\documentclass[hyperpdf,bindnopdf]{hepthesis}

\usepackage{lmodern}

\usepackage[T1]{fontenc}
\usepackage{charter}
\usepackage[expert]{mathdesign}

\usepackage{tikz}
\usepackage{morefloats,subfig,afterpage}
\usepackage{mathrsfs} 
\usepackage{verbatim}
\usepackage{amsmath}

\usepackage[english]{babel}
\usepackage[backend=bibtex,style=numeric,sorting=none]{biblatex}
\DeclareSourcemap{
 \maps[datatype=bibtex,overwrite=true]{
  \map{
    \step[fieldsource=collaboration, final=true]
    \step[fieldset=usera, origfieldval, final=true]
  }
 }
}

\renewbibmacro*{author}{%
  \iffieldundef{usera}{%
    \printnames{author}%
  }{%
    \textbf{\printfield{usera}}, \printnames{author}%
  }%
}%
\usepackage{capt-of}
\usepackage{array}
\usepackage{booktabs}
\newcolumntype{C}[1]{>{\centering\let\newline\\\arraybackslash\hspace{0pt}}m{#1}}
\usepackage{enumerate}
\usepackage[shortlabels]{enumitem}

\usepackage{stackrel}

\definecolor{cerulean}{rgb}{0.0, 0.48, 0.65}

\definecolor{ao(english)}{rgb}{0.0, 0.5, 0.0}

\definecolor{goldenrod}{rgb}{0.85, 0.65, 0.13}

\definecolor{blue-violet}{rgb}{0.54, 0.17, 0.89}

\definecolor{fandango}{rgb}{0.71, 0.2, 0.54}

\usepackage{anyfontsize}

\usepackage[ddmmyyyy]{datetime}
\usepackage{braket}
\usepackage{slashed}
\usepackage{graphicx}
\usepackage[export]{adjustbox}

\usepackage[subfigure]{tocloft}
\addtocontents{toc}{\vspace{-4ex}} 
\addtocontents{preface}{\vspace{-4ex}} 
\addtocontents{publication}{\vspace{-4ex}} 
\setkomafont{subsubsection}{\normalfont}
\makeatletter
\renewcommand\subsubsection{\@startsection{subsubsection}{3}{\z@}%
                                     {-3.25ex\@plus -1ex \@minus -.2ex}%
                                     {-1.5ex \@plus -.2ex}
                                     {\normalfont\normalsize\bfseries}}
\makeatother

\usepackage{amssymb}
\newcommand{\gsim}{\raisebox{-0.13cm}{~\shortstack{$>$ \\[-0.07cm]
      $\sim$}}~}
\newcommand{\lsim}{\raisebox{-0.13cm}{~\shortstack{$<$ \\[-0.07cm]
$\sim$}}~}

\usepackage{bibentry}

\usepackage{booktabs}
\usepackage{multirow}
\usepackage{tabularx}

\usepackage{comment}
\usepackage{wrapfig}
\usepackage{makecell}

\DeclareOldFontCommand{\rm}{\normalfont\rmfamily}{\mathrm}
\DeclareOldFontCommand{\sf}{\normalfont\sffamily}{\mathsf}
\DeclareOldFontCommand{\tt}{\normalfont\ttfamily}{\mathtt}
\DeclareOldFontCommand{\bf}{\normalfont\bfseries}{\mathbf}
\DeclareOldFontCommand{\it}{\normalfont\itshape}{\mathit}
\DeclareOldFontCommand{\sl}{\normalfont\slshape}{\@nomath\sl}
\DeclareOldFontCommand{\sc}{\normalfont\scshape}{\@nomath\sc}
\newcommand{\be}{\begin{equation}}
\newcommand{\ee}{\end{equation}}
\newcommand{\bea}{\begin{eqnarray}}
\newcommand{\eea}{\end{eqnarray}}
\newcommand{\bi}{\begin{itemize}}
\newcommand{\ei}{\end{itemize}}
\newcommand{\ben}{\begin{enumerate}}
\newcommand{\een}{\end{enumerate}}
\newcommand{\la}{\left\langle}
\newcommand{\ra}{\right\rangle}
\newcommand{\lc}{\left[}
\newcommand{\rc}{\right]}
\newcommand{\lp}{\left(}
\newcommand{\rp}{\right)}
\DeclareMathOperator{\Tr}{Tr}

\usepackage{floatrow}


\usepackage{caption}
\usepackage{sidecap}

\usepackage{colortbl}
\usepackage{pifont}
\newcommand{\xmark}{\ding{55}}

\newcommand{\grokcell}{\cellcolor{green!25}\checkmark}

\newcommand{\ylcell}{\cellcolor{yellow!25}}

\newcommand{\rdxcell}{\cellcolor{red!25}\xmark}
\newcommand{\ftm}[1]{\footnotemark[#1]}

\newcounter{mysym}

\newcommand{\eko}{{\ttfamily EKO}}
\newcommand{\ekos}{{\ttfamily EKOs}}
\newcommand{\yadism}{{\ttfamily Yadism}}
\newcommand{\pineline}{{\ttfamily Pineline}}
\newcommand{\pineko}{{\ttfamily Pineko}}
\newcommand{\pinefarm}{{\ttfamily Pinefarm}}
\newcommand{\apfel}{{\ttfamily APFEL}}
\newcommand{\apfelpp}{{\ttfamily APFEL++}}
\newcommand{\qcdnum}{{\ttfamily Qcdnum}}
\newcommand{\pegasus}{{\ttfamily PEGASUS}}
\newcommand{\lhapdf}{{\ttfamily LHAPDF}}
\newcommand{\hoppet}{{\ttfamily HOPPET}}
\newcommand{\xfitter}{{\ttfamily xFitter}}
\newcommand{\pineappl}{{\ttfamily PineAPPL}}
\newcommand{\madgraph}{{\ttfamily Mg5aMC@NLO}}
\newcommand{\Matrix}{{\ttfamily MATRIX}}
\newcommand{\nnlojet}{{\ttfamily nnlojet}}
\newcommand{\nnpdf}{{\ttfamily NNPDF}}

\makeatletter
\@ifpackageloaded{hyperref}{%
\hypersetup{%
  pdftitle = {A high-resolution imaging of the collinear substructure of the proton},
  pdfsubject = {Giacomo Magni's PhD thesis},
  pdfkeywords = {proton parton distribution functions, nPDFs, LHC, QCD, machine learning, neural networks},
  pdfauthor = {\textcopyright\ Giacomo Magni}
}}{}
\makeatother

\begin{document}

\begin{frontmatter}

\thispagestyle{empty}
\begin{center}
\Huge{\textbf{A high-resolution imaging of the collinear substructure of the proton}}\\
\vspace*{\fill}
\Large{Giacomo Magni}
\end{center}

\newpage
\thispagestyle{empty}

\vspace*{\fill}
\noindent This work is financed by the Netherlands Organization for Scientific Research (NWO) and carried out at Nikhef and Vrije Universiteit Amsterdam.
\begin{figure*}[!h]
  \begin{center}
  \includegraphics[width=0.35\textwidth]{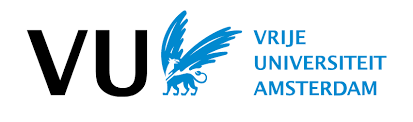}
  \includegraphics[width=0.25\textwidth]{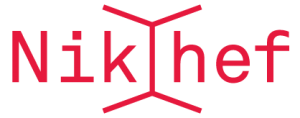}
  \includegraphics[width=0.38\textwidth]{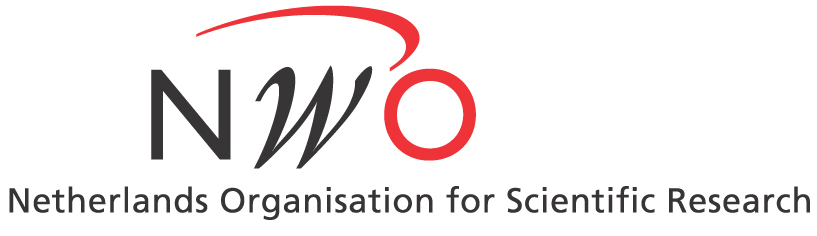}
  \end{center}
\end{figure*}

\newpage
\thispagestyle{empty}
\begin{center}
VRIJE UNIVERSITEIT
\\ \vspace{2cm}
\textbf{A HIGH-RESOLUTION IMAGING OF THE COLLINEAR SUBSTRUCTURE OF THE PROTON}
\\ \vspace{5cm}
ACADEMISCH PROEFSCHRIFT
\\ \vspace{1cm}
ter verkrijging van de graad Doctor of Philosophy aan\\
de Vrije Universiteit Amsterdam,\\
op gezag van de rector magnificus\\
prof.dr. J.J.G. Geurts,\\
volgens besluit van de decaan\\
van de Faculteit der Bètawetenschappen\\
in het openbaar te verdedigen\\
op vrijdag 4 april 2025 om 11.45 uur\\
in de universiteit \\
\end{center}
\vspace*{\fill}
\begin{center}
door \\
Giacomo Magni \\
geboren te Milaan, Italië
\end{center}

\newpage
\thispagestyle{empty}
\begin{table}
  \centering
  \begin{tabular}{ll} 
promotor: & prof.dr. J. Rojo \\
& \\
& \\
copromotor: & dr. M. Senghi Soares \\ 
& \\
& \\
promotiecommissie:
& prof.dr. K.S.E. Eikema \\
& dr. M. van Beekveld  \\
& prof.dr. D. Boer   \\
& prof.dr. T. du Pree  \\
& dr. W.D. Hulsbergen \\

\end{tabular}
\end{table}


\begin{abstract}
  \thispagestyle{empty}
  Accurate Standard Model predictions of proton-proton collisions are essential for interpreting 
  the current and forthcoming experimental measurements from high-energy colliders. The quest for 
  physics beyond the Standard Model is in fact strongly related to a reliable determination of 
  systematic uncertainties, among which the theoretical ones have nowadays become a significant 
  component.
  In particular, an important role is played by the universal Parton Distribution Functions (PDFs).
  This thesis presents a collection of studies aimed at improving the accuracy and precision 
  of collinear PDF determinations.
  By employing the NNPDF methodology, we present recent polarized and unpolarized PDF fits 
  including higher order QCD corrections and theoretical uncertainties. Additionally, we examine
  the presence of non-vanishing intrinsic charm and its phenomenological implications.
\end{abstract}

\clearpage

\tableofcontents
\markboth{}{}

\begin{preface}
\phantomsection
\addcontentsline{toc}{chapter}{Preface}

\paragraph{Introduction.} 

Protons are the building blocks of the universe, and describing their
dynamics has been one of the main challenges in modern physics.
The interpretation of the proton as a composite state made of quarks and gluons 
is now firmly established and provides one of the key elements behind 
the understanding of high energy physics measurements. 
The main subject of this thesis are the collinear Parton Distribution Functions 
(PDFs), which enable an effective description of the proton's content in terms 
of the momentum carried by its constituents.
Their study lies at the intersections of theory and data analysis, requiring both 
a precise understanding of Quantum Chromodynamics (QCD) and an accurate 
treatment of experimental data.
If beyond Standard Model phenomena will show up in collider experiments only via 
indirect effects through complex patterns, having a clear understanding 
of the proton substructure is indispensable.
Hence, PDFs constitute a necessary ingredient to improve the accuracy of high energy 
phenomenology and eventually detecting tiny but consistent deviations from the Standard Model. 

This thesis presents a collection of works aimed at providing more precise PDF 
determinations by improving the accuracy of the theoretical 
calculations employed.

\paragraph{Outline of the thesis.}
The thesis can be divided into two main parts: in the first part, 
\cref{chap:introduction,chap:methodology}, I introduce the fundamental tools adopted 
in the studies I carried out during my PhD, some of which are presented 
in the second part, \cref{chap:ic,chap:an3lo,chap:pol}.
The latter are all based on different projects I worked on within the NNPDF 
collaboration.

In \cref{chap:introduction}, I review the theoretical formalism which is currently 
used to describe the high energy scattering of lepton-proton and proton-proton systems. 
This is mainly based on QCD and on the Factorization theorem, which allow the separation of
quantities computable with perturbative methods from non-perturbative objects encoding 
the long-distance dynamics between the proton components.
Collinear PDFs are an example of such objects: they are currently not computable 
a priori but have to be fitted from high energy scattering data.

\cref{chap:methodology} focuses on the computational tools and methodological aspects 
of the NNPDF PDF fitting framework. Specifically, I summarize the features of two main 
tools used to compute PDF evolution and Deep Inelastic Scattering (DIS) cross-sections.
I also discuss the treatment of uncertainties and the fitting methodology, which employs 
an artificial neural network to model the PDF functional form.

\cref{chap:ic} contains a summary of my studies about intrinsic charm~\cite{Ball:2022qks,NNPDF:2023tyk}.
The phenomenon of intrinsic charm is purely a quantum mechanics effect that entails 
a non-vanishing contribution of non-perturbative generated charm quarks
to the proton wave function. 
Ref.~\cite{Ball:2022qks} provides a first evidence of this effect starting from the NNPDF4.0 PDF analysis, 
and validates the result with a comparison to the most recent data, from the LHCb 
experiment, which could potentially be sensitive to this phenomenon.
In the last part of the chapter, based on Ref.~\cite{NNPDF:2023tyk}, I investigate 
the possibility of finding an intrinsic charm asymmetry in the proton,
which could definitively shed light on this topic.

In \cref{chap:an3lo}, I present the most up-to-date global PDF analysis from the 
NNPDF collaboration~\cite{NNPDF:2024nan} which includes the known four-loop and three-loop QCD corrections 
to PDF evolution and DIS predictions, respectively.
This allows the extraction of PDFs at approximate next-to-next-to-next-to-leading order 
(N$^3$LO) with a consistent treatment of underlying theoretical uncertainties
and provides a fundamental input to improve the accuracy of both future and 
experimental high energy physics analysis.

\cref{chap:pol} outlines my most recent work~\cite{Cruz-Martinez:2025ahf} about a new 
determination of helicity dependent proton PDFs.
The study improves previous results, both from the theoretical and 
methodological points of view, and it will be beneficial to upcoming 
interpretations of polarized scattering measurements, for instance, from the EIC.
Finally, \cref{chap:Conclusion} contains a summary of the work.

All the presented results are provided along with open source codes and corresponding 
testing suites, which facilitate their development and maintenance.

\end{preface}

\newpage
\phantomsection
\addcontentsline{toc}{chapter}{Publications}
\pagestyle{plain}

\section*{Publications}

The following list include all the publications 
which are discussed in the thesis:

\subsection*{Journal papers}
\begin{itemize}
\item \fullcite{Barontini:2024xgu}
\item \fullcite{NNPDF:2024nan}
\item \fullcite{NNPDF:2024dpb}
\item \fullcite{Candido:2024rkr}
\item \fullcite{NNPDF:2023tyk}
\item \fullcite{Ball:2022qks}
\item \fullcite{Candido:2022tld}
\end{itemize}

\subsection*{Conference Proceedings and Reports}
\begin{itemize}
\item \fullcite{Hekhorn:2023gul}
\item \fullcite{Barontini:2022jci}
\end{itemize}

\subsection*{Open source codes} 

\begin{description}
    \item[\textbf{EKO}] DGLAP evolution kernel operators:
    \href{https://github.com/NNPDF/eko}{https://github.com/NNPDF/eko}
    \item[\textbf{Yadism}] Yet Another DIS Module:
    \href{https://github.com/NNPDF/yadism}{https://github.com/NNPDF/yadism}
    \item[\textbf{Pineko}] PineAPPL + EKO = fast theories:
    \href{https://github.com/NNPDF/pineko}{https://github.com/NNPDF/pineko}
    \item[\textbf{NNPDF}] A machine learning framework for global analyses of parton distributions: \\
    \href{https://github.com/NNPDF/nnpdf}{https://github.com/NNPDF/nnpdf} 
\end{description}

\vspace{1cm}

Other works, carried during the PhD and not presented in this manuscript, include:

\subsection*{Journal papers}
\begin{itemize}
\item \fullcite{NNPDF:2024djq}
\item \fullcite{Hekhorn:2024tqm}
\item \fullcite{terHoeve:2023pvs}
\item \fullcite{Giani:2023gfq}
\item \fullcite{Candido:2023utz}
\item \fullcite{Ethier:2021bye}
\item  \fullcite{Ethier:2021ydt}
\end{itemize}

\subsection*{Conference Proceedings and Reports}
\begin{itemize}
\item \fullcite{Andersen:2024czj}
\item \fullcite{Cooper-Sarkar:2024crx}
\item \fullcite{Amoroso:2022eow}
\item \fullcite{Magni:2021rno}
\end{itemize}

\subsection*{Open source codes} 

\begin{description}
    \item[\textbf{SMEFiT}] A standard model effective field theory fitter: \\
    \href{https://github.com/LHCfitNikhef/smefit_release}{https://github.com/LHCfitNikhef/smefit$\_$release}
    \item[\textbf{NNSF$\nu$}] Predictions for neutrino structure functions: \\
    \href{https://github.com/NNPDF/nnusf} {https://github.com/NNPDF/nnusf}
\end{description}

\clearpage
\pagestyle{headings}
\thispagestyle{empty}

\end{frontmatter}

\RedeclareSectionCommand[
  beforeskip=0pt,
  afterskip=2\baselineskip]{chapter}
\RedeclareSectionCommand[
  beforeskip=\baselineskip,
  afterskip=.5\baselineskip]{section}
\RedeclareSectionCommand[
  beforeskip=.75\baselineskip,
  afterskip=.5\baselineskip]{subsection}

\begin{mainmatter}

  \chapter{Scattering Protons}
\label{chap:introduction}
\vspace{-0.5cm}
\begin{center}
\begin{minipage}{1.\textwidth}
    \begin{center}
    \textit{This chapter is based on Refs.~\cite{Ellis:1996mzs,Schwartz:2014sze,Borsa:2022irn,Davies:2016bwb,Hekhorn:2019nlf,ParticleDataGroup:2022pth}} 
    \end{center}
\end{minipage}
\end{center}

Collisions of nuclei at high center-of-mass energies have
been one of the richest sources of experimental data 
in particle physics.
Starting from the 1960s this activity has led to the discovery of
the theory describing strong interactions, 
known as Quantum Chromodynamics 
(QCD) and, nowadays, the extensive information 
that can be extracted from hadron-hadron collisions continues 
to motivate physicists to operate the largest collider ever built, 
the Large Hadron Collider (LHC).
In order to formulate accurate predictions of such scattering 
processes the knowledge of the initial state conditions is 
an essential theoretical input.
However, this task is non-trivial, as hadrons do not behave 
like point-like particles and their constituents, quarks and 
gluons, become weakly interacting only in the high energy limit.

\paragraph{Outline.}
We begin this chapter with a review of the main concepts 
of the current formulation of perturbative QCD (\cref{sec:pert_qcd}),
we then describe how it is possible to make predictions 
about Deep Inelastic Scattering (DIS), i.e. the lepton-hadron 
scattering at high center-of-mass energies (\cref{sec:dis}). 
In \cref{sec:pdfs}, we introduce the concepts of Factorization
and Parton Distribution Function (PDF) which are the main tools
used to describe how the hadron constituents are distributed 
inside the bound states allowing the computation of cross-sections.  
In the following \cref{sec:heavy_quarks}, we discuss how perturbative 
QCD can be used to improve the accuracy of DIS predictions 
and how heavy quark mass effects can be accounted.
Finally, in \cref{sec:double_hadronic}, we sketch a  
generalization of the above concepts for multi hadron scattering,
especially for the proton-proton case.

\section{Perturbative QCD}
\label{sec:pert_qcd}

QCD is the theory describing the strong interaction which is the 
dominant force binding quarks and gluons.
Its formulation has been a result 
of many ideas and experimental data collected from the end of 
the '60s. Below, we summarize the key findings that steered the 
development of current QCD theory. 

The existence of quarks was first observed at SLAC \cite{Breidenbach:1969kd}
in high-energy electron-proton collisions, 
where many hadrons were produced in the final state.
Their classification was compatible with the description
given by Gell-Mann and Zweig, the eightfold way \cite{Gell-Mann:1961omu}. 
The model is based on spin-$\frac{1}{2}$ constituent particles,
the quarks, \textit{up}, \textit{down}, \textit{strange} which 
obey an $SU(3)$ flavor symmetry and carry a 
fractional electric charge. 
The necessity to have antisymmetric wave-functions in spin-$\frac{3}{2}$
baryons, led to the idea that quarks were carrying an additional
charge, called \textit{color} charge.

Slightly later in mid '70s, data from electron-positron collisions 
\cite{Hanson:1975fe}, showed the presence of two sprays of collimated 
particles in the final state, which were interpreted as
"jets" initiated by quark fragmentation. 
Moreover, the existence of events in which a third jet was observed, 
lead to the discovery of the gluon, a bosonic particle with integer spin, 
acting as the strong force mediator.
By measuring the cross-section ratio $\sigma(e^+ e^- \to \text{jet} + \text{jet})/\sigma(e^+ e^- \to \mu^+ + \mu^-)$,
physicists determined the number of color charges, which was 
compatible with $N_C=3$. 
Finally, the fact that quarks and gluons were not observed as 
free particles, or better are only asymptotically free, as 
visible in the DIS data, motivates that the underlying 
symmetry beyond strong interaction was coming from a non-abelian group.

Due to its complexity, a full analytical solution of a QCD scattering process 
is not feasible. Although other approaches are available, here we will deal only with
perturbative QCD (pQCD), i.e. all the quantities will be expressed in terms
of a perturbative series in the strong coupling. 
The larger the number of perturbative corrections included, 
the more precise will be the prediction.
In order to maintain a physical description of the theory, 
the inclusion of higher order corrections requires 
to redefine the \textit{bare} quantities appearing in the Lagrangian 
in terms of the renormalized physical quantities.
This procedure, as we shall see in \cref{sec:alpha_s,sec:dglap},
allows the definition of some Renormalization Group Equations (RGE)
which are again solvable in perturbation theory and fix
the running of physical input parameters at different scales.
As a matter of fact, the value of the strong coupling is a very good 
expansion parameter only at large scales, and blows up at 
$\approx 300~\text{MeV}$, where the approach of pQCD is 
no longer reliable.
Thus, pQCD is an excellent tool for describing hadronic scattering 
at high energy colliders, but might fail for instance in providing 
a description of low energy cross-sections. 
Another major downside of the pQCD method is that the complexity of 
the computation increases dramatically at higher orders, 
for instance due to the large number of Feynman diagrams 
that must be taken into account.

In \cref{sec:qcd_lagrangian}, starting from the QCD Lagrangian, we 
set up the notation that will be used in all the rest of the work, 
including the electroweak couplings,
while in \cref{sec:alpha_s} we examine how the 
running of the strong coupling can be used to explain the 
observed asymptotic freedom. 
   
\subsection{QCD Lagrangian}
\label{sec:qcd_lagrangian}

The derivation of the QCD Lagrangian is a generalization of Quantum Electrodynamics (QED)
to a more complicated symmetry group, $SU(3)$.
The full QCD Lagrangian is given by the terms
\begin{equation}
    \mathcal{L}_{QCD} = \mathcal{L}_{classical} + \mathcal{L}_{gauge} + \mathcal{L}_{ghost}
    \label{eq:lqcd}
\end{equation}
where the classical part describes the dynamics of the quark and gluon fields, 
while $\mathcal{L}_{gauge}$ contains the gauge fixing 
terms and $\mathcal{L}_{ghost}$ add the necessary ghost fields~\cite{Faddeev:1967fc} 
needed to remove unphysical gluon polarizations.
The first part takes the form
\begin{equation}
    \mathcal{L}_{classical} = \frac{1}{4}  G^{a}_{\mu \nu} G_{a}^{\mu \nu} + \sum_{q} \bar{\psi}_q (i \slashed{D} - m_q) \psi_q \, ,
    \label{eq:lqcd_classical}
\end{equation}
$G^{a}_{\mu \nu}$ is the gluon field strength tensor
$G^{a}_{\mu \nu} = \partial_\mu A^a_\nu - \partial_\nu A^a_\mu + g_s f^{a}_{bc} A^b_\mu A^{c}_\nu$, 
where the gluon field vector $A^a_\nu$ is supplemented 
with the color index $a$. 
The constants $g_s$ and $f^{a}_{bc}$ are respectively 
the QCD coupling and the QCD structure constants.
As the underlying QCD symmetry group is non-abelian, the gluon self interactions
are present and, both triple and four gluon vertices are present.

The second right-handed part of \cref{eq:lqcd_classical} set the propagation of
the quark fields $\psi_q$ and their interaction with gluons through the 
covariant derivative $D$.
Also, quarks carry a color index which has been omitted here.
The sum runs over the different flavor ($up$, $down$, $strange$,
$charm$, $bottom$ and $top$) and $m$ indicates their masses. 
In the rest of the work we assume $u$, $d$ and $s$ 
to be massless, as we aim to describe only high energy scattering processes, 
while $c$, $b$, $t$ will be treated separately as their mass effect is not 
always negligible, and it often requires ad-hoc prescriptions (see \cref{sec:heavy_quarks}).
The covariant derivative is given by $D_\mu = \partial_\mu + i g_s t^a A^a$, 
where the matrices $t^a$ are the $SU(3)$ symmetry generators.
The entire Lagrangian is invariant under $SU(3)$ transformation 
which ensure color charge conservation.
%
In particular, by setting the number of colors $N_C=3$ we fix the normalization 
constant as
\begin{align}
    {\rm \Tr} \left(t^a t^b \right) = T_R \delta_{ab}, & \quad T_R = \frac{1}{2} \,, \\ 
    \sum_a t^a t^a = \frac{N_C^2-1}{2 N_C} = C_F, & \quad C_F =  \frac{4}{3} \,, \\
    {\rm \Tr} \left(T^a T^b \right) = N \delta_{ab} = C_A \delta_{ab}, & \quad C_A = 3 \, .
    \label{eq:qcd_const}
\end{align} 
These factors appear during the computation of Feynman diagrams, as one needs to sum 
over the possible color states. 
Higher order calculations generate sums on more complicated 
topologies and other algebraic constants might be required.
However, formally their value can always be computed a priori from the 
properties of the $SU(3)$ representations.

\subsection{The strong coupling}
\label{sec:alpha_s}

In \cref{eq:lqcd_classical} we have introduced the strong coupling, 
which enters explicitly in the gluon strength tensor definition 
and in the covariant derivative.  
The inclusion of one loop gluon propagator corrections
lead to integrals over the momentum flowing inside the loop, which 
diverge as this quantity increases. 
This is the simplest example of well-known behavior
which is then reconciled with the physical states
through a procedure called \textit{renormalization}~\cite{tHooft:1973mfk,Weinberg:1973xwm}.
In doing so, all the fields and the parameters (the coupling $g_s$ 
and quark masses) entering in \cref{eq:lqcd} are redefined in terms of \textit{bare}
quantities which are not physical and absorbs all the infinite ultraviolet 
divergences, leaving the one-loop corrections to the physical
parameter finite.
The price to pay is that one need to introduce a new set of
\textit{renormalized} quantities which now depend on one 
additional scale and coincide with the physical parameter 
for a precise boundary condition. 

This procedure entails two custom choices:
first, the inclusion of finite terms in either the \textit{bare}
or \textit{renormalized} quantity is not constrained.
Here, as common in literature we will adopt the $\overline{MS}$ scheme, 
where in addition to the divergent pieces, 
only the term $\gamma_E - \ln(4 \pi)$
\footnote{Here $\gamma_E = 0.577 \dots$ is the Eulero-Mascheroni constant.} 
are kept in the \textit{bare} parameters.
Let us stress that, in order to obtain consistent results, 
all the different quantities needs to be defined with the same 
subtraction scheme. 
The second choice regards the definition of the renormalization scale $\mu_R$, 
at which the parameter redefinition is performed with respect to the physical 
scale $Q$ at which the degree of freedom is probed.
Noticing that the renormalized quantity depends only on the 
ratio $Q/\mu_R$ and, imposing the scaling invariance of the \textit{renormalized} 
quantity upon the choice of $\mu_R$ it leads to a renormalization group 
equation (RGE), which fixes the running of the parameter. 
Focusing on the strong coupling $a_s = \alpha_s / (4 \pi) = g_s^2 / (4 \pi)^2$, one obtains
\begin{equation}
    \mu_R \frac{d a_s}{d \mu_R} = \beta(a_s) = - \sum_{i=2}^{\infty} a_s^{i+2} \beta_i\, ,
    \label{eq:as_running}
\end{equation}
where the beta function is now known up to five-loop $\mathcal{O}(a_s^5)$~\cite{Baikov:2016tgj,Herzog:2017ohr}.
In the r.h.s we explicitly factor out the global minus sign with the leading coefficient 
positive definite $\beta_0 > 0$ and $n_f=3,\dots,6$.
In the following, for all the pQCD expansions, we assume 
the boundary condition to be $a_s(\mu_0 = M_Z = 91.18~\text{GeV}) = 0.118 / (4 \pi)$.
The leading-order solution
\footnote{
    See \href{https://eko.readthedocs.io/en/latest/theory/pQCD.html}{https://eko.readthedocs.io/en/latest/theory/pQCD.html}
    for RGE solution up to four-loop.
}
of \cref{eq:as_running} is given by
\begin{equation}
    a_s(\mu_R) = \frac{a_s(\mu_0)}{ 1 + \beta_0 a_s(\mu_0) \ln(\frac{\mu^2_R}{\mu^2_0})} \,.
    \label{eq:lo_as}
\end{equation}
The actual values of the QCD $\beta_i$ coefficients (and in particular of $\beta_0$) 
depend not only on the number of colors $N_C=3$, but also on 
the number of active flavors $n_f$ running in the loops, which can be 
at most $6$. 

All these observations, which seem quite simple at first sight, 
have a huge phenomenological consequence which distinguishes
the behavior of the QED and QCD interactions.
First, a negative beta function implies that the strong
coupling becomes smaller at high scales and 
colored particles are then asymptotically free as observed
for instance in DIS (see \cref{sec:parton_model}).  
On the other hand, at low energies the strong coupling 
can blow up, and it becomes large. From the leading order
solution (\cref{eq:lo_as}) one can read out the scale 
at which this happens, finding
$\Lambda_{QCD} = M_Z \exp(- 1 / ( 2 a_s(M_z) \beta_0)) \approx 10~\text{MeV}$,
although a more reasonable value, including higher order corrections,
is of the order $300~\text{MeV}$ \cite{ParticleDataGroup:2022pth}.
This can explain why the strong interaction is dominant for particles
with low momentum and why quarks are only observed inside bound 
colorless states, i.e. mesons and baryons (often referred ad QCD confinement).
%

In principle also the mass of the \textit{heavy} quarks, 
i.e. the ones with $m_q > \Lambda_{QCD}$, should depend 
on the scale. However, in a realistic scenario,
this effect is much smaller than the strong coupling running
thus, for our purposes, it will be neglected.
In the following we assume $m_c = 1.51~\text{GeV}$, 
$m_b = 4.92~\text{GeV}$, $m_t = 172.5~\text{GeV}$
as pole masses~\cite{LHCHiggsCrossSectionWorkingGroup:2016ypw}.


\paragraph{Other Standard Model parameters.} 
For completeness, we also list the convention used for the electroweak 
coupling of quarks and leptons to the $\gamma, Z$ and $W$.
We define the interaction vertex to be on the form $-i e \Gamma_{b,f}^\mu$,
with $e$ the electric charge unit.
The couplings can then be split for each fermion $f$ into axial-vector and vector-vector as
\begin{align}
    \Gamma_{\gamma,f}^\mu & = e_f \gamma^\mu \label{eq:gamma_coupling},\\
    \Gamma_{Z,f}^\mu & = \frac{1}{2 \sin(\theta_W) \cos(\theta_W)} ( V_f \gamma^\mu + A_f \gamma_5 \gamma^\mu \label{eq:Z_coupling} ), \\
    \Gamma_{W,f}^\mu & = \frac{1}{\sqrt{2} \sin(\theta_W)} \gamma^\mu  \frac{1 - \gamma_5}{2} \label{eq:W_coupling},
\end{align} 
where we assume the shorthand notation for the $Z$ coupling
\begin{equation}
    V_{f} = I_3^f - 2 e_f \sin^2(\theta_W) \, , \quad  A_{f} = I_3^f \,,
    \label{eq:fZ_coupling}
\end{equation}
with $\sin^2(\theta_W) = 0.2315$ and the corresponding electroweak charges given in \cref{tab:electroweak_charges}.
For the QED coupling we assume a fixed expansion parameter 
$a_{em} = e^2 / (4 \pi) = 1 / 137.04$.
%

\begin{table}[!t]
    \centering
    \begin{tabularx}{0.21\textwidth}{c|c|c}
            & $e_f$ & $I_3$ \\
        \toprule
            $\ell^{-}$  & $-1$ & $-\frac{1}{2}$ \\
            $\nu$ & $0$ & $\frac{1}{2}$ \\
        \midrule
            $u,c,t$  & $\frac{2}{3}$ & $\frac{1}{2}$ \\
            $d,s,b$ & $-\frac{1}{3}$ & $-\frac{1}{2}$ \\
        \bottomrule
\end{tabularx}



    \vspace{0.3cm}
    \caption{Electroweak charges. $e_f$ denotes the electric charge, $I_3$ 
    the third component of the electroweak isospin.
    Note that for antiparticles all the charges are reversed. }
    \label{tab:electroweak_charges}
\end{table}
Finally, note that we treat the charged leptons electrons and muons 
to be massless and, we neglect tau effects.

\section{Deep Inelastic scattering}
\label{sec:dis}

The simplest type of interaction involving hadrons in the 
initial state is the electron-proton scattering. 
The spectrum of the cross-section for energies much smaller
than the proton mass ($m_p = 0.938~\text{GeV}$) has been well known
even before quarks were discovered, and can be computed with 
standard quantum field theory assuming that protons behave as point-like 
particles.
On the other hand, already from the '50 \cite{Hofstadter:1953zjy} 
data in the $\text{MeV}$ region showed an incompatible 
behavior with the above assumption.
Surprisingly, at even higher energies, where the inelastic
scattering dominates (DIS regime) and the proton breaks up, the 
energy of the final state lepton decreases, and data are 
again compatible with a point-like nucleon interaction.
This is the regime we are interested in and where pQCD
can be used.

In the following \cref{sec:had_sf}, we show how the computation of 
a DIS cross-section reduces to determining certain scalar functions 
describing nucleon properties. 
Then, in \cref{sec:parton_model}, we illustrate a first simple model, 
which provides a description of the proton in terms of quarks, 
introducing the concept of Parton Distribution Functions (PDFs).

\subsection{Hadronic Structure Functions}
\label{sec:had_sf}

Let us consider the following lepton-hadron scattering process
\begin{equation}
    \ell(k) + p(P) \to \ell'(k') + X \,,
    \label{eq:dis_def} 
\end{equation}
where the brackets denote the 4-momentum associated with the particle.
The scattering charged, or neutral, leptons are assumed to be massless with $k^\mu = (E, k)$ 
and ${k'}^\mu = (E', k')$ and $X$ is a generic final state consisting of the proton remnants. 
We define the following kinematics variables:
\begin{align}
    \nu = P \cdot q &= M (E-E'), \label{eq:nu_def} \\
    Q^2 = - q^2 & = - (k'-k)^2, \label{eq:q2_def}\\
    M^2 & = P^2 , \label{eq:m2_def} \\
    x = \frac{Q^2}{2 (q \cdot P)} & = \frac{Q^2}{2 M (E - E')} , \label{eq:x_def}\\
    y = \frac{q \cdot P }{k \cdot P} & = 1 - \frac{E'}{E} . \label{eq:y_def}
\end{align} 
$M$ denotes the hadron invariant mass, while 
the variables $x$ and $y$, called respectively Bjorken-x and inelasticity,
are two dimensionless variables bounded in the range $(0,1]$. 
The center-of-mass energy can be expressed from the above quantities as
\begin{equation}
    s = (P + k)^2 = M^2 + \frac{Q^2}{x y} \ ,
    \label{eq:s2_def}
\end{equation}
and final state invariant mass $W^2$ can be defined through
\begin{equation}
    W^2 = (P + q)^2 = M^2 - Q^2 (1 - \frac{1}{x}) \ .
    \label{eq:w2_def}
\end{equation}
In the elastic scattering regime the hadron does not break, and the final invariant
mass coincides with $M$, thus it corresponds to $x = 1$ limit. 
On the other hand, the high energy limit is given by $E \to \infty$ 
and thus coincides with $x \to 0$.
\begin{figure}[!t]
    \includegraphics[trim={0 6cm 0 4cm},clip]{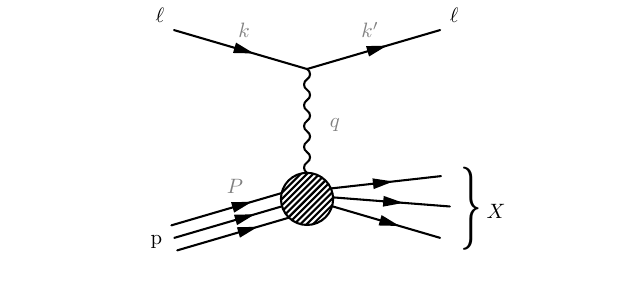}
    \caption{Feynman diagram associated to the high energy lepton-proton scattering 
    $\ell(k) + p(P) \to \ell'(k') + X$}
    \label{fig:dis_hadronic}
\end{figure}
Without loss of generality, we can write the cross-section,
associated to the squared amplitude of the diagram in \cref{fig:dis_hadronic}
and representing the probability of the processes, as
\begin{equation}
    \frac{d^2\sigma_{i}}{d x d y} = \frac{2 \pi y a_{em}^2}{Q^4} \sum_j \eta_j L_{j}^{\mu \nu} W^{j}_{\mu \nu} \, .
    \label{eq:dis_xs}
\end{equation}
Depending on whether the final-state lepton coincides with the incoming one, 
we distinguish between neutral current (NC) processes ($i = NC$), 
where the incoming and outgoing leptons are identical, and charged current (CC) 
processes ($i = CC$), where a scattering charged lepton results in a neutrino 
in the final state, or vice versa.
In the former case we need to sum over photon $\gamma$ and $Z$ boson 
contributions, taking into account also the interference term. Thus, 
for $i = NC$ we have 3 different contributions $j \in \{ \gamma\gamma, \gamma Z, ZZ \}$ with
corresponding normalization:
\begin{align}
    \eta_{\gamma \gamma} = 1 \, , \quad
    \eta_{\gamma Z} = \frac{1}{4 \sin^2(\theta_W) \cos^2(\theta_W)} \frac{Q^2}{Q^2 + M_Z^2} \, , \quad
    \eta_{Z Z} =  \eta_{\gamma Z}^2 \, ,
    \label{eq:etaNC_def}
\end{align} 
which are needed to account for the different propagators and coupling constants.

%
In the CC case only $W^{\pm}$ bosons can be exchanged, thus the sum on $j$ is trivial
and the normalization $\eta_{WW}$ is given by
\begin{equation}
    \eta_{WW} = \left ( \frac{\eta_{\gamma Z}}{2} \frac{1 + Q^2/M_Z^2}{1 + Q^2/M_W^2} \right ) ^2.
    \label{eq:etaCC_def}
\end{equation}

In \cref{eq:dis_xs} we have introduced two tensors which describe respectively 
the lepton-boson ($L_{j}^{\mu \nu}$) and the hadron-boson ($W_{j}^{\mu \nu}$) 
interactions. As we are dealing with massless leptons and no QED corrections,
the structure of the leptonic tensor is quite simple and fully determined by the
electroweak coupling. It is custom to normalize the different contributions 
in terms of the photon exchange
\begin{equation}
    L_{\gamma\gamma}^{\mu \nu} = 2 ( 
        k^\mu {k'}^\nu + k^\nu {k'}^\mu 
        - (k \cdot k') g^{\mu \nu} 
        - i \lambda_{\ell} \epsilon_{\mu \nu \alpha \beta} k^\alpha {k'}^\beta
        ) \,,
        \label{eq:Lgg_def}
\end{equation}
for which the antisymmetric part is fully proportional to the lepton
helicity $\lambda_{\ell}$. 
The other tensors are then:
\begin{align}
    L_{\gamma Z}^{\mu \nu} & = (V_\ell + \lambda_{\ell} A_{\ell}) L_{\gamma\gamma}^{\mu \nu} \, , \label{eq:LgZ_def} \\
    L_{ZZ}^{\mu \nu} & = (V_\ell + \lambda_{\ell} A_{\ell})^2 L_{\gamma\gamma}^{\mu \nu} \, , \label{eq:LZZ_def} \\
    L_{WW}^{\mu \nu} & = ( 1 + \lambda_{\ell} I_{W} )^2L_{\gamma\gamma}^{\mu \nu} \, , \label{eq:LWW_def} 
\end{align} 
with the vector-vector $V_\ell$ and axial-vector $A_\ell$ coupling defined as in \cref{eq:fZ_coupling}.
In the case of electroweak interaction, the part proportional to the lepton
helicity does not correspond anymore to the full antisymmetric part (or parity violating);
however the antisymmetric terms proportional to the helicity, 
have exactly the same tensor structure as the ones independent of $\lambda_\ell$, 
and similar for the symmetric ones.
In the limit $Q^2 \ll M_Z^2$ the contributions from the $Z$ boson 
are suppressed. We refer to this case as electromagnetic (EM) DIS. 

The last and more involving part of \cref{eq:dis_xs} is the hadronic tensors $W_{j}^{\mu \nu}$.
For the unpolarized case the kinematic constraints suggest us to decompose it as 
a linear combination of terms $g^{\mu \nu}, P^\mu P^\nu, q^\mu q^\nu, P^\mu q^\nu, q^\mu P^\nu$ and 
$\epsilon_{\mu \nu \alpha \beta} q^\alpha P^\beta$. 
A slightly more complicated structure is foreseen for the hadron spin $S^{\mu}$ dependent part;
\footnote{
    The spin 4-vector of a fermion field is generally defined as $S^\mu = (0, \mathbf{S})$, with $\mathbf{S} = \frac{1}{2} \gamma_5 \mathbf{\alpha}$
    and $\alpha_i$ the spinor representation of the Pauli matrices.
} 
however, since we have assumed massless leptons, we can use the contraction $L^{\mu \nu} q_{\mu} = 0$, 
to simplify the $W^{\mu \nu}$ decomposition.  
Following Ref.~\cite{ParticleDataGroup:2022pth}, we can write
\begin{align}
\begin{split}
    W_{j}^{\mu \nu} = 
        & - g_{\mu \nu} \left [ F_1^j(x,Q^2) + \frac{S \cdot q}{ P \cdot q} g_5^j (x, Q^2) \right ] \\
        & + \frac{\hat{P}^\mu \hat{P}^\nu}{p \cdot q} \left [ F_2^j(x,Q^2) + \frac{S \cdot q}{ P \cdot q} g_4^j (x, Q^2) \right ] \\
        & - i \epsilon_{\mu \nu \alpha \beta} \frac{q^\alpha}{2 P \cdot q} \left [ F_3^j(x,Q^2) P^\beta - 2 g_1^j(x, Q^2) S^\beta  \right ] \\
        & + i \epsilon_{\mu \nu \alpha \beta} \frac{q^\alpha}{P \cdot q} \left [ S^\beta - \frac{S \cdot q}{ P \cdot q} P^\beta \right ] g_2^j(x, Q^2) \\
        & + \frac{1}{P \cdot q} \left [ \frac{1}{2} \left ( \hat{P}^\mu \hat{S}^\nu + \hat{P}^\nu \hat{S}^\mu \right ) -  \frac{S \cdot q}{ P \cdot q}  \hat{P}^\mu \hat{P}^\nu \right ] g_3^j(x, Q^2) \, ,
    \label{eq:Wmunu_def}
\end{split}
\end{align} 
having defined the hat vectors $\hat{X} \in \{ \hat{P}^\mu, \hat{S}^\mu \}$ as:
\begin{equation}
    \hat{X}^\mu = X^{\mu} - \frac{X \cdot q}{q^2} q^\mu \, .
    \label{eq:hat_vactors} 
\end{equation}
The functions $F_k, g_k$ appearing in \cref{eq:Wmunu_def} are called
hadronic structure functions and encode all the information about how 
the partons are distributed inside the hadrons.
First, we observe that terms linearly proportional to the hadron spin 
$S^\mu$ vanish when averaging over the hadron polarizations. 
Thus, for unpolarized observables only structure functions $F_k$ survive.
The opposite holds for observables in which the target is polarized, where
$g_k$, called polarized structure functions, are the only contribution.

Second, we note that some structure functions are fully 
antisymmetric and thus are parity violating. 
This is manifest for $F_3$ which has to vanish for EM DIS only.
For polarized structure functions, the spin operator is also odd under parity, 
thus we refer to $g_1,g_2$ as parity conserving and to $g_3,g_4,g_5$ as parity violating 
(see \cref{tab:sf_parity}).
\begin{table}[!t]
    \centering
    \begin{tabularx}{0.5\textwidth}{c|c|c}
            Parity & Unpolarized & Polarized \\
        \toprule
            conserving  & $F_1,F_2$ & $g_1,g_2$ \\
        \midrule
            violating   & $F_3$ & $g_3,g_4,g_5$ \\
        \bottomrule
\end{tabularx}
    \vspace{0.3cm}
    \caption{Structure functions classification.}
    \label{tab:sf_parity}
\end{table}

Third, for longitudinal hadron polarizations, after performing all the contractions 
with $L_{j}^{\mu \nu}$, the structure functions $g_2$ and $g_3$ turn out to be
 fully proportional to $M^2/Q^2$~\cite{ParticleDataGroup:2022pth}, thus are suppressed 
in the high energy limit, $M^2/Q^2  \to 0$ and, they are not be considered further in this work.
In this limit the symmetry between the unpolarized and polarized is fully visible, and 
we obtain the relations:
\begin{equation}
    F_1 \leftrightarrow - g_5, \quad F_2 \leftrightarrow - g_4, \quad x F_3 \leftrightarrow 2 x g_1 \, .
    \label{eq:pol_unpol_symmetry}
\end{equation}
For reasons which will be clear later, it is convenient to introduce the 
linear combinations:
\begin{equation}
    F_L = F_2 - 2 x F_1, \quad g_L = g_4 - 2 x g_5,
    \label{eq:fl_gl_def}
\end{equation}
Finally, inserting \cref{eq:Lgg_def,eq:LgZ_def,eq:LZZ_def,eq:LWW_def,eq:Wmunu_def}, in \cref{eq:dis_xs},
we get the expressions for unpolarized and double (longitudinally) polarized cross-section:
\begin{align}
    \frac{d^2\sigma_{i}}{d x d y} = \frac{1}{4} \sum_{\lambda_{\ell},h = \pm 1} \frac{d^2\sigma_{i}}{d x d y} 
        & = \frac{4 \pi a_{em}^2 }{x y Q^2} \xi_i  \left [ Y_+ F^i_2 - y^2 F^i_L\ \mp Y_{-} x F^i_3 \right ] \,, 
    \label{eq:dis_unpol_sx} \\
        \frac{d^2 \Delta \sigma_{i}}{d x d y} = \frac{1}{4} \sum_{\lambda_{\ell},h = \pm 1} \lambda_{\ell} h \frac{d^2\sigma_{i}}{d x d y}
        & = \frac{4 \pi a_{em}^2 }{x y Q^2} \xi_i \left [- Y_{+} g^i_4 + y^2 g^i_L\ \mp Y_{-} 2 x g^i_1 \right ] \,.
    \label{eq:dis_pol_sx}
\end{align} 
Here $h$ denotes the hadron polarization, the factor $Y_{\pm} = 1 \pm (1 - y)^2$ and the sign $\mp$ refers to positive or negative 
polarized charged leptons $\ell^{\pm}$.
To define the normalization factors $\xi_i$, with $i \in \{ NC, CC\}$, 
we have to sum over the different contributions given by each boson.
For CC, this is trivial and, we get:
\begin{equation}
    \xi_{CC} = (1 + \lambda_\ell I_{W}) \eta_{WW}
    \label{eq:xi_CC}
\end{equation}
while for neutral currents we obtain, for parity conserving structure functions $H_k \in \{F_2, F_L, g_1\}$:
\begin{equation}
    H_k^{NC} = H^{\gamma \gamma}_k - \left [ V_\ell \pm \lambda_\ell A_\ell \right ] \eta_{\gamma Z} H^{\gamma Z}_k + \left [ V^2_\ell + A^2_\ell \pm 2 \lambda_\ell V_\ell A_\ell \right ] \eta_{ZZ} H^{ZZ}_k \, ,
    \label{eq:sf_pc_nc}
\end{equation}
and for the parity violating $H_k \in \{F_3, g_4, g_5\}$:
\begin{equation}
    H_k^{NC} = - \left [ A_\ell \pm \lambda_\ell v_\ell \right ] \eta_{\gamma Z} H^{\gamma Z}_k + \left [ \pm \lambda_\ell (V^2_\ell + A^2_\ell) + 2 V_\ell A_\ell \right ] \eta_{ZZ} H^{ZZ}_k \, . 
    \label{eq:sf_pv_nc}
\end{equation}
Depending on the experimental setup, which can assume different spin configurations
polarized DIS cross-sections are often reported as asymmetries or normalized to 
the unpolarized counterparts \cite{Ball:2013lla} (cf. \cref{subsec:pol_data}).

Eventually, in order to show how it is possible to compute predictions about the behavior
of structure functions using perturbative QCD, we now have to introduce the
parton model and collinear factorization.


\subsection{The parton model and QCD}
\label{sec:parton_model}

At energies exceeding the proton mass ($E \approx 2 m_p~\text{GeV}$), 
the measured cross-section of electron-proton scattering indicates 
that the structure of the proton is composed of point-like particles.
Based on this observation a naive parton model was derived by Feynman~\cite{Feynman:1969ej,Feynman:1973xc}
even before QCD became a well adopted framework. In this section we show 
how the leading-order (LO) parton model can explain the observed Bjorken 
scaling of the DIS structure functions and provide an intuitive definition 
of Parton Distribution functions (PDFs).

Let's suppose that hadrons are made up of constituents, called partons~\cite{Feynman:1969wa},
which become somehow weakly interacting at high energy scales $Q \to \infty$, 
and that the probing leptons scatter on them incoherently.
We can then modify the scattering process of \cref{eq:dis_def} to be 
\begin{equation}
    \ell(k) + q(\xi P) \to \ell'(k') + X \,,
    \label{eq:partonic_dis_def}
\end{equation}
where now we have substituted the full proton by a parton $q$ with momentum
$p_q = \xi P$ with $\xi \in (0, 1]$. 
In analogy to the derivation of \cref{sec:had_sf} define the 
partonic momentum fraction $z$
\begin{equation}
    z =  \frac{Q^2}{2 p_q \cdot q} = \frac{x}{\xi} \, ,
    \label{eq:partonic_x}
\end{equation}
and we can now compute the cross-section of the process by substituting 
the tensors $W_i^{\mu \nu}$, with its partonic counterpart $\hat{W}_i^{\mu \nu}$,
implicitly defined via
\begin{align}
\begin{split}
    W_i^{\mu \nu}(x, Q) & = \sum_{j} \int_0^1 d z \int_0^1 d \xi \hat{W}_i^{\mu \nu}(z, Q) q_j(\xi) \delta(x - \xi z) \\
                        & = \sum_{j} \int_x^1 \frac{d\xi}{\xi} \hat{W}_i^{\mu \nu}(\frac{x}{\xi}, Q) q_j(\xi) \,,
    \label{eq:Wmunu_factorization}
\end{split}
\end{align} 
where $q_i(\xi)$ is a functional distribution describing the probability 
to extract a parton $q$ of flavor $i$ from the colliding 
hadron. The sum runs over all possible flavors, and we have to integrate 
over all possible momentum fractions and probability configurations.
Effectively, energy-momentum conservation implies that the integration bounds
in \cref{eq:Wmunu_factorization} are limited by $x$ and 1.
Note that the object $q_i(\xi)$ contains a spin index, which is here omitted, 
but follows automatically once the PDF is defined in terms of field operators
\cite{Manohar:1990jx,Manohar:1990kr}.
As before we can decompose the partonic tensors according to \cref{eq:Wmunu_def}
\begin{align}
    \begin{split}
        \frac{1}{2 z} \hat{W}_{j}^{\mu \nu} = 
            & - g_{\mu \nu} \left [ \hat{F}_1^j(z,Q^2) + \frac{S \cdot q}{ p_q \cdot q} \hat{g}_5^j (z, Q^2) \right ] \\
            & + \frac{p_q^\mu p_q^\nu}{p_q \cdot q} \left [ \hat{F}_2^j(z,Q^2) + \frac{S \cdot q}{ p_q \cdot q} \hat{g}_4^j (z, Q^2) \right ] \\
            & - i \epsilon_{\mu \nu \alpha \beta} \frac{q^\alpha}{2 p_q \cdot q} \left [ \hat{F}_3^j(z,Q^2) p_q^\beta - 2 \hat{g}_1^j(z, Q^2) S^\beta  \right ] \, .
        \label{eq:Wmunu_part_def}
\end{split}
\end{align}
The major difference is that now the partonic structure functions $\hat{F}_i,\hat{g}_i$ 
are computable directly from the Feynman diagrams associated to the partonic scattering,
multiplied by a suitable projector (see Ref.~\cite{Hekhorn:2019nlf} for a complete list).

At LO, only diagrams with quarks in the initial state can contribute and,
the computation of \cref{eq:Wmunu_part_def} reduces to the diagram $q(p_q) + V^{*}(q) \to q'(p'_q)$.
For example, in the case of photon exchange only, after averaging over the spin, its amplitude 
leads to the integral
\begin{equation}
    \frac{1}{2z} \hat{W}_{j}^{\mu \nu} & = e_q^2 \int \frac{d^3 p'_q}{(2 \pi)^3} \frac{1}{2 E'_q} {\rm \Tr} \left[ \gamma^{\mu} \slashed{p}_q \gamma^{\nu} \slashed{p'}_q \right] (2 \pi)^4 \delta^4(p_q + q - p'_q) \\
                          & = 2 \pi e_q^2 \left[ \left( - g_{\mu \nu} + \frac{q^\mu q^\nu}{q^2} \right) + \frac{4z}{Q^2} \left( p_q^\mu - \frac{p_q \cdot q}{q^2} q^\nu \right) \left( p_q^\nu - \frac{p_q \cdot q}{q^2} q^\mu \right) \right] \delta(1-z) \,. \nonumber
    \label{eq:Wmunu_part_LO}
\end{equation}
By comparing to \cref{eq:Wmunu_def} we can identify the 2 non-vanishing partonic structure functions, 
$F_1$ and $F_2$, which are proportional to Dirac-delta.
More in general, after applying the projectors, summing or subtracting on the parton spin 
and performing the convolution with the PDF, one finds the relations between hadronic structure function 
and PDFs. For the NC, we get:
\begin{allowdisplaybreaks} 
\begin{align}
    \left [ F_2^{\gamma\gamma}, F_2^{\gamma Z}, F_2^{ZZ} \right ] &= x \sum_q \left[ e_q^2, 2 e_q V_q, V_q^2 + A_q^2 \right] (q + \bar{q}) \,, \\ 
    x \left [ F_3^{\gamma\gamma}, F_3^{\gamma Z}, F_3^{ZZ} \right ] &= x \sum_q \left[ 0 , 2 e_q A_q, 2 V_q A_q \right] (q - \bar{q}) \,, \\
    2 x \left [ g_1^{\gamma\gamma}, g_1^{\gamma Z}, g_1^{ZZ} \right ] &= x \sum_q \left[ e_q^2, 2 e_q V_q, V_q^2 + A_q^2 \right] (\Delta q + \Delta \bar{q}) \,, \\ 
    \left [ g_4^{\gamma\gamma}, g_4^{\gamma Z}, g_4^{ZZ} \right ] &= x \sum_q \left[ 0, 2 e_q A_q, 2 V_q A_q \right] (\Delta q - \Delta\bar{q}) \,,
    \label{eq:sf_NC_LO}
\end{align} 
\end{allowdisplaybreaks}
\noindent where $V_q, A_q$ are defined in \cref{eq:fZ_coupling} for up and down type quarks,
while the PDFs $q_k,\ \Delta q_k$ read:
\begin{align}
    q_k(x) = q_k^{\uparrow\uparrow}(x) &+ q_k^{\uparrow\downarrow}(x) \\
    \Delta q_k(x) = q_k^{\uparrow\uparrow}(x) - q_k^{\uparrow\downarrow}(x), & \quad
    \Delta \bar{q}_k(x) = \bar{q}_k^{\downarrow\uparrow}(x) - \bar{q}_k^{\uparrow\uparrow}(x)\,,
    \label{eq:pdf_decomposition}
\end{align}
where for each PDF the first arrow indicates the direction of
the proton spin and the second the partonic helicity.
In the case of CC, the average the sum $W^+ + W^-$ has exactly the same 
flavor decomposition as of the NC but with a different coupling,
while for the combination $W^+ - W^-$ up and down quark types 
contribute with opposite sign:
%
\begin{align}
    F_2^{W^+ \pm W^{-}} &= 2 x  \sum_{d_j} \sum_{u_i} |V_{u_i,d_j}|^2 d_j^\pm \pm  \sum_{u_j} \sum_{d_k} |V_{d_k,u_j}|^2 u_j^\pm \,, \\
    x F_3^{W^+ \pm W^{-}} &= 2 x \sum_{d_j} \sum_{u_i} |V_{u_i,d_j}|^2 d_j^\mp \pm \sum_{u_j} \sum_{d_k} |V_{d_k,u_j}|^2 u_j^\mp \,, \\
    2 x g_1^{W^+ \pm W^{-}} &= 2 x \sum_{d_j} \sum_{u_i} |V_{u_i,d_j}|^2 \Delta  d_j^\pm \pm  \sum_{u_j} \sum_{d_k} |V_{d_k,u_j}|^2 \Delta u_j^\pm \,, \\
    g_4^{W^+ \pm W^{-}} &= 2 x \sum_{d_j} \sum_{u_i} |V_{u_i,d_j}|^2 \Delta d_j^\mp \pm  \sum_{u_j} \sum_{d_k} |V_{d_k,u_j}|^2 \Delta u_j^\mp \,,
    \label{eq:sf_CC_LO}
\end{align} 
where $V_{u_i,d_j}$ are the Cabibbo-Kobayashi-Maskawa (CKM) matrix and the sum is performed
in the first addend on all the active up-types given a down-type quark (CKM columns), 
and vice versa in the second case (CKM rows). We have adopted the shorthand notation $q^\pm = (q \pm \bar{q})$.

The longitudinal structure functions $F_L,g_L$ are vanishing at $\mathcal{O}(a_s)$.
This is known as the \textit{Callan-Gross} \cite{Callan:1969uq} (or \textit{Dicus} \cite{Dicus:1972pq}) relation and follows directly 
from the quark being a spin-$\frac{1}{2}$ field, which cannot absorb a longitudinally
polarized vector boson.

Historically, the greater success of the parton model was to predict 
two major observed features of the DIS structure functions, the above-mentioned 
\textit{Callan-Gross} relation and the so-called \textit{Bjorken scaling} \cite{Bjorken:1968dy}. 
In the limit of $Q,\nu \to \infty$ and at fixed $x$, the structure functions 
become scale independent, i.e. $F(x, Q) \to F(x)$; 
this seemed compatible with the idea that partons behaves like a free particle
only in that limit where are asymptotically free. 
However, as explained in the next sections, QCD interactions cause 
the scaling violation for mid-range $Q$ as well as deviation from the structure 
functions sum rules.
Before describing how such QCD correction can be included in the DIS structure
functions (\cref{sec:dis_coeff}), let us introduce more formally the PDFs and 
their renormalization group equations.

\section{Parton Distribution Functions}
\label{sec:pdfs}

In \cref{sec:factorization}, we generalize the concept of PDFs, 
and relate them to the DIS parton model using the factorization theorem.
Then in \cref{sec:dglap} we introduce the PDF renormalization group equations, 
which allow for the description of PDF scale dependence taking into
account the QCD splitting. 
Finally, in \cref{sec:dis_coeff} we explain how the higher
order QCD corrections are organized inside the DIS 
structure functions.

\subsection{Collinear factorization}
\label{sec:factorization}
%
%
\paragraph{PDF operator definition.}

In the previous section we have introduced the concept of Parton Distribution
Function (PDF) from a phenomenological point of view.
Starting from the quantized quarks $\psi_q$ and gluon fields $A^\mu$ of the QCD Lagrangian 
(\cref{eq:lqcd_classical}), one can construct the PDFs as 
Fourier transform of the operator matrix element that counts the number of quarks and gluons 
in the hadron state $\ket{P}$ and at given momentum fraction.
For quark fields, we define
\begin{equation}
    f_q (x) = \int \frac{d \xi^{-}}{4 \pi} e^{-i x P^{+} \xi^{-}} \bra{P} \bar{\psi}_q(\xi^{-}) \gamma^+ U(\xi^-, 0) \psi_q(0)  \ket{P} \, ,
    \label{eq:pdf_def}
\end{equation}
where the hadron momentum is $P^\mu = (P^0,0,0,P^z)$ and $P^\pm = \frac{P^0 \pm P^z}{\sqrt{2}}$ are the light-cone 
coordinates, with the extracted parton carrying a momentum $xP^\mu$ with $x \in [-1, 1]$.
$U$ is the parallel transport operator of the gauge field given by the path ordering 
\begin{equation}
    U(\xi^{-},0) = \mathcal{P} \exp \left[ -i a_s \int_0^{\xi^-} d \eta A(\eta^-) \right]
    \label{eq:wilson_line}
\end{equation}
The connection between \cref{eq:pdf_def} and the phenomenological quantities in \cref{sec:parton_model} is then
\begin{equation}
    f_q(x) = 
        \begin{cases}
        q(x)    \quad \text{for} \quad x > 0 \\
        - \bar{q}(-x) \quad \text{for} \quad x < 0
        \end{cases}
    \label{eq_pdf_formal_mapping}
\end{equation}
assuming the caveat that, to obtain the correct helicity combinations, 
one now has to sum or subtract the spin projection 
as in \cref{eq:pdf_decomposition}.
The gluon PDF can be defined in an analogous way. We refer to Refs.~\cite{Collins:1981uw,Collins:1989gx} and 
Refs.~\cite{Manohar:1990jx,Manohar:1990kr} for a more detailed discussion on the unpolarized or 
polarized case.
Finally, let us notice that PDFs, as defined in \cref{eq:pdf_def} must 
undergo the renormalization procedure which subtracts the ultraviolet divergences 
of the bare field.

\paragraph{Factorization theorem.}

The operator PDF definition is valid for any hadron (including nuclei)
implying that PDFs are really a characteristic property of the hadron 
and do not depend on the scattering process we are looking at.
%
%
But most importantly, it allows us to prove 
the factorization theorem~\cite{Collins:1989gx}: any inclusive DIS observable can be 
computed as convolution between the PDF $f_i$, describing the non-perturbative 
(i.e. long distance) dynamics and the partonic matrix element $\hat{\sigma}_i$, associated
with the hard interaction process (i.e. short distance)
\begin{equation}
    \sigma(x, Q) = \sum_{i=q,\bar{q},g} \int_x^1 \frac{dy}{y} \hat{\sigma}_i \left(\frac{x}{y}, Q, \mu_F \right) f_i\left( y, \mu_F\right)  + \mathcal{O}\left(\frac{\Lambda^2}{Q^2} \right) \, .
    \label{eq:dis_factorization}
\end{equation}
This formula is well justified only if the scale at which the short and long
distance interaction occurs are well separated.
If this is no longer the case, then multiple parton interaction (higher-twist)
can occur and spoil the factorization. 
Higher-twist are then process dependent and \cref{eq:dis_factorization} ensures that they 
are suppressed by powers of $Q^2$. 
Partonic cross-sections can be computed using pQCD, for every given 
partonic process, while PDFs are universal and their numerical 
value have to be extracted by fitting to experimental data.
Similarly to the case of renormalization, in \cref{eq:dis_factorization} 
we have introduced an additional scale $\mu_F$, called factorization scale, 
which is not physical and plays as similar role as the renormalization scale.

\paragraph{NLO corrections to $q + \gamma \to q$.}

To explain the physical origin of this factorization scale let us look again at the 
simple process $q + \gamma \to q$. Here $\hat{\sigma}_i$ can be identified 
with a partonic structure function $\hat{F}_{2,i}$.
\begin{equation}
    F_2(x, Q) = \sum_{i=q,\bar{q},g} \int_x^1 \frac{dy}{y} \hat{F}_{2,i} \left(\frac{x}{y}, Q^2 \right) f_i\left( y \right) \,,
    \label{eq:dis_factorization_2}
\end{equation}
Next-to-leading (NLO) order QCD corrections, as the real gluon emission from the 
initial state quark, induce integrations over the quark propagator momentum $k$
\begin{equation}
    \hat{F}_{2}\left |_{q\gamma \to q g}(x, Q^2) \right . \propto \int_{Q_T^2}^{Q^2} \frac{dk^2}{k^2} a_s x P(x) =  a_s \ln\frac{Q^2}{Q_T^2} x P(x) \, ,
    \label{eq:partonic_nlo}
\end{equation}
with $P(x)$ a characteristic function describing the quark to gluon splitting.
This leads to a logarithmic divergence for $Q_T \to 0$, i.e. when the emitted gluon 
is \textit{collinear} to the quark, and has vanishing transverse momentum.
Again one can identify such limit as low-energy interaction and thus reabsorb
the divergence into the PDF definition, leaving the partonic structure function $\hat{F}_{2,i}$
finite.
To do so, we introduce a cut-off scale $\mu_F$ which acts as regulator of the
divergent logarithmic term separating $\ln\frac{Q^2}{Q_T^2} = \ln\frac{Q^2}{\mu_F^2} + \ln\frac{\mu_F^2}{Q_T^2}$.
The first contribution is finite and can be kept inside the partonic structure function, 
while the second can be arbitrary large and must be retained inside the PDFs.
Note that to get the full NLO divergent contribution to $q + \gamma \to q$,
one would need to account also for the virtual corrections to the quark 
leg and photon vertex.
Before subtraction, the full quark initiated contribution to $F_2$ up to NLO then is
\begin{equation}
    F_{2,q} (x, Q^2) \propto x \sum_{q,\bar{q}} \left( q(x) + a_s \int_x^1 \frac{dy}{y}  \left[ P \left (\frac{x}{y} \right ) \ln\frac{Q^2}{Q_T^2} + C \left (\frac{x}{y} \right ) \right] q(y) \right) \, .
    \label{eq:f2_lo_nlo}
\end{equation}
where $C$ is again a finite function typical of this process. We redefine the quark PDF to be
\begin{equation}
    q (x, \mu_F^2) =  q(x) + a_s \int_x^1 \frac{dy}{y}  \left[ P \left (\frac{x}{y} \right ) \ln\frac{\mu_F}{Q_T^2} \right] q(y) \,,
    \label{eq:ren_pdf}
\end{equation}
and obtain the physical finite structure function via
\begin{equation}
    F_{2,q} (x, Q^2) \propto x \sum_{q,\bar{q}} \int_x^1 \frac{dy}{y} \left [ P \left (\frac{x}{y} \right ) \ln\frac{Q^2}{\mu_F^2} + C \left( \frac{x}{y}  \right) \right ] q( y , \mu_F) \,.
    \label{eq:ren_f2_lo_nlo}
\end{equation}
This corresponds to say that the PDFs introduced in \cref{eq:Wmunu_factorization,eq:dis_factorization_2}
are unmeasurable (bare) quantities, which absorb corrections of type
$\left (a_s \ln\frac{\mu_F^2}{Q_T^2} \right )$ for each collinear gluon 
emission.
As the final structure function in l.h.s of \cref{eq:ren_f2_lo_nlo} must be independent 
on the choice of the factorization scale, it is possible to derive a differential 
equation which properly resums the collinear emissions inside the physical PDF. 

It is possible to show that the singularities arising in the parton model correspond precisely 
to the infrared divergences of the PDF, defined with the operator point of view, 
when these are evaluated for on-shell massless partonic states \cite{Collins:2021vke,Candido:2023ujx}.


\subsection{DGLAP equations}
\label{sec:dglap}

The dependency of the PDFs on the factorization scale is governed by the RGEs, 
known as the Dokshitzer-Gribov-Lipatov-Altarelli-Parisi 
(DGLAP) equations~\cite{Gribov:1972ri,Altarelli:1977zs,Dokshitzer:1977sg}
\begin{equation}
    \mu^2_F \frac{f_i(x, \mu^2_F)}{d \mu^2_F} =  P_{ij} \otimes f_j = \sum_{j=q,\bar{q},g} \int_x^1 \frac{dy}{y} P_{ij}\left(y, a_s(\mu^2_F)\right) f_j(x/y, \mu^2_F)
    \label{eq:dglap}
\end{equation}
where $P_{ij}$ are the Altarelli-Parisi splitting functions, 
$\otimes$ denotes the Mellin convolution and the 
sum runs over all the active flavors.
As the splitting kernels are not diagonal, \cref{eq:dglap} 
is a system of coupled equations. However, the gluon distribution
has to be flavor blind and couples only with the total quark
PDF. Thus, it is possible to maximally disentangle the system,
rotating to the evolution basis
\begin{equation}
    \mathcal{F}_{ev} = \text{span}(g, \Sigma, V, T_3, T_8, T_{15}, T_{24}, V_3, V_8, V_{15}, V_{24})
    \label{eq:evol_basis}
\end{equation}
\begin{alignat}{3}
    \label{eq:pm_to_evol_basis}
    \Sigma &= \sum_i^{n_f}f_{i}^+ \qquad && V &&= \sum_i^{n_f}f_{i}^- \nonumber \\
    T_{3} &= u^+ - d^+ \qquad && V_3 &&= u^- - d^-\nonumber \\
    T_{8} &= u^+ + d^+ -2s^+ \qquad && V_8 &&= u^- + d^- - 2s^- \\
    T_{15} &= u^+ + d^+ + s^+ - 3c^+ \qquad && V_{15} &&= u^- + d^- + s^- - 3c^- \nonumber\\
    T_{24} &= u^+ + d^+ + s^+ + c^+ - 4b^+ \qquad && V_{24} &&= u^- + d^- + s^- + c^- - 4b^- \nonumber
\end{alignat}
where the $q_k^{\pm} = q_k \pm \bar{q}_k$. The basis
elements can be separated in to two categories: 
$\Sigma$ (total) singlet, gluon $g$ and $T_i$
form the \textit{singlet} sector; while we refer to $V$ as (total) valence 
distribution and to $V_i$ \textit{non-singlet} distributions.
Here, we are considering the case of pure QCD evolution 
and neglecting any photon PDF contribution, 
although a generalization is possible, see Ref.~\cite{NNPDF:2024djq} for details. 
For any phenomenological application also the top $t$, $\bar{t}$ PDFs are always
neglected.
In the evolution basis only $\Sigma$ and $g$ are coupled via 
\begin{equation}
    \mu^2_F \frac{d}{d \mu^2_F} \begin{pmatrix} g \\ \Sigma \end{pmatrix} = 
    \begin{pmatrix} P_{gg} & P_{gq} \\ P_{qg} & P_{qq} \end{pmatrix}
    \otimes
    \begin{pmatrix} g \\ \Sigma \end{pmatrix}
    \label{eq:singlet_dglap}
\end{equation}
while the distributions $V$, $T_i$, $V_i$ evolve all independently.
The polarized DGLAP evolution is analogue with all the quantities $f_i$ and $P_{ij}$ replaced by 
$\Delta f_i$ and $ \Delta P_{ij}$ respectively.

The splitting functions can be expanded in perturbation theory as
\begin{equation}
    (\Delta ) P_{ij}(x, \mu) = \sum_{k=0}^{\infty} a_s^{k+1}(\mu) (\Delta ) P^{(k)}_{ij}(x)
    \label{eq:p_expansion}
\end{equation}
In principle 7 different splitting functions combinations are possible, 
4 in the singlet sector \cref{eq:singlet_dglap} and 3 for the non-singlet
$P_{ns,-}, P_{ns,+}, P_{ns,s}$ with the total valance and singlet-to-singlet 
splitting given by
\begin{align}
    P_{ns,v} &= P_{ns,+} +  P_{ns,s}\, , \label{eq:Pnsv} \\
    P_{qq} &= P_{ns,+} +  P_{qq,ps} \, . \label{eq:Pqq}
\end{align} 
This separation facilitates to isolate perturbative suppressed 
contributions as $P_{ns,s}$, which starts at NNLO (i.e. $\mathcal{O}(a_s^3)$), while
$P_{qq,ps}$ is $\mathcal{O}(a_s^2)$. Moreover, at LO one finds 
that $ P_{qq} = P_{ns,-} = P_{ns,+}$.
Symmetry considerations imply that polarized non-singlet splitting 
functions coincide with the spin-averaged ones to all orders after 
they are swapped as follows:
\begin{equation}
    \Delta P_{ns,\pm} = P_{ns,\mp}
    \label{eq:deltap_ns}
\end{equation}
The interpretation of $P_{ij}(x,\mu)$,
as the probability to find the parton $i$ inside the parton of type $j$
with a given momentum fraction $x$ and energy less than $\mu$,
allows us to formulate some constraints on their integrals and conserved 
quantities:
\begin{align}
    \int_0^1 dx \ P_{ns,-}(x) & = 0 \quad \text{Quark number conservation}\, , \label{eq:quark_num_cons} \\
    \int_0^1 dx \ x [P_{gg}(x) + P_{qg}(x)] & = 0 \quad \text{Gluon momentum conservation}\, , \label{eq:quark_mom_cons} \\
    \int_0^1 dx \ x [P_{qq}(x) + P_{gq}(x)] & = 0 \quad \text{Quark momentum conservation}\, , \label{eq:gluon_mom_cons} \\ 
    \int_0^1 dx \ \Delta P_{qg}(x) & = 0 \quad \text{Helicity conservation}\, . \label{eq:helicity_cons}
\end{align} 
For completeness, note that splitting functions carry a dependency on 
the number of active flavors $n_f$, and $(\Delta)P_{qg}$ as well as 
$(\Delta)P_{qq,ps}$, $(\Delta)P_{ns,s}$ are fully proportional to $n_f$.
All the above listed sum rules are valid at every order and for every 
$n_f$ component in perturbation theory.

The actual expressions of the splitting kernels can be obtained directly
from the Feynman diagrams of the quark-to-gluon and gluon-to-gluon splitting, 
although these kernels appear also explicitly during the calculation of higher 
order QCD cross-sections as we have sketched in \cref{sec:factorization}.
%
%
The unpolarized analytic $P_{ij}$ expressions are known up to NNLO~\cite{Moch:2004pa,Vogt:2004mw,Blumlein:2021enk},
while we will discuss explicitly the different N$^3$LO approximations 
in \cref{sec:an3lo_dglap}.
The NNLO helicity dependent $\Delta P_{ij}$ were presented in Refs.~\cite{Moch:2014sna,Moch:2015usa} 
and benchmarked independently in Refs.~\cite{Blumlein:2021enk,Blumlein:2021ryt}.
%

\paragraph{DGLAP solution.}
Having simplified the system of the DGLAP equations with respect to the flavor space, 
we can now attempt to solve it.
However, we see that \cref{eq:dglap} contains a convolution which makes the 
differential equations more complicated. Therefore, it is convenient to introduce 
the Mellin transformation:
\begin{equation}
    \tilde{g}(N) = \mathcal{M}\left[ g(x) \right](N) = \int_0^1 dx \ x^{(N-1)} g(x)\, .
    \label{eq:mellin_def}
\end{equation}
In Mellin space a convolution becomes a simple product
\begin{align}
\begin{split}
    \mathcal{M}\left[ c \otimes f \right](N) 
    & = \int_x^1 dx \ x^{(N-1)} \int_{x}^{1} \frac{dy}{y} c(y) f(x/y) \\
    & = \int_x^1 dx \int_{x}^{1} \frac{du}{u} \ x^{(N-1)} c(x/u) f(u) \\
    & = \int_x^1 u \ dt \int_{0}^{1} \frac{du}{u} \ u^{(N-1)}t^{(N-1)} c(t) f(u)
    = \tilde{c}(N) \tilde{f}(N) \, ,
\end{split}
\end{align} 
thus we can define the anomalous dimension (note the additional minus sign) as
\begin{equation}
    \gamma_{ij}\left(N, a_s\right) =  - \mathcal{M}\left[ P_{ij}(x, a_s) \right](N) \, ,
    \label{eq:ad_def}
\end{equation}
and rewrite \cref{eq:dglap} leaving implicit the sum over flavors
\begin{equation}
    \mu^2_F \frac{\tilde{f}_i(x, \mu^2_F)}{d \mu^2_F} =  
    - \gamma_{ij}\left(N, a_s\right) \tilde{f}_j
    \label{eq:dglap_mellin} \, .
\end{equation}
By changing the evolution variable to $a_s$ and using \cref{eq:as_running} we get
\begin{equation}
    \frac{d \tilde{f}_i(x, a_s)}{d a_s} =  
    \frac{d \mu^2_F}{ d a_s} \frac{\tilde{f}_i(x, a_s)}{d \mu^2_F} =  
    - \frac{\gamma_{ij}\left(N, a_s\right)}{\beta(a_s)} \tilde{f}_j
\end{equation}
which admits a formal solution in terms of Evolution Kernel Operators
(EKO)~\cite{Bonvini:2012sh} and a given boundary condition $\{ a_{s,0}, {f}_j(a_{s,0}) \}$
\begin{align}
\begin{split}
    \tilde{f}_i(a_s) 
    & = E_{ij}(a_s \leftarrow a_{s,0}) \tilde{f}_j(a_{s,0})  = \mathcal{P} \exp\left[ - \int_{a_{s,0}}^{a_s} dt \frac{\gamma(t)}{\beta(t)} \right] \tilde{f}_j(a_{s,0}) \,,
    \label{eq:eko}
\end{split}
\end{align} 
being $\mathcal{P}$ the path-ordering operator.
For the non-singlet sector it is possible to find an analytic solution,
while in the singlet case different type of approximations are possible.
We refer to \cref{sec:eko} for a more detailed discussion of how these methods 
are implemented at different perturbative orders and how the final result 
is then converted back into $x$-space. 
The LO the solution is
\begin{equation}
    E_{ij}(a_s \leftarrow a_{s,0}) = \exp \left[ \ln \left( \frac{a_s}{a_{s,0}} \right) \frac{\gamma_{ij}^{0}}{\beta_{0}} \right] \,.
    \label{eq:eko_lo}
\end{equation}
Phenomenologically, at higher scales the DGLAP splitting induces a raise in 
the small-$x$ tails of the singlet and gluon PDFs, while it decrease the large-$x$ 
PDF.
At low energy the proton is dominated by the valence PDFs,
and the higher we go in energy the more the quark and gluon sea 
becomes relevant (see also \cref{fig:pdg}).

\subsection{PDF and DIS coefficients}
\label{sec:dis_coeff}

We are now equipped with all the necessary ingredients to analyze how higher 
order QCD correction are included into DIS structure functions.
By using the factorization theorem of \cref{eq:dis_factorization} we can write 
\begin{align}
\begin{split}
    F^{j}_i(x, Q) = C^{j}_{i,k} \otimes f_{k} 
    & = \sum_{k=q,\bar{q},g} \int_x^1 \frac{dy}{y} C^{j}_{i,k} \left(y, Q, \mu_F \right) f_k\left( \frac{x}{y}, \mu_F \right) \\
    & =  C^{j}_{i,g} \otimes x g  + \sum_{k=\bar{q},q} C^{j}_{i,q} \otimes x q_k \, ,
    \label{eq:coefficient_functions}
\end{split}
\end{align} 
where the index $j$ distinguish the type of electroweak interaction $j\in \{\gamma\gamma,\gamma Z,Z\gamma,ZZ,WW\}$, 
and $i \in \{2,L,3\}$. 
The objects $C^{j}_{i,k}$ are called DIS coefficient functions,
which can be expanded as
\begin{equation}
    C^{j}_{i,k}(y, Q) = \sum_{l=0}^{\infty} a_s^l(Q^2) \ C^{j,(l)}_{i,k}(y) \, ,
    \label{eq:pert_dis_coeff}
\end{equation}
having assumed $\mu_F=Q$, with the dependency on the scale $Q$ 
fully contained in the $a_s$ running. 
Their expressions are known up to three-loop ($l=3$) both for NC
\cite{Moch:2004xu,Vermaseren:2005qc,Blumlein:2022gpp} and 
CC \cite{Moch:2007rq,Moch:2008fj,Davies:2016ruz} coefficients, 
with some partial results at four-loop ($l=4$) \cite{Basdew-Sharma:2022vya}.
Here and in the following we call LO coefficient function 
the contributions $\mathcal{O}(a_s^0)$ irrespective of the first 
non-vanishing order.

\begin{figure}[!t]
    \includegraphics[width=0.18\textwidth]{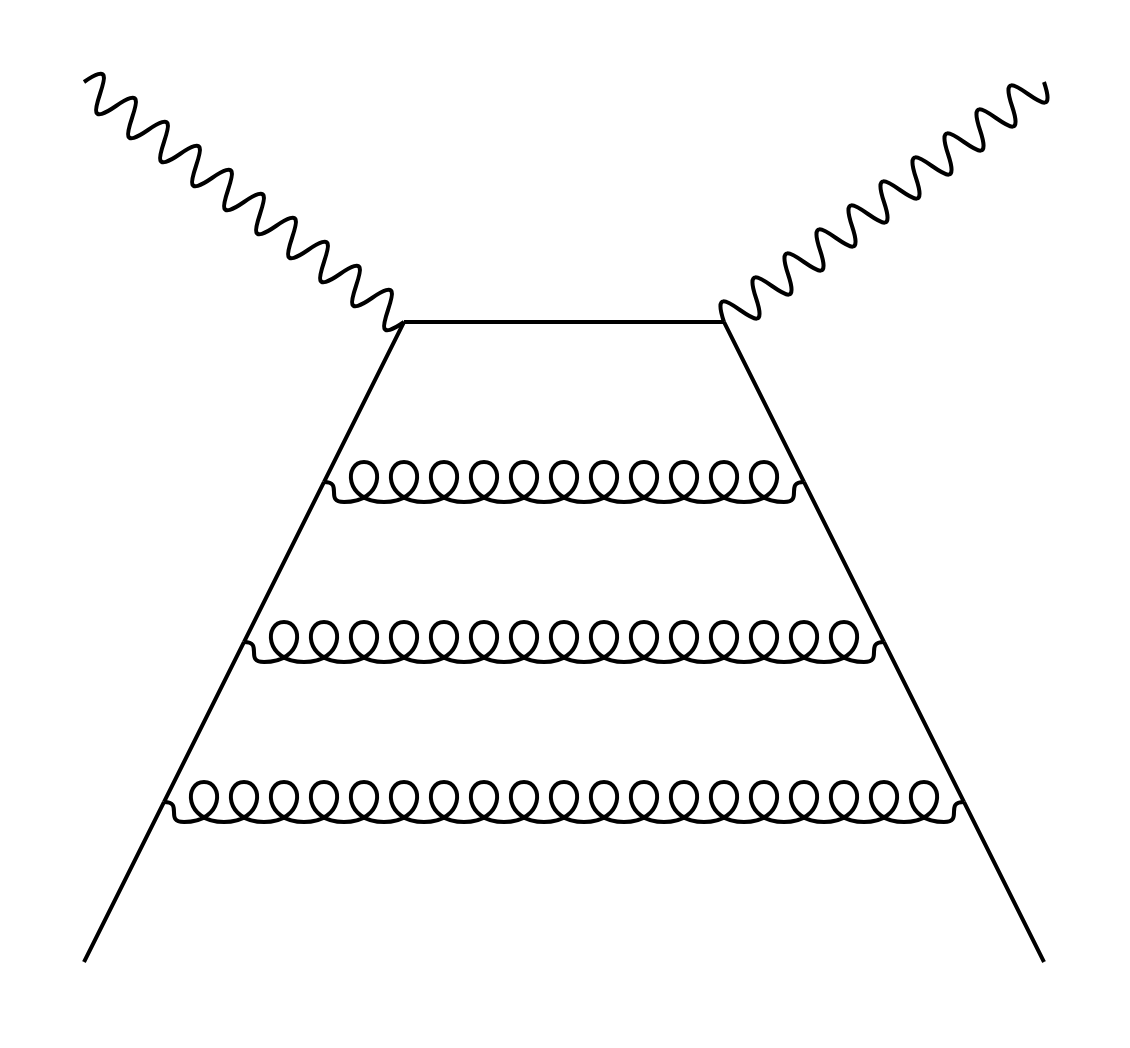}
    \includegraphics[width=0.18\textwidth]{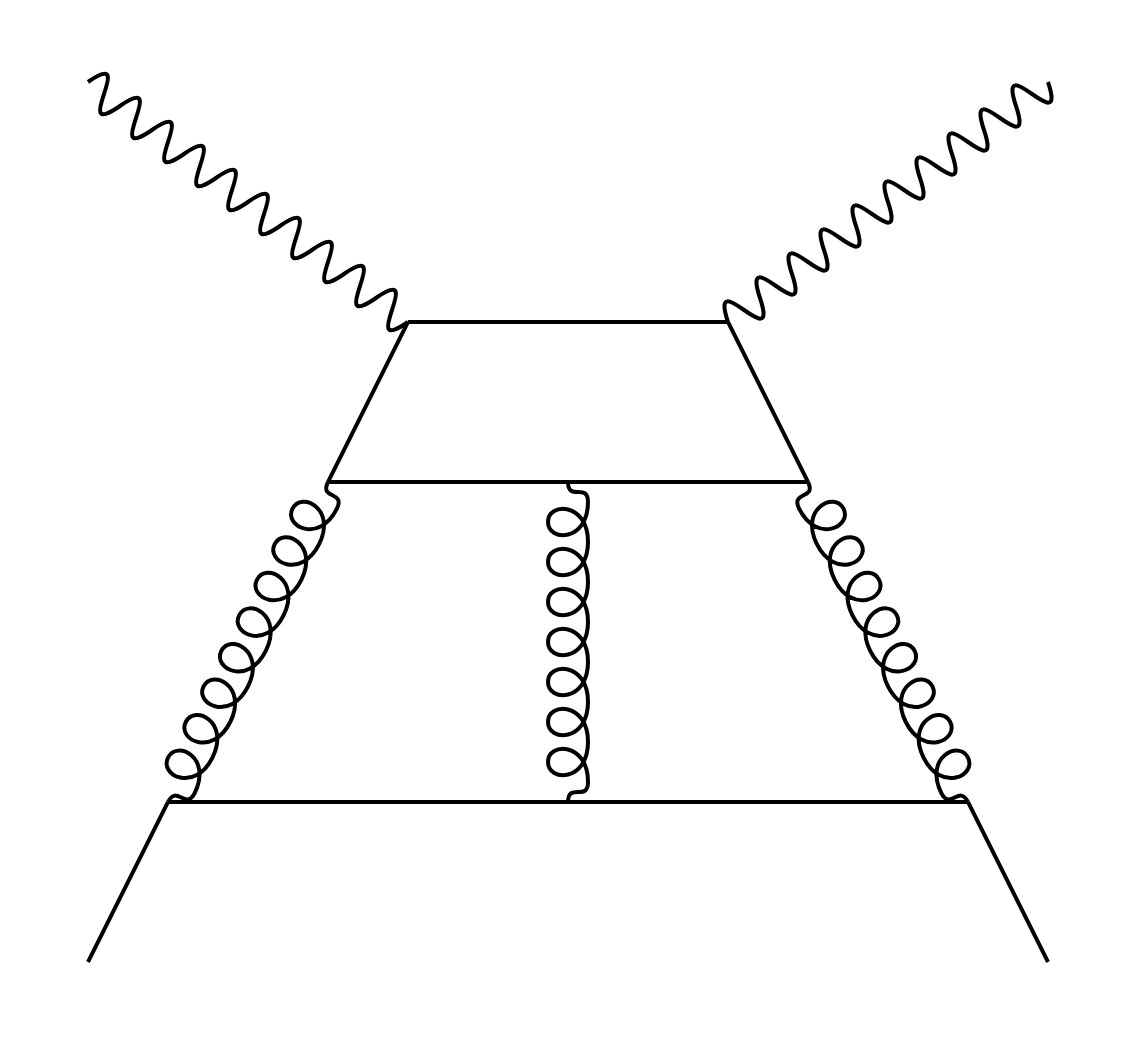} 
    \includegraphics[width=0.18\textwidth]{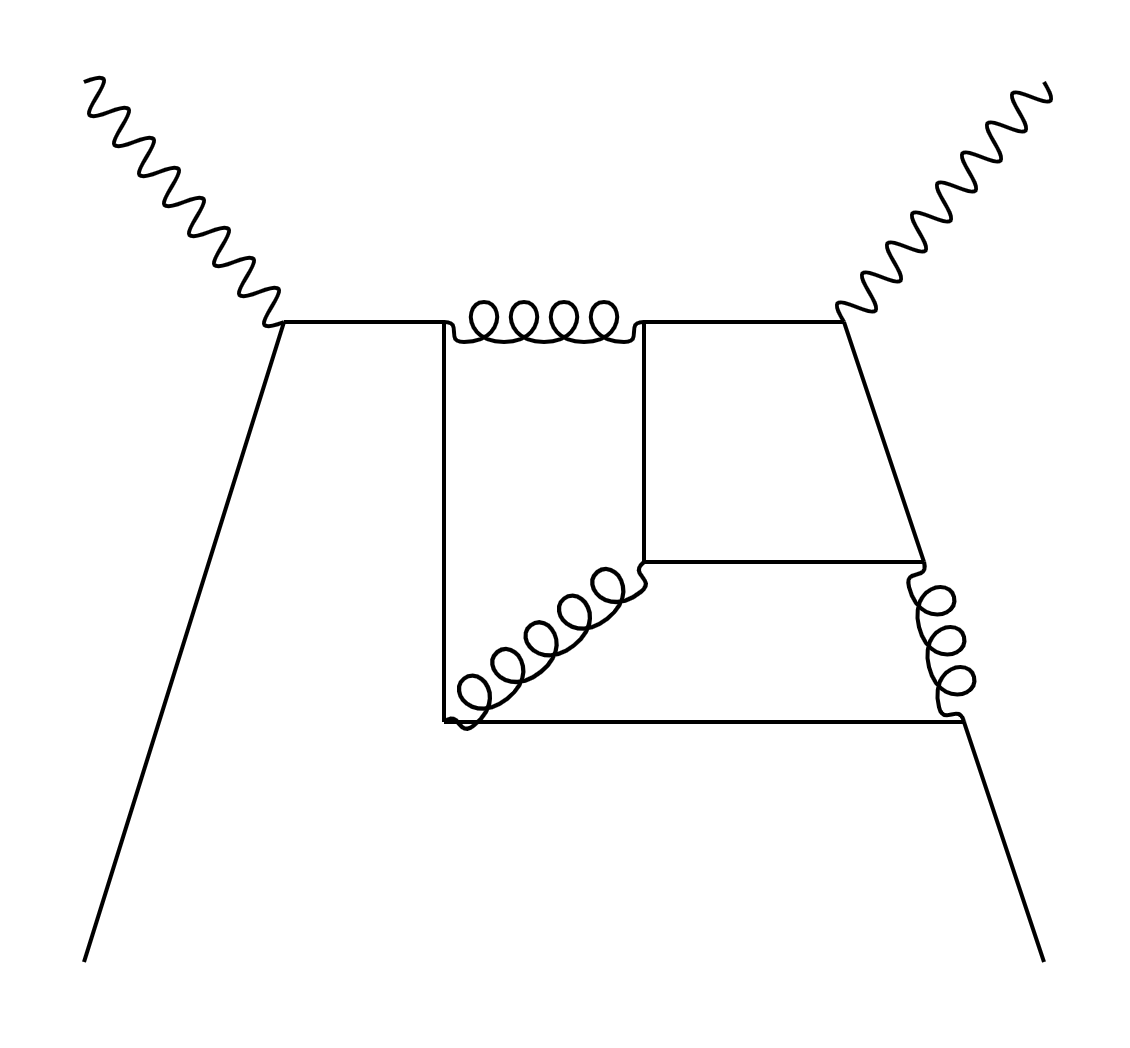}
    \includegraphics[width=0.18\textwidth]{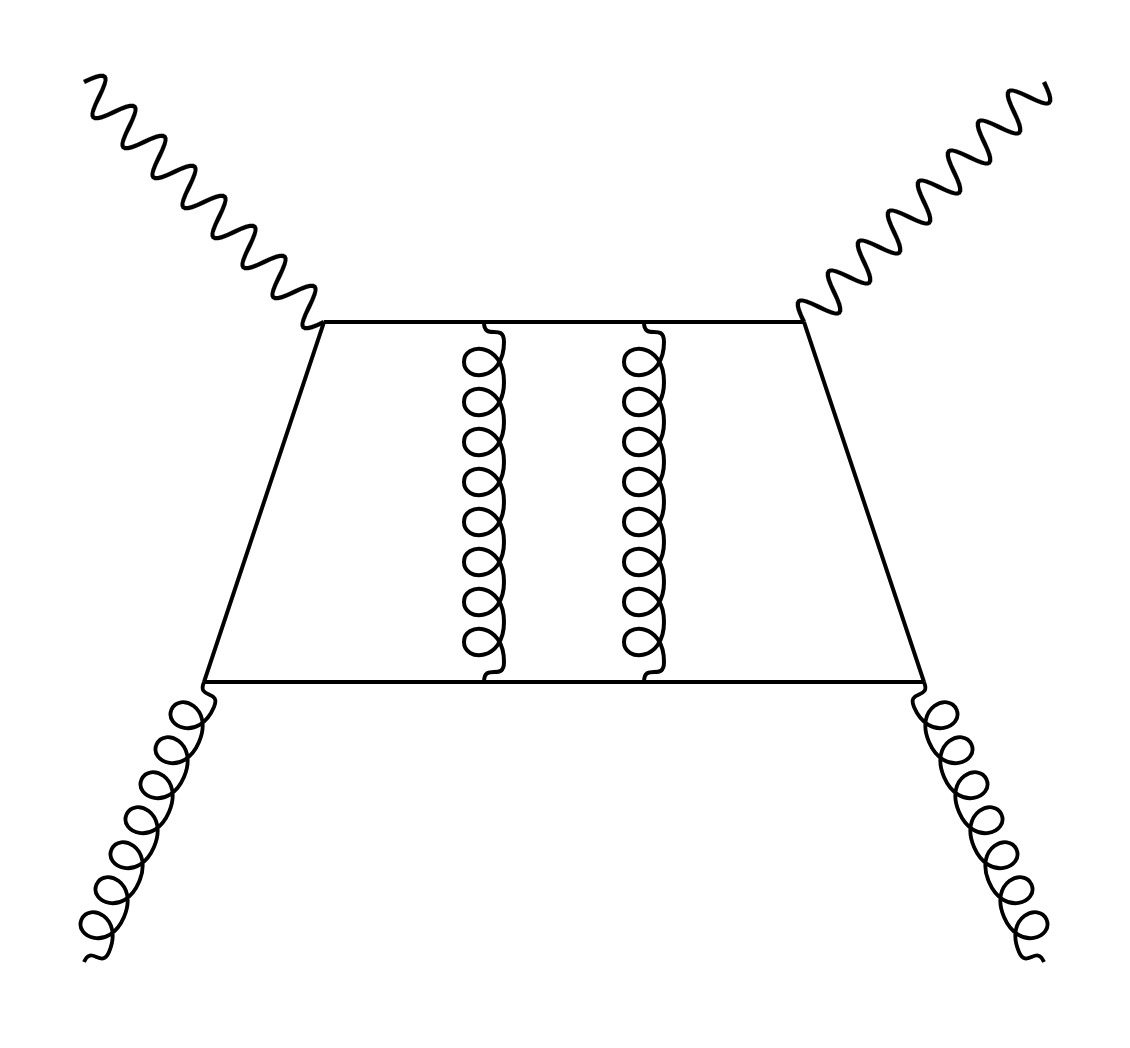}
    \includegraphics[width=0.18\textwidth]{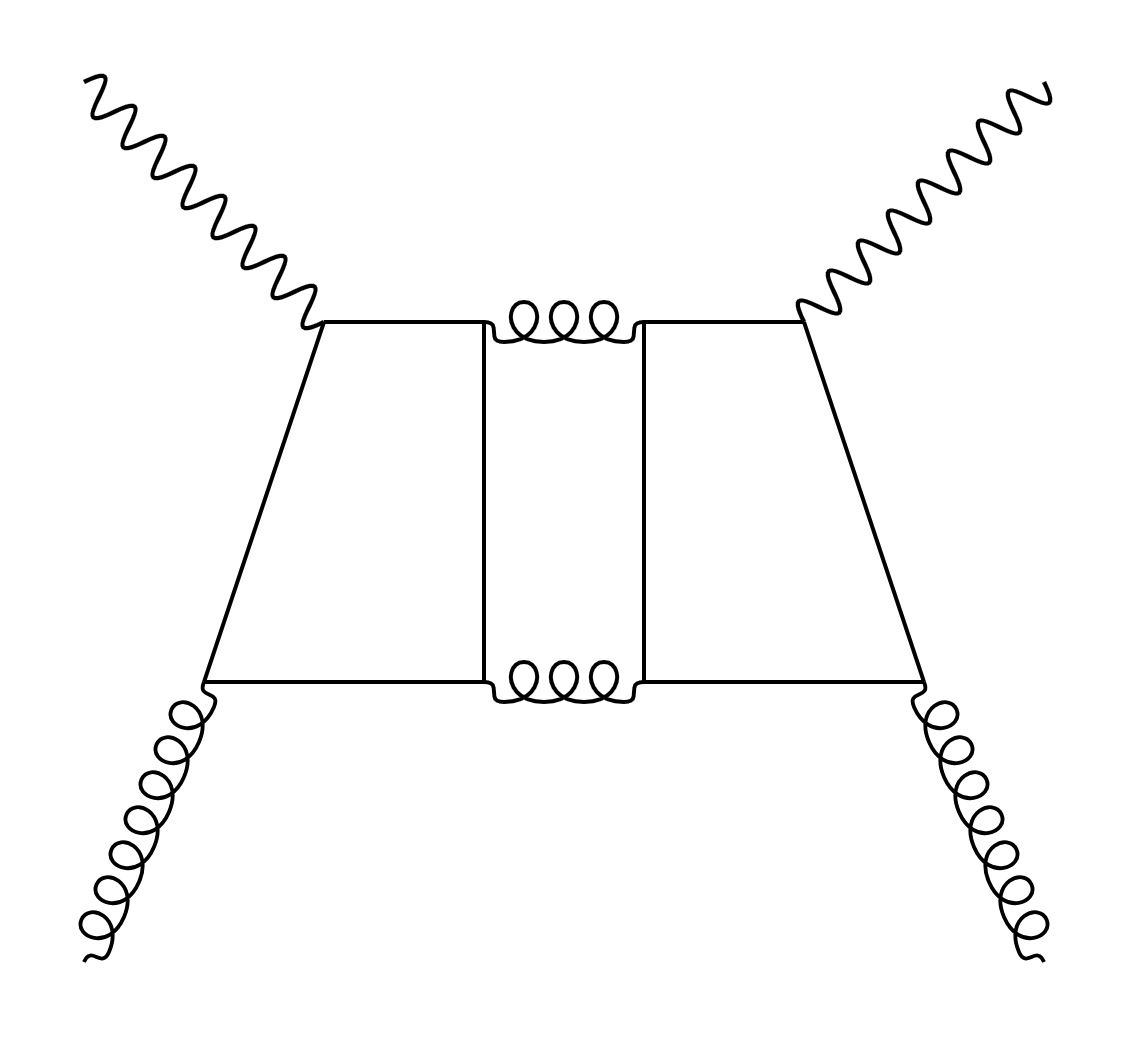}
    \caption{
        Representative three-loop squared DIS diagrams of the flavor class 
        $FC_{2}$, $FC_{02}$, $FC_{11}$, $FC^{g}_{2}$ and $FC^{g}_{11}$
        for quark and gluon DIS scattering.
        All diagrams can contribute to NC, while $FC_{11}$ and $FC^{g}_{11}$ are 
        not present for CC as single $W^\pm$ boson cannot couple to a fermion loop. 
        From Refs.~\cite{Vermaseren:2005qc,Davies:2016ruz}.
    }
    \label{fig:dis_n3lo_ns}
\end{figure}

\paragraph{Flavor decomposition.}
To simplify further the quark sector in \cref{eq:coefficient_functions}, 
and factorize the electroweak coupling $g^{(j)}_{k}$ from the coefficient function, 
one can observe that massless diagrams with $n_f$ active flavors obeys 
a $SU(n_f)$ flavor symmetry \cite{Larin:1996wd},
and distinguish different contributions.
Up to one-loop all the squared diagrams belong to the same flavor class,
$FC_{2}$, where all the coefficients are proportional
to the quark-electroweak boson coupling squared 
$g^{(j),2}_{k} = \text{diag}(g^{(j)}_{d},g^{(j)}_{u},g^{(j)}_{s}, \dots)^2$.
Two-loop corrections can introduce diagrams of the flavor class $FC_{02}$,
where both electroweak boson legs are attached to the same internal quark
loop. These diagrams are then proportional to ${\rm Tr}(g^{(j),2}_{k}) \mathbf{1}$.
From three loops the situation is more complex, as visible in \cref{fig:dis_n3lo_ns} 
for some representative diagrams.
At this order, another flavor class, $FC_{11}$, can be present.
The latter collects all the diagrams where one boson is attached to a 
quark loop and the other one to an open quark line.
These diagram contribute with a weight given by ${\rm Tr}(g^{(j)}_{k}) g^{(j)}_{k}$. 
Finally, gluon initiated contributions always couple with the average coupling of the
quarks running in the loop, so $FC^{g}_{2}$ and $FC^{g}_{11}$ diagrams
weight ${\rm Tr}(g^{(j),2}_{k})$ and ${\rm Tr}(g^{(j)}_{k})^2$ respectively. 

The $FC_{2}$ diagrams can contribute to the
projections along the $\lambda_a$ generators of $SU(n_f)$
symmetry, and their associated coefficient functions are usually 
called \textit{non-singlet}, $C^{j}_{i,ns}$; 
instead, the loop suppress contributions from $FC_{02}$ are called 
\textit{pure-singlet}, $C^{j}_{i,ps}$ 
(or \textit{pure-valence}, $C^{j}_{i,pv}$ for parity violating structure functions),
as they contribute the same for all quark flavor lines. 
By contrast, $FC_{11}$ diagrams do not follow this classification. 
Eventually these diagrams are fully proportional to the QCD constants 
$(d^{a}_{bc})^2$ and can be easily isolated.
Since they contain loops coupled with a single electroweak boson, 
this flavor class cannot contribute to CC DIS.

Starting from the NC structure functions, $j\in \{\gamma\gamma,\gamma Z,Z\gamma,ZZ\}$, we find the following decomposition
\begin{align}
    F^{j}_i = 
        & C^{FC_2}_{i,ns} \otimes \sum_{k=1}^{n_f} g^{(j),2}_{k} \ xq^+_k 
            + \braket{g^{(j),2}} \left[ C^{FC_{02}}_{i,ps} \otimes x\Sigma + C^{FC_2}_{i,g} \otimes \ xg \right] \nonumber \\
        & + C^{FC_{11}}_{i,q} \otimes \braket{g^{(j)}} \sum_{k=1}^{n_f} g^{(j)}_{k} \ xq^+_k 
            + \braket{g^{(j)}}^2 C^{FC_{11}}_{i,g} \otimes \ xg 
        \, ,\quad i \in \{2,L\} 
        \label{eq:F2L_NC_decomposition} \\
    x F^{j}_3(x) = 
        & C^{FC_2}_{i,ns} \otimes \sum_{k=1}^{n_f} g^{(j),2}_{k} \ xq^-_k 
            + \braket{g^{(j),2}} C^{FC_{02}}_{i,pv} \otimes xV \nonumber \\
        & + C^{FC_{11}}_{3,q} \otimes  \braket{g^{(j)}} \sum_{k=1}^{n_f} g^{(j)}_{k} \ xq^-_k \,,
        \label{eq:F3_NC_decomposition}
\end{align} 
%
where we have normalized the structure functions such that they are always convolved with 
$x f(x)$ and for the averaged couplings are summed over all the active flavors.
Lastly, we note that $FC_{11}$ diagrams are not symmetric (see \cref{fig:dis_n3lo_ns} middle), 
thus when summing up all the different electroweak channels 
in the NC case one has to consider the case $j = \gamma Z$ 
and $j=Z \gamma$ separately.

For CC, the combination $W^{+} + W^{-}$ leads to diagrams which have the same topology
as the NC $FC_{2}$. It is then convenient to categorize the coefficient functions for the combinations
$C^{+}_{i,k}=C^{W^+}_{i,k}+ C^{W^-}_{i,k}$ and $C^{-}_{i,k}=C^{W^+}_{i,k}-C^{W^-}_{i,k}$
\cite{Davies:2016bwb}.
We obtain
\begin{align}
    C^{+}_{i,k} &= C^{FC_{2}}_{i,k} \quad \text{for} \quad k = ns,\ g,\ ps,\ pv\,, \label{eq:CCp} \\
    C^{+}_{i,k} &= C^{FC_{02}}_{i,k} \quad \text{for} \quad k = ps,\ pv\,, \label{eq:CCp_02} \\
    C^{-}_{i,k} &= 0 \quad \text{for} \quad k = g,\ ps,\ pv\,, \label{eq:CCm}
\end{align} 
while $C^{-}_{i,ns}$ are genuinely different. 
\cref{eq:CCm} originates from $g$, $\Sigma$ and $V$ being coupled the same 
to $W^{\pm}$.
In the $W^{+} + W^{-}$ case, the CC structure functions have the same flavor decomposition 
as \cref{eq:F2L_NC_decomposition,eq:F3_NC_decomposition} with different couplings 
for up and down quark types; while in $W^{+} - W^{-}$ the parity conserving structure
functions depends on $d_k^{-}$ and $u_k^{-}$, and the parity violating 
on $d_k^{+}$ and $u_k^{+}$.
\begin{allowdisplaybreaks} 
\begin{align}
    F^{W^+ + W^{-}}_i 
        & = C^{+}_{i,ns} \otimes \left( \sum_{d_j} \sum_{u_k} |V_{u_k,d_j}|^2 x d_j^{+} + \sum_{u_j} \sum_{d_k} |V_{d_k,u_j}|^2 x u_j^{+} \right) \nonumber \\
        & + \sum_{d_j} \sum_{u_k} |V_{u_k,d_j}|^2 \left[ C^{+}_{i,ps} \otimes x\Sigma + C^{+}_{i,g} \otimes \ xg \right] \, ,
    \quad i \in \{2,L\} \label{eq:F2L_CCp_decomposition} \\
    x F^{W^+ + W^{-}}_3(x) 
        & = C^{+}_{3,ns} \otimes \left( \sum_{d_j} \sum_{u_k} |V_{u_k,d_j}|^2 x d_j^- + \sum_{u_j} \sum_{d_k} |V_{d_k,u_j}|^2 x u_j^- \right)  \nonumber \\
        & + \sum_{d_j} \sum_{u_k} |V_{u_k,d_j}|^2 C^{+}_{3,pv} \otimes xV \, , \label{eq:F3_CCp_decomposition}
\end{align} 
\end{allowdisplaybreaks}
\begin{allowdisplaybreaks} 
\begin{align}
    F^{W^+ - W^{-}}_i & = C^{-}_{i,ns} \otimes \left( \sum_{d_j} \sum_{u_k} |V_{u_k,d_j}|^2 x d_j^- -  \sum_{u_j} \sum_{d_k} |V_{d_k,u_j}|^2 x u_j^- \right) 
    \quad i \in \{2,L\} \label{eq:F2L_CCm_decomposition} \\
    x F^{W^+ - W^{-}}_3(x) & = C^{-}_{3,ns} \otimes \left( \sum_{d_j} \sum_{u_k} |V_{u_k,d_j}|^2 x d_j^+ - \sum_{u_j} \sum_{d_k} |V_{d_k,u_j}|^2 x u_j^+ \right)
    \label{eq:F3_CCm_decomposition}
\end{align} 
\end{allowdisplaybreaks}
%

\paragraph{Polarized coefficient functions.}

The case of polarized structure functions follows exactly 
the same factorization of \cref{eq:coefficient_functions}, where coefficient and 
PDFs are replaced by the suitable polarized counterparts: 
$C^{j}_{i,k} \to \Delta C^{j}_{i,k}$ and $f_k \to \Delta f_k$.
The flavor decomposition can be divided in again according to the 
parity type, for instance $2x g_1$ has the same decomposition as 
$F_2$, while $g_4$ follows $xF_3$.
The coefficients of $g_1$ are known up to three loops \cite{Blumlein:2022gpp}.
Symmetry considerations lead to the map 
between the $FC_{2}$ \textit{non-singlet} coefficients \cite{Zijlstra:1993zs,Borsa:2022irn}
\begin{align}
    \Delta C^{b}_{1,ns} & = C^{b}_{3,ns}\,, \label{eq:delta_1ns} \\
    \Delta C^{b}_{4,ns} & = C^{b}_{2,ns}\,, \label{eq:delta_4ns} \\
    \Delta C^{b}_{L,ns} & = C^{b}_{L,ns}\,, \label{eq:delta_Lns}
\end{align} 
which holds both for NC $b \in FC_{2}$ and CC currents combinations, $b \in \{+,-\}$.
However, these relations are not valid for the other flavor classes 
and in particular for $FC_{11}$ which at N$^3$LO spoils the
\textit{non-singlet} sector symmetry. The full three-loop corrections
to $g_4$ and $g_L$ are not yet known.

Finally, we note that the NC structure function $g_1$ is the only one
of phenomenological relevance and for which experimental measurements are available.


\section{Heavy Quark treatment}
\label{sec:heavy_quarks}

So far we have considered a scattering process where 
all the initial and final state quarks were massless. 
However, at scales $Q \approx m_c, m_b$ the heavy quark mass 
effects cannot be neglected when describing proton interactions.
Again focusing on DIS we sketch how coefficient functions 
have to be modified to take into account of mass effects (\cref{sec:heavy_quarks_dis}),
and then in \cref{sec:fns} we show how different QCD schemes 
can be combined to properly describe the real case of multiple heavy quarks (charm and bottom).

\subsection{Mass effects in DIS}
\label{sec:heavy_quarks_dis}
%

%

Heavy quark contributions to DIS may be treated in a decoupling scheme~\cite{Collins:1978wz}, 
in which heavy quarks do not contribute, neither to the running of $a_s$, and neither to PDF evolution, 
but coefficient functions acquire a dependence on the heavy quark mass $m_h$.
We define $H^{j}_{i,k}=H^{j}_{i,k}\left(y,Q,\mu_f,m^2_h\right)$ 
as the coefficient function originating from diagrams where the heavy quark couples 
to the virtual gauge boson.
This definition is infrared-safe. However, a naive definition 
of $H^{j}_{i,k}$ based on the tag of heavy final state  would not be theoretically 
sound, due to gluon splitting.
Depending on whether one considers massive partons in the initial or final state,
there are several modifications that need to be introduced in the DIS factorization 
\cref{eq:coefficient_functions}.

First let us consider the process, often called heavy quark open production,
\begin{equation}
  \ell(k) + p(P) \to \ell'(k') + h + X \,,
  \label{eq:massive_dis_def}
\end{equation}
where we require to have a massive quark $h$ with mass $m_h$ in the final state. 
The presence of a massive particle requires to impose a kinematic cut in the 
convolution range with the PDF in \cref{eq:coefficient_functions}, which 
now reads for massless to massive structure functions
\begin{equation}
  \left.\hat{F}^{j}_i(x, Q)\right|_{h} 
    = \sum_{k=q,\bar{q},g} \int_x^{x_{max}} \frac{dy}{y} 
      H^{j}_{i,k}\left(y,Q, \mu_f,m^2_h \right) 
      f_k\left( \frac{x}{y}, \mu_F \right)
  \label{eq:massive_coefficient_functions}
\end{equation}
with $x_{max} = (1 + 4 m_h^2 / Q^2)^{-1}$ for NC, where the heavy quarks are created 
in pairs and $x_{max} = (1 + m_h^2 / Q^2)^{-1}$ for CC.
In the former case, the process cannot occur at LO
and at NLO is driven by the gluon channel, while the latter starts at LO.
\footnote{
  Recall that we adopt an \textit{absolute} definition of perturbative order, 
  i.e., LO $=O(a_s^0)$ irrespective of the first non-zero order, 
  e.g.\ for $F_L$ or $F_2^{(c)}$.
}
The massive coefficient functions are known exactly up to NNLO for 
photon~\cite{Laenen:1992zk,Alekhin:2003ev}, Z~\cite{Hekhorn:2018ywm,Hekhorn:2019nlf} 
and W~\cite{Gluck:1996ve,Gao:2017kkx} exchange while at N$^3$LO only
partial results are available in Ref.~\cite{Laenen:1998kp,Catani:1990eg,Kawamura:2012cr,bbl2023},
or in the $Q^2 \gg m_h^2$ limit~\cite{
  Bierenbaum:2009mv,Ablinger:2010ty,Ablinger:2014vwa,Ablinger:2014nga,Behring:2014eya}
(see also \cref{sec:an3lo_dis}).
Massive coefficients differ from the massless ones also because 
of the parity structure. In fact, in the case of Z boson exchange
and parity conserving structure functions, the axial vector-axial vector 
coefficient function is no longer equal to the vector-vector piece 
which contributes both to the photon and Z exchange.

The case of initial massive partons is more involved; the factorization 
formula requires also to sum over the PDF of the massive quark
which is then called \textit{intrinsic}. 
In this case, we add a further contribution to \cref{eq:massive_coefficient_functions} 
given by
\begin{equation}
  \left.\hat{F}^{j}_i(x, Q)\right|_{f_h} 
    = \sum_{k=h,\bar{h}} \int_\chi^{1} \frac{dy}{y} 
      H^{j}_{i,k}\left(y,Q, \mu_f,m^2_h \right) 
      f_k\left( \frac{\chi}{y}, \mu_F \right)
  \label{eq:ic_massive_coefficient_functions}
\end{equation}
with the convolution point now shifted to
\begin{equation}
  \chi_{NC} = \frac{x(1 + \sqrt{ 1 + 4 m_h^2 /Q^2})}{2}, 
  \quad 
  \chi_{CC} = \frac{x(1 + \sqrt{ 1 + m_h^2 /Q^2})}{2}, 
  \label{eq:chi_def}
\end{equation}
for NC and CC respectively.
The intrinsic coefficient functions are available only up to NLO \cite{Kretzer:1998ju}, 
with the CC part computed very recently \cite{Candido:2024rkr,Kirill}.
Intrinsic polarized coefficient functions are not yet known.
In \cref{tab:coef-funcs,tab:pol-coeff-funcs} we collect a summary of 
the coefficient functions as currently available in literature and
implemented in \yadism~(see \cref{sec:yadism}),
both for the polarized and unpolarized case.
For each perturbative order and process, we distinguish contributions from light-to-light (light), 
light-to-heavy (heavy) and heavy-to-light or heavy-to-heavy (intrinsic) coefficients.

\begin{table}[!t]
  \centering
  \renewcommand{\arraystretch}{1.60}
  
\begin{tabular}{c|c c c}
    \toprule
    NLO $O(a_s)$ 
    & light 
    & heavy 
    & intrinsic
    \\
    \hline
    NC & \grokcell\cite{Vermaseren:2005qc,Moch:2004xu,Moch:1999eb} 
       & \grokcell\cite{Hekhorn:2019nlf} 
       & \grokcell\cite{Kretzer:1998ju}
    \\
    CC 
      & \grokcell \cite{Moch:2007rq,Moch:2008fj} 
      & \grokcell \cite{Gluck:1996ve} 
      & \grokcell \cite{Candido:2024rkr,Kirill} 
    \\
    \midrule
    NNLO  $O(a_s^2)$  & & &\\
    \hline
    NC 
      & \grokcell\cite{Vermaseren:2005qc,Moch:2004xu,Moch:1999eb} 
      & \grokcell\cite{Hekhorn:2019nlf} 
      & \rdxcell 
    \\
    CC 
      & \grokcell\cite{Moch:2007rq,Moch:2008fj} 
      & \ylcell \cite{Gao:2017kkx}
      & \rdxcell 
    \\
    \midrule
    N$^3$LO $O(a_s^3)$ & & &\\
    \hline
    NC 
      & \grokcell\cite{Vermaseren:2005qc,Moch:2004xu,Moch:1999eb} 
      & \grokcell\cite{Kawamura:2012cr,Laurenti:2024anf,bbl2023} 
      & \rdxcell 
    \\
    CC 
      & \grokcell\cite{Moch:2007rq,Moch:2008fj} 
      & \rdxcell 
      & \rdxcell 
    \\
    \bottomrule
\end{tabular}

  \vspace{0.3cm}
  \caption{
    Overview of the unpolarized DIS coefficients currently available in literature
    at the corresponding order in perturbative QCD.
    In the columns we distinguish between light, heavy and intrinsic. 
    We mark in green coefficient function that are implemented in \yadism~(see \cref{sec:yadism}), 
    in red the ones which are not yet known and in yellow the ones 
    which are not yet implemented in \yadism, but available in literature.
     }
  \label{tab:coef-funcs}
\end{table}
\begin{table}[!t]
  \centering
  \renewcommand{\arraystretch}{1.60}
  
\begin{tabular}{c|c c c}
    \toprule
    & light 
    & heavy 
    & intrinsic 
    \\
    \hline
    NLO  $O(a_s)$ 
      & \grokcell\cite{Zijlstra:1993sh,deFlorian:1994wp,Anselmino:1996cd} 
      & \grokcell\cite{Hekhorn:2019nlf} 
      & \rdxcell 
    \\
    \midrule
    NNLO  $O(a_s^2)$ 
      & \grokcell\cite{Zijlstra:1993sh,Borsa:2022irn} 
      & \grokcell\cite{Hekhorn:2019nlf} 
      & \rdxcell
    \\
    \midrule
    N$^3$LO $O(a_s^3)$ 
      & \ylcell\cite{Blumlein:2022gpp}\ftm{1} 
      & \rdxcell 
      & \rdxcell
    \\
    \bottomrule
\end{tabular} \\
{
    \footnotesize
    \renewcommand{\arraystretch}{1.2}
    \begin{tabular}{r l}
      \ftm{1} & Only for the $g_1$ structure function.\\
    \end{tabular}
}

  \vspace{0.3cm}
  \caption{
    Same as \cref{tab:coef-funcs} for NC polarized coefficients.
  }
  \label{tab:pol-coeff-funcs}
\end{table}

Lastly, we can note that when including NNLO or higher corrections,
massive quarks can also contribute to \cref{eq:massive_dis_def}
with diagrams where the electroweak boson is coupling to 
a light fermionic line.
Such contributions, are not included in the heavy coefficient 
function, but need to be taken into account in the total structure
function.
Currently, they are known only up to NNLO \cite{Hekhorn:2019nlf}.
In summary when considering heavy quarks effects the total inclusive structure
functions are given by an incoherent sum over all the light, 
heavy and intrinsic partons contributions. 
Tagging a final state heavy quark is in principle not sufficient to disentangle the 
heavy contributions from the massless ones.


\subsection{Flavor Number Schemes}
\label{sec:fns}

As mentioned in the previous section a fully massive DIS 
coefficient functions contains terms $\mathcal{O}(m_h^2/Q^2)$ 
which are neglected in the massless case.
In particular, these terms are relevant to describe experimental measurements 
in the threshold region $Q^2 \approx m_h^2$, where the heavy quark mass can 
be either charm or bottom. 
On the other hand in this scheme, the quasi-collinear divergences of the heavy
quark splitting generates terms $\ln(Q^2/m_h^2)$, which can become large for 
$Q^2 \gg m_h^2$ and are not resummed through the DGLAP evolution spoiling perturbation convergence. 

The choice of the number of active flavor $n_f$ and the heavy quarks thresholds
$\mu_h$ at which a heavy quark becomes active define a so-called flavor number 
scheme (FNS).
In principle the scale $\mu_h$ can be different from the actual quark mass,
although it is common to set them to the same value.
Different FNS choices are possible depending upon the heavy quarks being treated as light 
($\mu_h=0$), heavy ($\mu_h$ finite) or decoupled ($\mu_h=\infty$).
We refer to the massive scheme with $n_f$ active light flavor and one heavy 
quark as Fixed Flavor Number Scheme (henceforth denoted by FFNS$n_f$).
To properly include massive effects, without spoiling the
high-$Q^2$ limit we need to introduce a variable flavor number scheme 
(VFNS), which combines different FNS depending on the considered scale.

\paragraph{Matching conditions.}
Since all the anomalous dimensions associated to the running of renormalized 
quantities depends explicitly on $n_f$, when crossing a heavy quark threshold,
there can be discontinuities. In order to recover the proper transition between
FNS one needs to introduce matching conditions~\cite{Collins:1986mp} 
connecting quantities in the $n_f+1$ scheme to $n_f$ one.
Such matching conditions can also be computed in perturbation theory and have to be
included consistently. 
Starting from the strong couplings $a_s$ one finds the relation
\begin{allowdisplaybreaks} 
\begin{align}
    \begin{split}
    a_s^{(n_f+1)}(\mu_h^2) &= d^{(n_f+1)}\left( a_s^{(n_f+1)}(\mu_h^2), \ln(\frac{\mu_h^2}{m_h^2}) \right) a_s^{(n_f)}(\mu_h^2)  \\
        &=\sum_{k=0}^{\infty} \left(a_s^{(n_f+1)}(\mu_h^2)\right)^k d^{(k),(n_f+1)} \left(\ln^k(\frac{\mu_h^2}{m_h^2}) \right) a_s^{(n_f)}(\mu_h^2), 
   \label{eq:asmatching}    
    \end{split}
\end{align} 
\end{allowdisplaybreaks}
where the decoupling constants $d^{(k),(n_f)}$ are known up to 4-loop~\cite{Chetyrkin:2005ia,Schroder:2005hy},
and their dependency on the quark mass is only through $\ln(\frac{\mu_h^2}{m_h^2})$.

Similarly, for the PDFs it holds
\begin{allowdisplaybreaks} 
\begin{align}
    \begin{split}
    f_i^{(n_f + 1)}(\mu_h^2) &= A_{ij}^{(n_f+1)}\left( a_s^{(n_f+1)}(\mu_h^2), \ln(\frac{\mu_h^2}{m_h^2}) \right) \otimes f_i^{(n_f)}(\mu_h^2) \\
     &= \sum_{k=0}^{\infty} \left(a_s^{(n_f+1)}(\mu_h^2)\right)^k A_{ij}^{(k),(n_f+1)}\left(\ln(\frac{\mu_h^2}{m_h^2})\right) \otimes f_j^{(n_f)}(\mu_h^2) 
     \label{eq:pdfmatching}
    \end{split}
\end{align} 
\end{allowdisplaybreaks}
where the coefficients of the matching matrix $A_{ij}^{(k),(n_f)}$ 
are known up to 3-loop order~\cite{Buza:1996wv,
    Bierenbaum:2009zt,Bierenbaum:2009mv,
    Ablinger:2010ty,Ablinger:2014vwa,Ablinger:2014uka,Behring:2014eya,
    Ablinger_2014,Ablinger:2014nga,Blumlein:2017wxd,Ablinger_2015,Ablinger:2022wbb,Ablinger:2023ahe,
    Ablinger:2024xtt} 
for $j\in \{q,g\}$ and 1-loop for $j = h$ (heavy-initial state)~\cite{Ball:2015tna}
These matrix elements are the partonic expectation values of the 
renormalized local twist-2 operators
\begin{equation}
    A_{ij} = \bra{j(p)} O_i \ket{j(p)}, \quad i,j = q,g
    \label{eq:ome_def}
\end{equation}
%
and relate to how the massive partonic coefficient functions
of \cref{eq:massive_coefficient_functions} factorize in terms of the massless one 
in the $Q^2 \gg \mu_h^2$ limit
\begin{equation}
    H_{k,i}\left(a_s(\mu_h^2), \ln(\frac{\mu_h^2}{m_h^2})\right) =  C_{k,i}\left(a_s(\mu_h^2)\right) \otimes A_{ij}\left(a_s(\mu_h^2),\ln(\frac{\mu_h^2}{m_h^2})\right), \quad k = 2,L,3.
    \label{eq:coeff_asy}
\end{equation}
Their flavor decomposition can be expressed in terms of the evolution basis PDFs 
as
\begin{equation}
    \begin{pmatrix} g^{(n_f+1)} \\ \Sigma^{(n_f+1)} \\ h^{+,(n_f+1)} \end{pmatrix} = 
    \begin{pmatrix} A_{gg,h} & A_{gq,h} & A_{gh} \\ 
                    A_{qg,h} & A_{qq,h}^{ps} + A_{qq,h}^{ns} & A_{qh}^{ps} \\ 
                    A_{hg} & A^{ps}_{hq} & A_{hh}^{ps} + A_{hh}^{ns} \end{pmatrix} \otimes
    \begin{pmatrix} g^{(n_f)} \\ \Sigma^{(n_f)} \\ h^{+,(n_f)} \end{pmatrix}
    \label{eq:singlet_matching}
\end{equation}
\begin{equation}
    \begin{pmatrix} V^{(n_f+1)} \\ h^{-,(n_f+1)} \end{pmatrix} = 
    \begin{pmatrix} A_{qq,h}^{ns} & 0 \\ 0 & A_{hh}^{ns} \end{pmatrix} \otimes
    \begin{pmatrix} V^{(n_f)} \\ h^{-,(n_f)} \end{pmatrix}
    \label{eq:ns_matching}
\end{equation}
where for the light-to-light element the underscore $h$ denotes that at least
one heavy fermion line is present in the corresponding diagram.
At LO only the diagonal element are present, 
the NLO corrections contribute to $A_{gg,h},A_{gh},A_{hg},A_{hh}^{ns}$, 
while $A_{gq,h},A^{ps}_{hq},A_{qh}^{ps},A_{hh}^{ps}$ start at NNLO, with also
$A_{qq,h}^{ns}$ receiving $\mathcal{O}(a^2)$ corrections.
The other entries $A_{qq,h}^{ps},A_{qg,h}$ are $\mathcal{O}(a^3)$.
\footnote{
    Since the heavy initial state matrix elements are known 
    only up to NLO, we always set $A_{hh}^{ps},A_{qh}^{ps}$ to $0$.
}

Polarized matching follows the same structure, with all the 2-loop \cite{Bierenbaum:2022biv} and 
3-loop light initiated contributions known.

\paragraph{Zero-Mass VFNS.}
The simplest VFNS construction is the Zero-Mass VFNS 
(ZM-VFNS): quarks are assumed decoupled below their respective threshold and
light above. 
%
%
The scheme resums all logarithmic corrections as they are provided by DGLAP evolution,
but it does not contain any power-like heavy quark corrections $m^2/Q^2$ which 
may be phenomenological important in certain regions of the kinematic phase space.
ZM-VFNS usually generates unphysical discontinuities where the scale $Q$ 
equals any of the heavy quark masses. However, as all the power suppressed 
corrections are neglected all the perturbative calculations 
simplify significantly.

\paragraph{The FONLL scheme.}
A more refined prescription is the FONLL scheme~\cite{Cacciari:1998it},
presented in Ref.~\cite{Forte:2010ta} for DIS observables, 
with a more recent implementation provided in~\cite{Barontini:2024xgu}.
The FONLL scheme enhances the fixed order calculation by the resummation of 
the pseudo-collinear logs, which can become arbitrary large.
The procedure can be applied either to one or more heavy quark thresholds, 
and generalization to the polarized case are possible, provided that the massive 
calculation are available~\cite{Hekhorn:2024tqm}.
In practice, one combines FFNS$n_f$ and FFNS$n_1+1$ while taking care of the 
double counting and define
\begin{equation}
    F^\mathrm{FONLL}(Q^2,m_h^2) = F^{(n_f)}(Q^2,m_h^2) + F^{(n_f+1)}(Q^2) - F^{(n_f\cap n_f+1)}(Q^2,m_h^2) 
    \label{eq:FONLL_m1}
\end{equation}
where the intersection operation $\cap$ indicate the overlap between the two schemes.

Calculations performed in a decoupling scheme with $n_f$ light quarks retain the 
full dependence on the heavy quark mass and include the contribution 
of heavy quarks at a fixed perturbative order (FFNS$n_f$). 
Calculations performed in a scheme in which the heavy quark is treated as massless 
(FFNS$n_1+1$), and endowed with a PDF that satisfies perturbative matching conditions,
resums logarithms of $Q^2/m_h^2$ to all orders through the running of
the coupling and the evolution of PDFs, but does not include terms that
are suppressed as powers of $\frac{m_h^2}{Q^2}$. 
In the r.h.s of \cref{eq:FONLL_m1}, each component obeys factorization and 
is thus given by a convolution between a PDF $f$ and a coefficient function $C$, 
given in the corresponding scheme.
The intersection coefficient is given by \cref{eq:coeff_asy}
and contains only the massive pseudo-collinear terms and has to be 
convolved with an $n_f$ PDF set.
This construction reduces to the decoupling calculation for $Q^2 \approx m_h^2$
and to the massless one for $Q^2 \gg m_h^2$.

When computing \cref{eq:FONLL_m1} multiple approaches are possible:
in the original publication \cite{Forte:2010ta}, all the terms
are rewritten analytically in the FFNS$n_f+1$ scheme such that a 
single PDF convolution is needed. However, this approach is not optimal 
at higher orders where the matching condition expressions become 
more involved and their explicit inversion can be complex;
in Ref.~\cite{Barontini:2024xgu} we show how \cref{eq:FONLL_m1} can be implemented
more easily evolving the different PDFs in their respective 
schemes and joining the different pieces only at structure function level.
The latter method is also more straightforward to extend in 
case of multiple thresholds (charm and bottom), where the procedure 
is applied iteratively, while the method of Ref.~\cite{Forte:2010ta} would 
require re-expressing the massive scheme PDFs into massless 
scheme PDFs twice.

\section{Proton-proton collisions}
\label{sec:double_hadronic}

DIS is the simplest scattering process that can be used to study the
structure of the nuclei. 
However, for various experimental and theoretical reasons, 
such as reaching higher center of mass energies or trying to probe Higgs boson
production, physicist started to investigate also hadron-hadron collisions.
In particular, measurements from Tevatron and LHC, on $p\bar{p}$ and $pp$ scattering
respectively, provide nowadays a vast amount of accurate data that can be
described through the pQCD and PDFs formalism.
Thanks to their universality, we can relate PDFs to the inclusive cross-section 
of double hadron scattering through a generalization of the DIS factorization 
theorem (cf. \cref{eq:dis_factorization}).
In this case, the formula entails a convolution of the process dependent partonic matrix 
element with two PDFs 
\begin{equation}
    \sigma(x, Q) = \sum_{i,j=q,\bar{q},g} \int_{x_1}^1 \frac{dy_1}{y_1} \int_{x_2}^1 \frac{dy_2}{y_2} 
        \hat{\sigma}_{ij} \left(y_1, y_2, Q, \mu_F \right) 
        f_i\left( \frac{x_1}{y_1}, \mu_F\right) f_j\left( \frac{x_2}{y_2}, \mu_F\right)  
        + \mathcal{O}\left(\frac{\Lambda^p}{Q^p} \right) \, .
    \label{eq:hadronic_factorization}
\end{equation}

Simple hadronic processes as single gauge boson production 
(often called Drell-Yan or DY) or inclusive jet production can provide essential constraints 
on the flavor separation and or the gluon PDF.
%
%
Recently, also other hadronic processes like single-$t$, $t\bar{t}$, and
prompt photon have been used during PDF fits, but their impact is not as competitive 
as DIS, jets and DY.

From the theoretical point of view, the computation of double hadronic cross-sections 
is more demanding than the DIS case, and it is usually available only at NLO or NNLO 
in QCD.
The presence of multi particle phase space complicates further the calculations.
Typically, fully analytical computations are not feasible beyond NLO and thus 
Monte Carlo methods are used to sample complex integrals.

Up to know, for technical limitations, PDF independent computations
of hadronic observables $\hat{\sigma}$, which are crucial in PDF 
fitting (see \cref{sec:theory_methodology}), are only available at NLO. 
Therefore, a $K$-factors approximation it is used to grasp the NNLO effect.
This is achieved by rescaling the NLO cross-section as follows
\begin{equation}
    \sigma_{\text{NNLO}} \approx K_{\text{NNLO}}^{\text{QCD}} \sum_{ab} 
        \hat{\sigma}^{\text{NLO}}_{ab} \mathcal{L}_{ab} \,.
    \label{eq:nnlo_qcd_pred}
\end{equation}
$\mathcal{L}_{ab}$ indicates the parton luminosity defined following 
Ref.~\cite{Mangano:2016jyj} as 
\begin{equation}
    \mathcal{L}_{ab}(m_X)= \frac{1}{s}\int_{\tau}^1 
        \frac{dx}{x}f_a \lp x,m_X^2\rp f_b \lp \tau/x,m_X^2\rp 
        \, ,\qquad 
        \tau=\frac{m_X^2}{s} 
        \, ,
    \label{eq:lumi_def}
\end{equation}
where $a,b$ label the species of incoming partons, $s$ is the center-of-mass energy 
of the hadronic collision, and $m_X$ is the final state invariant mass.
Finally, the $K$-factor is given computing the ratio
\begin{equation}
    K_{\text{NNLO}}^{\text{QCD}} = \frac{
        \sum_{ab} \hat{\sigma}^{\text{NNLO}}_{ab} \otimes \mathcal{L}^{\text{NNLO}}_{ab}
    }{
        \sum_{ab} \hat{\sigma}^{\text{NLO}}_{ab} \otimes \mathcal{L}^{\text{NNLO}}_{ab}
    }
    \label{eq:qcd_kf_def}
\end{equation}
where the luminosities are evaluated with NNLO PDFs both in the numerator and 
in the denominator.
This method is computationally advantageous, and it is motivated by the 
factorization of the QCD correction into the hadronic matrix element and 
the PDF evolution.
However, as a downside, different partonic channels $ab$, in \cref{eq:nnlo_qcd_pred}, 
are weighted in with the same $K$-factor, although this is assumption is not 
always well justified.
Current effort are ongoing in the HEP phenomenology community to overcome 
this issue and make full NNLO, PDF independent, computations public. 

We now conclude with a brief recap of the different kinematic variables 
for Drell-Yan and Jet production.

\paragraph{Single Electroweak bosons production.}

\begin{figure}[!t]
    \includegraphics[trim={0 4cm 0 4cm},clip]{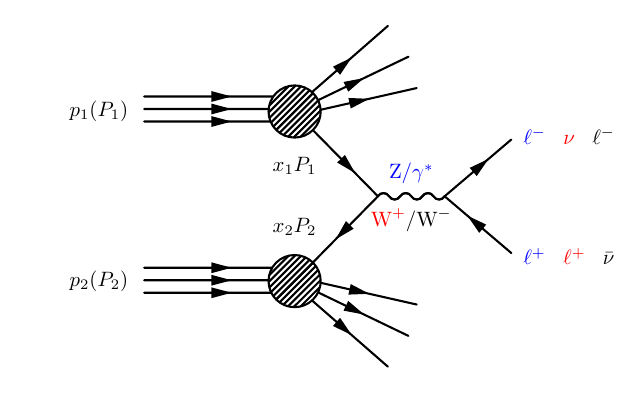}
    \caption{
        Representative Feynman diagram associated to Drell-Yan 
        production $p_1(P_1) + p_2(P_2) \to \ell + \ell' + X$. 
        Neutral bosons decays into $\ell^+,\ell^-$ pairs while
        $W^+ \to \ell^+ + \nu$ and  $W^- \to \ell^- + \bar{\nu}$-
    }
    \label{fig:dy_hadronic}
\end{figure}

Very precise observables in hadron-hadron collision are given by the 
Drell-Yan (DY) process. 
Similarly to DIS, single electroweak boson production can be classified 
in two different channels depending on exchanged boson which are then observed 
mainly through the leptonic decays. 
\begin{align}
    p_1(P_1) + p_2(P_2) & \to Z / \gamma \to \ell^+ + \ell^- + X
    \label{eq:NC_dy} \\
    p_1(P_1) + p_2(P_2) & \to W^{\pm} \to \ell + \nu + X
    \label{eq:CC_dy}
\end{align} 
The inclusive cross-section are usually measured as a function of the invariant 
mass or the rapidity of the charged lepton (for CC, $W^\pm$) 
or lepton pairs (for NC, $Z/\gamma$) reconstructed in the final state. 
At LO these kinematic variables are related to the PDFs by:
\begin{align}
    x_{1,2} & = \frac{M_{\ell\ell}}{\sqrt{s}} e^{\pm y_{\ell \ell}} , \label{eq:dy_x1x2_def} \\
    M_{\ell \ell}^2 & = x_1 x_2 s , \label{eq:dy_m2_def} \\
    s & = (P_1 + P_2)^2 , \\
    y_{\ell \ell} & = \frac{1}{2} \ln \left( \frac{E_{\ell \ell} + p_{{\ell \ell},z}}{E_{\ell \ell} - p_{{\ell \ell},z}} \right) .
    \label{eq:dy_rap_def}
\end{align} 
A representative QCD LO Feynman diagram is shown in \cref{fig:dy_hadronic};
we observe that this process is mainly driven by the $q \bar{q}$ or $q q$ channels for
NC and CC providing additional sensitivity to the quark flavor separation.

\paragraph{Jet and Dijet production.}

\begin{figure}[!t]
    \includegraphics[trim={0 4cm 0 4cm},clip]{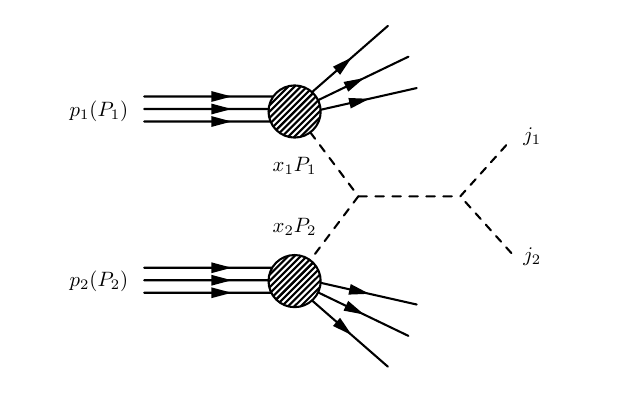}
    \caption{        
        Representative Feynman diagram associated to
        dijet production $p_1(P_1) + p_2(P_2) \to j_1 + j_2 + X$.
        The dashed line indicates any possible strong interaction
        diagram leading to $q$,$\bar{q}$ or $g$ final states.
    }
    \label{fig:jets_hadronic}
\end{figure}

Inclusive single jet and dijet production, \cref{fig:jets_hadronic}, are
among the processes with the largest cross-section in hadron-hadron
collisions.
They involve the reconstruction of one or two jets originated by any 
hard emission of $q$,$\bar{q}$ or $g$ in the partonic matrix element 
\begin{align}
    p_1(P_1) + p_2(P_2) & \to j_1 + X,
    \label{eq:jet} \\
    p_1(P_1) + p_2(P_2) & \to j_1 + j_2 + X .
    \label{eq:dijet}
\end{align} 
Here the characteristic scale of the process is given by the transverse momentum 
of the single jet or dijet system which is related to the proton momentum fraction 
via:
\begin{allowdisplaybreaks} 
\begin{align}
    x_{1,2} & = \frac{p_T}{\sqrt{s}} e^{\pm y} , \label{eq:jet_x1x2_def} \\
    p_T^2 & = x_1 x_2 s , \label{eq:jet_m2_def} \\
    s & = (P_1 + P_2)^2 , \\
    y & = \frac{1}{2} \ln \left( \frac{E + p_{z}}{E - p_{z}} \,. \right)
    \label{eq:jet_rap_def}
\end{align} 
\end{allowdisplaybreaks}
Since quark and gluon initiated jets are experimentally indistinguishable, 
this process at high energy is dominated by the $gg$ and $gq$ channels, providing important
constraints to the large-$x$ gluon PDF.

  \chapter{Tools and methodology}
\label{chap:methodology}
\vspace{-0.5cm}
\begin{center}
\begin{minipage}{1.\textwidth}
    \begin{center}
        \textit{
            This chapter is based on Ref.~\cite{NNPDF:2021njg} and my result presented in Refs.~\cite{Barontini:2022jci,Candido:2022tld,Candido:2024rkr,NNPDF:2024dpb}.}
    \end{center}
\end{minipage}
\end{center}


Perturbative QCD can be used to describe the behavior of high energy 
scattering processes involving hadrons as we have outlined in the previous
chapter.
This approach relies on the factorization theorem which allows 
to separate the perturbative from non-perturbative contributions by 
absorbing the latter into quantities such as Parton Distribution Functions (PDFs). 
Unfortunately, beyond the framework of lattice QCD, the functional form of
PDFs is not computable from first principles. The most common approach to overcome this 
issue entails parametrizing PDFs at a common given scale $Q_0$ and, by using DGLAP 
evolution, fitting them to experimental high energy data.
Historically, the first parametrizations of PDFs relied on fixed functional forms, 
while in the NNPDF approach, adopted in this thesis, PDFs are obtained by training a neural 
network, which can approximate any continuous function as dictated by the Universal 
Approximation Theorem \cite{balazs:2001}. 
During a PDF fit, different sources of uncertainties arise and have to be 
consistently propagated. These uncertainties, which can have both experimental 
and theoretical origins, will provide a bound on our PDFs estimate, 
which can systematically limit the accuracy of further computations
that make use of such PDFs. 
It is then clear that, to achieve accurate predictions for future colliders, 
as HL-LHC or EIC, both accurate pQCD calculations and reliable statistical tools
have to be adopted for PDF extraction.

In this chapter, we review the tools, assumptions and settings used within the NNPDF 
framework to fit PDFs and which constitute the starting setup used 
to derive the results presented in the next \cref{chap:ic,chap:an3lo,chap:pol}. 
Let us remark that the theoretical tools presented in the first section
are independent of and decoupled from the NNPDF methodology, which is outlined 
in the second and third parts of the chapter.
In particular, the former have also been adopted in other works which 
are not directly related to PDF fitting.

\paragraph{Outline.}

This chapter covers three rather independent topics, which 
are collected here to give a unitary description of the common 
working environment adopted during the studies of \cref{chap:ic,chap:an3lo,chap:pol}.

In \cref{sec:theory_methodology}, we describe the tools used 
to produce theory predictions of DIS and hadronic observables, able 
to constrain PDFs. In particular, we describe
two open-source code \eko~\cref{sec:eko} and \yadism~\cref{sec:yadism}
adopted respectively to perform DGLAP evolution and computing DIS coefficients.
In \cref{sec:nnpdf_methodology}, we summarize the treatment of 
uncertainties (both experimental and theoretical) during our PDF fits and 
outline the common aspects of the fitting methodology adopted both for polarized 
and unpolarized fits.
Finally, we conclude, in \cref{sec:nnpdf40}, by highlighting some features 
of the NNPDF4.0~\cite{NNPDF:2021njg} set including 
the recently published MHOU set \cite{NNPDF:2024dpb}. 
These sets constitute the baseline upon which the results of the 
following chapters are built.

\section{Theory predictions for PDF fitting}
\label{sec:theory_methodology}



Addressing the problem of PDF fitting requires integrating several 
elements from different sources: data from experiments - ranging 
over multiple decades and formats - and competitive theory predictions, 
coming from different providers. Moreover, a fitting methodology has to be 
selected and engineered to implement theory constraints, and to limit not 
physically motivated bias.
While data are a static component in the fit, the theory predictions depend 
on the candidate PDF, since, through the factorization theorem (cf. \cref{eq:hadronic_factorization}), 
they constitute the mapping that connects the unobserved PDF space, to the observed data space.

During a PDF fit, this map is evaluated repeatedly (at least 
once for every minimization step), so it is paramount to have an efficient way 
to evaluate it, otherwise it can become a serious bottleneck.
This issue has been solved by introducing an interface that is able to produce 
PDF independent theory predictions. Few examples are present in literature
\cite{Carli:2010rw,Britzger:2012bs,Britzger:2022lbf,Carrazza:2020gss} 
and, they are being used in various context. 
In particular, the convolution of \cref{eq:hadronic_factorization} is performed 
in three different spaces that needs be factored out: the flavor space, 
and the kinematic space of $x$ and $Q^2$.
The main concept of such interfaces is to split the prediction generator (usually a Monte Carlo generator
for hadronic processes) output into different luminosity components, perturbative orders, 
and observables binning. 
Essentially, this recast the partonic cross-sections predictions as a theory array (celled grid), 
for which the Mellin convolution is replaced by a linear algebra contraction over 
a single or multiple PDF sets.

However, this step is not exhaustive for the implementation of the PDF factorization.
In fact, while discretization on the luminosity and bins takes care of the flavor and $x$-space
dependency of the PDF, it leaves the dependence on the energy scale untouched. 
The latter dependence is not fitted, since it is only determined by perturbative QCD
through the DGLAP equation (\cref{eq:dglap}) and can be computed a priori.
Being DGLAP a set of integro-differential equations linear in the PDF, its solution can be 
converted in the application of a suitable evolution operator (\cref{eq:eko}).
It is possible to combine the two ingredients (the operator and the grid) in a single fast array 
interface, that directly produce the required theory predictions once contracted on the PDF 
candidate.
This way, the map from PDF space to data space discussed above, is reduced to a linear algebra product.
Such an interface is called a “Fast Kernel table” (shortened to FK-table) in the context of the 
NNPDF collaboration.

For example, given any observable $\mathcal{O}(Q)$, evaluated at the scale $Q$ and which depends 
linearly on the PDF (i.e. a DIS observable), the theory prediction is achieved via
\begin{equation}
    \mathcal{O}(Q) =
        \sum_{i} {\rm FK}_{i}\left (a_s(Q), a_{em}(Q), \frac{\mu_R}{Q}, \frac{\mu_F}{Q} \right ) 
        \otimes f_i(Q_0) \,, 
    \label{eq:dis_fks}
\end{equation}
\begin{allowdisplaybreaks} 
\begin{align}
\begin{split}
    {\rm FK}_{i} \left (a_s(Q), a_{em}(Q), \frac{\mu_R}{Q}, \frac{\mu_F}{Q} \right ) = \
        & {\rm G}_{i} \left (a_s(Q), a_{em}(Q), \frac{\mu_R}{Q} \right ) \\ 
        & \otimes \sum_{j} {\rm E}_{ij} \left (a_s(Q),a_s(Q_0), \frac{\mu_F}{Q} \right ) \,,
    \label{eq:fks_def}
\end{split}
\end{align} 
\end{allowdisplaybreaks}
where the convolutions are performed in the discretized $x$-space and the sums run over the flavor
space. 
${\rm FK}_i$ denotes the FK-table, ${\rm G}_i$ denotes the grid computed with a suitable generator, ${\rm E}_{ij}$ the DGLAP 
operator and $f_i$ the PDF evaluated at the arbitrary fitting scale $Q_0$. $\mu_F$ and $\mu_R$ 
are the unphysical factorization and renormalization scales.
By doubling all the flavor and $x$-space indices \cref{eq:dis_fks} can be generalized for the 
hadronic observables which depends quadratically on PDFs. 
During this procedure, there is no loss of generality, if the interpolation grid used for the 
conversion of the analytic convolutions is sufficiently dense. 
Moreover, since PDFs are non-perturbative objects, they are usually represented in terms of 
discrete grids~\cite{Buckley:2014ana}, called \lhapdf{} grids.

With the definition of \cref{eq:dis_fks,eq:fks_def} in mind, in the following
sections we explain the features of the two codes used to compute 
the DGLAP operators, \eko{} (\cref{sec:eko}), and the DIS grids, \yadism{} (\cref{sec:yadism}).
Finally, in \cref{sec:pineline} we outline how these programs are integrated
in a unique framework, called \pineline{} which is able to produce all the theoretical
calculation needed for a PDF fit. 

\subsection{EKO}
\label{sec:eko}
In this section, we highlight the most relevant features of \eko{} the
PDF evolution library adopted to produce the result of \cref{chap:ic,chap:an3lo,chap:pol}
as well as a number of additional works \cite{NNPDF:2024dpb,NNPDF:2024djq,Cooper-Sarkar:2024crx}. 
We refer to \cite{Candido:2022tld} for a more extensive presentation.

\eko{} solves the DGLAP evolution equations~\cref{eq:dglap} in Mellin space
allowing for simpler solution algorithms (both iterative and approximated).
Yet, it provides result in momentum fraction space to allow an easy 
interface with existing generator codes.

\eko{} computes evolution kernel operators (EKO) which are independent of
the initial boundary conditions but only depend on the given theory settings.
The operators can thus be computed once, stored on disk and then reused in the
actual application. This method can significantly speed-up when PDFs
are repeatedly being evolved, as it is customary in PDF fits.

\eko{} is open-source, allowing easy interaction with users and developers.
The project comes with a clean, modular, and maintainable codebase that guarantees
easy inspection and ensures it is future-proof.
We provide both a user and a developer documentation. 

\eko{} currently implements solution up to approximate N$^3$LO QCD, which are provided
with two different approximations (see also \cref{sec:an3lo_dglap}), 
and up to NNLO in QED with mixed QED$\otimes$QCD corrections.

\eko{} correctly treats intrinsic heavy quark distributions,
required for studies of the heavy quark content of the nucleon~(cf. \cref{chap:ic}).
While the treatment of intrinsic distributions in the evolution equations is
mathematically simple, as they decouple in a specific basis, their integration
into the full solution, including matching conditions, is non-trivial.
We also implement backward evolution, again including the non-trivial
handling of matching conditions.


\eko{} adopts Python as a programming language opting for a high-level 
language which is easy to understand for newcomers.
In particular, with the advent of Data Science and Machine Learning, Python has
become the language of choice for many scientific applications, mainly driven
by the large availability of packages and frameworks.
%
The code is developed mainly as a library, that contains
physics, math, and algorithmic tools, such as those needed for managing or
storing the computed operators. 


The full code documentation can be accessed at:
\begin{center}
    \url{https://eko.readthedocs.io/en/latest/}
\end{center}
This document is also regularly updated and extended upon the 
implementation of new features.
The code has been extensively benchmarked against the tables of Refs.~\cite{Giele:2002hx,Dittmar:2005ed} 
and with the program \apfel{}~\cite{Bertone:2013vaa} and \pegasus{}~\cite{Vogt:2004ns} 
as reported in \cite[Sec.~3.1]{Candido:2022tld} finding agreement
up to $\mathcal{O}(10^{-4})$ relative accuracy.
The following paragraphs describe some interesting options of the code.

\paragraph{Interpolation.}
Mellin space has the theoretical advantage that the analytical solution of the
DGLAP equations becomes simpler, but the practical disadvantage that it requires
PDFs in Mellin space. 
This constraint is in practice a serious limitation since most matrix element
generators~\cite{Buckley:2011ms} as well as the various generated coefficient
function grids are not using Mellin space, but rather $x$-space.


We are bypassing this limitation by introducing a Lagrange-interpolation~\cite{LagrangeInterpol,suli2003introduction} of the
PDFs in $x$-space on arbitrarily user-chosen grids $\mathbb G$:
\begin{equation}
    f(x) \sim \bar f(x) = \sum_{j} f(x_j) p_j(x),  \quad \text{with}\,x_j\in \mathbb G
\end{equation}
For the usage inside the library we do an analytic Mellin transformation of 
the polynomials $\tilde p_j(N) = \mathcal{M}[p_j(x)](N)$. For the interpolation polynomials 
$p_j$ we are choosing a subset with $N_{degree} + 1$ points of the interpolation 
grid $\mathbb G$ to avoid Runge's phenomenon~\cite{zbMATH02662492,suli2003introduction} 
%
and to avoid large cancellation in the Mellin transform.
\footnote{
    More details of the implementation are available at
    \href{https://eko.readthedocs.io/en/latest/theory/Interpolation.html}{https://eko.readthedocs.io/en/latest/theory/Interpolation.html}.
}

As standard setting we adopt a grid of at least 50 points with linear scaling in the large-$x$ 
region ($x \geq 0.1$) and with logarithmic scaling in the small-$x$ region 
and an interpolation of degree four.

\begin{figure}[!t]
    \begin{center}
    \includegraphics[width=0.8\textwidth]{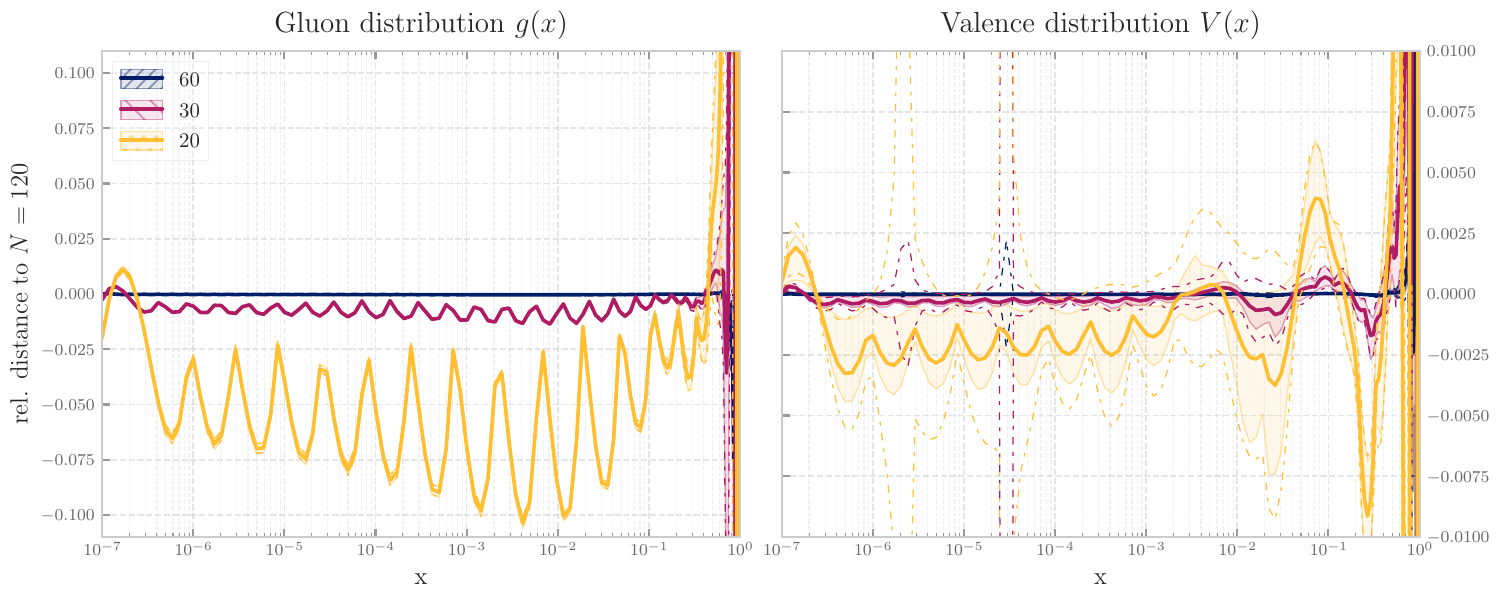}
    \end{center}
    \caption{
        Relative differences between the outcome of NNLO QCD evolution
        as implemented in \eko{} with 20, 30, and 60 points to 120
        interpolation points respectively.
        We used NNPDF4.0 as input PDF, the upper and lower borders of the envelope 
        correspond respectively to the $0.16$ and $0.84$ quintiles of the replicas 
        set, while the dashed lines correspond to one standard deviation.
        The distributions are evolved in the range $\mu_F=1.65 \to 10^2~\text{GeV}$.
    }
    \label{fig:interpolation} 
\end{figure}

For a first qualitative study, we show in \cref{fig:interpolation} a comparison between 
an increasing number of interpolation points distributed according to \cite[Eq.~2.12]{Carrazza:2020gss}.
The separate configurations are converging to the solution with the largest number 
of points. Using 60 interpolation points is almost indistinguishable from using 
120 points (the reference configuration in the plot). In the singlet sector (gluon) 
the convergence is significantly slower due to the more involved solution strategies 
and, specifically, the oscillating behavior is caused due to these difficulties.
The spikes for $x\to 1$ are not relevant since the PDFs are intrinsically small 
in this region ($f(x)\to 0$) and thus small numerical differences are enhanced.




\paragraph{Solution strategies.}
The formal solution of \cref{eq:dglap_mellin} in terms of evolution kernel operators
is given by \cref{eq:eko}.
If the anomalous dimension $\gamma_{ij}$ is diagonal in flavor space, i.e.\ it is 
in the non-singlet sector, it is always possible to find an analytical solution 
to \cref{eq:eko}. In the singlet sector, however, this is only true at 
LO and to obtain a solution beyond, we need to apply different approximations 
and solution strategies, on which \eko{} offers currently eight implementations,
which may differ only by the strategy in a specific sector. 
All provided strategies agree at fixed order, but differ by higher order terms.



\begin{figure}[!t]
    \centering
    \includegraphics[width=0.8\textwidth]{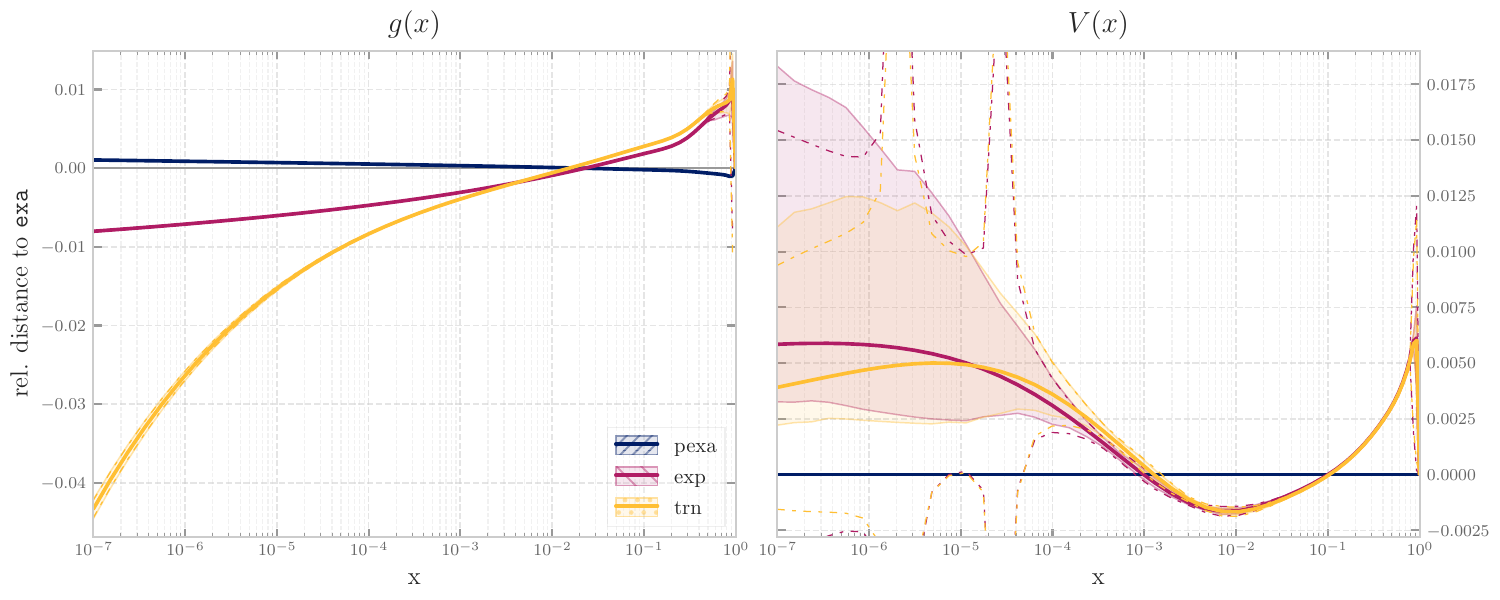}
    \caption{
        Comparison of selected solutions strategies, with respect to the
        \texttt{iterated-exact} (called \texttt{exa} in label) one. 
        In particular: \texttt{perturbative-exact} (\texttt{pexa}) 
        (matching the reference in the non-singlet sector),
        \texttt{iterated-expanded} (\texttt{exp}), and \texttt{truncated}
        (\texttt{trn}).
        The distributions are evolved in the range $\mu_F=1.65 \to 10^2~\text{GeV}$,
        with same plotting settings as in \cref{fig:interpolation}.
    }
    \label{fig:dglap_solutions}
\end{figure}

In \cref{fig:dglap_solutions} we show a comparison of a selected list of
solution strategies:
\footnote{
    For the full list of available solutions and a detailed description
    see \href{https://eko.readthedocs.io/en/latest/theory/DGLAP.html}{https://eko.readthedocs.io/en/latest/theory/DGLAP.html}.
}

\begin{itemize}
    \item \texttt{iterate-exact}:
        In the non-singlet sector we take the analytical solution of \cref{eq:dglap_mellin} 
        up to the order specified. In the singlet sector we split the evolution path 
        into segments and linearize the exponential in each segment~\cite{Bonvini:2012sh}.
        This provides effectively a straight numerical solution of \cref{eq:dglap_mellin}.
        In \cref{fig:dglap_solutions} we adopt this strategy as a reference.
    \item \texttt{perturbative-exact}:
        In the non-singlet sector it coincides with \texttt{iterate-exact}.
        In the singlet sector we make an ansatz to determine the solution as a
        transformation $U(a_s)$ of the LO solution (see \cite[Eq.~2.23]{Vogt:2004ns}). 
        We then iteratively determine the perturbative coefficients of $U$.
    \item \texttt{iterate-expanded}:
        In the singlet sector we follow the strategy of \texttt{iterate-exact}.
        In the non-singlet sector we expand \cref{eq:dglap_mellin} first to the order
        specified, before solving the equations.
    \item \texttt{truncated}: 
        In both sectors, singlet and non-singlet, we make an ansatz to determine 
        the solution as a transformation $U(a_s)$ of the LO solution 
        and then expand the transformation $U$ up to the order specified (see \cite[Eq.~2.24]{Vogt:2004ns}).
        Note that for programs using $x$-space this strategy is difficult
        to pursue as the LO solution is kept exact and only the transformation
        $U$ is expanded.
\end{itemize}

The strategies differ mostly in the small-$x$ region where the PDF evolution 
is enhanced and the treatment of sub-leading corrections become relevant.
This feature is seen prominently in the singlet sector between
\texttt{iterate-exact} (the reference strategy) and \texttt{truncated}.
On contrary, in the non-singlet sector the distributions vanish for small-$x$
and so the difference can be artificially enhanced.
This is eventually the source of the spread visible for the valence distribution $V(x)$
making it more sensitive to the initial PDF.

\paragraph{Matching at thresholds.}
\eko{} can perform calculations in a fixed flavor number scheme where
the number of active or light flavors $n_f$ is constant and 
in a variable flavor number scheme (VFNS) where the number of active flavors 
changes when the scale $\mu_F^2$ crosses a threshold $\mu_h^2$.
The latter requires a matching procedure as explained in \cref{sec:fns}.
Although the value of $\mu_h$ usually coincides with the respective
quark mass $m_h$, \eko{} implements the explicit expressions when the two 
scales do not match. This variation can be used to estimate MHOU.

\begin{figure}[!t]
    \centering
    \includegraphics[width=0.8\textwidth]{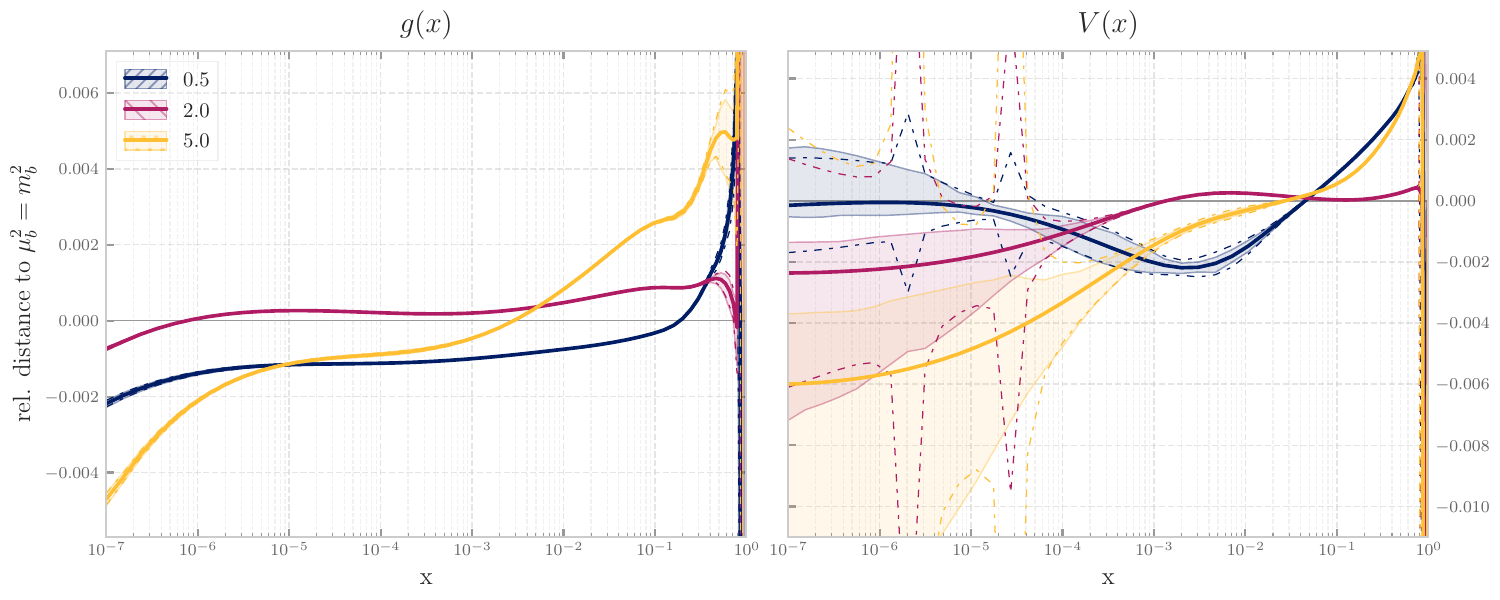}
    \caption{
        Difference of PDF evolution with the bottom matching $\mu_b^2$ at $1/2, 2,$ 
        and $5$ times the bottom mass $m_b^2$ relative to $\mu_b^2 = m_b^2$. 
        Note the different scale for the two distributions.  
        All evolved in $\mu_F=1.65\to 10^2~\text{GeV}$.
    }
    \label{fig:pdfmatching}
\end{figure}

In \cref{fig:pdfmatching} we show the relative difference for the PDF evolution 
with threshold values $\mu_h^2$ that do not coincide with the respective heavy
quark masses $m_h^2$. When matching at a lower scale the difference is significantly 
more pronounced as the evolution includes a region where the strong coupling 
varies faster. When dealing with $\mu_h^2 \neq m_h^2$ the PDF matching conditions 
become discontinuous already at NLO. 

\paragraph*{Backward matching.}
For backward evolution the PDF matching has to be applied in the reversed order.
In \eko{} we have implemented two different strategies to perform the inverse
matching: the first one is a numerical inversion, called \texttt{exact}, 
where the operator matrix elements of \cref{eq:singlet_matching,eq:ns_matching} 
are inverted exactly in Mellin space; in the second method, called \texttt{expanded},
the matching matrices are inverted through a perturbative expansion in $a_s$
in Mellin space, given by:
\begin{align}
\begin{split}    
    \left(A^{(n_f)}_{exp}\right)^{-1}(\mu_{h}^2) & = 
        I - a_s(\mu_{h}^2) A^{(1),(n_f)} 
        + a_s^2(\mu_{h}^2) \left[ - A^{(2),(n_f)} -  (A^{(1),(n_f)})^2 \right] \\ 
        & + a_s^3(\mu_{h}^2) \left[ - A^{(3),(n_f)} +  A^{(1),(n_f)} A^{(2),(n_f)} +  A^{(2),(n_f)} A^{(1),(n_f)} -  (A^{(1),(n_f)})^3\right] \,,
    \label{eq:invmatchingexp}
\end{split}
\end{align} 
with $I$ the identity matrix in flavor space.

\begin{figure}[!t]
    \centering
    \includegraphics[width=0.8\textwidth]{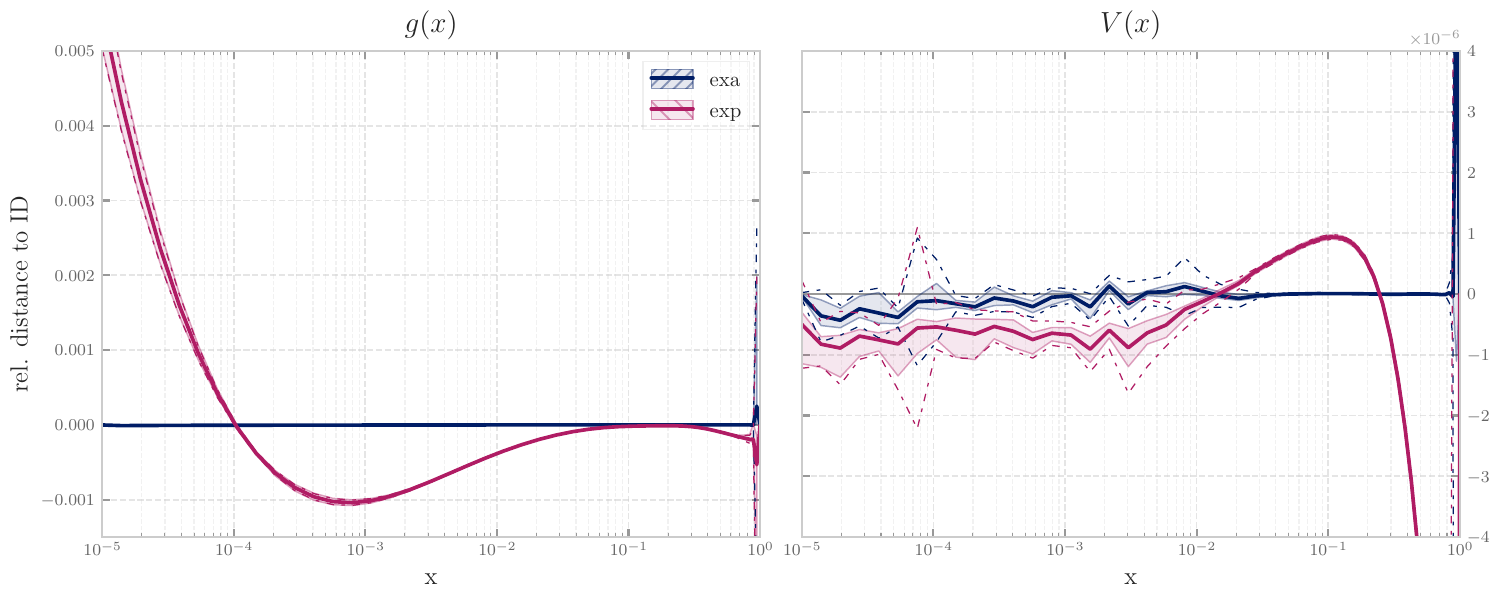}
    \caption{
        Relative distance of the product of two opposite NNLO \ekos{}
        and the identity matrix, in case of exact inverse and expanded
        matching (see \cref{eq:invmatchingexp}) when crossing the bottom
        threshold scale $\mu_{b}=4.92~\text{GeV}$.
        The plot setting are as in \cref{fig:interpolation}.
    }
    \label{fig:closure_test}
\end{figure}

As a consistency check we have performed a closure test verifying that after
applying two opposite \ekos{}s to a custom initial condition
we are able to recover the initial PDF. Specifically, the product of the
two kernels is an identity both in flavor and momentum space up to
the numerical precision. The results are shown in \cref{fig:closure_test} in 
case of NNLO evolution crossing the bottom threshold scale $\mu_{F}=m_{b}$. 
The differences between the two inversion methods are more visible for 
singlet-like quantities, because of the non-commutativity of the matching 
matrix $A_{S}^{(n_f)}$.  

Special attention must be given to the heavy quark distributions which are 
always treated as intrinsic, when performing backward evolution.
In fact, if the initial PDF (above the mass threshold) contains an intrinsic 
contribution, this has to be evolved below the threshold otherwise momentum 
sum rules can be violated.
This intrinsic component is then scale independent and fully decoupled 
from the evolving (light) PDFs. On the other hand, if the initial PDF 
is purely perturbative, it vanishes naturally below the mass threshold 
scale after having applied the inverse matching.


\subsection{Yadism}
\label{sec:yadism}
In this section, we present \yadism{}, the software library developed 
for the calculation of DIS observables. 
To date, \yadism{} has already been used in various papers. Specifically, it has been used 
to evaluate neutrino structure functions in Refs.~\cite{Candido:2023utz,Cruz-Martinez:2023sdv}, 
and to compute of polarized structure functions in Ref.~\cite{Hekhorn:2024tqm} (cf. \cref{chap:pol}).
Furthermore, \yadism{} has been adopted by the NNPDF collaboration, who has used it in their 
most recent papers~\cite{NNPDF:2024djq,NNPDF:2024dpb,NNPDF:2024nan} (cf. \cref{chap:ic,chap:an3lo}).
We refer to \cite{Candido:2024rkr} for a more extensive presentation.

%
\yadism{} differs from other QCD codes such as \apfel~\cite{Bertone:2013vaa},
\apfelpp~\cite{Bertone:2017gds}, \hoppet~\cite{Salam:2008qg}, and \qcdnum~\cite{Botje:2010ay}
in several ways.

\yadism{} includes most of the currently available results in literature, specifically 
it allows for the computation of polarized and unpolarized
structure functions up to N$^3$LO in QCD (cf. \cref{sec:an3lo_dis}).
Thanks to its modular design, the library can be easily extended as the results
of new computations become available.

\yadism{} provides both renormalization and factorization scale variations consistently~\cite{NNPDF:2024dpb} 
and both can be implemented at any order. The currently implemented coefficients 
allow performing renormalization scale variations up to N$^3$LO and factorization scale 
variations up to NNLO. Instead, N$^3$LO factorization scale variations can be included 
through the \eko{} evolution code (\cref{sec:eko}).

\yadism{} can, together with \eko, be used to construct general-mass variable flavor number 
schemes using coexisting PDFs with different numbers of active flavors~\cite{Barontini:2024xgu}. This can avoid  
the perturbative expansion of the evolution kernel as is done in the construction 
of the FONLL scheme~\cite{Forte:2010ta}.

\yadism{} has a uniform treatment of all heavy quarks, i.e., all features that are available 
for charm are also available for bottom and top. This strategy opens up the possibility 
for computations with an intrinsic bottom quark~\cite{Brodsky:2015fna,Lima:2024ilk}. 
We provide both the fixed-flavor number scheme (FFNS$n_f$) 
and zero-mass variable-flavor number scheme (ZM-VFNS) calculation, as well as the asymptotic limit, $Q^2 \gg m^2$, of the 
FFNS (FFNS$n_f^{(0)}$), which is required in the construction of the FONLL scheme~\cite{Forte:2010ta}.

%
The \pineappl{} grid output format allows \yadism{} to be integrated into the 
\xfitter{} framework~\cite{Alekhin:2014irh,xFitterDevelopersTeam:2017xal,xFitter:2022zjb}
and the \pineline{} framework~(see \cref{sec:pineline}). 

\yadism{} is written in the \texttt{Python} programming language, which is known f
or its ease of use, and thus reduces the threshold for potential new contributors. 
For these reasons, development of new functionality can be quick to, e.g., rapidly 
adopt new computations. 
%

The up-to-date code documentation is available at:
\begin{center}
  \url{https://yadism.readthedocs.io/en/latest/}.
\end{center}

Benchmarks to other available libraries, as \apfelpp{} and \qcdnum{}, have been performed, 
both at coefficient function and structure function level, finding good agreement 
as shown in \cite[Sec.~3.1]{Candido:2024rkr}.

If one wishes to actually compute a structure function one needs to define a number 
of theory parameters and parameters of the experimental setup.
Such input settings are passed to \yadism{} through \textit{runcards} in YAML format,
\footnote{\url{https://yaml.org/}} and they are divided into two parts:
an \textit{observable runcard} describing the experimental setup 
(such as scattering particles, kinematic bins, or helicity settings) and a 
\textit{theory runcard} describing the parameters of the theory framework 
(such as coupling strength, perturbative orders, or quark masses).
While observable runcards are usually tailored to a given experiment, theory 
parameters are usually shared by multiple runs. 
Below, we describe the most important options for the configuration of the observables 
and theories that can be defined in 
the respective runcard. We conclude analyzing differences between flavor 
schemes.

\subsection*{Observable configuration options}
\paragraph{Projectile.}
\yadism{} supports computations of DIS coefficients with massless charged leptons 
and their associated neutrinos as projectiles in the scattering process.
Specifically, to describe, e.g., the HERA~\cite{H1:2009pze,H1:2018flt} data one needs 
both electrons and positrons and, e.g., for the CHORUS~\cite{CHORUS:2005cpn} data 
neutrinos as well as anti-neutrinos are needed. Charged leptons can interact both 
electromagnetically and weakly with the scattered nuclei, whereas neutrinos only 
carry weak charges. Recently, together with a machine-learning parametrization 
of experimental data, CC neutrino DIS predictions computed with \yadism{} have 
been used to extend predictions for neutrino structure functions~\cite{Candido:2023utz}.

\paragraph{Target.}
\yadism{} supports computations with nuclei with mass number $A$ and $Z$ protons as 
targets in the scattering process.
By acting on the coefficients associated to up and down partons \yadism{} implements 
the isospin symmetry of the form:
\begin{equation}
  \begin{pmatrix} c'_u \\ c'_d \end{pmatrix} =
  \frac{1}{A}
  \begin{pmatrix} Z & A - Z \\ A - Z & Z \end{pmatrix}
  \begin{pmatrix} c_u \\ c_d \end{pmatrix}
  \label{eq:isospin}
\end{equation}
where $c'_i$ and $c_i$ are the effective and the proton coefficient associated with 
the parton $i$. This rotation is particularly useful in the context of proton PDF fitting
where it can be used to relate neutron, deuteron, and heavier nuclear structure functions 
to the proton ones. In this way, isospin is used as a first approximation of nuclear 
correction by just swapping up and down contribution for the amount specified by the 
target nuclei. In particular for:
\begin{itemize}
  \item \texttt{proton targets} \textit{($A=1, Z=1$)}: up and down are kept as they are.
  \item \texttt{neutron targets} \textit{($A=1, Z=0$)}: up and down components are fully swapped,
  such that the up coefficient function is matched to the down PDF and conversely.
  \item \texttt{isoscalar targets, i.e. deuteron} \textit{($A=2, Z=1$)}:
  the effective coefficient functions are mixed such that
  $c'_{u}$ is half the original $c_{u}$ and half the original $c_{d}$.
\end{itemize}
\yadism{} is completely general with respect to the nuclear target allowing a user to provide 
values for $A$ and $Z$ as input to the computation. Alternatively, for a number of targets, 
the name itself can also be provided as input.
The readily available targets are: iron ($A=49.618,Z=23.403$), used to describe NuTeV data;
lead ($A=208,Z=82$), used to describe CHORUS data; neon and marble ($CaCO_{3}$) with both $A=20, Z=10$,
used to describe respectively the BEBCWA59~\cite{BEBCWA59:1987rcd} and CHARM~\cite{CHARM:1984ikt} data.


\paragraph{Cross-sections.}
\yadism{} supports the computation of both structure functions and (reduced) cross-sections.
In particular, we implement the structure functions:
\begin{equation}
  & F_2, \ F_L, \ x F_3, \ g_4, \ g_L, \ 2 x g_1 \,,
\end{equation}
where the normalization is chosen such that at LO, all the structure functions are 
proportional to different PDF combinations of the form $xf(x)$.
%
Generally, we can write the (reduced) cross-sections for a DIS process 
in terms of the structure functions as
\begin{equation}
  \sigma(x,Q^2,y)=N\left(F_2(x,Q^2)-d_L F_L(x,Q^2)+d_3 x F_3(x,Q^2)\right),
\end{equation}
where $N$, $d_L$, and $d_3$ may depend on the experimental setup or the scattered lepton. 
The different reduced cross-sections implemented in \yadism{}, and their definitions in 
terms of $N$, $d_L$, and $d_3$ can be found in the online documentation.
\footnote{\url{https://yadism.readthedocs.io/en/latest/theory/intro.html\#cross-sections}}
The implemented definitions can be used to describe data from
HERA, CHORUS, NuTeV, CDHSW~\cite{Berge:1989hr}, and FPF~\cite{Cruz-Martinez:2023sdv}.
Finally, we provide the linearly dependent structure functions $2x F_1$ and $2x g_5$.

\paragraph{Flavor tagging.}
In general, any total DIS structure function $F$ can be decomposed in three 
different components, according to the type of quark coupling to the exchanged EW boson
(see \cref{sec:heavy_quarks}):
\begin{equation}
  F = F^{(l)} + F^{(h)} + F^{(hl)},
  \label{eq:sf_decomposition}
\end{equation}
where $F^{(l)}$ denotes the contribution coming from diagrams where all the fermion 
lines are massless, $F^{(h)}$ is the contribution due to heavy quarks coupling to the 
EW boson and $F^{(hl)}$ originates from higher order diagrams where a light quark is coupling 
to the boson, but heavy quarks lines are present.

Given \cref{eq:sf_decomposition}, we support the calculation of fully inclusive (total) 
observables, where only the lepton is observed in the final state, and flavor tagged 
final state, where we require a specific heavy quark (charm, bottom, or top) to couple 
with the mediating boson.
%
For completeness, also light structure functions $F^{(l)}$ are available,
in isolation, although they do not correspond to any physical observable.

\subsection*{Theory configuration options}
\paragraph{Renormalization and factorization scale variations.}
In perturbative QCD the DIS coefficients of \cref{eq:pert_dis_coeff},
are expanded in powers of $a_s$.
The estimate of the error introduced by the truncation of such series 
is quite relevant in multiple precision applications.
Some information about the missing higher orders, and the related uncertainty 
(MHOU), can be extracted from the Callan-Symanzyk equations violation.
In this sense, a practical approach to obtain a numerical estimate consists 
in varying the relevant unphysical scales (see \cref{sec:th_err}).

In DIS, the two involved unphysical scales are the \textit{renormalization scale}, 
arising from the subtraction of ultraviolet divergences, and the 
\textit{factorization scale}, from the subtraction of collinear logarithms in 
the PDF definition.

The explicit expressions of the $C_i$ expansion upon scale variations can be found, 
in \cite[Sec.~2]{vanNeerven:2000uj}.
Generally, these depend, order by order in perturbation theory, on the derivatives
of $a_s$ and the PDFs with respect to the mentioned scales.
The former are the $\beta$-function coefficients and the latter the splitting functions.
In \yadism, necessary $\beta$-function coefficients are taken from the \eko{} package, 
while the $x$-space splitting functions are directly implemented.

At the level of structure function, scale variations can be cast into an additional
convolution with a kernel $K$:
\begin{equation}
  F(x,\mu\neq Q) = \left(K \otimes C \otimes f\right)(x)
\end{equation}
It can be shown that the transformation can be applied a posteriori to an already 
computed interpolation grid.

\paragraph{Target mass corrections.}
While the DIS factorization is usually derived for the scattering of two massless particles,
it is possible to account for the finite mass of the scattering target through target mass 
corrections~\cite{Schienbein:2007gr,Goharipour:2020gsw,Hekhorn:2024tqm}.
These corrections become relevant for either small virtualities or large Bjorken-$x$.
They can be implemented as an additional convolution, and we provide several approximations 
(corresponding to higher twist expansions) following Ref.~\cite{Schienbein:2007gr}.

\paragraph{Flavor Number scheme.}
Flavor number schemes provide a prescription to resolve the ambiguous treatment 
of heavy quark masses (see also \cref{sec:fns}).
Generally, to achieve a faithful description of experimental data at scales 
roughly around the heavy quarks mass $Q\sim m_h$, quarks should be treated fully massive.
However, in the region where $Q \gg m_h$, quarks should be considered massless.
In \yadism{} we allow for 3 different schemes. Only one single heavy quark is allowed 
at each time.

\begin{itemize}
  \item \texttt{Fixed flavor number scheme (FFNS$n_f$).}
  The FFNS$n_f$, is defined as a configuration with a fixed number of flavors at all scales,
  i.e.\ all quark masses are fixed to be either light (up to $n_f$), heavy ($n_f+1$) or 
  decoupled (above $n_f+1$).

  \item \texttt{Zero mass-variable flavor number scheme (ZM-VFNS).}
  In the ZM-VFNS all quark masses in the calculations are either light or decoupled.
  The number of light quarks $n_f$ is not fixed, but instead varies with the number of active flavors depending on the scale of the process, i.e.\ $n_f(Q^2)$. Specifically, $n_f=3$ below $m_c$ and this increases as the heavy quark thresholds are crossed, i.e.\ $Q>m_h$, after which the corresponding heavy quark is treated to be light.

  \item \texttt{Asymptotic fixed flavor number scheme (FFNS$n_f^{(0)}$).}
  The FFNS$n_f{(0)}$ is similar to the FFNS$n_f$, but retains only the logarithmic corrections,
  i.e.\ it does not contain any power-like heavy quark corrections $m^2/Q^2$.
  The FFNS$n_f{(0)}$ is constructed as the overlap between FFNS$n_f$ and ZM-VFNS and can be used to
  construct a VFNS flavor number schemes.
\end{itemize}

\yadism{} does not provide explicitly the FONLL scheme, but all the necessary 
ingredients FFNS$n_f$, FFNS$n_f^{(0)}$, and ZM-VFNS are available.

\begin{figure}[!t]
  \includegraphics[width=\textwidth]{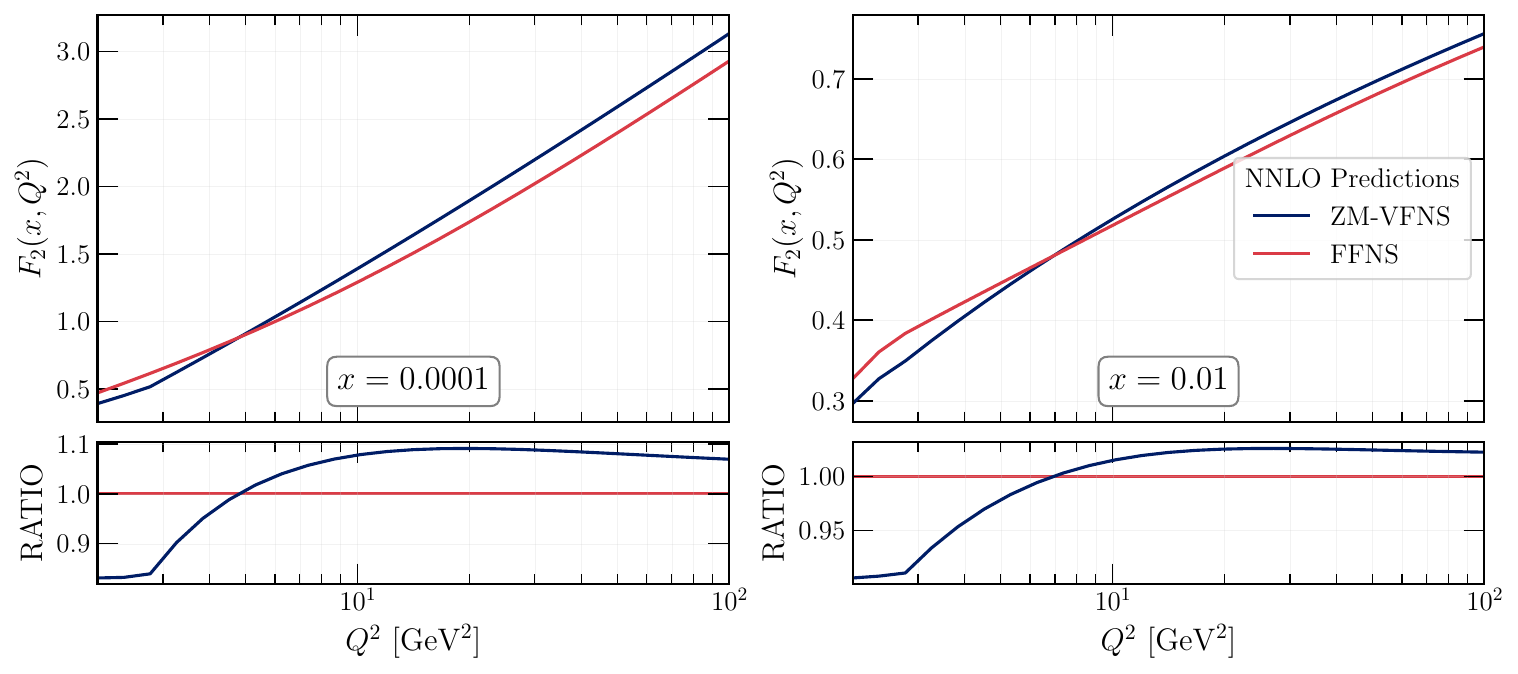}
  \includegraphics[width=\textwidth]{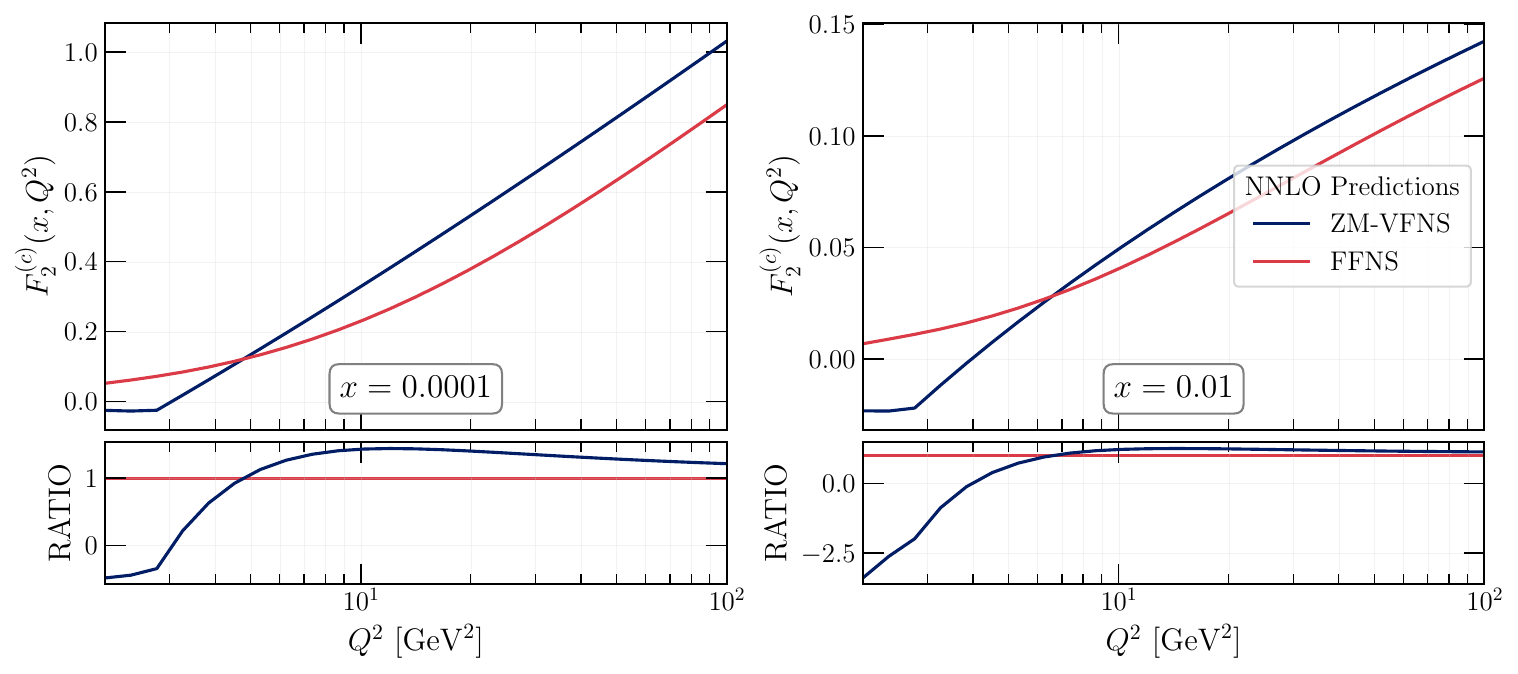}
  \caption{
        Comparison of the structure functions $F_2$ (top) and $F^{(c)}_2$ (bottom) 
        using FFNS3 and ZM-VFNS at NNLO accuracy.
        The top panels show the absolute comparisons while the bottom ones show the ratio 
        w.r.t.\ ZM-VFNS.
  }
  \label{fig:zm_vs_ffns}
\end{figure}

\begin{figure}[!t]
  \includegraphics[width=\textwidth]{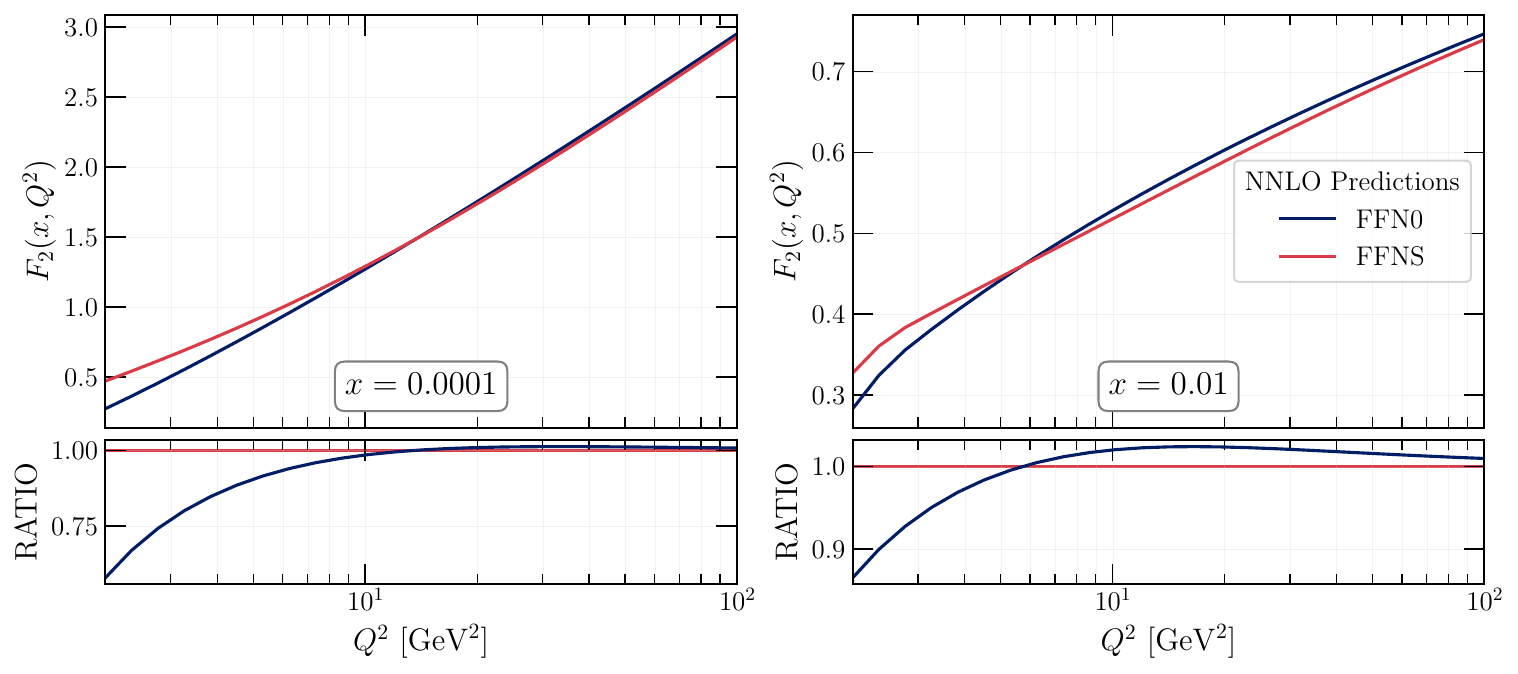}
  \includegraphics[width=\textwidth]{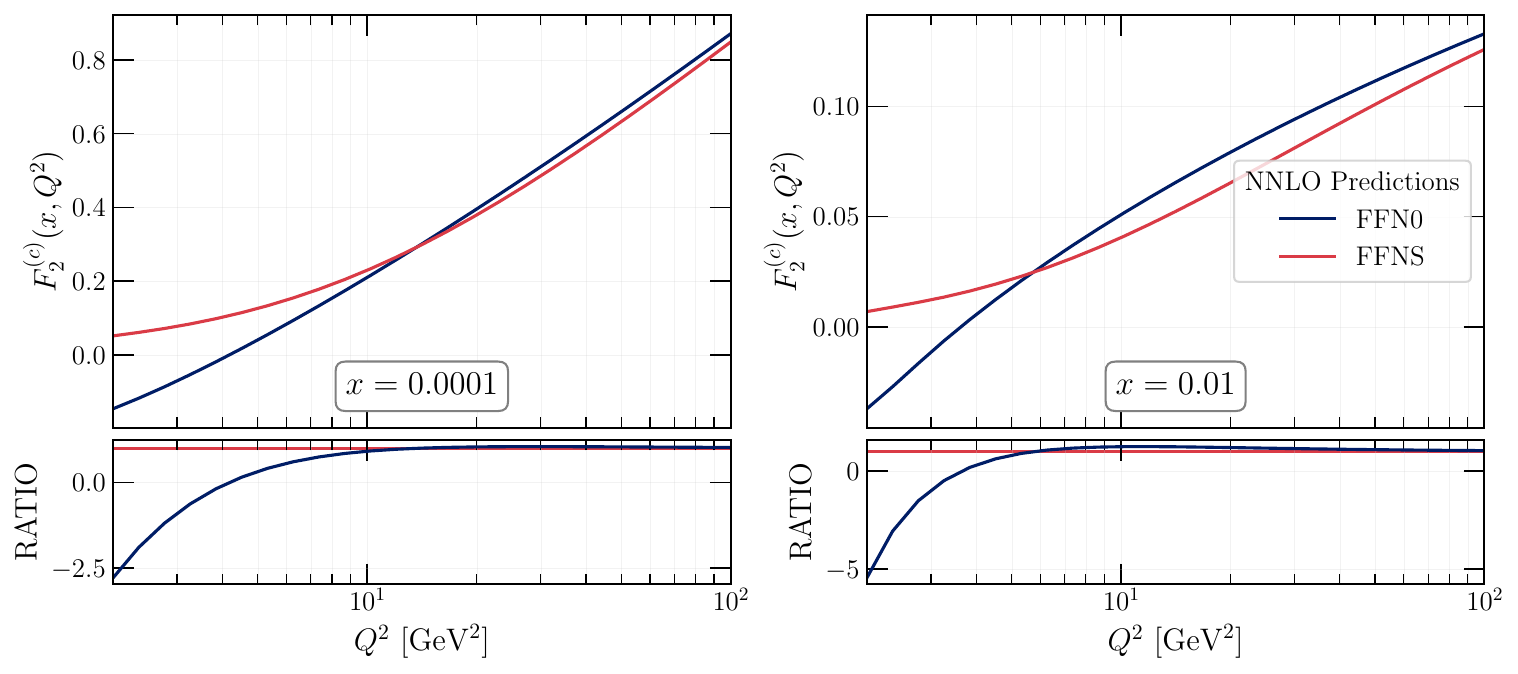}
  \caption{Same as \cref{fig:zm_vs_ffns} but now comparing FFNS3 and FFNS3$^{(0)}$.}
  \label{fig:asy_vs_ffns}
\end{figure}

We now provide representative comparisons on the different 
prescriptions used to treat heavy quark masses in order to underline their 
relevance in the different kinematic regions.
In all the subsequent comparisons, we adopt a fixed boundary condition defined 
as a PDF set at a given scale $\mu_F=\mu_0$. Evolution of the boundary condition, 
including changing of the number of active flavors, is performed using \eko.
First, in \cref{fig:zm_vs_ffns}, we compare the ZM-VFNS and FFNS3 coefficient functions 
as a function of $Q^2$. 
We expect both calculations to differ more in the low-$Q^2$ region and progressively 
reach better agreement towards the large-$Q^2$ region. However, while ZM-VFNS 
fully resums all (collinear) logarithms $\log(m^2/Q^2)$, FFNS$n_f$ is a fixed order 
calculation which only collects a finite number of (collinear) logarithms and 
hence a finite difference between the two calculations remains.
We indeed observe for both structure functions $F_2$ and $F_2^{(c)}$ this
expected pattern, thus confirming a consistent implementation.
Next, in \cref{fig:asy_vs_ffns}, we compare FFNS3 and FFNS3$^{(0)}$ coefficient functions as a function 
of $Q^2$.
While we can indeed observe this convergence at large-$Q^2$, we also find a
relevant region at mid to low $Q^2$ where mass effects can grow up to $25\%$.
This latter region can reach up to $\mathcal{O}(100)$ times the heavy quark mass 
and clearly demonstrates the need for a VFNS to improve the accuracy of the prediction.

\subsection{The Pineline framework}
\label{sec:pineline}

As mentioned above, the determination of PDFs from high energy experimental data, 
requires FK-tables i.e. a theory map given by \cref{eq:fks_def}, made of two main 
components: a PDF independent interpolation \textit{grid} and an \textit{evolution operator}.
The latter are computed with \eko{}, while the former have to be produced by 
different generators in order to cover the full variety of available processes.
For this reason, we have constructed a unique infrastructure, \pineline{} with the following 
targets: standardize the input/output format and make the results reproducible.

\begin{figure}
  \centering
  \includegraphics[width=0.9\textwidth]{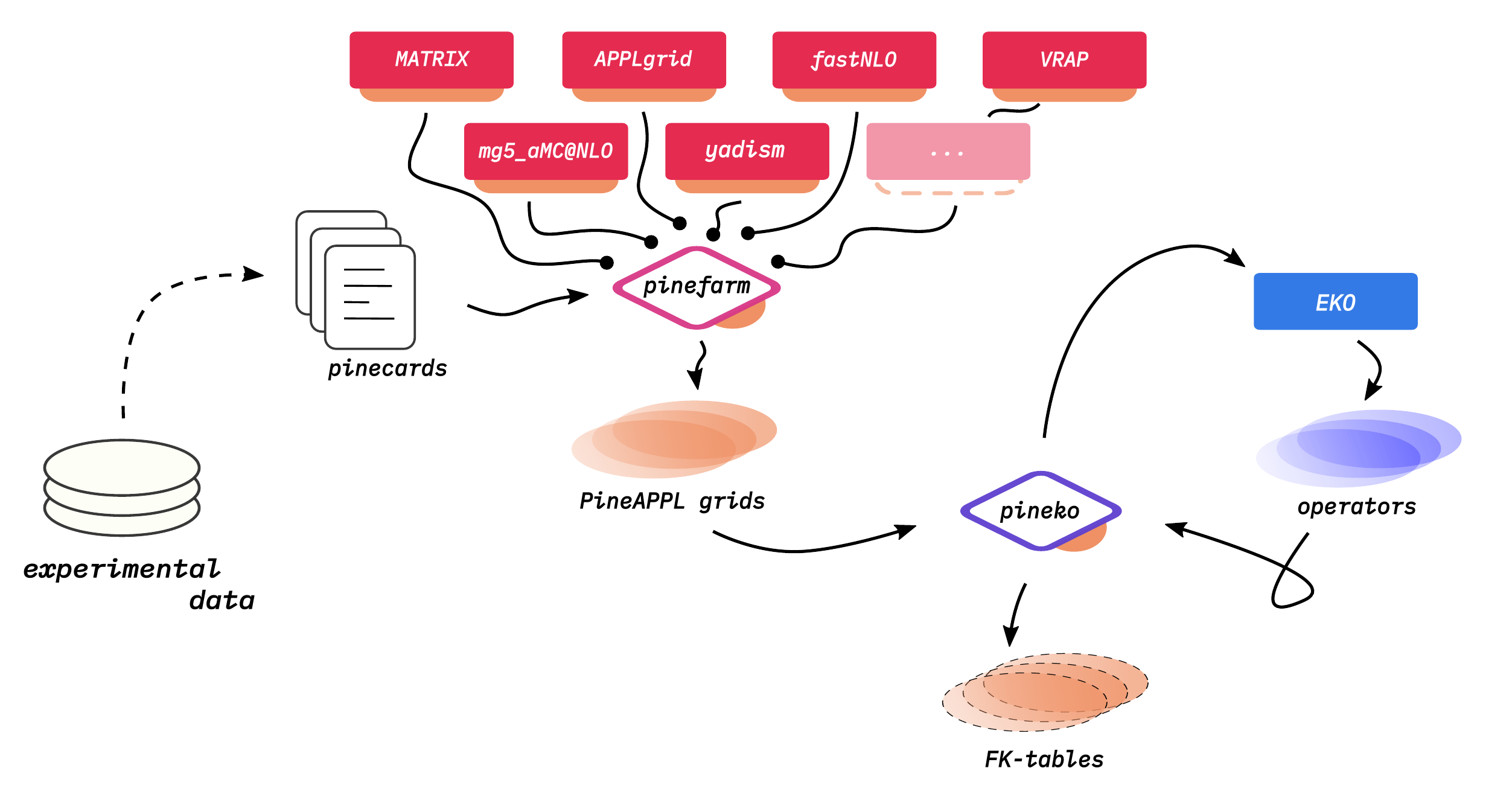}
  \caption{
    The \pineline{} architecture.
    Arrows in the picture indicate the flow of information (together with the
    execution order), and the orange insets on other elements indicate an
    interface to \pineappl{} (notice \eko{} not having it).
    In particular, magenta blocks above \pinefarm{} are the grid providers 
    \cite{Gehrmann:2018szu,Grazzini:2017mhc,Alwall:2014hca,Frederix:2018nkq,Candido:2024rkr,Carli:2010rw,Britzger:2012bs,Anastasiou:2003ds}.
  }
  \label{fig:pineline}
\end{figure}

The framework adopts \pineappl{} as back-end and bridges the output of different generators 
to an FK-table.
In particular, \pineappl{} exposes APIs to different languages: it is natively
written in Rust, but has an API to C/C++, that can be consumed also by a
FORTRAN application, and a Python API. Different \textit{grid} providers can thus 
interface directly to \pineappl{} when filling grids.
The\pineline{} architecture is visible in \cref{fig:pineline}.
Starting with the experimental data, we standardize 
them into a common format, called \textit{pinecard}.
These are used together with the theory parameters as inputs for various grid generators 
which are manged by \pinefarm{} a unique Python package working as a front-end.
Such interface is still work in progress, nevertheless,
among others \pinefarm{} is able to run \madgraph~\cite{Alwall:2014hca,Frederix:2018nkq}, 
\Matrix~\cite{Grazzini:2017mhc}, \nnlojet~\cite{Gehrmann:2018szu} allowing
to compute predictions for numerous scattering processes.
Once the grid is available, \pineko{}, another package dedicated to the final
construction of FK-tables, can extract the details of the needed DGLAP operator and 
run the \eko{} library. Finally, it will also take care of combining the grid and the operator 
into the final FK-table.

All the components of the pipeline are open source and the code is available in
on GitHub at:
\begin{itemize}
  \item \pineappl: \url{https://github.com/NNPDF/pineappl}
  \item \pineko: \url{https://github.com/NNPDF/pineko}
  \item \pinefarm: \url{https://github.com/NNPDF/pinefarm} 
\end{itemize}
It is important to emphasize that this set of tools does not depend on the \nnpdf fitting methodology and can 
be used in general for any (polarized) hadronic function fitting.

\section{The NNPDF methodology}
\label{sec:nnpdf_methodology}

The problem of determining PDFs can be seen
from the statistical point of view as a regression problem. 
In fact, from a finite set of experimental datapoints, we are trying to 
reconstruct a set of real functions defined in the domain $f(x, Q_0): (0,1] \to \mathcal{R}^{+}$
at the scale $Q_0$ fixed.
The Fast Kernel technology, allows us to efficiently map the dependent variables,
i.e. the cross-section theoretical predictions, to the independent
degrees of freedom the PDFs and overcome all the complications 
arising due to the flavor mixing and $Q^2$ dependency.
However, to solve the PDF regression problem we are left with at least 
3 major obstacles that have to be taken into account. 
First, the problem might seem mathematically ill-defined, as
the number of datapoints is finite, while PDFs have a continuous
domain. This is practically solved by restricting the region
of validity of the fitted PDFs to a specific $x$ range (approximately $[10^{-4},0.7]$), 
called data region, where the behavior of PDFs can be really determined 
by the data, and adopting a discrete grid parametrization. 
In the very small-$x$ and large-$x$ regions our analysis
can only be interpreted as an extrapolation, which might be possible to 
validate with future measurements or so-called future tests~\cite{Cruz-Martinez:2021rgy}.
Second, both the experimental data and the theory predictions
used during the fit, have a finite accuracy. Thus, their uncertainties,
need to be taken into account. In this work we adopt a Monte Carlo
method to sample the experimental distributions (described in \cref{sec:exp_err})
and a covariance matrix formalism to propagate different theoretical
errors such as nuclear uncertainties or missing higher order corrections
(cf. \cref{sec:th_err}) to the PDF fit.
Finally, there is an arbitrariness in the choice of the parametrization
to adopt for the initial scale PDFs. Various approaches have been proposed in literature, 
in the thesis we mainly present results 
based on the NNPDF methodology where the PDFs are given as an output of
feed forward neural network. 
The solutions adopted to overcome these issues together with the
setting used to define a cost function and its minimization, define the 
so-called fitting methodology, which is described in \cref{sec:nnpdf_fitting}. 

\subsection{Treatment of experimental uncertainties}
\label{sec:exp_err}
In particle physics, measurements are typically presented as binned 
and fiducial cross-sections.
In a simplistic way, given a scattering process,
defined by the kinematics of its final state and, the possible background processes,
we can define a fiducial cross-section $\sigma$ as
\begin{equation}
    \sigma = \frac{N_c - N_b}{\epsilon L}
    \label{eq:xs_def}
\end{equation}
where $N_c$ and $N_b$ are respectively the total number of observed events satisfying
the selection criteria and the estimate of background events. 
$L$ corresponds to the integrated luminosity which acts as a normalization by total number of 
collision happening in a certain time and detected by the experimental apparatus.
$\epsilon$ instead, called acceptance, is used to model possible systematic effects
such as, detector acceptance, trigger efficiency, reconstruction efficiency, 
which are typical of the experimental setup.
Subsequently, one can divide the phase space regions into different bins and 
obtain differential cross-sections. 
Often, in order to reduce uncertainties, measurements are presented as ratios
of cross-sections: for instance, polarized asymmetries are computed from the ratio of 
the polarized to unpolarized cross-sections.  

The experimental uncertainties can thus be classified into:
\begin{itemize}
    \item 
    \textit{statistical}, which arise due to random fluctuation of the finite 
    number of samples collected during a measurement. 
    The statistical fluctuations in repeated observations of the same process, 
    are, by definition, uncorrelated in different kinematic bin.
    \item  
    \textit{systematics}, originating from the procedure adopted during the data 
    taking, as well as due to the design and calibration of the apparatus. 
    Examples of systematic uncertainties include trigger efficiency,
    the signal selection efficiency, the beam polarization uncertainty, 
    the luminosity uncertainty or the jet reconstruction uncertainties.
    Such uncertainties are generally correlated in all the different kinematic 
    bins. 
\end{itemize}

While statistical uncertainties are are generally additive and depend on the number of observed 
events (a naive scaling suggest $\approx 1/\sqrt{N_c}$), the estimation of systematics 
require ad-hoc analysis and, generally, it is a non-trivial procedure. 
Depending on the setup used for their determination, systematics can be either 
additive either multiplicative.
Given a set of measurements $i = 1 \dots N_{dat}$ obtained with $k= 1 \dots N_{rep}$ observations, 
of the same phenomenon $x_i^{(k)}$, we assume the data to follow a Gaussian distribution given by
\begin{equation}
    \mathcal{G}(x_i^{(k)}) \propto \text{exp} \left [ - (x_i^{(k)} - \sigma_i) \ {\rm cov}_{ij} \ (x_j^{(k)} - \sigma_j) \right ] \,,
    \label{eq:gauss_def}
\end{equation}
where the expectation value $\sigma_i$ is given by the average of all the observation $N_{rep}$
and ${\rm cov}_{ij}$ is the total experimental covariance matrix defined via
\begin{equation}
    {\rm cov}_{ij} = \frac{1}{N_{rep}} \sum x_i^{(k)} x_j^{(k)} - \sigma_i \sigma_j \, .
    \label{eq:cov_def}
\end{equation}
Specifically, measurements are presented in terms of the best values $\sigma_i$  
accompanied by a number of uncertainties $s_i^{m}$ for each bin. 
As \cref{eq:cov_def} is not practically useful, we have to reconstruct 
the experimental covariance matrix via: 
\begin{equation}
    C^{exp}_{ij} = \delta_{ij} s_i^{stat} s_j^{stat} + \sum s_i^{sys,add} s_j^{sys,add} + \sum s_i^{sys,mult} s_j^{sys,mult} \sigma_i \sigma_j, 
    \label{eq:cov_by_bin_bin}
\end{equation}
where $s_i^{m}$ are split in: statistical $s_i^{stat}$, additive systematic 
$s_i^{sys,add}$ and multiplicative systematics $s_i^{sys,mult}$.
In case bin uncertainties are not symmetric, they are symmetrized with a standard
procedure and shifting the corresponding central value by  
\begin{equation}
    s_i = \frac{s_i^{+} + s_i^{-}}{2}, \quad \sigma_i  \to \sigma_i + \frac{s_i^{+} - s_i^{-}}{2} \, . 
    \label{eq:asy_unc}
\end{equation}

Finally, we can note that in the l.h.s of \cref{eq:cov_by_bin_bin}, multiplicative uncertainties
are multiplied with the best value of the measurements $\sigma_i$.
This can originate the D'Agostini bias~\cite{DAgostini:1993arp}, which 
might lead to a systematic underestimation of the underlying best 
fitting theory (PDF) estimate. 
To avoid this, we substitute in \cref{eq:cov_by_bin_bin}, $\sigma_i$ with 
a theoretical prediction of the value $T_i$ obtained with consistent parameters and 
a pre-determined PDF \cite{Ball:2009qv}.

\subsection{Treatment of theoretical uncertainties}
\label{sec:th_err}


Completely orthogonal to the experimental uncertainties 
are the uncertainties arising during the computation of theory predictions.
In the following, we delineate how scale variations can be used to 
estimate the missing higher order uncertainties (MHOU) arsing in pQCD
and propagated to a PDF fit by means of a covariance matrix.
Scale variation are justified by RGE invariance
and have the advantage that can be performed in a coherent way for all the very 
different theoretical predictions entering in a PDF fit.
Other independent types of theoretical uncertainties such as due to nuclear 
corrections \cite{Ball:2018twp} or the approximate N$^3$LO incomplete higher 
order uncertainties (cf. \cref{sec:an3lo_ihou}) can also be added to the
theory covariance matrix whenever needed.

\paragraph{MHOU from scale variations.}
As we have sketched in \cref{sec:theory_methodology}, theoretical predictions 
for high energy proton-proton scattering depend on two quantities that are
 computed perturbatively by expanding in $a_s(Q^2)$: the partonic cross-sections 
or coefficient functions, \cref{eq:pert_dis_coeff}, and the anomalous dimensions,
\cref{eq:ad_def}, that determine the scale dependence, of the PDF. 
The MHOU on the predictions is due to the truncation of these perturbative 
expansions at a given order.

In principle, if a VFNS (see \cref{sec:fns}) is used, a further MHOU is
introduced by the truncation of the perturbative expansion of the
matching conditions that relate PDFs in schemes with a different number
of active flavors. If one is interested in precision LHC phenomenology, 
then physics predictions are produced in an $n_f=5$ scheme, but PDFs are 
also determined by comparing to data predictions whose vast majority is 
computed in the $n_f=5$ scheme. Hence, the matching uncertainties only affect 
the small amount of data below the bottom threshold or charm threshold. 
The MHOU related to the matching conditions are thus subdominant 
and we neglect them here.

We thus focus on MHOUs for the hard cross-sections and anomalous dimensions.
For each perturbative result MHOUs are obtained by producing various expansions,
that differ by the subleading terms that are generated when varying the 
scale at which the strong coupling is evaluated. 
Given a perturbative quantity, we construct a scale-varied N$^k$LO 
coefficient function
\begin{equation}
    \label{eq:renscalvar}
    \bar C(a_s(Q^2),\rho) = a_s^m(Q^2) \sum_{j=0}^k \left(a_s(Q^2)\right)^{j} \bar C_j(\rho)
\end{equation}
by requiring that
\begin{equation}
  \bar C(a_s(\rho Q^2),\rho)=C(a_s(Q^2))\left[1+{\cal O}(a_s)\right],
    \label{eq:Cbarren}
\end{equation} 
which fixes the scale-varied coefficients $\bar C_j(\rho)$ in terms of the starting $C_j$. 
Here $\rho$ denotes the ratio between the physical scale $Q$ and the unphysical 
scale $\mu_F,\mu_R$ that appears upon imposing RGE invariance of the perturbative quantity. 
At any given order the difference between $C$ and $\bar C$ is taken as an estimate of
the missing higher orders, and  it may be  used for the construction of a covariance matrix.
Explicit expressions of the scale varied coefficient functions and anomalous dimension, 
up to N$^3$LO, can be found in \cite[App.~A]{NNPDF:2024dpb}.
In particular, we refer to renormalization scale variation when varying $\rho = \rho_r$
inside the partonic coefficient, while we consider variation of $\rho = \rho_f$ inside the
anomalous dimensions as factorization scale variations.
It is possible to prove that scale variations factorize during the DGLAP evolution \cite{NNPDF:2024dpb} 
and the scale-varied EKO can be constructed as
\begin{equation}
    \bar E(Q^2 \leftarrow Q_0^2,\rho_f )=K\left(a_s(\rho_f Q^2),\rho_f\right) E( \rho_f Q^2 \leftarrow Q_0^2 ) \, ,
    \label{eq:ekok}
\end{equation}
where at N$^k$LL (i.e.\ with the anomalous dimension computed at N$^k$LO) the additional 
evolution kernel $K(a_s(\rho_f Q^2),\rho_f)$ is obtained by imposing 
\begin{equation}
  \bar E(Q^2 \leftarrow Q_0^2,\rho_f )=E(Q^2 \leftarrow Q_0^2)\left[1+O(a_s)\right] \, ,
  \label{eq:ekoscale}
\end{equation}
and expanding
\begin{equation}
  K\left(a_s(\rho_f Q^2),\rho_f \right)= \sum_{j=0}^{k} \left(a_s(\rho_f Q^2)\right)^{j}K_j(\rho_f) \, .
  \label{eq:ks}
\end{equation}
\cref{eq:ekoscale,eq:ekok} mean that the effective scale-varied evolution kernel evolves 
from $Q_0^2$ to $\rho_f Q^2$, and then from $\rho_f Q^2$ back to $Q^2$, but 
with the latter evolution expanded out to fixed N$^k$LO.
Let us mention that the procedure to perform scale variations is not unique and 
different scale varied terms can either be included during the PDF evolution 
or in the coefficient function, as pointed out in Refs.~\cite{NNPDF:2024dpb,NNPDF:2019ubu}. 
In particular, in this work, we adopt the scheme B of Ref.~\cite{NNPDF:2019ubu}, 
taking the advantage that, both during renormalization and factorization scale variation, 
PDFs are always evaluated at the initial unvaried scale $Q_0^2$ which facilitate
the fitting procedure.

\paragraph{Construction of the covariance matrix.}

To construct a theory covariance matrix from scale varied predictions 
we follow Refs.~\cite{NNPDF:2019vjt,NNPDF:2019ubu}.
First, we define the shift in theory prediction for the $i$-th datapoint due
to renormalization and factorization scale variation
\begin{equation}
    \label{eq:delta}
    \Delta_{i}(\rho_{f}, \rho_{r}) \equiv T_{i}(\rho_f, \rho_r) - T_{i}(0, 0),
\end{equation}
where $T_{i}(\rho_f, \rho_r)$ is the prediction for the $i$-th datapoint obtained 
by varying the renormalization and factorization
scale by a factor $\rho_r$, $\rho_f$ respectively. 
%
Next, we choose a correlation pattern for scale variation, as
follows:
\begin{itemize}
  \item factorization scale variation is correlated for all datapoints, 
  because the scale dependence of PDFs is universal;
  \item renormalization scale variation is correlated for all datapoints 
  belonging to the same category, i.e.\ either the same observable 
  (such as, for instance, fully inclusive DIS cross-sections) or to 
  different observables for the same process (such as, different distributions
  of DY production).
\end{itemize}
Renormalization scale variations require a categorization of processes and, 
in this thesis, we adopt nine categories, namely: neutral-current deep-inelastic 
scattering (DIS NC), charged-current deep-inelastic scattering (DIS CC), and the 
following seven hadronic production processes: top-pair; $Z$, i.e.\ 
neutral-current Drell-Yan (DY NC);  $W^\pm$, i.e.\ charged current Drell-Yan (DY CC); 
single top; single-inclusive jets; prompt photon and dijet.

These choices correspond to the assumption that missing higher order terms are of 
a similar nature and thus of a similar size in all processes included in a 
given process category.
Different assumptions are consequently possible, for instance decorrelating 
the renormalization scale variation from contributions to the same process 
from different partonic sub-channels, or introducing a further variation 
of the scale of the process on top of the renormalization and factorization 
scale variation discussed above.

We then define a MHOU covariance matrix, whose matrix element between 
two datapoints $i$, $j$ is
\begin{equation}
    \label{eq:thcovmat}
    C^{MHOU}_{ij} = n_{m}\sum_{V_{m}} \Delta_{i}(\rho_{f}, \rho_{r_{i}})\Delta_{j}(\rho_{f}, \rho_{r_{j}}),
\end{equation}
where the sum runs over the space $V_m$ of the $m$ scale variations
that are included; the factorization scale $\rho_f$ is always varied
in a correlated way, the renormalization scales $\rho_{r_i}$, $\rho_{r_j}$ 
are varied in a correlated way ($\rho_{r_i}=\rho_{r_j}$) if datapoints 
$i$ and $j$ belong to the same category, but are varied independently if $i$ and $j$
belong to different categories, and $n_m$ is a normalization
factor.
The computation of the normalization factor is nontrivial
because it must account for the mismatch between the dimension of the
space of scale variations when two datapoints are in the same category
(so there is only one correlated set of renormalization scale
variations) and when they are not (so there are two independent sets
of variations). These normalization factors were computed for various
choices of the space $V_m$ of scale variations and for various values
of $m$ in Ref.~\cite{NNPDF:2019ubu}, to which we refer for details.

As custom in literature, we consider scale variation by a factor 2, so we set
\begin{equation}
 \label{eq:scales}
  \kappa_f=\ln \rho_f=\pm\ln 4\qquad
  \kappa_r=\ln \rho_r=\pm\ln 4.
\end{equation}
Also in this case, different choices for the space of allowed variations can be considered,
among others: the 9-point prescription, in which $\kappa_r$, $\kappa_f$ 
are allowed to both take all values in $\{- \ln4,0,\ln 4\}$, with
$m=8$ (eight variations about the central value); and 
the commonly used 7-point prescription, with $m=6$, which is obtained from the
former by discarding the two outermost variations, in which
$\kappa_r=+\ln 4$, $\kappa_f=-\ln 4$ or $\kappa_r=-\ln 4$,
$\kappa_f=+\ln 4$. 
In Ref.~\cite{NNPDF:2024dpb} we provide the explicit expression of 
the covariance matrices with the 7-point and 9-point prescription 
showing that, the two prescriptions lead to a similar behavior.

The set of assumptions including the correlation patterns of
renormalization and factorization scale variations, the process
categorization, the range of variation of the scales, and the specific
choice of variation points involves a certain degree of arbitrariness.
This is inevitable given that the MHOU is the estimate of the probability 
distribution for the size of an unknown quantity which has a unique true value, 
and thus it is intrinsically Bayesian. The only way to validate this kind of
estimate is by comparing its performance to cases in which the true value is known.
Finally, we acknowledge that there exist cases where scale variations are known 
to fail in the estimate of MHOU. By construction, scale varied terms only include 
ingredients that are available at previous perturbative orders, so they will never 
be able to predict effects due to new partonic channels or due to higher logarithmic 
divergences appearing in the small or large-$x$ regions, which can also spoil 
the pQCD expansion.
This boundary can be seen as the theoretical counterpart of the limitation that 
we have on the finite kinematic coverage of experimental data and, it constrains 
the validity of the result we shall derive solely to the region where pQCD is 
a descriptive tool. 

\subsection{Fitting methodology}
\label{sec:nnpdf_fitting}


We now review the main aspects of the PDF fitting methodology 
adopted through this work: the Monte Carlo replica method and the neural 
network workflow along with its implementation.

\paragraph{The Monte Carlo replica method.}
Given an ensemble of data ($D_i, \ i = \{1 \dots N_{\rm dat}\}$) and the corresponding PDF dependent theoretical 
predictions, in order to extract the best fitting PDFs, we begin by defining 
the likelihood function $\mathcal{L}(D_i | \boldsymbol{\theta})$.
This describes the probability of observing the given sample of data for a given set of parameters 
$\boldsymbol{\theta}$, which, in our case, are any complete set of parameters able to 
describe a PDF.
Under the assumption that data are Gaussian distributed $\mathcal{G}(\sigma,C)$ 
around the expected values $\sigma_i$ with a covariance ${\rm cov}_{ij}$, we can 
write the likelihood as
\begin{equation}
    \mathcal{L}(D_i | \boldsymbol{\theta})
    \propto  \text{exp} \left [- \frac{1}{2} (T_i(\boldsymbol{\theta}) - \sigma_i) \ {\rm cov}_{ij} \ (T_j(\boldsymbol{\theta}) - \sigma_j) \right ]
    = \text{exp} \left [- \frac{1}{2} \chi^2(\boldsymbol{\theta}) \right ] \,,
    \label{eq:likihood}
\end{equation}
where $T_i(\boldsymbol{\theta})$ are the theoretical predictions evaluated with the PDF 
we aim to determine.
By the application of Bayes theorem, we can see that the posterior 
distribution describing the parameters $\boldsymbol{\theta}$ given the data can be obtained 
by maximizing \cref{eq:likihood}. 
For practical reasons, it is more convenient to minimize the argument of the exponential, 
i.e. the $\chi^2$, which, in summary, describes how well the theoretical predictions model the data. 

Since experimental data and theory prediction are not 
exact we adopt a Monte Carlo replica method to propagate uncertainties to the PDF parameter space
\cite{Ball:2008by}. 
The Monte Carlo replica method proceed as follows: for all the datapoints, we construct 
an artificial replica $\sigma_i^{(r)}$ from the distribution $\mathcal{G}(\sigma,C)$
by fluctuating the central data via 
\begin{equation}
    \sigma_i^{(r)} = \sigma_i + a_i^{(r)} L_{ij}, \quad {\rm cov}_{ij} = L_{ik} L_{kj}^T \, ,
    \label{eq:replica_def}
\end{equation}
where $a_i^{(r)}$ is a normally distributed random array and $L_{ik}$ are obtained 
form the Cholesky decomposition of the covariance matrix. ${\rm cov}_{ij}$ can include both 
experimental and theoretical uncertainties. 
We substitute these fluctuated values in the $\chi^2$ definition \cref{eq:likihood} 
obtaining the loss function $\chi^{(r),2}$, which is then evaluated during the minimization.
We then repeat the sampling procedure of \cref{eq:replica_def} until the distribution 
of the best fitting parameters is sufficiently populated, 
typically 100 replicas are adequate to describe the full PDFs distributions.

As we will see later, since we are parametrizing our unknown PDF parameter space
with a neural network, to avoid over-fitting, we introduce a cross validation 
technique, splitting the data into two categories: the validation and training set
\cite{Ball:2014uwa}.
During the minimization we seek for the minimum of $\chi^{(r),2}$ over 
the training set, while evaluating the stopping criteria 
at subsequent iterations on the validation set (look-back stopping).
Note that, the training and validation split is preformed independently 
for each replica and, datapoints of the same datasets can also be split.

\paragraph{Feed forward neural networks.}
The constraints imposed by QCD do not allow us to determine a precise 
functional for of the PDFs on the all $x$ domain and fixed scale $Q$.
Thus, to perform a fit, it is important to select a sufficiently
flexible PDF parametrization and avoid introducing a parametrization bias 
which can distort the result drastically. 
Feed forward neural networks provide a useful tool to address this problem.
From a Machine Learning prospective the task of PDF fitting
can be classified as a supervised learning problem, 
where the input data are labelled and, the goal is to reconstruct a mapping 
which has to accurate if the output is known (data region), and unbiased 
as possible where the output is not known (extrapolation region). 

Neural networks are inspired by the biological mechanism behind the human brain. 
They consist in a relation graph, where each basic unit, called \textit{artificial neuron}, 
is connected to others by a precise quantitative rule.
In a feed forward network, neurons are organized in layers where each element 
is connected to the previous by an activation function and provide as output, 
a real number which is then used to weight the response of subsequent neuron layer.
Thanks to their flexibility the neural networks are able to continuously update their 
response and learn hidden patters present in the data used for training. 
In particular, it has been proven that a single layer is sufficient to represent any 
function within the range of  the given inputs \cite{balazs:2001}.
The way in which information flows inside the network is specified by a set of 
hyperparameters that control, among others, the activation function, 
the number of neurons and layers, the learning rate.

As mentioned, the forward propagation is given by the recursive evaluation 
of the activation function. 
Starting from an array of inputs point $\boldsymbol{x}$ (the PDF $x$-grid in our case),
for each layer $l$ the weight of the subsequent neurons ${\rm NN}^{(l+1)}_j$ is given by 
\begin{equation}
    {\rm NN}^{(l+1)}_j \left( \boldsymbol{x}; \boldsymbol{w}^{(l)}, \boldsymbol{\xi}^{(l)} \right) = \phi \left( 
        \sum_{i=1}^{n_l} w^{(l)}_i {\rm NN}^{(l)}_i \left( \boldsymbol{x}; \boldsymbol{w}^{(l-1)}, \boldsymbol{\xi}^{(l-1)} \right) 
        + \xi^{(l)}_i \right) \, ,
    \label{eq:neuron} 
\end{equation}
where $\phi$, the activation function, contains two degrees of freedom: 
a multiplicative weight ($w^{(l)}_i$) and a linear bias ($\xi^{(l)}_i$), 
which are usually normalized in a unit interval.
The sum, in \cref{eq:neuron}, runs over the number of connected nodes, 
and the recursive evaluation requires and initial boundary condition.

On the other hand, backward propagation corresponds to the moment in which the network 
trains and learns the pattern of the target data. This is fixed by the optimization 
of a loss function, i.e. the $\chi^2$ in our case.
At each minimization step, starting from the final layer, it is possible 
to compute weight and bias values ($\boldsymbol{w}^{(l)}, \boldsymbol{\xi}^{(l)}$) 
that leads to lower loss and, by a chain rule, update all the previous layers of the network.

The specific SGD optimizer, its settings, the training and validation fraction, the neural network 
architecture are all tunable parameters of the fitting methodology that can be determined via the 
hyperoptimization.
In order to assess the independence of the result on the choice of these hyperparameters while 
adopting an efficient methodology, one needs to scan many hyperparameter combinations
and test their performance on different subset of the data.

In this thesis, for the unpolarized fits, we adopt the same hyperparameters selected with the 
$K$-folding procedure described in \cite[Sec.~3.3]{NNPDF:2021njg}; while for the polarized 
fits of \cref{chap:pol} we perform a different hyperoptimization as described in \cref{subsec:pol_hyperoptimization}.
The former, adopts hyperbolic tangent activation function and 
a stochastic gradient descent (SGD) algorithm, called \texttt{nADAM} 
implemented in \texttt{TensorFlow}~\cite{tensorflow2015:whitepaper}, 
for which the backward propagation reduces essentially to a computation of 
derivatives $\chi^2$ in terms of $\boldsymbol{w}^{(l)}, \boldsymbol{\xi}^{(l)}$.
\texttt{nADAM} being based on a numerical minimization, ensures good efficiency
with respect to other minimizer as genetic algorithms and, by adapting the learning rate 
based on the previous iterations, it reduces the possibility of being trapped in local minima.

\begin{figure}[!t]
    \centering
    \includegraphics[width=0.75\textwidth]{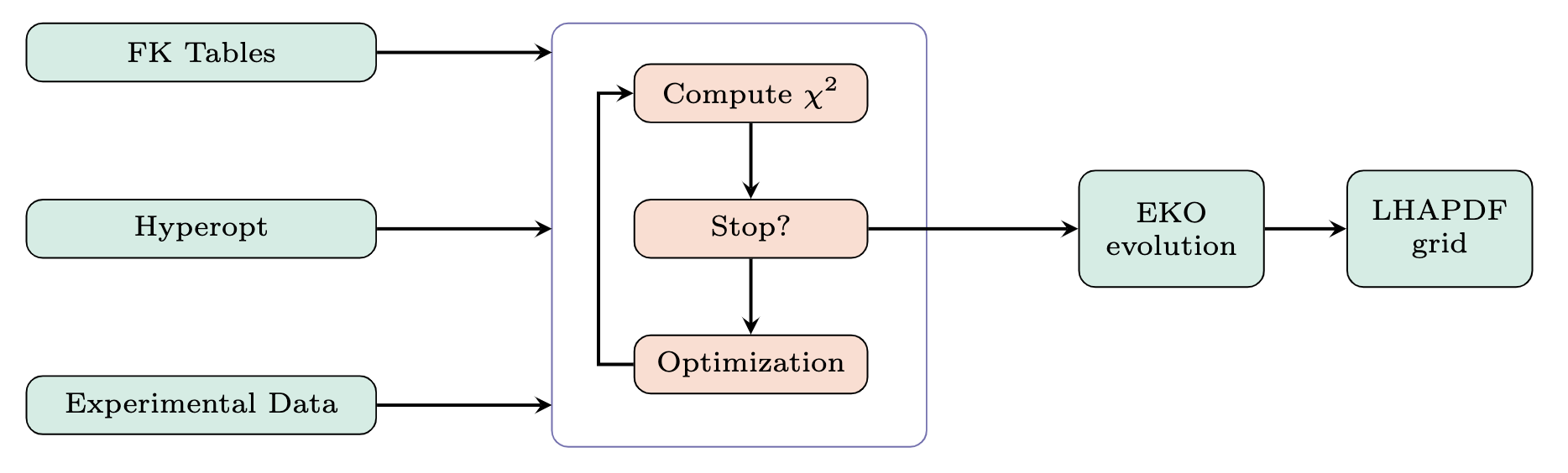}
    \caption{
        Diagrammatic representation of the NNPDF fitting framework. 
        The blue box contains the minimization of the $\chi^2$ figure of merit, 
        whose computation is illustrated in \cref{fig:nnpdf_chi2_eval}.
        From Ref.~\cite{NNPDF:2021njg}.
    }
    \label{fig:nnpdf_workflow}
\end{figure}

\paragraph{Fitting strategy.}
\cref{fig:nnpdf_workflow} summarizes the workflow of the NNPDF fitting 
framework. 
%
%
The main three inputs are given by the theoretical calculation encoded in the FK-tables, 
the experimental data and the set of optimal hyperparameters which determine 
the neural network and the optimization algorithm. 
These are used to compute (and optimize) the figure of merit $\chi^2$ over different
Monte Carlo replicas.
The algorithm make use of the cross validation technique to avoid over-fitting, 
impose necessary constraints and, provide as output
a set of best fitting PDF grids for each replica.
Finally, we can obtain the all scale PDFs running the DGLAP evolution with the program 
\eko{} and dump the final result as a standard \lhapdf{} set.

\begin{figure}[!t]
    \centering
    \includegraphics[width=0.90\textwidth]{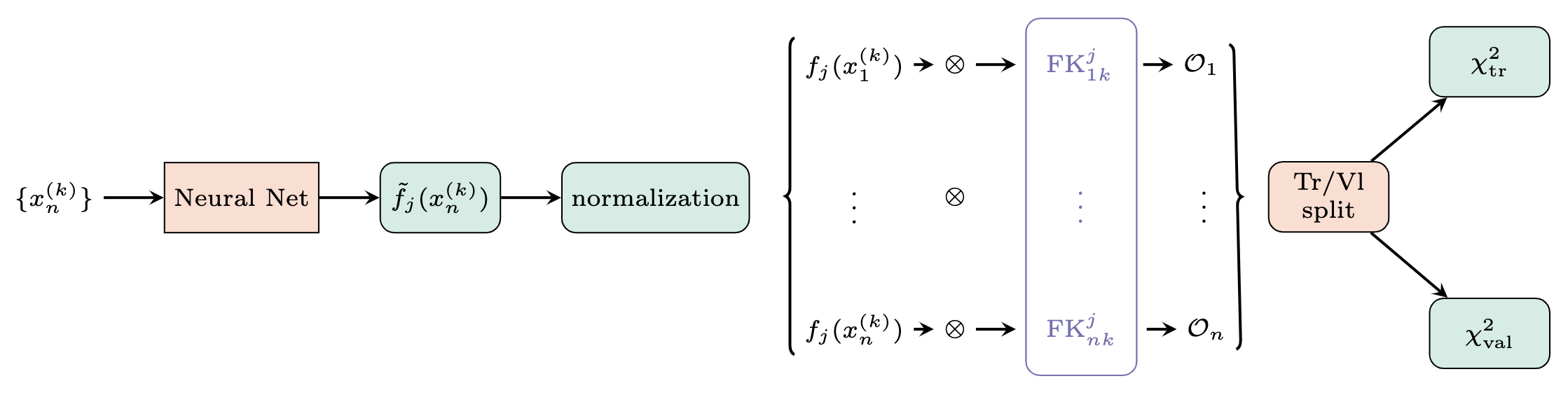}
    \caption{
        Diagrammatic representation of the calculation of the $\chi^2$ 
        in the NNPDF fitting framework as a function of the values of $\{x^{(k)}_n\}$ 
        for the different datasets. Each block indicates an independent component. 
        From Ref.~\cite{NNPDF:2021njg}.
    }
    \label{fig:nnpdf_chi2_eval}
\end{figure}

More specifically, see \cref{fig:nnpdf_chi2_eval}, the $\chi^2$ computation is preformed  
starting from the initial scale PDFs which are parametrized in  
the evolution basis (\cref{eq:evol_basis}) at a given scale $Q_0$ and related to the 
initial $x$-grid through the neural network.
PDF parametrizations are normalized to match theoretical constraints, such as sum rules and 
integrability and then convolved with the FK-tables to match the experimental observables.
Finally, the datapoints are split in training and validation sets and we proceed to update 
the neural network parameters until the stopping criteria are matched.

\section{The NNPDF 4.0 PDF set}
\label{sec:nnpdf40}


The previous sections summarize the key concepts of the NNPDF 
methodology. In principle, these apply to a generic polarized and 
unpolarized PDF determination of the NNPDF family but, the result
that we shall present in \cref{chap:ic,chap:an3lo,chap:pol} are build
upon a specific unpolarized PDF set, the NNPDF4.0 PDF set.
In particular, in \cref{chap:ic,chap:an3lo} we assume the NNLO set 
with fitted charm to be our baseline, while in \cref{chap:pol} 
we adopt the NNLO set with perturbative charm as boundary condition.
This section, by describing a concrete example of a PDF fit,
aims to bridge the introductory part of the thesis,
with the chapters focused on the actual results.

NNPDF4.0~\cite{NNPDF:2021njg} has been a major release that 
has improved the previous NNPDF3.1~\cite{Ball:2017nwa}, 
both upon kinematic coverage, including systematically LHC data,
and upon fitting methodology which has been supported by validation 
tools as closure tests and hyperoptimization. 
In Ref.~\cite{NNPDF:2024dpb} this determination has been complemented
with the inclusion of theoretical error, while a photon PDF 
has been determined in Ref.~\cite{NNPDF:2024djq} together with mixed
QED$\otimes$QCD effects.


We now proceed to \cref{sec:nnpdf40_param} specifying the PDF parametrization 
and the theoretical constraint adopted, then we recap the 
experimental data used in our framework (\cref{sec:nnpdf40_kinematic}) 
and we conclude in \cref{sec:nnpdf40_pdfs} with an overview of 
NNPDF4.0 PDF up to NNLO with and without MHOU.

\subsection{PDF parametrization and theoretical constrain}
\label{sec:nnpdf40_param}

\begin{figure}[!t]
    \centering
    \includegraphics[width=0.75\textwidth]{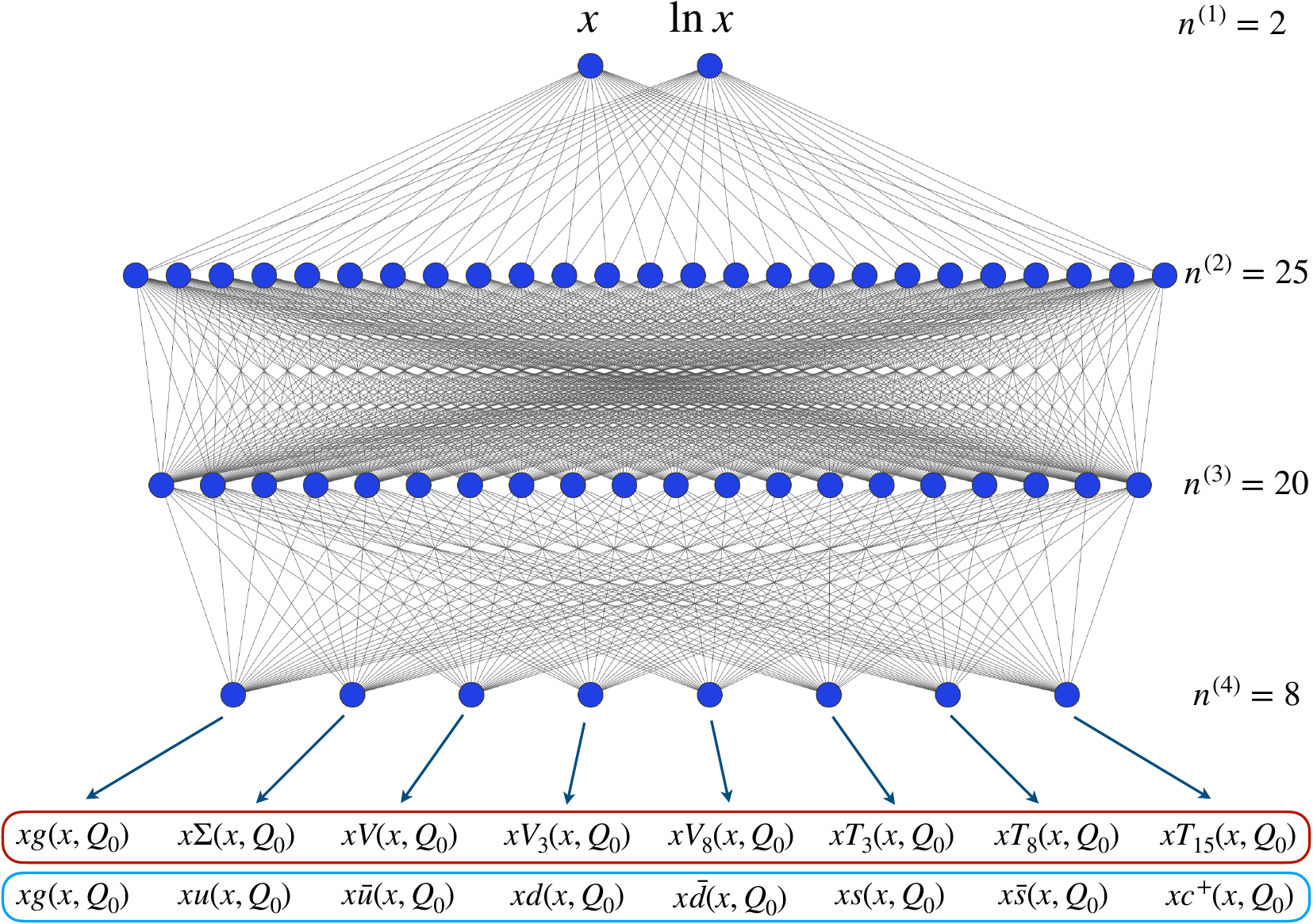}
    \caption{
        The neural network architecture adopted for unpolarized NNPDF4.0 PDF set. 
        A single network is used, whose  eight output values are the PDFs 
        in the evolution (red, default) or the flavor basis (blue box). 
        The architecture displayed corresponds to the optimal choice of the 
        hyperparameters in the evolution basis.
        From Ref.~\cite{NNPDF:2021njg}.
    }
    \label{fig:NNPDF_arch}
\end{figure}

Any PDF analysis requires to select a fitting scale $Q_0$, a PDF parametrization 
and a flavor basis, namely the choice of different flavor combinations that are 
determined independently during the fit. 
In NNPDF4.0 methodology we fit PDFs at the initial scale $Q_0 = 1.65~\text{GeV}$,
and we assume the evolution basis of \cref{eq:evol_basis} as the default, considering
8 different linear combinations
\begin{equation}
	\label{eq:fitted_basis}
    {f}_{k}=\{V,\, V_3,\, V_8,\, T_3,\, T_8,\, T_{15},\, \Sigma,\, g\} \, .
\end{equation}
This basis facilitates the implementation of sum rules and integrability constrain
but, fits in the standard flavor basis can also be performed and used to cross-check 
the final result.
Each of the flavor combinations is related to the output of the neural network via
\begin{equation}
	xf_k\left( x,Q_0; {\boldsymbol \theta} \right) = A_k\,x^{1-\alpha_k}(1-x)^{\beta_k}{\rm NN}_k(x;
        {\boldsymbol\theta}) \, ,\quad k=1,\ldots,8\,,
    \label{eq:evolution_basis_param}
\end{equation}
where ${\rm NN}_k(x;{\boldsymbol\theta})$ is the $k$-th output of a neural network, 
whose architecture is shown in \cref{fig:NNPDF_arch}, and ${\boldsymbol\theta}$ 
collectively indicates the full set of neural network parameters.

The polynomial part in \cref{eq:evolution_basis_param}, referred as \textit{preprocessing}, 
ensures that the PDF has the correct behavior in the large-$x$ and small-$x$ limit, speeding 
up the training of the neural network.
In particular, the factor $(1-x)^{\beta_k}$ guarantees the convergence at $x=1$;
$x^{1-\alpha_k}$ is based instead on Regge theory arguments~\cite{Roberts:1990ww},
%
%
which imply that the singlet and non-singlet have a different small-$x$ behavior: 
the non-singlet has a finite integral, while the singlet first moment diverges.
The small-$x$ preprocessing is not imposed in a flavor basis fit.
The numerical values of the parameters $\alpha_k$ and $\beta_k$ are determined
with an iterative procedure~\cite{Ball:2014uwa}, but are not fitted during the
neural network training.

Finally, the normalization constants $A_k$ are needed to impose the constraints
coming from sum rules.
At any scale $Q$, momentum conservation impose a constraint on the first moment 
of unpolarized quark and gluon total content
\begin{equation}
    \label{eq:mom_sum_rule}
    \int_0^1 dx\,x \left [ g\left(x, Q\right) + \Sigma\left(x, Q\right) \right ] = 1 \, , 
\end{equation}
while from proton charge conservation we can infer the three valence sum rules:
\begin{equation}
    \label{eq:valence_sum_rules}
    \int_0^1 dx\, V\left(x, Q\right) = \int_0^1 dx\, V_8\left(x, Q\right) = 3\,.  
    \quad
    \int_0^1 dx\, V_3\left(x, Q\right) = 1\,.
\end{equation}

\paragraph{Integrability and Positivity.}

Additional constraints on the allowed PDFs space, such as positivity and integrability
can be imposed during the fit, by means of Lagrange multipliers as explained below.
In this case the cost function is supplemented by 
\begin{equation}
	\label{eq:chi2pos_k}
	\chi^2_{\rm tot} (\boldsymbol{\theta}) \to \chi^2_{\rm tot} (\boldsymbol{\theta}) + \sum_{i=1}^{N_C} \lambda_i h_i (\boldsymbol{\theta}) \,,
\end{equation}
where the sum runs over the number of constraints $N_{C}$, $\lambda_i$ are a set constant 
and $h_i (\boldsymbol{\theta})$ are real functions of the PDFs parameters.
If the term $\lambda_i h_i (\boldsymbol{\theta})$ is allowed to be much larger than 
$\chi^2_{\rm tot} (\boldsymbol{\theta})$ for some specific PDF configurations,
this effectively act as a penalty and make such configurations less
favorable upon minimization.

We can distinguish two types of positivity conditions. 
First, the PDFs we aim to determine should have a physical meaning, 
i.e. for each flavor they should correspond to positive definite observables.
Following Ref.~\cite{NNPDF:2021njg}, one imposes positivity of the 
structure functions $F_2^u$, $F_2^d$, $F_2^s$, $F_2^c$, and $F_{L}$
and of the flavor-diagonal Drell-Yan rapidity
distributions $\sigma_{{\rm DY},u\bar{u}}$, $\sigma_{{\rm DY},d\bar{d}}$,
$\sigma_{{\rm DY},s\bar{s}}$.
\footnote{
    A positivity condition on the physical cross-section corresponds
    to a direct positivity condition on the PDF only at LO. 
    In this case, partonic matrix elements are $\delta(1-x)$,
    thus the physical cross-sections can be directly proportional to the PDF.
}
Second, in Ref.~\cite{Candido:2020yat} it has been shown that 
in the $\overline{MS}$ scheme at sufficiently large $Q^2$
massless PDFs are positive defined.
Thus, one can require the gluon, the up, down and strange quark 
and antiquark at $Q^2 = 5\, \text{GeV}^2$ to be positive
(see \cite[Sec.~3.1.3]{NNPDF:2021njg} for more details).
This additional constraints help the convergence of the fit 
especially in the very large-$x$ region.

Regarding the integrability conditions,
the momentum sum rule and valence sum rule imply respectively a vanishing small-$x$
limit of the second moment of $g,\Sigma$ and the first moment of $V, V_3, V_8$.
As observed in Ref.~\cite{NNPDF:2021njg}, to fulfill these 
conditions, it is sufficient to restrict the range of the small-$x$ preprocessing exponent 
\cref{eq:evolution_basis_param} to: $\alpha_i < 2$ for $g, \Sigma$ and 
$\alpha_i < 1$ for the other non-singlet combinations.
However, Regge theory arguments suggest also 
\begin{equation}
    \label{eq:cond_integ_Tx}
    \lim_{x\rightarrow 0}\,  xf_k(x,Q)= 0 \, ,\quad \forall~Q \, ,\qquad f_k=T_3,\,T_8 \, ,
\end{equation}
which are then imposed using Lagrange multipliers, evaluated at $x = 10^{-9}$ and 
penalize configurations with $T_3,T_8$ non-vanishing moments.
\subsection{Kinematic coverage}
\label{sec:nnpdf40_kinematic}

The data used in the NNPDF4.0 analysis \cite{NNPDF:2021njg}, in the 
subsequent updates \cite{NNPDF:2024dpb,NNPDF:2024nan,NNPDF:2024djq}
and in the related studies \cite{Ball:2022qks,NNPDF:2023tyk}
are discussed in details in \cite[Sec.~2]{NNPDF:2021njg}.
Here, we limit ourselves to list the type of processes, 
the experiments and the tools used to compute the corresponding predictions.

\begin{itemize}
    \item \texttt{Fixed-target DIS.}
        We include neutral current (NC) structure function data from NMC~\cite{Arneodo:1996kd,Arneodo:1996qe},
        SLAC~\cite{Whitlow:1991uw} and BCDMS~\cite{Benvenuti:1989rh}, fixed-target inclusive and 
        dimuon charged current (CC) cross-section data from CHORUS~\cite{Onengut:2005kv} 
        and NuTeV~\cite{Goncharov:2001qe,MasonPhD}.
        Theoretical predictions are computed with \yadism{} 
        at NNLO in QCD with massive corrections included as in the FONLL scheme 
        (cf.~\cref{sec:fns}).
    \item \texttt{Collider DIS.}
        We consider collider NC and CC cross-section data from HERA~\cite{Abramowicz:2015mha} 
        together with the reduced charm and bottom cross-section from H1 and 
        ZEUS~\cite{Abramowicz:1900rp,Aaron:2009af,Abramowicz:2014zub}.
        To compute the theoretical predictions, we adopt the same setting as fixed-target DIS data.
    \item \texttt{Fixed-target DY.}
        Among all the Fermilab data we select measurements from E605~\cite{Moreno:1990sf} 
        and E866,E906~\cite{Webb:2003ps,Towell:2001nh,Dove:2021ejl},
        as well as rapidity distributions from Tevatron CDF~\cite{Aaltonen:2010zza} and 
        D0~\cite{Abazov:2007jy,Abazov:2013rja,D0:2014kma}.
        Corresponding FK-tables are computed with {\tt Vrap}~\cite{Anastasiou:2003ds} at NLO 
        with the inclusion of  NNLO $K$-factors. 
    \item \texttt{Collider gauge boson production.}
        We encompass inclusive cross-sections, differential distributions 
        in the gauge boson invariant mass or rapidity 
        from ATLAS~\cite{Aad:2011dm,Aaboud:2016btc,Aad:2014qja,Aad:2013iua},
        CMS~\cite{Chatrchyan:2012xt,Chatrchyan:2013mza,Chatrchyan:2013tia,Khachatryan:2016pev}
        and LHCb~\cite{Aaij:2012mda,Aaij:2015gna,Aaij:2015vua,Aaij:2015zlq}.
        Data include central rapidity regions as well as more forward production; 
        in all the selected measurements the electroweak boson decays leptonically.
        We also include $Z$-boson transverse momentum production data from ATLAS~\cite{Aad:2015auj} 
        and CMS~\cite{Khachatryan:2015oaa}.
        Data are described with NLO calculations from \madgraph~\cite{Alwall:2014hca,Frederix:2018nkq}
        supplemented with NNLO $K$-factors from {\tt FEWZ}~\cite{Gavin:2010az,Gavin:2012sy,Li:2012wna} 
        and {\tt DYNNLO}~\cite{Catani:2007vq,Catani:2009sm}.
    \item \texttt{Collider gauge boson plus jets.}
        Among the available measurements, we consider differential distributions 
        of $W$-boson production with $N_{\rm jets}\geq 1$ from 
        ATLAS~\cite{Aaboud:2017soa}. We select the distribution differential in the transverse
        momentum of the $W$ boson, $p_T^W$.
        Theoretical predictions are determined at NLO, with {\tt MCFM}, 
        while NNLO, QCD corrections are implemented by means of 
        $K$-factors~\cite{Boughezal:2015dva,Ridder:2015dxa}.
    \item \texttt{Single inclusive jet and dijet production.}
        Measurements of single inclusive jet production
        comprehend differential distributions in the transverse momentum, 
        $p_T^{\rm jet}$, and of the rapidity, $y^{\rm jet}$, of the jet
        from ATLAS~\cite{Aaboud:2017dvo} and CMS~\cite{Khachatryan:2016mlc}.
        For dijet production instead we consider double differential in 
        the dijet invariant mass $m_{jj}$ and in the absolute difference of the 
        rapidities of the two jets $y^*$ from ATLAS~\cite{Aad:2011fc,Aad:2013tea} 
        and CMS~\cite{Chatrchyan:2012bja,Sirunyan:2017skj}.
        The accompanying predictions are computed with \nnlojet~\cite{Gehrmann:2018szu}
        at full NLO with NNLO $K$-factors.
    \item \texttt{Top quark pair production.}
        We include differential cross-section in the top pair rapidity and/or 
        invariant mass from ATLAS~\cite{Aaboud:2016iot,Aad:2020tmz,Aaboud:2016pbd,Aad:2015mbv} 
        and CMS~\cite{Sirunyan:2017azo,Sirunyan:2018wem,Sirunyan:2018ucr,Khachatryan:2015oqa}. 
        We use \madgraph{} to compute NLO predictions and NNLO
        corrections are determined from publicly available {\tt FastNLO}
        tables~\cite{Czakon:2017dip,Czakon:2019yrx}.
    \item \texttt{Single top quark production.}
        We consider data from ATLAS~\cite{Aad:2014fwa,Aaboud:2017pdi,Aaboud:2016ymp} 
        and CMS~\cite{Chatrchyan:2012ep,Khachatryan:2014iya,Sirunyan:2016cdg}.
        These encompass ratio of the top to antitop inclusive cross-sections,
        differential distributions in the top or antitop quark rapidity
        and sum of top and antitop inclusive cross-sections.
        Similarly to $t \bar{t}$ data we compute predictions with \madgraph{}
        at NLO and account for NNLO effects with $K$-factors 
        as in Refs.~\cite{Berger:2016oht,Berger:2017zof}.
    \item \texttt{Direct Photon production.}
        In this category, we select the ATLAS measurements~\cite{Aad:2016xcr,Aaboud:2017cbm}. 
        The measurements are provided for the cross-section differential
        in the photon transverse energy $E_T^\gamma$ in different bins of the photon 
        pseudorapidity $\eta_\gamma$ and compared to theoretical
        predictions from {\tt MCFM}. NNLO QCD corrections are
        incorporated by means of the $K$-factors computed in~\cite{Campbell:2016lzl}
\end{itemize}

Whenever possible, all the experimental correlations are taken into account 
and we select observables that are less affected by higher order corrections.
For DIS data, we apply the kinematic cuts on $Q^2 \geq 3.5~\text{GeV}^2$
and $W^2 \geq 12.5~\text{GeV}^2$, assuming that outside 
this region, higher-twist effects might become relevant.
For the hadronic dataset, the precise adopted cuts are listed in \cite[Tab.~4.1]{NNPDF:2021njg}.
These aim to remove specific points for which the $K$-factor 
approach is poorly reliable due to numerical instabilities, 
the electroweak corrections are relevant, for instance in the large mass tails
of the DY distributions or, resummation effects are large, 
as in the small-$p_T$ regions.
If correlations are not available, the dijet data are preferred over the 
single jet measurements. 
The total number of datapoints included in the baseline is 
therefore $4426$ at NLO and $4618$ at NNLO.

Approximately $30~\%$ of the data points listed above involve measurements on
deuterium and heavy nuclear targets, but in our analysis we aim 
to determine a proton PDF.
In order to take this effect into account, we supplement the experimental uncertainties with 
a covariance matrix, computed as in \cref{sec:th_err}, now starting from the shifts between predictions
evaluated on a proton and on a nuclear PDF from the {\tt nNNPDF2.0} set \cite{AbdulKhalek:2020yuc}.

\begin{figure}[!p]
    \centering
    \includegraphics[width=0.88\textwidth]{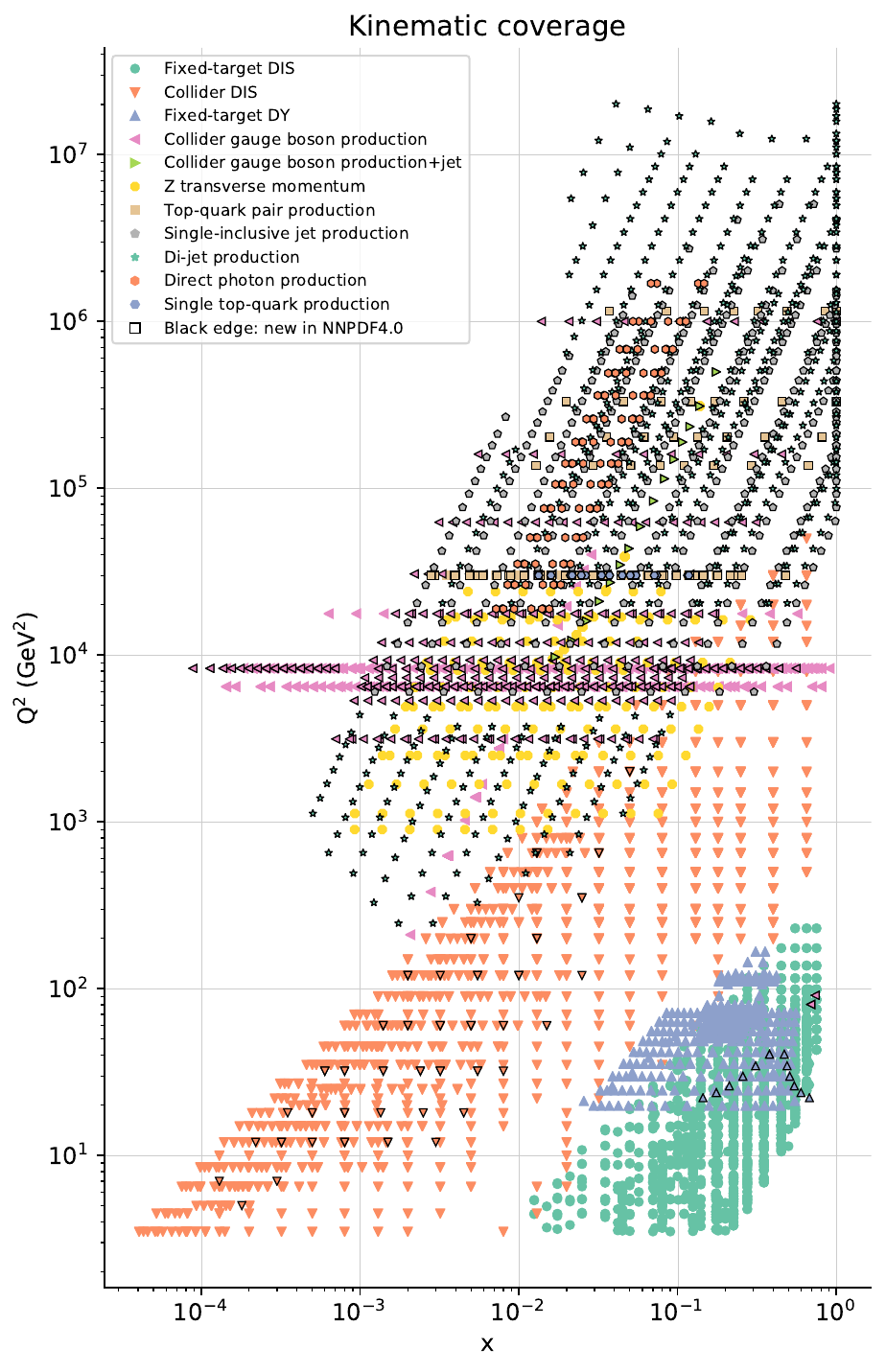}
    \caption{
      The kinematic coverage in the $(x,Q)$ plane covered by the 
      4618 cross-sections used in the NNPDF4.0 PDF set.
      These cross-sections have been classified into the main 
      different types of processes entering the global analysis.
      From Ref.~\cite{NNPDF:2021njg}.
    }
    \label{fig:kinplot}
\end{figure}

The NNPDF4.0 analysis comprehends a vast number of new datapoints
from LHC measurements with respect to previous PDF analysis.
This is visible in \cref{fig:kinplot}, where we display the kinematic coverage 
in the $(x,Q^2)$ plane.
%
%
%
While the majority of datapoints still belongs to the DIS data 
(roughly $50~\%$), the hadronic processes are essential to constrain
specific PDFs flavor combinations or kinematic regions.
Drell-Yan data provide a handle on the quark-antiquark flavor separation 
and allow the determination of the valence distributions, especially in 
the peak region at $x \geq 0.05$. 
In particular, LHC forward measurements are sensitive both to high $Q^2$ 
small-$x$ and large-$x$ regions, providing an independent constrain
with respect to the information carried by small-$x$, low-$Q^2$ HERA data and 
the fixed target DY data.
Jets data are crucial to shape the gluon PDF, 
with dijet being more constraining also on the small-$x$ region.
Top data are found to have a mild impact of the up and down PDFs,
while being potentially sensitive also to a bottom quark PDF.
Finally, the direct photon measurements can affect the mid-$x$
gluon PDF.

\subsection{NNLO baseline and MHOU set}
\label{sec:nnpdf40_pdfs}

\begin{figure}[!t]
  \centering
  \includegraphics[width=0.49\linewidth]{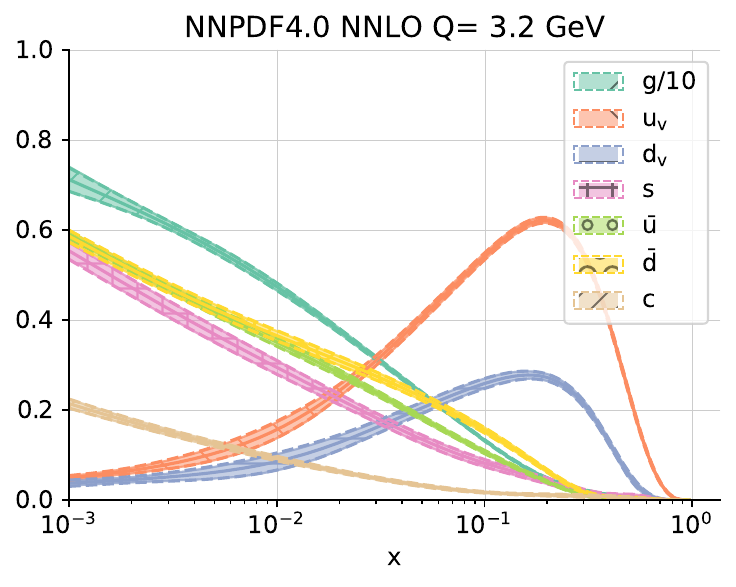}
  \includegraphics[width=0.49\linewidth]{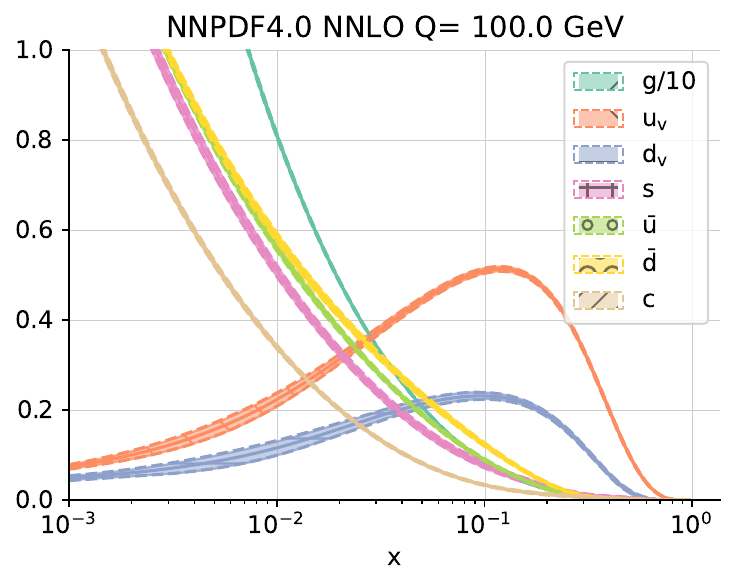}\\
  \caption{
      The NNPDF4.0 NNLO PDFs at $Q=3.2$~GeV (left) and $Q=10^2$~GeV (right).
      We display PDFs without MHOU.
      From Ref.~\cite{NNPDF:2021njg}.
  }
  \label{fig:pdg}
\end{figure}

We now turn to the description of the NNPDF4.0 PDFs set. 
For simplicity, we focus on the NNLO PDF
set with fitted charm, discussing quality of the fit and then the impact 
of the inclusion of MHOU. 
The final NNPDF4.0 NNLO PDFs are shown in \cref{fig:pdg} both at a low 
($Q=3.2~\text{GeV}$) and a high ($Q=100~\text{GeV}$) scale.
The relative uncertainty of almost all the NNLO baseline PDFs is of the 
order of 1-2 \% in the region probed by experimental data.
This underscores the importance of treating theoretical errors appropriately 
and studying the N$^3$LO effects.
The NNPDF4.0 set is consistent with the previous NNPDF3.1 set, but it 
improves the PDF accuracy by a factor of $30-50~\%$ in most of the 
kinematic regions probed by the data. 

\paragraph{Fit quality.}

In \cref{tab:chi2_TOTAL_nnlo_mhou} we report the number of
data points and the $\chi^2$ per data point in the NLO and NNLO NNPDF4.0
PDF determinations before and after inclusion of MHOUs. 
%
When MHOUs are included, the theory covariance matrix is computed with a 
7-point prescription. 
Note that the MHOU contribution is respectively excluded or included
both in the definition of the $\chi^2$ used by the NNPDF
algorithm (i.e.\ for pseudodata generation and in training and validation
loss functions), and in the covariance matrix used to compute the values given 
in \cref{tab:chi2_TOTAL_nnlo_mhou}.
Datasets are aggregated according to the process categorization of \cref{sec:th_err}: 
correlations among different groups are lost when showing $\chi^2$ values for data subsets, 
thus, the total $\chi^2$ shown in the last row is not the weighted average 
of individual values.

Fit quality is generally good, with the total $\chi^2$ being closer to the unity 
for the NNLO fits.
One can notice the clear improvement in the description of the 
data once NNLO corrections are included. 
This is visible in particular for the NC DIS, DY and top pair which 
are high precision measurements; on the other hand the
low $\chi^2$ of single top and prompt photon data can be explained 
with a larger experimental uncertainty.
The biggest impact of MHOU is visible in the inclusive and dijet data.

\begin{table}[!t]
  \scriptsize
  \centering
  \renewcommand{\arraystretch}{1.4}
  \begin{tabularx}{\textwidth}{Xrcccc}
  \toprule
    \multirow{2}{*}{Dataset}
  & \multirow{2}{*}{$N_{\rm dat}$ }
  & \multicolumn{2}{c}{NLO}
  & \multicolumn{2}{c}{NNLO}  \\
  & 
  & $C+S^{\rm (nucl)}$ & $C+S^{\rm (nucl)}+S^{\rm (7pt)}$
  & $C+S^{\rm (nucl)}$ & $C+S^{\rm (nucl)}+S^{\rm (7pt)}$ \\
  \midrule
  DIS NC
  & 2100
  & 1.30 & 1.22
  & 1.23 & 1.20 \\
  DIS CC
  &  989 
  & 0.92 & 0.87
  & 0.90 & 0.90 \\
  DY NC
  &  736 
  & 2.01 & 1.71
  & 1.20 & 1.15 \\
  DY CC
  &  157 
  & 1.48 & 1.42
  & 1.48 & 1.37 \\
  Top pairs
  &   64 
  & 2.08 & 1.24
  & 1.21 & 1.43 \\
  Single-inclusive jets
  &  356 
  & 0.84 & 0.82
  & 0.96 & 0.81 \\
  Dijet
  &  144 
  & 1.52 & 1.84
  & 2.04 & 1.71 \\
  Prompt photons 
  &   53 
  & 0.59 & 0.49
  & 0.75 & 0.67 \\
  Single top
  &   17 
  & 0.36 & 0.35
  & 0.36 & 0.38 \\
  \midrule
  Total
  & 4616 
  & 1.34 & 1.23
  & 1.17 & 1.13 \\
\bottomrule
\end{tabularx}

  \vspace{0.3cm}
  \caption{
    The number of data points and the $\chi^2$ per data
    point for the NLO and NNLO NNPDF4.0 PDF sets without and
    with MHOUs. Datasets are grouped according to
    the process categorization of \cref{sec:th_err}.
  }
  \label{tab:chi2_TOTAL_nnlo_mhou}
\end{table}

\cref{tab:chi2_TOTAL_nnlo_mhou} show that, upon inclusion of the MHOU covariance matrix, 
the total $\chi^2$ decreases for both the NLO and NNLO fits but, the decrease 
is more substantial at NLO. Even after inclusion of the MHOU, the
NLO $\chi^2$ remains somewhat higher than the NNLO one. Inspection of
specific datasets shows that this is in fact due to a few number of datasets 
(for e.g. ATLAS low-mass Drell-Yan), for which NNLO corrections are
substantially underestimated by scale variation. 
However, for the majority of datapoints and of process categories, 
the MHOU covariance matrix correctly accounts for the mismatch between 
data and theory predictions at NLO due to missing NNLO terms.

\begin{figure}[!t]
  \centering
  \includegraphics[width=\textwidth]{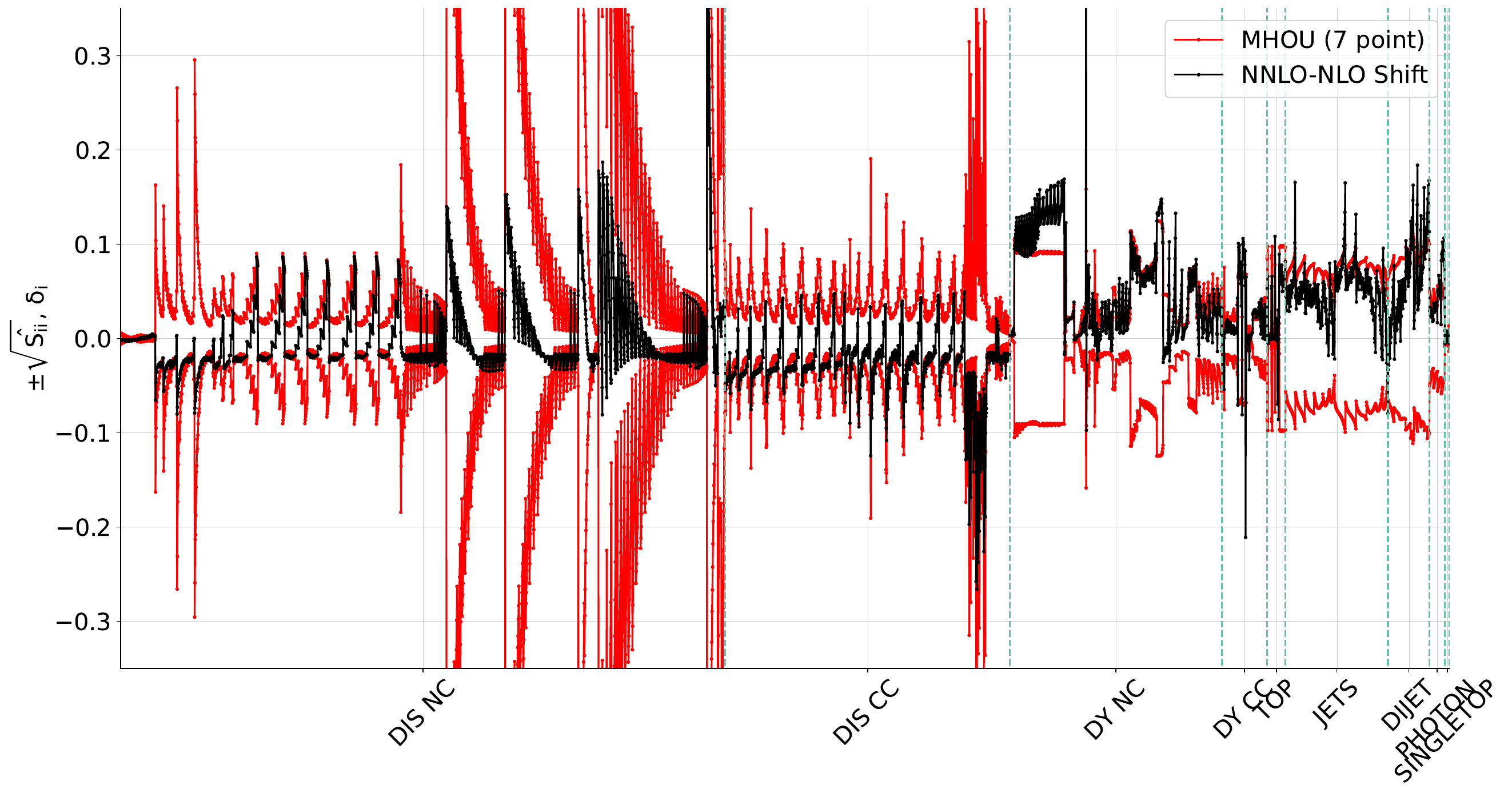}
  \caption{
      Comparison of the symmetrized NLO MHOU $\pm{\sqrt{\hat{S}_{ii}}}$, 
      defined as the square root of the diagonal element of the covariance 
      matrix normalized to the value of the theory prediction (red), 
      and the normalized NNLO-NLO shift $\delta_i$ of \cref{eq:shift} (black) 
      for all datapoints. 
      Results are obtained with the 7-point prescription.
  }
  \label{fig:shift_diag}
\end{figure}

This picture can be validated with the plot of \cref{fig:shift_diag}, 
where we compare the shifts
\begin{equation}
  \label{eq:shift}
  \delta_{i} = \frac{T_{i}^{\text{NNLO}} - T_{i}^{\text{NLO}}}{T_{i}^{\text{NLO}}},
\end{equation}
to the theoretical uncertainty on individual points (also normalized). 
The latter is given by the square root of the diagonal entries of the normalized NLO MHOU 
covariance matrix 
\begin{equation}
  \label{eq:normcov}
  \hat{S}_{ij}^{\text{NLO}} = \frac{S_{ij}^{\text{NLO}}}{T_{i}^{\text{NLO}}T_{j}^{\text{NLO}}}\, . 
\end{equation}

It is clear that, for DIS, 7-point scale variations at NLO provide 
a very conservative uncertainty estimate that significantly overestimates the 
NNLO-NLO shift. 
On the other hand, for hadronic processes the shift and scale variation estimate 
are generally comparable in size.
Only for DY, scale variations perform less well, with instances of underestimation of
the shift. Whereas this may suggest adjusting the range of scale
variation on a process-by-process basis, it is unclear to which extent the NLO
behavior could be generalized to higher orders.

\begin{figure}[!t]
  \centering
  \includegraphics[width=0.45\textwidth]{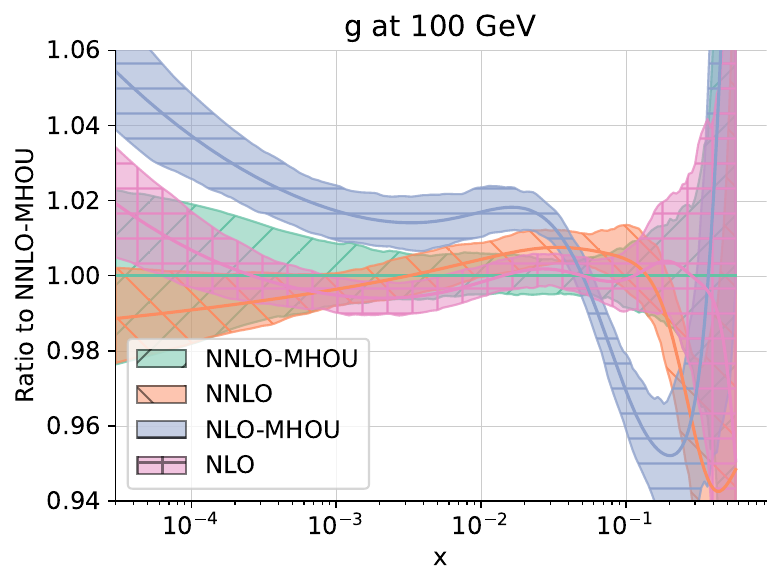}
  \includegraphics[width=0.45\textwidth]{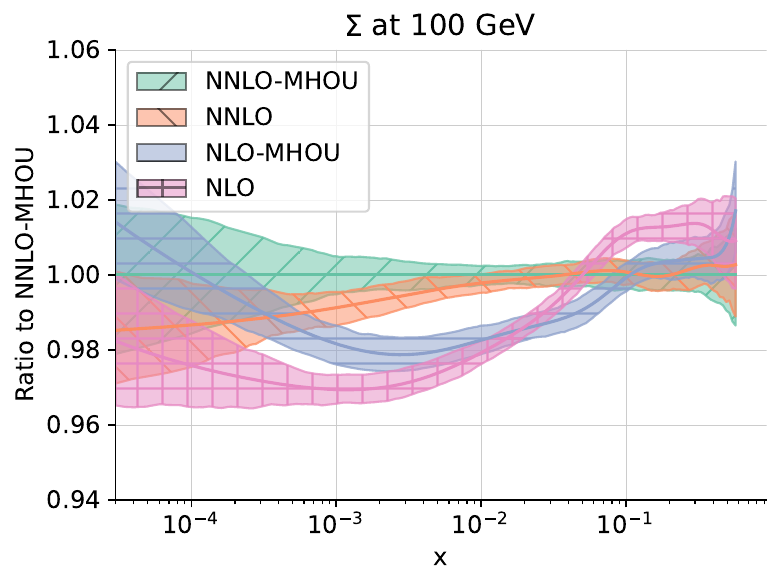} \\
  \includegraphics[width=0.45\textwidth]{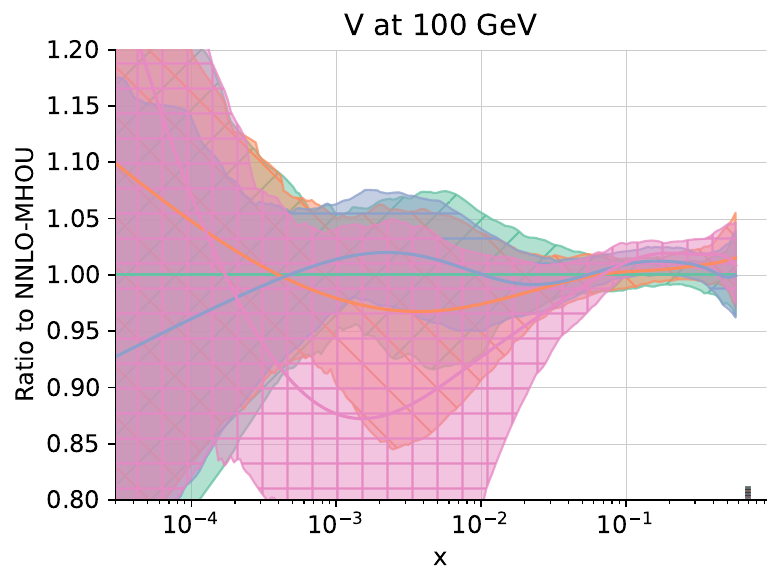}
  \caption{
    The NLO and NNLO NNPDF4.0 PDFs with and without MHOUs at $Q=100$~GeV. 
    We display the gluon, singlet, and valence PDFs. 
    All curves are  normalized to the NNLO with MHOUs. The
    bands correspond to one sigma uncertainty.
  }
  \label{fig:nnlo_nlo_pdfs}
\end{figure}

\paragraph{Perturbative convergence and MHOU uncertainties.}
Individual PDFs at NLO and NNLO, with and without MHOUs, are compared 
in \cref{fig:nnlo_nlo_pdfs} at $Q=100$~GeV. 
We show the gluon, singlet and valence distributions (reference 
for the other non singlet quantities), 
all shown as a ratio to the  NNLO PDFs with MHOUs. 
%
The change in central value due to the inclusion of MHOUs is generally 
moderate at NNLO; at NLO it is significant for the
gluon and singlet, but quite moderate for all other PDF
combinations.

\begin{figure}[!t]
  \centering
  \includegraphics[width=0.45\textwidth]{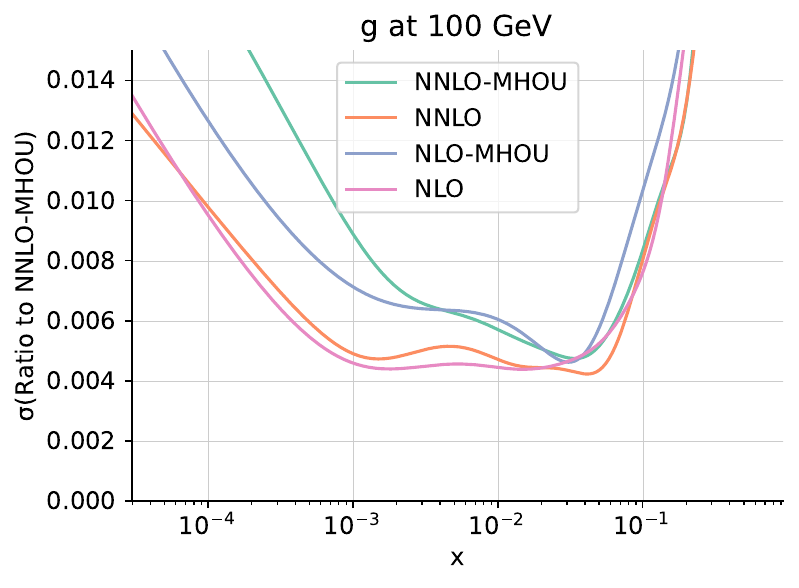}
  \includegraphics[width=0.45\textwidth]{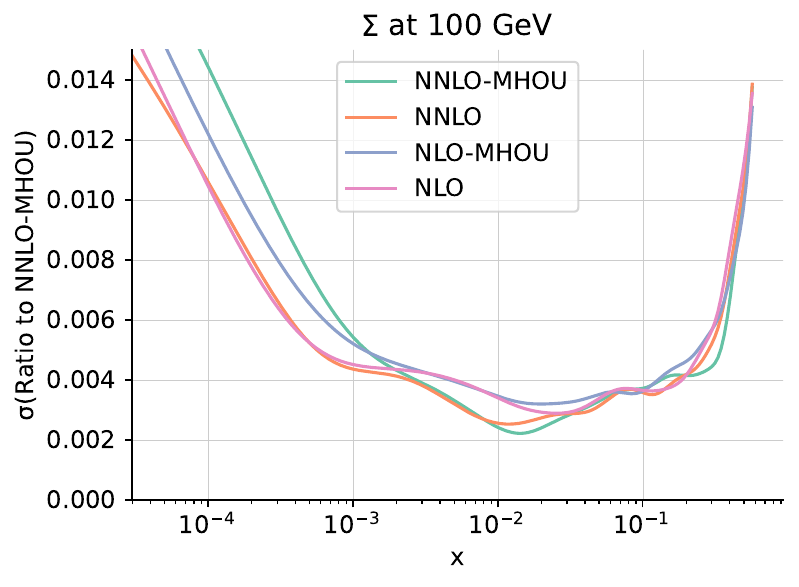}\\
  \includegraphics[width=0.45\textwidth]{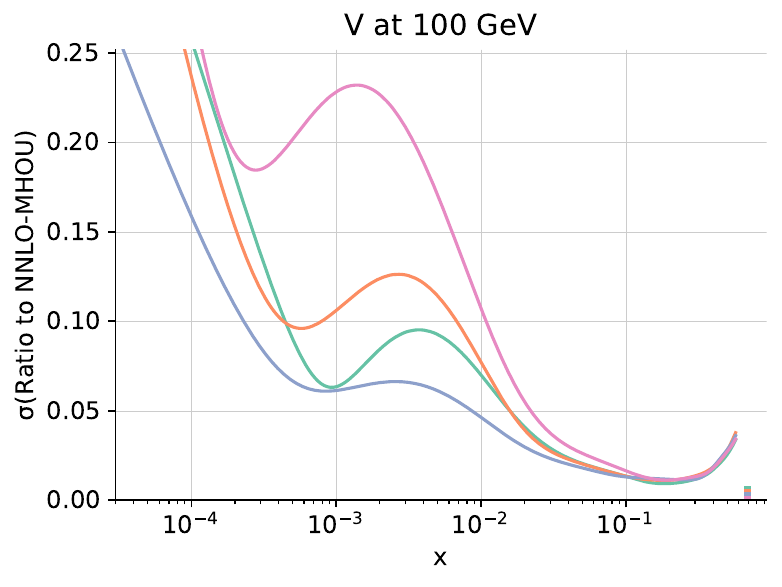}
  \caption{
    Relative one sigma uncertainties for the PDFs shown in
    \cref{fig:nnlo_nlo_pdfs}. All uncertainties are normalized 
    to the corresponding central NNLO PDFs with MHOUs.
  }
  \label{fig:nnlo_nlo_pdf_uncertainties}
\end{figure}

Inspection of \cref{fig:nnlo_nlo_pdf_uncertainties} shows that the PDF
uncertainty at NNLO in the data region remains on average unchanged 
upon inclusion of MHOUs, though in the singlet
sector it increases at small $x$, especially for the gluon where the
increase is up to $x\sim 10^{-2}$.
At NLO the uncertainty is generally reduced in the non-singlet sector, 
while in the singlet sector the uncertainty increases for all $x$, 
especially for the gluon. 
This is consistent with the observation of that at NLO the MHOU from 
scale variation does not fully account for the large shift from NLO to 
NNLO for some datasets. 

The somewhat counter-intuitive fact that the uncertainty on the PDF does 
not increase and may even  be reduced upon inclusion of an extra source of 
uncertainty in the $\chi^2$ was already observed in
Refs.~\cite{Ball:2018twp,Ball:2020xqw} and demonstrates the increased
compatibility of the data due to the MHOU. 

  \chapter{Evidence for Intrinsic charm in the proton}
\label{chap:ic}
\begin{center}
\begin{minipage}{1.\textwidth}
    \begin{center}
        \textit{
            This chapter is based my result presented in Refs.~\cite{Ball:2022qks,NNPDF:2023tyk}.
            In these works my contribution has focused to develop the theoretical framework 
            allowing us to extract the $n_f=3$ charm PDF and to the phenomenology computations.   
        } 
    \end{center}
\end{minipage}
\end{center}

\paragraph{Motivation.}


The description of electron-proton and proton-proton collisions at high
momentum transfers in terms of collisions between partons is now rooted 
in the theory of QCD, and it provides the basis of modern-day precision 
phenomenology at proton accelerators such as the LHC as well as for future 
facilities including the EIC~\cite{AbdulKhalek:2021gbh,AbdulKhalek:2022hcn}, the FPF~\cite{Feng:2022inv},
and neutrino telescopes~\cite{IceCube-Gen2:2020qha}.

Knowledge of the structure of the proton, which is necessary in order to 
obtain quantitative prediction for physics processes at the LHC and other
experiments, is encoded in the distribution of momentum carried by partons
of each type (gluons, up quarks, down quarks, up antiquarks, etc.): parton 
distribution functions (PDFs).
These PDFs could be in principle computed from first principles, but in practice 
even their determination from numerical simulations~\cite{Constantinou:2020hdm} 
is extremely challenging.
Consequently, the only strategy currently available for obtaining the reliable 
determination of the proton PDFs which is required to evaluate LHC predictions 
is empirical, through the global analysis of data for which precise theoretical 
predictions and experimental measurements are available, so that the PDFs are 
the only unknown~\cite{Gao:2017yyd}.

While the successful framework of PDF has by now been worked through in great 
detail, several key open questions remain open.
One of the most controversial of these concerns the treatment of so-called 
heavy quarks, i.e.\ those whose mass is greater than that of the proton 
($m_p=0.94$~GeV). Indeed, virtual quantum effects and energy-mass considerations 
suggest that the three light quarks and antiquarks (up, down, and strange) 
should all be present in the proton wave-function.
Their PDFs are therefore surely determined by the low-energy dynamics that controls 
the nature of the proton as a bound state.
However, it is a well-known fact~\cite{DeRoeck:2011na,
    Kovarik:2019xvh,Gao:2017yyd,Rojo:2019uip}
that in high enough energy collisions all species of quarks can be excited 
and hence observed inside the proton, so their PDFs are nonzero.
This excitation follows from standard QCD radiation and it can be computed 
accurately in perturbation theory.

But then the question arises: do heavy quarks also contribute to the proton 
wave-function? Such a contribution is called ``intrinsic'', to distinguish 
it from that computable in perturbation theory, which originates from QCD 
radiation.
Already since the dawn of QCD, it was argued that all kinds of intrinsic 
heavy quarks must be present in the proton wave-function~\cite{Brodsky:1984nx}.
In particular, it was suggested~\cite{Brodsky:1980pb} that the intrinsic 
component could be non-negligible for the charm quark, whose mass 
($m_c=1.51$~GeV) is of the same order of magnitude as the mass of 
the proton.
This question has remained highly controversial, and indeed recent dedicated studies 
have resulted in disparate claims, from excluding momentum fractions carried 
by intrinsic charm larger than $0.5~\%$ at the $4\sigma$ level~\cite{Jimenez-Delgado:2014zga} 
to allowing up to a $2~\%$ charm momentum fraction~\cite{Hou:2017khm}.
A particularly delicate issue in this context is that of separating the 
radiative component: finding that the charm PDF is nonzero at a low scale 
is not sufficient to argue that intrinsic charm has been identified.
In the following we present a resolution of this four-decades-long conundrum 
by providing a first evidence for intrinsic charm in the proton.

We provide also a first quantitative indication that the proton wave functions 
contains unequal distributions of charm quarks and antiquarks, i.e.\ a non-vanishing 
intrinsic valence charm.
A significant non-vanishing valence component cannot be perturbatively generated, 
hence our results reinforce evidence that the proton contains an intrinsic 
(i.e., not radiatively generated) charm quark component.

\paragraph{Outline.}



This chapter is structured as follows: first we review how the charm PDF 
is determined in the NNPDF framework (\cref{sec:ic_fitting}), in particular 
discussing and comparing with the case of perturbative charm only (\cref{sec:ic_pch})
in the four-flavor-number scheme (4FNS).
Then, in \cref{sec:ic_ic}, we show how it is possible to disentangle the 
perturbative component of the charm PDF and isolate a possible intrinsic
component by isolating the three-flavor-number scheme (3FNS) charm. 
We discuss the stability of such procedure, and we validate the result by 
comparing to some recent measurement of LHCb (\cref{sec:ic_lhcb}).
Finally, in \cref{sec:ic_icasy} we show how this method can be extended to 
probe a non-vanishing charm PDF asymmetry, and we propose some dedicated 
observables, which can be measured at future colliders such as HL-LHC or 
EIC and can further constrain a proton intrinsic charm component (\cref{sec:icasy_pheno}).


\section{Fitting charm PDF}
\label{sec:ic_fitting}


In this section, by using the methodology described in \cref{sec:nnpdf_methodology},
we discuss some feature of the charm NNPDF4.0 PDF~\cite{NNPDF:2021njg}.

%
This fitted charm PDF will be the boundary condition of the studies presented
then in \cref{sec:ic_ic,sec:ic_icasy}, its determination is performed 
at NNLO in an expansion in powers of the strong coupling, $\alpha_s$, 
which represents the precision frontier for collider phenomenology.

The charm PDF determined in this manner includes a radiative component, 
and indeed it depends on the resolution scale: it is given in a 
four-flavor-number scheme (4FNS), in which up, down, strange and charm 
quarks are subject to perturbative radiative corrections and mix with 
each other and the gluon as the resolution is increased.
In the following, we review the parametrization of the fitted charm 
PDF (\cref{sec:ic_nnpdf40charm}), we compare it with the alternative
scenario of the perturbative charm only in \cref{sec:ic_pch}:
discussing its stability at PDF level (\cref{sec:ic_stability_4FNS}).

\subsection{The NNPDF4.0 charm PDF}
\label{sec:ic_nnpdf40charm}
The 4FNS charm PDF and its associated uncertainties is determined by means of 
a  global QCD analysis within the NNPDF4.0 framework.
All PDFs, including the charm PDF, are parametrized at $Q_0=1.65$~GeV in a 
model-independent manner using a neural network, which is fitted to data using 
supervised machine learning techniques. The Monte Carlo replica method is deployed 
to ensure a faithful uncertainty estimate (cf \cref{sec:nnpdf_fitting}).
Specifically, we express the 4FNS total charm PDF ($c^+=c+\bar{c}$) in terms of 
the output neurons associated to the quark singlet $\Sigma$ and non-singlet $T_{15}$ 
distributions, as 
\begin{equation}
  xc^+(x,Q_0;{\boldsymbol \theta}) = 
    \left( x^{\alpha_{\Sigma}}(1-x)^{\beta_{\Sigma}} {\rm NN}_{\Sigma}(x,{\boldsymbol \theta}) 
      - x^{\alpha_{T_{15}}}(1-x)^{\beta_{T_{15}}} {\rm NN}_{T_{15}}(x,{\boldsymbol \theta}) 
    \right)/4 \, ,
    \label{eq:fitted_charm_param}
\end{equation}
where ${\rm NN}_{i}(x,{\boldsymbol \theta})$ is the $i$-th output neuron of a neural 
network with input $x$ and  parameters ${\boldsymbol \theta}$, and 
$\left( \alpha_i,\beta_i\right)$ are preprocessing exponents.
A crucial feature of \cref{eq:fitted_charm_param} is that no {\it ad hoc} 
specific model assumptions are used: the shape and size of $xc^+(x,Q_0)$ are entirely 
determined from experimental data. Hence, our determination of the 4FNS fitted 
charm PDF, and thus of the intrinsic charm, is unbiased.
%


The neural network parameters ${\boldsymbol \theta}$ in \cref{eq:fitted_charm_param}
are determined by fitting an extensive global dataset that consists of 4618 
cross-sections from a wide range of different processes, measured over the years 
in a variety of fixed-target and collider experiments (see~\cite{NNPDF:2021njg} 
for a complete list).
The kinematic coverage of these cross-sections, is displayed in \cref{fig:kinplot}.
%
%
Many of these processes provide direct or indirect sensitivity to the charm 
content of the proton.
Particularly important constraints come from $W$ and $Z$ production from ATLAS, 
CMS, and LHCb as well as from neutral and charged current DIS structure functions 
from HERA.
The 4FNS PDFs at the input scale $Q_0$ are related to experimental measurements 
at $Q \neq Q_0$ by means of NNLO QCD calculations, including the FONLL-C 
general-mass scheme for DIS~\cite{Forte:2010ta} generalized to allow for 
fitted charm~\cite{Ball:2015tna} (cf. \cref{sec:fns}).

As we shall explain in \cref{sec:ic_stability_4FNS}, we have 
verified that the determination of 4FNS charm PDF \cref{eq:fitted_charm_param} 
and the ensuing three-flavor-number-scheme (3FNS) intrinsic charm PDF are stable upon variations of methodology 
(PDF parametrization basis), input dataset, and values of Standard Model parameters 
(the charm mass). 
We have also studied the stability of our results upon replacing the current 
NNPDF4.0 methodology~\cite{NNPDF:2021njg} with the previous NNPDF3.1 methodology
~\cite{NNPDF:2017mvq}. It turns out that results are perfectly consistent. 
Indeed, the old methodology leads to somewhat larger uncertainties, corresponding 
to a moderate reduction of the local statistical significance for intrinsic charm, 
and to a central value which is within the smaller error band of our current result.

A determination in which the vanishing of intrinsic charm is imposed has also been 
performed (see also \cref{sec:ic_pch}). In this case, the fit quality 
significantly deteriorates: the values of the $\chi^2$ per data point of 1.162, 
1.26, and 1.22 for total, Drell-Yan, and NC DIS data respectively, found when fitting 
charm, are  increased to 1.198, 1.31, 1.28 when the vanishing of intrinsic charm is imposed.
The absolute worsening of the total $\chi^2$ when the vanishing of intrinsic charm 
is imposed is therefore of 166 units, corresponding to a $2\sigma$ effects 
in units of $\sigma_{\chi^2}= \sqrt{2n_{\rm dat}}$.

\subsection{The perturbative charm}
\label{sec:ic_pch}

In the absence of intrinsic charm, the charm PDF is fully determined by
perturbative matching conditions, i.e.\ by the matrix $\mathbf{A}^{(n_f)}(Q_{c}^2)$ 
in \cref{eq:singlet_matching,eq:ns_matching}.
We will denote the charm PDF thus obtained as ``perturbative charm PDF'', 
for short. The PDF uncertainty on the perturbative charm PDF is directly 
related to that of the light quarks and especially the gluon, and is typically 
much smaller than the uncertainty on our fitted charm PDF, that includes
a possible intrinsic charm. Here and in the following we will refer to our 
result of fitted 4FNS charm as ``default''.
It should be noticed that the matching conditions for charm are nontrivial 
starting from NNLO: while at NLO the perturbative charm PDF vanishes at threshold.
Hence, having implemented in \eko{} also the N$^3$LO matching conditions, we are 
able to assess the MHOU of the perturbative charm at the matching scale $Q_c$, 
by comparing results obtained at the first two non-vanishing perturbative
orders.

\begin{figure}[!t]
    \centering
    \includegraphics[width=0.49\textwidth]{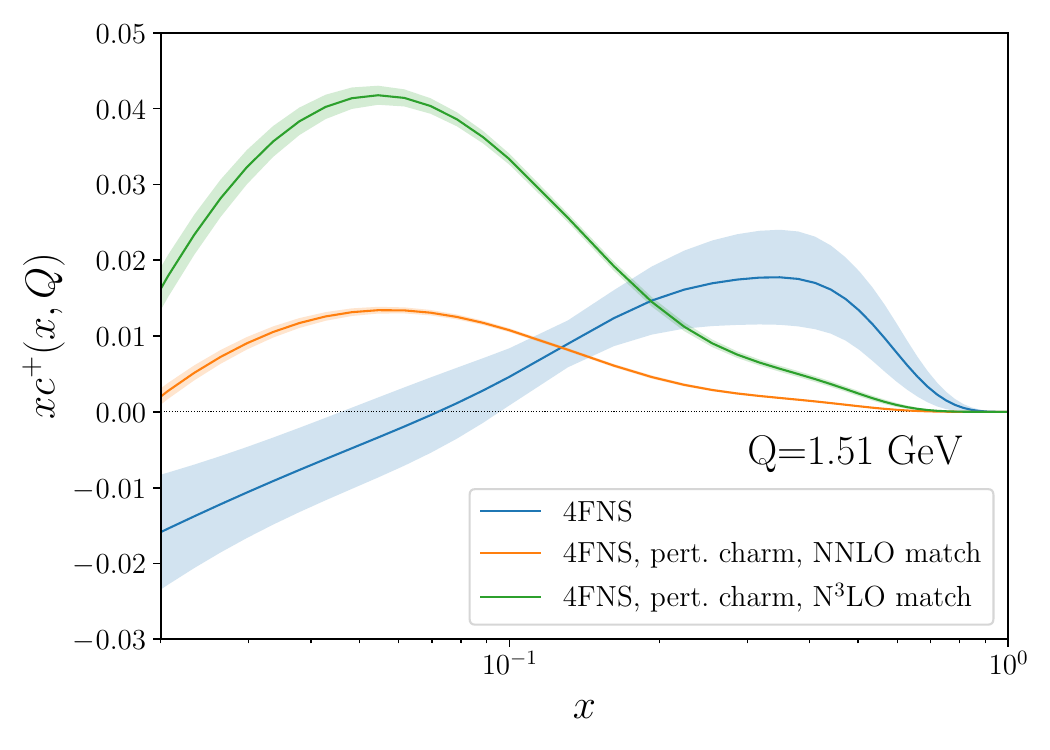}
    \includegraphics[width=0.49\textwidth]{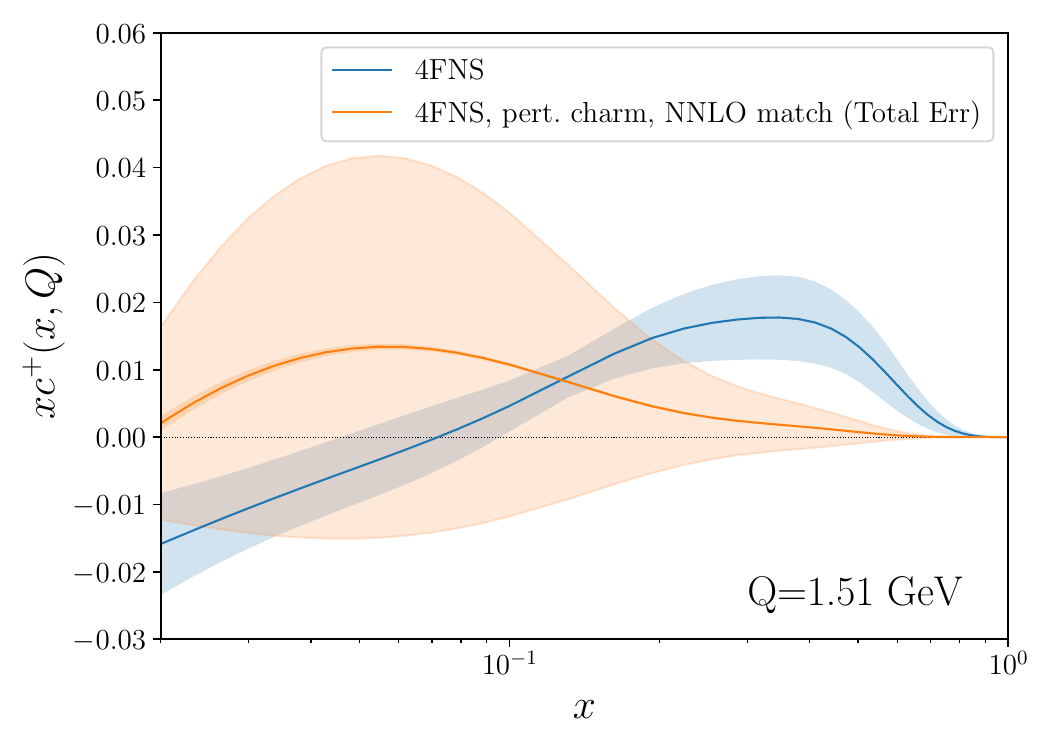}
    \caption{
        Left: the perturbative charm PDF at $Q=1.51$~GeV obtained from NNLO 
        PDFs using NNLO and N$^3$LO matching conditions.
        Right: the NNLO perturbative charm PDF including the MHOU computed 
        as the difference between NNLO and N$^3$LO matching. In both plots our 
        default (fitted) charm PDF is also shown for comparison.
   }
   \label{fig:charm_fitted_vs_perturbative_mhous}
\end{figure}

We construct a PDF set with perturbative charm, in which the full PDF determination 
from the global dataset leading to the NNPDF4.0 PDF set is repeated, but now 
with the assumption that l.h.s of \cref{eq:fitted_charm_param} is fixed by 
the matching condition at $Q_c$ and the following DGLAP evolution from $Q_c \to Q_0$.
This perturbative charm PDF is compared to our default result in 
\cref{fig:charm_fitted_vs_perturbative_mhous}~(left), where the 4FNS
perturbative charm PDF at scale $Q_c=m_c$ obtained using either NNLO or N$^3$LO
under the assumption of no intrinsic charm are shown, together with our default 
result allowing for possible intrinsic charm.
It is clear that while on the one hand, the PDF uncertainty on the perturbative 
charm PDF is indeed tiny, on the other hand the difference between the result 
for perturbative charm obtained using NNLO or N$^3$LO matching is large, and in
fact larger at small-$x$ than the difference between perturbative charm and our
default (fitted) result.

We may use the difference between the 4FNS perturbative charm obtained from 
NNLO and N$^3$LO matching as an estimate of the MHOU on perturbative charm 
at the scale $Q_c$.
The total uncertainty is found by adding this in quadrature to the PDF 
uncertainty (which however in practice is negligible).
The result is shown in \cref{fig:charm_fitted_vs_perturbative_mhous}~(right).
Within this total uncertainty there is now good agreement between our fitted charm 
result and perturbative charm for all $x\lsim 0.2$. On the other hand, there 
is a clear deviation for larger $x$. We may view the difference between the 
4FNS default result and the 4FNS perturbative charm as the intrinsic component 
in the 4FNS, and indeed it is clear from \cref{fig:charm_fitted_vs_perturbative_mhous} 
that the 4FNS intrinsic component is sizable and positive at large $x$.
%

\subsection{Stability of the 4FNS charm}
\label{sec:ic_stability_4FNS}

The main input to our determination 
of intrinsic charm is the 4FNS charm PDF extracted from high-energy data. 
While this determination comes with an uncertainty estimate, it is important 
to verify that this adequately reflects the various sources of uncertainty, 
and that there are no further sources of uncertainty that may be unaccounted for.
To this purpose, here we assess the stability of our boundary condition PDF 
first, upon the choice of underlying dataset, next upon changes in methodology, 
and finally, upon variation of Standard Model parameters.
In each case we verify stability upon the most important possible source 
of instability: respectively, the use of collider vs.\ fixed target and 
deep-inelastic vs.\ hadronic data (dataset); the choice of parametrization basis 
(methodology); and the value of the charm quark mass.
%
%
In all comparisons we focus on the large-$x$ region in which the fitted charm 
displays a valence-like peak. As we shall see in \cref{sec:ic_ic} this is the 
$x$-region where intrinsic charm could be localized.
In this section, the 4FNS charm PDF is displayed at the scale $Q = 1.65$~GeV 
so that results for all fit variants, including those with different 
$m_c$ values, can be shown at a common scale.

\begin{figure}[!t]
    \centering
    \includegraphics[width=0.99\textwidth]{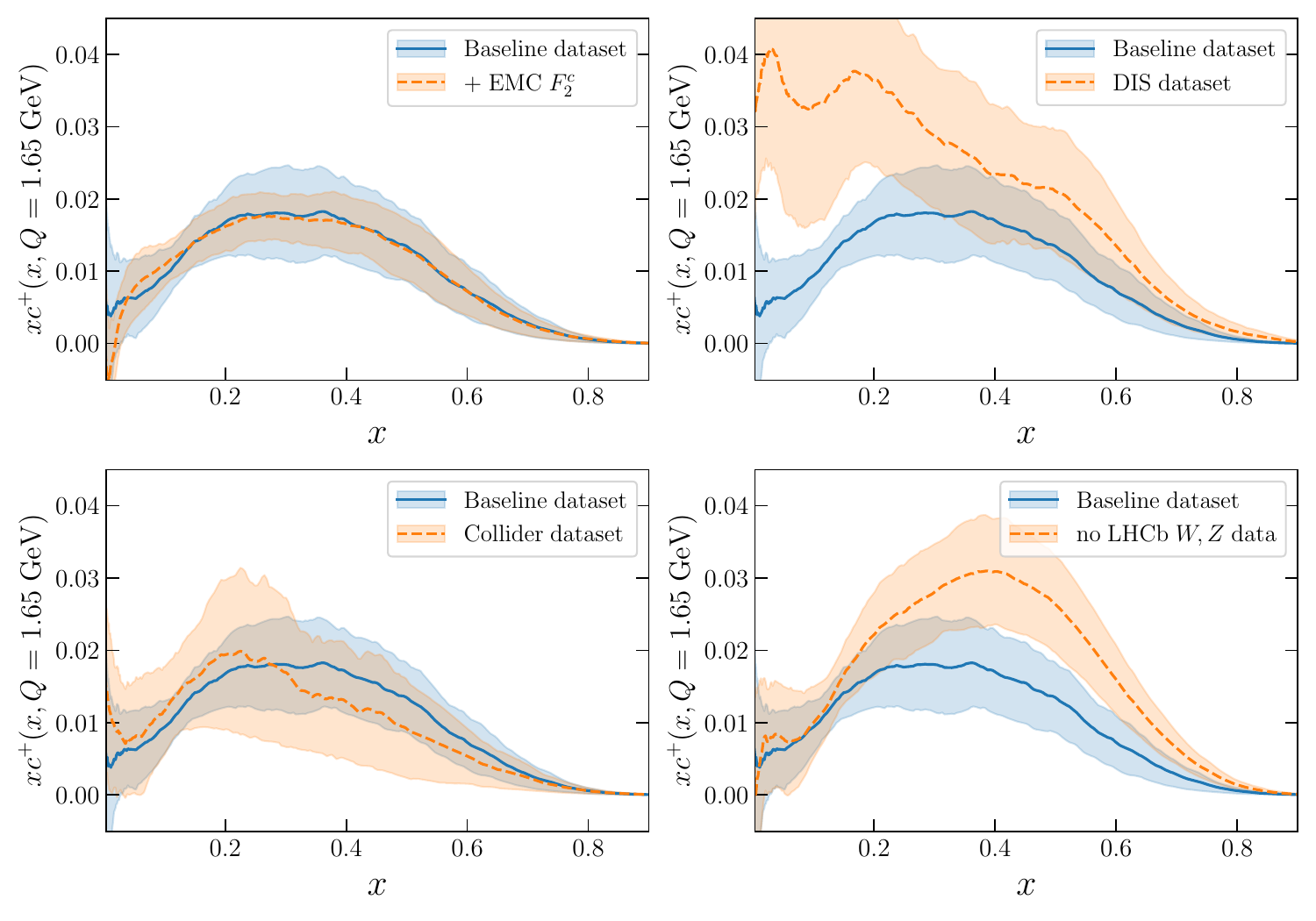}
    \caption{
        The dependence of the 4FNS charm PDF at $Q=1.65$~GeV on 
        the input dataset. We compare the baseline result with 
        that obtained by also including EMC $F_2^c$ data (top left), 
        only including DIS data (top right), only including collider
        data (bottom left) and removing LHCb gauge boson production 
        data (bottom right). 
  }
  \label{fig:charm_dataset_dep} 
\end{figure}

\paragraph{Dependence on the choice of dataset.}
We now study the stability of the 4FNS charm determination upon variation 
of the underlying data, which also allows us to identify the datasets or 
groups of processes that provide the leading constraints on fitted charm.
To this purpose, we have repeated our PDF determination using a variety of 
subsets of the global dataset used for our default determination. Results 
are shown in \cref{fig:charm_dataset_dep}, where we compare the result using
the baseline dataset to determinations performed by adding to the baseline
the EMC charm~\cite{Aubert:1982tt} structure function data; by only including 
DIS data; by only including collider data (HERA, Tevatron and LHC); and by 
removing the LHCb $W$ and $Z$ production data.

The EMC data are relatively imprecise as they were taken at relatively low 
scales, where radiative corrections are large and their accuracy has often 
been questioned. For these reasons, we have not included them in our baseline
analysis. However, it is interesting to assess the impact of their inclusion.
We find that the extra information provided by these $F_2^c$ data is subdominant 
in comparison to that from the global dataset. The result is stable and only a 
moderate uncertainty reduction at the peak is observed. It is interesting to
contrast this with the previous NNPDF study~\cite{Ball:2016neh}, in which the 
global fit provided only very loose constraints on the charm PDF, which was 
then determined mostly by the EMC data.
Indeed, a DIS-only fit, for which most data were already available at the 
time of the previous determination, determines charm with very large
uncertainties. On the other hand, both the central value and uncertainty 
found in the collider-only fit are quite similar to the baseline result.
This shows that the dominant constraint is now coming from collider, and 
specifically hadron collider data. Among these, LHCb data (which are taken 
at large rapidity and thus impact PDFs at large and small $x$) are especially 
important, as demonstrated by the increase in uncertainty when they are removed.

In all these determinations, the charm PDF at $x\simeq 0.4$ remains consistently 
nonzero and positive, thus emphasizing the stability of our results.

\paragraph{Dependence on the parametrization basis.}
Among the various methodological choices, a possibly critical one is the choice of 
fitting basis functions. Specifically, in our default analysis, the output of the 
neural network does not provide the individual quark flavor and anti-flavor PDFs, 
but rather linear combinations corresponding to the so-called evolution basis
(cf. \cref{sec:dglap}). Our charm PDF is given in \cref{eq:fitted_charm_param} as 
the linear combination of the two basis PDFs $\Sigma$ and $T_{15}$.
One may thus ask whether this choice may influence the final results for individual 
quark flavors, specifically charm. Given that physical results are basis independent, 
the outcome of a PDF determination should not depend on the basis choice.

\begin{figure}[!t]
    \centering
    \includegraphics[width=0.49\textwidth]{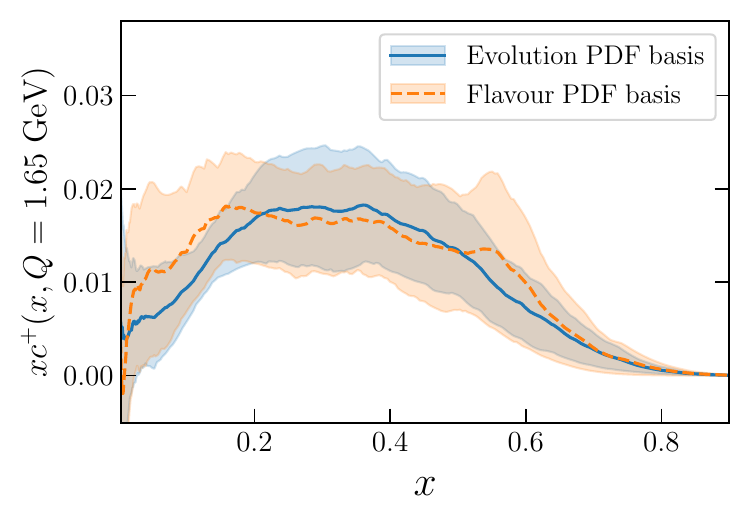}
    \caption{
        The default 4FNS charm PDF at $Q=1.65$~GeV compared to a result 
        obtained by parametrizing PDFs in the flavor basis instead 
        of the evolution basis. 
    }
    \label{fig:charm_basisdep}
\end{figure}

In order to check this, we have repeated the PDF determination, but now using the 
flavor basis, in which each of the neural network output neurons now correspond to 
individual quark flavors, so in particular, instead of \cref{eq:fitted_charm_param}, 
one has
\begin{equation}
    xc^+(x,Q_0;{\boldsymbol \theta}) = (1-x)^{\beta_{c^+}} {\rm NN}_{c^+}(x,{\boldsymbol \theta}) \, ,
    \label{eq:fitted_charm_param_flavour}
\end{equation}
where ${\rm NN}_{c^+}(x,{\boldsymbol \theta})$ indicates the value of the output neuron 
associated to the charm PDF $c^+$.
The 4FNS charm PDFs determined using either basis are compared in \cref{fig:charm_basisdep} 
at $Q=1.65$~GeV. We find excellent consistency, and in particular the valence-like 
structure at high-$x$ is independent of the choice of parametrization basis.

\paragraph{Dependence on the charm mass.}
The kinematic threshold for producing charm perturbatively depends on the value 
of the charm mass. Therefore, the perturbative contribution to the 4FNS charm PDF, 
and thus the whole charm PDF if one assumes perturbative charm, depends strongly 
on the value of the charm mass.
On the other hand, the intrinsic charm PDF is of non-perturbative origin, so it 
should be essentially independent of the numerical value of the charm mass that 
is used in perturbative computations employed in its determination (though it will 
of course depend on the true underlying physical value of the charm mass).

\begin{figure}[!t]
    \centering
    \includegraphics[width=0.99\textwidth]{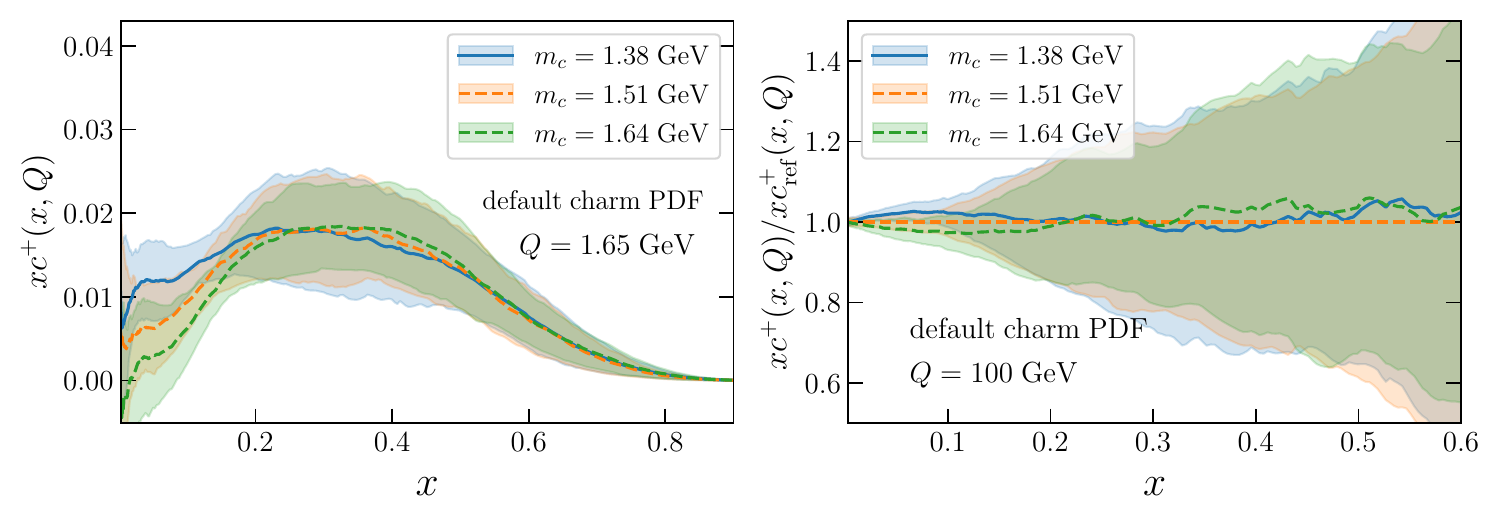}
    \caption{
        The 4FNS charm PDF determined using three different values 
        of the charm mass. The absolute result (left) is shown at 
        $Q=1.65$~GeV, while the ratio to the default value $m_c=1.51$~GeV 
        (right) used elsewhere in this paper is shown at $Q=100$~GeV.
   }
   \label{fig:charm_fitted_mcdep} 
\end{figure}

In order to study this mass dependence, we have repeated our charm PDF determination 
using different values for the charm mass. The definition of the charm mass which is 
relevant for kinematic thresholds is the pole mass, for which we adopt the value 
recommended by the Higgs cross-section working group~\cite{deFlorian:2016spz} based 
on the study of~\cite{Bauer:2004ve}, namely $m_c = 1.51 \pm 0.13$~GeV. 
Results are shown in \cref{fig:charm_fitted_mcdep}, where our default charm PDF 
determination with $m_c = 1.51$~GeV is repeated with $m_c=1.38$~GeV and $m_c=1.64$~GeV.
In order to understand these results note that this is the 4FNS PDF, so it includes 
both a non-perturbative and a perturbative component. The latter is strongly dependent 
on the charm mass, but of course the data correspond to the unique true value of the 
mass and the mass dependence of the perturbative component is present only due to our
ignorance of the actual true value. When determining the PDF from the data, as we do, 
we expect this spurious dependence to be to some extent reabsorbed into the fitted PDF. 
So we expect results to display a moderate dependence on the charm mass.

\begin{figure}[!t]
    \centering
    \includegraphics[width=0.99\textwidth]{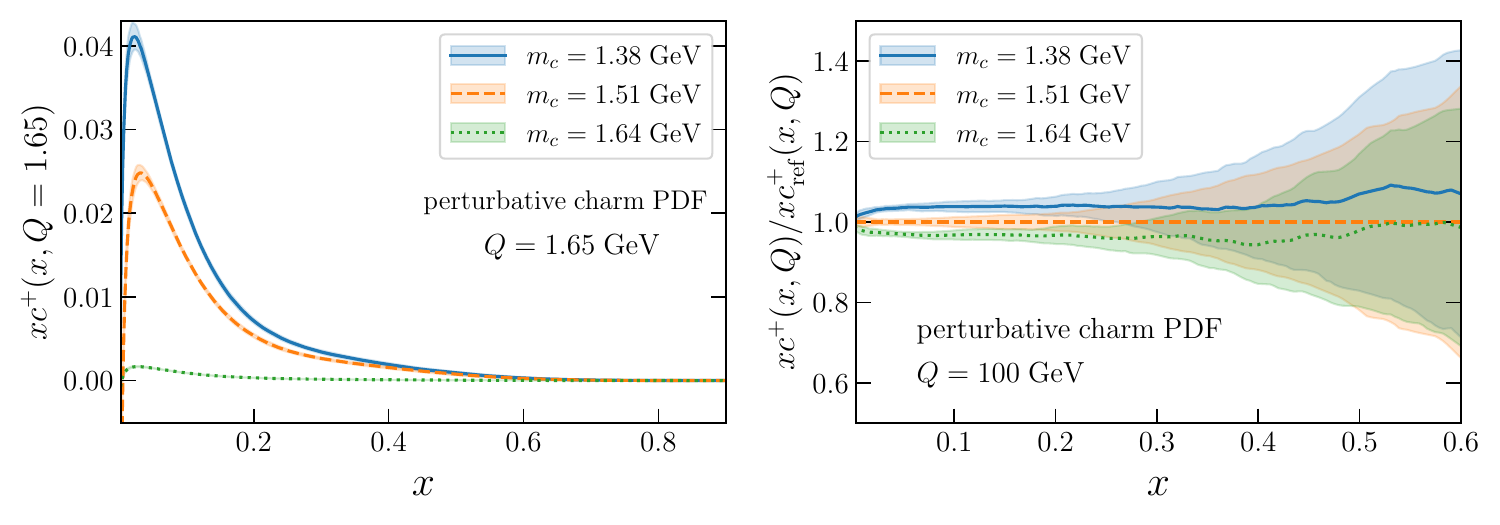}
    \caption{
        The same as \cref{fig:charm_fitted_mcdep} but now for 
        the perturbative charm PDF.
    }
    \label{fig:charm_pert_mcdep}
\end{figure}

In \cref{fig:charm_pert_mcdep} the same result is shown, but now for the perturbative 
charm PDF discussed in \cref{sec:ic_pch}, so the charm PDF is of purely perturbative 
origin and fully determined by the strongly mass-dependent matching condition. This 
dependence is clearly seen in the plots. Furthermore, comparison with \cref{fig:charm_fitted_mcdep} 
shows that indeed this spurious dependence is partly reabsorbed in the fit when the 
charm PDF is determined from the data, so that the residual mass dependence is moderate.
In particular, the large-$x$ valence peak, is very stable.

\section{Evidence of intrinsic charm}
\label{sec:ic_ic}

Here we provide for the first time evidence for intrinsic charm 
by exploiting a high-precision determination of the  quark-gluon 
content of the nucleon~\cite{NNPDF:2021njg} based on machine 
learning and the largest experimental dataset ever.
We disentangle the intrinsic charm component from charm-anticharm pairs 
arising from high-energy radiation~\cite{Ball:2015tna} (\cref{sec:ic_methods}).

In \cref{sec:ic_results} we establish the existence of intrinsic 
charm at the $3\sigma$ level, with a momentum distribution in remarkable 
agreement with model predictions~\cite{Brodsky:1980pb,Hobbs:2013bia}.

Later, in we analyze this evidence at the level of momentum fraction (\cref{sec:ic_momentum}).
%
Finally, in \cref{sec:ic_lhcb,sec:ic_lumi} we confirm these findings by comparing to a recent 
data on $Z$ production with charm jets in the forward region from the 
LHCb experiment~\cite{LHCb:2021stx}. 

\subsection{The intrinsic charm determination}
\label{sec:ic_methods}

The intrinsic charm component can be disentangled from it as follows.
First, we note that in the absence of an intrinsic component, the initial
condition for the charm PDF is determined using perturbative matching
conditions, computed up to NNLO, and recently extended up to N$^3$LO 
(cf. \cref{sec:fns}).
%
%
%
The assumption that there is no intrinsic charm amounts to the assumption
that all 4FNS PDFs are determined using perturbative matching conditions 
in terms of 3FNS PDFs that do not include a charm PDF.
These matching conditions determine the charm PDF in terms of the PDFs of 
the three-flavor-number-scheme (3FNS), in which only the three lightest 
quark flavors are radiatively corrected.
So the assumption of no intrinsic charm amounts to the assumption that 
if the 4FNS PDFs are transformed back to the 3FNS, the 3FNS charm PDF 
is found to vanish. In this context, intrinsic charm is by definition 
the deviation from zero of the 3FNS charm PDF~\cite{Ball:2015dpa}. 
Hence, this perturbative charm PDF is entirely determined in terms of the 
three light quarks and antiquarks and the gluon.
However, these perturbative matching conditions are actually given by a 
square matrix that also includes a 3FNS charm PDF and this does not need 
to vanish (\cref{eq:singlet_matching}): in fact, if the charm quark PDF in the 
4FNS is freely parametrized and thus determined from the data~\cite{Ball:2015tna}, 
the matching conditions can be inverted.
This possible 3FNS charm PDF, is then by definition the intrinsic charm PDF: 
indeed, in the absence of intrinsic charm it would vanish~\cite{Ball:2015dpa}. 
Unlike the 4FNS charm PDF, that includes both an intrinsic and a radiative
component, the 3FNS charm PDF is purely non-perturbative and is not equal 
to the 4FNS charm PDF, since matching conditions reshuffle all PDFs among 
each other.

\begin{figure}[!t]
  \centering
  \includegraphics[width=0.8\textwidth]{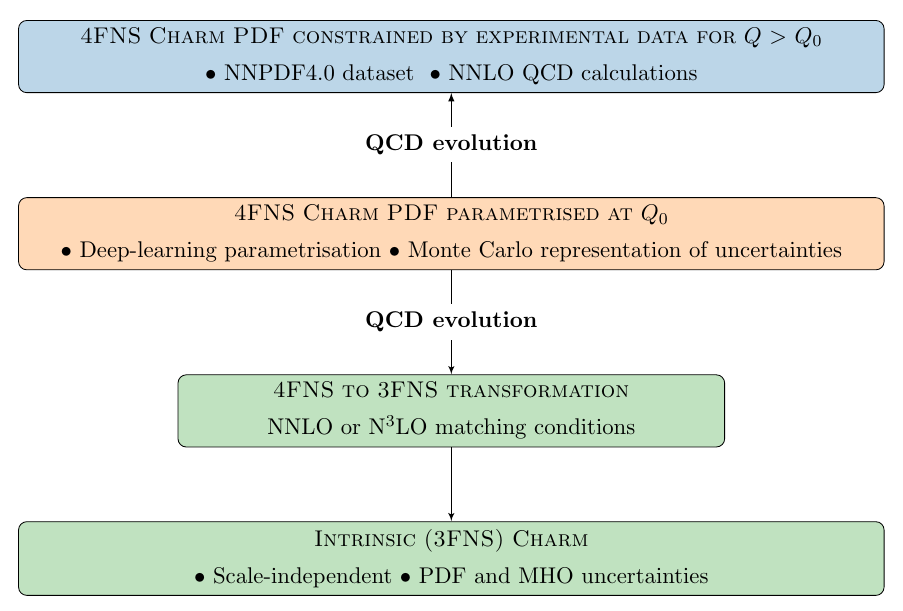}
  \caption{
    The 4FNS charm PDF is parametrized  at $Q_0$ and evolved to all $Q$, 
    where it is  constrained by the NNPDF4.0 global dataset. 
    Subsequently, it is transformed to the 3FNS where (if nonzero) it provides 
    the intrinsic charm component.
  }
  \label{fig:ic_strategy}
\end{figure}

In summary, intrinsic charm can then be determined through the following 
two steps, summarized in \cref{fig:ic_strategy}.
First, all the PDFs, including the charm PDF, are parametrized in the 4FNS 
at an input scale $Q_0$ and evolved using NNLO perturbative QCD to $Q \not = Q_0$.
These evolved PDFs can be used to compute physical cross-sections, also 
at NNLO, which then are compared to a global dataset of experimental 
measurements.
The result of this first step in our procedure is a Monte Carlo (MC) representation 
of the probability distribution for the 4FNS PDFs at the input parametrization 
scale $Q_0$. 
Next, this 4FNS charm PDF is transformed to the 3FNS at some scale matching 
scale $Q_c$ by inverting \cref{eq:singlet_matching}. 
The choice of both $Q_0$ and $Q_c$ are immaterial.
The former because perturbative evolution is invertible, so results for the 
PDFs do not depend on the choice of parametrization scale $Q_0$. 
The latter because the 3FNS charm is scale independent, so it does not depend 
on the value of $Q_c$.
Both statements of course hold up to fixed perturbative accuracy, and are 
violated by MHO corrections.
In practice, we parametrize PDFs at the scale $Q_0=$~1.65~GeV and perform 
the inversion at a scale chosen equal to the charm mass $Q_c=m_c=1.51$~GeV.
The scale-independent 3FNS charm PDF is then the sought-for intrinsic charm.

\paragraph{Calculation of the 3FNS charm PDF.}
The Monte Carlo representation of the probability distribution associated to the 
4FNS charm PDF determined by the global analysis contains an intrinsic component 
mixed with a perturbatively generated contribution, with the latter becoming 
larger in the $x\lsim 0.1$ region as the scale $Q$ is increased.
In order to extract the intrinsic component, we transform PDFs to the 
3FNS at the scale $Q_c=m_c=1.51$~GeV using \eko{} (\cref{sec:eko}).
In this study we have performed this inversion at NNLO as well as at N$^3$LO 
which as we shall see provides a handle on the perturbative uncertainty of the 
NNLO result.
This work has performed when not all the N$^3$LO matching conditions were fully 
known. In particular $A^{(3)}_{gg}$~\cite{Ablinger:2022wbb} and $A^{(3)}_{Qg}$
~\cite{Ablinger:2023ahe,Ablinger:2024xtt} were published in a later stage.
The impact of the N$^3$LO corrections on intrinsic charm is discussed further
in \cref{sec:anl3o_ic}.

Therefore, our results have NNLO accuracy and we can only use the N$^3$LO 
contributions to the $\mathcal{O}(\alpha_s^3)$ corrections to the heavy 
quark matching conditions as a way to estimate the size of the 
missing higher orders. Indeed, these corrections have a very significant 
impact on the perturbatively generated component, see \cref{sec:ic_pch}.
They become large for $x \lsim 0.1$, which coincides with the region
dominated by the perturbative component of the charm PDF, and are 
relatively small for the valence region where intrinsic charm dominates.

\subsection{Results: the intrinsic charm PDF}
\label{sec:ic_results}



\begin{figure}[!t]
    \centering
    \includegraphics[width=0.49\linewidth]{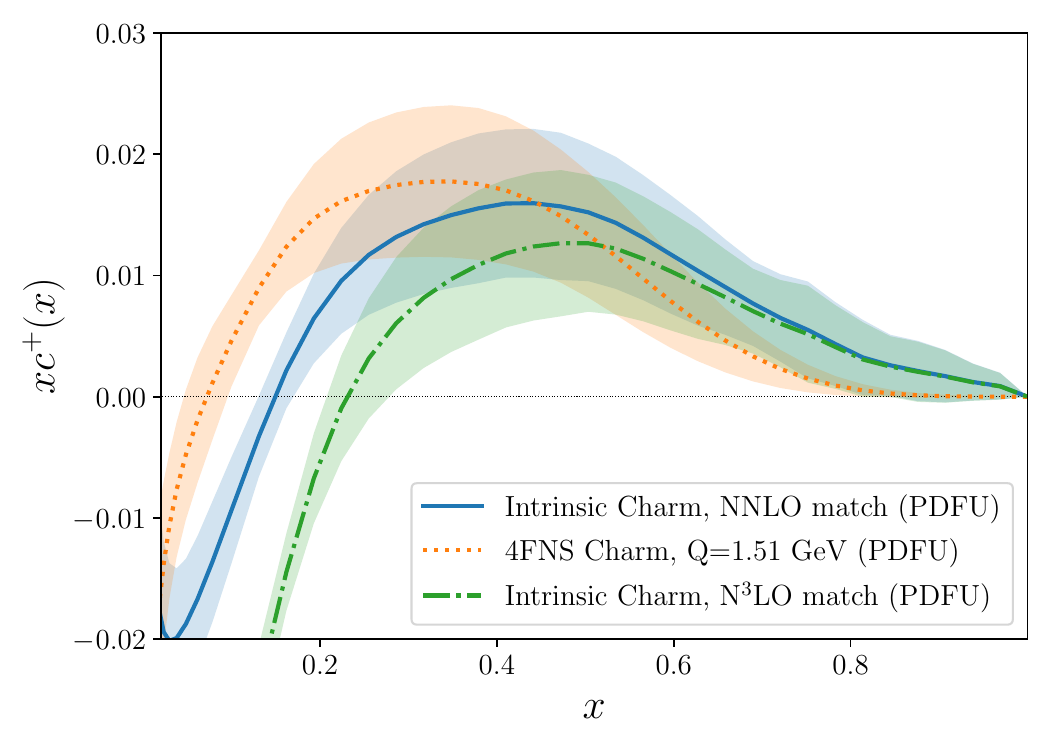}
    \includegraphics[width=0.49\linewidth]{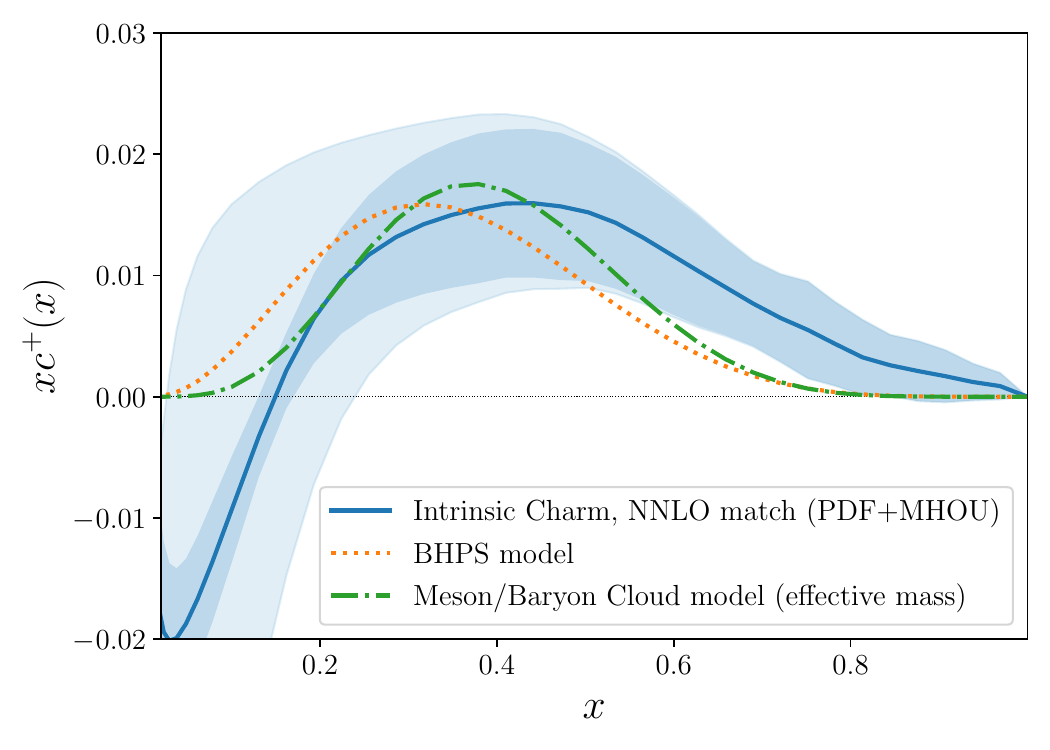}
    \caption{
      The intrinsic charm PDF and comparison with models.
      Left: the purely intrinsic (3FNS) result (blue)
      with PDF uncertainties only, compared to the 4FNS PDF, that
      includes both an intrinsic and radiative
      component, at $Q=m_c=1.51$~GeV (orange). 
      The purely intrinsic (3FNS) result obtained using N$^3$LO 
      matching is also shown (green).
      Right: the purely intrinsic (3FNS) final result with 
      total uncertainty (PDF+MHOU), with the PDF uncertainty 
      indicated as a dark shaded band; the predictions from the original 
      BHPS model~\cite{Brodsky:1980pb} and from the more recent meson/baryon
      cloud model~\cite{Hobbs:2013bia} are also shown for comparison
      (dotted and dot-dashed curves respectively).
    }
    \label{fig:charm_content_3FNS} 
\end{figure}

As summarized in \cref{sec:ic_fitting}, our starting point is the NNPDF4.0 
global analysis, which provides a determination of the sum of the charm 
and anticharm PDFs, namely $c^+(x,Q) = c(x,Q)+\bar c(x,Q)$, in the 4FNS. 
This can be viewed as a probability density in $x$, the fraction of the 
proton momentum carried by charm, in the sense that the integral over all 
values of $0\le x\le1$ of $xc^+(x)$ is equal to the fraction of the proton 
momentum carried by charm quarks, though note that PDFs are generally 
not necessarily positive-definite. 
Our initial 4FNS $xc^+(x,Q)$ at the charm mass scale, $Q=m_c$ with 
$m_c=1.51$~GeV, is displayed in \cref{fig:charm_content_3FNS}~(left).
The ensuing intrinsic charm is determined from it by transforming to the 3FNS 
using NNLO matching.
This result is also shown in \cref{fig:charm_content_3FNS}~(left).
The bands indicate the $68~\%$ confidence level (CL) interval associated with 
the PDF uncertainties (PDFU) in each case. Henceforth, we will refer to the 
3FNS $xc^+(x,Q)$ PDF as the intrinsic charm PDF. 

The intrinsic (3FNS) charm PDF displays a characteristic valence-like
structure at large-$x$ peaking at $x\simeq 0.4$.
While intrinsic charm is found to be small in absolute terms
(it contributes less than $1~\%$ to the proton total momentum),
it is significantly different from zero.
Note that the transformation to the 3FNS has little effect on the peak region,
because there is almost no charm radiatively generated at such large values of $x$: 
in fact, a very similar valence-like peak is already found in the 4FNS calculation.

Because at the charm mass scale the strong coupling $\alpha_s$ is rather
large, the perturbative expansion converges slowly.
In order to estimate the effect of missing higher order uncertainties (MHOU), 
we have also performed the transformation from the 4FNS NNLO charm PDF
determined from the data to the 3FNS charm PDF at one order higher, namely at N$^3$LO. 
The result is also shown \cref{fig:charm_content_3FNS} (left). 
Reassuringly, the intrinsic valence-like structure is unchanged.
On the other hand, it is clear that for $x\lsim 0.2$ perturbative uncertainties 
become very large.
We can estimate the total uncertainty on our determination of intrinsic charm 
by adding in quadrature the PDF uncertainty and a MHOU estimated from the shift 
between the result found using NNLO and N$^3$LO matching.

This procedure leads to our final result for intrinsic charm and its total
uncertainty, shown in \cref{fig:charm_content_3FNS} (right).
The intrinsic charm PDF is found to be compatible with zero for
$x\lsim 0.2$: the negative trend seen in \cref{fig:charm_content_3FNS} with 
PDF uncertainties only becomes compatible with zero upon inclusion of theoretical
uncertainties. However, at larger $x$ even with theoretical uncertainties
the intrinsic charm PDF differs from zero by about 2.5 standard deviations 
($2.5\sigma$) in the peak region.
This result is stable upon variations of dataset, methodology (in
particular the PDF parametrization basis) and Standard Model parameters 
(specifically the charm mass), as demonstrated in the 
\cref{sec:ic_stability_4FNS}. 

Our determination of intrinsic charm can be compared to theoretical expectations.
Subsequent to the original intrinsic charm model of~\cite{Brodsky:1980pb} 
(BHPS model), a variety of other models were proposed~\cite{
  Hoffmann:1983ah,Pumplin:2005yf,Paiva:1996dd,Steffens:1999hx,Hobbs:2013bia},
see~\cite{Brodsky:2015fna} for a review.
Irrespective of their specific details, most models predict a valence-like
structure at large $x$ with a maximum located between $x\simeq 0.2$ and 
$x\simeq 0.5$, and a vanishing intrinsic component for $x\lsim 0.1$.
In \cref{fig:charm_content_3FNS}~(right) we compare our result to the original BHPS 
model and to the more recent meson/baryon cloud model of~\cite{Hobbs:2013bia}.

As these models predict only the shape of the intrinsic charm distribution, 
but not its overall normalization, we have normalized them by requiring
them to reproduce the same charm momentum fraction as our determination.
We find agreement between the shape of our determination and the model predictions.
In particular, we reproduce the presence and location of the large-$x$ valence-like peak
structure (with better agreement, of marginal statistical significance, with
the meson/baryon cloud calculation), the vanishing of intrinsic charm at small-$x$.
The fraction of the proton momentum carried by charm quarks that we obtain from our 
analysis, used in this comparison to models, is $\lp 0.62 \pm 0.28\rp \%$ including 
PDF uncertainties only (see \cref{sec:ic_momentum} for details).
However, the uncertainty upon inclusion of MHOU greatly increases, and we obtain 
$\lp 0.62 \pm 0.61\rp \%$, due to the contribution from the small-$x$ region, 
$x\lsim 0.2$, where the MHOU is very large, see \cref{fig:charm_content_3FNS}~(right).
Note that in most previous analyses~\cite{Hou:2017khm} 
intrinsic charm models (such as the BHPS model) are fitted to the data, with only 
the momentum fraction left as a free parameter.
A comparison between our result and Ref.~\cite{Hou:2017khm} is available in \cite[App.~F]{Ball:2022qks}. 

We emphasize that in our analysis the charm PDF is entirely determined by the 
experimental data included in the PDF determination. The data with the most 
impact on charm are from recently measured LHC processes, which are both accurate 
and precise.
Since these measurements are made at high scales, the corresponding hard cross-sections 
can be reliably computed in QCD perturbation theory.

\subsection{The charm momentum fraction}
\label{sec:ic_momentum}

The fraction of the proton momentum carried by charm quarks is given by
\begin{equation}
  \lc c\rc = \int_0^1dx\, x c^+(x,Q^2) \, .
  \label{eq:charm_momentum_fraction}
\end{equation}

Model predictions, as mentioned, are typically provided up to an overall 
normalization, which in turn determines the charm momentum fraction.
Consequently, the momentum fraction is often cited as a characteristic parameter 
of intrinsic charm. It should however be borne in mind that, even in the absence 
of intrinsic charm, this charm momentum fraction is nonzero due to the perturbative 
contribution.

In \cref{tab:momfrac_lowQ} we indicate the values of the charm momentum fraction 
in the 3FNS for our default charm determination and in the 4FNS (at $Q=1.65$~GeV) 
both for the default result and for perturbative charm PDF.
We provide results for three different values of the charm mass $m_c$ and indicate 
separately the PDF and the MHO uncertainties. The 3FNS result is scale-independent, 
it corresponds to the momentum fraction carried by intrinsic charm, and it vanishes 
identically by assumption in the perturbative charm case.
The 4FNS result corresponds to the scale-dependent momentum fraction that combines 
the intrinsic and perturbative contribution, while of course it contains only the
perturbative contribution in the case of perturbative charm.
As discussed in \cref{sec:ic_pch}, the large uncertainty associated to higher order 
corrections to the matching conditions affects the 3FNS result (intrinsic charm) 
in the default case, in which the charm PDF is determined from data in the 4FNS scheme, 
while it affects the 4FNS result for perturbative charm, that is determined assuming 
the vanishing of the 3FNS result.

For our default determination, the charm momentum fraction in the 4FNS at low scale
differs from zero at the $3\sigma$ level.
However, it is not possible to tell whether this is of perturbative or intrinsic origin, 
because, due to the large MHOU in the matching condition, the intrinsic (3FNS) charm 
momentum fraction is compatible with zero. This large uncertainty is entirely due to the
small $x\lsim 0.2$ region, see \cref{fig:charm_content_3FNS}~(right).
Accordingly, for perturbative charm the low-scale 4FNS momentum fraction is 
compatible with zero.
Consistently with the results of \cref{sec:ic_stability_4FNS}, the 4FNS result is 
essentially independent of the value of the charm mass, but it becomes strongly dependent 
on it if one assumes perturbative charm.

\begin{table}[!t]
    \footnotesize
    \centering
    \renewcommand{\arraystretch}{1.30}
    \begin{tabularx}{\textwidth}{C{2.0cm}C{2.3cm}C{2.2cm}C{2.2cm}C{5.6cm}}
  \toprule
  Scheme  & $Q$ & Charm PDF & $m_c$  &  $\lc c\rc~\lp\%\rp$ \\
  \midrule
  \midrule
  3FNS  & --  &default  &  1.51~GeV  &   $ 0.62\pm0.28_{\rm pdf}\pm 0.54_{\rm mhou} $ \\
  3FNS  & --  &default  &  1.38~GeV  &   $ 0.47\pm0.27_{\rm pdf}\pm 0.62_{\rm mhou} $ \\
  3FNS  & --  &default  &  1.64~GeV  &    $ 0.77\pm0.28_{\rm pdf}\pm 0.48_{\rm mhou} $ \\
  \midrule
  4FNS  & 1.65~GeV  & default  &  1.51~GeV  &   $0.87 \pm 0.23_{\rm pdf}$  \\
  4FNS  & 1.65~GeV  & default &  1.38~GeV  &   $0.94 \pm 0.22_{\rm pdf}$  \\
  4FNS  & 1.65~GeV  & default   &  1.64~GeV  &  $0.84 \pm 0.24_{\rm pdf}$  \\
  \midrule
  \midrule
  4FNS  & 1.65~GeV   & perturbative  &  1.51~GeV  &   $0.346\pm 0.005_{\rm pdf}\pm 0.44_{\rm mhou}$ \\
  4FNS  & 1.65~GeV   & perturbative  &  1.38~GeV  &    $0.536\pm 0.006_{\rm pdf}\pm 0.49_{\rm mhou}$ \\
  4FNS  & 1.65~GeV   & perturbative  &  1.64~GeV  &    $0.172\pm 0.003_{\rm pdf}\pm 0.41_{\rm mhou}$ \\
  \bottomrule
\end{tabularx}
    \vspace{0.3cm}
    \caption{
        The charm momentum fraction, \cref{eq:charm_momentum_fraction}.
        We show results both in the 3FNS and the 4FNS (at $Q=1.65$~GeV)
        for our default charm, and also in the 4FNS for perturbative charm.
        We provide results for  three different values of the charm mass $m_c$ 
        and indicate separately the PDF and the MHO uncertainties.
    }
\label{tab:momfrac_lowQ}
\end{table}

\begin{figure}[!t]
    \centering
    \includegraphics[width=0.49\textwidth]{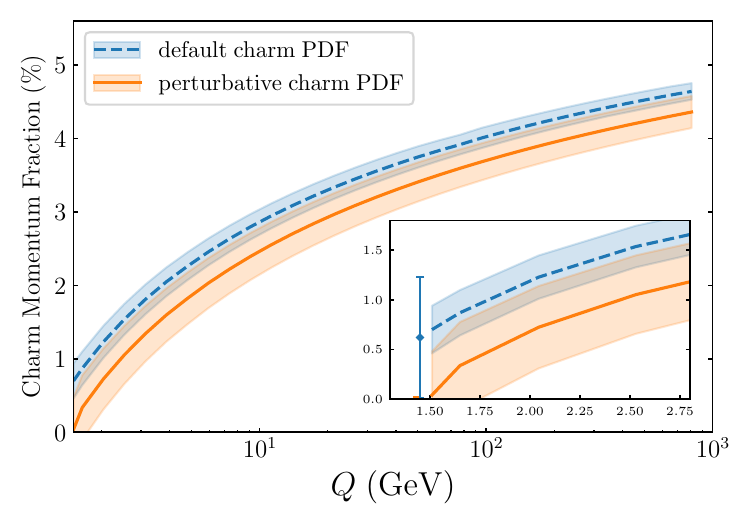}
    \caption{ 
        The 4FNS charm momentum fraction in NNPDF4.0 as a function of scale $Q$,
        both for the default and perturbative charm cases, for a charm mass value 
        of $m_c=1.51$~GeV.
        The inset zooms on the low-$Q$ region and includes the 3FNS (default) 
        result from \cref{tab:momfrac_lowQ}. 
        Note that the uncertainty includes the MHOU for the 3FNS default
        and 4FNS perturbative charm cases, while it is the PDF uncertainty 
        for the 4FNS default charm case.
    }
    \label{fig:comparison_IC_models} 
\end{figure}

\begin{figure}[!t]
    \centering
    \includegraphics[width=0.49\textwidth]{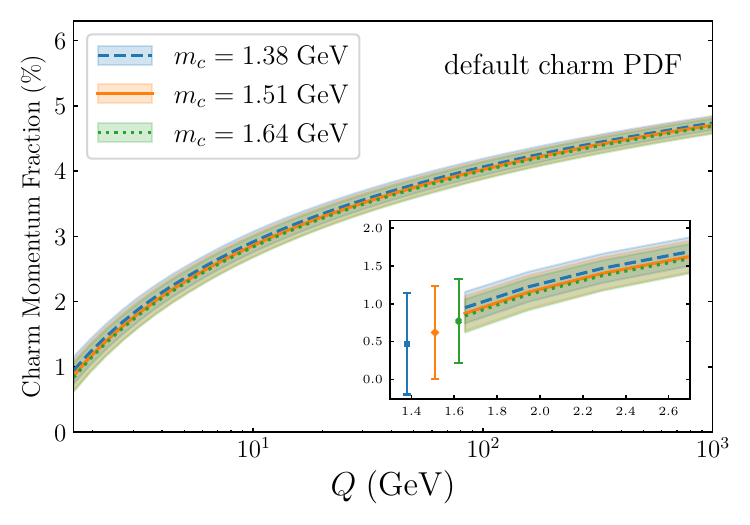}
    \includegraphics[width=0.49\textwidth]{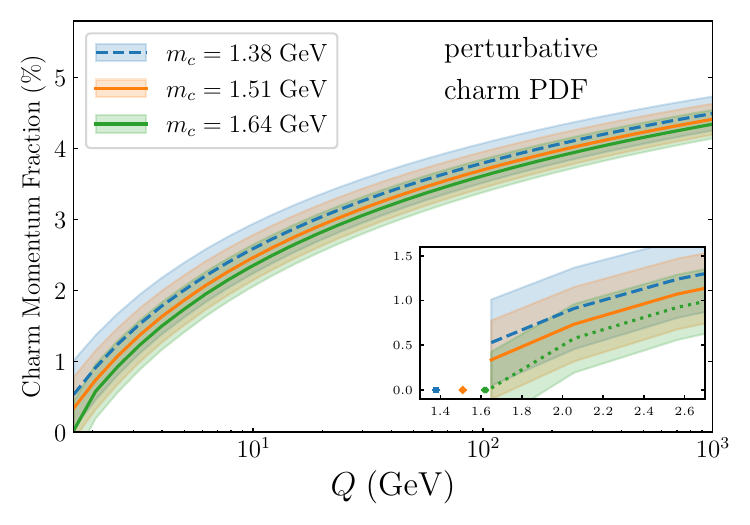}
    \caption{
      Same as \cref{fig:comparison_IC_models} for different values of the charm mass. 
      Note that the 3FNS momentum fraction for perturbative charm vanishes identically 
      by assumption.
    }
    \label{fig:charm_momfrac_qdep_mc} 
\end{figure}

The 4FNS charm momentum fraction is plotted as a function of scale in 
\cref{fig:comparison_IC_models}, both in the default case and for perturbative charm, 
with the 3FNS values and the detail of the low-$Q$ 4FNS results shown in an inset.
The dependence on the value of the charm mass is shown in \cref{fig:charm_momfrac_qdep_mc}.
The large MHOUs on the 3FNS result, and on the 4FNS result in the case of perturbative 
charm, are apparent. The stability of the default result upon variation of the value of
$m_c$, and the strong dependence of the perturbative charm result on $m_c$, are also 
clear. Both the PDF uncertainty, and the strong dependence on the value of $m_c$
for perturbative charm are seen to persist up to large scales.

\begin{figure}[!t]
    \centering
    \includegraphics[width=0.49\textwidth]{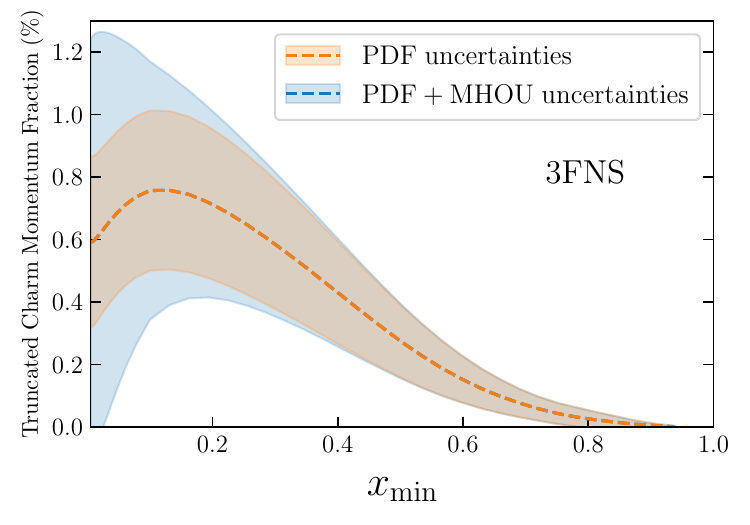}
    \caption{
      The value of the truncated charm momentum integral, \cref{eq:charm_momentum_fraction_truncated}, 
      as a function of the lower integration limit $x_{\rm min}$ for our baseline 
      determination of the 3FNS intrinsic charm PDF. We display separately the PDF 
      and the total (PDF+MHOU) uncertainties.
    }
    \label{fig:charm_momfrac_xmin_dep} 
\end{figure}

It is interesting to understand in detail the impact of the MHOU on the momentum 
fraction carried by intrinsic charm. To this purpose, we have computed the truncated 
momentum integral, i.e. \cref{eq:charm_momentum_fraction} but only integrated down to
some lower integration limit $x_{\rm min}$:
\begin{equation}
  \lc c\rc_{\rm tr}(x_{\rm min}) = \int_{x_{\rm min}}^1dx\, x c^+(x,Q^2) \, .
  \label{eq:charm_momentum_fraction_truncated}
\end{equation}
Note than in the 3FNS $x c^+(x)$ does not depend on scale, so this becomes a scale-independent 
quantity.
The result for our default intrinsic charm determination is displayed in 
\cref{fig:charm_momfrac_xmin_dep}, as a function of the lower integration limit $x_{\rm min}$.
It is clear that for $x_{\rm min} \gtrsim 0.2$ the truncated momentum fraction differs 
significantly from zero, thereby providing evidence for intrinsic charm with similar 
statistical significance as the local pull shown in \cref{fig:Zc_bottomrow_pull}.
For $x \lsim 0.2$ this significance is then washed out by the large MHOUs.

Hence, while the total momentum fraction has been traditionally adopted as a measure of 
intrinsic charm, our analysis shows that, once MHOUs are accounted for, the information
provided by the total momentum fraction is limited, at least with current data and theory.

\subsection{$Z+$charm production in the forward region}
\label{sec:ic_lhcb}
The production of $Z$ bosons in association with charm-tagged jets (or alternatively,
with identified $D$ mesons) at the LHC is directly sensitive to the charm content
of the proton via the dominant $gc \to Zc$ partonic scattering process.
Measurements of this process at the forward rapidities covered by the LHCb acceptance
~\cite{LHCb:2021stx} provide access to the large-$x$ region where the intrinsic contribution 
is expected to dominate.
This is in contrast with the corresponding measurements from ATLAS and CMS, which only 
become sensitive to intrinsic charm at rather larger values of $p_T^Z$ than those
currently accessible experimentally.

Following~\cite{Boettcher:2015sqn,LHCb:2021stx}, we have obtained theoretical predictions 
for $Z$+charm production at LHCb with NNPDF4.0, based on NLO QCD calculations using 
{\sc\small POWHEG-BOX}~\cite{Alioli:2010xd} interfaced to {\sc\small Pythia8}
with the Monash 2013 tune for showering, hadronization, and underlying event~\cite{Sjostrand:2007gs}.
Acceptance requirements and event selection follow the LHCb analysis, where in particular 
charm jets are defined as those anti-$k_T$ $R=0.5$ jets containing a reconstructed charmed 
hadron.
The details of the calculation are reported below in a dedicated paragraph.
The ratio between $c$-tagged and untagged $Z$+jet events can then be compared with the 
LHCb measurements of
\begin{equation}
  \mathcal{R}_j^c(y_Z) = 
    \frac{N(c~{\rm tagged~jets};y_Z)}{N({\rm jets};y_Z)} =
    \frac{\sigma(pp\to Z+{\rm charm~ jet};y_Z)}{\sigma(pp \to Z+{\rm jet};y_Z)} \, .
    \label{eq:Rcj}
\end{equation}
Here $N(c~{\rm tagged~jets};y_Z)$ and $N({\rm jets};y_Z)$ are, respectively, the number 
of charm-tagged and untagged jets, for a $Z$ boson rapidity interval that satisfies the 
selection and acceptance criteria. 
The more forward the rapidity $y_{Z}$, the higher the values of the charm momentum $x$ 
being probed.


\begin{figure}[!t]
    \centering
    \includegraphics[width=0.49\textwidth]{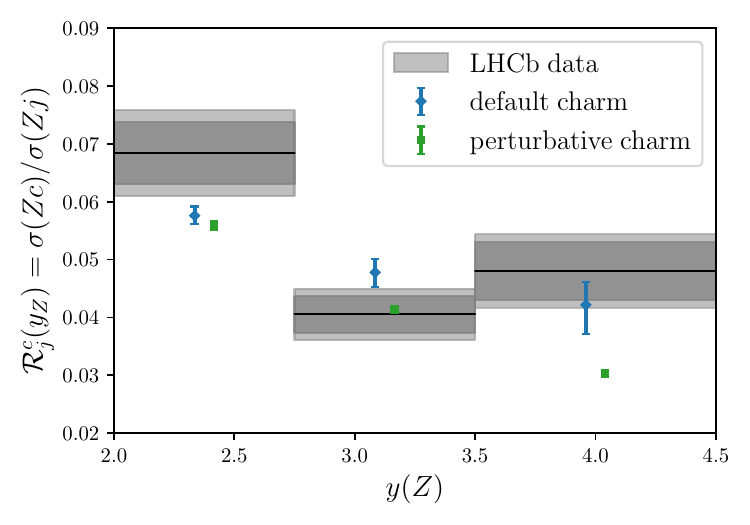}
    \includegraphics[width=0.49\textwidth]{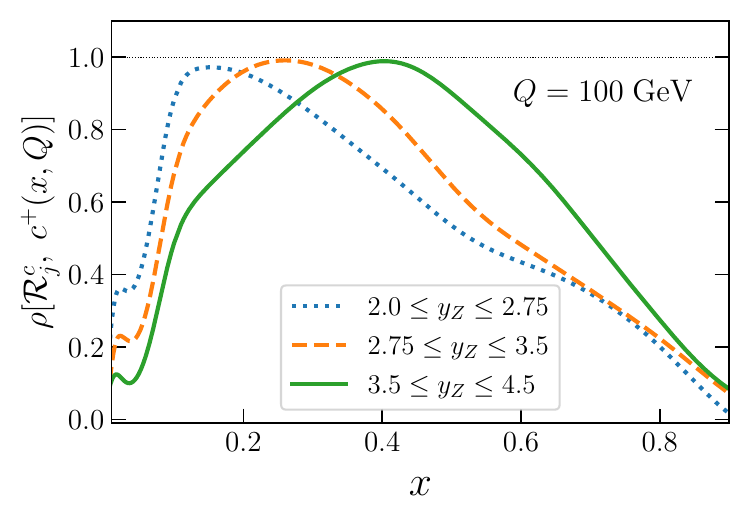}
    \caption{
      Intrinsic charm and $Z+$charm production at LHCb.
      Left: the LHCb measurements of $Z$ boson production in association 
      with charm-tagged jets, $\mathcal{R}_j^c$, at $\sqrt{s}=13$ TeV, compared 
      with our default prediction which includes an intrinsic charm component,
      as well as with a variant in which we impose the vanishing of the 
      intrinsic charm component.
      The thicker (thinner) bands in the LHCb data indicate the statistical
      (total) uncertainty, while the theory predictions include both PDF and 
      MHOU.
      Right: the correlation coefficient between the charm PDF at $Q=100$~GeV 
      in NNPDF4.0 and the LHCb measurements of $\mathcal{R}_j^c$ for the three 
      $y_Z$ bins.
    }
    \label{fig:Zc_toprow} 
\end{figure}

In \cref{fig:Zc_toprow}~(left) we compare the LHCb measurements of $\mathcal{R}_j^c$, 
provided in three bins of the $Z$-boson rapidity $y_Z$, with the theoretical 
predictions based on both our default PDFs and the PDF set in which 
we impose the vanishing of intrinsic charm.
In \cref{fig:Zc_toprow}~(right) we also display the correlation coefficient between 
the charm PDF at $Q=100$~GeV and the observable $\mathcal{R}_j^c$, demonstrating 
how this observable is highly correlated to charm in a localized $x$ region that 
depends on the rapidity bin.
It is clear that our prediction is in excellent agreement with the LHCb measurements, 
while in the highest rapidity bin, which is highly correlated to the charm PDF 
in the region of the observed valence peak $x\simeq 0.45$, the prediction obtained 
by imposing the vanishing of intrinsic charm undershoots the data at the $3\sigma$ level.
Hence, this measurement provides independent direct evidence in support of our result.


We have also determined the impact of these LHCb $Z$+charm measurements on the charm 
PDF by means of the Bayesian reweighting.


\begin{table}[!t]
  \small
  \renewcommand{\arraystretch}{1.45}
  \begin{tabularx}{\textwidth}{C{3.5cm}C{2.5cm}C{2.5cm}C{2.5cm}C{2.5cm}}
    \toprule
    \multirow{2}{*}{ $\chi^2/N_{\rm dat}$}  &
    \multicolumn{2}{c}{ default charm}  & 
    \multicolumn{2}{c}{perturbative charm} \\
    &  $\rho_{\rm sys}=0$   & $\rho_{\rm sys}=1$ &  $\rho_{\rm sys}=0$ &   $\rho_{\rm sys}=1$ \\
    \midrule
    Prior        &  1.85   &  3.33      &   3.54  & 3.85      \\
    \midrule
    Reweighted   &  1.81   &  3.14      &   $-$   &  $-$     \\
    \bottomrule
\end{tabularx}
  \vspace{0.3cm}
  \caption{
    The values of $\chi^2/N_{\rm dat}$ for the LHCb $Z$+charm data before (prior)
    and after (reweighted) their inclusion in the PDF fit. Results are given for 
    two experimental correlation models, denoted as $\rho_{\rm sys}=0$ and 
    $\rho_{\rm sys}=1$. We also report values before inclusion for the perturbative 
    charm PDFs.
  }
  \label{tab:chi2_zcharm}
\end{table}

We first compare the quality of the description of the LHCb data before their inclusion. 
In \cref{tab:chi2_zcharm} we show the values of $\chi^2/N_{\rm dat}$ for the LHCb 
$Z$+charm data both with default and perturbative charm.
Since the experimental covariance matrix is not available for the LHCb data we determine 
the $\chi^2$ values assuming two limiting scenarios for the correlation of experimental 
systematic uncertainties.
Namely, we either add in quadrature statistical and systematic errors ($\rho_{\rm sys}=0$),
or alternatively we assume that the total systematic uncertainty is fully correlated 
between $y_Z$ bins ($\rho_{\rm sys}=1$). Fit quality is always significantly better 
in our default intrinsic charm scenario than with perturbative charm.
As it is clear from \cref{fig:Zc_toprow} (left), the somewhat poor fit quality is mostly due 
to the first rapidity bin, which is essentially uncorrelated to the amount of intrinsic 
charm (see \cref{fig:Zc_toprow}, right).

The LHCb $Z$+charm data are then included in the PDF determination through Bayesian 
reweighting~\cite{Ball:2010gb,Ball:2011gg}. The $\chi^2/N_{\rm dat}$ values obtained 
using the PDFs found after their inclusion are given in \cref{tab:chi2_zcharm}.
They are computed by combining the PDF and experimental covariance matrix so both sources 
of uncertainty are included --- MHOUs from the hadronic matrix element are negligible, 
see computation the details below.
The fit quality is seen to improve only mildly, and the effective number of replicas
~\cites{Ball:2010gb,Ball:2011gg} after reweighting is only moderately reduced, from the 
prior $N_{\rm rep}=100$ to $N_{\rm eff}=92$ or $N_{\rm eff}=84$ in the $\rho_{\rm sys}=0$ 
and $\rho_{\rm sys}=1$ scenarios respectively.
This demonstrates that the inclusion of the LHCb $Z$+charm measurements affects the PDFs 
only weakly. 



\begin{figure}[!t]
  \centering
  \includegraphics[width=0.49\textwidth]{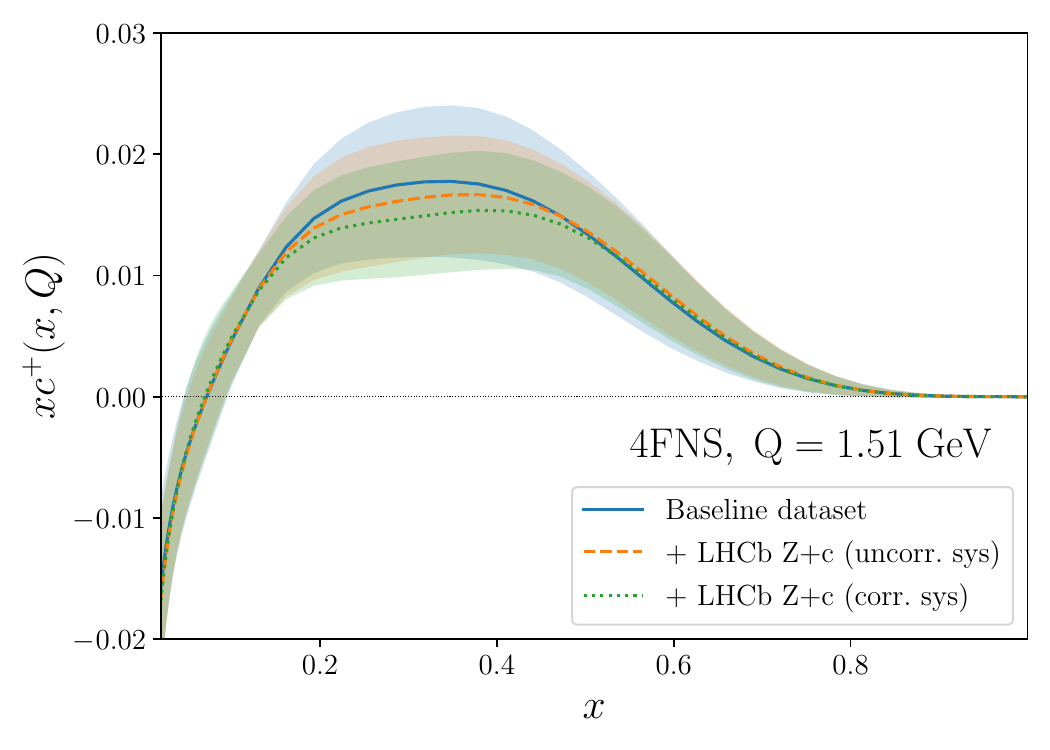}
  \includegraphics[width=0.49\textwidth]{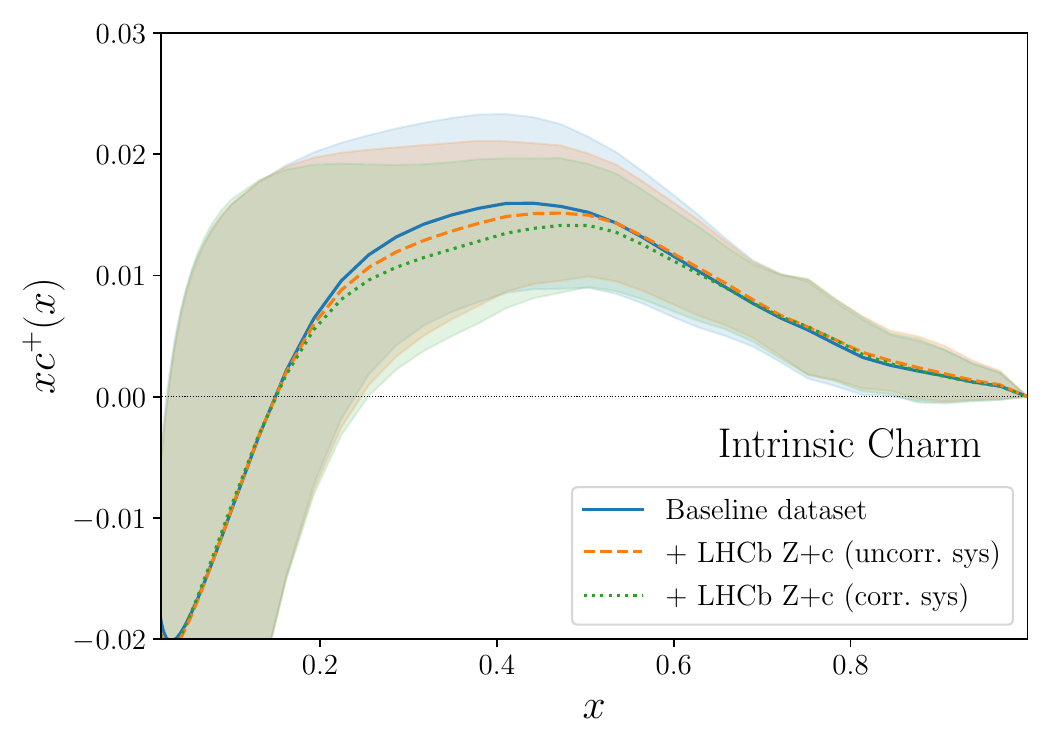}
  \caption{
    Impact of $Z$+charm LHCb data on the charm PDF.
    The charm PDF in the 4FNS (left) and the intrinsic (3FNS) 
    charm PDF (right) before and after inclusion of the LHCb $Z$+charm data.
    Results are shown for both experimental correlation models discussed in the text.
  }
  \label{fig:Zc_midrow} 
\end{figure}
%

The charm PDF in the 4FNS before and after inclusion of the LHCb data 
(with either correlation model), and the intrinsic charm PDF obtained from it, 
are displayed in \cref{fig:Zc_midrow}~(left and right respectively).
The bands account for both PDF and MHOU.
The results show full consistency: inclusion of the LHCb $\mathcal{R}_j^c$ data leaves
the intrinsic charm PDF unchanged, while moderately reducing the uncertainty on it.

We can summarize our results through their so-called local statistical significance, 
namely, the size of the intrinsic charm PDF in units of its total uncertainty.

\begin{figure}[!t]
  \centering
  \includegraphics[width=0.49\textwidth]{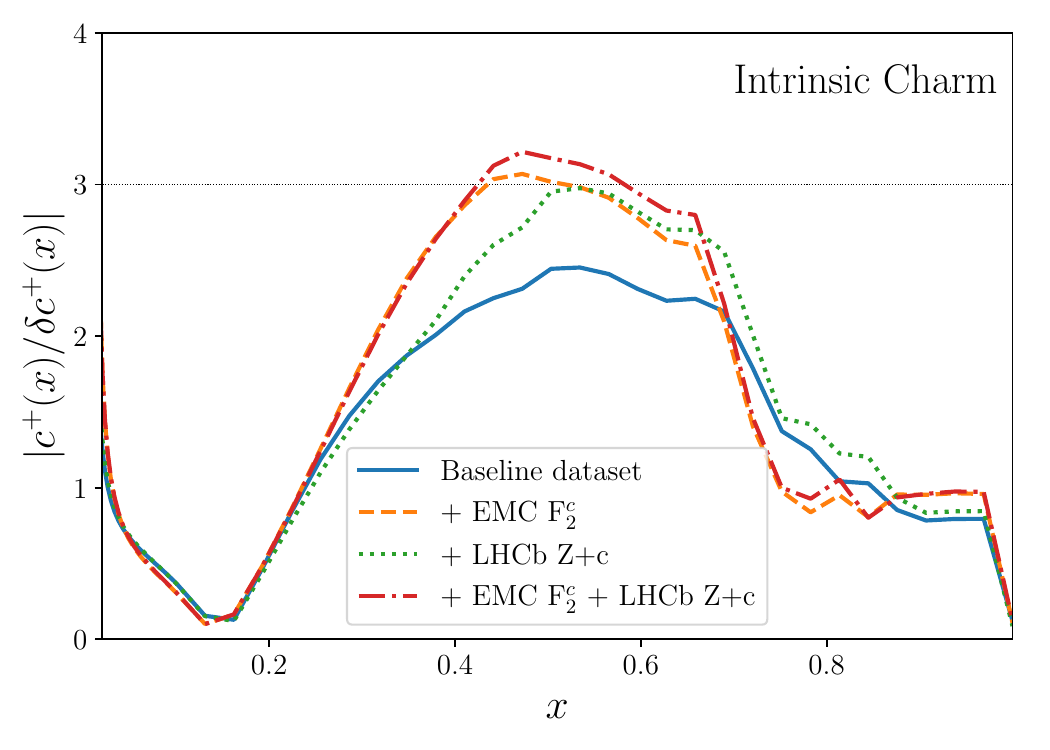}
  \caption{
    The statistical significance of the intrinsic charm 
    PDF in our baseline analysis, compared to the results obtained 
    also including either the LHCb $Z$+charm (with uncorrelated systematics) 
    or the EMC structure function data, or both.
  }
  \label{fig:Zc_bottomrow_pull} 
\end{figure}

%
This is displayed in \cref{fig:Zc_bottomrow_pull} for our default determination of
intrinsic charm, as well as after inclusion of either the LHCb $Z$+charm or the EMC data, 
or both.
We find a local significance for intrinsic charm at the $2.5\sigma$ level in the region 
$0.3 \lsim x \lsim 0.6$.
This is increased to about $3\sigma$ by the inclusion of either the EMC or the LHCb data, 
and above if they are both included.
The similarity of the impact of the EMC and LHCb measurements is especially remarkable 
in view of the fact that they involve very different physical processes and energies.


\paragraph{$\mathcal{R}_j^c$ computational settings.}

Here we provide full details on our computation of $Z$+charm production 
and on the inclusion of the LHCb data for this process in the determination 
of the charm PDF shown in \cref{fig:Zc_toprow}.
We follow the settings described in~\cite{Boettcher:2015sqn}.
$Z$+jet events at NLO QCD theory are generated for $\sqrt{s}= 13$ TeV using the $Zj$ 
package of the {\sc\small POWHEG-BOX}~\cite{Alioli:2010xd}.
The parton-level events produced by {\sc\small POWHEG} are then interfaced to 
{\sc\small Pythia8}~\cite{Sjostrand:2007gs} with the Monash 2013 tune~\cite{Skands:2014pea} 
for showering, hadronization, and simulation of the underlying event and multiple
parton interactions.
Long-lived hadrons, including charmed hadrons, are assumed stable and not decayed.

Selection criteria on these particle-level events are imposed to match the LHCb 
acceptance.
$Z$ bosons are reconstructed in the dimuon final state by requiring 
$60~{\rm GeV}\le m_{\mu\mu} \le 120~{\rm GeV}$, and only events where these muons 
satisfy $p_T^\mu \ge 20~{\rm GeV}$ and $2.0 \le \eta_{\mu}\le 4.5$ are retained.
Stable visible hadrons within the LHCb acceptance of $2.0 \le \eta \le 4.5$ are 
clustered with the anti-$k_T$ algorithm with radius parameter of $R=0.5$~\cite{Cacciari:2008gp}.
Only events with a hardest jet satisfying $20~{\rm GeV} \le p_T^{\rm jet} \le 100~{\rm GeV}$
and $2.2 \le \eta_{\rm jet}\le 4.2$ are retained.
Charm jets are defined as jets containing a charmed hadron, specifically jets 
satisfying $\Delta R(j, c{\rm-hadron})\le 0.5$ for a charmed hadron with 
$p_T(c{\rm-hadron})\ge 5~{\rm GeV}$.
Jets and muons are required to be separated in rapidity and azimuthal angle, 
so we require $\Delta R(j, \mu)\ge 0.5$.
The resulting events are then binned in the $Z$ bosom rapidity $y_Z = y_{\mu \mu}$.

The physical observable measured by LHCb is the ratio of the fraction of $Z$+jet events 
with and without a charm tag, given by \cref{eq:Rcj}, where the denominator of \cref{eq:Rcj} 
includes all jets, even those containing heavy hadrons.
%
%
The charm tagging efficiency is already accounted for at the level of the experimental 
measurement, so it is not required in the theory simulations.

Predictions for \cref{eq:Rcj} are produced using our default PDF determination 
(NNPDF4.0 NNLO), as well as the corresponding PDF set with perturbative charm \cref{sec:ic_pch}.
We have explicitly checked that our results are essentially independent of the value of 
the charm mass.
We have evaluated MHOUs and PDF uncertainties using the output of the {\sc\small POWHEG+Pythia8} 
calculations. We have checked that MHOUs, evaluated with the standard seven-point 
prescription, essentially cancel in the ratio \cref{eq:Rcj}. Note that this is not 
the case for PDF uncertainties, because the dominant partonic sub-channels in the 
numerator and denominator are not the same.

\subsection{Parton luminosities}
\label{sec:ic_lumi}

The impact of intrinsic charm on hadron collider observables can be assessed 
by studying parton luminosities. Indeed, the cross-section for hadronic processes 
at leading order is typically proportional to an individual parton luminosity 
or linear combination of parton luminosities.
Comparing parton luminosities determined using our default PDF set to those obtained 
imposing perturbative charm (\cref{sec:ic_pch}) provides a qualitative estimate of 
the measurable impact of intrinsic charm. Of course this is then modified by 
higher-order perturbative corrections, which generally depend on more partonic
sub-channels and thus on more than a single luminosity.
In this section we illustrate this by considering the parton luminosities 
that are relevant for the computation of the $Z$+charm process in the LHCb 
kinematics, see \cref{sec:ic_lhcb}.

The parton luminosity without any restriction on the rapidity $y_X$ of the final 
state can be defined as in \cref{eq:lumi_def}.
For the more realistic situation where the final state rapidity is restricted, 
$y_{\rm min}\le y_X\le y_{\rm max}$, \cref{eq:lumi_def} is modified as
\begin{equation}
    \mathcal{L}_{ab}(m_X)= \frac{1}{s}\int_{\tau}^1 
        \frac{dx}{x}f_a \lp x,m_X^2\rp f_b\lp \tau/x,m_X^2\rp 
        \theta\lp y_X-y_{\rm min} \rp \theta\lp y_{\rm max}-y_X \rp\, ,
    \label{eq:lumi1D_restricted}
\end{equation}
where $y_X = \lp \ln x^2/\tau \rp /2$. For this observable we consider in 
particular the quark-gluon and the charm-gluon luminosities, defined as
\begin{equation}
    \mathcal{L}_{qg}(m_X) = \sum_{i=1}^{n_f} 
        \lp 
            \mathcal{L}_{q_ig}(m_X)+ \mathcal{L}_{\bar{q}_ig}(m_X) 
        \rp\, , 
    \quad
    \mathcal{L}_{cg}(m_X) = 
        \lp 
            \mathcal{L}_{cg}(m_X)+ \mathcal{L}_{\bar{c}g}(m_X)
        \rp\, ,
    \label{eq:lumis}
\end{equation}
where $n_f$ is the number of active quark flavors for a given value of $Q=m_X$
with a maximum value of $n_f=5$.
These are the flavor combinations that provide the leading contributions respectively 
to the numerator ($\mathcal{L}_{cg}$) and the denominator ($\mathcal{L}_{qg}$) 
of $\mathcal{R}_j^c$ in \cref{eq:Rcj}. 

\begin{figure}[!t]
    \centering
    \includegraphics[width=0.99\textwidth]{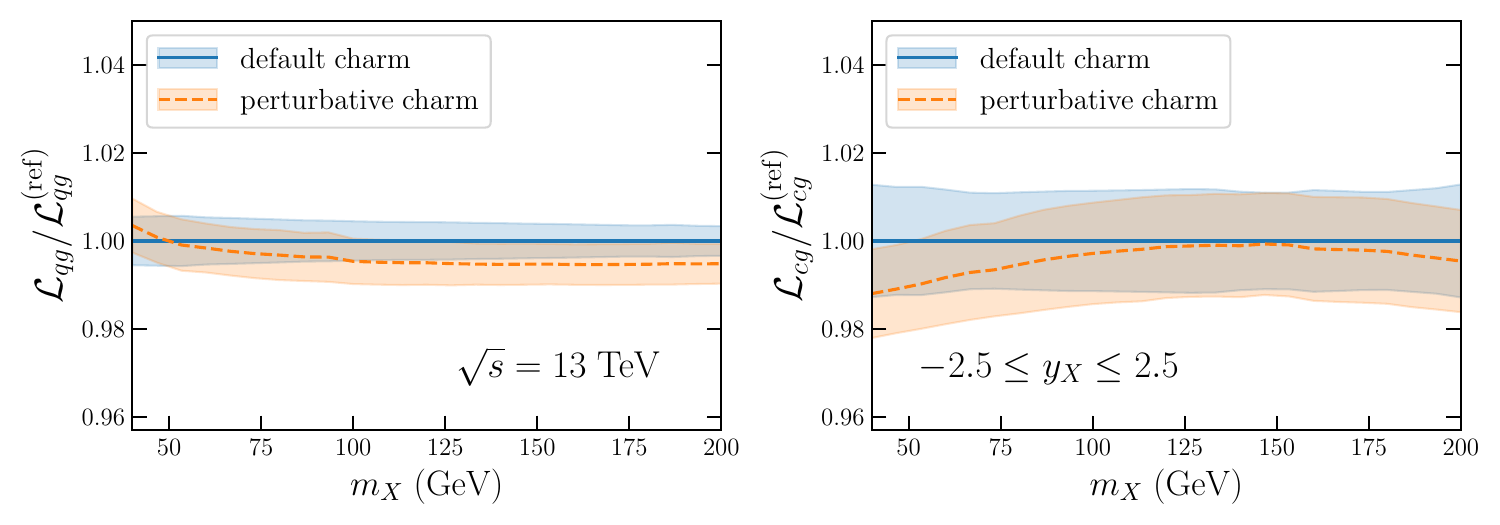}
    \includegraphics[width=0.99\textwidth]{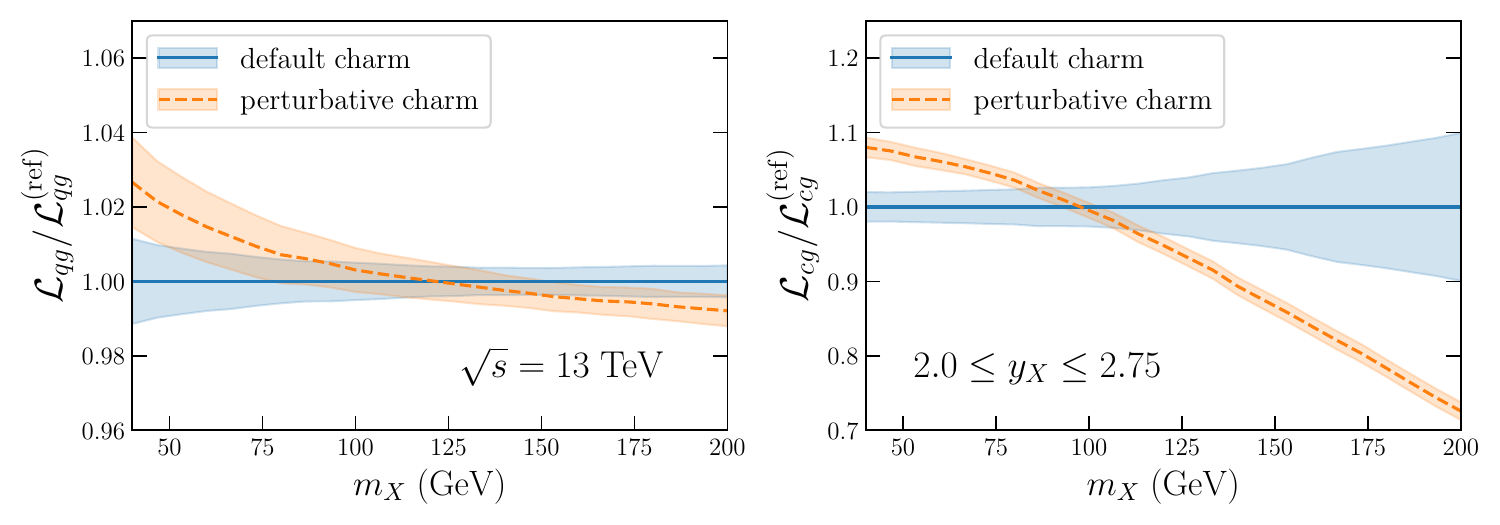}
    \includegraphics[width=0.99\textwidth]{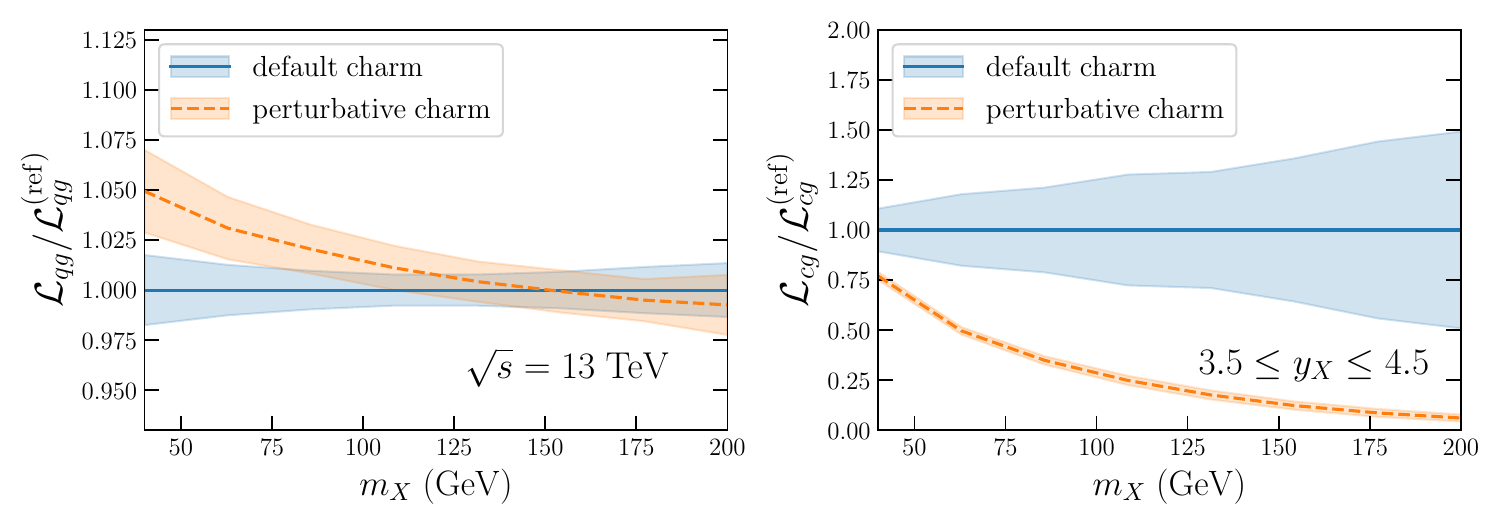}
    \caption{
        The quark-gluon (left) and charm-gluon (right) parton luminosities 
        in the $m_X$ region relevant for $Z$+charm production and three 
        different rapidity bins (see text). Results are shown both for our 
        default charm PDFs and for the variant with perturbative charm. 
    }
    \label{fig:charm_luminosities} 
\end{figure}

The luminosities are displayed in \cref{fig:charm_luminosities}, in the invariant 
mass region, $40~{\rm GeV}\le m_X \le 200~{\rm GeV}$ which is most relevant for
$Z$+charm production.
Results are shown for three different rapidity bins, $-2.5 \le y_X \le 2.5$ 
(central production in ATLAS and CMS), $2.0 \le y_X \le 2.75$ (forward production, 
corresponding to the central bin in LHCb), and $3.5 \le y_X \le 4.5$ 
(highly boosted production, corresponding to the most forward bin in the 
LHCb selection), as a ratio to our default case.

For central production it is clear that both the quark-gluon and charm-gluon 
luminosities with our without intrinsic charm are very similar. This means 
that central $Z$+charm production in this invariant mass range is insensitive 
to intrinsic charm.
For forward production, corresponding to the central LHCb rapidity bin, 
$2.0 \le y_X \le 2.75$, in the invariant mass region $m_X\simeq 100$~GeV again 
there is little difference between results with or without intrinsic charm, but 
as the invariant mass increases the charm-gluon luminosity with intrinsic charm 
is significantly enhanced.
For very forward production, such as the highest rapidity bin of LHCb, 
$3.5 \le y_X \le 4.5$, the charm-gluon luminosity at $m_X \simeq 100$~GeV is enhanced 
by a factor of about 4 in our default result in comparison to the perturbative 
charm case, corresponding to a $\simeq 3\sigma$ difference in units of the 
PDF uncertainty, consistently with the behavior observed for the $\mathcal{R}_j^c$ 
observable in \cref{fig:Zc_toprow}~(left) in the most forward rapidity bin.
This observation provides a qualitative explanation of the results of \cref{sec:ic_lhcb}.

\section{Intrinsic charm asymmetry}
\label{sec:ic_icasy}




%
In the previous \cref{sec:ic_ic} we have presented a determination 
of intrinsic charm in the proton from a global analysis of parton distribution 
functions (PDFs)~\cite{NNPDF:2021njg,Ball:2022qks}.
This study found evidence for intrinsic charm at the $3\sigma$ level, and was 
supported by independent constraints from forward $Z$ production  with charm 
jets at the LHCb experiment~\cite{LHCb:2021stx}.

In doing so we determined the distributions of charm quarks and antiquarks assuming 
equality of the intrinsic (scale-independent) charm and anticharm PDFs, i.e.\ the 
vanishing of the charm valence PDF
\begin{equation}
    c^-(x,Q^2)=c(x,Q^2)-\bar c(x,Q^2)\, .
    \label{eq:cmin}
\end{equation}
Nevertheless, the valence charm PDF $c^-(x,Q^2)$ must have vanishing integral over 
$x$ at all scales $Q^2$, because the proton does not carry the charm quantum number, 
but the PDF itself may well be nonzero, as it happens for the strange valence 
PDF $s^-=s-\bar s$.
Indeed, a non-vanishing charm valence component is always generated, like for 
any other quark flavor, by perturbative QCD evolution~\cite{Catani:2004nc}.
However, any perturbatively generated valence charm component is tiny in comparison 
to all other PDFs, including those of heavy quarks. Hence, any evidence of a sizable 
valence charm PDF is a definite sign of its intrinsic nature.
Model calculations~\cite{Brodsky:2015fna,Hobbs:2013bia}, while in broad agreement 
on the shape of total intrinsic charm PDF, widely differ in predictions for the shape 
and magnitude of the intrinsic valence charm component.
Model calculations of intrinsic charm complemented with input from lattice QCD
~\cite{Sufian:2020coz} also predict a non-vanishing valence component.

Here we investigate this issue by performing a data-driven determination of the 
intrinsic valence charm PDF of the proton, based on the same methodology as 
in~\cite{Ball:2022qks} (cf. \cref{sec:ic_nnpdf40charm,sec:ic_ic}).
We generalize the NNPDF4.0 PDF determination by introducing an independent 
parametrization of the charm and anticharm PDFs, determine them from a global QCD 
analysis (\cref{sec:icasy_pdf}), and subtract the perturbatively generated contributions 
by transforming all PDFs to the 3FNS in which perturbative charm vanishes so any 
residual charm PDF is intrinsic.

We find a non-zero charm valence PDF, with a positive valence peak for $x\sim 0.3$,
whose local significance is close to two sigma. 
We demonstrate the stability of this result with respect to theoretical, dataset, 
and methodological variations (\cref{sec:icasy_results}).
In \cref{sec:icasy_pheno} we conclude proposing two novel experimental probes to 
further scrutinize this asymmetry between charm and anticharm PDFs: $D$-meson 
asymmetries in $Z$+c-jet production at LHCb, and flavor-tagged structure functions 
at the upcoming EIC.

\subsection{The valence charm PDF}
\label{sec:icasy_pdf}

Also in this study we follow the NNPDF4.0 methodology, theory settings 
and dataset~\cite{NNPDF:2021njg} (\cref{sec:nnpdf_methodology}), 
the only modifications being related to the independent parametrization 
of the charm valence PDF. 
Firstly, the neural network architecture is extended with an additional 
neuron in the output layer in order to independently parametrize $c^-(x,Q_0)$, 
\cref{eq:cmin}, at the PDF parametrization scale $Q_0=1.65$~GeV.
In the default PDF basis (``evolution basis'', see \cref{eq:evol_basis}) this 
extra neuron is taken to parametrize the valence non-singlet combination 
$V_{15}=(u^- + d^- + s^- - 3c^-)$. In an alternative basis (``flavor basis'')
it instead parametrizes $\bar{c}$: so in both cases the valence component 
is obtained by taking linear combinations of the neural network outputs.
In our previous analysis, the assumption of vanishing intrinsic valence was 
enforced by setting $V_{15}=V=\sum_i q_i^-$ in the evolution basis or $\bar{c}=c$ 
in the flavor basis at the scale $Q_0$.

In addition to experimental constraints, a non-zero charm valence must, 
as mentioned, satisfy the sum rule
\begin{allowdisplaybreaks} 
\begin{align}
    Q_{15} = \int_0^1 dx\, V_{15}(x,Q_0)&=3 \, ,
    \label{eq:vsr_evol} \\
    Q_c = \int_0^1 dx\, (c-\bar{c})(x,Q_0)&=0  \,,
    \label{eq:vsr_flav}
\end{align} 
\end{allowdisplaybreaks}
in the evolution or flavor basis respectively.
This sum rule is enforced in the same manner as that of the strange valence sum rule.
Finally, to ensure cross-section positivity (at $Q^2=5$~GeV$^2$) separately for charm- 
and anticharm initiated processes, we replace the neutral current $F_2^c$ positivity
observable (sensitive only to $c^+$) with its charged current-counterparts $F_2^{c,W^-}$ 
and $F_2^{\bar{c},W^+}$.
The charm PDFs $xc$ and $x\bar{c}$ themselves are not required to be positive-definite
~\cite{Collins:2021vke,Candido:2023ujx}.
Integrability and preprocessing are imposed as in NNPDF4.0.
We have verified that results are stable upon repeating the hyperoptimization of all
parameters defining the fitting algorithm, and thus we keep the same settings as in
~\cite{NNPDF:2021njg}.


As explained in \cref{sec:ic_methods}, intrinsic charm is the charm PDF in the 3FNS, 
where charm is treated as a massive particle that does not contribute to the running 
of the strong coupling or the evolution of PDFs.
In the absence of intrinsic charm (i.e. ``perturbative charm'' only) the charm and anticharm 
PDFs in the 3FNS vanish identically.
In the 4FNS, in which charm is treated as a massless parton, these PDFs are determined by
perturbative matching conditions between the 3FNS and the 4FNS.
In our data-driven approach, the charm and anticharm PDFs, instead of being fixed by 
perturbative matching conditions, are determined from data on the same footing as the light 
quark PDFs.
The deviation of data-driven charm from perturbative charm, i.e., in the 3FNS the deviation 
of the charm and anticharm PDFs from zero, is identified with the intrinsic component.
In practice, we parametrize PDFs at $Q_0=1.65$~GeV in the 4FNS, and then invert the matching 
conditions of \cref{eq:singlet_matching,eq:ns_matching} to determine the intrinsic component 
in the 3FNS.

\begin{figure}[!t]
  \centering
  \includegraphics[width=0.99\textwidth]{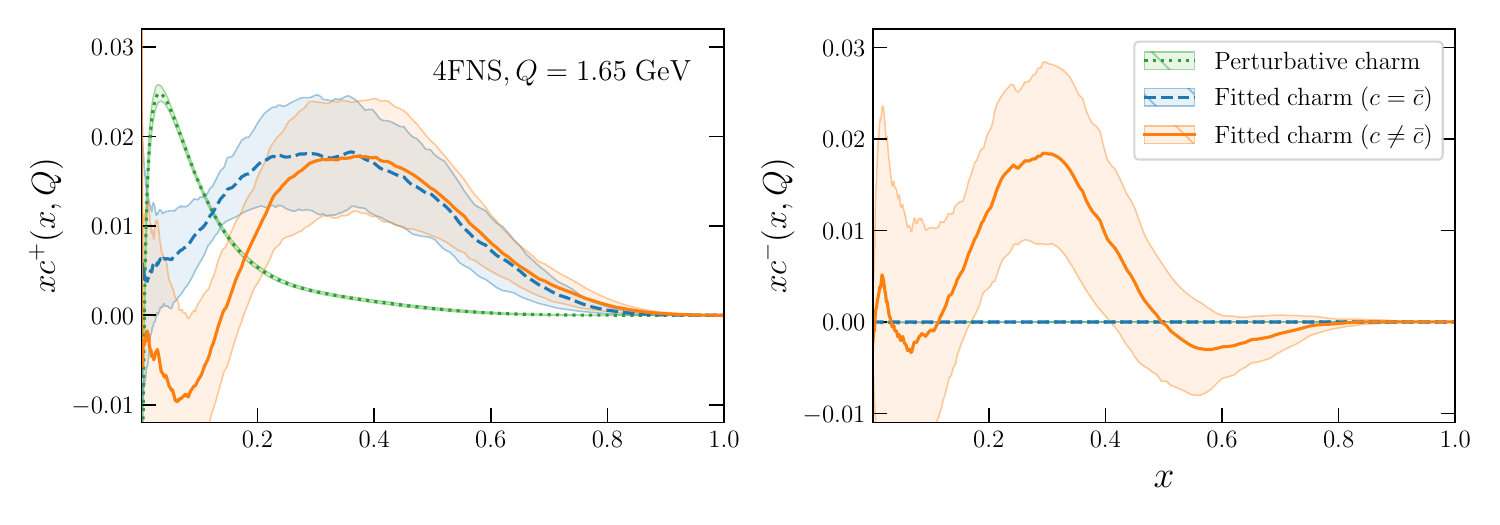}
  \caption{
    The charm total $xc^+$ (left) and valence $xc^-$ (right) PDFs in the 4FNS 
    at $Q=1.65$~GeV. The perturbative and data-driven results are compared, 
    in the latter case either assuming $c^-=0$ or $c^-$ determined from data.
  }
  \label{fig:CharmAsymmetry-q1p65gev-Fit1Main}
\end{figure}

In \cref{fig:CharmAsymmetry-q1p65gev-Fit1Main} we show $xc^+$ and $xc^-$ in the 4FNS at
$Q=1.65$~GeV, i.e.\ just above the charm mass that we take to be $m_c=1.51$~GeV, determined 
using NNLO QCD theory. The bands are $68~\%$ confidence level (CL) PDF uncertainties. We show 
both the purely perturbative and data-driven results, in the latter case both for $c=\bar{c}$
and  $c \ne \bar{c}$.
Note that the purely perturbative valence PDF vanishes at $Q=m_c$ at NNLO, and only develops 
a tiny component at one extra perturbative order (N$^3$LO), or at higher scales. Hence, a
non-vanishing valence component in the 4FNS provides already evidence for intrinsic charm.

Upon allowing for a vanishing valence $xc^-$ component, the total charm $xc^+$ is quite stable, 
especially around the peak at $x\sim0.4$. This total charm PDF is also somewhat suppressed for
smaller $x\lesssim 0.2$ as compared to the baseline result. In terms of fit quality, the $\chi^2$ 
per data point for the global dataset decreases from 1.162 to 1.151, corresponding
to an improvement by about 50 units in absolute $\chi^2$. The main contributions to this decrease 
comes from neutral current deep-inelastic scattering and LHC gauge boson production data.

The valence component is nonzero and positive at more than one sigma level in the $x \in [0.2, 0.4]$ 
region, and consistent with zero within the large PDF uncertainties elsewhere.
All other PDFs are mostly left unaffected by having allowed for a non-vanishing valence charm.

\begin{figure}[!t]
    \centering
    \includegraphics[width=0.99\textwidth]{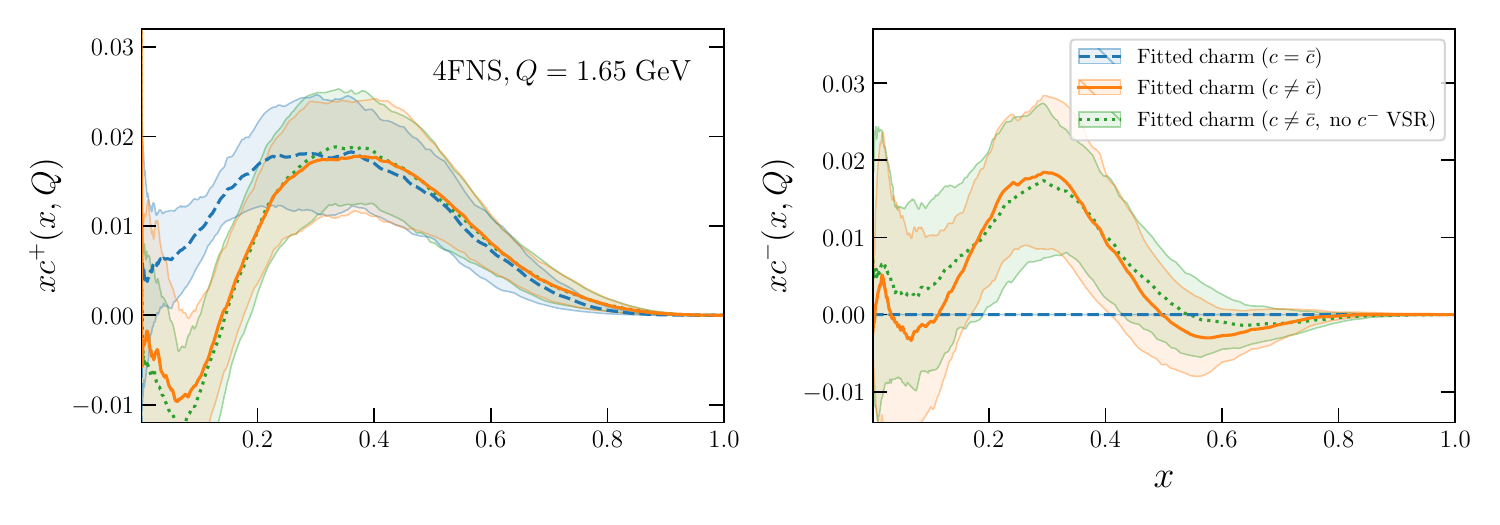}
    \caption{
      Same as \cref{fig:CharmAsymmetry-q1p65gev-Fit1nsr}, now without imposing 
      the charm valence sum rule \cref{eq:vsr_evol} when $c\not=\bar c$.
    }
    \label{fig:CharmAsymmetry-q1p65gev-Fit1nsr}
\end{figure}

Whereas in our default determination we have imposed the charm valence sum rule \cref{eq:vsr_evol}, 
we have also repeated our determination without imposing this theoretical constraint.
We then obtain $Q_c=0.07\pm 0.14$ and the resulting charm PDFs are shown in \cref{fig:CharmAsymmetry-q1p65gev-Fit1nsr}.
This result demonstrates that the valence sum rule is actually enforced by the data, 
and our result is data-driven.

In the following paragraphs we prove that the size and shape of the valence charm PDF 
seen in \cref{fig:CharmAsymmetry-q1p65gev-Fit1Main}  are stable upon variations of PDF 
parametrization basis, the value of $m_c$, the input dataset, and the kinematic cuts 
in $W^2$ and $Q^2$.


\paragraph{Fit quality and data impact.}

We compare the fit quality for the PDF determination presented here with $c\ne \bar{c}$ 
to the published NNPDF4.0 determination with $c = \bar{c}$, by showing in \cref{tab:chi2-baseline} 
the experimental $\chi^2$ per data point for different groups of processes and 
for the total dataset.
We refer to~\cite{NNPDF:2021njg} for the definition of the dataset and of the 
$\chi^2$ and the process categories (see in particular \cite[Tab.~5.1]{NNPDF:2021njg}).

The largest reduction in absolute $\chi^2$ upon allowing for a non-vanishing charm 
valence component is in collider DIS (i.e. HERA), and the largest reduction in $\chi^2$ 
in charged-current collider DIS. There the largest impact is seen in the large $Q^2$, 
large $x$ bins, 
consistent with 
the observation that the intrinsic charm PDFs are localized at large $x$.
Note that HERA data for the $F_2^c$ charm structure function, that are included in the fit, 
have no impact on intrinsic charm because they are in the medium-to-low $x$ region
where the charm PDF is dominated by the perturbative component.

\begin{table}[!t]
  \centering
  \small
  \renewcommand{\arraystretch}{1.40}
  \begin{tabularx}{\textwidth}{Xrcc}
\hline
Dataset  &  $n_{\rm dat}$
         & $\qquad \chi^{2}/n_{\rm dat}$ ($c \ne \bar{c}$) $\qquad$
         & $\qquad\chi^{2}/n_{\rm dat}$ ($c = \bar{c}$) $\qquad$\\
\hline
DIS NC (fixed-target)                     &  973 & 1.24 & 1.26 \\
DIS CC (fixed-target)                     &  908 & 0.86 & 0.86 \\
DIS NC (collider)                         & 1127 & 1.18 & 1.19 \\
DIS CC (collider)                         &   81 & 1.23 & 1.28 \\
Drell-Yan (fixed-target)                  &  195 & 1.02 & 1.00 \\
Tevatron $W$, $Z$               &   65 & 1.06 & 1.09 \\
LHC $W$, $Z$                    &  463 & 1.35 & 1.37 \\
LHC $W$, $Z$  ($p_T$ and jets)  &  150 & 0.99 & 0.98 \\
LHC top-quark pair              &   64 & 1.28 & 1.21 \\
LHC jet                         &  520 & 1.25 & 1.26 \\ 
LHC isolated $\gamma$           &   53 & 0.76 & 0.77 \\
LHC single $t$                  &   17 & 0.36 & 0.36 \\
\hline
{\bf Total}                               & {\bf 4616} & {\bf 1.151} & {\bf 1.162}\\
\hline
\end{tabularx}

  \caption{
    The values of the experimental $\chi^2$ per data point for the different groups 
    of processes entering the NNPDF4.0 determination as well as for the total dataset. 
    We compare the results of the baseline NNPDF4.0 fit ($c = \bar{c}$) with the 
    results of this work ($c \ne \bar{c}$).
  }
  \label{tab:chi2-baseline}
\end{table}


\paragraph{Parametrization basis dependence.}

Because here we are determining a difference between PDFs, it is especially important 
to check stability upon choice of basis.
In the evolution basis, the charm PDFs are parametrized through the two combinations
$T_{15}$ and $V_{15}$ of \cref{eq:evol_basis}, at a scale $Q=Q_0=1.65$~GeV. 
In the flavor basis, they are parametrized as $c(x,Q)$ and $\bar{c}(x,Q)$. Note that 
in neither case the total and valence combinations $c^\pm$ are elements of the basis.

\begin{figure}[!t]
  \centering
  \includegraphics[width=0.99\textwidth]{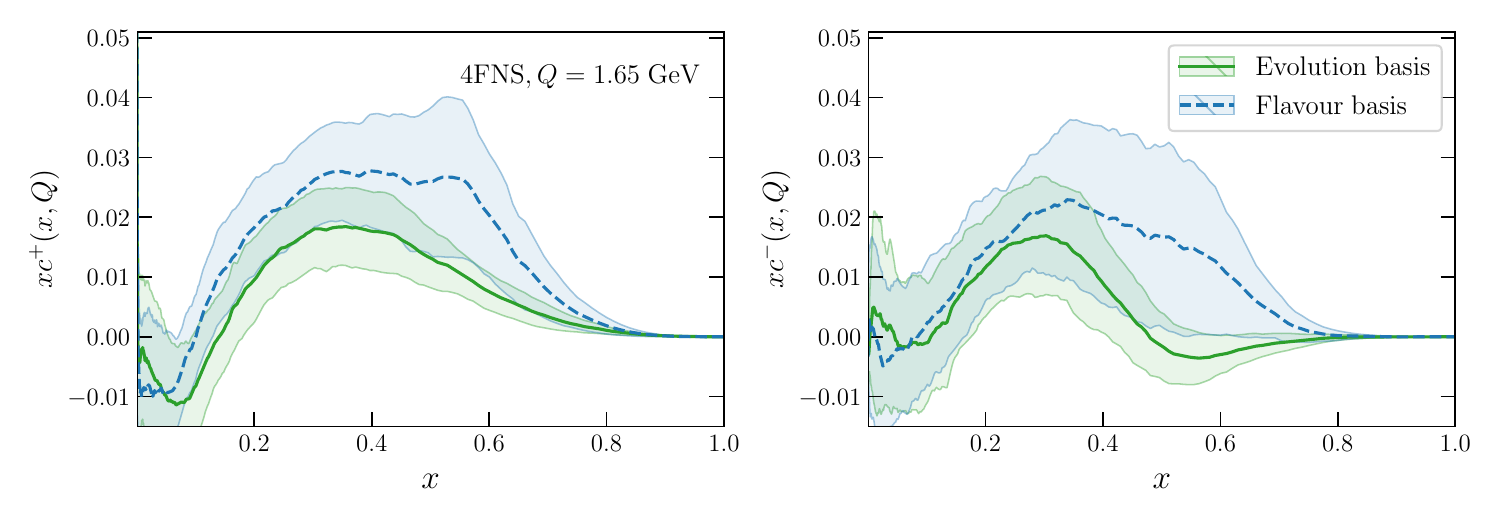}
  \caption{
    Comparison between the total  $xc^+$ (left) and valence $xc^-$ 
    (right) charm PDFs in the 4FNS at $Q=1.65$~GeV, obtained parametrizing 
    PDFs in the evolution basis (default) and in the flavor basis.
  }
  \label{fig:ICasym-SI-basisdep}
\end{figure}

In \cref{fig:ICasym-SI-basisdep} the $xc^\pm$ PDFs found using either basis are compared. 
Agreement at the one sigma level is found for all $x$.
The main qualitative features are independent of the basis choice, specifically the presence 
of a positive valence peak around $x\sim0.3$ for $xc^-$.
Results in the flavor basis display larger PDF uncertainties, possibly because the 
flavor basis has not undergone the same extensive hyperoptimization as the fits 
in the evolution basis.

\paragraph{Dependence on the value of the charm mass.}

\begin{figure}[!t]
    \centering
    \includegraphics[width=0.99\textwidth]{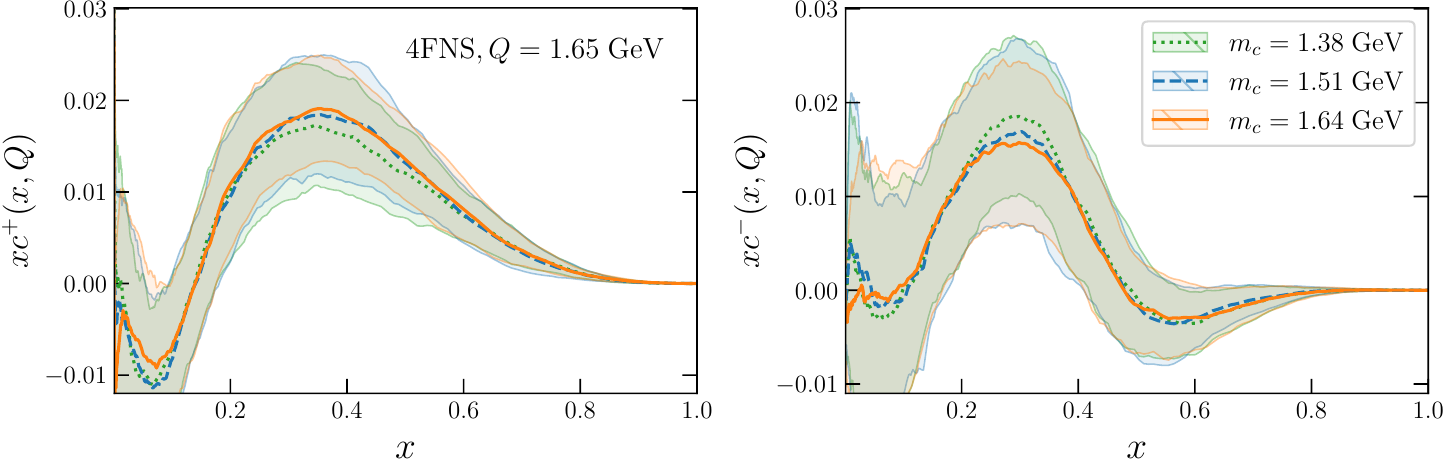}
    \caption{
      Comparison of the total $xc^+$ (left) and valence charm $xc^-$ (right) PDFs 
      in the 4FNS at $Q=1.65$~GeV  as the charm pole mass is varied about
      the default central value $m_c=1.51$~GeV by $\pm\,0.13$~GeV.
  }
  \label{fig:ICasym-SI-mcdep}
\end{figure}

We verify the independence of our results on the value of the charm quark mass, 
by repeating our determination as the (pole) charm mass is varied from the default 
$m_c=1.51$~GeV to $m_c=1.38$~GeV and 1.64~GeV.
Note that we always choose the scale $\mu_c=m_c$ as matching scale between the 
4FNS and 3FNS, hence this is also varied alongside $m_c$.
The total and valence charm PDFs $xc^\pm$ at $Q=1.65$~GeV in the 4FNS are displayed 
in \cref{fig:ICasym-SI-mcdep}.
The result is found to be essentially independent of the charm mass value, 
in agreement with the corresponding result of \cref{sec:ic_stability_4FNS} 
(\cref{fig:charm_fitted_mcdep}).

\paragraph{Dataset dependence.}

\begin{figure}[!t]
    \centering
    \includegraphics[width=0.99\textwidth]{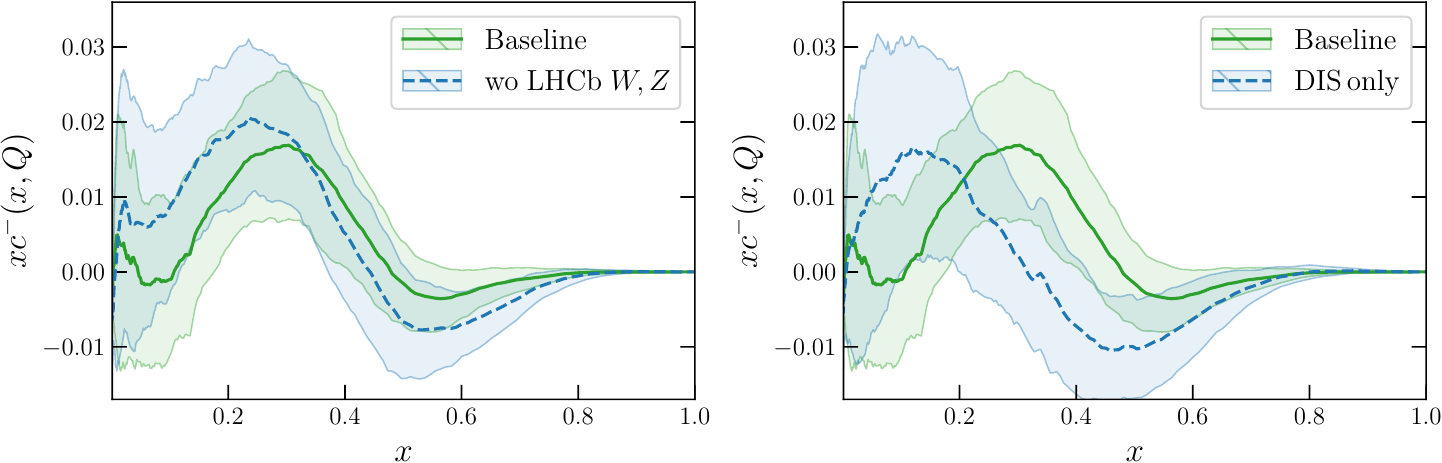}
    \caption{
        The charm valence PDF in the default determination compared to a
        determination in which LHCb $W,Z$ inclusive production data are excluded (left)
        and a determination based on DIS structure functions only (right).
    }
    \label{fig:ICasym-SI-dataset}
\end{figure}

We study the effect of removing some datasets from our determination, with the dual goal 
of checking the stability of our results, and investigating which data mostly determine 
the valence charm PDF.
Specifically, we remove the LHCb $W,Z$ data, which was found in \cref{sec:ic_stability_4FNS}
to dominate the constraints on the total charm PDF from all collider measurements, 
and we determine the PDFs only using DIS structure function data.
The valence charm PDF found in either case is compared to the default in 
\cref{fig:ICasym-SI-dataset}.
Removing the LHCb electroweak data leaves $xc^-$ mostly unchanged, hence the valence 
PDF appears to be less sensitive to this data than the total charm.
When only including DIS data a nonzero valence component is still found but now 
with a reduced significance: the result is consistent with zero at the 
one sigma level.

\paragraph{Kinematic cuts.}

\begin{figure}[!t]
    \centering
    \includegraphics[width=0.99\textwidth]{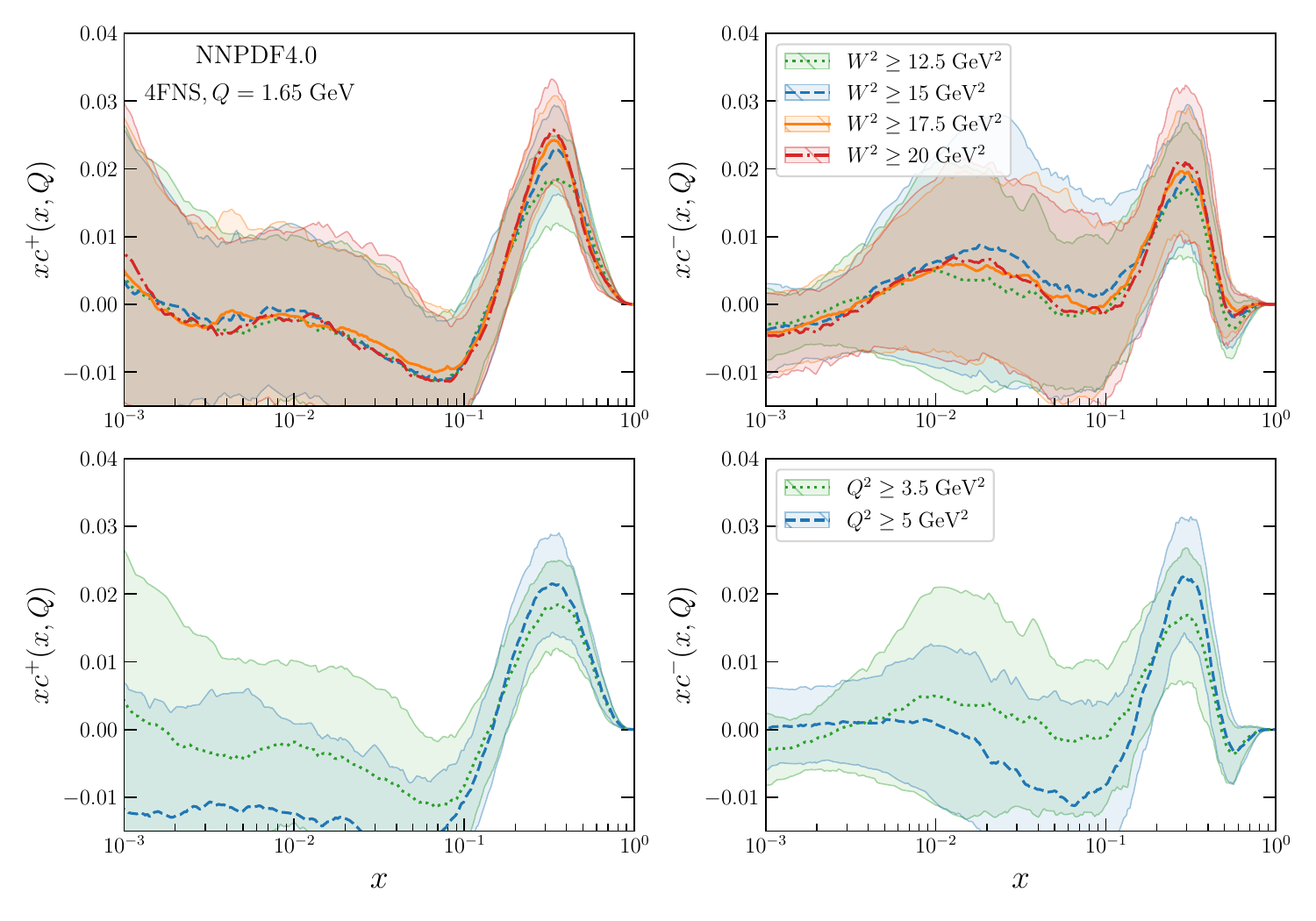}
    \caption{
      The variation in the 4FNS total (left) and valence (right) charm
      PDFs at $Q=1.65$~GeV as the $W^2$ cut is raised to 20~GeV$^2$ in steps 
      of 2.5~GeV$^2$ (top) and the $Q^2$ cut is raised to 5~GeV$^2$ (bottom).
      The kinematic cuts in the baseline fit are $Q^2\ge 3.5$~GeV$^2$ and 
      $W^2\ge 12.5$~GeV$^2$.
    }
    \label{fig:ICasym-SI-KinCuts}
\end{figure}

The NNPDF4.0 dataset only includes data with $Q^2 \ge 3.5$~GeV$^2$ and $W^2 \ge 12.5$~GeV$^2$, 
in order to ensure the reliability of the leading-twist, fixed-order perturbative approximation.
It is important to verify that results for intrinsic charm are stable upon variation 
of these cuts, as this checks that intrinsic charm is not contaminated by possible 
non-perturbative corrections not accounted for in the global PDF fitting framework.
To this purpose, we have raised the $W^2$ cut in steps of 2.5~GeV$^2$ up to 20~GeV$^2$, 
and the $Q^2$ cut up to 5~GeV$^2$.
Results are displayed in \cref{fig:ICasym-SI-KinCuts}, and prove satisfactory stability: 
upon variation of the $W^2$ cut nothing changes, and upon variation of the $Q^2$ cut 
(which removes a sizable amount of data) the central value is stable and the uncertainty 
only marginally increased.

\subsection{The intrinsic valence charm PDF}
\label{sec:icasy_results}

\begin{figure}[!t]
  \centering
  \includegraphics[width=0.99\columnwidth]{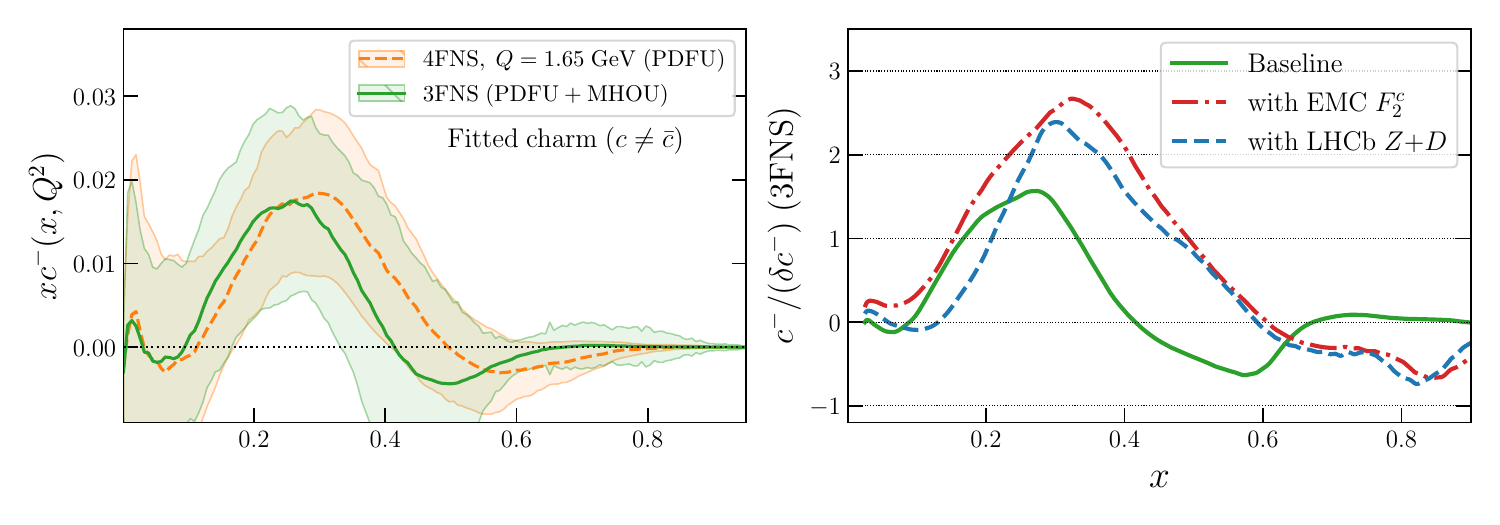}
  \caption{
    Left: the 3FNS (intrinsic) valence charm PDF $xc^-$, compared to the 
    4FNS result (same as \cref{fig:CharmAsymmetry-q1p65gev-Fit1Main} right).
    The 3FNS also includes MHOU due to the inversion from the 4FNS to the 3FNS.
    Right: the pull for valence $xc^-$ charm PDF in the 3FNS.
    Results are shown both for the default fit and also when including the EMC 
    $F_2^c$ and LHCb $Z+c$ data.
  }
  \label{fig:3FNS_ICasy}
\end{figure}

The intrinsic valence charm PDF is now determined by transforming back to the 
3FNS scheme, and is displayed in \cref{fig:3FNS_ICasy} (upper panel), together 
with its 4FNS counterpart already shown in \cref{fig:CharmAsymmetry-q1p65gev-Fit1Main}. 
An estimate of the missing higher order uncertainties (MHOU) related
to the truncation of the perturbative expansion is also included. 
This, as in~\cref{sec:ic_results}, is estimated as the change in the 3FNS PDF 
when the transformation from the 4FNS to the 3FNS is performed to one higher 
perturbative order, i.e.\ N$^3$LO, as this is estimated to be the dominant 
missing higher order correction.

The 3FNS and 4FNS valence PDFs turn out to be quite close, implying that for 
the valence PDF, unlike for the total charm PDF, the theory uncertainty is smaller 
than the PDF uncertainty. 
We thus find that the intrinsic (3FNS) charm valence is nonzero and positive
roughly in the same $x$ region as its 4FNS counterpart.

The statistical significance of the non-vanishing valence is quantified
by the pull, defined as the median PDF in units of the total uncertainty,
shown in \cref{fig:3FNS_ICasy} (right).
The local significance of the intrinsic valence is slightly below two sigma,
peaking at $x\sim 0.3$. The significance of the total intrinsic component 
is similar to that found in Ref.~\cite{Ball:2022qks} (cf. \cref{fig:Zc_bottomrow_pull}), 
namely about three sigma for $x\sim 0.5$. As in the previous sections, we 
also show the results found in fit variants including the EMC $F_2^c$ and 
LHCb $Z+c$ data~\cite{LHCb:2021stx}, both of which increase the local significance.

The results of \cref{fig:CharmAsymmetry-q1p65gev-Fit1Main,fig:3FNS_ICasy} suggest 
that the intrinsic valence component may be nonzero, but their significance falls 
below the three sigma evidence level.
In the next section, we thus propose two novel experimental observables engineered 
to probe this valence charm component.
\subsection{Charm asymmetries at LHCb and EIC}
\label{sec:icasy_pheno}

The LHCb LHC Run~2 data, which, as shown in \cref{sec:ic_lhcb}, 
reinforce the evidence for an intrinsic total charm component, correspond 
to measurements of forward $Z$ production in association with charm-tagged 
jets~\cite{LHCb:2021stx}. 
They are presented as a measurement of $\mathcal{R}_j^c(y_Z)$, the ratio 
between $c$-tagged and untagged jets in bins of the $Z$-boson rapidity $y_Z$, 
and they are obtained from tagging $D$-mesons from displaced vertices.
The higher statistics available first at Runs~3 and 4 and later at the HL-LHC 
will enable the reconstruction of the exclusive decays of $D$-mesons, and thus 
the separation of charm and anticharm tagged final states. 
We thus define the asymmetry
\begin{equation}
  \mathcal{A}_c(y_Z) = \frac{N_j^c(y_Z) - N_j^{\bar{c}}(y_Z)}{N_j^c(y_Z) + N_j^{\bar{c}}(y_Z)} \, ,
  \label{eq:charm_lhcb_asym}
\end{equation}
where $N_j^c$~($N_j^{\bar{c}}$) is defined in the same manner as $\mathcal{R}_j^c$
(\cref{eq:Rcj}), but now restricted to events with $D$-mesons containing a 
charm quark (antiquark).
This asymmetry is directly sensitive to a possible difference between the charm
and anticharm PDFs in the initial state.

\begin{figure}[!t]
  \centering
  \includegraphics[width=0.49\textwidth]{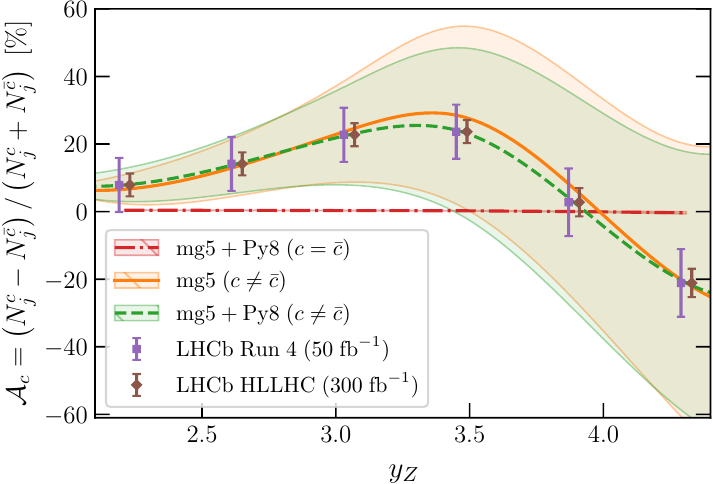}
  \caption{
    The charm asymmetry $\mathcal{A}_c(y_Z)$, \cref{eq:charm_lhcb_asym}, 
    in $Z$+$c$-jet production at LHCb ($\sqrt{s}=13$ TeV) evaluated at LO matched 
    to parton showers with the non-vanishing valence PDF determined here.
    The pure LO result and the result with vanishing charm valence are also 
    shown for comparison.
    The bands correspond to one-sigma PDF uncertainties.
    Projected statistical uncertainties for LHCb measurements at Run~4 
    ($\mathcal{L}=50~$fb$^{-1}$) and the HL-LHC ($\mathcal{L}=300~$fb$^{-1}$) 
    are also shown.
  }
  \label{fig:LHCb_zcham_HLLHC}
\end{figure}

In \cref{fig:LHCb_zcham_HLLHC} we display the asymmetry $\mathcal{A}_c(y_Z)$,
\cref{eq:charm_lhcb_asym}, computed for $\sqrt{s}=13$~TeV using the PDFs determined 
here, that allow for a non-vanishing valence component, as well as the default 
NNPDF4.0 with $c=\bar{c}$. Results are computed using {\sc\small mg5\_aMC@NLO}~\cite{Alwall:2014hca} 
at leading order (LO) matched to {\sc\small Pythia8}~\cite{Sjostrand:2007gs,Skands:2014pea},
with the same $D$-meson tagging and jet-reconstruction algorithm 
as in~\cite{Boettcher:2015sqn,LHCb:2021stx}.
The leading order parton-level result is also shown.

It is apparent from \cref{fig:LHCb_zcham_HLLHC} that, even though the forward-backward 
asymmetry of the $Z$ decay generates a small asymmetry $\mathcal{A}_c\ne 0$ even 
when $c=\bar{c}$~\cite{Gauld:2015qha,Gauld:2019doc}, the LO effect due to an asymmetry 
between $c$ and $\bar c$ PDFs is much larger, and stable upon showering and hadronization 
corrections. Indeed, higher-order QCD corrections largely cancel in the ratio 
$\mathcal{A}_c(y_Z)$.

In \cref{fig:LHCb_zcham_HLLHC} we also display projected uncertainties for the LHCb 
measurement of this asymmetry at Run~3 and at the HL-LHC, showing that a valence 
component of the same size as our central prediction could be detected respectively 
at about a two sigma or four sigma level.


The projected statistical uncertainties for the future LHCb measurement of 
$\mathcal{A}_c$ shown in \cref{fig:LHCb_zcham_HLLHC} are obtained extrapolating 
from those of the Run~2 data by correcting both for the higher luminosity and 
for the acceptance associated to the different charm-tagging procedure required 
in this case.
The uncertainties obtained in the Run~2 measurement~\cite{LHCb:2021stx} and based 
on an integrated luminosity of $\mathcal{L}=6$ fb$^{-1}$ are rescaled to the expected 
luminosity to be accumulated by LHCb by the end of Run~4, $\mathcal{L} \sim 50$ fb$^{-1}$,
and at the HL-LHC, $\mathcal{L} \sim 300$ fb$^{-1}$.
Furthermore, the Run~2 measurement was based on charm-meson tagging with displaced 
vertices, with a charm-tagging efficiency of $\epsilon_c \sim 25\%$.
The asymmetry $\mathcal{A}_c$ (\cref{eq:charm_lhcb_asym}) requires separating charm 
from anticharm in the final states, which in turn demands reconstructing the 
$D$-meson decay products.
The associated efficiency is estimated by weighting the $D$-meson branching fractions
to the occurrence of each species in the LHCb $Z$+charm sample, resulting in an efficiency 
of $\epsilon_c \sim 3\%$.
The uncertainty on the asymmetry is then determined by using error propagation with 
$N_j^c=N^{\bar{c}}_j$, neglecting the dependence of the uncertainty on the value of 
the asymmetry itself.


Let us know turn to another observable that can be used as probe 
of the charm component of the proton i.e. the deep-inelastic charm structure function 
$F_2^c$~\cite{Aubert:1982tt,Forte:2010ta,Ball:2015dpa,H1:2018flt} and the associate 
deep-inelastic reduced charm production cross-section $\sigma_{\rm red}^{c\bar{}c}$
(\cref{eq:dis_xs}).

Correspondingly, the charm valence can be determined from the reduced cross-section 
asymmetry 
\begin{equation}
  \mathcal{A}_{\sigma^{c\bar{c}}}(x,Q^2) = 
  \frac{\sigma_{\rm red}^{c}(x,Q^2)-\sigma_{\rm red}^{\bar{c}}(x,Q^2)}{\sigma_{\rm red}^{c\bar{c}}(x,Q^2)} \, .
  \label{eq:EIC_asy_F2c}
\end{equation}
A measurement of this observable requires reconstructing final-state $D$-mesons by 
identifying their decay products.
At the future EIC this will be possible with good precision using the proposed 
ePIC detector~\cite{Khalek:2021ulf,Armesto:2023hnw,Kelsey:2021gpk}.

\begin{figure}[!t]
    \centering
    \includegraphics[width=0.49\textwidth]{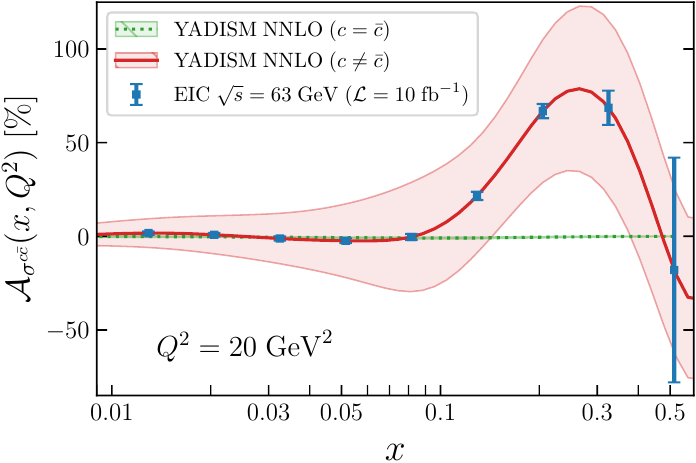}
    \caption{
      The reduced charm-tagged cross-section asymmetry $\mathcal{A}_{\sigma^{c\bar{c}}}$, 
      \cref{eq:EIC_asy_F2c}, at $Q^2=20$~GeV$^2$ computed at NNLO QCD using the 
      non-vanishing valence PDF determined here.
      The result with vanishing charm valence is also shown for comparison.
      The bands correspond to one-sigma PDF uncertainties.
      The projected statistical uncertainties at the EIC~\cite{Kelsey:2021gpk} 
      (running at $\sqrt{s}=63$~GeV for $\mathcal{L}=10$ fb$^{-1}$) are also shown.
  }
  \label{fig:F2c-EIC}
\end{figure}

The predicted asymmetry $\mathcal{A}_{\sigma^{c\bar{c}}}$ at $Q^2=20$~GeV$^2$ is shown 
in \cref{fig:F2c-EIC}; results are shown at the reduced charm (parton) cross-section level, 
evaluated with \yadism~(\cref{sec:yadism}) at NNLO accuracy.
As in \cref{fig:LHCb_zcham_HLLHC}, we show results obtained using the PDFs determined 
here, that allow for a non-vanishing valence component, as well as the default NNPDF4.0 
with $c=\bar{c}$.
We also display the projected statistical uncertainties~\cite{Kelsey:2021gpk} at the EIC 
running at $\sqrt{s}=63$~GeV for $\mathcal{L}=10$ fb$^{-1}$. It is clear that a non-vanishing 
charm valence component can be measured at the EIC to very high significance even for 
a moderate amount of integrated luminosity.

\begin{figure}[!t]
    \centering
    \includegraphics[width=0.49\textwidth]{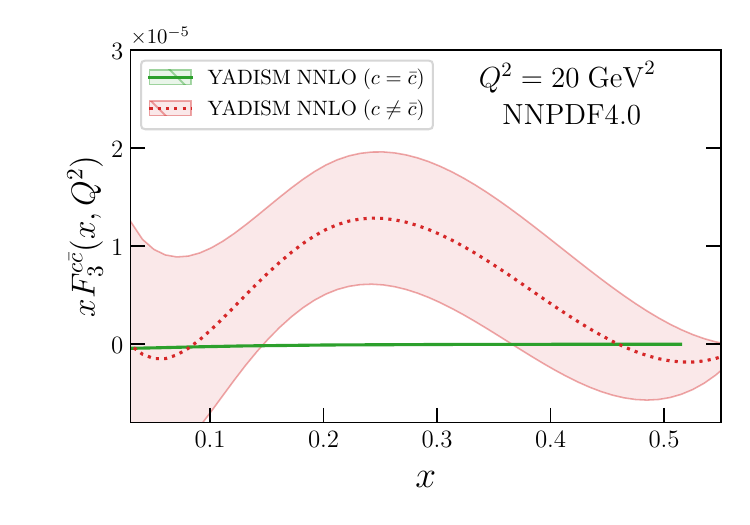}
    \caption{
      Same as \cref{fig:F2c-EIC} for the charm-tagged parity-violating structure 
      function $xF_3^{c\bar{c}}(x,Q^2)$ at the EIC (no projection for the statistical 
      accuracy of the EIC measurement is available).
    }
    \label{fig:ICasym-SI-EIC-xF3c}
\end{figure}

In addition to the charm-tagged structure function $F_2^{c\bar{c}}$, at the EIC
complementary sensitivity to the charm valence content of the proton would be 
provided by the charm-tagged parity-violating structure function $xF_3^{c\bar{c}}(x,Q^2)$.
This observable has the advantage that at LO is already proportional to $xc^-$ 
(see \cref{eq:sf_NC_LO}), hence provides a direct constraint on valence charm.
Predictions for this observable, are presented in \cref{fig:ICasym-SI-EIC-xF3c}. 
Even in the absence of detailed predictions for prospective EIC measurements of 
this observable, it is clear that its measurement would significantly constrain 
the charm valence PDF.
Similarly, if we consider the CC DIS a possible charm asymmetry can be probed 
by the measurements of $F_2^{c\bar{c}}$ (see \cref{eq:sf_CC_LO}) in 
neutrino, anti-neutrino DIS, although the experimental uncertainties are expected
to be larger.


\paragraph{$\mathcal{A}_{\sigma^{c\bar{c}}}$ computational settings.}

The projected statistical uncertainties for the future EIC measurement of the 
charm-tagged asymmetry $\mathcal{A}_{\sigma^{c\bar{c}}}$ shown in \cref{fig:F2c-EIC} 
are obtained as follows.
We adopt the projections from~\cite{Kelsey:2021gpk} for the kinematic coverage 
in the $(x,Q^2)$ plane and the expected statistical precision, based on running 
at a center-of-mass energy of $\sqrt{s}=63$~GeV for $\mathcal{L}=10~{\rm fb}^{-1}$.
These projections entail that measurements of charm production at the EIC 
will cover the region $1.3~{\rm GeV}^2\lesssim Q^2\lesssim 120$~GeV$^2$ 
and $5\times 10^{-4}\lesssim x\lesssim 0.5$.
Charm production is tagged from the reconstruction of $D^0$ and $\bar{D}^0$ 
exclusive decays, and a detailed estimate of experimental uncertainties would 
require a full detector simulation. Here, however, we limit ourselves to 
estimating the statistical accuracy on the asymmetry \cref{eq:EIC_asy_F2c}, 
which is expressed in terms of reduced cross-sections, defined as in 
Ref.~\cite{H1:2018flt} in terms of charm structure functions.
For this, we take the statistical uncertainties provided in~\cite{Kelsey:2021gpk} 
and increase them by a factor $\sqrt{2}$ since the measured sample has to be 
separated into $D$- and $\bar{D}$-tagged subsamples.

\section{Summary}
\label{sec:ic_summary}


In this chapter we have presented a first evidence for intrinsic 
charm quarks in the proton.
%
%
By carefully disentangling the perturbative component, we obtain 
unambiguous evidence for total intrinsic charm $c^+(x)$, which turns out to be 
in qualitative agreement with the expectations from model calculations.
Our determination of the charm PDF, driven by indirect constraints 
from the latest high-precision LHC data, is perfectly consistent 
with direct constraints both from EMC charm production data taken 
forty years ago, and with more recent $Z$+charm production data in 
the forward region from LHCb.
Combining all data, we find local significance for total intrinsic charm 
in the large-$x$ region just above the $3\sigma$ level.


Regarding a possible intrinsic charm asymmetry $c^-(x)$, 
our main conclusion is that current experimental data provide support
for the hypothesis that the valence charm PDF may be nonzero, even 
though with the NNPDF4.0 dataset it is not possible to reach three-sigma evidence.
Whereas the situation may improve somewhat with future PDF determinations 
based on the full LHC Run-3 dataset, dedicated observables such as the LHCb
charm asymmetry \cref{eq:charm_lhcb_asym} as well as charm production
at the EIC \cref{eq:EIC_asy_F2c} will be needed in order to achieve firm 
evidence or discovery. 


Our results motivate further dedicated studies of intrinsic charm through 
a wide range of nuclear, particle and astroparticle physics experiments.
These can include: the High-Luminosity LHC~\cite{Azzi:2019yne} with the fixed-target 
programs of LHCb~\cite{LHCb:2018jry,LHCb:2022cul} and ALICE~\cite{QCDWorkingGroup:2019dyv} 
in the forward region; the Forward Physics Facility~\cite{Feng:2022inv,Cruz-Martinez:2023sdv}
with the proposed experiments FASER$\nu$~\cite{FASER:2023zcr}, SND@LHC~\cite{SNDLHC:2023pun}; 
the Electron Ion Collider and neutrino telescopes~\cite{Halzen:2016thi}.

We refer to \cref{sec:anl3o_ic} for a discussion about the impact of 
N$^3$LO corrections ind DGLAP evolution, on intrinsic charm, as well as 
the effect of additional N$^3$LO matching conditions not available by 
the time of writing \cite{Ablinger:2022wbb}.
These have mainly the effect of reduce some of the theory uncertainties
affecting the current determination.


  \chapter{The Path to aN$^3$LO Parton Distributions}
\label{chap:an3lo}
\begin{center}
\begin{minipage}{1.\textwidth}
    \begin{center}
        \textit{
          This chapter is based my result presented in Refs.~\cite{NNPDF:2024nan}.
          In particular, I have worked on the approximation of the aN$^3$LO splitting 
          functions, on the computation of the necessary theoretical calculations 
          for the PDF fit and on the fit running.
          } 
    \end{center}
\end{minipage}
\end{center}

\paragraph{Motivation.}
Calculations of hard-scattering cross-sections at fourth perturbative order in
the strong coupling (N$^3$LO),
have been available for a long time for massless in DIS~(see \cref{sec:dis_coeff}),
and have more recently become available for a rapidly growing set of hadron
collider processes. These include inclusive Higgs production in
gluon-fusion~\cite{Anastasiou:2015ema,Mistlberger:2018etf},
bottom-fusion~\cite{Duhr:2019kwi}, in association with vector
bosons~\cite{Baglio:2022wzu}, and in
vector-boson-fusion~\cite{Dreyer:2016oyx},
Higgs pair production~\cite{Chen:2019lzz}, inclusive
Drell-Yan (DY) production~\cite{Duhr:2020sdp,Duhr:2021vwj}, differential Higgs
production~\cite{Dulat:2017prg,Dulat:2018bfe,Chen:2021isd,Billis:2021ecs,
  Camarda:2021ict}, and differential DY
distributions~\cite{Chen:2021vtu,Chen:2022lwc}, 
see~\cite{Caola:2022ayt} for an overview.

In order to obtain predictions for hadronic observables with this
accuracy, these  partonic cross-sections must be combined with parton
distribution functions (PDFs) determined  at the same perturbative order.
These, in turn, must be determined by comparing to experimental data theory
predictions computed at the same accuracy. The main bottleneck in
carrying out this programme is the lack of exact expressions for
the N$^3$LO splitting functions
that govern the scale dependence of the PDFs: for these only partial information
is available~\cite{Davies:2016jie,Moch:2017uml,Davies:2022ofz,Henn:2019swt,
  Duhr:2022cob,Moch:2021qrk,Soar:2009yh,Falcioni:2023luc,Falcioni:2023vqq,
  Moch:2023tdj,Falcioni:2023tzp,Falcioni:2024xyt}. This information includes a set of integer
$N$-Mellin moments, terms proportional to $n_f^k$ with $k\ge 1$, and the large-
and small-$x$ limits. By combining these partial results it is possible
to attempt an approximate determination of the N$^3$LO splitting
functions~\cite{Moch:2017uml,Falcioni:2023luc,Falcioni:2023vqq,Moch:2023tdj,McGowan:2022nag}, 
as was successfully done in the past at NNLO~\cite{vanNeerven:2000wp}.

At present a global PDF determination at N$^3$LO must consequently be
based on incomplete information: the approximate knowledge of
splitting functions, and full knowledge of partonic cross-sections
only for a subset of processes. 
Here we use a theory covariance matrix formalism
in order to account for such the missing perturbative information,
as well as nuclear uncertainties and missing higher-order uncertainties.
Equipped with such theory covariance matrices, we can
perform a determination of PDFs at ``approximate N$^3$LO''
(hereafter denoted aN$^3$LO), in
which the theory covariance matrix accounts both for incomplete
knowledge of  N$^3$LO  splitting functions and massive coefficient
functions (IHOUs), and for missing N$^3$LO corrections to 
the partonic cross-sections for hadronic processes (MHOUs).

We will thus present the  aN$^3$LO NNPDF4.0 PDF determination, to be added to
the existing LO, NLO and NNLO sets~\cite{NNPDF:2021njg}, as well as the
more recent NNPDF4.0 MHOU PDFs~\cite{NNPDF:2024dpb} that also include
MHOUs in the PDF uncertainty. 
With PDFs determined from the same global dataset and using the same
methodology at four consecutive perturbative orders it is now possible to
assess carefully perturbative stability and provide a reliable uncertainty
estimation.

\paragraph{Outline.}
In \cref{sec:an3lo_dglap} we construct an approximation to the N$^3$LO splitting functions 
based on all known exact results and limits. We compare it with the MSHT
approximation~\cite{McGowan:2022nag} as well as with the
more recent approximation of Refs.~\cite{Moch:2017uml,Falcioni:2023luc,Falcioni:2023vqq,Moch:2023tdj}. 
In \cref{sec:an3lo_coefffun} we discuss available and approximate N$^3$LO 
corrections to hard cross-sections: specifically, DIS coefficient functions, 
including a generalization to this order of the FONLL~\cite{Forte:2010ta,Ball:2015tna} 
method for the inclusion of heavy quark mass effects, 
and the DY cross-section.
In \cref{sec:an3lo_results} we present the main results
of this chapter, namely the aN$^3$LO  NNPDF4.0 PDF set. 
We discuss in detail perturbative convergence before and after 
the inclusion of MHOUs, and results are compared to those
of the MSHT group~\cite{McGowan:2022nag}. 
Finally, a first assessment of the impact of aN$^3$LO PDFs on DY and
Higgs production is presented in \cref{sec:an3lo_pheno}.

\section{Approximate DGLAP N$^3$LO evolution}
\label{sec:an3lo_dglap}
Having introduced DGLAP evolution in \cref{sec:dglap},
in this section we proceed to the construction and implementation of aN$^3$LO evolution.
We first describe our strategy to approximate the N$^3$LO
evolution equations, the way this is used to construct aN$^3$LO
anomalous dimensions and splitting functions, and to estimate the
uncertainty in the approximation and its impact on theory predictions.
We then use this strategy to construct an approximation in the non-singlet
sector, where accurate results have been available for a
while~\cite{Moch:2017uml}, and benchmark it against these results. We
then present our construction of aN$^3$LO singlet splitting functions,
examine  our results, their uncertainties and their
perturbative behavior. We next
describe our implementation of aN$^3$LO evolution and study the
impact of aN$^3$LO on the perturbative evolution of PDFs. Finally,
we compare our aN$^3$LO singlet splitting functions to 
those of the MSHT group and to the recent
results of~\cite{Falcioni:2023luc,Falcioni:2023vqq,Moch:2023tdj}.

\subsection{Construction of the approximation}
\label{sec:an3lo_general_strategy}

The approximation of N$^3$LO DGLAP splitting functions (\cref{eq:dglap})
is more conveniently performed in Mellin space, 
where the kernels are called anomalous dimensions (\cref{eq:ad_def}).
At N$^3$LO there are seven independent contributions: 
three in the non-singlet sector,
$\gamma_{{\rm ns},\pm}$ and $\gamma_{\rm ns,s}$, and four in the singlet
sector, $\gamma_{{qq},{\rm ps}}$, $\gamma_{{qg}}$,
$\gamma_{{gq}}$, and $\gamma_{{gg}}$.
In turn, each of these anomalous dimensions can be expanded
according to 
\begin{equation}
  \gamma_{ij}(N,a_s(\mu^2)) = a_s \gamma_{ij}^{(0)}(N)
    + a_s^2 \gamma_{ij}^{(1)}(N)
    + a_s^3 \gamma_{ij}^{(2)}(N)
    + a_s^4 \gamma_{ij}^{(3)}(N)
  + \mathcal{O}\lp a_s^5\rp  \, .
  \label{eq:ad_expansion}
\end{equation}
Our goal is to determine an approximate expression for the corresponding seven $\gamma_{ij}^{(3)}(N)$ N$^3$LO
terms.
The information that can be exploited in order to achieve this goal
comes from three different sources: (1) full analytic knowledge of
contributions to the anomalous dimensions proportional to the highest
powers of the number of flavors $n_f$; (2) large-$x$ and small-$x$ resummations
provide all-order information on terms that are logarithmically enhanced by
powers of
$\ln (1-x)$ and $\ln x$ respectively; (3) analytic knowledge of a finite
set of integer moments. We construct an approximation based on this
information by first separating off the analytically known terms
(1-2),  then expanding the remainder on a set of basis functions and
using the known moments to  determine the expansion coefficients.
Finally, we vary the set of basis functions in order to obtain an
estimate of the uncertainties. 

Schematically, we proceed as follows:
\begin{enumerate}

\item We include all terms in the expansion
  \begin{equation}\label{eq:gammaf}
  \gamma_{ij}^{(3)}(N) = \gamma_{ij}^{(3,0)}(N) + n_f
  \gamma_{ij}^{(3,1)}(N) + n_f^2
  \gamma_{ij}^{(3,2)}(N) + n_f^3
  \gamma_{ij}^{(3,3)}(N) \, ,
  \end{equation}
  of the anomalous dimension in powers of $n_f$ 
  that are fully or partially known analytically.
  We collectively denote such terms as $\gamma_{ij,n_f}^{(3)}(N)$.

\item We include all terms from large-$x$ and small-$x$ resummation,
  to the highest known logarithmic accuracy, including all known
  subleading power corrections in both limits.
  We  denote these terms as $\gamma_{ij,N\to \infty}^{(3)}(N)$
  and $\gamma_{ij,N\to 0}^{(3)}(N)$, $\gamma_{ij,N\to 1}^{(3)}(N)$ respectively.
  Possible double counting coming from the overlap of these terms with
  $\gamma_{ij,n_f}^{(3)}(N)$
  is removed.

\item We write the approximate anomalous dimension matrix element
  $\gamma_{ij}^{(3)}(N)$ as the sum of the
  terms which are known exactly and a remainder 
  $\widetilde{\gamma}_{ij}^{(3)}(N)$ according to
  \begin{equation}
  \label{eq:ad_expansion_terms}
  \gamma_{ij}^{(3)}(N) = \gamma_{ij,n_f}^{(3)}(N)
  + \gamma_{ij,N\to \infty}^{(3)}(N)
  + \gamma_{ij,N\to 0}^{(3)}(N)
  + \gamma_{ij,N\to 1}^{(3)}(N)
  + \widetilde{\gamma}_{ij}^{(3)}(N) \,.
  \end{equation}
  We determine the remainder  as a linear
  combination of a set of  $n^{ij}$
  interpolating functions $G^{ij}_\ell(N)$ (kept fixed)
  and $H^{ij}_\ell(N)$ (to be varied)
  \begin{equation}
  \label{eq:gamma_residual_basis}
  \widetilde{\gamma}_{ij}^{(3)}(N) = \sum_{\ell=1}^{n^{ij }-n_H}
  b^{ij}_\ell G^{ij}_\ell(N) +\sum_{\ell=1}^{n_H}
  b^{ij}_{n^{ij} -2+\ell} H^{ij}_{\ell}(N) \, ,
  \end{equation}
  with $n^{ij}$ equal to the number of known
  Mellin moments of 
  $\gamma_{ij}^{(3)}(N)$. We determine the coefficients
  $b^{ij}_\ell$ by equating the evaluation of
  $\widetilde{\gamma}_{ij}^{(3)}(N)$ to the known moments of the splitting
  functions.

\item In the singlet sector, we take $n_H=2$ and we make  $\widetilde{N}_{ij}$
  different choices for the two functions $H^{ij}_\ell(N)$, by selecting them
  out of a list of distinct  basis
  functions (see \cref{sec:an3lo_singlet} below). Thereby, we obtain
  $\widetilde{N}_{ij}$  expressions for the
  remainder $\widetilde{\gamma}_{ij}^{(3)}(N)$ and accordingly for the
  N$^3$LO anomalous dimension matrix element ${\gamma}_{ij}^{(3)}(N)$ through
  \cref{eq:ad_expansion_terms}. These are used to construct the
  approximate anomalous dimension matrix and the uncertainty on it, in
  the way discussed in \cref{sec:an3lo_ihou} below. In the non-singlet
  sector instead, we take  $n_H=0$, i.e.\ we take a unique answer as our
  approximation, and we neglect the uncertainty on it, for reasons to
  be discussed in greater detail at the end of \cref{sec:an3lo_non_singlet}.  
\end{enumerate}

\subsection{The approximate anomalous dimension matrix and its uncertainty}
\label{sec:an3lo_ihou}

The procedure described in \cref{sec:an3lo_general_strategy} provides us with an
ensemble of $\widetilde{N}_{ij}$ different approximations 
to the N$^3$LO anomalous dimension, 
denoted $\gamma_{ij}^{(3),\,(k)}(N)$, $k=1,\dots\widetilde{N}_{ij}$. 
Our best estimate for the approximate anomalous dimension is then their average
\begin{equation}
  \label{eq:centralgij}
  \gamma_{ij}^{(3)}(N)=
  \frac{1}{\widetilde{N}_{ij}}\sum_{k=1}^{\widetilde{N}_{ij}} \gamma_{ij}^{(3),\,(k)}(N).
\end{equation}

We include  the uncertainty on the approximation, and the
ensuing uncertainty on N$^3$LO theory predictions,
using the general formalism for the treatment of theory uncertainties
developed in Refs.~\cite{NNPDF:2019vjt,NNPDF:2019ubu}. Namely, we
consider the uncertainty on each anomalous dimension matrix element
due to its incomplete knowledge as a source of
uncertainty on  theoretical predictions, uncorrelated from
other sources of uncertainty, and neglecting possible correlations
between our incomplete knowledge of each individual matrix element
$\gamma_{ij}^{(3)}$. This uncertainty on the incomplete higher
(N$^3$LO) order terms (incomplete higher order uncertainty, or IHOU)
is then treated in the same way as the uncertainty due to missing
higher order terms (missing higher order uncertainty, or MHOU).

We construct the shift of theory prediction for the $m$-th
data point due to replacing the central anomalous dimension matrix
element $\gamma_{ij}^{(3)}(N)$, \cref{eq:centralgij}, with each of
the instances $\gamma_{ij}^{(3),\,(k)}(N)$, viewed as an independent
nuisance parameter:
\begin{equation}
  \label{eq:deltaij}
  \Delta_m(ij,k)=  T_{m}(ij,k) - \bar T_{m},
\end{equation}
where  $\bar T_{m}$ is the prediction for the $m$-th datapoint
obtained using the best estimate \cref{eq:centralgij} for the
full anomalous dimension matrix, while $T_{m}(ij,k)$ is the prediction
obtained when the $ij$ matrix element of our best estimate is replaced
with the $k$-th instance $\gamma_{ij}^{(3),\,(k)}(N)$.

We then construct the covariance matrix over theory predictions for
individual datapoints due to the IHOU on the $ij$
N$^3$LO matrix element as the covariance of the shifts
$\Delta_m(ij,k)$ over all $\widetilde{N}_{ij}$ instances:
\begin{equation}
  \text{cov}^{(ij)}_{mn} = \frac{1}{\widetilde{N}_{ij}-1} \sum_{k=1}^{\widetilde{N}_{ij} }\Delta_m(ij,k)\Delta_n(ij,k).
  \label{eq:covihouij}
\end{equation}

We recall that we do not associate an IHOU to the non-singlet anomalous
dimensions, and we assume conservatively that there is no correlation
between the different singlet anomalous dimension matrix elements.
Thus, we can write the total contribution to the theory covariance matrix due to IHOU as
\begin{equation}
  \label{eq:covihou}
  \text{cov}^\text{IHOU}_{mn} = \text{cov}^{(gg)}_{mn}  +\text{cov}^{(gq)}_{mn}  +\text{cov}^{(qg)}_{mn}  +\text{cov}^{(qq)}_{mn}.  
\end{equation}

The mean square uncertainty on the anomalous dimension matrix element itself is
then determined, by viewing it as a pseudo-observable, as the variance 
\begin{equation}
  \label{eq:sigihou}
  (\sigma_{ij}(N))^2=  \frac{1}{\widetilde{N}_{ij}-1}
\sum_{k=1}^{\widetilde{N}_{ij}}
\left(\gamma_{ij}^{(3),\,(k)}(N)-\gamma_{ij}^{(3)}(N)\right)^2.
\end{equation}

\subsection{The non-singlet sector}
\label{sec:an3lo_non_singlet}

Information on the Mellin moments of non-singlet anomalous dimensions is
especially abundant, in that eight moments of
$\gamma_{{\rm ns},\pm}^{(3)}$ and nine moments of $\gamma_{{\rm ns,s}}^{(3)}$
are known. An approximation based on this knowledge  was 
given in Ref.~\cite{Moch:2017uml}. More recently, further information
on the small-$x$ behavior of
$\gamma_{{\rm ns},\pm}^{(3)}$ was derived in Ref.~\cite{Davies:2022ofz}.
While for $\gamma_{{\rm ns},s}^{(3)}$ we directly rely on the approximation of
Ref.~\cite{Moch:2017uml}, which already includes all the available
information, we construct an approximation to $\gamma_{{\rm ns},\pm}^{(3)}$ 
based on the procedure described in \cref{sec:an3lo_general_strategy}, 
in order to include also this more recent information, 
and also as a warm-up for the construction of our approximation 
to the singlet sector anomalous dimension that we present in the next section.

Contributions to $\gamma_{{\rm ns},\pm}^{(3)}$ proportional to $ n_f^2$ and $ n_f^3$
are known exactly~\cite{Davies:2016jie} (in particular the $n_f^3$
contributions to  $\gamma_{{\rm ns},\pm}^{(3)}$ coincide), while
$\mathcal{O}( n_f^0)$ and $\mathcal{O}( n_f)$ terms
\footnote{The $n_f C^3_F$
  terms have also been published very recently~\cite{Gehrmann:2023iah},
  but are not yet included in our study.
}
are known in the large-$N_c$ limit~\cite{Moch:2017uml} and we include these in
$\gamma_{{\rm ns},\pm ,n_f}^{(3)}(N)$.

Small-$x$ contributions to  $\gamma_{{\rm ns},\pm}$
are double logarithmic, i.e.\ of the form
$a_s^{n+1} \ln^{2n-k}(x)$, corresponding in Mellin space
to poles of order $2n-k+1$ in
$N=0$, i.e.\ $\frac{1}{N^{2n-k+1}}$, so at N$^3$LO we have $n=3$ and thus
\begin{equation}
  P_{\rm ns,\,\pm}^{(3)}(x) = \sum_{k=1}^{6} c^k_{{\rm ns}, \,N\to 0} \ln^k(1/x)+ O(x)\, .
  \label{eq:ns_smallxx}
\end{equation}

The coefficients $c^k_{{\rm ns},\,N\to 0}$ are known~\cite{Davies:2022ofz} exactly up to
NNLL accuracy ($k=4,\,5,\,6)$, and approximately  up to N$^6$LL ($k=1,\,2,\,3$).
Hence, we let
\begin{equation}
    \label{eq:ns_smallx}
    \gamma_{{\rm ns},\pm, \,N\to 0}^{(3)}(N)=\sum_{k=1}^{6} c^k_{{\rm ns}, \,N\to 0} (-1)^k \frac{k!}{N^{k+1}}.
\end{equation}

Large-$x$ logarithmic contributions in the $\overline{\rm MS}$
scheme only appear in coefficient functions~\cite{Albino:2000cp},
and so the $x\to 1$ behavior of splitting functions is provided by the
cusp anomalous dimension $\sim\frac{1}{(1-x)_+}$, corresponding to a
single $\ln(N)$ behavior in Mellin space as $N\to\infty$. 
This behavior is common to the pair of anomalous dimensions $\gamma_{{\rm ns},\pm}^{(3)}(N)$.
Furthermore, several subleading power corrections as $N\to\infty$ can also be
determined and we set
\begin{equation}
  \gamma_{{\rm ns},\pm,\,N\to \infty}^{(3)}(N) =A^{q}_4 S_1(N) + B_4^{q} + C_4^{q} \frac{S_1(N)}{N} + D_4^{q} \frac{1}{N} ,
\label{eq:ns_largen}
\end{equation}

where $S_1$ denotes the first harmonic sum, 
which is usually analytically continued in terms of the polygamma function
\begin{equation}
  S_1(N) = \sum_{j=1}^{N} \frac{1}{j} = \psi(N+1) + \gamma_E, 
  \label{eq:s1}
\end{equation}
The coefficient of the $\mathcal{O}(\ln(N))$ term  $A^{q}_4$, is the quark cusp
anomalous dimension~\cite{Henn:2019swt}. The constant coefficient
$B^{q}$ is determined by the integral of the non-singlet splitting
function, which was originally computed in Ref.~\cite{Moch:2017uml} 
in the large-$N_c$ limit and recently updated to the full color
expansion~\cite{Duhr:2022cob}  
as a result of computing different N$^3$LO cross-sections in the soft limit.
The coefficients of the terms suppressed by $1/N$ in the large-$N$ limit,
$C^{q}$ and $ D^{q}$,
can be obtained directly from lower-order anomalous dimensions
by exploiting  large-$x$ resummation techniques~\cite{Davies:2016jie}.
For completeness, the explicit expressions of $\gamma_{{\rm ns}\, \pm,\,N\to \infty}^{(3)}(N)$ and
$\gamma_{{\rm ns}\, \pm,\,N\to 0}^{(3)}(N)$ are given in
Appendix A of~\cite{NNPDF:2024nan}.

\begin{table}[!t]
  \centering
  \small
  \renewcommand{\arraystretch}{1.50}
  \begin{tabularx}{0.6\textwidth}{Xcc}
\midrule
$G^{{\rm ns}, \pm}_1(N)$ &   1 \\
$G^{{\rm ns}, \pm}_2(N)$ & $\mathcal{M}[(1-x)\ln(1-x)](N)$ \\
$G^{{\rm ns}, \pm}_3(N)$ & $\mathcal{M}[(1-x)\ln^2(1-x)](N)$ \\
$G^{{\rm ns}, \pm}_4(N)$ &  $\mathcal{M}[(1-x)\ln^3(1-x)](N)$ \\
$G^{{\rm ns}, \pm}_5(N)$ & $\frac{S_1(N)}{N^2}$  \\
$G^{{\rm ns}, \pm}_6(N)$ &  $\frac{1}{(N+1)^2}$  \\
$G^{{\rm ns}, \pm}_7(N)$ & $\frac{1}{(N+1)^3}$ \\
$G^{{\rm ns}, +}_8(N)$, $G^{{\rm ns}, -}_8(N)$  &  $\frac{1}{(N+2)}$, $\frac{1}{(N+3)}$ \\
\bottomrule
\end{tabularx}
  \vspace{0.3cm}
  \caption{The Mellin space interpolating functions $G^{{\rm ns},\pm}_\ell(N)$ 
    entering the parametrization of the remainder term
    $\widetilde{\gamma}^{(3)}_{{\rm ns}\, \pm}(N)$
    for the non-singlet anomalous dimension expansion
    of \cref{eq:gamma_residual_basis}.}
  \label{tab:functions_interpolating}
\end{table}

The remainder terms, $\widetilde{\gamma}_{\rm ns, \pm}^{(3)}(N)$, are
expanded over the set of eight functions $G^{{\rm ns},\pm}_\ell(N)$ listed in
\cref{tab:functions_interpolating}. 
The coefficients $b^{\rm ns,\pm}_\ell$ (defined in \cref{eq:gamma_residual_basis})
are determined by imposing  that the values of the eight moments given in
Ref.~\cite{Moch:2017uml} be reproduced. The set of functions
$G^{{\rm ns},\pm}_\ell(N)$ is chosen to adjust the overall constant ($\ell=1$),
model the large-$N$ behavior ($2\le \ell\le 5$) and model the small-$N$ behavior
($\ell=6,7$), consistent with the general analytic
structure of fixed order anomalous dimensions.
Specifically, the large-$N$ functions are chosen as the
logarithmically enhanced next-to-next-to-leading power terms ($\ln^k(N)/N^2$,
$\ell=2,3,4,5$) and the small-$N$ functions are chosen as
logarithmically enhanced subleading poles ($1/(N+1)^k$, $\ell=6,7$)
and sub-subleading poles ($1/(N+2)$ or $1/(N+3)$,  $\ell=8$).
The last element, $\ell = 8$, is chosen
at a fixed distance from the lowest known moment, $N=2$ for
$\gamma_{{\rm ns},+}^{(3)}(N)$ and $N=1$ for $\gamma_{{\rm ns},-}^{(3)}(N)$.

\begin{figure}[!t]
  \centering
  \includegraphics[width=.49\textwidth]{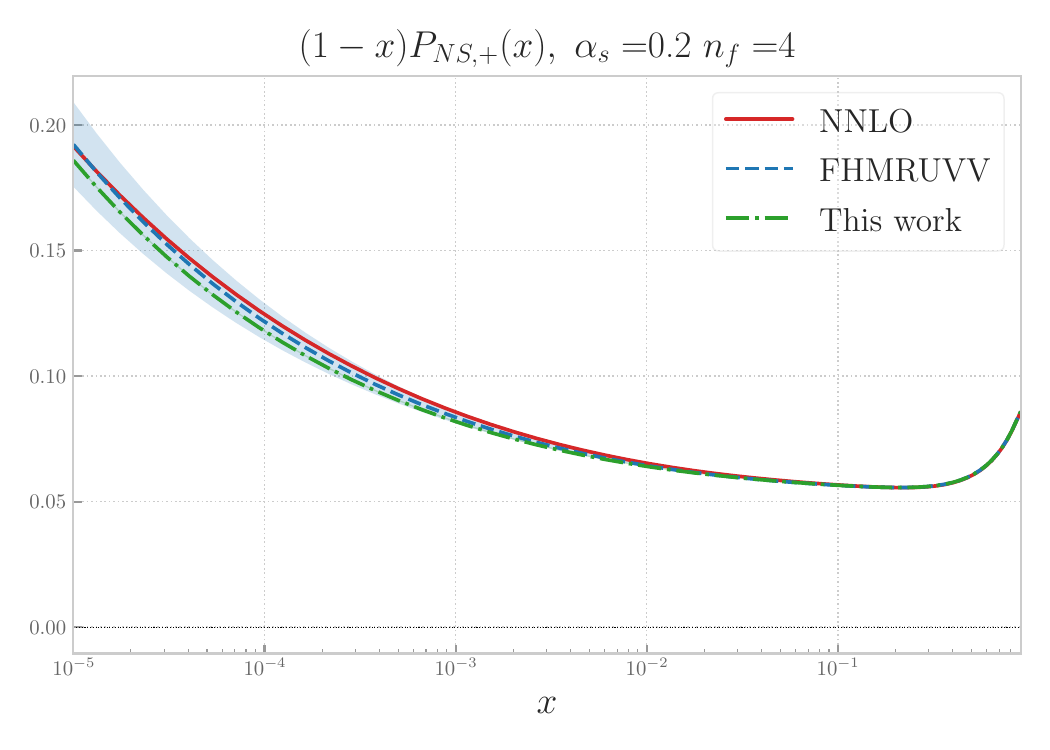}
  \includegraphics[width=.49\textwidth]{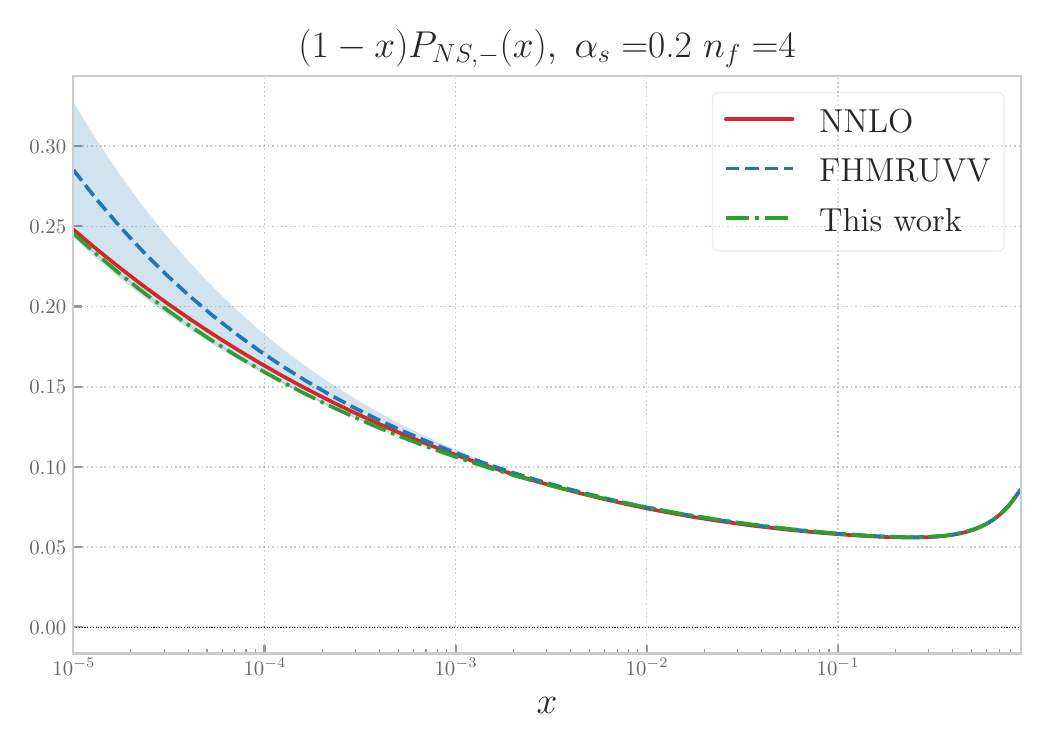}
  \caption{The aN$^3$LO non-singlet splitting functions $(1-x)
    P_{{\rm ns},+}(x,\alpha_s)$ 
    and $(1-x) P_{{\rm ns},-}(x,\alpha_s)$, evaluated 
    as a function of $x$  for $n_f=4$ and $\alpha_s=0.2$ in our
    approximation compared to the previous  approximation of
    Ref.~\cite{Moch:2017uml} (denoted FHMRUVV), for which the approximation
    uncertainty, as estimated by its authors, is also displayed. 
    For comparison, the (exact) NNLO result is also shown.}
  \label{fig:nsgamma} 
\end{figure}

In \cref{fig:nsgamma} we plot the resulting splitting functions
$P_{{\rm ns},\pm}^{(3)}(x)$, obtained by Mellin inversion of the anomalous
dimension. We compare our approximation to the approximation of
Ref.~\cite{Moch:2017uml}, for $\alpha_s=0.2$ and $n_f=4$, and also show 
the (exact) NNLO result for reference. Because
the splitting function is a distribution at $x=1$ we plot $(1-x)P(x)$.
The result of Ref.~\cite{Moch:2017uml} also provides an estimate
of the uncertainty related to the approximation, shown in the
figure as a band, and we observe that this uncertainty
is negligible except at very small $x$.
As we include further constraints on the small-$x$ behavior,
the uncertainty on the approximation becomes negligible, as it can be
checked by comparing results obtained by including increasingly more
information in the construction of the approximation. 
Consequently, as mentioned in \cref{sec:an3lo_general_strategy} above, we take
$n_H=0$ in \cref{eq:gamma_residual_basis}.

\subsection{The singlet sector}
\label{sec:an3lo_singlet}

In order to determine the singlet-sector anomalous dimension
matrix, we must determine $\gamma_{{qq},{\rm ps}}$ that, 
together with the previously determined non-singlet anomalous dimension, 
contributes to the $qq$ entry, 
and then also the three remaining matrix elements
$\gamma_{{qg}}$, $\gamma_{{gq}}$, and $\gamma_{{gg}}$.

For all matrix elements, the leading large-$n_f$ $\mathcal{O}(n_f^3)$
contributions in \cref{eq:gammaf} are 
known analytically~\cite{Davies:2016jie}, while for
$\gamma_{{qq},{\rm ps}}$~\cite{Gehrmann:2023cqm}
and $\gamma_{{gq}}$~\cite{Falcioni:2023tzp} the $\mathcal{O}(n_f^2)$
contributions are also known and we include all of them in
$\gamma_{ij,n_f}^{(3)}(N)$.

Small-$x$ contributions in the singlet sector include, 
on top of the double-logarithmic contributions $a_s^{n+1} \ln^{2n-k}(x)$,
also single-logarithmic contributions $a_s^{n+1} \frac{1}{x} \ln^{n}(x)$. 
In Mellin space, this means that on top of order $2n-k+1$ subleading poles in
$N=0$, there are also leading poles in $N=1$ of order $n-k+1$, 
i.e.\ $\frac{1}{(N-1)^{n-k+1}}$.
The leading-power single logarithmic contributions can be extracted from the 
high-energy resummation at LL$x$~\cite{Jaroszewicz:1982gr,Ball:1995vc,Ball:1999sh,} 
and NLL$x$~\cite{Bonvini:2018xvt} accuracy. 
This allows for a determination of the coefficients of the leading
$\frac{1}{(N-1)^4}$ and next-to-leading $\frac{1}{(N-1)^3}$ contributions to
$\gamma^{(3)}_{gg}$ and of the next-to-leading $\frac{1}{(N-1)^3}$ contributions
to $\gamma^{(3)}_{qg}$. The remaining entries can be obtained from these
by using the color-charge (or Casimir scaling) relation
$\gamma_{iq}=\frac{C_F}{C_A}\gamma_{ig}$~\cite{Catani:1994sq,Bonvini:2018xvt}.
Hence, we set
\begin{allowdisplaybreaks} 
\begin{align}
 \label{eq:smallxgg}
 \gamma_{gg,\,N\to 1}^{(3)}(N) & = c^4_{gg, \,N\to 1}\frac{1}{(N-1)^4}+c^3_{gg, \,N\to 1}\frac{1}{(N-1)^3}; \\
 \label{eq:smallxqg}
 \gamma_{qg,\,N\to 1}^{(3)}(N) & = c^3_{qg, \,N\to 1}\frac{1}{(N-1)^3}; \\ 
 \label{eq:smallxiq}
 \gamma_{iq,\,N\to 1}^{(3)}(N) & = \frac{C_F}{C_A} \gamma_{ig,\,N\to 1}^{(3)}(N),\quad i= q,\,g.
\end{align} 
\end{allowdisplaybreaks}

Although only the leading pole of $\gamma_{gq}$ satisfies
\cref{eq:smallxiq} exactly, at NNLO this relation is only violated 
at the sub-percent level~\cite{Vogt:2004mw}, so this is likely to be an adequate
approximation also at this order: this approximation is also adopted in
Ref.~\cite{Moch:2023tdj,Falcioni:2024xyt}. An important observation is
that both NLO and NNLO coefficients of
the leading poles, $\frac{1}{(N-1)^2}$ and $\frac{1}{(N-1)^3}$
respectively, vanish accidentally. Hence, at N$^3$LO the leading poles
contribute for the first time beyond leading order.
The subleading poles can be determined up to NNLL accuracy~\cite{Davies:2022ofz} 
and, thus, fix the coefficients of the $\frac{1}{N^7}$, $\frac{1}{N^6}$ and
$\frac{1}{N^5}$ subleading poles for all entries of the singlet
anomalous dimension matrix. 
All these contributions are included in
$\gamma_{ij,N\to 1}^{(3)}(N)$ and $\gamma_{ij,N\to 0}^{(3)}(N)$. 

In the singlet sector, large-$x$ contributions, whose Mellin transform
is not suppressed in the large-$N$ limit, only appear in the diagonal $qq$ and $gg$
channels. 
In the quark-to-quark channel these are already included 
in $\gamma_{{\rm ns},+,\,N\to \infty}^{(3)}(N)$,
according to \cref{eq:ns_largen}, while  $\gamma^{(3)}_{qq,{\rm ps}}$
is suppressed in this limit. 
In the gluon-to-gluon channel they take the same form as in the 
non-singlet and diagonal quark channel. 
Hence, we expand, as in \cref{eq:ns_largen},
\begin{equation}
  \label{eq:largexgg}
  \gamma_{gg,\,N\to \infty}^{(3)}(N) = A^{g}_4 S_1(N) + B^{g}_4 +  C^{g}_4 \frac{S_1(N)}{N} + D^{g}_4 \frac{1}{N} \, .
\end{equation}
  
The coefficients $A^{g}_4$, $ B^{g}_4$, $C^{g}_4$
and $D^{g}_4$ are the  counterparts of those of
\cref{eq:ns_largen}: the gluon cusp anomalous dimension was determined
in Ref.~\cite{Henn:2019swt} and the constant in
Ref.~\cite{Duhr:2022cob}, while the  $C^{g}_4$ and $D^{g}_4$
coefficients can be determined using results from
Refs.~\cite{Dokshitzer:2005bf,Moch:2023tdj}.  

Off-diagonal $qg$ and $qg$ splitting functions have logarithmically
enhanced next-to-leading power behavior at large-$x$:
\begin{equation}
      P_{ij}^{(3)}(x) = \sum_{k=0}^{6} \sum_{l=0}^{\infty} c^{k,l}_{ij,\,N\to \infty} (1-x)^l \ln^k(1-x) . 
      \label{eq:largex_expansion}
\end{equation}
For $l=0$ the coefficients of the higher logs $k=4,5,6$ can be determined from 
N$^3$LO coefficient functions, based on a conjecture~\cite{Soar:2009yh,Almasy:2010wn} 
on the large-$x$ behavior of the physical evolution kernels that give the scale
dependence of structure functions.
The coefficient with the highest power $k=6$ cancels and thus we let
\begin{allowdisplaybreaks} 
\begin{align}
      \label{eq:largexgq}
      \gamma_{gq,\,N\to \infty}^{(3)}(N) = \sum_{k=4}^5 c^{k,0}_{gq, \,N\to \infty} & L_{k,0}(N), \\
      \label{eq:largexqg}
      \gamma_{qg,\,N\to \infty}^{(3)}(N) = \sum_{k=4}^5 c^{k,0}_{qg, \,N\to \infty} L_{k,0}(N) & + c^{k,1}_{qg, \,N\to \infty} L_{k,1}(N), 
\end{align} 
\end{allowdisplaybreaks}
where in $\gamma_{qg,\,N\to \infty}^{(3)}$ we have retained also the
$l=1$ terms \cite{Falcioni:2023vqq} and used the shorthand notation
\begin{equation}
  L_{k,m}(N) = \mathcal{M}\left[(1-x)^m \ln^k(1-x)\right]
  \label{eq:mellin_lm1kmk}
\end{equation}
Finally, the pure singlet quark-to-quark splitting function starts at 
next-to-next-to-leading power as $x\to 1$,
i.e.\ it behaves as  $(1-x)\ln^k(1-x)$, with $k\le 4$.
The coefficients of the higher logs $k=3,4$  can be extracted by expanding the
$x=1$ expressions from Refs.~\cite{Soar:2009yh,Falcioni:2023luc}. 
Hence, we let
 \begin{equation}
 \label{eq:largexps}
 \gamma_{qq,{\rm ps},\,N\to \infty}^{(3)}(N)= \sum_{k=3}^4 \left[c^{k,1}_{qq,{\rm ps},\,N\to \infty} L_{k,1}(N) + c^{k,2}_{qq,{\rm ps},\,N\to \infty} L_{k,2}(N) \right] 
 \end{equation}
Note that for the $qq$ and $qg$ entries we also include the (known) 
next-to-leading power contributions, while we do not include them for $gq$ and
$gg$ because for these anomalous dimension matrix elements a
significantly larger number of higher Mellin moments is known, hence
there is no risk that the inclusion of these contributions could
contaminate the intermediate $x$ region where they are not
necessarily dominant.
The explicit expressions of
$\gamma_{ij\,N\to \infty}^{(3)}(N)$,
$\gamma_{ij\,N\to 0}^{(3)}(N)$ and $\gamma_{ij\,N\to 1}^{(3)}(N)$ 
are given in Appendix A of~\cite{NNPDF:2024nan}.

As discussed in \cref{sec:an3lo_general_strategy}, the remainder
contribution $\widetilde{\gamma}_{ij}^{(3)}(N)$,
\cref{eq:gamma_residual_basis}, is determined by expanding each
of its matrix elements  over
a set of  $n^{ij}$ basis functions, where  $n^{ij}$ is the number of
known Mellin moments,  and determining the expansion coefficients by
demanding that the known moments be reproduced.
Specifically, the known moments  are the four moments
computed in Ref.~\cite{Moch:2021qrk}, the six additional moments
for $\gamma_{qq,{\rm ps}}$ and $\gamma_{qg}$ computed in
Ref.~\cite{Falcioni:2023luc} and Ref.~\cite{Falcioni:2023vqq} respectively,
and the additional moment $N=10$ for $\gamma_{gg}$ and $\gamma_{gq}$
evaluated in Ref.~\cite{Moch:2023tdj}.
These constraints automatically implement momentum conservation:
\begin{allowdisplaybreaks} 
\begin{align}
\begin{split}
    \gamma_{qg}(N=2) + \gamma_{gg}(N=2) &= 0 \, , \\
    \gamma_{qq}(N=2) + \gamma_{gq}(N=2) &= 0 \, .
    \label{eq:singlet_scaling}
\end{split}
\end{align} 
\end{allowdisplaybreaks}
The additional 5 moments $N=12,\dots,20$ of $\gamma_{qg}$~\cite{Falcioni:2024xyt}
were not available by the time of writing, and are not included in the study. 
However, we refer to \cref{app:new_an3lo_pgq} for a discussion on the impact
of this newer constraints.

\begin{table}[!t]
  \centering
  \small
  \renewcommand{\arraystretch}{1.70}
  \begin{tabularx}{\textwidth}{Xcc}
    \toprule
    \multirow{6}{*}{$\gamma_{gg}^{(3)}(N)$} 
    & $ G^{gg}_{1}(N) $ &	$\mathcal{M}[(1-x)\ln^3(1-x)](N)$\\
    & $ G^{gg}_{2}(N) $  &	$\frac{1}{(N-1)^2} $\\
    & $ G^{gg}_{3}(N) $  &	$\frac{1}{N-1}$ \\
    & \multirow{2}{*}{$\{ H^{gg}_{1}(N), \ H^{gg}_{2}(N) \}$} &
    $\frac{1}{N^4},\ 
    \frac{1}{N^3},\ 
    \frac{1}{N^2},\ 
    \frac{1}{N+1},\ 
    \frac{1}{N+2}$,\ \\
    & & $\mathcal{M}[(1-x)\ln^2(1-x)](N),\ 
    \mathcal{M}[(1-x)\ln(1-x)](N)$ \\
    \midrule
    \multirow{5}{*}{$\gamma_{gq}^{(3)}(N)$} 
    & $ G^{gq}_{1}(N) $  &	$\mathcal{M}[\ln^3(1-x)](N)$ \\ 
    & $ G^{gg}_{2}(N) $  &	$\frac{1}{(N-1)^2} $ \\
    & $ G^{gq}_{3}(N) $  &	$\frac{1}{N-1}$ \\
    & $\{ H^{gq}_{1}(N), \ H^{gq}_{2}(N) \}$ &	
    $\frac{1}{N^4},\
    \frac{1}{N^3},\ 
    \frac{1}{N^2},\ 
    \frac{1}{N+1},\ 
    \frac{1}{N+2},\ 
    \mathcal{M}[\ln^2(1-x)](N),\ 
    \mathcal{M}[\ln(1-x)](N)$ \\
    \midrule
    \multirow{6}{*}{$\gamma_{qg}^{(3)}(N)$}  
    & $ G^{qg}_{1}(N) $ &	$\mathcal{M}[\ln^3(1-x)](N)$ \\
    & $ G^{qg}_{2}(N) $ &	$\frac{1}{(N-1)^2}$ \\
    & $ G^{qg}_{3}(N) $ &	 $\frac{1}{N-1}-\frac{1}{N}$ \\
    & $ G^{qg}_{4,\dots,8}(N) $  &	
    $\frac{1}{N^4},\
    \frac{1}{N^3},\
    \frac{1}{N^2},\
    \frac{1}{N},\ 
    \mathcal{M}[\ln^2(1-x)](N)$ \\
    &  \multirow{2}{*}{$\{ H^{qg}_{1}(N), \ H^{qg}_{2}(N) \}$}  &	
    $\mathcal{M}[\ln(x) \ln(1-x)](N),\
    \mathcal{M}[\ln(1-x)](N),\
    \mathcal{M}[(1-x)\ln^3(1-x)](N)$ \\
    & & $\mathcal{M}[(1-x)\ln^2(1-x)](N),\ 
    \mathcal{M}[(1-x)\ln(1-x)](N),\ 
    \frac{1}{1+N}$ \\
    \midrule
    \multirow{6}{*}{$\gamma_{qq,{\rm ps}}^{(3)}(N)$}
    & $ G^{qq,{\rm ps}}_{1}(N) $  &	$\mathcal{M}[(1-x)\ln^2(1-x)](N)$  \\ 
    & $ G^{qq,{\rm ps}}_{2}(N) $  &	$-\frac{1}{(N-1)^2} + \frac{1}{N^2}$ \\
    & $ G^{qq,{\rm ps}}_{3}(N) $  &	$-\frac{1}{(N-1)} + \frac{1}{N}$ \\
    & \multirow{2}{*}{$ G^{qq,{\rm ps}}_{4,\dots,8}(N) $}  & 
    $\frac{1}{N^4},\
    \frac{1}{N^3},\
    \mathcal{M}[(1-x)\ln(1-x)](N)$ \\
    & & $\mathcal{M}[(1-x)^2\ln(1-x)^2](N),\ 
    \mathcal{M}[(1-x)\ln(x)](N)$ \\
    & \multirow{2}{*}{$\{ H^{qq,{\rm ps}}_{1}(N), \ H^{qq,{\rm ps}}_{2}(N) \}$} &  
    $\mathcal{M}[(1-x)(1+2x)](N),\ 
    \mathcal{M}[(1-x)x^2](N)$, \\
    & & $\mathcal{M}[(1-x) x (1+x)](N),\ 
    \mathcal{M}[(1-x)](N)$ \\
    \bottomrule
\end{tabularx}

  \vspace{0.3cm}
  \caption{The set of basis functions $G^{ij}_\ell (N)$ and $H^{ij}_\ell (N)$
    used to parametrize the singlet sector remainder 
    anomalous dimensions matrix elements $\widetilde{\gamma}_{ij}^{(3)}(N)$
    according to \cref{eq:gamma_residual_basis}.}
  \label{tab:functions_interpolating_ihou}
\end{table}

The set of basis functions is chosen based on the idea of constructing
an approximation that reproduces the singularity structure of the Mellin
transform of the anomalous dimension viewed as analytic functions in
$N$ space, hence corresponding to the leading and subleading (i.e.\ rightmost)
$N$-space poles with unknown coefficients as well as the leading
unknown large-$N$ behavior.  
As mentioned in \cref{sec:an3lo_general_strategy}, the uncertainty 
on the parametrization is then estimated by varying the set of basis functions, 
specifically by varying two out of the $n^{ij}$ basis functions. 
The way the basis functions are partitioned between the fixed functions $G^{ij}$
and  the varying functions $H^{ij}$ is by always including in the
fixed set the most leading unknown contributions, and in the  $H^{ij}$
further subleading ones. 

Specifically, the functions $G^{ij}$ are chosen as follows.
\begin{enumerate}
\item The function $G^{ij}_1(N)$ reproduces the leading unknown
  contribution in the large-$N$ limit, i.e.\ the unknown term in
  \cref{eq:largex_expansion} with highest $k$ and lowest $l$.
\item The functions $G^{ij}_2(N)$ and $G^{ij}_3(N)$
  reproduce the first two
  leading  unknown contributions in the small-$N$ limit, i.e.\ the unknown
  $\frac{1}{(N-1)^k}$ leading poles with highest and next-to-highest
  values of $k$, i.e.\ $k=2$ and  $k=1$. For $\gamma_{qq,{\rm ps}}$
  and  $\gamma_{qg}$ a 
  subleading small-$x$ pole 
  with the same power and opposite sign is added to the leading pole
  with respectively  $k=1,\,2$ and $k=1$, so as to leave unaffected the
  respective large-$x$ leading power behavior \cref{eq:largexqg,eq:largexps}.
\item For $\gamma_{qq,{\rm ps}}$ and  $\gamma_{qg}$, 
  for which an additional five moments are known, the
  functions $G^{qj}_{4,\dots,8}(N)$ reproduce subleading small- and
  large-$N$ terms.
\end{enumerate}
Note that a larger number of basis functions is chosen to describe the
small-$N$ poles  rather than the large-$N$ behavior because  less
exact information  is available in the former case: so for instance
only the leading pole \cref{eq:smallxqg} is known for
$\gamma_{qg}^{(3)}(N)$, while the first two logarithmically enhanced
large-$N$ contributions to it \cref{eq:largexqg} are known. 

As mentioned, the functions $H^{ij}$ are chosen to reproduce further
subleading contributions:
\begin{enumerate}
\item The functions  $H^{gj}_1(N),H^{gj}_2(N)$ in the gluon sector,
  where only five moments are known exactly, are chosen to reproduce subleading
  small- and large-$N$ terms, i.e.\ similar to  $G^{qj}_{4,\dots,8}(N)$.
\item The functions  $H^{qg}_1(N),H^{qg}_2(N)$ are chosen as
  subleading and next-to-leading power large-$x$ terms and the
  remaining unknown leading small-$N$ pole.
\item The functions  $H^{qq,\,{\rm ps}}_1(N),H^{qq,\,{\rm ps}}_2(N)$
  are chosen as low-order polynomials, i.e., sub-subleading small-$x$
  poles.
\end{enumerate}
The number of basis functions is greater for anomalous
dimension matrix elements for which less exact information is
available: 7 in the gluon sector (i.e.\ $gg$ and $gq$), 6 for the $qg$ entry
and 4 for the pure singlet entry. For the $gg$ entry
two combinations are discarded as they lead to unstable
(oscillating) results and we thus end up with $\widetilde{N}_{gg}=19$,
$\widetilde{N}_{gq}=21$, $\widetilde{N}_{qg}=15$, and $\widetilde{N}_{qq}=6$
different parametrizations.
The full set of basis functions $G^{ij}$ and $H^{ij}$ is listed in 
\cref{tab:functions_interpolating_ihou}. We have checked that
results are stable upon variation of these choices, so for instance
including a larger number of $H^{ij}$ functions does not lead to
significantly larger uncertainties. 

Upon combining the exactly known contributions with the
$\widetilde{N}_{ij}$ remainder terms according to
\cref{eq:ad_expansion_terms} we end up with an ensemble of
$\widetilde{N}_{ij}$ instances of $\gamma_{ij}^{(3),\,(k)}(N)$ for each singlet
anomalous dimension matrix element and the final matrix elements
$\gamma_{ij}^{(3)}(N)$ and their uncertainties $\sigma_{ij}(N)$ are computed using
\cref{eq:centralgij,eq:sigihou} respectively.

\subsection{Results: aN$^3$LO splitting functions}
\label{sec:an3lo_splittings}

\begin{figure}[!t]
  \centering
  \includegraphics[width=.49\textwidth]{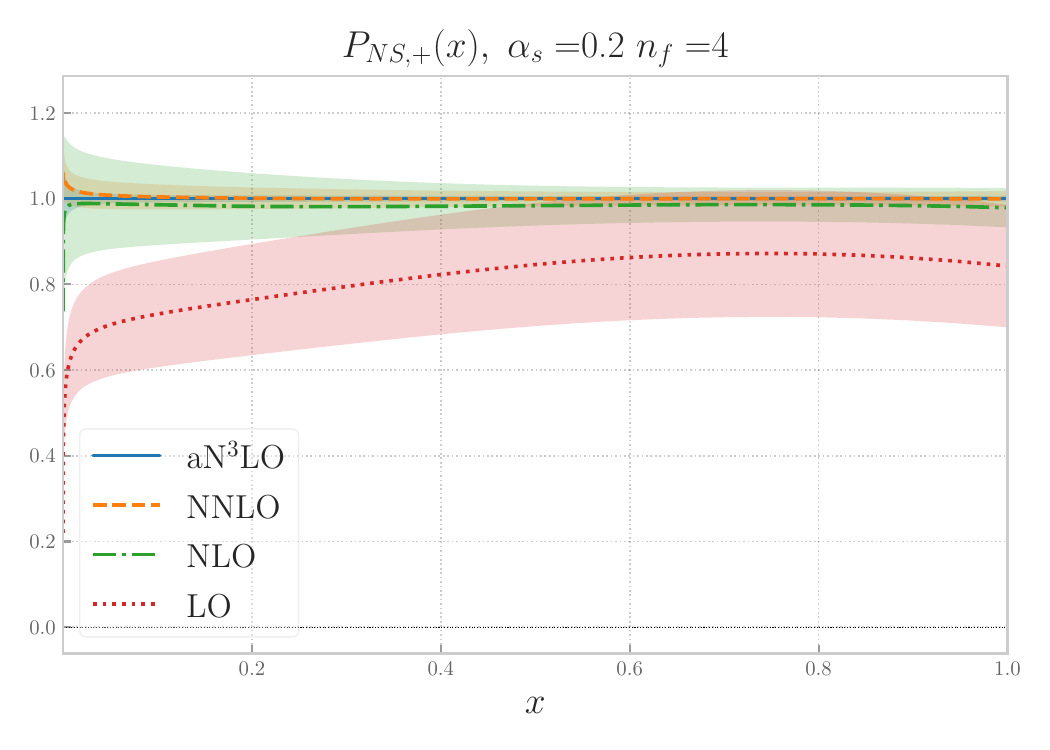}
  \includegraphics[width=.49\textwidth]{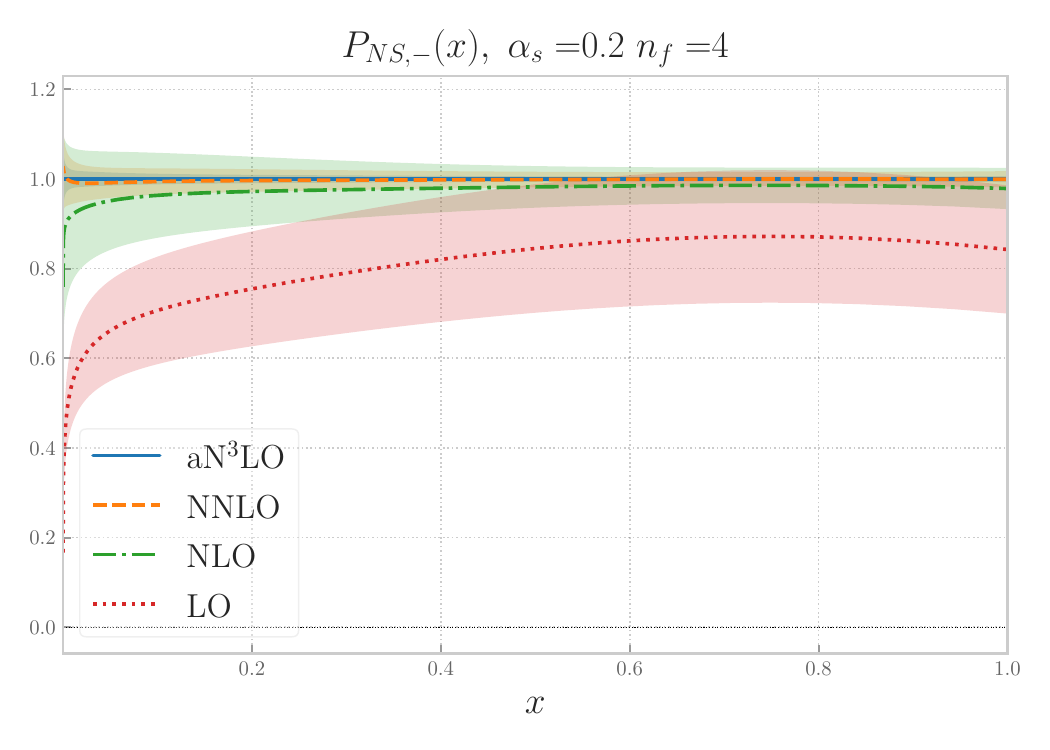}
  \caption{\small The non-singlet splitting functions at
    LO, NLO, NNLO, and aN$^3$LO, normalized to the aN$^3$LO
    central value and with a linear scale on the $x$ axis.
    In each case we show also the 
    uncertainty due to missing higher orders (MHOU) estimated by scale
    variation according to Refs.~\cite{NNPDF:2019vjt,NNPDF:2019ubu}.
  }
  \label{fig:splitting-functions-ns} 
\end{figure}

\begin{figure}[!t]
  \centering
  \includegraphics[width=.49\textwidth]{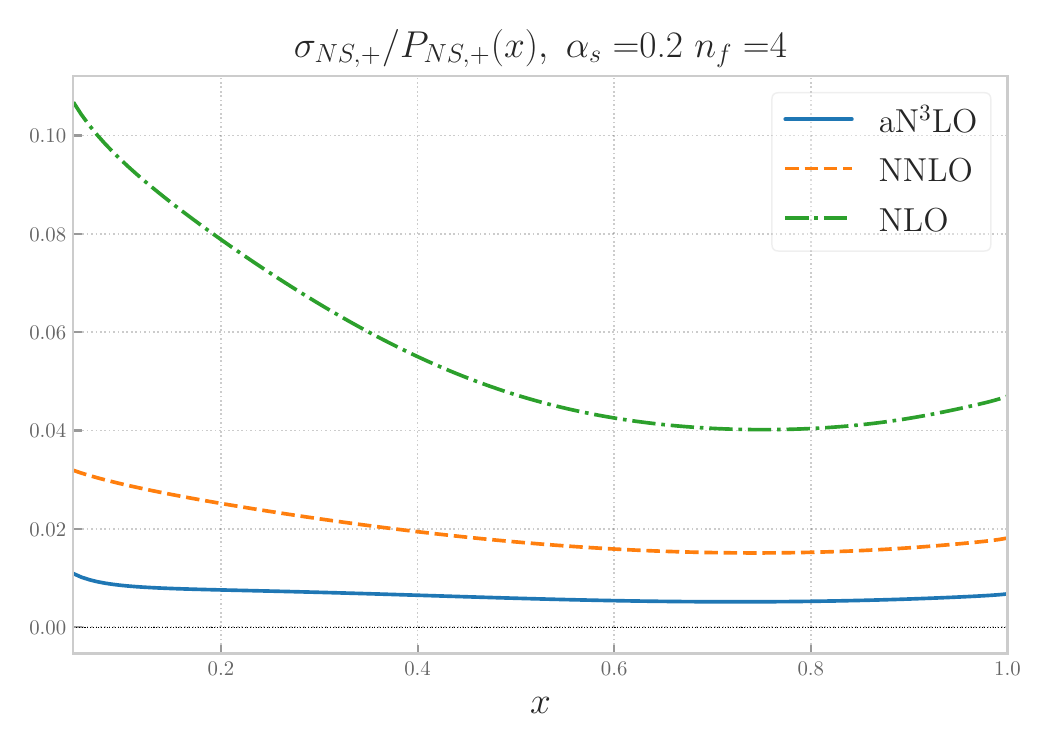}
  \includegraphics[width=.49\textwidth]{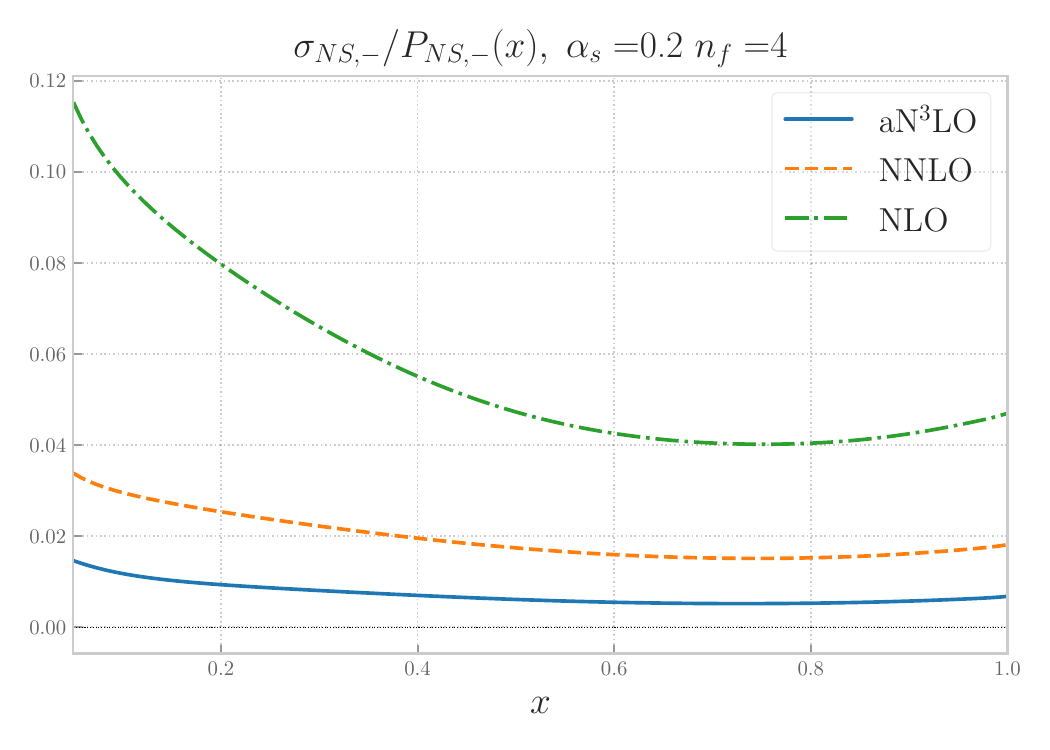}
  \caption{\small The relative size of the uncertainty due to missing higher
    orders (MHOU) on the splitting functions
    of \cref{fig:splitting-functions-ns}.}
  \label{fig:splitting-functions-ns-unc} 
\end{figure}

\begin{figure}[!t]
  \centering
  \includegraphics[width=.49\textwidth]{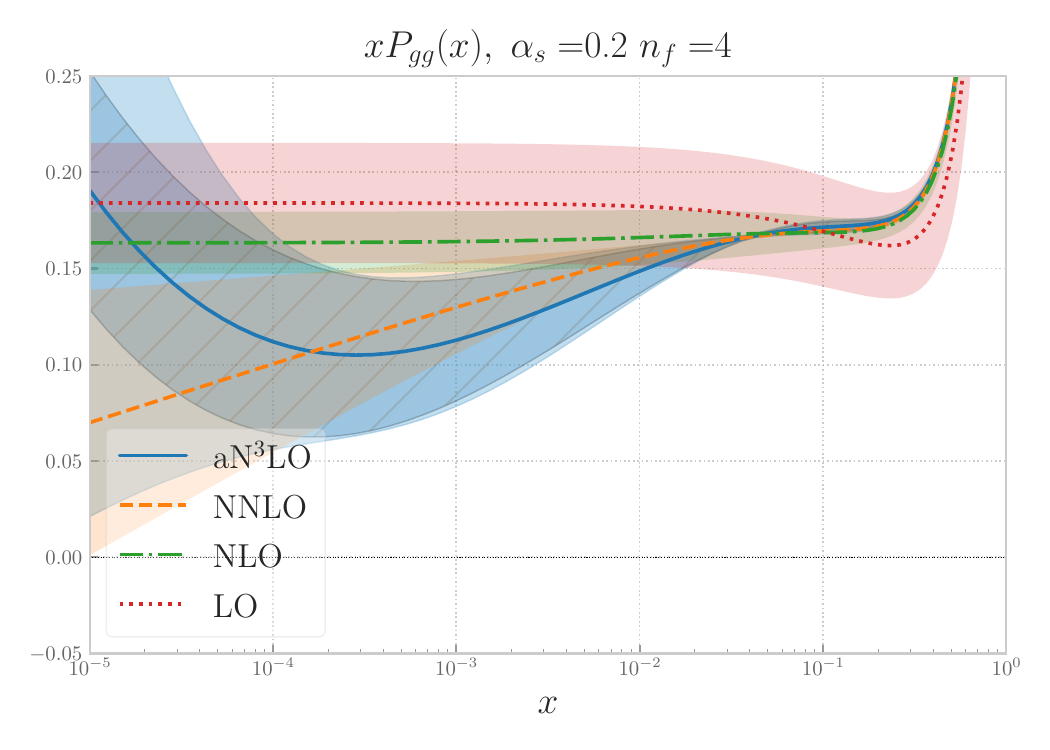}
  \includegraphics[width=.49\textwidth]{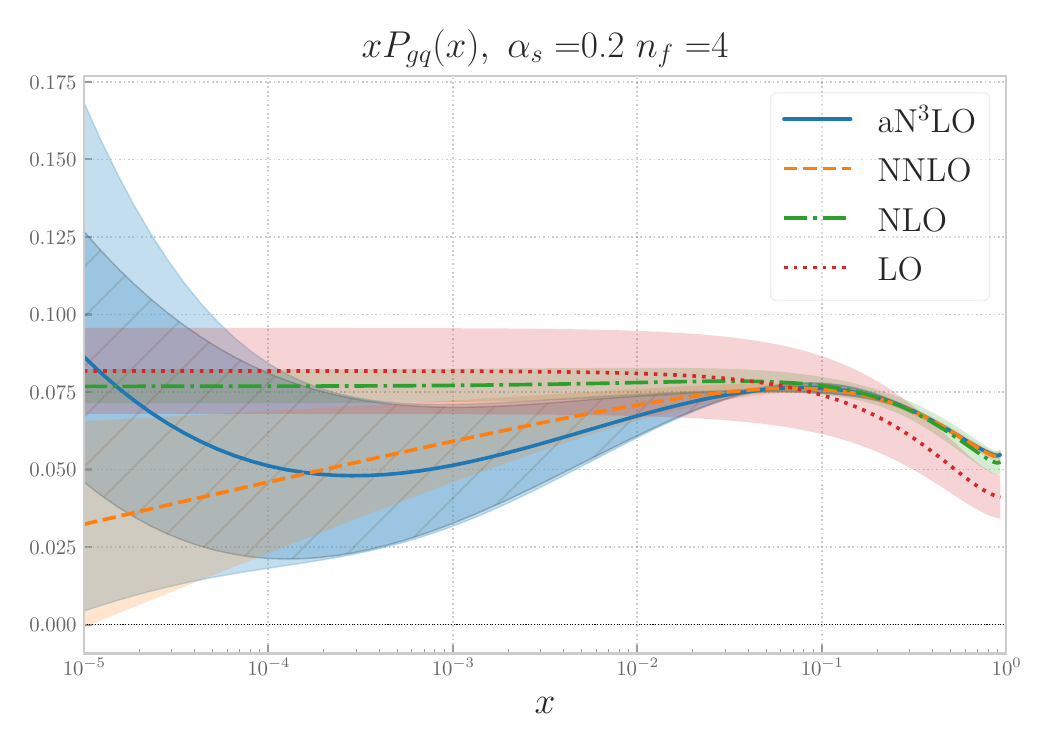} \\
  \includegraphics[width=.49\textwidth]{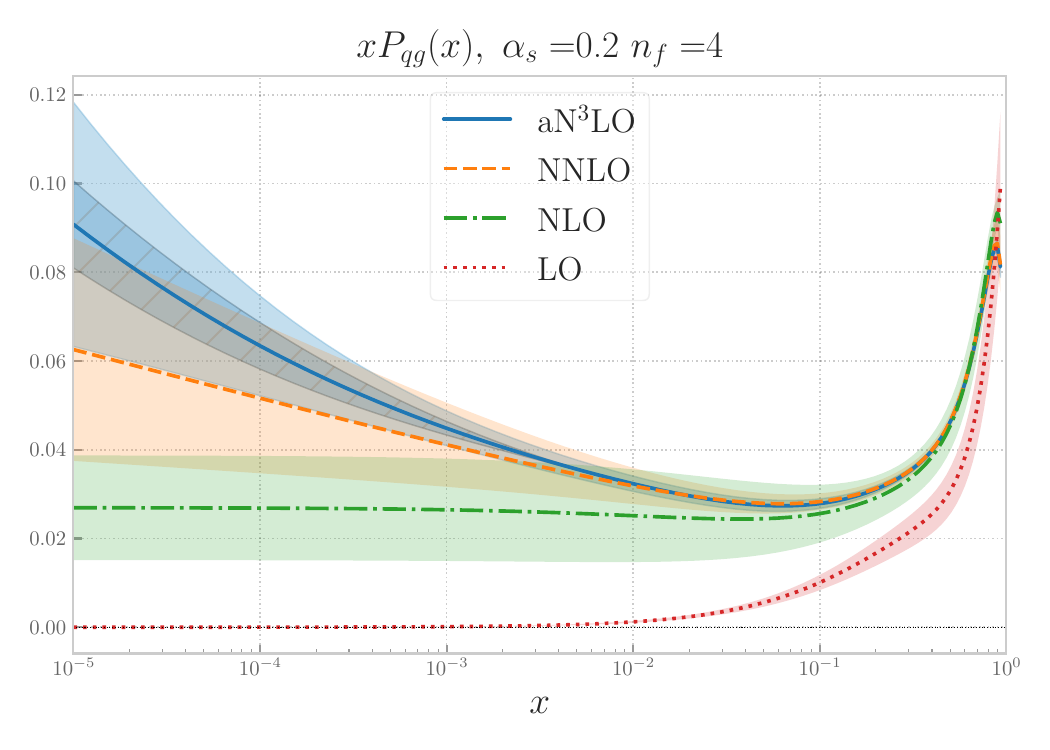}
  \includegraphics[width=.49\textwidth]{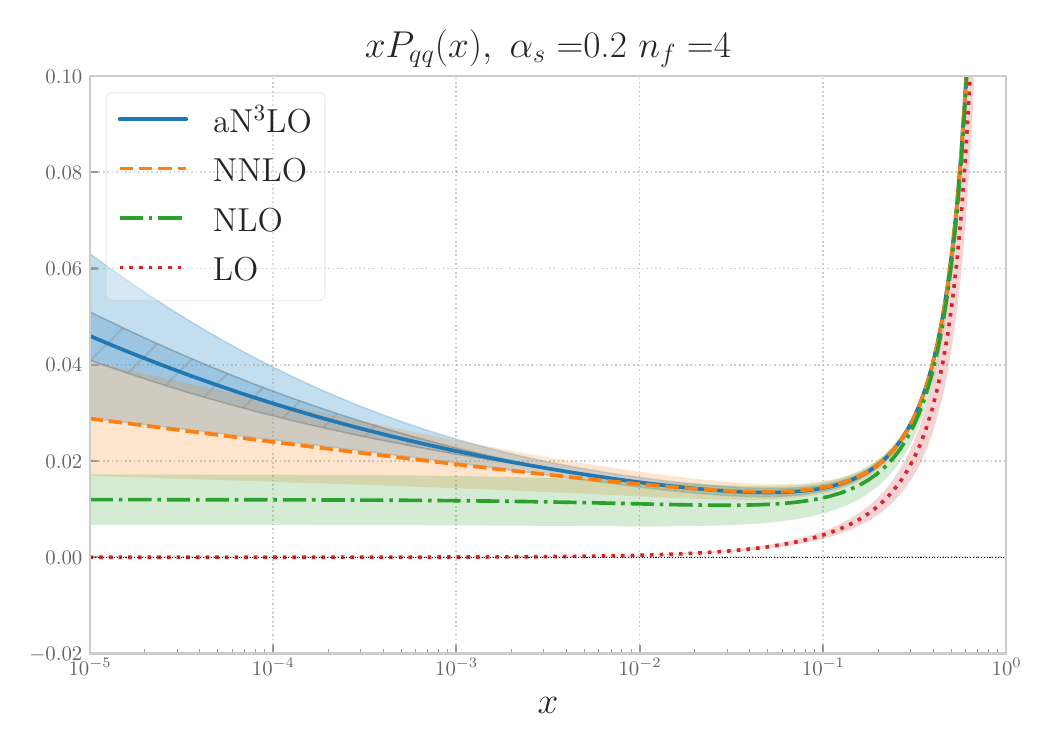}
  \caption{\small The singlet matrix of splitting functions $xP_{ij}$  at
    LO, NLO, NNLO  and aN$^3$LO. From left to
    right and from top to bottom the $gg$, $gq$, $qg$ and $qq$ entries
    are shown. The MHOU estimated by scale
    variation is shown to all orders. At aN$^3$LO
    the dark blue band corresponds to IHOU only, 
    while the light blue band is the sum in 
    quadrature of IHOU and MHOU.}
  \label{fig:splitting-functions-singlet-logx} 
\end{figure}

\begin{figure}[!t]
  \centering
  \includegraphics[width=.49\textwidth]{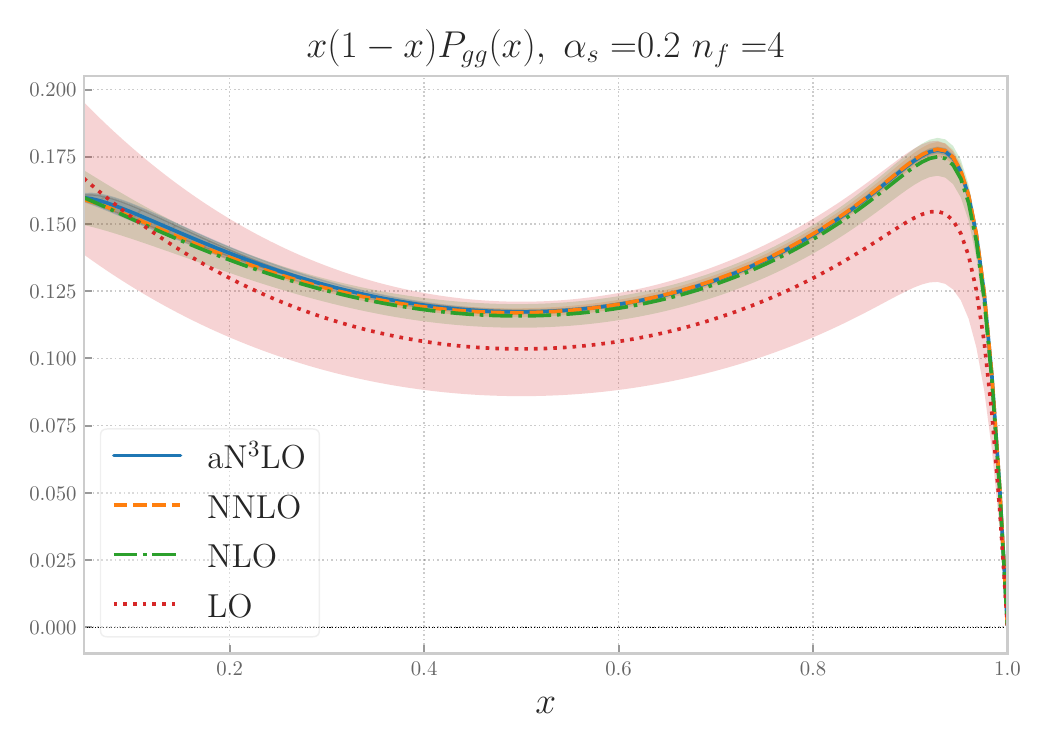}
  \includegraphics[width=.49\textwidth]{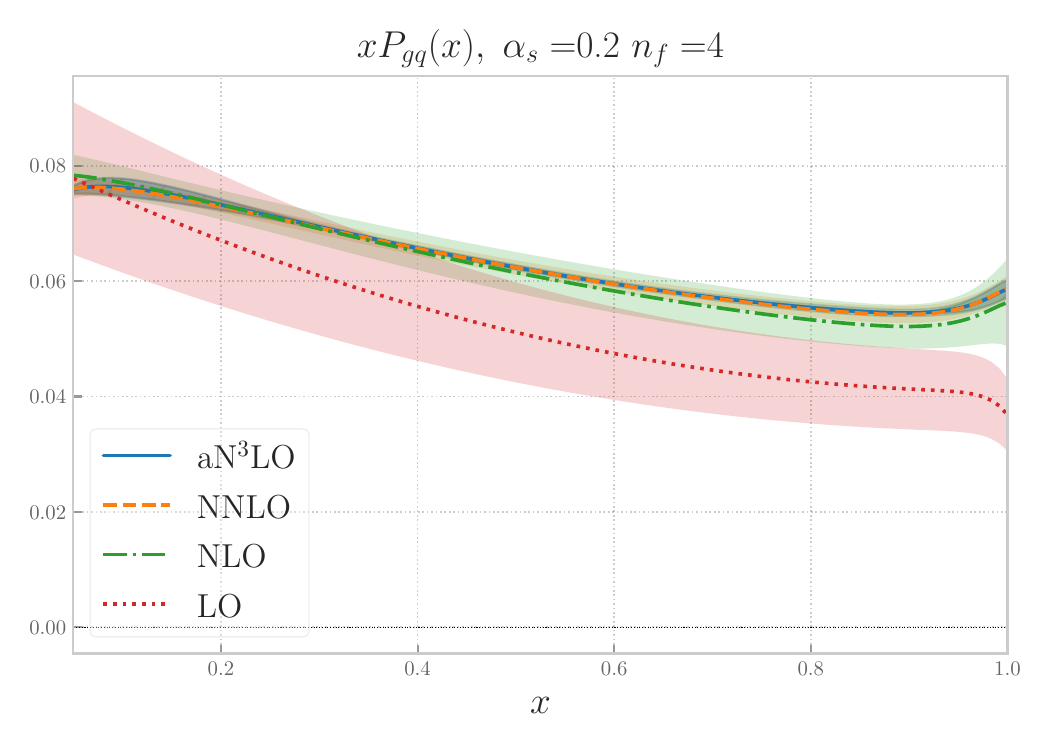} \\
  \includegraphics[width=.49\textwidth]{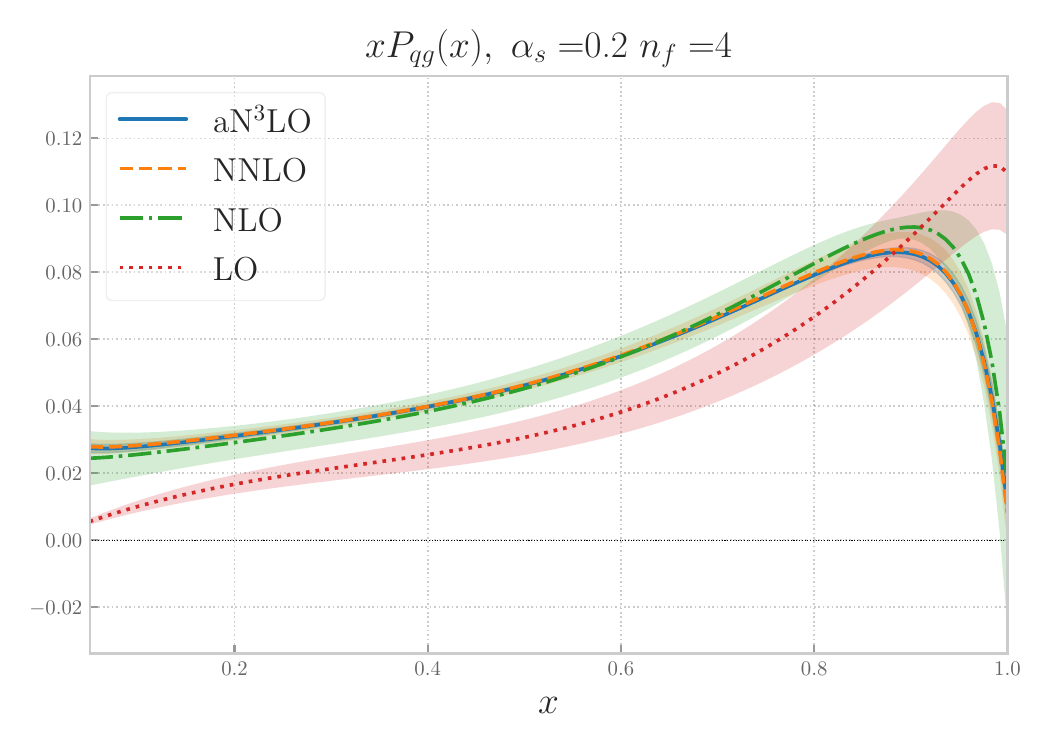}
  \includegraphics[width=.49\textwidth]{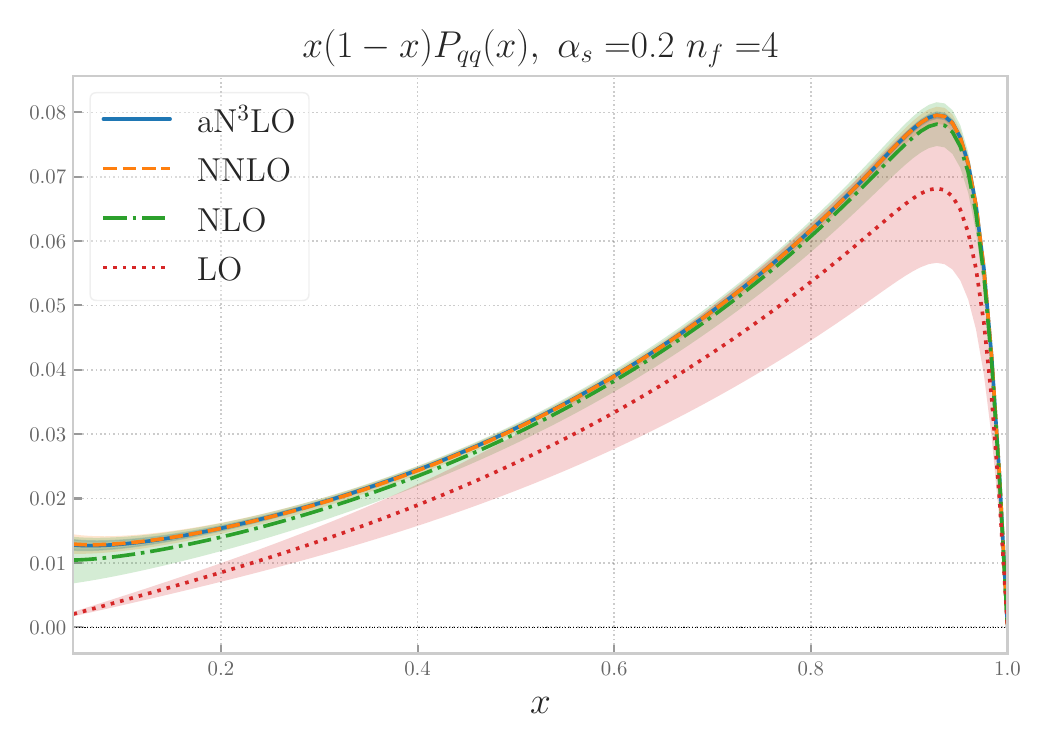}
  \caption{\small Same as \cref{fig:splitting-functions-singlet-logx}
    with a  linear scale on the $x$ axis, and plotting
    $(1-x)x P_{ii}$ for diagonal entries.}
  \label{fig:splitting-functions-singlet-linx} 
\end{figure}

\begin{figure}[!t]
  \centering
  \includegraphics[width=.49\textwidth]{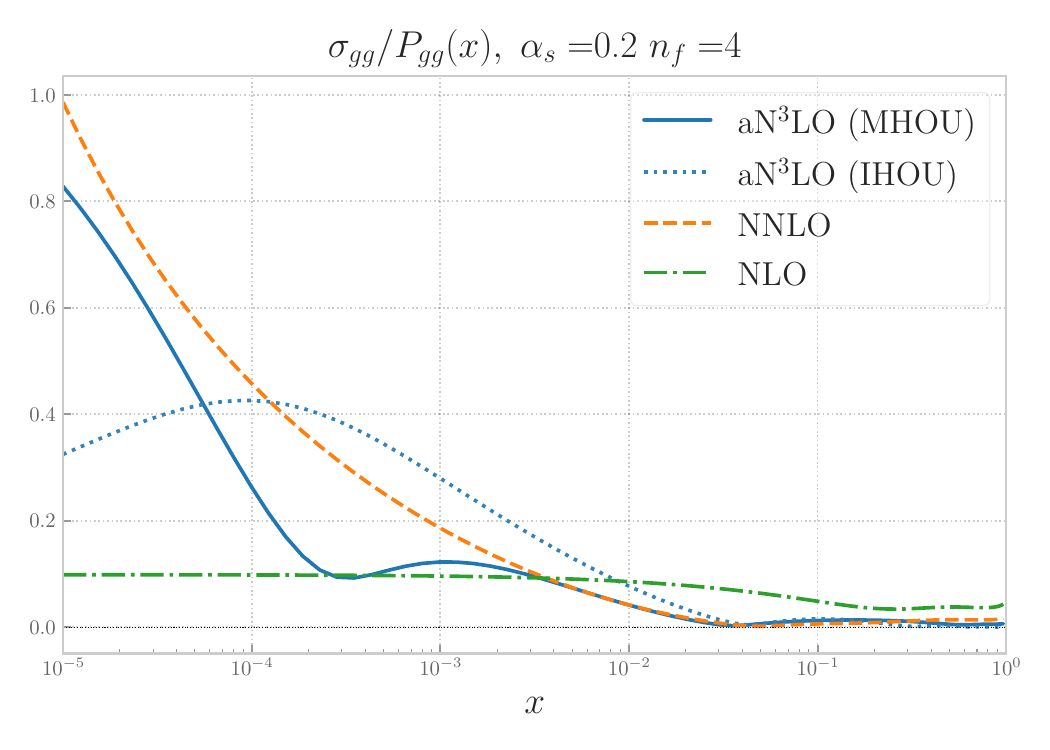}
  \includegraphics[width=.49\textwidth]{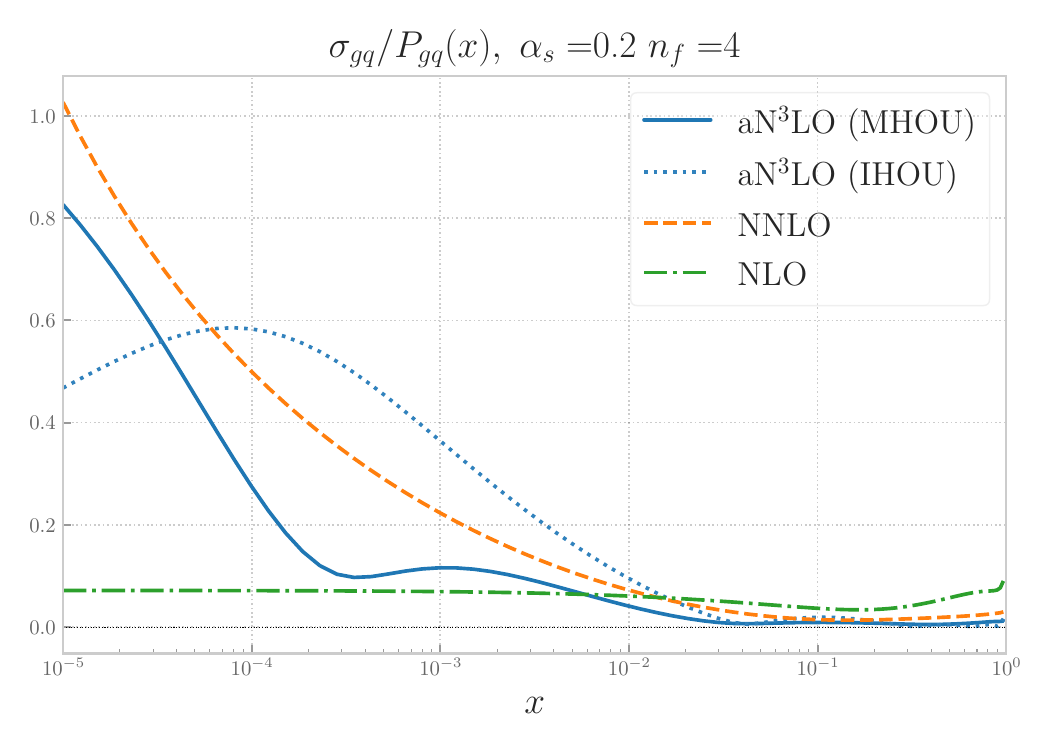} \\
  \includegraphics[width=.49\textwidth]{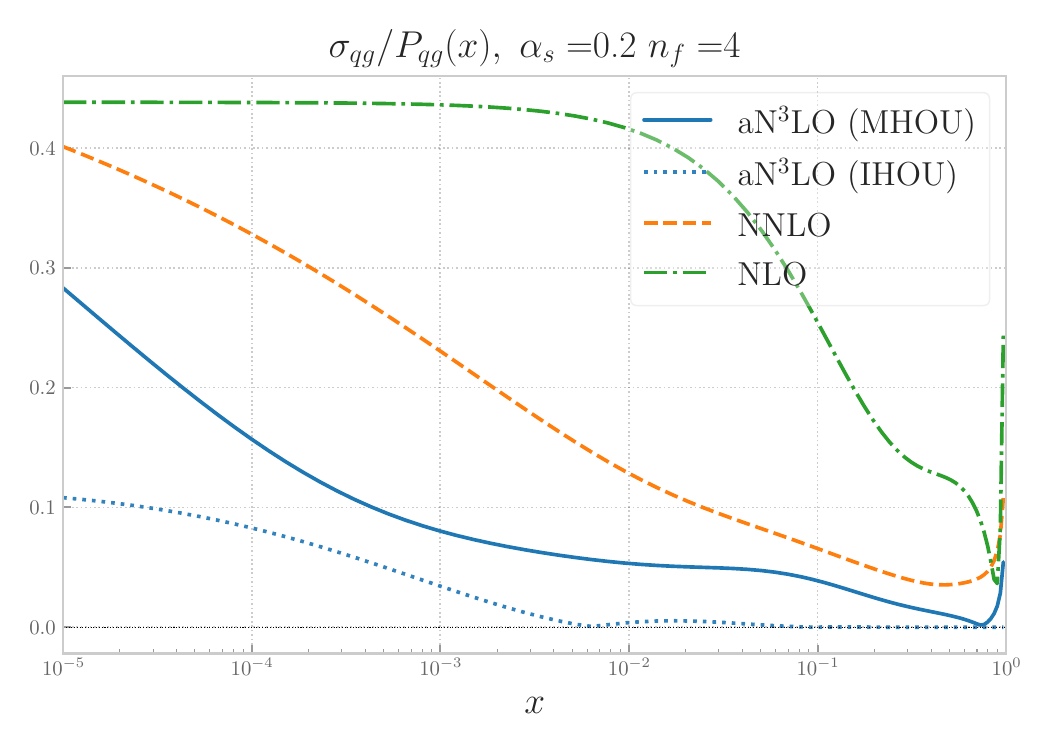}
  \includegraphics[width=.49\textwidth]{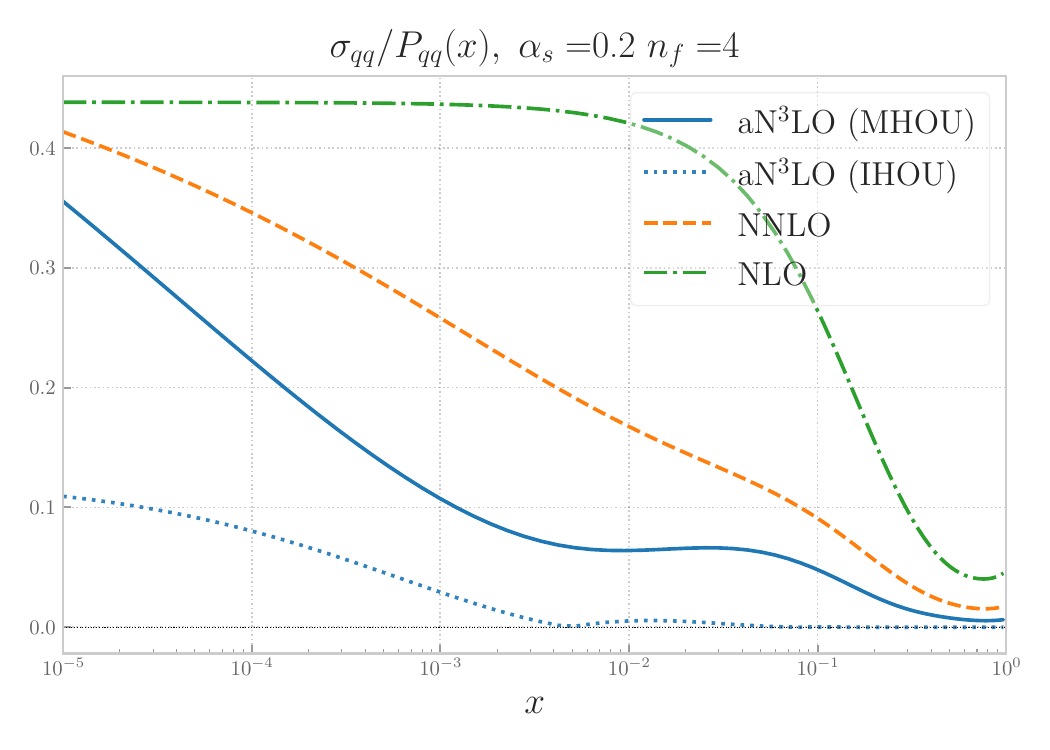}
  \caption{\small Same as \cref{fig:splitting-functions-ns-unc}
    for the singlet splitting function matrix elements. At NLO and NNLO we show
    the MHOU, while at aN$^3$LO we also show the IHOU.}
  \label{fig:splitting-functions-singlet-unc} 
\end{figure}

We now present the aN$^3$LO splitting functions constructed following the
procedure described in \cref{sec:an3lo_general_strategy,sec:an3lo_non_singlet,sec:an3lo_singlet}.
The non-singlet result, already compared in \cref{fig:nsgamma} to
the previous approximation of  Ref.~\cite{Moch:2017uml}, is shown 
in \cref{fig:splitting-functions-ns} at the first four
perturbative orders as a ratio to the aN$^3$LO result. 
For each order we include the MHOU determined by
scale variation according to Refs.~\cite{NNPDF:2019vjt,NNPDF:2019ubu} and recall
that there are no IHOU in the non-singlet sector.
As the non-singlet splitting function are subdominant at small
$x$ we only show the plot with a linear scale in $x$. The relative size of the
MHOU is shown in \cref{fig:splitting-functions-ns-unc}.

Inspecting \cref{fig:splitting-functions-ns,fig:splitting-functions-ns-unc}
reveals good perturbative convergence
\footnote{Here and henceforth by
``convergence'' we mean that the size of the missing N$^4$LO corrections is
negligible compared to the target accuracy of theoretical predictions,
i.e.\ at the sub-percent level.}
for all values of $x$.
Specifically, the differences between two subsequent perturbative orders are
reduced as the accuracy of the calculation increases, and, correspondingly, the
MHOUs associated to factorization scale variations decrease with the
perturbative accuracy. Indeed, the MHOU
appears to reproduce well the observed behavior of the
higher orders, with overlapping uncertainty bands between subsequent
orders except at LO at the smallest $x$ values. Hence, the behavior of
the perturbative series suggests that the MHOU estimate based on scale
variation at N$^3$LO is reliable. 

Based on these results it is clear that in the non-singlet
sector the N$^3$LO contribution to the splitting function is
essentially negligible except at the smallest $x$ values, as shown in
\cref{fig:nsgamma}. Consequently, for all practical purposes we
can consider the current approximation to the non-singlet anomalous
dimension to be essentially exact, and with negligible MHOU.

The situation in the singlet sector is more challenging.
The singlet matrix of splitting functions is shown in
\cref{fig:splitting-functions-singlet-logx,fig:splitting-functions-singlet-linx}, 
respectively with a logarithmic or linear scale on the $x$ axis.
Because the diagonal splitting functions are
distributions at $x=1$ in the linear scale plots we display $x(1-x)P_{ii}$.
The corresponding relative size of the MHOU is shown in
\cref{fig:splitting-functions-singlet-unc} for the first
four perturbative orders, along with the IHOU on the aN$^3$LO result,
determined using \cref{eq:sigihou}.

A different behavior is observed for the quark sector $P_{qi}$ and for
the gluon sector $P_{gi}$. In the quark sector, the MHOU decreases with
perturbative order for all $x$, but it remains sizable at aN$^3$LO
for essentially
all $x$, of order $5~\%$ for $10^{-2}\lesssim x\lesssim10^{-1}$. 
In the gluon sector instead for $x \gtrsim 0.03$ the MHOU is negligible, 
but at smaller $x$ it grows rapidly, and in fact at very small $x$
it becomes larger than the NLO MHOU. This is due to the presence of
leading small-$x$ logarithms, \cref{eq:smallxgg}, which are absent at NLO.
In fact the true gluon-sector MHOU at very small $x$ is likely to be
underestimated
by scale variation, because while it generates the fourth-order leading pole
present in the N$^4$LO (the fifth-order pole vanishes), it fails to generate 
the sixth-order pole known to be present in the N$^5$LO splitting function.

We now turn to the IHOU and find again contrasting behavior in the different
sectors. In the quark sector, thanks to the large
number of known Mellin moments and the copious information on the
large-$x$ limit, the IHOU are significantly smaller than the MHOU, by about a
factor three, and become negligible for $x\gtrsim 10^{-2}$. In the gluon
sector instead the IHOU, while still essentially negligible for
$x\gtrsim0.1$, is larger than the MHOU except at very small
$x\lesssim 10^{-4}$ where the MHOU dominates.

Consequently, for all matrix elements at large $x\gtrsim 0.1$ the behavior of
the singlet is similar to the behavior of the non-singlet: IHOU and MHOU are
both negligible, meaning that aN$^3$LO results are essentially
exact, and the perturbative expansion has essentially converged, see 
\cref{fig:splitting-functions-singlet-linx}. At smaller $x$, while
the aN$^3$LO and NNLO results agree within uncertainties, the
uncertainties on the aN$^3$LO are sizable, dominated by MHOUs in the
quark channel and by IHOUs in the gluon channel.

\subsection{Results: aN$^3$LO evolution}
\label{sec:an3lo_evol}

The aN$^3$LO anomalous dimensions discussed in the previous sections
have been implemented in the
Mellin space open-source evolution code \eko{} (\cref{sec:eko}). 
The parametrization is expressed in
terms of a basis of Mellin space functions 
which are numerically efficient to evaluate.
In order to achieve full aN$^3$LO accuracy, in addition to the anomalous dimensions,
we have also been implemented the four-loop running of 
the strong coupling constant $\alpha_s(Q)$ and the 
N$^3$LO matching conditions, as discussed in \cref{sec:fns}, 
dictating the transitions between schemes with different 
numbers of active quark flavor.
A special case is the $a_{Hg}^{(3)}$ entry of the matching condition matrix, 
which at the time of publication was still unknown and has been parametrized 
using the first 5 known moments~\cite{Bierenbaum:2009mv} and the LL$x$ contribution.
\footnote{ 
  Results of Ref.~\cite{Ablinger:2024xtt}, were not available when this study was originally presented.
}

\begin{figure}[!t]
  \centering
  \includegraphics[width=0.49\textwidth]{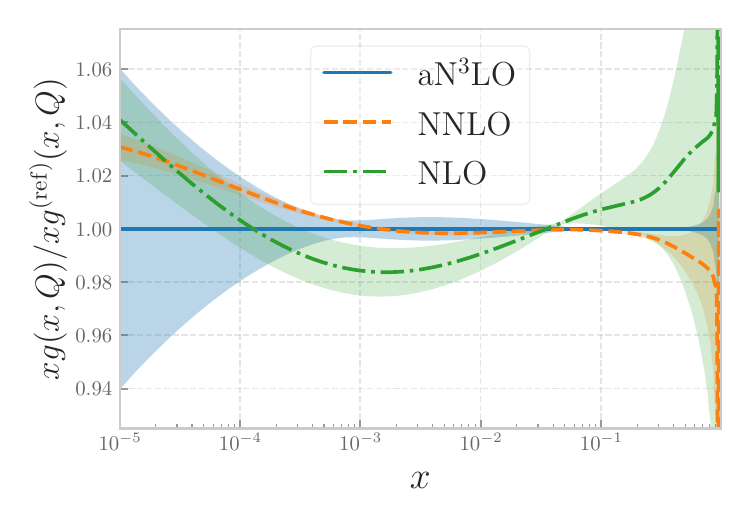}
  \includegraphics[width=0.49\textwidth]{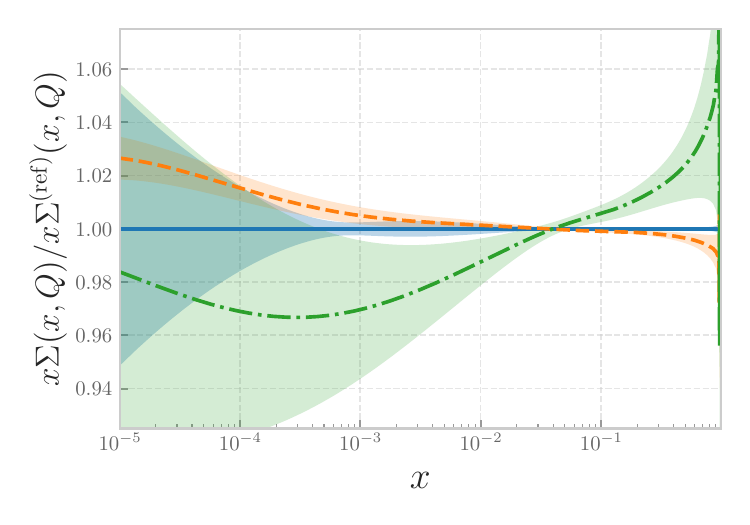}\\
  \includegraphics[width=0.49\textwidth]{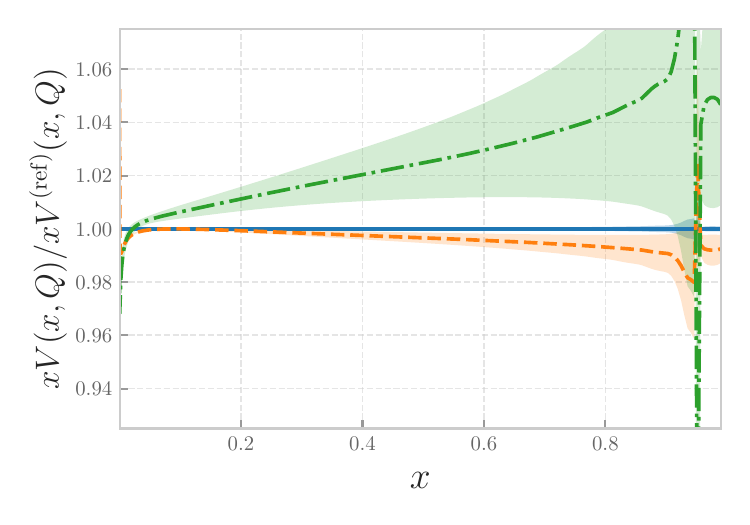}
  \includegraphics[width=0.49\textwidth]{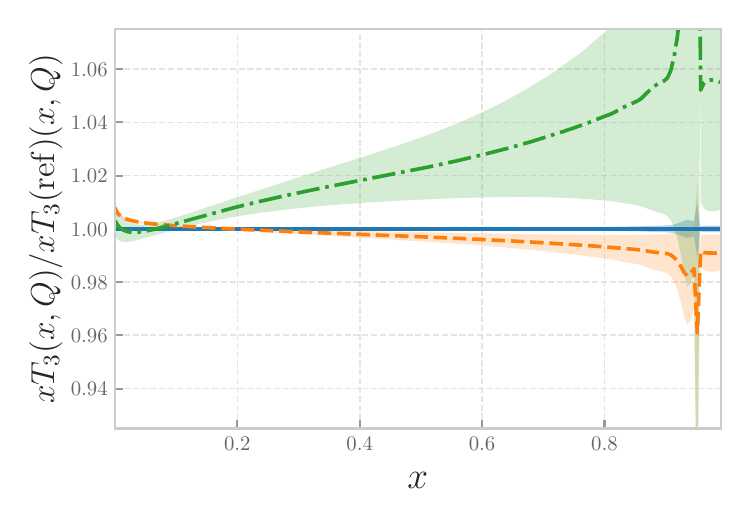}\\ 
  \includegraphics[width=0.49\textwidth]{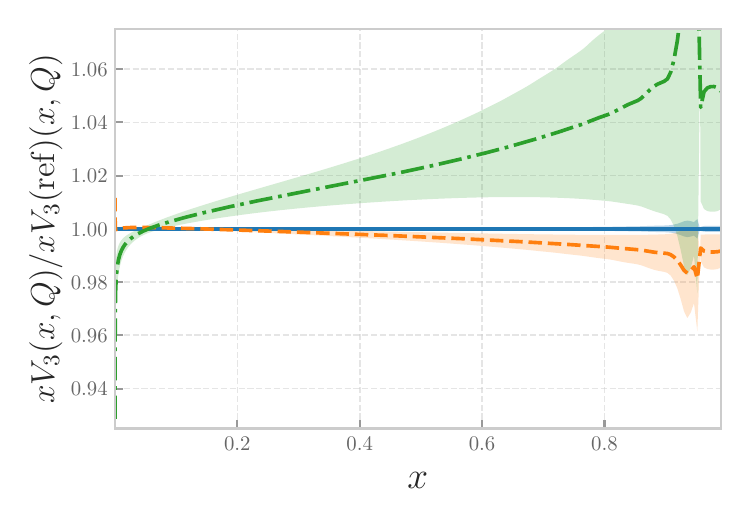} 
  \caption{Comparison of the result obtained evolving from  $Q_0=1.65$~GeV
    to $Q=100$~GeV at NLO, NNLO, and aN$^3$LO using NNPDF4.0 NNLO as fixed
    starting PDF. Results are shown as ratio to the aN$^3$LO (from left to right
    and from top to bottom) for the gluon and singlet $\Sigma$, and for the
    $V$, $V_3$ and $T_3$ quark eigenstates of perturbative evolution (see
    \cref{sec:an3lo_general_strategy}). The total theory uncertainty is
    shown in all cases, i.e.\ the MHOU at NLO and NNLO, and the sum in
    quadrature of MHOU and IHOU at aN$^3$LO.}
  \label{fig:N3LOevolution-q100gev-ratios} 
\end{figure}

In \cref{fig:N3LOevolution-q100gev-ratios} we compare the result of
evolving a fixed set of PDFs  from $Q_0=1.65$~GeV up to $Q=100$~GeV at
NLO, NNLO, and aN$^3$LO. We take as input the NNPDF4.0NNLO PDF set,
and show results normalized to the aN$^3$LO evolution. Results are shown for
all the combinations that evolve differently, as discussed in 
\cref{sec:an3lo_general_strategy}, namely the singlet, gluon, total
valence and non-singlet $\pm$ combinations, with a logarithmic scale
on the $x$ axis for the singlet sector and a linear scale for the
valence and non-singlet combinations.

In all cases the perturbative expansion appears to have converged
everywhere, with almost no difference between NNLO and aN$^3$LO except
at small $x\lesssim10^{-3}$, where singlet evolution is weaker at
aN$^3$LO than at NNLO due to the characteristic dip seen in the
gluon sector splitting functions of \cref{fig:splitting-functions-singlet-logx}. 
Because the gluon-driven small-$x$ rise dominates small-$x$ evolution this is a
generic feature of all quark and gluon PDFs in this small-$x$ region.
In fact, the total theory uncertainty at aN$^3$LO is at the
sub-percent level for all  $x\gtrsim10^{-3}$. Hence, not only has the
MHOU become negligible, but also the effect of IHOU on PDF evolution
is only significant at small-$x$. 

\subsection{Comparison to other groups}
\label{sec:an3lo_comp}

We finally compare our approximation of the N$^3$LO splitting
functions to other recent results from
Refs.~\cite{McGowan:2022nag,Falcioni:2023luc,Falcioni:2023vqq,Moch:2023tdj}.
While the approach of
Refs.~\cite{Falcioni:2023luc,Falcioni:2023vqq,Moch:2023tdj}
(FHMRUVV, henceforth) is very similar to our own,
with differences only due to details of the choice of basis functions,
a rather different approach is adopted in
Ref.~\cite{McGowan:2022nag} (MSHT20, henceforth). 
There, the approximation is constructed from similar theoretical 
constraints (small-$x$, large-$x$ coefficients and Mellin moments), 
but supplementing the parametrization with additional
nuisance parameters, which control the uncertainties arising from 
unknown N$^3$LO terms. However, these approximations are taken as a prior, 
and the nuisance parameters are fitted to the
data along with the PDF parameters. The best-fit values of the
parameters determine the posterior for the splitting function, and
their uncertainties are interpreted as the final IHOU on it. A consequence
of this procedure is that the posterior can reabsorb not only N$^3$LO
corrections, but any other missing contribution, of theoretical or
experimental origin. 

\begin{figure}[!t]
  \centering
  \includegraphics[width=.49\textwidth]{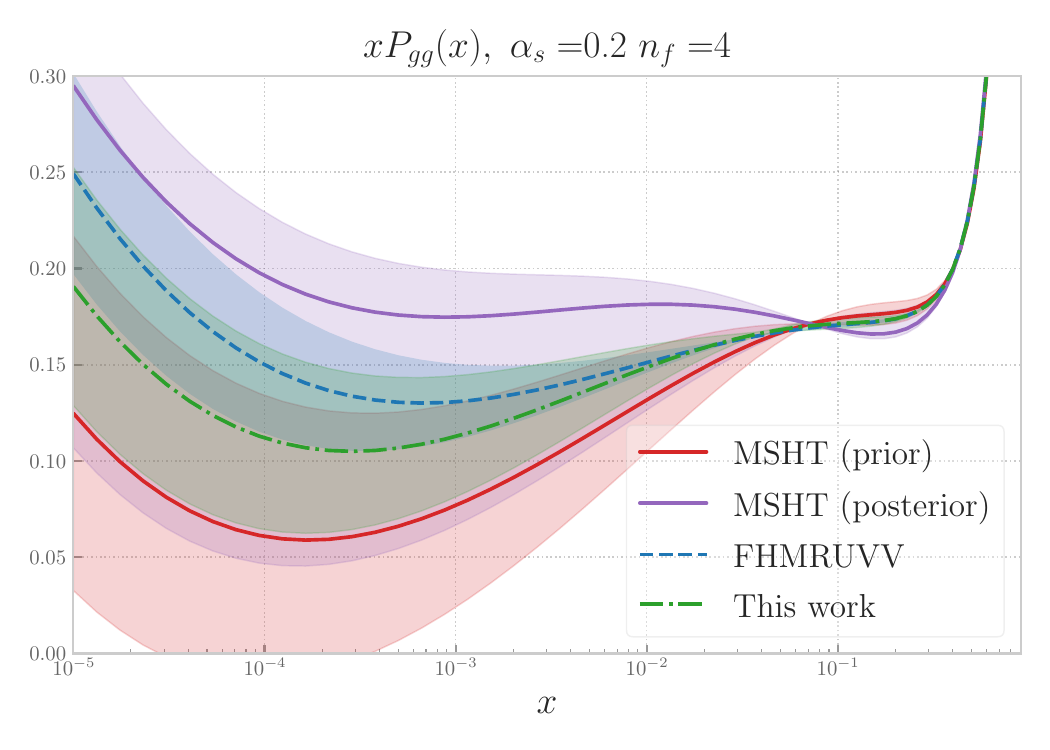}
  \includegraphics[width=.49\textwidth]{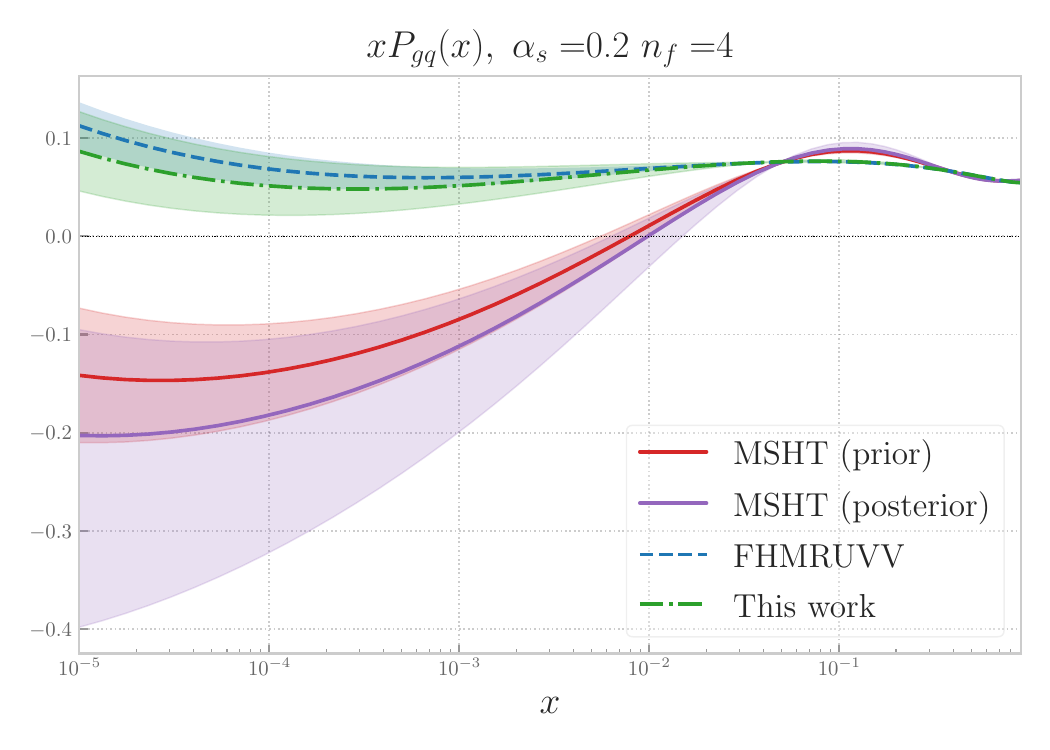} 
  \includegraphics[width=.49\textwidth]{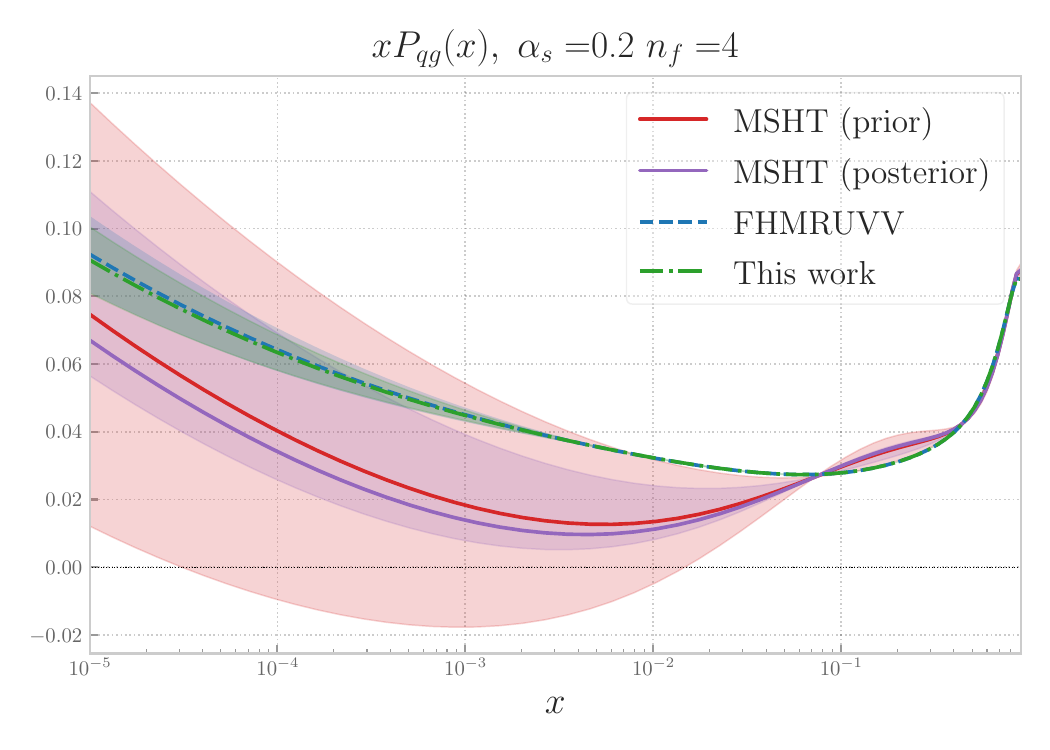}
  \includegraphics[width=.49\textwidth]{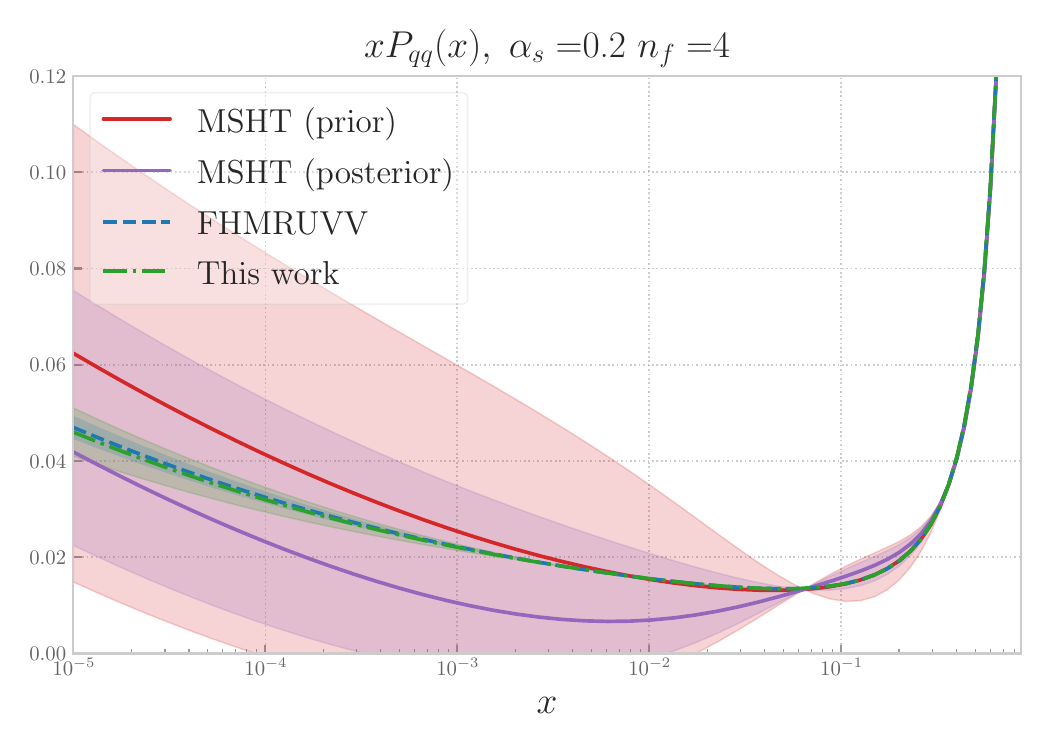}
  \caption{\small Same as \cref{fig:splitting-functions-singlet-logx}, now comparing
    our aN$^3$LO result to those of Ref.~\cite{McGowan:2022nag} (MSHT20) and
    Refs.~\cite{Falcioni:2023luc,Falcioni:2023vqq,Moch:2023tdj} (FHMRUVV).
    In all cases the uncertainty
    band correspond to the IHOU as estimated by the various groups.
    For the MSHT20 results, we display both the prior and the
    posterior parametrizations (see text).}
  \label{fig:splitting-functions-mhst} 
\end{figure}

The comparison is presented in \cref{fig:splitting-functions-mhst},
for all the four singlet splitting functions. For the MSHT20 results 
both prior and posterior are displayed. It should be noticed that even though
the uncertainty bands on the NNPDF4.0, FHMRUVV and MSHT20 prior are
all obtained by varying the set of basis functions, they are found
using somewhat different procedures, and their meaning is accordingly
somewhat different. Indeed, for NNPDF4.0 the is constructed out of the
covariance matrix according to \cref{eq:sigihou}. For FHMRUVV is instead the band between an
upper and lower estimates which are representative of the envelope of
all variations. Finally, for the MSHT20 prior it is the variance of
the probability distribution obtained assuming a  multi-Gaussian
distribution of suitable nuisance parameters. 

As expected, excellent agreement is found with the FHMRUVV result, for  
all  splitting functions and for all $x$, especially for the $P_{qg}$
and $P_{qq}$ splitting functions, for which the highest number of
Mellin moments is known. Good qualitative agreement is also found for $P_{gq}$
and $P_{gg}$, although at small $x$ IHOUs
are larger and consequently  central values differ somewhat more,
though still in agreement within uncertainties. Uncertainties are
qualitatively similar, except at small $x$, where less exact
information is available and both central values and uncertainties are
less constrained. In this region the NNPDF4.0 is generally somewhat
more conservative, possibly due to the fact that it is obtained by
adding individual shifts in quadrature, rather than taking their envelope.

Coming now to  MSHT20 results, good agreement is found with the
prior, except for $P_{gq}$, for which MSHT20 shows a small-$x$ dip
accompanied by a large-$x$ bump. The different small-$x$ behavior
is likely due to the fact that
MSHT20 do not enforce the color-charge relation \cref{eq:smallxiq}
at NLL$x$, with the large-$x$ bump then following from the constraints
\cref{eq:singlet_scaling}.  Also, in the quark sector the MSHT20
prior has significantly larger IHOUs due to the fact that it does not include 
the more recent information on Mellin moments from
Refs.~\cite{Davies:2022ofz,Falcioni:2023luc,Falcioni:2023vqq,Moch:2023tdj,
  Falcioni:2023tzp},
which were not available at the time of the MSHT20
analysis~\cite{McGowan:2022nag}.
At the level of posterior, however, significant differences appear
also for  $P_{gg}$, while persisting for $P_{gq}$. 
This means that the gluon evolution at aN$^3$LO is being modified
by the data entering the global fit, and it is not fully determined by the
perturbative computation. Further benchmarks of aN$^3$LO splitting functions
are presented in Ref.~\cite{Cooper-Sarkar:2024crx}.

\section{N$^3$LO partonic cross-sections}
\label{sec:an3lo_coefffun}
A PDF determination at N$^3$LO requires, in addition to the splitting
functions discussed in \cref{sec:an3lo_dglap}, hard cross-sections at the same
perturbative order. Exact N$^3$LO massless DIS coefficient functions have been
known for several years~\cite{Vermaseren:2005qc,Moch:2004xu,Moch:2007rq,
  Moch:2008fj,Davies:2016ruz,Blumlein:2022gpp}, 
while massive coefficient functions are only available in various
approximations~\cite{Kawamura:2012cr,Laurenti:2024anf,bbl2023}.
For hadronic processes, N$^3$LO results are available for inclusive
Drell-Yan (DY) production for the total
cross-section~\cite{Duhr:2021vwj,Duhr:2020sdp,Baglio:2022wzu} as well
as for rapidity~\cite{Chen:2021vtu} and transverse momentum
distributions~\cite{Chen:2022lwc}, though neither of these is publicly
available.

We now describe the implementation of these corrections in our fitting framework.
First, we review available results on DIS coefficient functions and summarize
the main features of the approximation that we will use for massive coefficient
functions~\cite{bbl2023,Laurenti:2024anf}.
Next we discuss how massless and massive DIS
coefficient functions are combined to 
extend the FONLL general-mass variable-flavor 
number scheme to $\mathcal{O}\lp \alpha_s^3\rp$.
Finally, we discuss 
N$^3$LO corrections for hadronic processes and different options for
their inclusion in PDF determination.

\subsection{N$^3$LO corrections to DIS structure functions}
\label{sec:an3lo_dis}

\begin{figure}[!t]
  \centering
  \includegraphics[width=.49\textwidth]{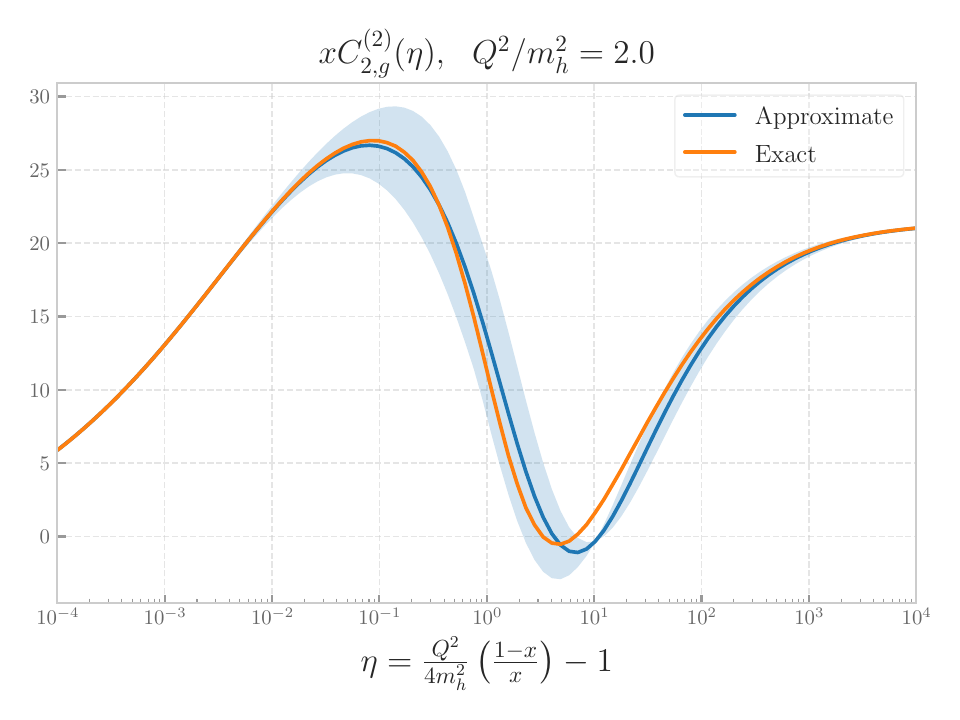}
  \includegraphics[width=.49\textwidth]{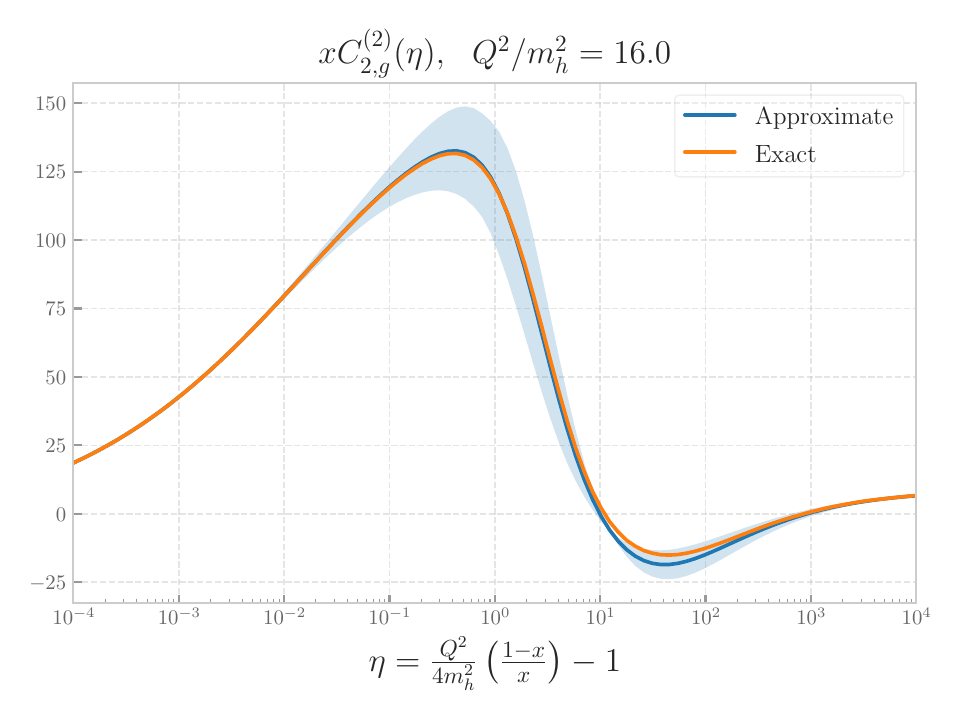}
  \caption{Comparison of the exact
    NNLO massive gluon-initiated coefficient function $x C_{2,g}^{(2)}(\eta)$
    to the approximation \cref{eq:massive_cf} from
    Ref.~\cite{Laurenti:2024anf}, plotted as a function of $\eta$,
    \cref{eq:etadef}, for fixed  $Q^2$. 
    Results are shown for two different values of $Q^2$, one
    close to threshold $Q^2=2 m_h^2$ (left) and one at high scales $Q^2=16 m_h^2$
    (right). The uncertainty on the approximate result
    is obtained by varying the interpolating functions
    $f_1(x)$ and $f_2(x)$ in \cref{eq:massive_cf}.}
  \label{fig:massive_nnlo_bench} 
\end{figure}

\begin{figure}[!t]
  \centering
  \includegraphics[width=.49\textwidth]{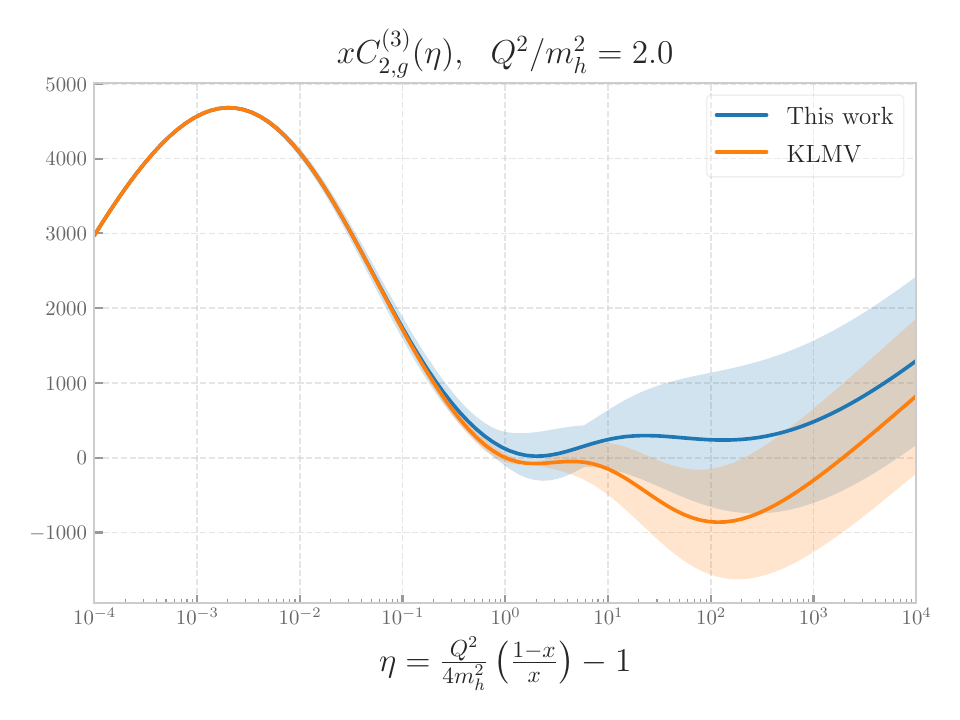}
  \includegraphics[width=.49\textwidth]{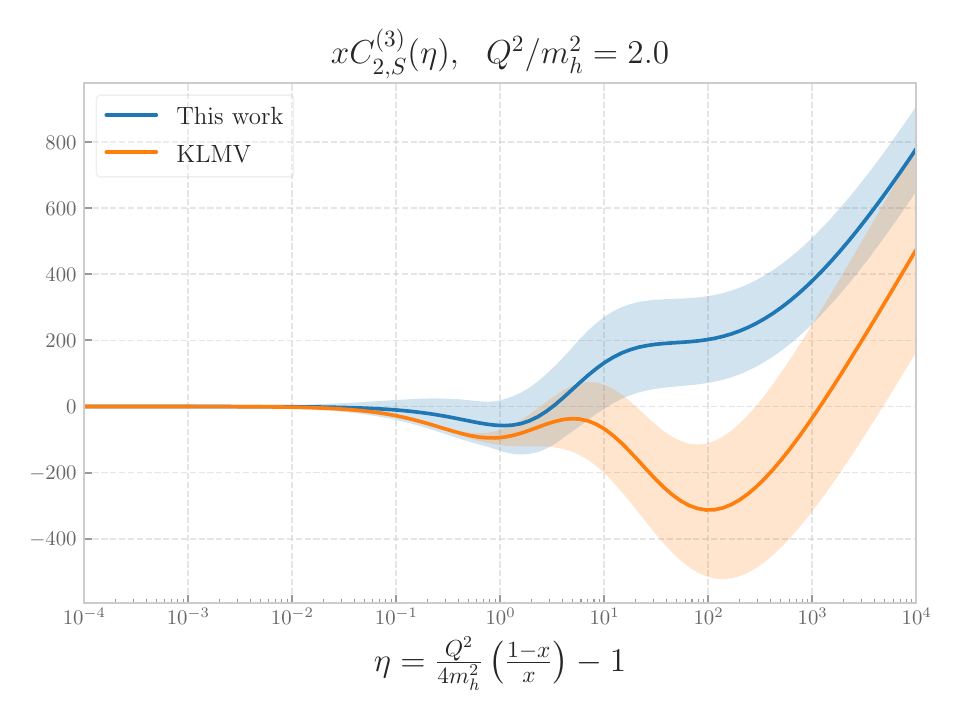}
  \caption{The approximate N$^3$LO massive gluon (left) and quark
    singlet (right) coefficient functions as a function of $\eta$
    for fixed  $Q^2=2 m_h^2$. Our result based on the approximation
    of Ref.~\cite{Laurenti:2024anf} is compared to the approximation of
    Ref.~\cite{Kawamura:2012cr} (KLMV).}
  \label{fig:massive_n3lo_bench} 
\end{figure}

\begin{figure}[!t]
  \centering
  \includegraphics[width=.85\textwidth]{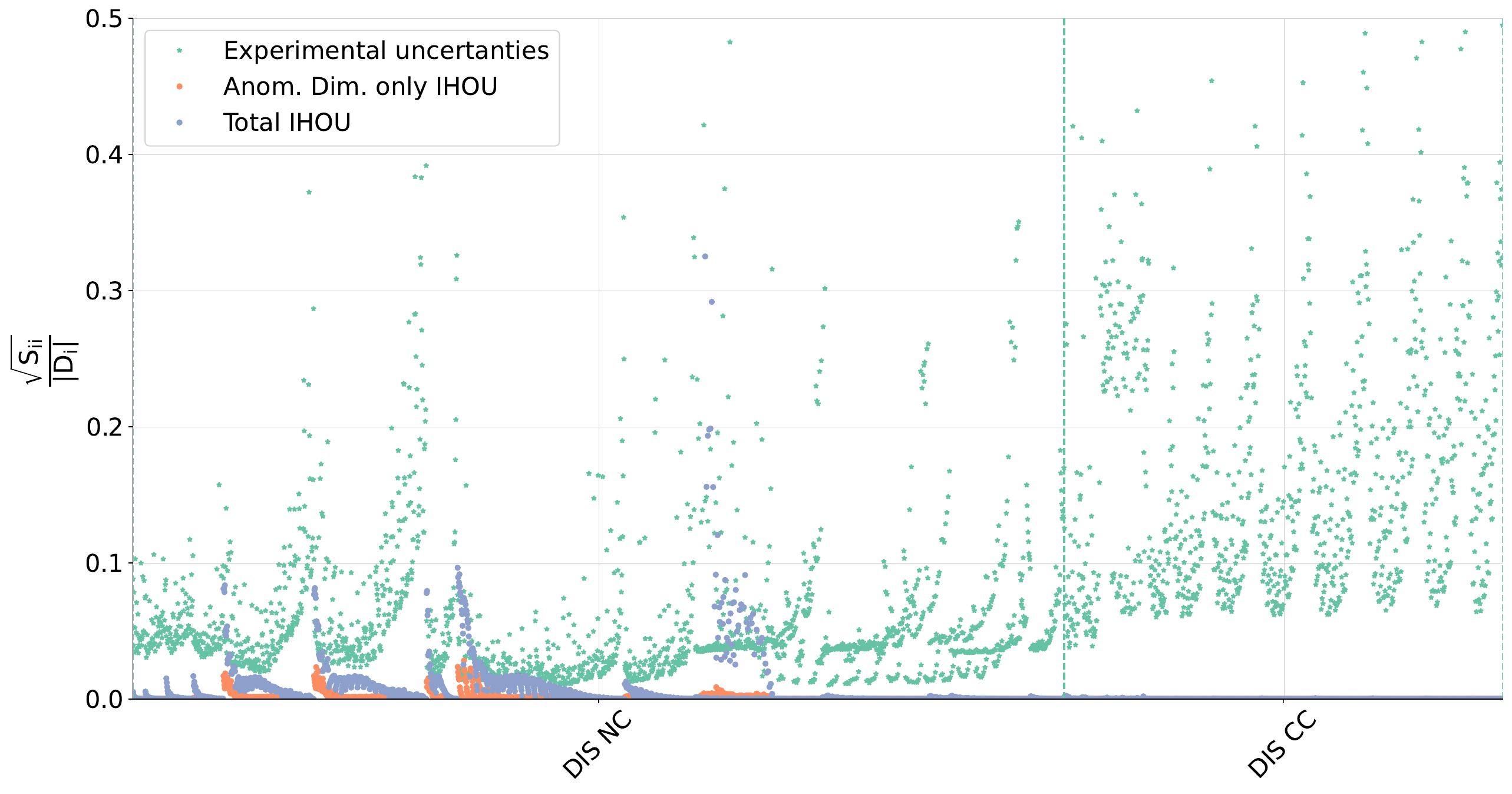} \\
  \caption{
    Square root of the diagonal entries of the IHOU
    covariance matrix for the DIS
    datasets normalized to the experimental central value $D_i$. We show the
    IHOU before and after adding to the covariance matrix 
    \cref{eq:covihou} that accounts for uncertainty on anomalous
    dimensions the extra component
    \cref{eq:covihouC} due to the  massive coefficient
    function. The experimental uncertainty is also shown for comparison.
  }    
  \label{fig:dis_ihou}
\end{figure}

In \cref{sec:dis_coeff} we have summarized how the DIS structure functions $F_i$ 
are evaluated from the convolution of PDFs and coefficient functions.
As mentioned, the N$^3$LO massless DIS coefficient are known,
while the massive corrections are not completely available, 
see also \cref{sec:dis_coeff,sec:heavy_quarks_dis}. 

Here, with the same spirit of \cref{sec:an3lo_evol}, 
we adopt an approximation for the N$^3$LO contribution
$C_{i,k}^{(3)}(x,\alpha_s,m^2_h/Q^2)$ to massive
coefficient functions for photon-induced DIS and neglect the axial-vector
coupling of the Z boson, while we treat heavy quarks in the massless
approximation for the W boson exchange.
Such an approximation, based on known partial results, has
been presented in Ref.~\cite{Kawamura:2012cr}, and recently revisited 
in Ref.~\cite{Laurenti:2024anf}. The approaches of these references rely on the
same known exact results, and differ in the details of the way they are
combined and interpolated. We will follow
Ref.~\cite{Laurenti:2024anf}, see also Ref.~\cite{bbl2023}, to which
we refer for further details. 
Exact results come from threshold resummation and high-energy
resummation, and are further combined with the asymptotic large-$Q^2$ limit,
thereby ensuring that the approximate massive coefficient function
reproduces the exact massless result in the $Q^2/m_h^2\to\infty$ limit.
In the approach of Refs.~\cite{Laurenti:2024anf,bbl2023} the massive
coefficient functions 
are written as
\begin{equation}
  C_{i,k}^{(3)}(x,m_h^2/Q^2) = C_{i,k}^{(3),{\rm thr}}(x,m_h^2/Q^2) f_1(x) + C_{i,k}^{(3),{\rm asy}}(x,m_h^2/Q^2) f_2(x) \, ,
  \label{eq:massive_cf}
\end{equation}
where $C_{i,k}^{(3),{\rm thr}}$ and $C_{i,k}^{(3),{\rm asy}}$
correspond to the contributions coming from differently resummations,
and  $f_1(x)$ and  $f_2(x)$ are 
two suitable matching functions.

For massive quarks the threshold limit is
$x\to x_{\rm max}$ with $x_{\rm max} = \frac{Q^2}{4m_h^2 + Q^2}$ or
$\beta \to 0$, with
$\beta\equiv \sqrt{1- \frac{4m_h^2}{s}}$ and $s=Q^2\frac{1-x}{x}$
the center-of-mass energy of the partonic cross-section.
In this limit, the coefficient
function contains logarithmically enhanced terms of the form
$\alpha_s^n \ln^m\beta$ with $m\le 2n$ due to soft gluon emission, which are
predicted by threshold resummation~\cite{Bonciani:1998vc}. 
Further contributions of the form $\alpha_s^n \beta^{-m}\ln^l\beta$, 
with $m\le n$, arise from Coulomb exchange between the heavy quark and 
antiquark, and can also be resummed using  non-relativistic QCD
methods~\cite{Pineda:2006ri}. 
At N$^3$LO all these contributions are
known and can be extracted from available resummed
results~\cite{Kawamura:2012cr}; they are included in
$C_{i,k}^{(3),{\rm thr}}$.

In the high-energy limit, the coefficient function contains logarithmically
enhanced terms of the form 
$\alpha_s^n \ln^m x$ with $m\le n-2$, which are determined at all
orders through small-$x$ resummation at the LL
level~\cite{Catani:1990eg}, from which the N$^3$LO expansion can be
extracted~\cite{Kawamura:2012cr}. This result can be further
improved~\cite{Laurenti:2024anf,bbl2023} by including a particular
class of NLL terms related to NLL perturbative evolution and the
running of the coupling. 
In the approach of Refs.~\cite{Laurenti:2024anf,bbl2023} the high-energy contributions
are combined into $C_{i,k}^{(3),{\rm asy}}$ with 
the asymptotic $Q^2 \gg m_h^2$ limit of the coefficient function  
in the decoupling scheme~\cite{
  Bierenbaum:2009mv,Ablinger:2010ty,
  Ablinger:2014vwa,Ablinger:2014nga,Behring:2014eya},
while subtracting overlap terms. 
This ensures that in the $Q^2 \gg m_h^2$ limit, the structure function,
computed from $C_{i,k}^{(3),{\rm asy}}$
combined with decoupling-scheme PDFs, coincides with the structure function
computed in the limit in which the heavy quark mass is
neglected and the heavy quark is treated as a massless parton.
However, by the time of writing, the asymptotic limit could
only be determined approximately since in particular some of the
matching conditions were not fully known. 

The interpolating functions, used to combine the two contributions in
\cref{eq:massive_cf}, are chosen to satisfy the requirements 
\begin{allowdisplaybreaks} 
\begin{align}
\begin{split}
    f_1(x) \xrightarrow[x \to 0]{} 0, & \quad  f_1(x) \xrightarrow[x \to x_{\rm max}]{} 1 \, , \\
    f_2(x) \xrightarrow[x \to 0]{} 1, & \quad  f_1(x) \xrightarrow[x \to x_{\rm max}]{} 0 \, , 
    \label{eq:dampingfunctions}
\end{split}
\end{align} 
\end{allowdisplaybreaks}
which ensure that the threshold contribution vanishes in the small-$x$
limit and conversely. This guarantees that the approximation
\cref{eq:massive_cf} is  reliable in a broad kinematic range in
the $(x,Q^2)$ plane:  $C_{i,k}^{(3),{\rm asy}}$
reproduces the massless limit for large $Q^2$ values and for all values of $x$,
including
the small-$x$ limit, while  $C_{i,k}^{(3),{\rm thr}}$ describes the threshold limit,
with $x$ close to $x_{\rm max}$. 
An uncertainty on the approximate coefficient
function can be constructed varying the functional form of the
interpolating functions, as well as that of terms which are not fully
known. This includes the NLL small-$x$ resummation and the matching
functions that enter the asymptotic high $Q^2$ limit.
This uncertainty vanishes in the $x\to x_{\rm max}$ limit, for which
the exact known limit is reproduced  (with a fixed choice for the
unknown constant $\beta$-independent terms), and becomes larger in the
intermediate $\eta$ region. The interpolating functions and their
uncertainties are optimized by using the same methodology at NNLO,
where the full result is known.
We refer to Ref.~\cite{Laurenti:2024anf,bbl2023} for a detailed discussion of this
construction. 

This optimized approximation is shown at NNLO in
\cref{fig:massive_nnlo_bench}, where  we compare it to
the  exact result for the
massive gluon-initiated coefficient function $x C_{2,g}^{(2)}(\eta)$,
expressed in terms of the variable
\begin{equation}
  \label{eq:etadef}
  \eta= \frac{Q^2(1-x)}{4m_h^2x}-1 \, .
\end{equation}
Results are shown for two different values of the $Q^2/m_h^2$ ratio,
close to threshold and at higher scales.
Note that 
 $\eta \to 0$ corresponds to $x\to x_{\rm max}$ (threshold limit),
while $\eta \to \infty$ corresponds to either $Q^2/m_h^2 \to \infty$ for fixed $x$ (asymptotic limit),
or $x\to 0$ for fixed $Q^2$ (high-energy limit). In this case the
uncertainty band is obtained by varying the interpolating functions only.

The results found using the same procedure for the gluon and quark singlet
coefficient functions at aN$^3$LO are displayed in
\cref{fig:massive_n3lo_bench}, compared to the
approximation of Ref.~\cite{Kawamura:2012cr}, each shown with the
respective uncertainty estimate.
Good agreement between the different approximations is found,
especially for  the dominant  gluon 
coefficient function. The approximations agree in the asymptotic
$\eta\to 0$ and $\eta\to \infty$ limits and in most of the $\eta$
range,  but differ somewhat in the sub-asymptotic large $\eta$ region
at fixed $Q^2$, which corresponds to the small $x$ limit at fixed
$Q^2$. These differences can be traced to the aforementioned inclusion
in the procedure of Ref.~\cite{Laurenti:2024anf,bbl2023} of a  particular
class of NLL terms related.

The uncertainty involved in the approximation can be included as a
further IHOU, alongside that discussed in \cref{sec:an3lo_ihou},
through an additional contribution to the theory  
covariance matrix. Namely, we define
\begin{equation}
  \label{eq:covihouC}
  \text{cov}^{C}_{mn} = \frac{1}{2} \left(\Delta_m(+)\Delta_n(+)+\Delta_m(-)\Delta_n(-)\right).
\end{equation}

Here $\Delta_m(\pm)$ is the shift in the prediction for the $m$-th
DIS data point obtained by replacing the central approximation to the
massive coefficient function with the upper or lower edge of the
uncertainty range determined in Ref.~\cite{Laurenti:2024anf} and shown
as an uncertainty band in \cref{fig:massive_n3lo_bench}. Note that
unlike in \cref{eq:covihou}, we divide by the number of
independent variations, without decreasing it by one, because the
central value is not the average of the variations, and thus is independent.
The contribution \cref{eq:covihouC} is then added to the IHOU covariance
matrix as a further term on the right-hand side of \cref{eq:covihou}.

The impact of this contribution to the IHOU is assessed in
\cref{fig:dis_ihou}, where 
the square root of the diagonal component of the covariance matrix is shown 
for all the DIS data points in our dataset, comparing the IHOU before
and after adding to \cref{eq:covihou} the extra component
\cref{eq:covihouC} due to
the IHOU on the massive coefficient function. It is clear that the
impact of IHOUs due to perturbative evolution is generally negligible,
in agreement with the results discussed in \cref{sec:an3lo_evol}
and shown in \cref{fig:N3LOevolution-q100gev-ratios}: IHOUs on
splitting functions  are
only significant at small $x$, but available small-$x$ data are at
relatively low scale where the evolution length is small. The impact
of IHOUs on massive coefficient functions is relevant for data on
tagged bottom and charm structure functions, but otherwise moderate
and only significant for structure function data close to the heavy
quark production thresholds. 

\subsection{A general-mass variable flavor number scheme at N$^3$LO}
\label{sec:an3lo_fonll_d}

\begin{figure}[!t]
    \centering
    \includegraphics[width=\textwidth]{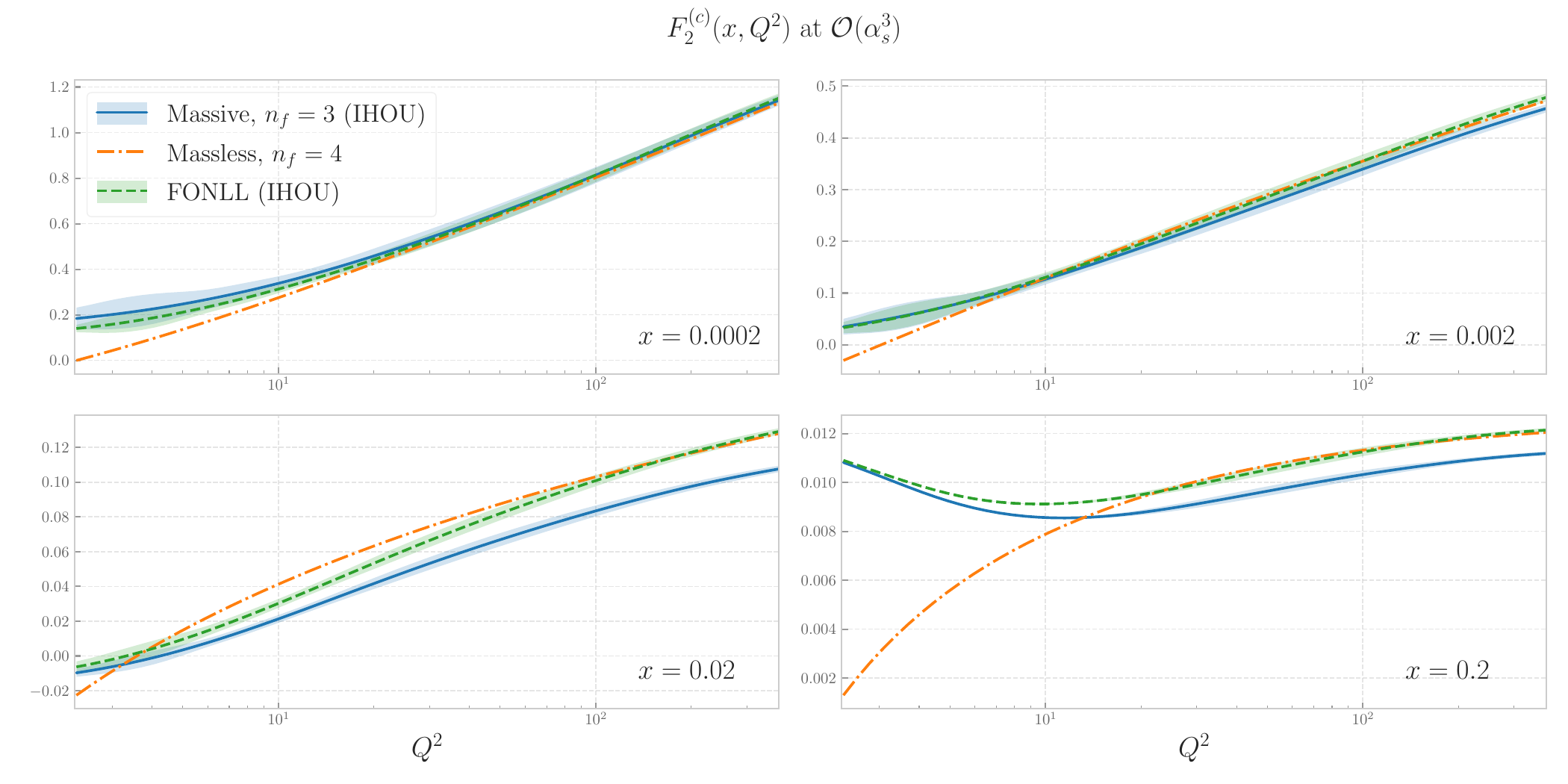}
    \caption{
      The charm structure function $F^{(c)}_{2}(x,Q^2,m_c^2)$ in the
      FONLL-E scheme, compared to the massive and massless scheme
      results (see text). Results are shown as a function of  $Q^2$
      for $x=2\times 10^{-4}$ (top left), $x=2\times 10^{-3}$ (top right),  
      $x=2\times 10^{-2}$ (bottom left), and $x=2\times 10^{-1}$ (bottom
      right). The uncertainty shown on the FONLL and massive curves is
      the IHOU on the heavy quark coefficient functions \cref{eq:covihouC}.}
    \label{fig:f2_charm_n3lo} 
\end{figure}
\begin{figure}[!t]
    \centering
    \includegraphics[width=\textwidth]{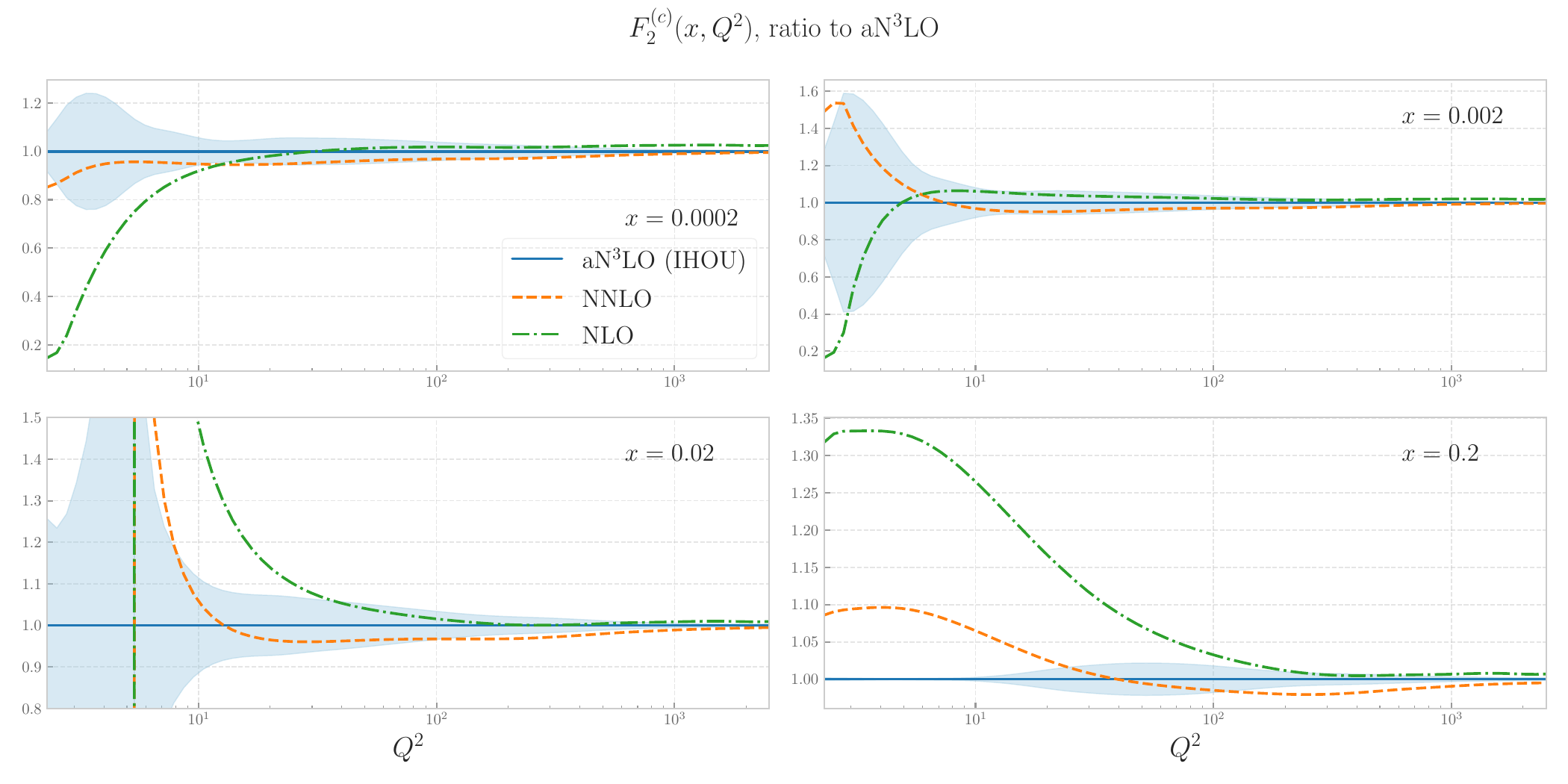}
    \caption{Same as \cref{fig:f2_charm_n3lo}, now comparing  the
      FONLL-A (used at NLO $\mathcal{O}\lp \alpha_s\rp$),
      FONLL-C (used at NNLO $\mathcal{O}\lp \alpha_s^2\rp$), 
      and FONLL-E (used at N$^3$LO $\mathcal{O}\lp \alpha_s^3\rp$), all
      shown as a ratio to FONLL-E. The FONLL-E result
      includes the IHOU on the heavy quark coefficient functions
      \cref{eq:covihouC}. 
    }
    \label{fig:f2_charm_pto} 
\end{figure}

The N$^3$LO DIS coefficients functions described in the previous 
section enable the extension to $\mathcal{O}\lp \alpha_s^3\rp$ 
of the FONLL general-mass variable flavor number scheme for DIS,
as discussed in \cref{sec:fns}.

The FONLL prescription of \cref{eq:FONLL_m1} was implemented in Ref.~\cite{Forte:2010ta,Ball:2015tna} 
for DIS to NNLO, by expressing all terms on the right-hand side
in terms of $\alpha_s$ and PDFs all defined in the massless scheme. 
This has the advantage of providing an expression that can used 
with externally provided PDFs, that are typically available only 
in a single factorization scheme for each value of the scale $Q$.
However, the recent \eko{} code (\cref{sec:eko}) 
allows, at any given scale, the coexistence of PDFs defined in
schemes with a different number of massless flavors. 
Furthermore, the recent \yadism{} program (\cref{sec:yadism}) 
implements DIS coefficient functions corresponding to all three contributions 
of the right-hand side of \cref{eq:FONLL_m1}.
It is then possible to implement the FONLL prescription
\cref{eq:FONLL_m1} by simply combining expressions
computed in different schemes~\cite{Barontini:2024xgu}. 
This formalism is especially advantageous at higher perturbative orders, 
where the analytic expressions relating PDFs in different scheme grow in 
complexity.

In the FONLL method, \cref{eq:FONLL_m1}, the first two terms
on the right-hand side may be computed at different perturbative
orders, provided one ensures that the third term correctly
includes only their common contributions. In Ref.~\cite{Forte:2010ta} 
some natural choices were discussed, based on the observation that in 
the massive scheme, the heavy quark contributes to the structure functions 
only at $\mathcal{O}\lp \alpha_s\rp$ and beyond, 
while in the massless scheme it already contributes at $\mathcal{O}\lp \alpha_s^0\rp$. 
Hence, natural choices are to combine both the massive and
massless contributions at $\mathcal{O}\lp \alpha_s\rp$ (FONLL-A), or
else the massive contribution at
$\mathcal{O}\lp \alpha_s^2\rp$ and the massless 
contribution at $\mathcal{O}\lp \alpha_s\rp$, i.e.\ both at
second nontrivial order (FONLL-B). The corresponding two options at the next
order are called FONLL-C and -D.

Here, we will consider FONLL-E, in which both the massless and
massive contributions are determined at $\mathcal{O}\lp\alpha_s^3\rp$. 
The charm structure function $F^{(c)}_{2}(x,Q^2)$, computed in this scheme,
is displayed in \cref{fig:f2_charm_n3lo} as a function of $Q^2$
for four values of $x$ (with $m_c=1.51$~GeV), and compared to the
massive and massless scheme results, with the IHOU on the massive
coefficient function shown for the first two cases. The structure
functions are computed using the NNPDF4.0 aN$^3$LO PDF set (to be
discussed in \cref{sec:an3lo_results} below) which satisfies aN$^3$LO evolution
equations, as is necessary for consistency with the massless scheme
result at high scale.
It is clear that the FONLL results interpolate between the massive and
massless calculations as the scale grows. 
The $Q^2$ value at which either of the massive or massless results
dominate depend strongly on $x$.
Except for the lowest $Q^2$ values, the IHOUs associated with the 
calculation remain moderate.

The perturbative convergence of the charm structure function is assessed in
\cref{fig:f2_charm_pto}, where we compare the FONLL-A, FONLL-C and
FONLL-E results, all shown as a ratio to FONLL-E, the latter also
including the IHOU 
as in \cref{fig:f2_charm_n3lo}.
Clearly, convergence is faster at higher scales due to asymptotic
freedom, and it appears that the perturbative expansion has
essentially converged for $Q^2\gtrsim 10$~GeV$^2$. On the other hand, the
impact of aN$^3$LO at low scale is sizable, up to $50~\%$ for small
$Q^2$ and $x = 2\times 10^{-3}$. The IHOUs are correspondingly
sizable at low scale, and in fact always larger than the difference
between the NNLO and aN$^3$LO results except at the highest $x$
values and the lowest scales, implying that for the charm
structure function aN$^3$LO may be more accurate, but possibly 
not more precise than NNLO.

\begin{figure}[!t]
  \centering
  \includegraphics[width=\textwidth]{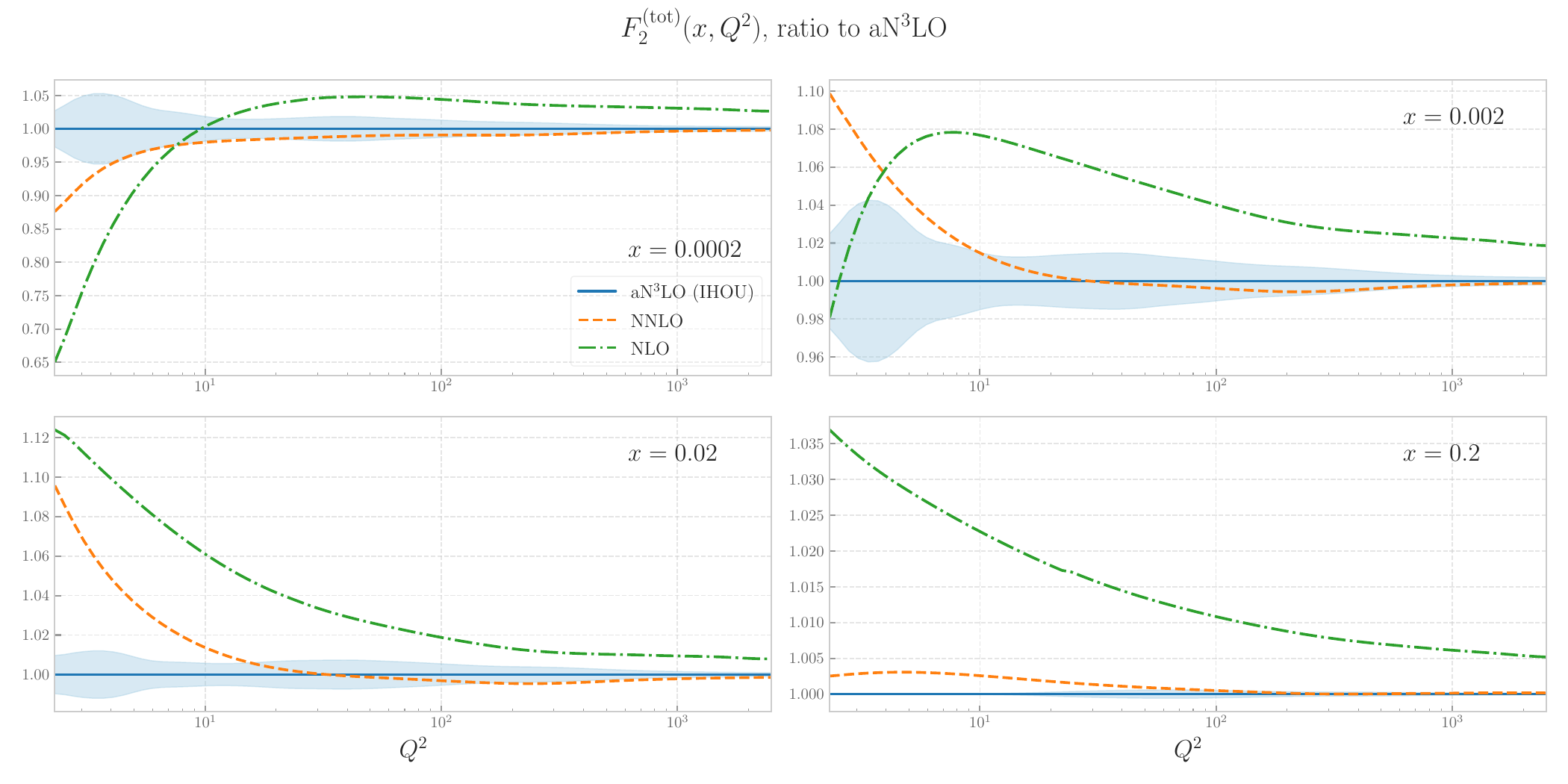}
  \caption{Same as \cref{fig:f2_charm_pto} for the inclusive structure
    function $F^{(\text{tot})}_{2}(x,Q^2)$. Note the different scale
    on the $y$ axis.
   }
  \label{fig:f2_total_pto} 
\end{figure}

An analogous study of perturbative convergence of the 
inclusive structure function is shown in \cref{fig:f2_total_pto}
(note the different scale on the $y$ axis). 
Interestingly, the effect of the aN$^3$LO corrections changes sign when going 
from $x=2\times 10^{-4}$ to larger values of $x$.
In general, N$^3$LO corrections are smaller at the inclusive level: specifically,
aN$^3$LO corrections to the inclusive structure function are below 
$2~\%$ for $Q^2\gsim 10$~GeV$^{2}$, and at most of the order $10~\%$ around the
charm mass scale. 
The impact of the IHOUs on the heavy coefficient is further reduced due to the fact 
that charm contributes at most one quarter of the total structure function. 
Consequently, the aN$^3$LO correction to the NNLO result is now larger than the IHOU 
in a significant kinematics region. 
This, together with the fact that aN$^3$LO corrections are comparable
or larger than typical experimental uncertainties on structure function data,
motivates their inclusion in a global PDF determination.

\subsection{N$^3$LO corrections to hadronic processes}
\label{sec:an3lo_hadronic_coeff}


N$^3$LO corrections to the total cross-section for inclusive NC and CC DY 
production~\cite{Duhr:2020sdp,Duhr:2021vwj} are available 
through the {\sc\small n3loxs} public code~\cite{Baglio:2022wzu}, both
for on-shell $W$ and $Z$ and as a function of the dilepton invariant
mass $m_{\ell \ell}$.
Differential distributions at the level of leptonic observables for
the same processes have also been computed~\cite{Chen:2021vtu,Chen:2022lwc},
but are not publicly available.
No N$^3$LO calculations are available for other processes included in
the NNPDF4.0 dataset.


Total cross-section data are obtained by extrapolating measurements
performed in a fiducial region. Whereas for NC DY production
in the central rapidity region and for dilepton invariant masses
around the $Z$-peak, the N$^3$LO/NNLO cross-section ratio
depends only mildly on the dilepton rapidity $y_{\ell\ell}$~\cite{Chen:2021vtu,Chen:2022lwc},
it is unclear whether this is the case also off-peak or at very large
and very small rapidities. Hence, the inclusion of N$^3$LO corrections for
hadronic processes is, at present, not fully reliable. 
We have consequently not included them in our default determination, but only
in a dedicated variant, with the goal of assessing their impact. 
For conciseness, we refer to Ref.~\cite{NNPDF:2024nan} for its discussion.

Despite the fact that we are not yet able to determine reliably 
N$^3$LO corrections for currently available LHC measurements, 
we wish to include the full NNPDF4.0 dataset in our aN$^3$LO PDF
determination. 
To this purpose, we endow all data for which N$^3$LO
are not included with an extra uncertainty that accounts for the
missing N$^3$LO terms. This is estimated using the methodology of
Refs.~\cite{NNPDF:2019vjt,NNPDF:2019ubu}, recently used in
Ref.~\cite{NNPDF:2024dpb} to produce a variant of the NNPDF4.0 PDF
sets that includes MHOUs, summarized in \cref{sec:nnpdf_methodology}. 

Thus, when not including N$^3$LO corrections to the hard
cross-section, the theory prediction is evaluated by combining
aN$^3$LO evolution with the NNLO cross-sections. 
The prediction is then supplemented with a theory covariance matrix,
computed varying the renormalization scale $\mu_R$ using a three-point
prescription~\cite{NNPDF:2019vjt,NNPDF:2019ubu}:
\begin{equation}
  \label{eq:covihouNNLO}
  \text{cov}^{\rm NNLO}_{mn} = \frac{1}{2} \left(\Delta_m(+)\Delta_n(+)+\Delta_m(-)\Delta_n(-)\right),
\end{equation}
analogous to \cref{eq:covihouC}, but now with $\Delta_m(\pm)$ the
shift in the prediction for the $m$-th data point obtained by
replacing the coefficient functions with those obtained by performing
upper or lower renormalization scale variation using the methodology 
of Ref.~\cite{NNPDF:2019ubu} (as implemented and
discussed in \cite[Eq. 2.9]{NNPDF:2024dpb}).
This MHOU covariance matrix is then added to the IHOU covariance matrix 
as a further term on the right-hand side of \cref{eq:covihou}.

\begin{figure}[!t]
  \centering
  \includegraphics[width=.7\textwidth]{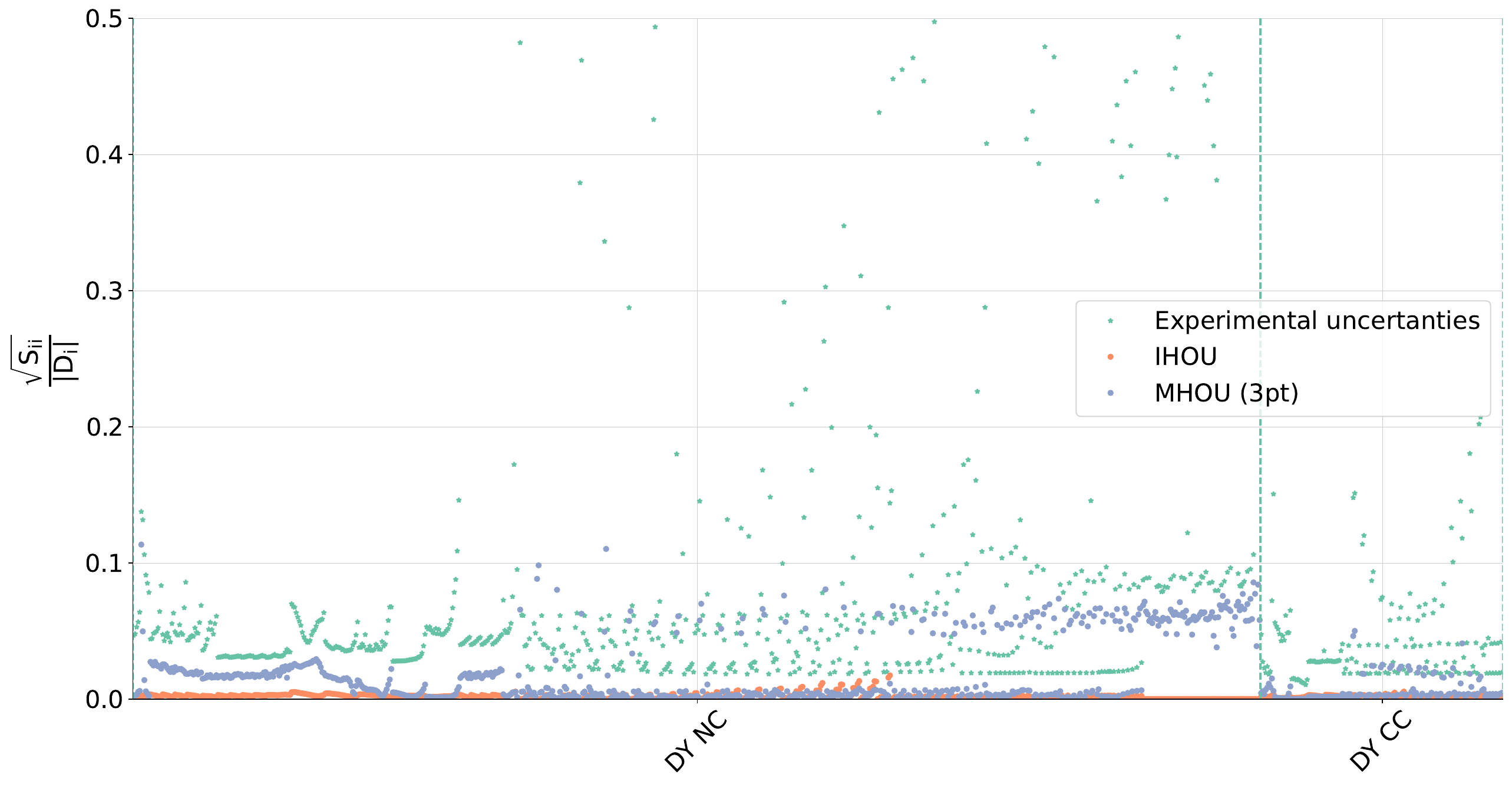} \\
  \includegraphics[width=.7\textwidth]{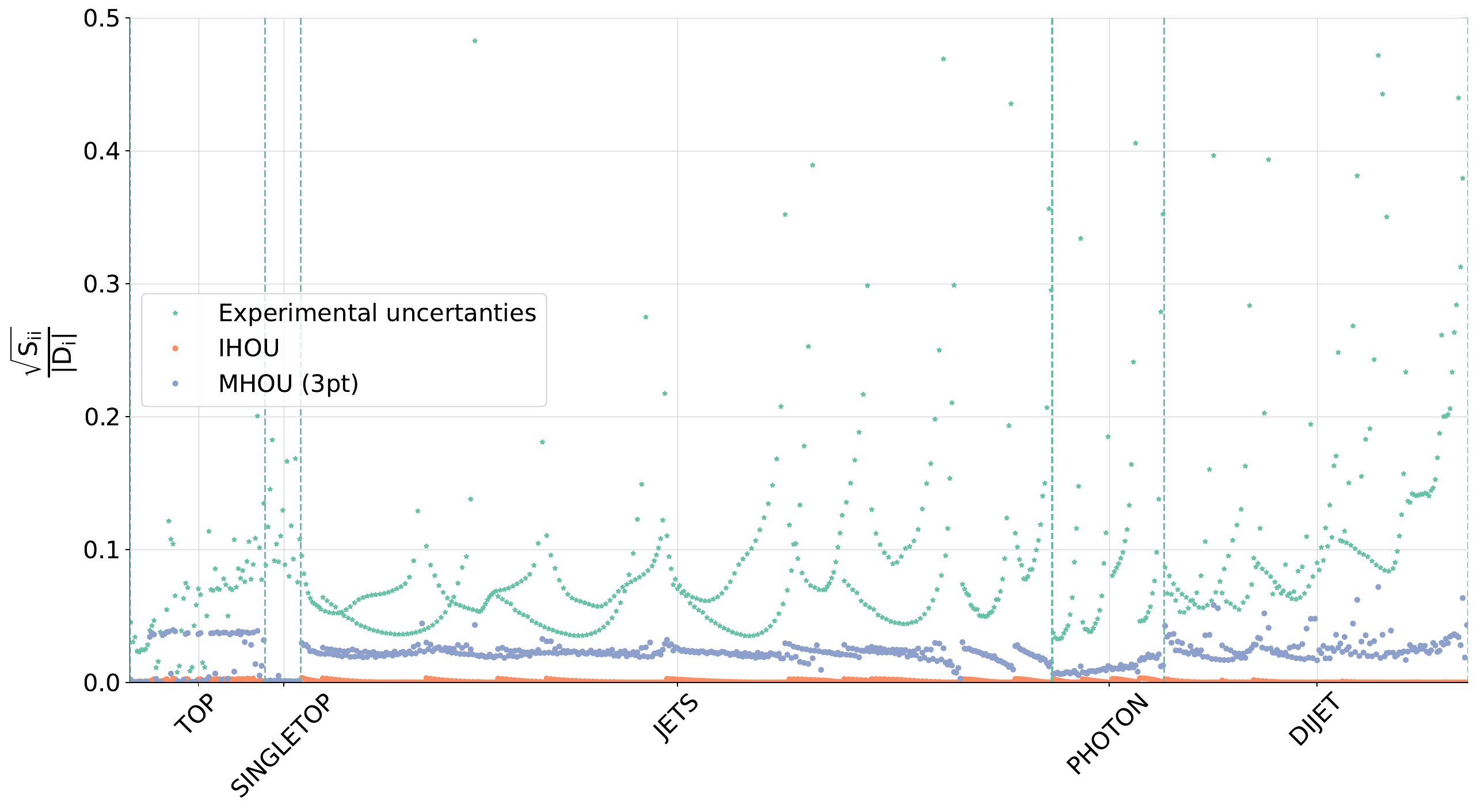}
  \caption{Same as \cref{fig:dis_ihou}
    now comparing the IHOU from
    \cref{eq:covihou} with the MHOU from
    \cref{eq:covihouNNLO} due to the missing N$^3$LO correction
    to the matrix element. Results are shown for all hadronic data in
    the NNPDF4.0 dataset: specifically DY (top) and top pair,
    single top, single-inclusive jet, prompt photon and dijet
    production (bottom). 
  }    
  \label{fig:hadronic_mhou}
 \end{figure}

The impact of this uncertainty is shown in \cref{fig:hadronic_mhou}, 
where we show for all hadronic datasets the square root of the 
diagonal entries of the MHOU covariance matrix \cref{eq:covihouNNLO}, 
compared to those of the IHOU covariance matrix \cref{eq:covihou},
and to the experimental uncertainties, all normalized to the central
theory prediction. The MHOU is generally larger than the IHOU,
indicating that the missing N$^3$LO terms in the hard cross-sections are
larger than the IHOU uncertainty in N$^3$LO perturbative evolution.
The experimental uncertainties are generally larger still.

In addition to the NNPDF4.0 aN$^3$LO baseline PDF set obtained in this
manner, we will also produce a NNPDF4.0 MHOU aN$^3$LO set, in analogy
to the NLO and NNLO MHOU. For this set, MHOUs on both perturbative
evolution and on the hard matrix elements are included using the
methodology of Refs.~\cite{NNPDF:2019vjt,NNPDF:2019ubu} with a theory
covariance matrix determined performing combined correlated
renormalization and factorization scale variations with a 7-point
prescription, as discussed in \cref{sec:nnpdf_methodology}
and with more details in Ref.~\cite{NNPDF:2024dpb}. In this case, we simply perform scale
variation on the expressions at the order at which they are being
computed, namely aN$^3$LO for 
anomalous dimensions and DIS coefficient functions and NNLO for
hadronic processes. The scale variation then is automatically larger
and suitable deweights processes for which N$^3$LO corrections are
not available. The possibility of simultaneously including in a PDF
determination processes for which theory predictions are only
available at different perturbative orders is an advantage of the
inclusion of MHOUs in the PDF determination, as already pointed out
in Refs.~\cite{Faura:2020oom,Amoroso:2022eow}.

\section{NNPDF4.0 at aN$^3$LO}
\label{sec:an3lo_results}
We now present the aN$^3$LO NNPDF4.0 PDF sets.
They have been obtained by using the dataset and methodology discussed
in~\cite{NNPDF:2021njg} and reviewed in \cref{sec:nnpdf_methodology}.
The aN$^3$LO results are obtained using the approximate N$^3$LO splitting
functions of \cref{sec:an3lo_dglap}, the exact massless and approximate massive
N$^3$LO coefficient functions of \cref{sec:an3lo_dis}, and NNLO
hadronic cross-sections supplemented by an extra uncertainty as per
\cref{sec:an3lo_hadronic_coeff}. 

Theoretical predictions are obtained using
the theory pipeline described in \cref{sec:theory_methodology}. 
As discussed in \cref{sec:an3lo_fonll_d}, this pipeline in particular 
includes a new FONLL implementation, that differs from the previous one by subleading
terms. A further small difference in comparison to Ref.~\cite{NNPDF:2021njg}
is the correction of a few minor bugs in the data implementation.
The overall impact of all these changes was assessed in Appendix~A of
Ref.~\cite{NNPDF:2024djq}, and was found to be very limited, so that
the new and old implementations can be considered equivalent,
and the PDF sets presented here can be considered the extension
to aN$^3$LO of the NNPDF4.0 PDF sets of Ref.~\cite{NNPDF:2021njg}.

In addition to the default NNPDF4.0 aN$^3$LO PDF determination, we also
present an aN$^3$LO PDF determination that includes MHOUs on all the
theory predictions used in the PDF determination. This is constructed
using the same methodology recently used to produce the
NNPDF4.0MHOU NNLO PDF set in Ref.~\cite{NNPDF:2024dpb} (see also \cref{sec:nnpdf40_pdfs}). In order to be
able to discuss perturbative convergence and the impact of MHOUs we
will also present a NNPDF4.0MHOU NLO PDF set constructed using the
same methodology, and exactly the same dataset as the default NNPDF4.0
NLO PDF set (which differs from the NNPDF4.0 NNLO dataset).


We first assess the fit quality, then present the PDFs and their
uncertainties, and study perturbative convergence and the
effect on it of the inclusion of MHOUs. We then specifically study the
impact of aN$^3$LO corrections on intrinsic charm. 
The comparison of our results to the recent MSHT20 aN$^3$LO PDFs~\cite{McGowan:2022nag}
is reported in \cref{app:an3lo_msht20}.

\subsection{Fit quality}
\label{sec:an3lo_fit_settings}

\begin{figure}[!t]
  \centering    
  \includegraphics[width=.60\textwidth]{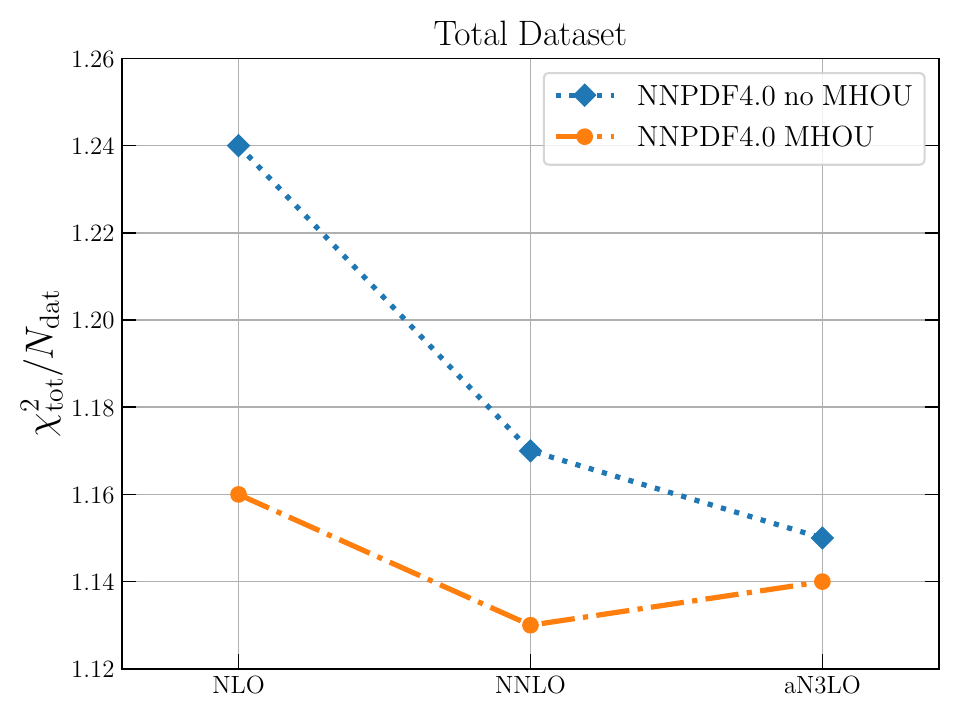}
  \caption{The values of the total $\chi^2$ per data point in the
    NNPDF4.0 NLO, NNLO, and aN$^3$LO fits without and with MHOUs.}
  \label{fig:chi2_n3lo_summary} 
\end{figure}

\cref{tab:chi2_TOTAL} display the number
of data points and the $\chi^2$ per data point obtained in the NLO, NNLO, and
aN$^3$LO NNPDF4.0 fits with and without MHOUs. In \cref{tab:chi2_TOTAL}
the datasets are grouped according to the process categorization used
in Ref.~\cite{NNPDF:2024dpb}. \footnote{
  Results for individual datasets are reported in \cite[Section~4.1.]{NNPDF:2024nan}, 
  
} 
The value of the total $\chi^2$ per data point is also shown as a function of the perturbative order in
\cref{fig:chi2_n3lo_summary}. 

\begin{table}[!t]
  \scriptsize
  \centering
  \renewcommand{\arraystretch}{1.4}
  \begin{tabularx}{\textwidth}{Xrccrccrcc}
  \toprule
  & \multicolumn{3}{c}{NLO}
  & \multicolumn{3}{c}{NNLO}
  & \multicolumn{3}{c}{aN$^3$LO} \\
  Dataset
  & $N_{\rm dat}$
  & no MHOU
  & MHOU
  & $N_{\rm dat}$
  & no MHOU
  & MHOU 
  & $N_{\rm dat}$
  & no MHOU
  & MHOU \\
  \midrule
  DIS NC
  & 1980 & 1.30 & 1.22
  & 2100 & 1.22 & 1.20
  & 2100 & 1.22 & 1.20 \\
  DIS CC
  &  988 & 0.92 & 0.87
  &  989 & 0.90 & 0.90 
  &  989 & 0.91 & 0.92 \\
  DY NC
  &  667 & 1.49 & 1.32
  &  736 & 1.20 & 1.15
  &  736 & 1.17 & 1.16 \\
  DY CC
  &  193 & 1.31 & 1.27
  &  157 & 1.45 & 1.37
  &  157 & 1.37 & 1.36 \\
  Top pairs
  &   64 & 1.90 & 1.24
  &   64 & 1.27 & 1.43
  &   64 & 1.23 & 1.41 \\
  Single-inclusive jets
  &  356 & 0.86 & 0.82
  &  356 & 0.94 & 0.81
  &  356 & 0.84 & 0.83 \\
  Dijet
  &  144 & 1.55 & 1.81
  &  144 & 2.01 & 1.71
  &  144 & 1.78 & 1.67 \\
  Prompt photons 
  &   53 & 0.58 & 0.47
  &   53 & 0.76 & 0.67
  &   53 & 0.72 & 0.68 \\
  Single top
  &   17 & 0.35 & 0.34
  &   17 & 0.36 & 0.38
  &   17 & 0.35 & 0.36 \\
  \midrule
  Total
  & 4462 & 1.24 & 1.16
  & 4616 & 1.17 & 1.13
  & 4616 & 1.15 & 1.14 \\
\bottomrule
\end{tabularx}

  \vspace{0.3cm}
  \caption{The number of data points and the $\chi^2$ per data point obtained
    in the NLO, NNLO, and aN$^3$LO NNPDF4.0 fits without and with MHOUs,
    see text for details. The datasets are grouped according to
    the same process categorization as that used in Ref.~\cite{NNPDF:2024dpb}.}
  \label{tab:chi2_TOTAL}
\end{table}

The NLO and NNLO results without MHOUs are obtained using the NLO and NNLO
NNPDF4.0 PDF sets~\cite{NNPDF:2021njg}. The NNLO result with MHOUs is
obtained using the NNPDF4.0MHOU NNLO set from Ref.~\cite{NNPDF:2024dpb},
while, as already mentioned, the NNPDF4.0MHOU NLO presented here for
the first time uses an identical methodology to NNPDF4.0MHOU
NNLO~\cite{NNPDF:2024dpb}, but the same dataset as NNPDF4.0
NLO~\cite{NNPDF:2021njg}. Hence, the datasets with and without MHOU are always
the same, but the NLO and NNLO datasets are not the same but rather follow
Ref.~\cite{NNPDF:2021njg}. The N$^3$LO dataset is the same as NNLO. 
The covariance matrix, whenever needed, is computed as described in
\cref{sec:nnpdf_methodology}, see also \cite[Section~4.1]{NNPDF:2024dpb} for more details. 

The N$^3$LO predictions are based on the same datasets and
kinematic cuts as the NNPDF4.0 NNLO PDF sets, use the theoretical predictions
discussed in \cref{sec:an3lo_dglap,sec:an3lo_coefffun}, and are
supplemented with a IHOU covariance matrix as discussed in
\cref{sec:an3lo_ihou,sec:an3lo_dis} and a MHOU for hadronic processes for
which N$^3$LO hard cross-sections are not available as discussed in
\cref{sec:an3lo_hadronic_coeff}.

\cref{tab:chi2_TOTAL} and \cref{fig:chi2_n3lo_summary}
show that without MHOUs fit quality improves as the perturbative
order increases. Note that this also happens when going from NNLO to
N$^3$LO, despite the fact that N$^3$LO corrections are only partially included,
with hadronic matrix elements still computed at NNLO. The latter indicates that
the impact of N$^3$LO corrections to evolution and DIS coefficient
functions is significant enough to affect fit quality in a way that is
qualitatively compatible with what one would expect when adding an
extra perturbative order.

On the other hand, when MHOUs are included, fit quality
becomes independent of perturbative order within uncertainties.
(note that, with $N_{\rm dat}=4462$, $\sigma_{\chi^2}=\sqrt{2n_{\rm dat}}=0.03$). 
This suggests that the MHOU covariance matrix estimated through scale variation is
correctly reproducing the observed shift between perturbative orders,
i.e. the true MHOU. Note that if true this also means that at
aN$^3$LO the missing N$^3$LO corrections to hadronic processes are
correctly accounted for by the corresponding MHOU which is always
included. Also, at aN$^3$LO
the fit quality is the same within uncertainties irrespective of whether MHOUs 
are included or not. This strongly suggests that inclusion of higher
order terms in perturbative evolution and DIS coefficient function
would not lead to further improvements, i.e. that in this respect, with
experimental uncertainties, current methodology and current dataset
the perturbative expansion has converged.

\subsection{Parton distributions}
\label{sec:an3lo_PDFs}

\begin{figure}[!p]
  \centering
  \includegraphics[width=0.43\textwidth]{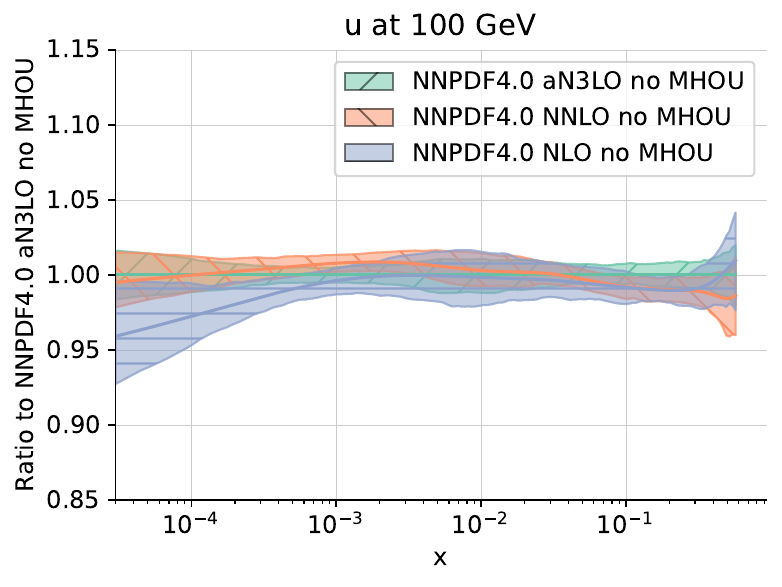}
  \includegraphics[width=0.43\textwidth]{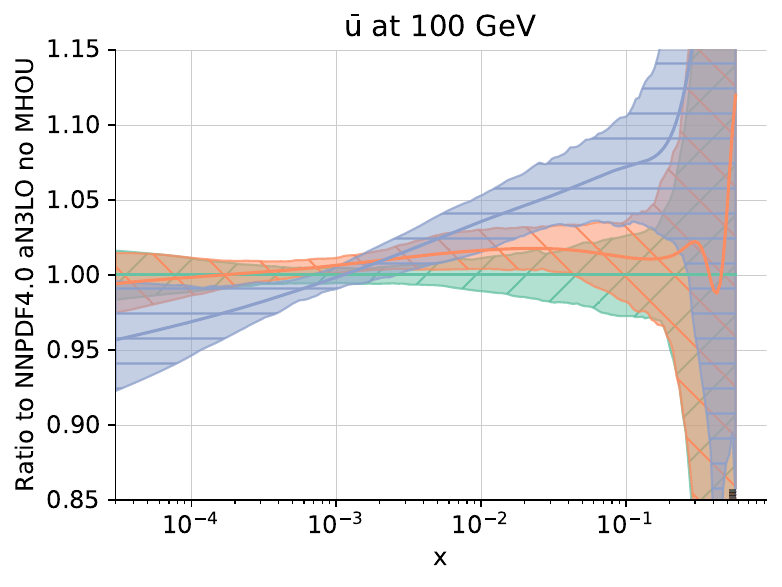} \\
  \includegraphics[width=0.43\textwidth]{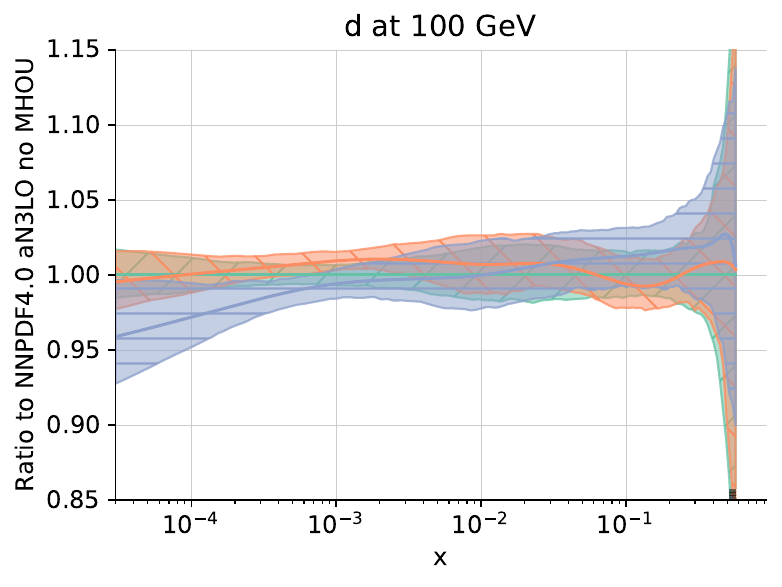}
  \includegraphics[width=0.43\textwidth]{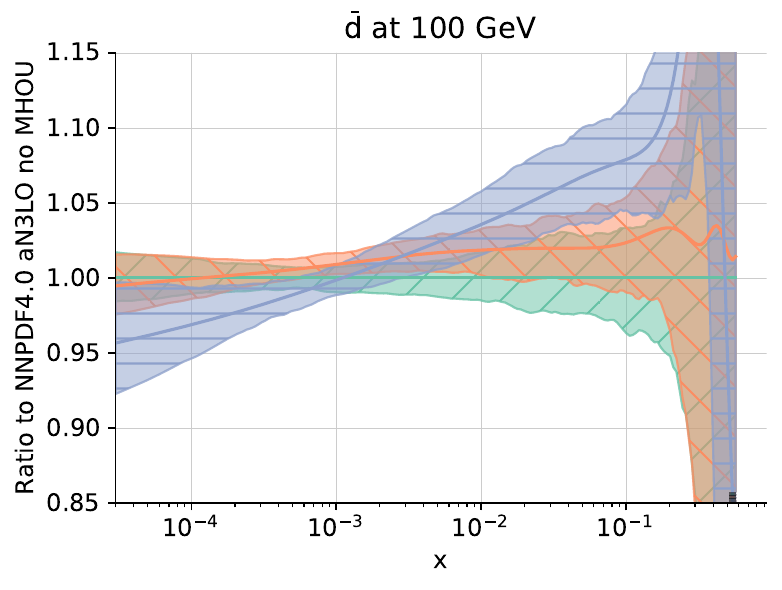} \\
  \includegraphics[width=0.43\textwidth]{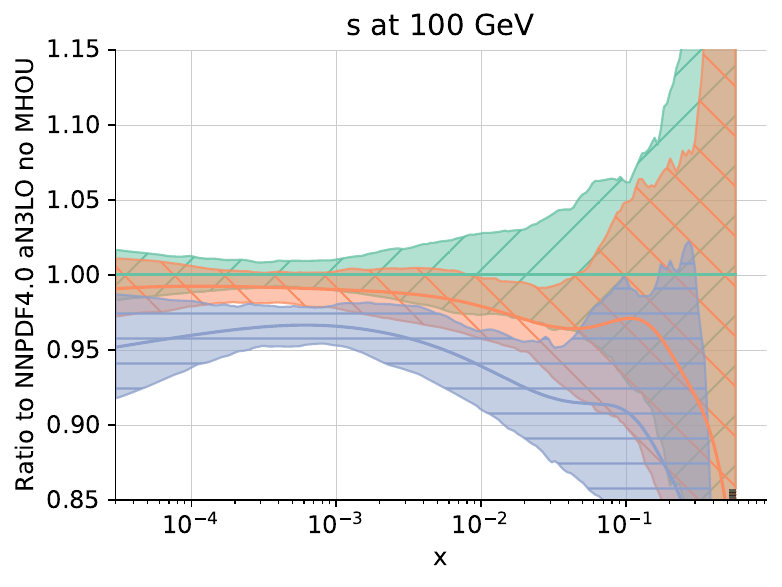}
  \includegraphics[width=0.43\textwidth]{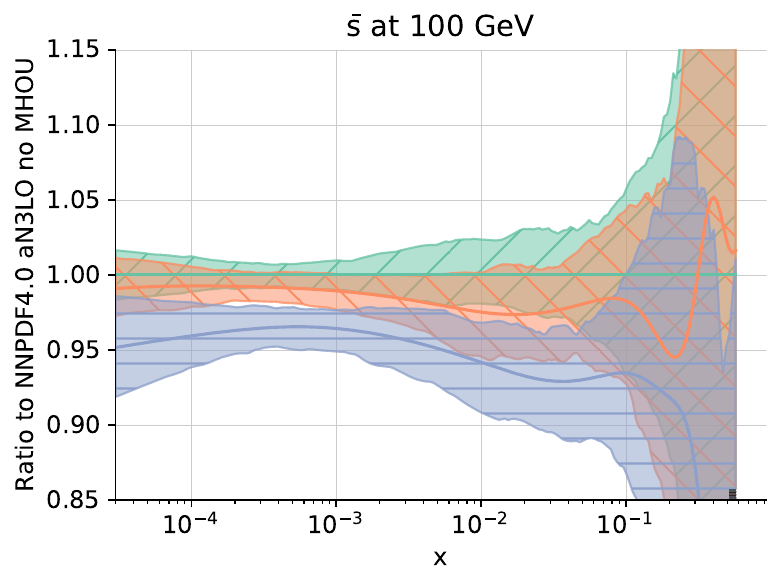} \\
  \includegraphics[width=0.43\textwidth]{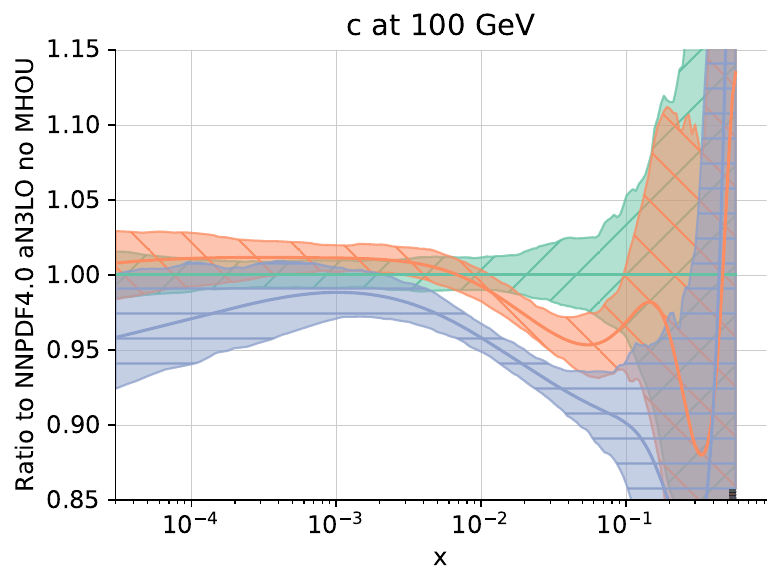}
  \includegraphics[width=0.43\textwidth]{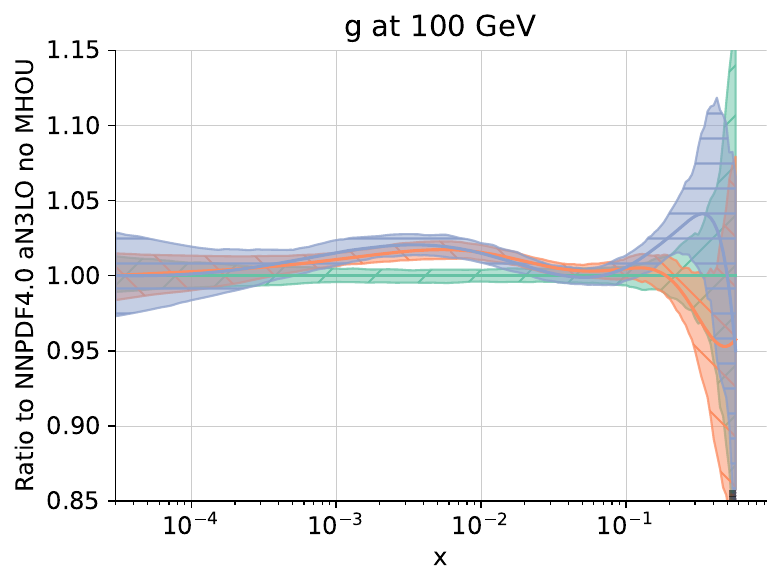}
  \caption{The NLO, NNLO and aN$^3$LO NNPDF4.0 PDFs at $Q=100$~GeV. We display
    the up, antiup, down, antidown, strange, antistrange, charm and gluon PDFs
    normalized to the aN$^3$LO result. Error bands correspond to one sigma
    PDF uncertainties, not including MHOUs on the theory predictions
    used in the fit.}
  \label{fig:pdfs_noMHOU_log}
\end{figure}

\begin{figure}[!p]
  \centering
  \includegraphics[width=0.45\textwidth]{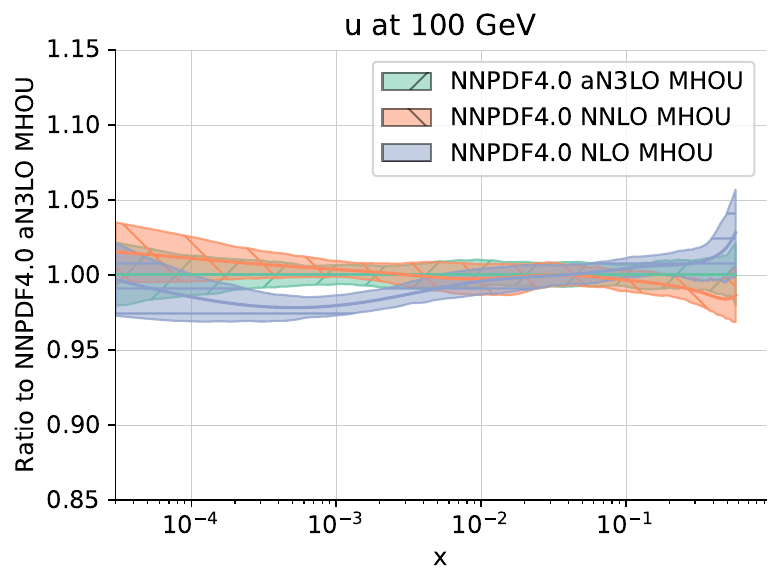}
  \includegraphics[width=0.45\textwidth]{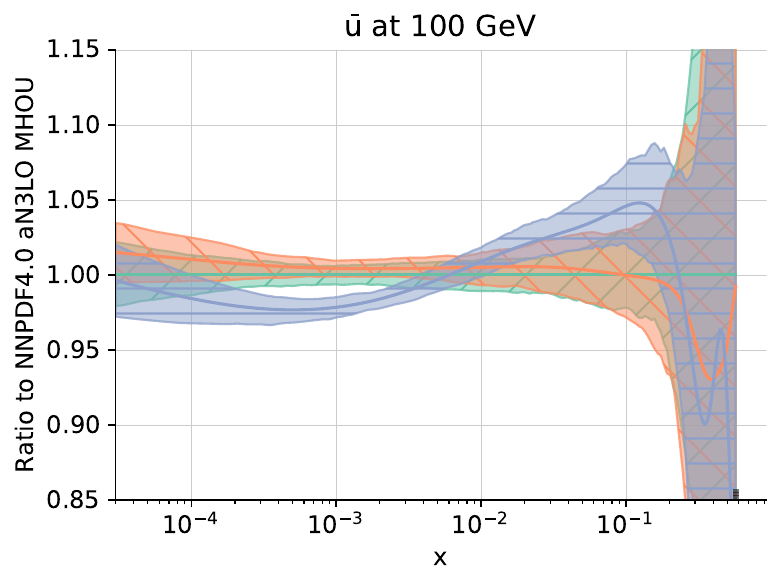}
  \includegraphics[width=0.45\textwidth]{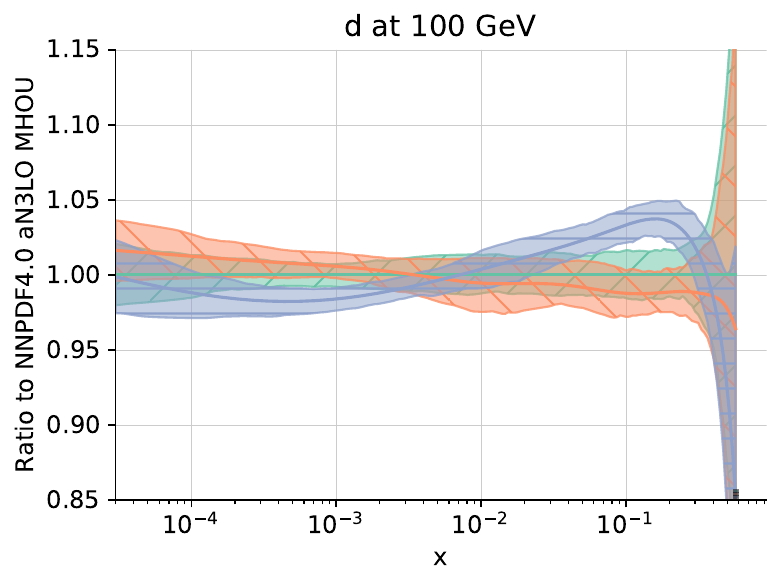}
  \includegraphics[width=0.45\textwidth]{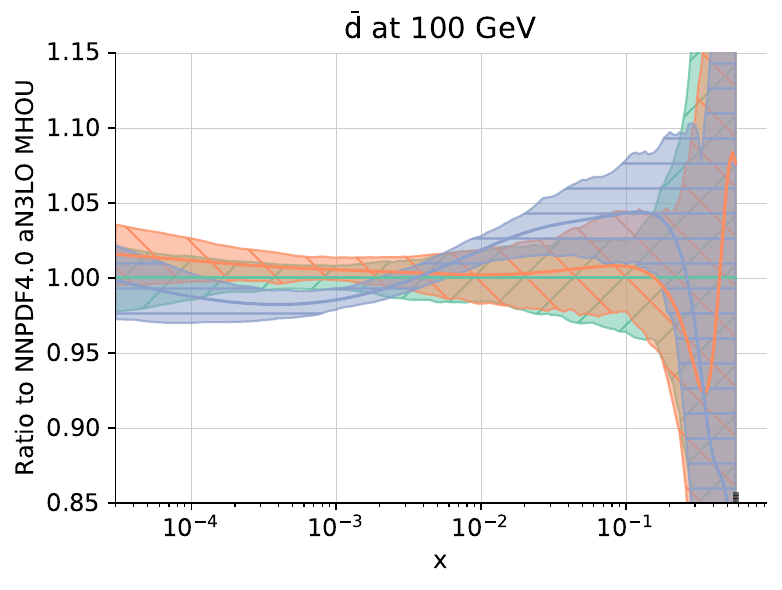}
  \includegraphics[width=0.45\textwidth]{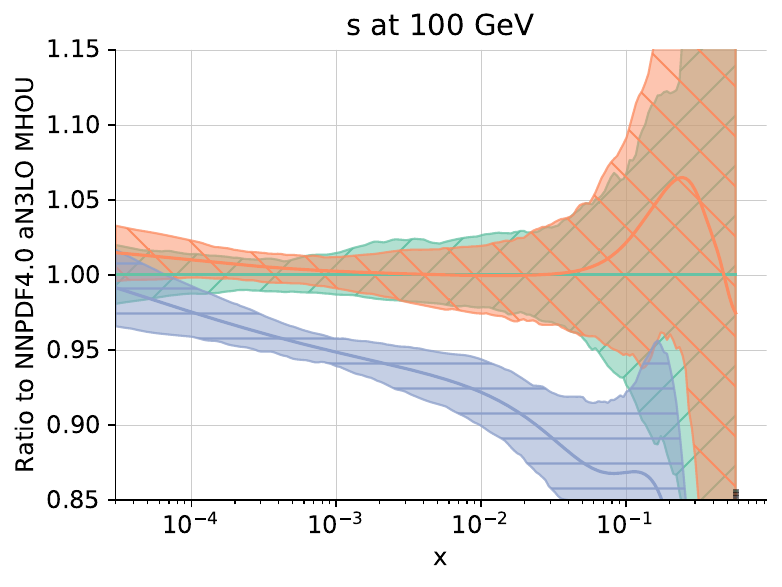}
  \includegraphics[width=0.45\textwidth]{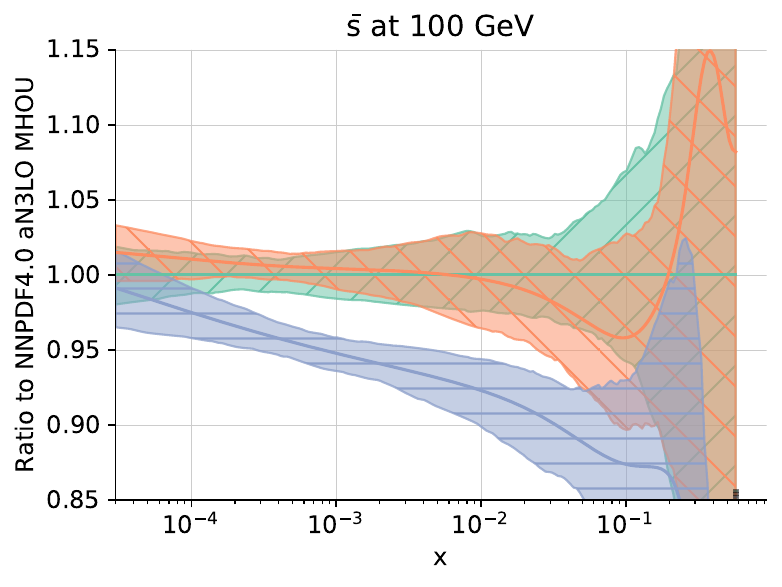}
  \includegraphics[width=0.45\textwidth]{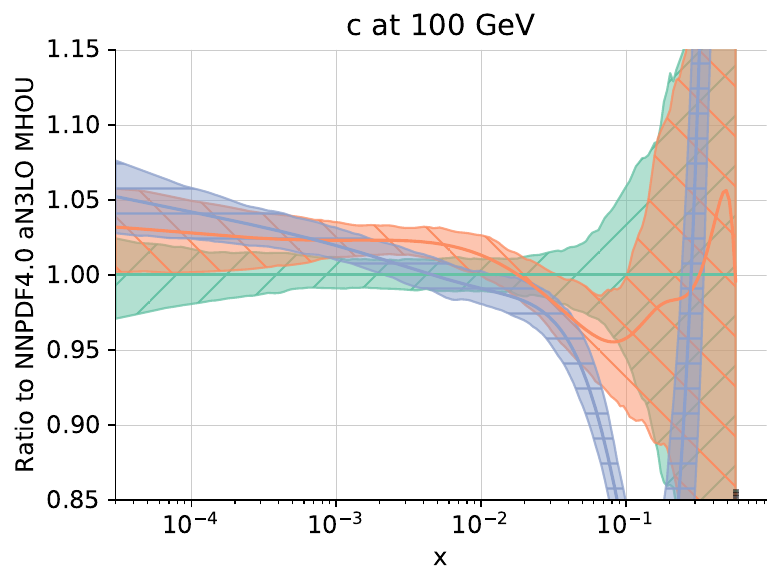}
  \includegraphics[width=0.45\textwidth]{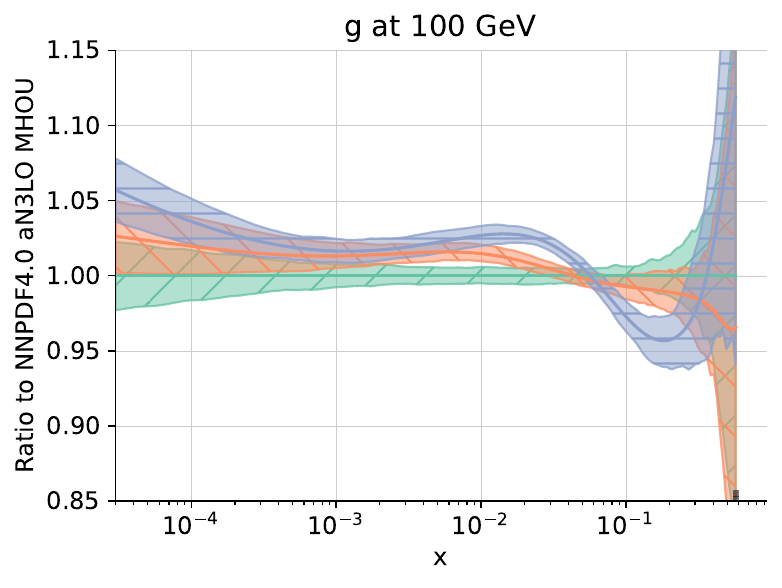}
  \caption{Same as \cref{fig:pdfs_noMHOU_log} for NNPDF4.0MHOU PDF
    sets. Error bands correspond to one sigma
    PDF uncertainties also including MHOUs on the theory predictions
    used in the fit.}
  \label{fig:pdfs_MHOU_log}
\end{figure}

We now examine the NNPDF4.0 aN$^3$LO parton distributions. We compare the NLO,
NNLO and aN$^3$LO NNPDF4.0 PDFs, obtained without and with inclusion of MHOUs,
in \cref{fig:pdfs_noMHOU_log} 
and in \cref{fig:pdfs_MHOU_log} 
respectively. Specifically, we show the up, antiup, down, antidown, strange,
antistrange, charm and gluon PDFs at $Q=100$~GeV, normalized to the aN$^3$LO
result, as a function of $x$ in logarithmic and linear scale. Error bands
correspond to one sigma PDF uncertainties, which do (MHOU sets) or do
not (no MHOU sets) include MHOUs on all theory predictions used in the fit. 
The PDF sets, with and without MHOUs, are the same used to compute the values 
of the $\chi^2$ in \cref{tab:chi2_TOTAL}. 

The excellent perturbative convergence seen in the fit quality is also manifest
at the level of PDFs. In particular, the NNLO PDFs
are either very close to or indistinguishable from their aN$^3$LO counterparts.
Inclusion of MHOUs further improves the consistency between NNLO and aN$^3$LO
PDFs, which lie almost on top of each other. This means that the NNLO
PDFs are made more accurate by the inclusion of MHOUs, and that the
aN$^3$LO PDFs have converged, in the sense discussed above.
Exceptions to this stability are the
charm and gluon PDFs, for which aN$^3$LO corrections have a sizable impact.
In the case of charm, they lead to an enhancement of the central value of
about $4\%$ for $x\sim 0.05$; in the case of gluon, to a suppression of
about $2-3\%$ for $x\sim 0.005$. In both cases, inclusion of MHOUs leads to an
increase in PDF uncertainties by about $1-2\%$. This makes the NNLO and
aN$^3$LO charm PDFs with MHOUs compatible within uncertainties, and the
NNLO and aN$^3$LO gluon PDFs with MHOU almost compatible. 

\begin{figure}[!p]
  \centering
  \includegraphics[width=0.40\textwidth]{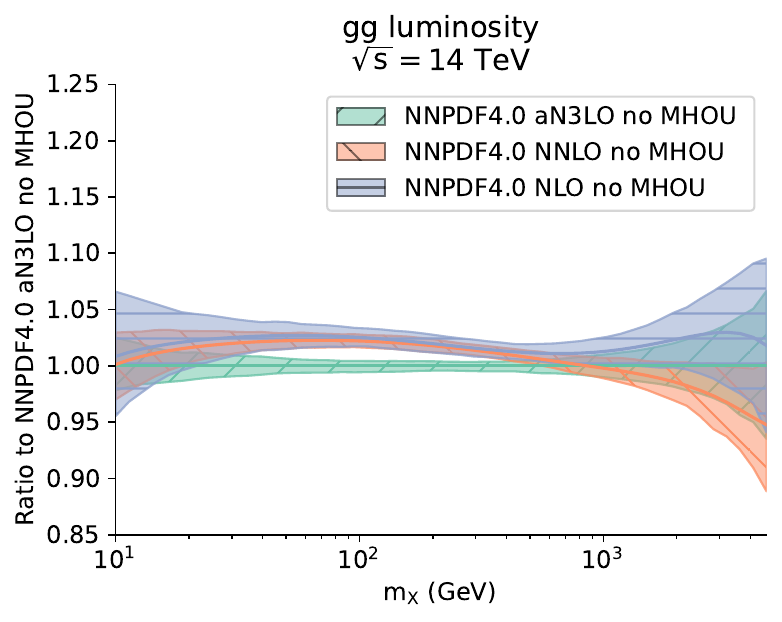}
  \includegraphics[width=0.40\textwidth]{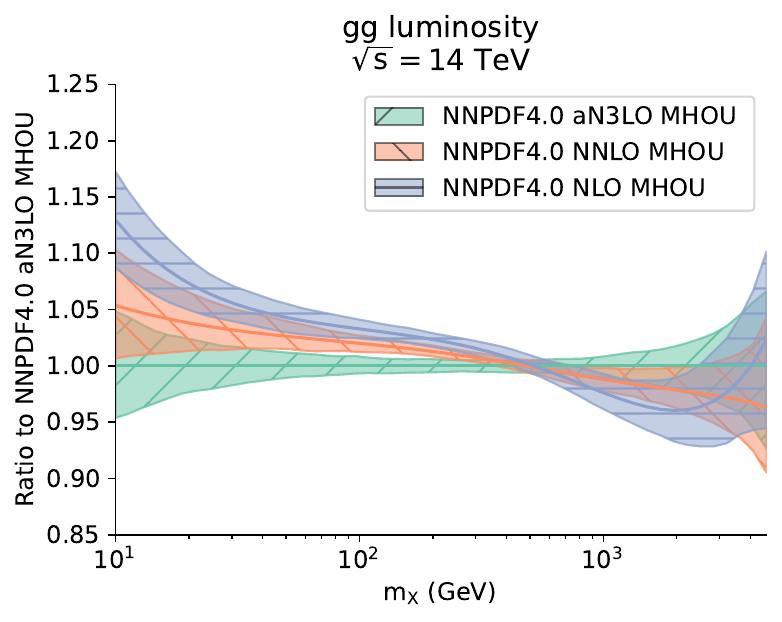}
  \includegraphics[width=0.40\textwidth]{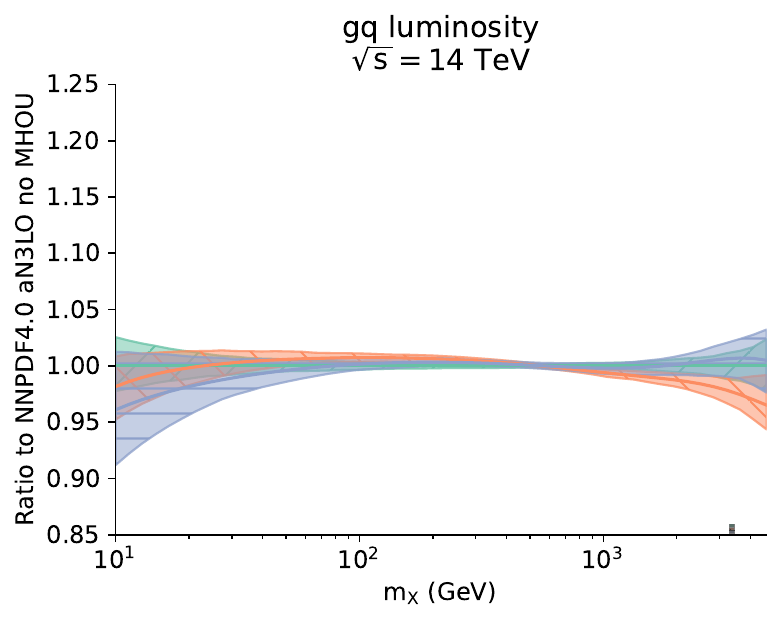}
  \includegraphics[width=0.40\textwidth]{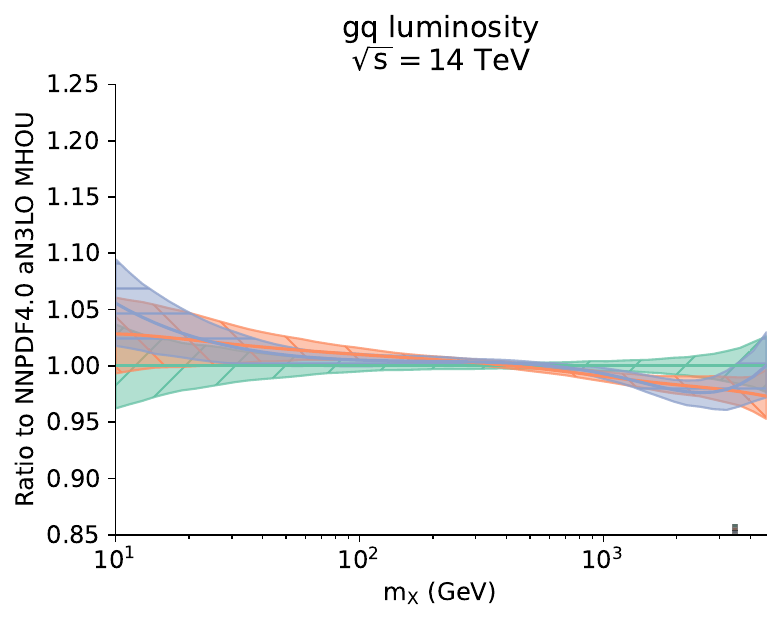}
  \includegraphics[width=0.40\textwidth]{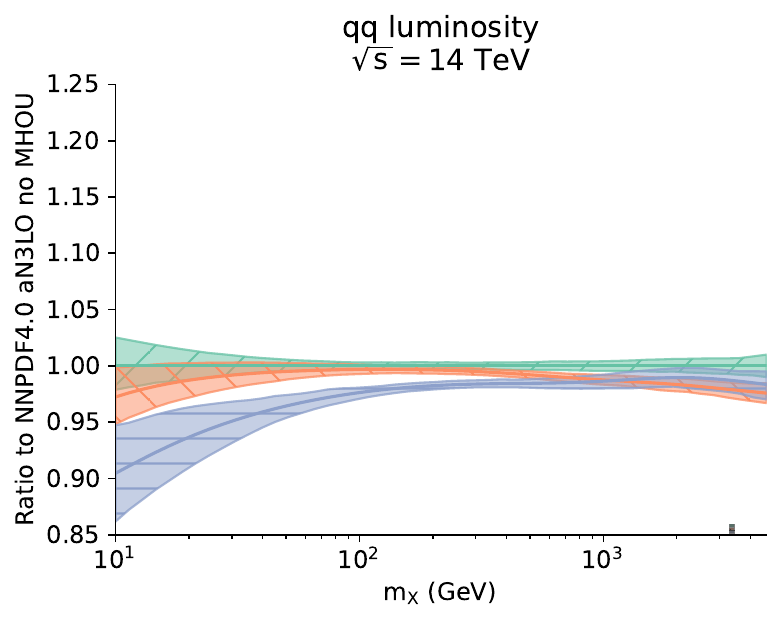}
  \includegraphics[width=0.40\textwidth] {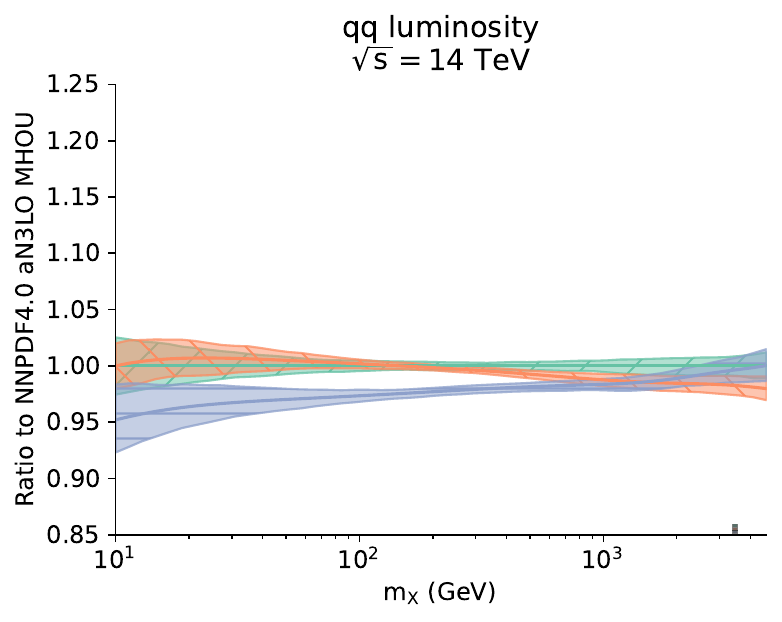}
  \includegraphics[width=0.40\textwidth]{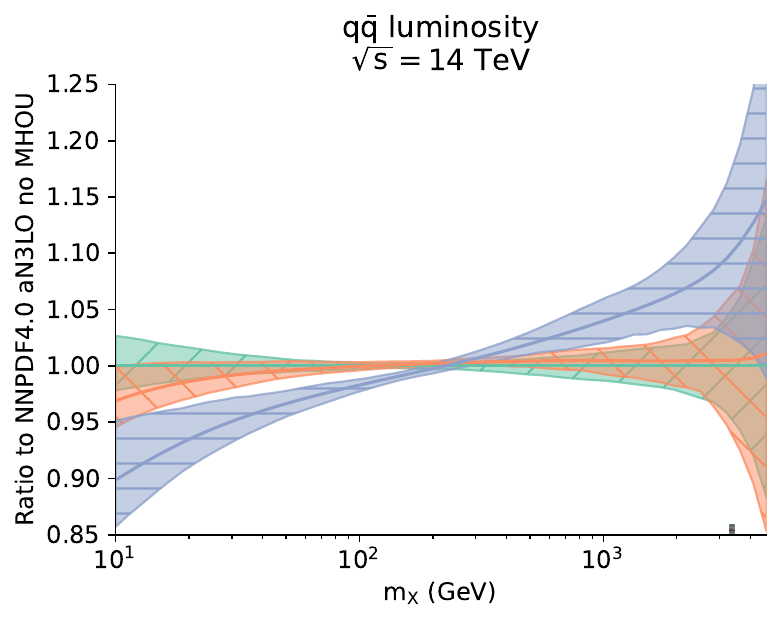}
  \includegraphics[width=0.40\textwidth]{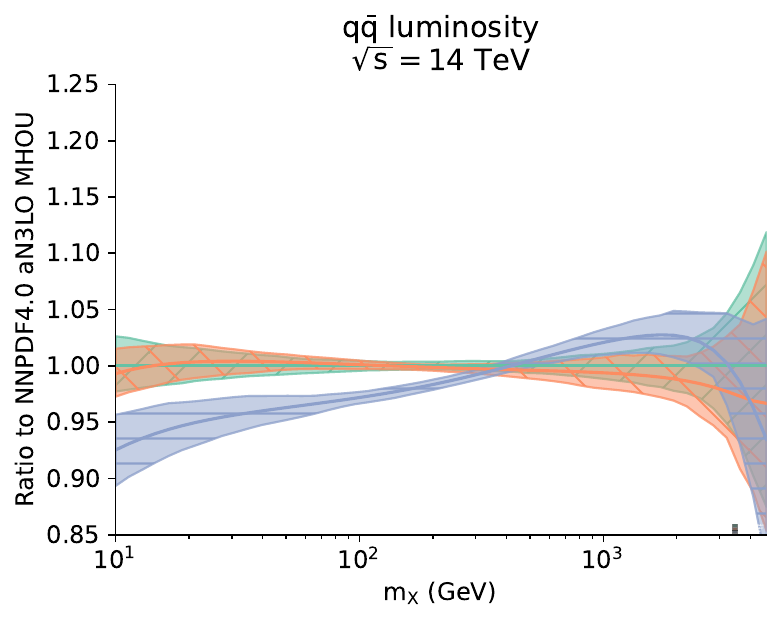}
  \caption{The gluon-gluon, gluon-quark, quark-quark, and quark-antiquark
    parton luminosities as a function of $m_X$ at $\sqrt{s}=14$~TeV,
    computed with NLO, NNLO and aN$^3$LO NNPDF4.0 PDFs without MHOUs (left) and
    with MHOUs (right), all shown as a ratio to the respective
    aN$^3$LO results. Uncertainties are as in \cref{fig:pdfs_noMHOU_log,fig:pdfs_MHOU_log}}
  \label{fig:lumis}
\end{figure}

\cref{fig:lumis} presents a comparison similar to that of
\cref{fig:pdfs_noMHOU_log,fig:pdfs_MHOU_log} for the gluon-gluon,
gluon-quark, quark-quark, and quark-antiquark parton luminosities.
These are shown integrated in rapidity as a function of the invariant mass of
the final state $m_X$ for a center-of-mass energy $\sqrt{s}=14$~TeV. Their
definition follows \cref{eq:lumi_def}.
%

As already observed for PDFs, perturbative convergence is excellent, and
improves upon inclusion of MHOUs. The NNLO and aN$^3$LO results are
compatible within uncertainties for the gluon-quark, quark-quark, and
quark-antiquark luminosities. Some are seen for the
gluon-gluon luminosity, consistent with the differences seen in
the gluon PDF. Specifically, the aN$^3$LO corrections lead to a
suppression of the gluon-gluon luminosity of $2-3~\%$ for
$m_X\sim 100$~GeV. This effect is somewhat compensated by an increase in
uncertainty of about $1~\%$ upon inclusion of MHOUs. Indeed,
the NNLO and aN$^3$LO gluon-gluon luminosities for $m_X\sim 100$~GeV
differ by about $2.5\sigma$ without MHOU, but become almost
compatible within uncertainties when MHOUs are included.

All in all, these results show that aN$^3$LO corrections are generally small,
except for the gluon PDF, and that at aN$^3$LO the perturbative expansion 
has all but converged, with NNLO and aN$^3$LO PDFs very close to each
other, especially upon inclusion of MHOUs. They also show that MHOUs
generally improve the accuracy of PDFs, though at aN$^3$LO they
have a very small impact.
The phenomenological consequences of this state of
affairs will be further discussed in \cref{sec:an3lo_pheno}.

\subsection{PDF uncertainties}
\label{sec:an3lo_uncertainties}

\begin{figure}[!p]
  \centering
  \includegraphics[width=0.45\textwidth]{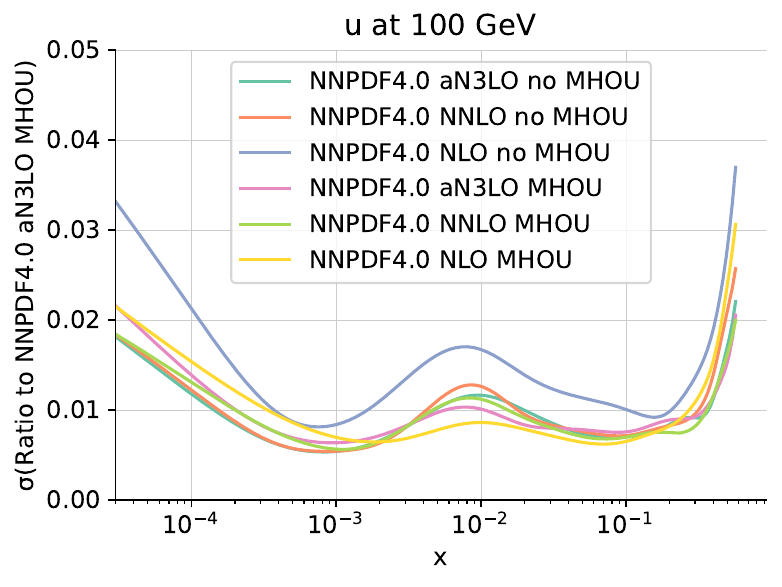}
  \includegraphics[width=0.45\textwidth]{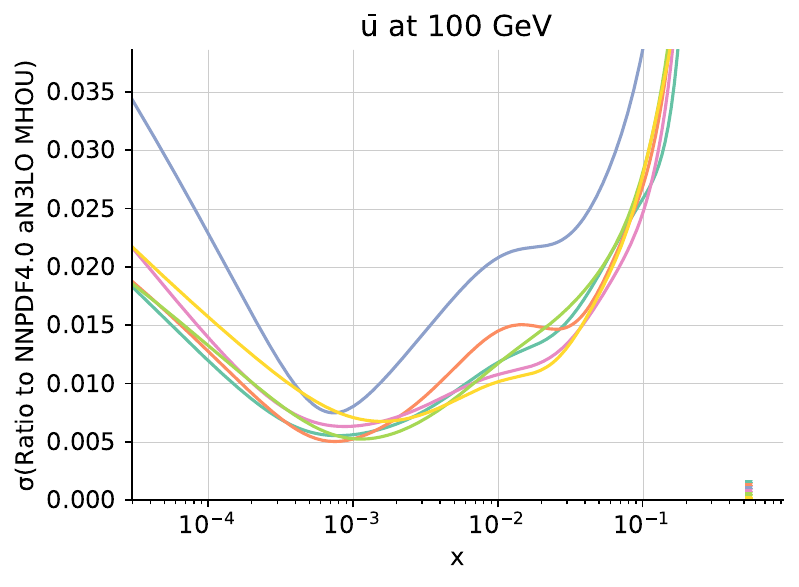}\\
  \includegraphics[width=0.45\textwidth]{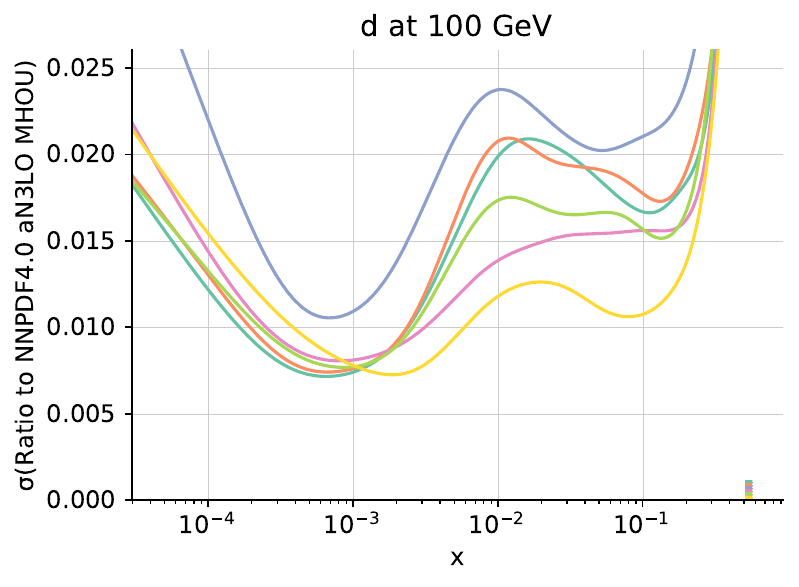}
  \includegraphics[width=0.45\textwidth]{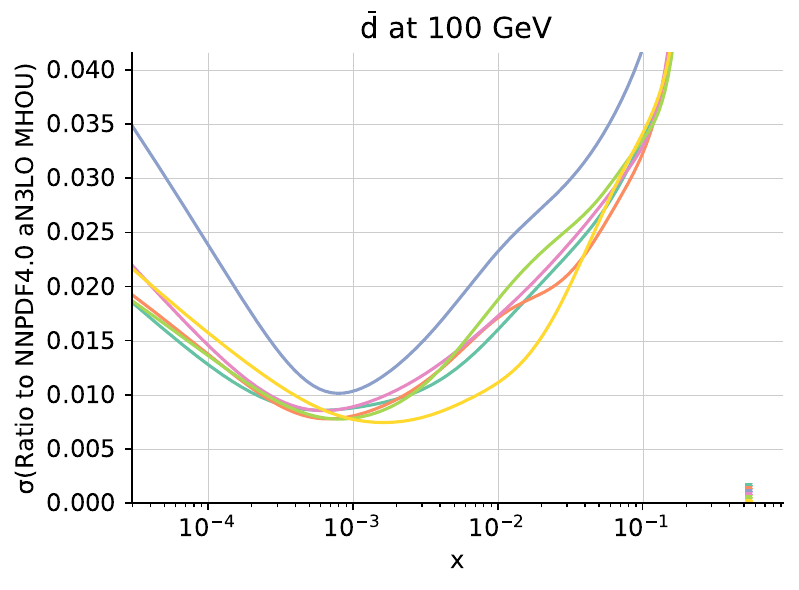}\\
  \includegraphics[width=0.45\textwidth]{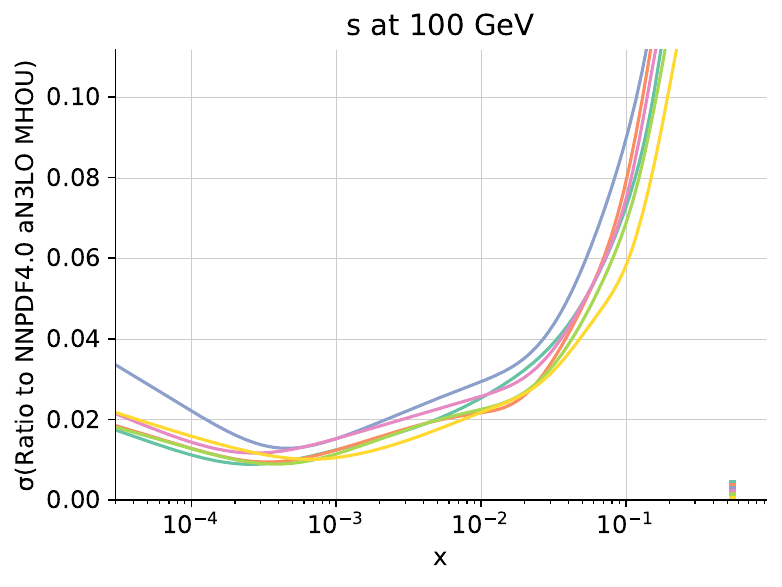}
  \includegraphics[width=0.45\textwidth]{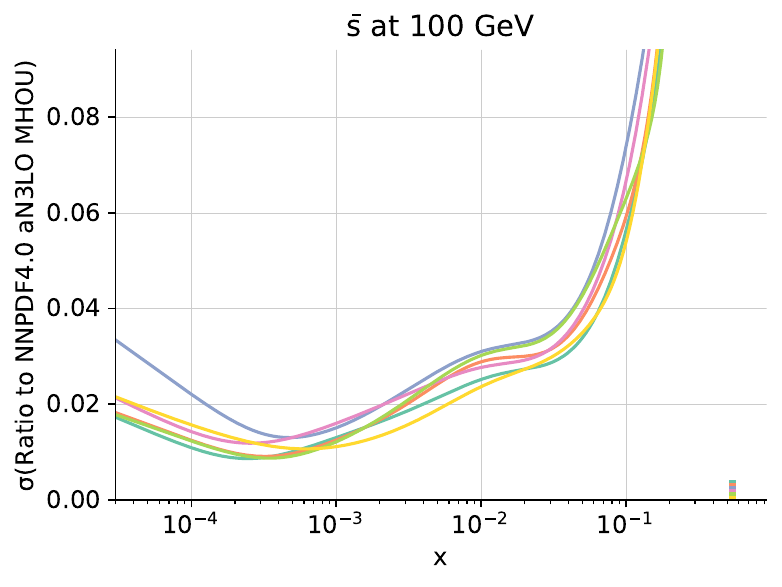}\\
  \includegraphics[width=0.45\textwidth]{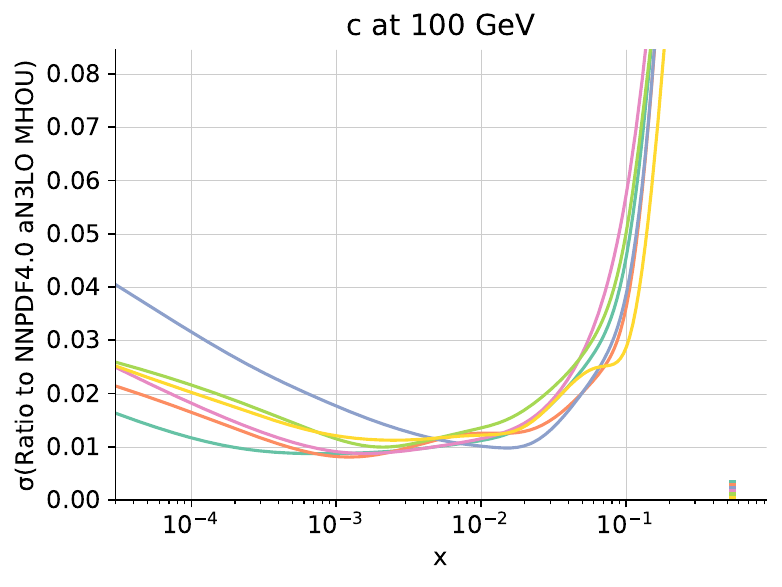}
  \includegraphics[width=0.45\textwidth]{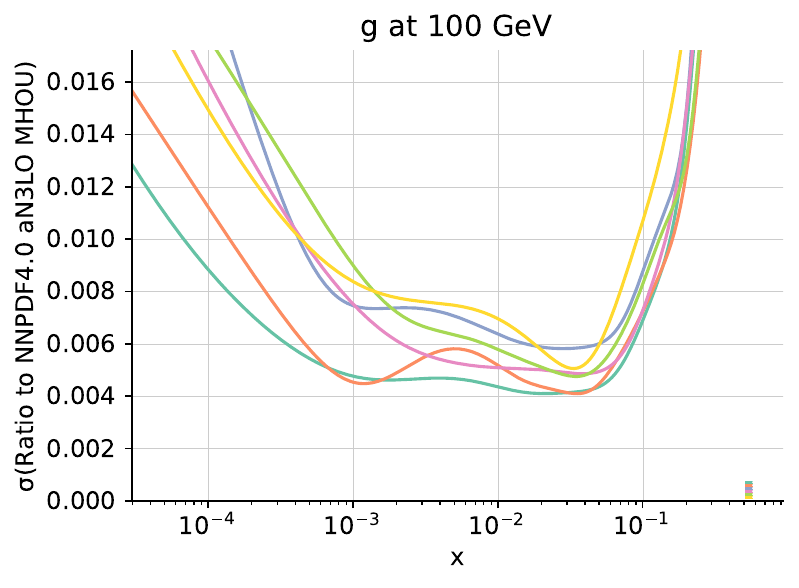}\\
  \caption{Relative one sigma uncertainties for the PDFs shown in
    \cref{fig:pdfs_noMHOU_log,fig:pdfs_MHOU_log}. All uncertainties
    are normalized to the central value of the NNPDF4.0 aN$^3$LO set with
    MHOUs.}
  \label{fig:pdf_uncs}
\end{figure}

We now take a closer look at PDF uncertainties.
In \cref{fig:pdf_uncs} we display one sigma uncertainties for the NNPDF4.0
NLO, NNLO, and aN$^3$LO PDFs with and without MHOUs at $Q=100$~GeV. All
uncertainties are normalized to the central value of the NNPDF4.0 aN$^3$LO PDF
set with MHOUs. The NLO uncertainty is generally the
largest of all in the absence of MHOUs, and for quark distributions
the smallest once MHOUs are included. All other uncertainties, at NNLO
and aN$^3$LO, with and without MHOUs, are quite similar to each other,
especially for quark PDFs. The fact that upon inclusion of an extra
source of uncertainty, namely the MHOU, 
PDF uncertainties are reduced (at NLO) or unchanged (at NNLO and
aN$^3$LO) may look counter-intuitive. However, as already pointed out in
Refs.~\cite{Ball:2018twp,Ball:2020xqw,NNPDF:2024dpb}, this can be
understood to be a consequence of the increased compatibility of the
data due to inclusion of MHOUs and of higher-order perturbative corrections.

\begin{figure}[!p]
  \centering
  \includegraphics[width=0.45\textwidth]{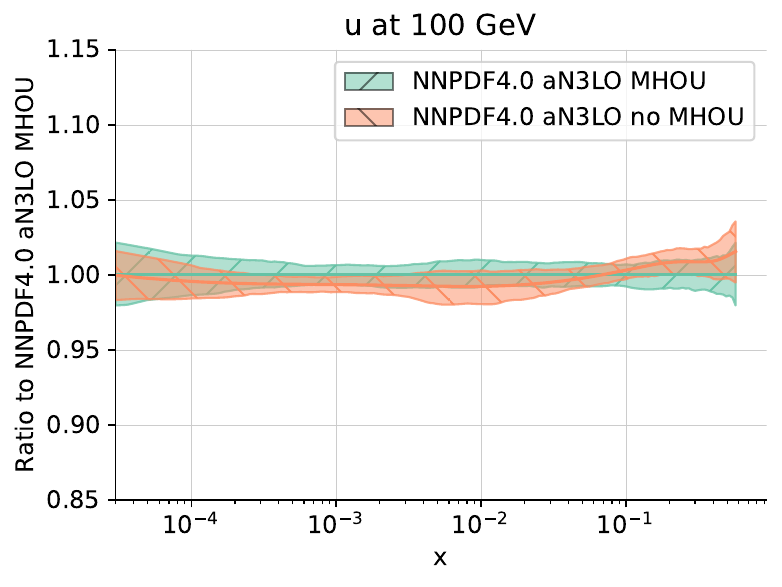}
  \includegraphics[width=0.45\textwidth]{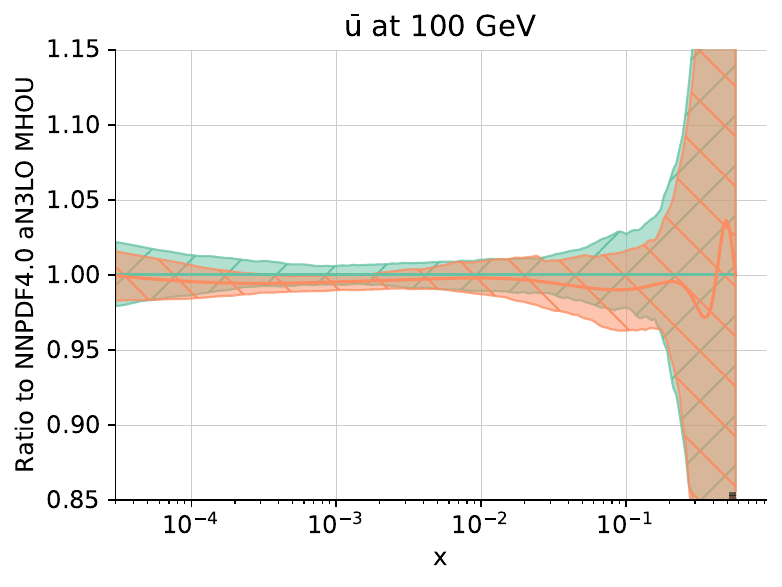}\\
  \includegraphics[width=0.45\textwidth]{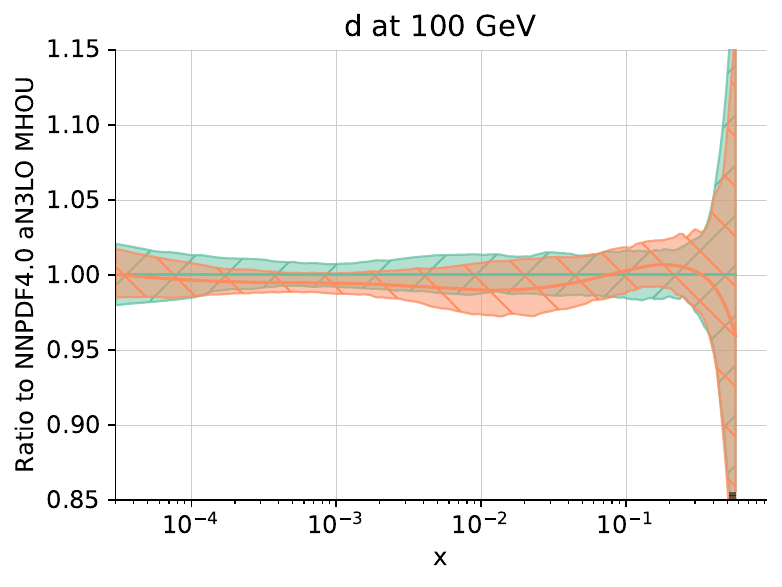}
  \includegraphics[width=0.45\textwidth]{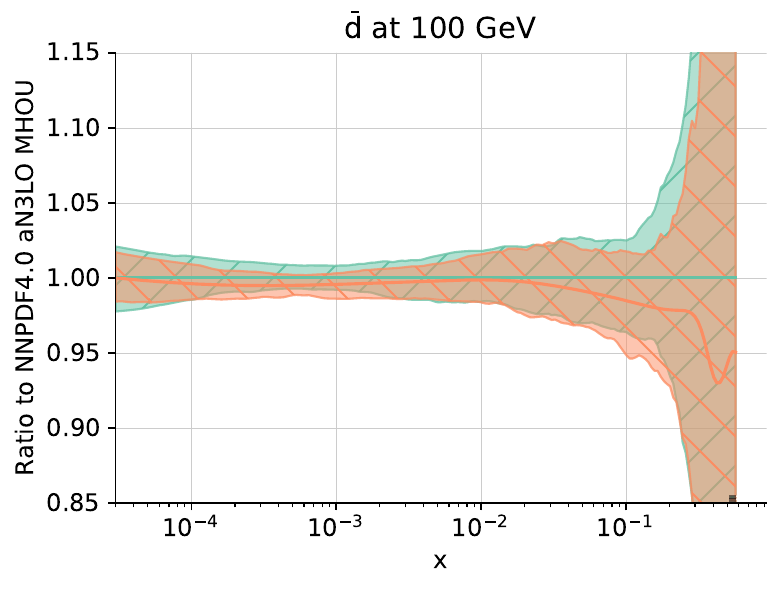}\\ 
  \includegraphics[width=0.45\textwidth]{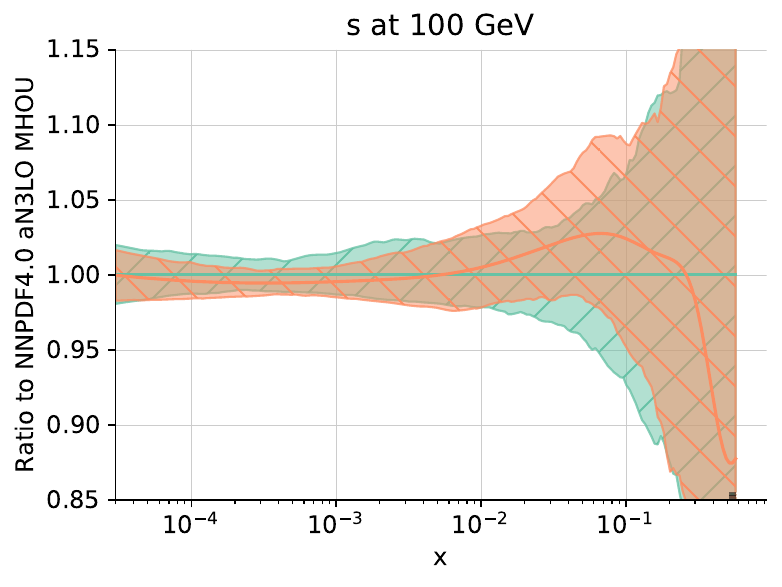}
  \includegraphics[width=0.45\textwidth]{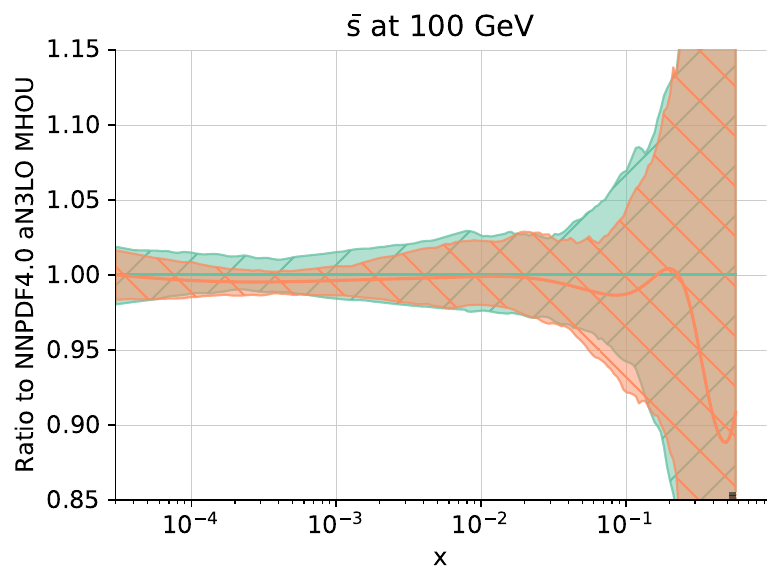}\\
  \includegraphics[width=0.45\textwidth]{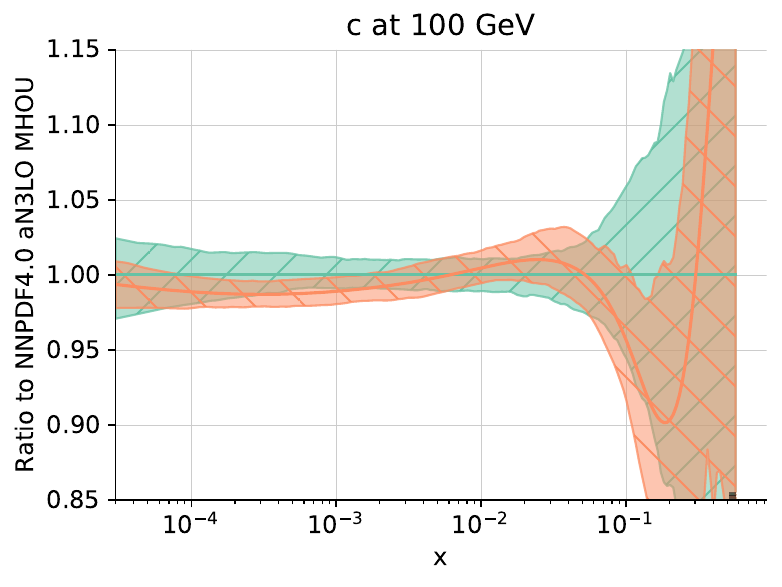}
  \includegraphics[width=0.45\textwidth]{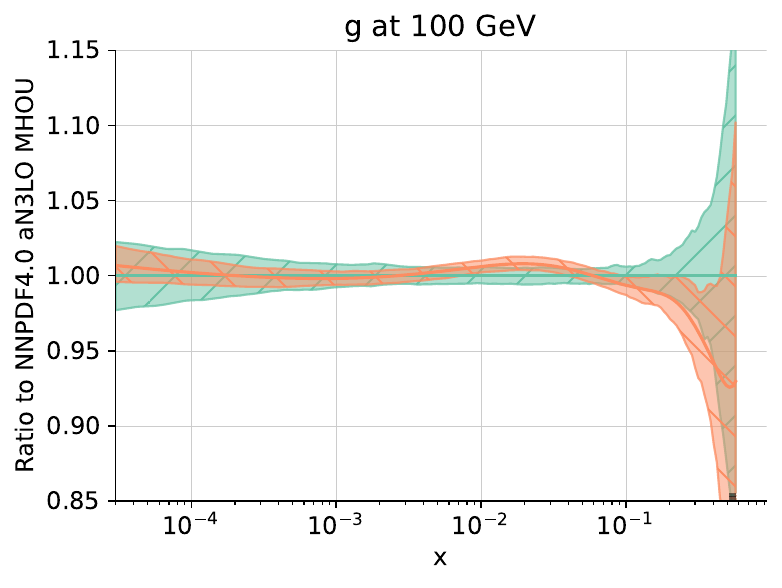}\\
  \caption{Same as
    \cref{fig:pdfs_noMHOU_log,fig:pdfs_MHOU_log}, now comparing NNPDF4.0 aN$^3$LO PDFs without and with
    MHOUs.}
  \label{fig:NNLO_vs_N3LO_pdfs}
\end{figure}

The impact of MHOUs on NLO and NNLO PDFs was extensively assessed in
Ref.~\cite{NNPDF:2024dpb} and summarized in \cref{sec:nnpdf40_pdfs}. 
In a similar vein, here we focus on the impact of MHOUs on aN$^3$LO PDFs. 
To this purpose, in \cref{fig:NNLO_vs_N3LO_pdfs} we compare the NNPDF4.0 aN$^3$LO PDFs
with and without MHOUs. 
Again, aN$^3$LO PDFs (and relative luminosities) with and without MHOU are very 
compatible with each other. This evidence reinforces the expectation that
perturbative corrections beyond N$^3$LO will not alter PDFs significantly,
at least with current data and methodology.

In analogy with Ref.~\cite{NNPDF:2024dpb}, we also compare the
$\phi$ estimator~\cite{NNPDF:2014otw}, which measures the standard deviation 
over the PDF replica sample in units of the data uncertainty
\begin{equation}
  \phi = \sqrt{\braket{\chi_{\rm exp}^2\left[ T[f], D \right]} - \chi_{\rm exp}^2 \left[ \braket{ T[f]}, D \right]},
  \label{eq:phi_def}
\end{equation}
and $T[f]$ and $D$ denotes the theory predictions and the dataset set
included in the fit and the average is taken over the PDF replicas, 
and solely the experimental covariance matrix is used in $\chi_{\rm exp}^2$.
$\phi$ provides an estimate of the consistency of the
data: consistent data are combined by the underlying theory and lead to an
uncertainty in the prediction which is smaller than that of the original data.
The value of $\phi$ obtained in the NLO, NNLO, and aN$^3$LO NNPDF4.0 fits
with and without MHOUs (as in \cref{tab:chi2_TOTAL}) is reported in
\cref{tab:phi_TOTAL}. It is clear that $\phi$ converges to very similar
values with the increase of the perturbative order and/or with inclusion of
MHOUs for both the total dataset and for most of the data categories.
This fact is further quantitative evidence of the perturbative
convergence of the PDF uncertainties.

\begin{table}[!t]
  \scriptsize
  \centering
  \renewcommand{\arraystretch}{1.4}
  \begin{tabularx}{\textwidth}{Xcccccc}
  \toprule
  & \multicolumn{2}{c}{NLO}
  & \multicolumn{2}{c}{NNLO}
  & \multicolumn{2}{c}{N$^3$LO} \\
  Dataset
  & no MHOU
  & MHOU
  & no MHOU
  & MHOU 
  & no MHOU
  & MHOU \\
  \midrule
  DIS NC
  & 0.14 & 0.13 
  & 0.15 & 0.13
  & 0.13 & 0.13 \\
  DIS CC
  & 0.11 & 0.11 
  & 0.12 & 0.12 
  & 0.12 & 0.12 \\
  DY NC
  & 0.19 & 0.17
  & 0.18 & 0.17
  & 0.17 & 0.18  \\
  DY CC
  & 0.33 & 0.27
  & 0.35 & 0.32
  & 0.31 & 0.32 \\
  Top pairs
  & 0.18 & 0.17
  & 0.17 & 0.17
  & 0.16 & 0.19 \\
  Single-inclusive jets
  & 0.13 & 0.13
  & 0.13 & 0.13
  & 0.13 & 0.13 \\
  Dijet
  & 0.10 & 0.10
  & 0.11 & 0.10
  & 0.10 & 0.10 \\
  Prompt photons 
  & 0.06 & 0.07
  & 0.06 & 0.06
  & 0.05 & 0.05 \\
  Single top
  & 0.04 & 0.04
  & 0.04 & 0.04
  & 0.04 & 0.04 \\
  \midrule
  Total
  & 0.18 & 0.15
  & 0.16 & 0.15
  & 0.15 & 0.15 \\
\bottomrule
\end{tabularx}

  \vspace{0.3cm}
  \caption{The $\phi$ uncertainty estimator for NNPDF4.0 PDFs at NLO, NNLO and
    aN$^3$LO without and with MHOUs for the process categories as in
    \cref{tab:chi2_TOTAL}.}
  \label{tab:phi_TOTAL}
\end{table}

\subsection{Implications for intrinsic charm}
\label{sec:anl3o_ic}

The availability of the aN$^3$LO PDFs discussed in
\cref{sec:an3lo_PDFs,sec:an3lo_uncertainties} allows us to revisit and
consolidate our results on intrinsic charm (cf.~\cref{chap:ic}). Specifically, based on the
NNPDF4.0 NNLO PDF determination, we have found evidence for intrinsic
charm~\cref{sec:ic_ic} and an indication for a non-vanishing valence charm
component~\cref{sec:ic_icasy}. In these analyses, the dominant source of theory
uncertainty was estimated to come from the matching conditions that are used in
order to obtain PDFs in a three-flavor charm decoupling scheme from
high-scale data, while MHOUs were assumed to be subdominant. The
uncertainty in the matching conditions was in turn estimated by comparing
results obtained using NNLO matching and the best available aN$^3$LO matching
conditions, both applied to NNLO PDFs.

It is now possible to improve these results on three counts. First, we
can now fully include MHOUs. Second, we can consistently combine
aN$^3$LO matching conditions and aN$^3$LO PDFs, and perform a consistent
comparison of NNLO and  aN$^3$LO results. Finally, knowledge of
aN$^3$LO matching conditions themselves is now improved thanks to
recent results~\cite{Ablinger:2022wbb} that were not available at the
time of the analysis of Ref.~\cite{Ball:2022qks}. We will specifically
discuss the determination of the total intrinsic charm component and we
do not consider the valence component,
because effects of MHOUs and of the flavor scheme transformation are
already very small at NNLO~\cite{NNPDF:2023tyk}.

\begin{figure}[!t]
  \centering
  \includegraphics[width=0.45\textwidth]{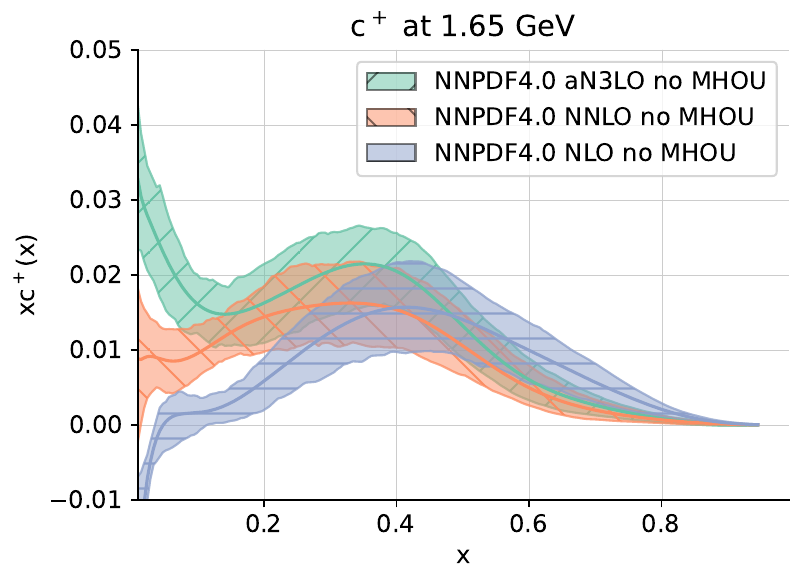}
  \includegraphics[width=0.45\textwidth]{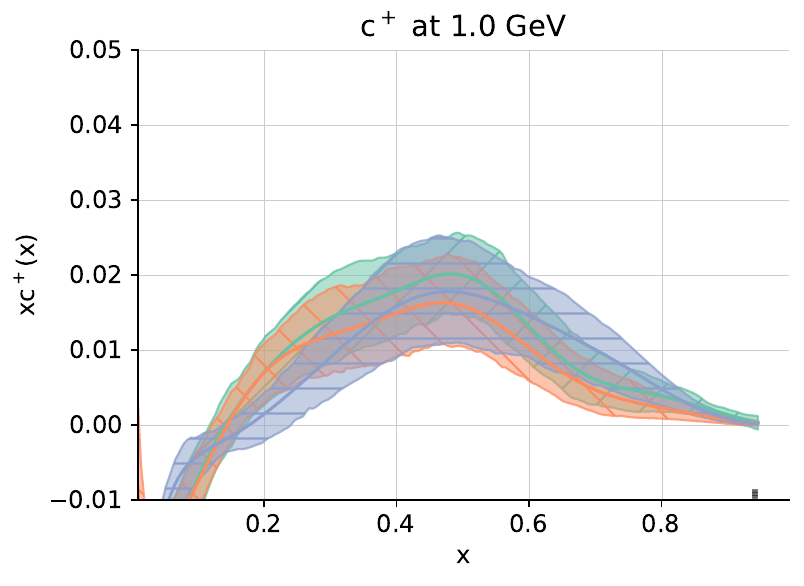}\\
  \includegraphics[width=0.45\textwidth]{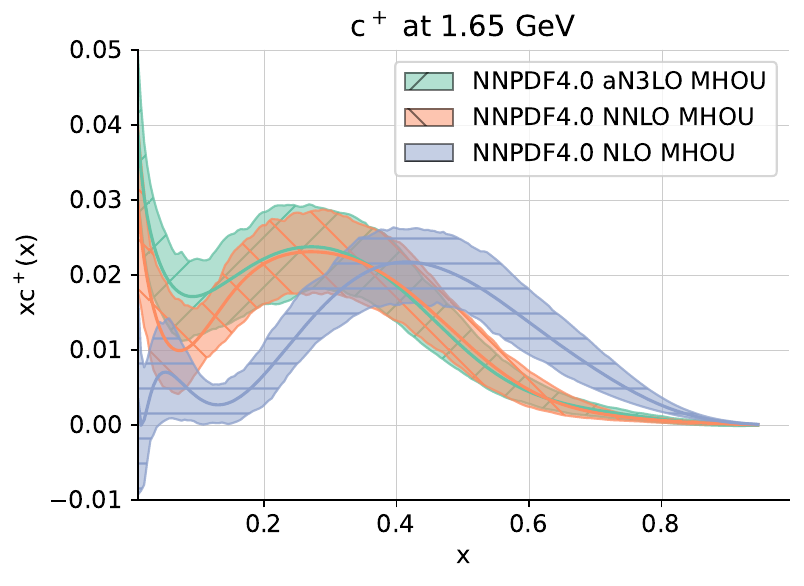}
  \includegraphics[width=0.45\textwidth]{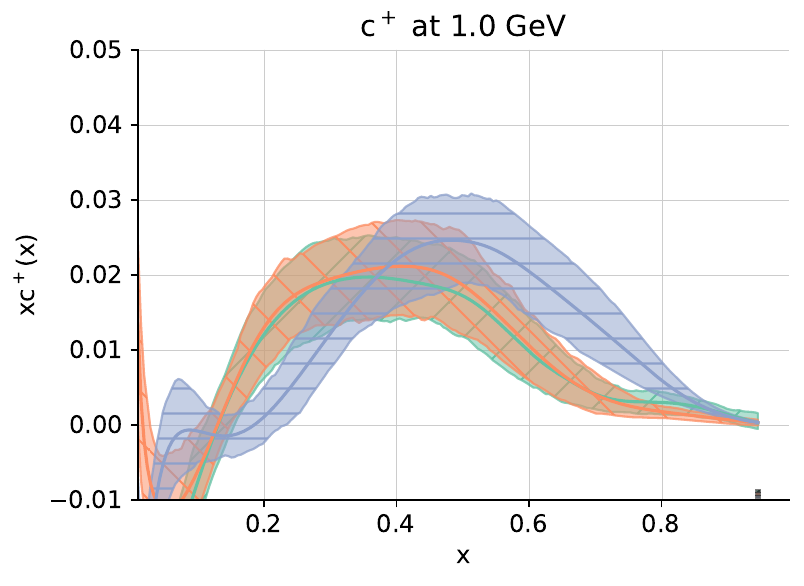}\\ 
  \caption{The total charm PDF, $xc^+(x,Q^2)$, in the 4FNS at $Q=1.65$~GeV
    (left) and 3FNS (right), as obtained from the NNPDF4.0 NLO, NNLO, and
    aN$^3$LO fits without (top) and with (bottom) MHOUs. Error bands correspond
    to one sigma PDF uncertainties. Note that in the 3FNS the charm
    PDF does not depend on scale.}
  \label{fig:PDFs-n3lo-q1p65gev-Charm-perturbative-convergence} 
\end{figure}

To this purpose, in
\cref{fig:PDFs-n3lo-q1p65gev-Charm-perturbative-convergence} we
show the total charm PDF, $xc^+(x,Q^2)$, in the 4FNS at $Q=1.65$~GeV and in the
3FNS, as obtained from using  NNPDF4.0 NLO, NNLO and aN$^3$LO without and with
MHOUs. Note that in the 3FNS the charm PDF does not depend on scale.
Error bands correspond to one sigma PDF uncertainties. The 4FNS
results share the general features discussed in \cref{sec:an3lo_PDFs}:
the perturbative expansion converges nicely, with the aN$^3$LO result
very close to the NNLO. The convergence is further improved by the
inclusion of MHOUs, which move the NNLO yet closer to the
aN$^3$LO. The 3FNS result is especially remarkable: whereas the
combination of aN$^3$LO matching with NNLO PDFs, used in
Ref.~\cite{Ball:2022qks} to conservatively estimate MHOUs, was somewhat
unstable, now results display complete stability, and in particular
the NNLO and aN$^3$LO results completely coincide.

\begin{figure}[!t]
  \centering
  \includegraphics[width=0.45\textwidth]{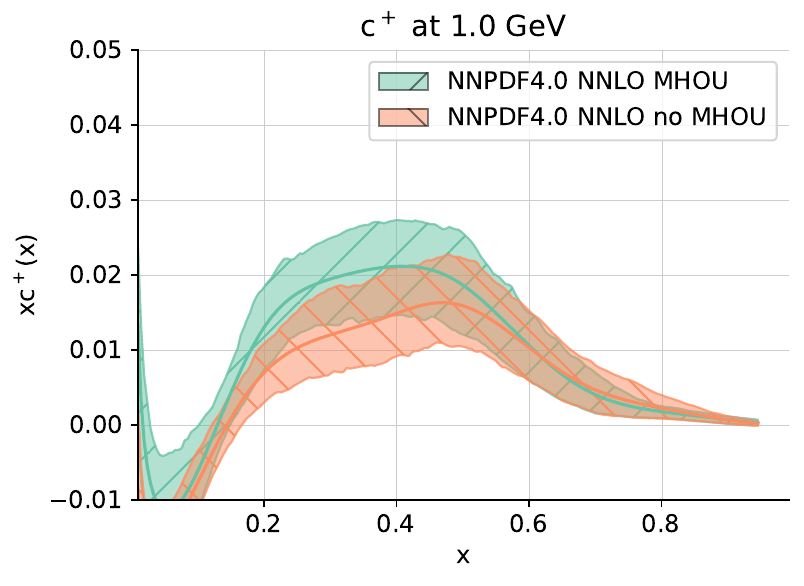}
  \includegraphics[width=0.45\textwidth]{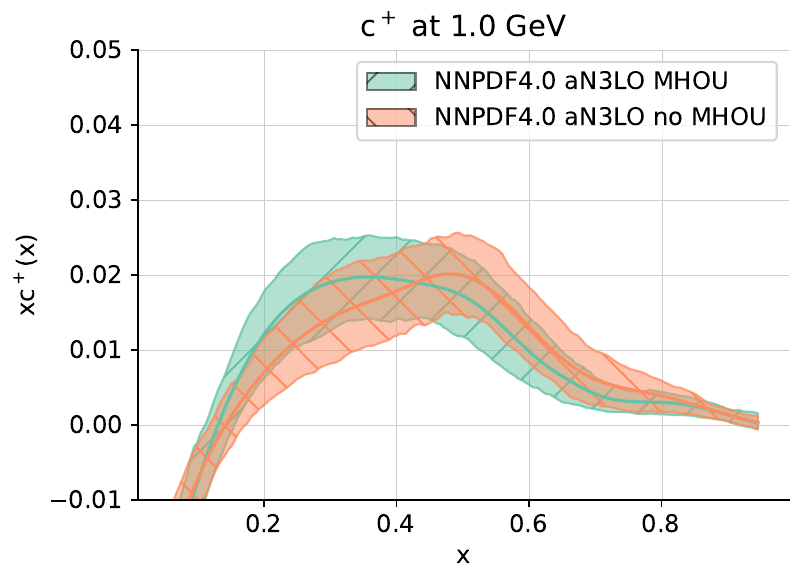}
  \caption{Same as
    \cref{fig:PDFs-n3lo-q1p65gev-Charm-perturbative-convergence}, now
    comparing the total charm PDF in the 3FNS with and without
    MHOUs, respectively at NNLO (left) and aN$^3$LO (right).}
  \label{fig:PDFs-n3lo-q1p65gev-Charm-impactMHOU} 
\end{figure}

In order to assess the impact of MHOUs more clearly, in
\cref{fig:PDFs-n3lo-q1p65gev-Charm-impactMHOU} we compare the total charm
PDF in the 3FNS with and without MHOUs, respectively at NNLO
and aN$^3$LO. At NNLO MHOUs have a small but non-negligible impact
on central values, with almost unchanged uncertainty, but at aN$^3$LO
they have essentially no impact, confirming the perturbative
convergence of the result.

\begin{figure}[!t]
  \centering
  \includegraphics[width=0.55\textwidth]{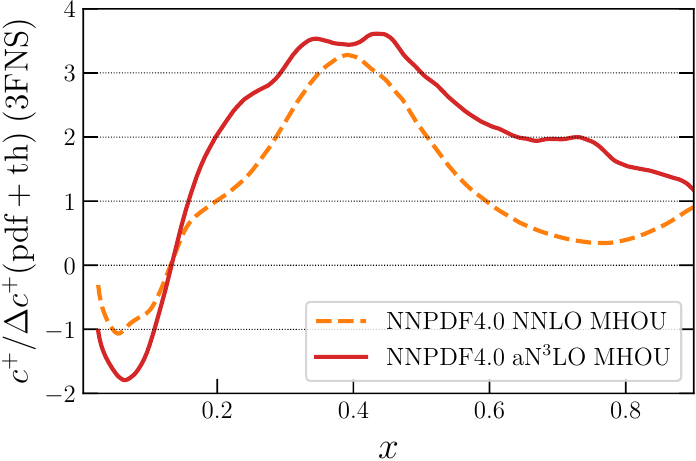}
  \caption{The pull (central value divided by total uncertainty)
    for the total charm PDF in the 3FNS obtained
    in the NNPDF4.0 NNLO and aN$^3$LO fits with MHOUs.}
  \label{fig:IC-Significance-N3LO} 
\end{figure}

We thus proceed to a final re-assessment of the significance of
intrinsic charm through the pull, defined as the central value divided
by total uncertainty, using NNPDF4.0MHOU NNLO and aN$^3$LO PDFs.
We estimate the total uncertainty by adding in quadrature to the PDF uncertainty
(which already includes the MHOU from the theory predictions used in the fit) a 
further  theory uncertainty, taken equal to the difference between the
central value at given perturbative order, and that at the previous perturbative
order (so at NNLO from the difference to NLO, and so on).
This now also includes the MHOU due to change in the 
matching from 4FNS to 3FNS, but also the shift in the 4FNS result that
is in principle already accounted for by the MHOU. Also, it 
conservatively assumes that the shift between the current order and
the next is equal to that from the previous order, rather than
smaller. Results obtained with this conservative error
estimate are shown in \cref{fig:IC-Significance-N3LO}.
It is clear that the
significance of intrinsic charm is increased somewhat when going from
NNLO to aN$^3$LO. It is now also somewhat increased already at NNLO in
comparison to the result of Ref.~\cite{Ball:2022qks}, despite the more
conservative uncertainty estimate, thanks to the  increased accuracy of MHOU
PDFs and the consistent and improved treatment of matching aN$^3$LO
conditions. Indeed, local significance at the peak is now more than
three sigma for the default fit.

\section{Impact on LHC phenomenology}
\label{sec:an3lo_pheno}
We present an assessment of the implications of aN$^3$LO PDFs
for LHC phenomenology, by looking at processes for which N$^3$LO
results are publicly available, namely the DY and Higgs total
inclusive cross-sections.
We present predictions at NLO, NNLO, and aN$^3$LO using both 
NNPDF4.0 and MSHT20 PDFs, consistently matching the perturbative order
of the PDF and matrix element. At N$^3$LO we also show results
obtained with the currently common approximation of using NNLO PDFs 
with aN$^3$LO matrix elements.

At each perturbative order, the uncertainty on the cross-section is
determined by adding in quadrature the PDF uncertainty to the MHOU on
the hard matrix element determined performing 7-point renormalization 
and factorization scale variation and taking the envelope of the results.
This is the procedure that is most commonly used for the estimation of 
the total uncertainty on hadron collider processes; here we follow it for 
ease of comparison with available results. 
\footnote{
  In a more refined treatment, MHOUs on the hard cross-section can be included 
  through a theory covariance matrix for the hard cross-section itself, 
  like the MHOUs and IHOUs on the PDF. 
  This would then make it possible to keep track of the correlation between 
  these different sources of uncertainty~\cite{Harland-Lang:2018bxd,Ball:2021icz}.
}

We display results with a total uncertainty obtained combining these
uncertainties in quadrature (both with and without MHOUs in the PDF fit), 
and we also tabulate this total uncertainty (without MHOUs in the PDF fit) 
along with the PDF uncertainty both with and without MHOUs.
Also, in order to assess the impact of the use of aN$^3$LO PDFs,
we plot N$^3$LO results obtained using NNLO and aN$^3$LO PDFs, 
we tabulate the shift between the N$^3$LO prediction obtained using NNLO 
and aN$^3$LO PDFs, and we compare it to previous estimate of this expected shift 
based on the differences between NNLO and NLO PDFs. 
Indeed, predictions for processes computed at N$^3$LO
accuracy are commonly obtained using NNLO PDFs, with an extra
uncertainty assigned to the result dues to this mismatch in
perturbative order between the PDF and the matrix element. 
A commonly used prescription in order to estimate this uncertainty
~\cite{Anastasiou:2016cez,Baglio:2022wzu} is to take it equal to
\begin{equation}
  \label{eq:PDFimpact_xsec_approx}
  \Delta^{\rm app}_{\rm NNLO} \equiv 
    \frac{1}{2}\Bigg| \frac{\sigma^{\rm NNLO}_{\rm NNLO-PDF} - \sigma^{\rm NNLO}_{\rm NLO-PDF}}{\sigma^{\rm NNLO}_{\rm NNLO-PDF}}\Bigg|,
\end{equation}
namely to assume that the same percentage shift, computed at one less
perturbative order, would be twice as large.
This prescription can now be compared to the exact result via
\begin{equation}
  \label{eq:PDFimpact_xsec_exact}
  \Delta^{\rm exact}_{\rm NNLO} \equiv \Bigg| \frac{\sigma^{\rm N^3LO}_{\rm N^3LO-PDF} - \sigma^{\rm N^3LO}_{\rm NNLO-PDF}}{\sigma^{\rm N^3LO}_{\rm N^3LO-PDF}}\Bigg| \, .
\end{equation}

\subsection{Inclusive Drell-Yan production}
\label{sec:gaugeboson}

\begin{figure}[!p]
  \centering
  \includegraphics[width=0.44\linewidth]{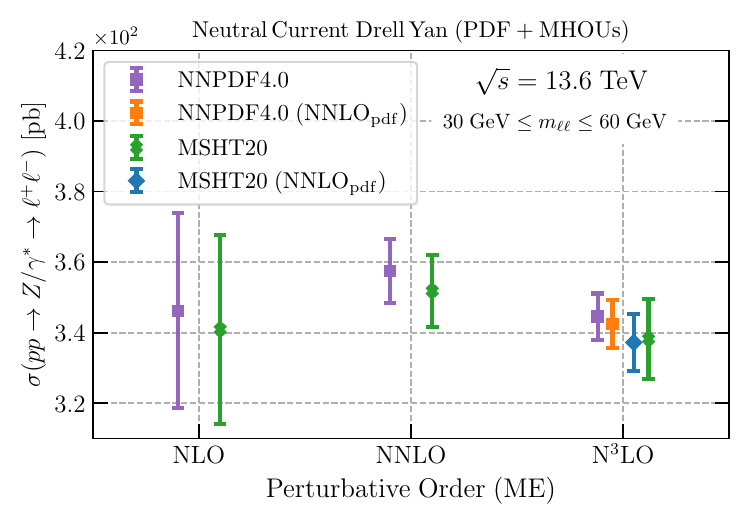}
  \includegraphics[width=0.44\linewidth]{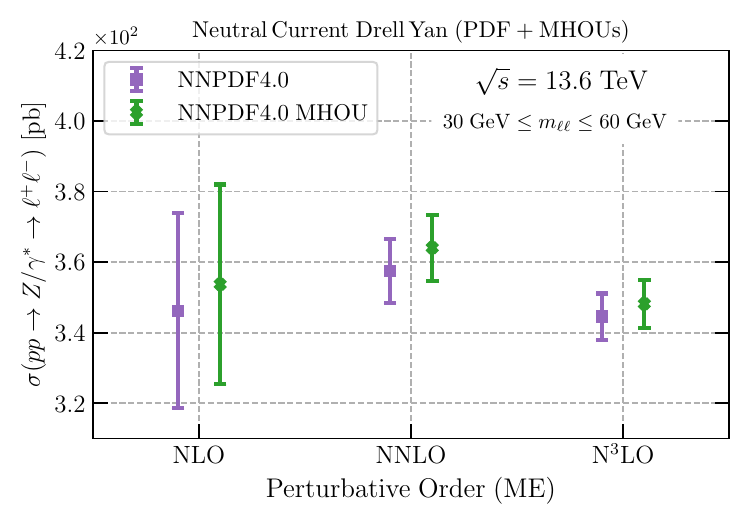}
  \includegraphics[width=0.44\linewidth]{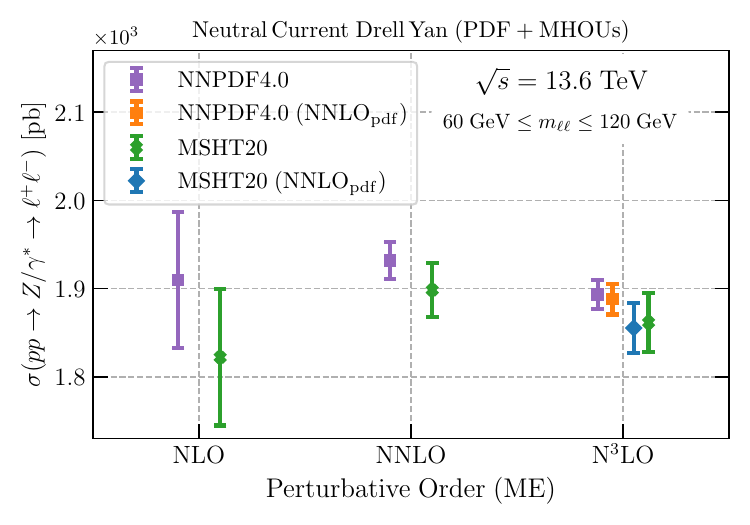}
  \includegraphics[width=0.44\linewidth]{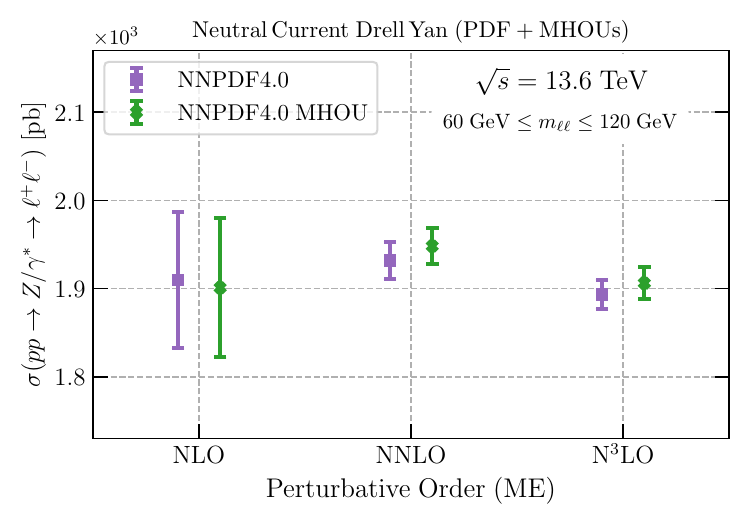}
  \includegraphics[width=0.44\linewidth]{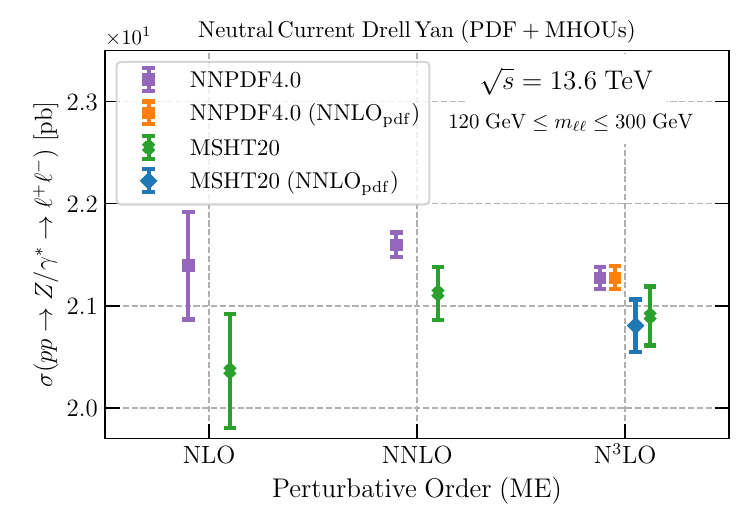}
  \includegraphics[width=0.44\linewidth]{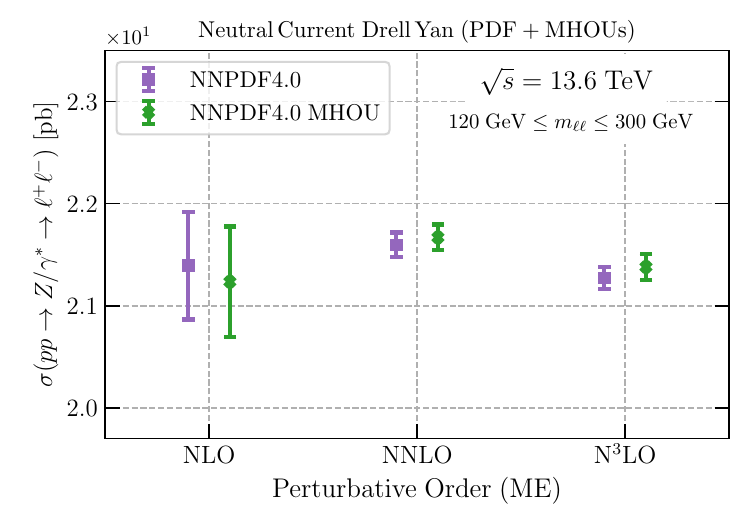}
  \includegraphics[width=0.44\linewidth]{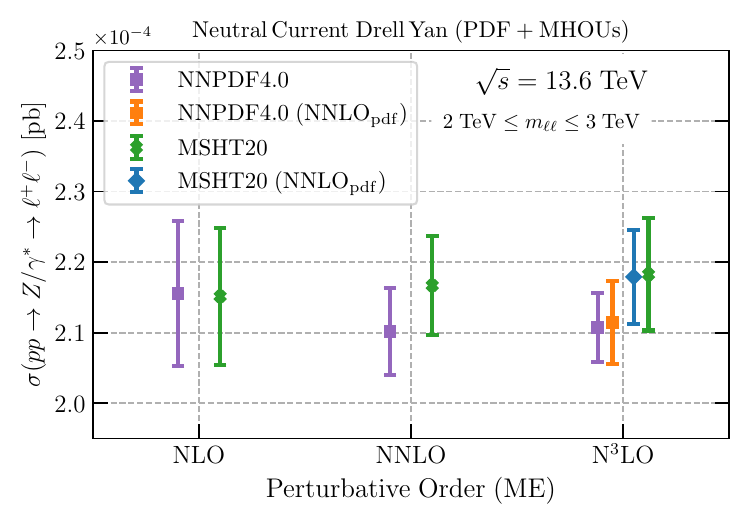}
  \includegraphics[width=0.44\linewidth]{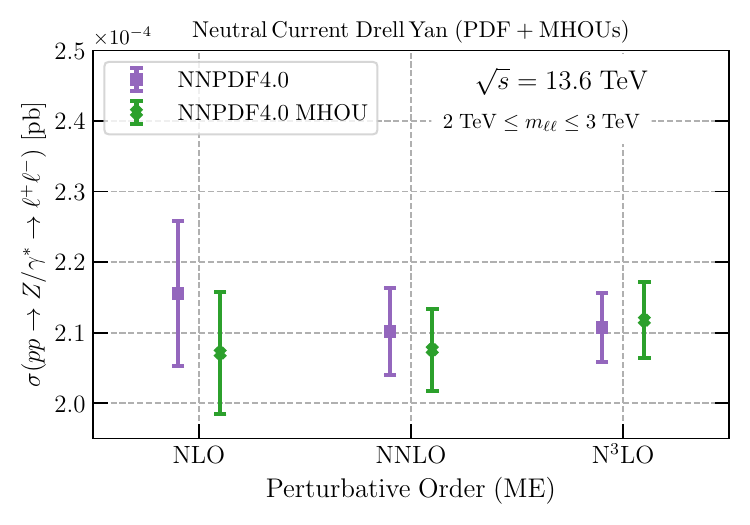}
  \caption{
    \small The inclusive NC DY production cross-section, $pp\to \gamma^*/Z \to \ell^+\ell^-$,
    for different ranges of the dilepton invariant mass $Q=m_{\ell\ell}$, from
    low to high invariant masses (top to bottom).
    In the left column results are shown comparing NLO, NNLO and aN$^3$LO
    with matched perturbative order in the matrix element and PDF, 
    and also at aN$^3$LO with NNLO PDFs using NNPDF4.0 and MSHT20 PDFs and at aN$^3$LO;
    in the right column, with PDFs without and with MHOUs.
  }
  \label{fig:nc-dy-pheno} 
\end{figure}

\begin{figure}[!p]
  \centering
  \includegraphics[width=0.48\linewidth]{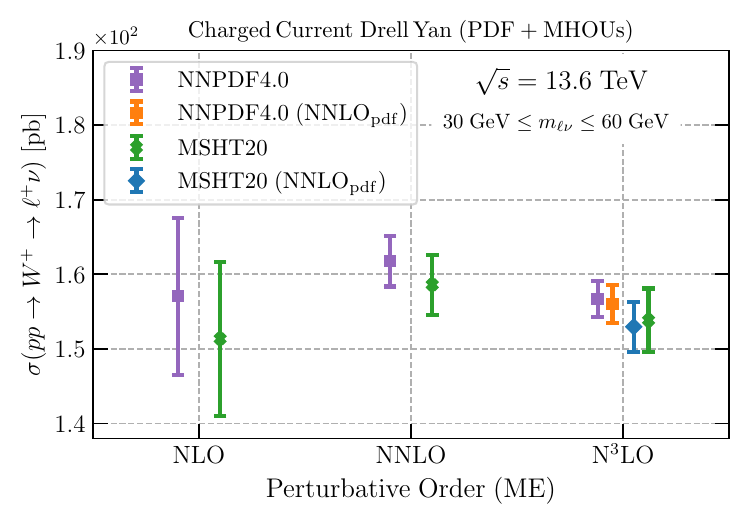}
  \includegraphics[width=0.48\linewidth]{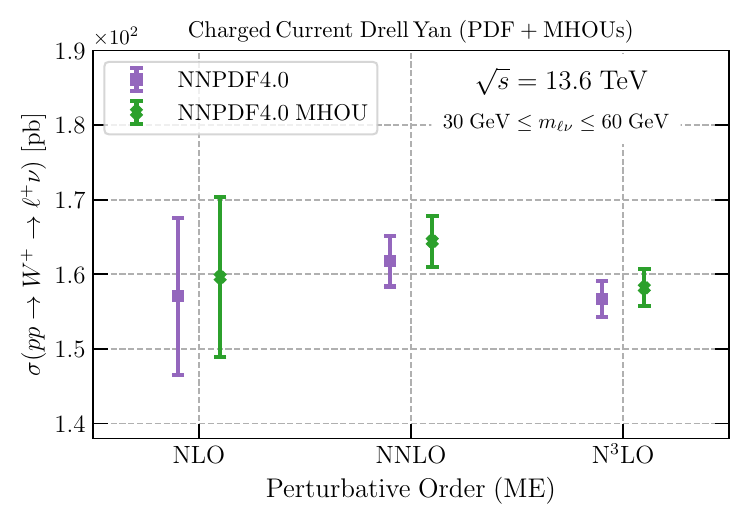}
  \includegraphics[width=0.48\linewidth]{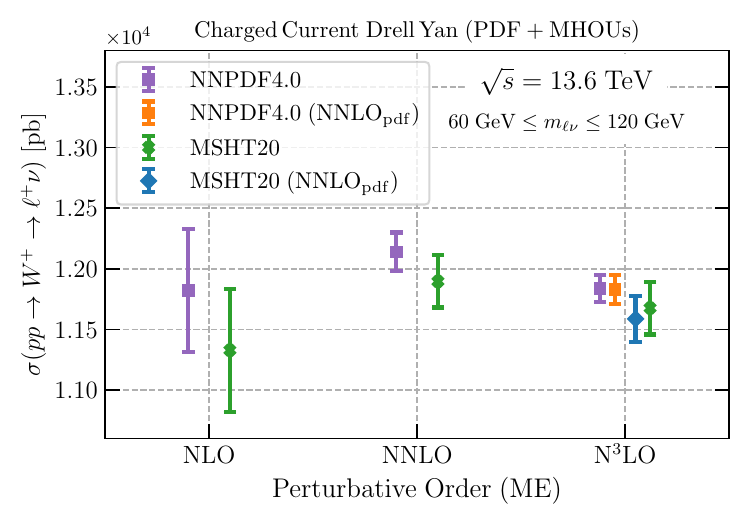}
  \includegraphics[width=0.48\linewidth]{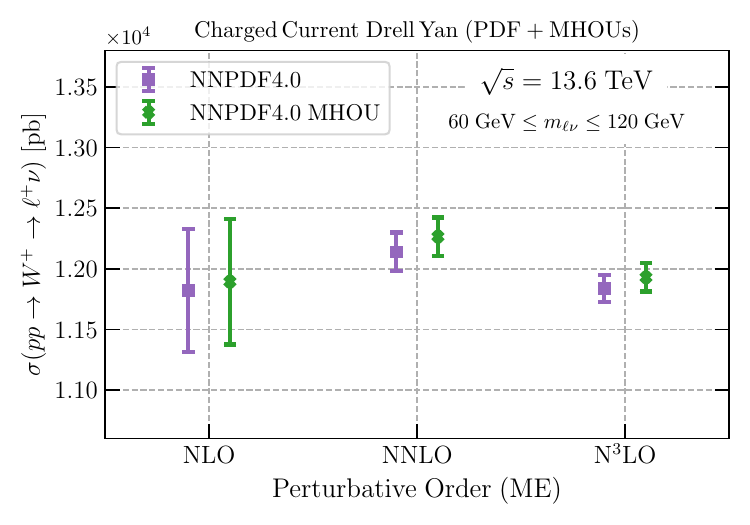}
  \includegraphics[width=0.48\linewidth]{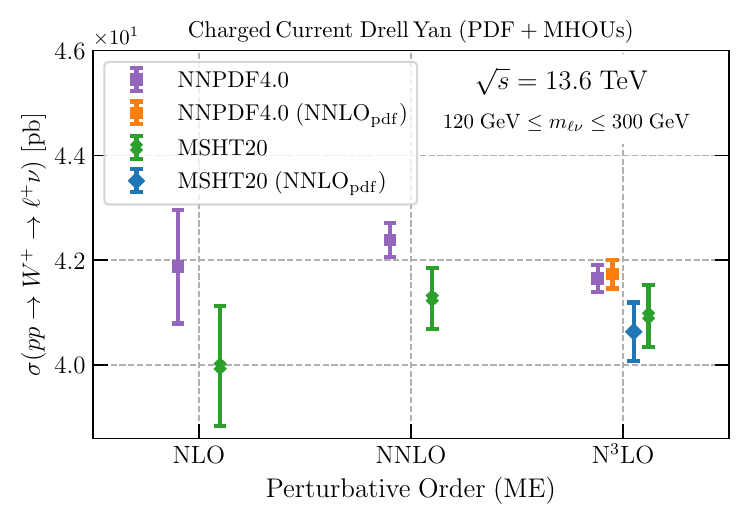}
  \includegraphics[width=0.48\linewidth]{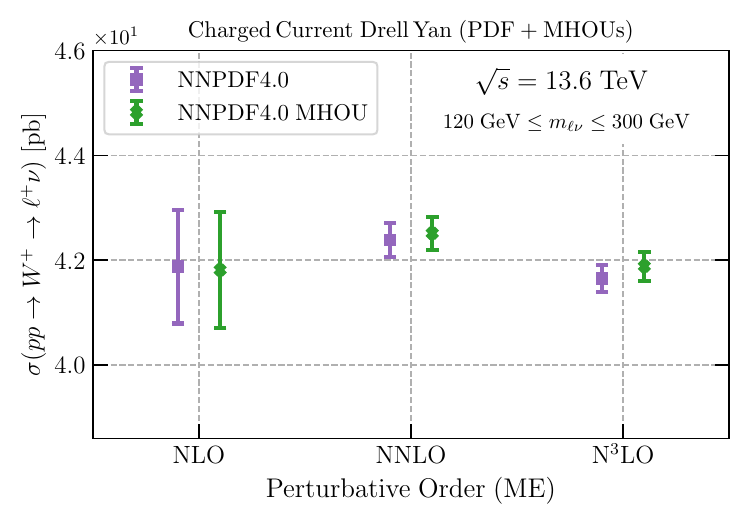}
  \includegraphics[width=0.48\linewidth]{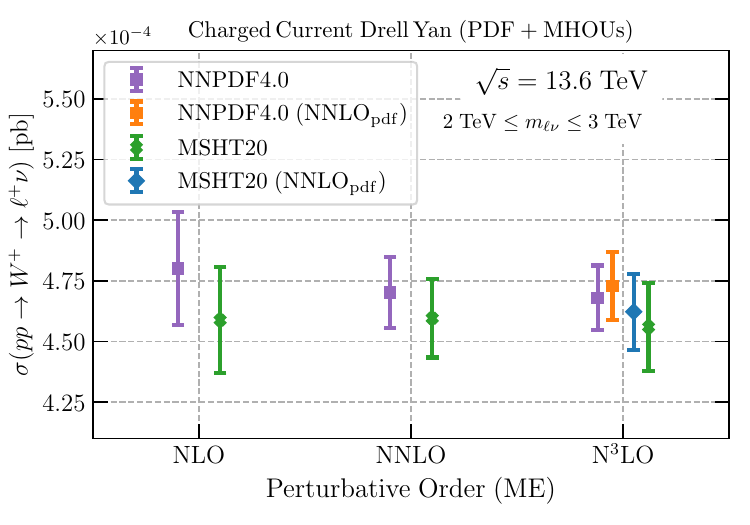}
  \includegraphics[width=0.48\linewidth]{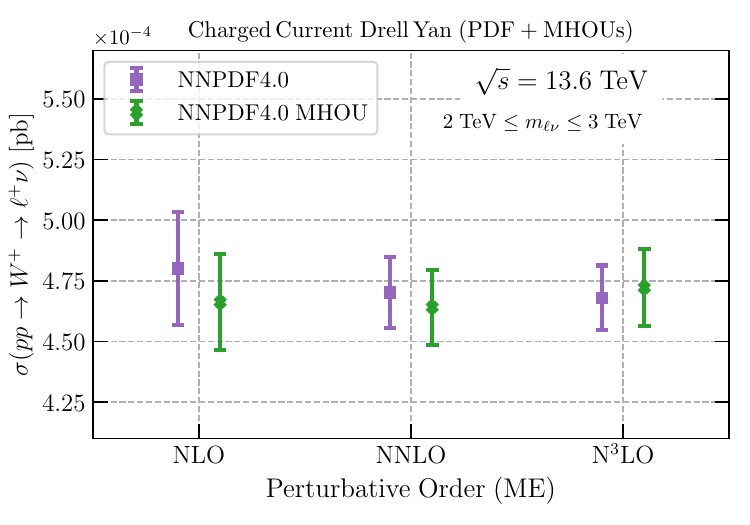}
  \caption{
    \small Same as \cref{fig:nc-dy-pheno} for the inclusive CC DY
    production cross-section, $pp\to W^+ \to \ell^+\nu_{\ell}$.
  }
  \label{fig:ccp-dy-pheno} 
\end{figure}

\begin{figure}[!p]
  \centering
  \includegraphics[width=0.48\linewidth]{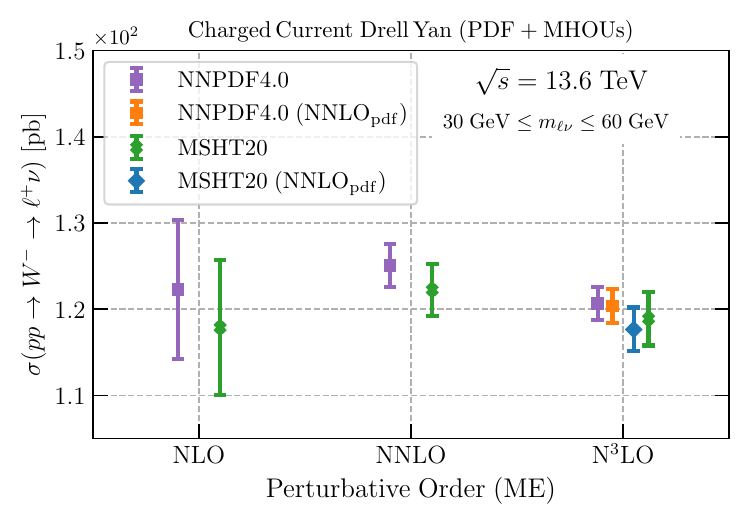}
  \includegraphics[width=0.48\linewidth]{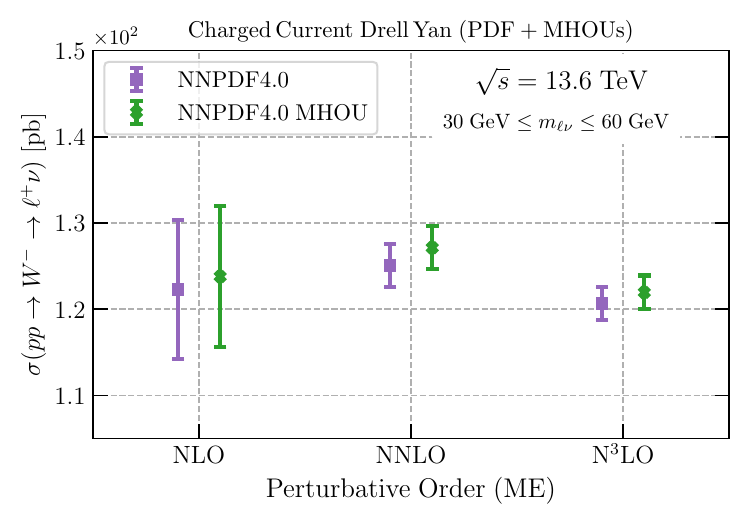}
  \includegraphics[width=0.48\linewidth]{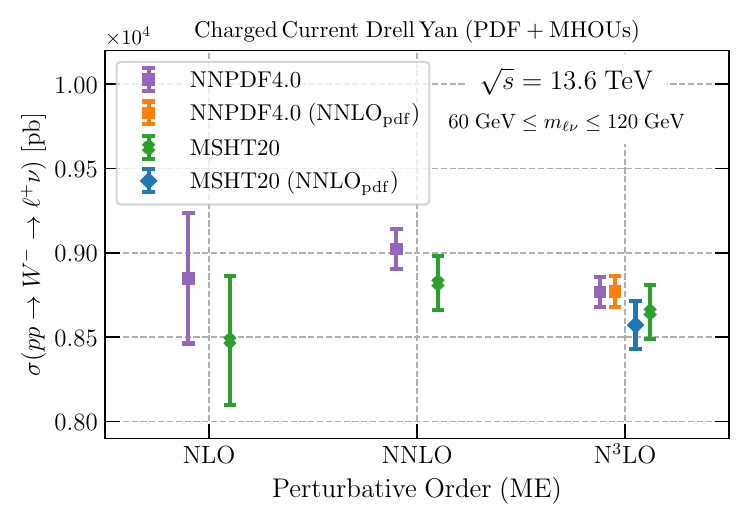}
  \includegraphics[width=0.48\linewidth]{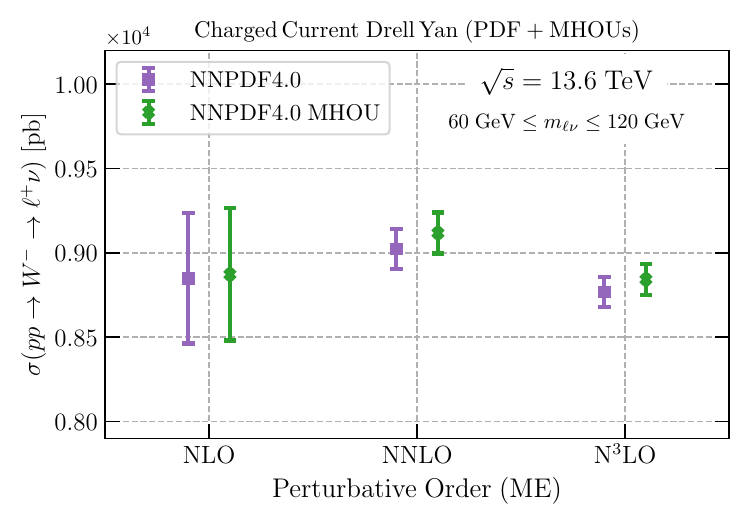}
  \includegraphics[width=0.48\linewidth]{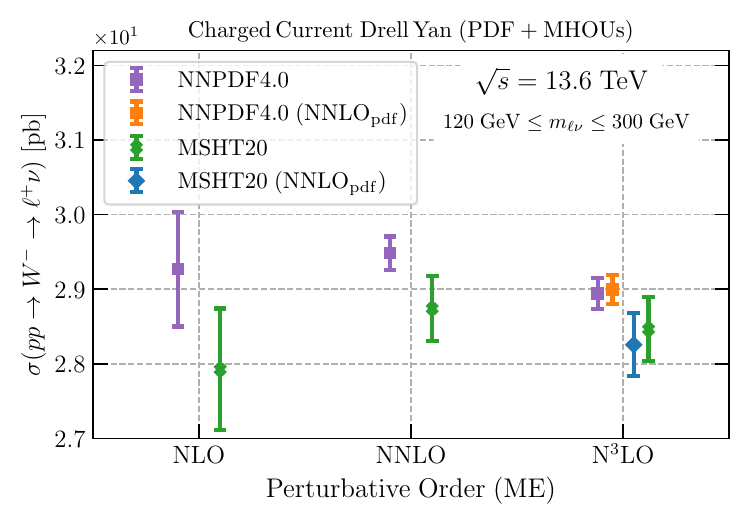}
  \includegraphics[width=0.48\linewidth]{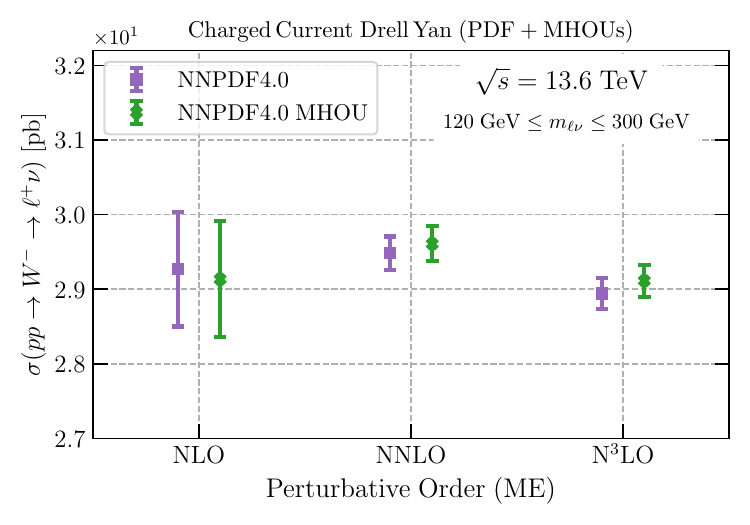}
  \includegraphics[width=0.48\linewidth]{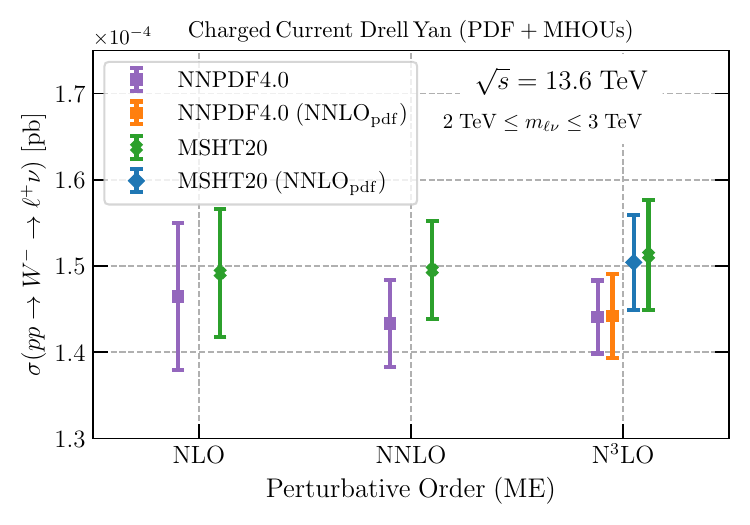}
  \includegraphics[width=0.48\linewidth]{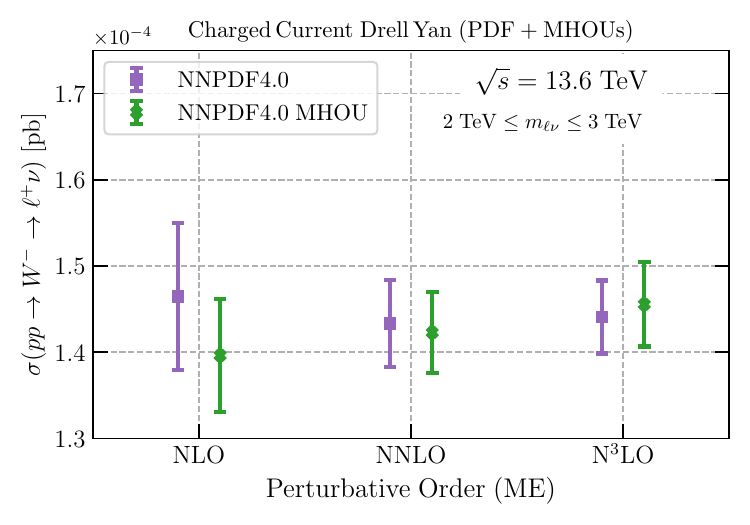}
  \caption{
    \small Same as \cref{fig:nc-dy-pheno} for the inclusive CC DY
    production cross-section, $pp\to W^- \to \ell^-\bar{\nu}_{\ell}$.
  }
  \label{fig:ccm-dy-pheno} 
\end{figure}

We start showing results for inclusive CC and NC gauge boson production 
cross-sections followed by their decays into the dilepton final state.
Cross-sections are evaluated using the {\sc\small n3loxs} code~\cite{Baglio:2022wzu}
for different ranges in the final-state dilepton invariant mass, 
$Q=m_{\ell\ell}$ for NC and $Q=m_{\ell \nu}$ for CC scattering.
\cref{fig:nc-dy-pheno} displays the inclusive NC DY cross-section 
$pp\to \gamma^*/Z \to \ell^+\ell^-$ and \cref{fig:ccp-dy-pheno,fig:ccm-dy-pheno} 
the CC cross-sections $pp\to W^\pm \to \ell^\pm\nu_{\ell}$.
We consider one low-mass bin ($30~{\rm GeV}\le Q \le 60~{\rm GeV}$),
the mass peak bin ($60~{\rm GeV}\le Q \le 120~{\rm GeV}$), and two high-mass
bins ($120~{\rm GeV}\le Q \le 300~{\rm GeV}$ and $2~{\rm TeV}\le Q\le 3~{\rm TeV}$), 
relevant for high-mass new physics searches.
In all cases, we compare the NLO, NNLO, and aN$^3$LO predictions using
NNPDF4.0 and MSHT20 PDFs determinations, with the same perturbative order 
in matrix element and PDFs, and also the aN$^3$LO result with NNLO PDFs, 
and then we compare the aN$^3$LO with NNPDF4.0 aN$^3$LO PDFs with and without MHOUs.
The values of cross-sections and uncertainties are collected in \cref{tab:DY_unc}.

\begin{table}[!t]
  \scriptsize
  \centering
  \renewcommand{\arraystretch}{1.7}
    \begin{tabularx}{\textwidth}{Xccccccccccc}
    \toprule
 \multirow{2}{*}{Process} & 
  \multicolumn{6}{c}{NNPDF4.0}
  & \multicolumn{5}{c}{MSHT20} \\
 &  $\sigma$ (pb) & $\delta_{\rm th}$
  &   $\delta^{\rm no MHOU}_{\rm PDF}$  &
  $\delta^{\rm MHOU}_{\rm PDF}$ &   $\Delta^{\rm app}_{\rm NNLO}$
  & $\Delta^{\rm exact}_{\rm NNLO}$   
 &  $\sigma$ (pb) & $\delta_{\rm th} \sigma$
  &   $\delta_{\rm PDF}$&   $\Delta^{\rm app}_{\rm
    NNLO}~ $   &$\Delta^{\rm exact}_{\rm NNLO}$    \\
  \midrule
  $W^+$  &  $1.2 \times 10^{4}$  &    1.0 & 0.5 & 0.5   & 1.1  &
  0.1    &  $1.2 \times 10^{4}$  &    1.9 & 1.7 & 2.3   & 0.8  \\
  $W^-$  &  $8.8 \times 10^{3}$  &    1.0 & 0.5 & 0.5   & 1.1  & 
  0.1    &  $8.7 \times 10^{3}$  &    1.9 & 1.6 & 2.1   & 0.0   \\
  $Z$    &  $1.9 \times 10^{3}$  &    0.9 & 0.4 & 0.5   & 1.1  & 
  0.3    &  $1.9 \times 10^{3}$  &    1.8 & 1.6 & 2.6   & 0.3  \\
\bottomrule
\end{tabularx}

  \vspace{0.3cm}
  \caption{
    The N$^3$LO cross-sections and uncertainties for the
    inclusive gauge boson production processes displayed in
    \cref{fig:nc-dy-pheno,fig:ccp-dy-pheno,fig:ccm-dy-pheno} and evaluated using the
    NNPDF4.0 and MSHT20 aN$^3$LO PDFs.
    We show the percentage total theory uncertainty $\delta_{\rm th}$,
    obtained adding in quadrature the 7-point scale variation
    MHOUs and the PDF uncertainty $\delta_{\rm PDF}$ (not including
    MHOUs in the fit), which is also separately provided.
    In the case of NNPDF4.0 the value of $\delta_{\rm PDF}$ with
    MHOUs in the fit is also listed. All uncertainties are
    expressed as percentage of the cross-section. 
    We finally show the error $\Delta^{\rm exact}_{\rm NNLO}$ 
    (\cref{eq:PDFimpact_xsec_exact}) due to using NNLO PDFs at N$^3$LO, 
    and the estimate of this error $\Delta^{\rm app }_{\rm NNLO}$
    (\cref{eq:PDFimpact_xsec_approx}), also expressed as a percentage. 
  }
  \label{tab:DY_unc}
\end{table}

In general, we observe a good perturbative convergence, with predictions at two 
subsequent orders in agreement within uncertainties, and generally improved 
convergence upon including MHOUs on the PDF.
Predictions based on NNPDF4.0 and MSHT20 are always consistent with
each other within uncertainties.
From \cref{fig:nc-dy-pheno,fig:ccp-dy-pheno,fig:ccm-dy-pheno,tab:DY_unc} we can draw three main conclusions. 
First, in many cases differences between the NNLO and N$^3$LO predictions tend to be
reduced when using consistently the appropriate PDFs at each order, rather than NNLO 
PDFs with N$^3$LO matrix elements (though in some cases the results are unchanged).
For instance, for the two lowest $m_{\ell\ell}$ bins for NC production
aN$^3$LO PDFs drive upwards the N$^3$LO prediction, making it
closer to the NNLO result. 
Second, the difference between PDFs with and without MHOUs, while moderate, 
remains non-negligible even at N$^3$LO, where it starts being comparable 
to the overall uncertainty, and thus it must be included in precision calculations. 
Third, the impact of using aN$^3$LO instead of NNLO PDFs is actually smaller than
the guess based on the estimate of \cref{eq:PDFimpact_xsec_approx}.

\subsection{Inclusive Higgs production}
\label{sec:higgsproduction}

\begin{figure}[!t]
  \centering
  \includegraphics[width=0.49\linewidth]{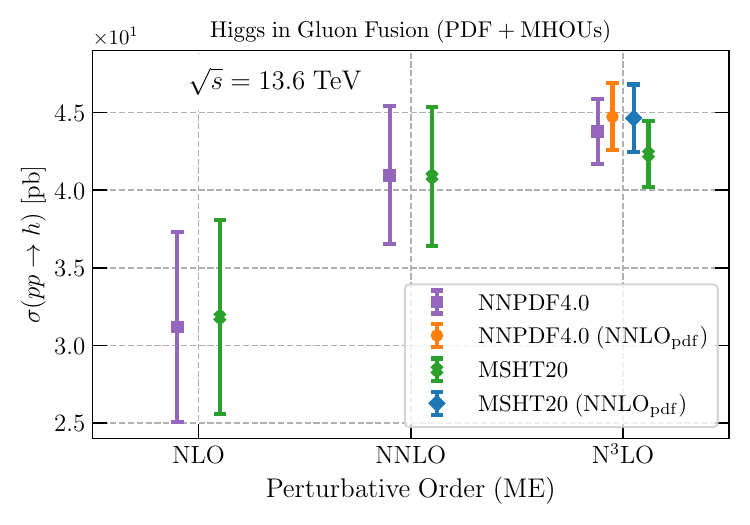}
  \includegraphics[width=0.49\linewidth]{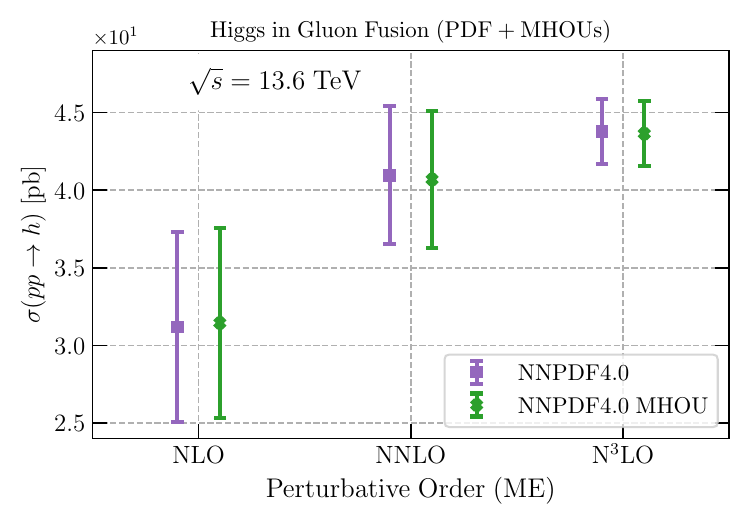}
  \includegraphics[width=0.49\linewidth]{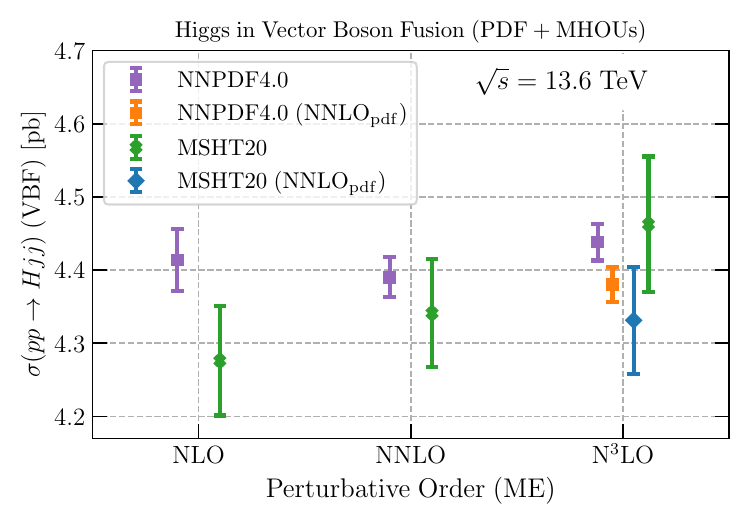}
  \includegraphics[width=0.49\linewidth]{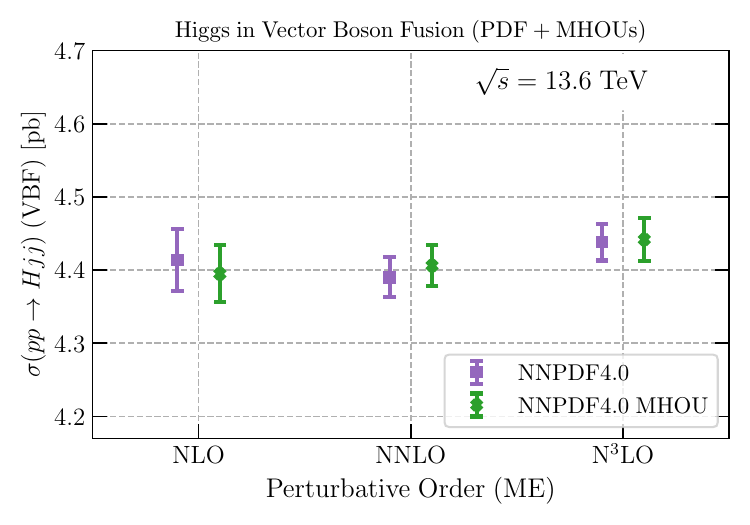}
  \caption{
    \small Same as \cref{fig:nc-dy-pheno} for Higgs production in gluon-fusion and 
    via vector-boson fusion.
  }
  \label{fig:higgs-pheno-1} 
\end{figure}

\begin{figure}[!t]
  \centering
  \includegraphics[width=0.49\linewidth]{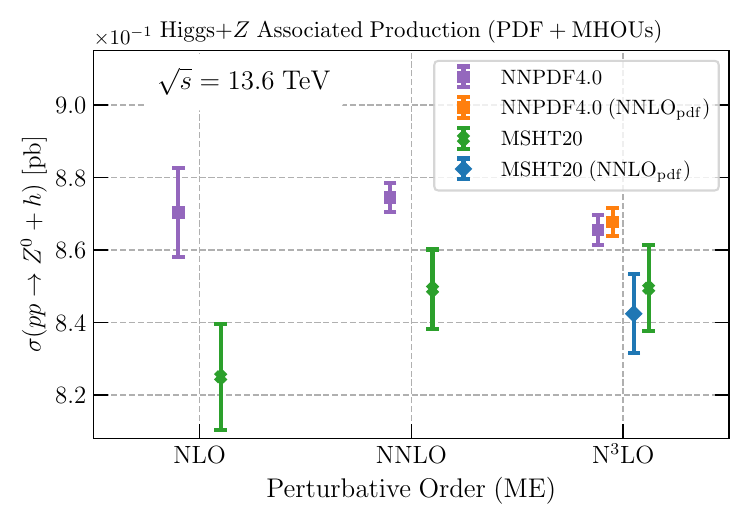}
  \includegraphics[width=0.49\linewidth]{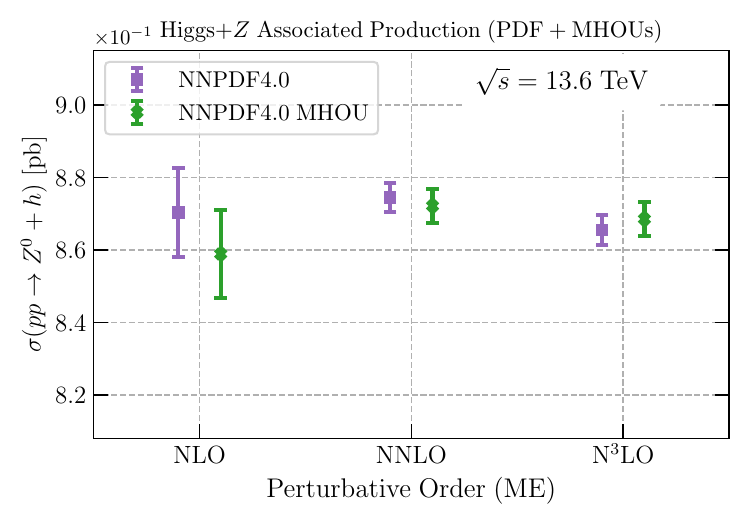}
  \includegraphics[width=0.49\linewidth]{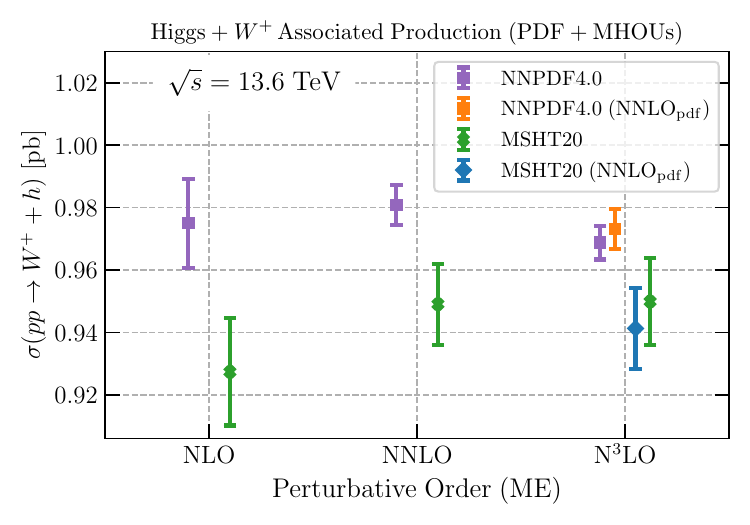}
  \includegraphics[width=0.49\linewidth]{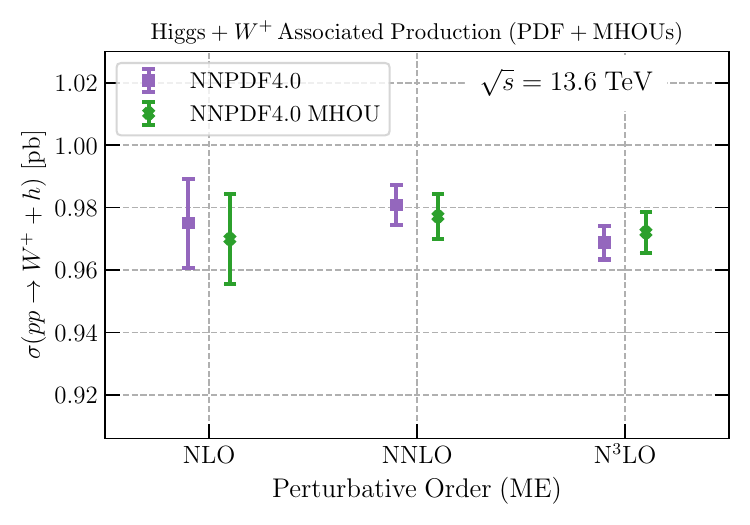}
  \includegraphics[width=0.49\linewidth]{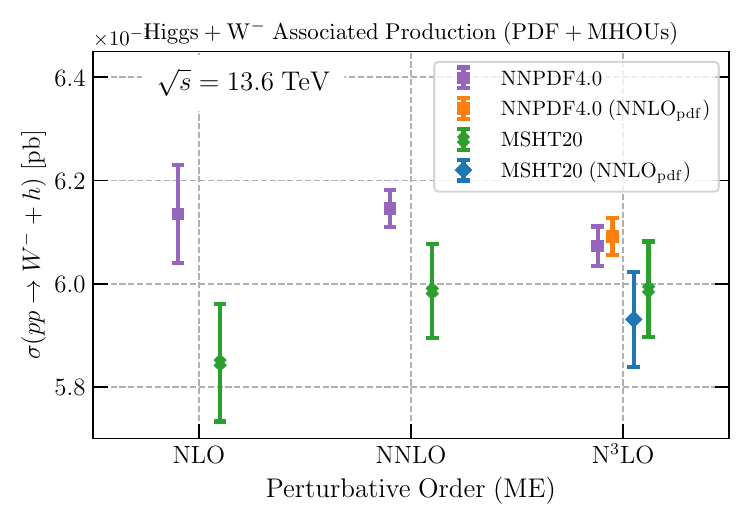}
  \includegraphics[width=0.49\linewidth]{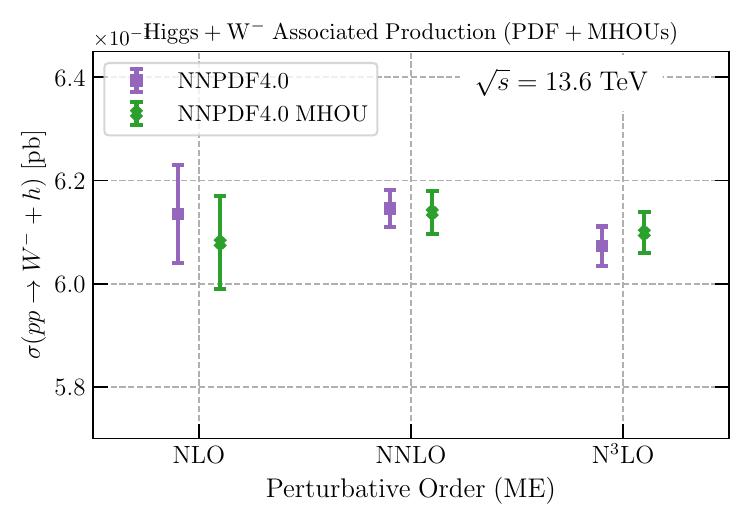}
  \caption{
    \small Same as \cref{fig:nc-dy-pheno} for Higgs production in association
    with $W^+$ and $Z$ gauge bosons: from top to bottom, $Zh$, $W^+h$, and $W^-h$.
  }
  \label{fig:higgs-pheno-2} 
\end{figure}

We now turn to Higgs production in gluon fusion, via vector-boson fusion (VBF) and 
in associated production with vector bosons. 
Predictions are obtained using the {\sc\small ggHiggs} code~\cite{Bonvini:2014jma} 
for gluon fusion, {\sc\small proVBFH} code~\cite{Dreyer:2018qbw} for VBF
and  {\sc\small n3loxs} for associate production.
Results are shown in \cref{fig:higgs-pheno-1} and \cref{tab:Higgs_unc}
for gluon-fusion and VBF, and \cref{fig:higgs-pheno-2} for associate
production with $W^+$ and $Z$.

\begin{table}[!t]
  \scriptsize
  \centering
  \renewcommand{\arraystretch}{1.4}
    \begin{tabularx}{\textwidth}{Xccccccccccc}
    \toprule
 \multirow{2}{*}{Process} & 
  \multicolumn{6}{c}{NNPDF4.0}
  & \multicolumn{5}{c}{MSHT20} \\
 &  $\sigma$ (pb) & $\delta_{\rm th}$
  &   $\delta^{\rm no MHOU}_{\rm PDF}$  &   $\delta^{\rm
    MHOU}_{\rm PDF}$ &   $\Delta^{\rm app}_{\rm NNLO}$
  & $\Delta^{\rm exact }_{\rm NNLO} $   
 &  $\sigma$ (pb) & $\delta_{\rm th} \sigma\,  $
  &   $\delta_{\rm PDF}~ $&   $\Delta^{\rm app}_{\rm
    NNLO}~ $   &$\Delta^{\rm exact }_{\rm NNLO} $    \\
  \midrule
  $gg\to h$  &  43.8   &    4.8  & 0.6  & 0.7   & 0.2     &   2.2   &    42.3  &  5.1   &  1.7    & 1.4    & 5.3  \\
  $h$ VBF  &   4.44  &   0.6    & 0.5  & 0.6    &  0.2    &  1.3    &    4.46    &    2.1      & 2.0   & 1.3   & 2.9  \\
  $hW^+$  &   0.97  &    0.6    &  0.5  & 0.6   &  0.2    &   0.5   &  0.95    &   1.5        & 1.4   &   0.8  &0.9   \\
  $hW^-$  &      0.61    &  0.6    & 0.6  &  0.6  & 0.2     & 0.3      &0.60      & 1.6     & 1.5  &   0.9  &  1.0 \\
  $hZ$  &     0.87  &  0.5    &  0.4 &   0.5   &  0.1    &  0.3    &    0.85    &    1.4    &  1.4   &  1.1  &   0.8 \\
\bottomrule
\end{tabularx}

  \vspace{0.3cm}
  \caption{
    Same as \cref{tab:DY_unc} for the Higgs production processes
    displayed in \cref{fig:higgs-pheno-1,fig:higgs-pheno-2}
  }
  \label{tab:Higgs_unc}
\end{table}

Here as well we observe generally good perturbative convergence, even for
gluon fusion, that notoriously has a very slowly converging expansion.
There is generally better agreement between NNPDF4.0 and MSHT20 
as the perturbative order increases, except for gluon fusion 
where the agreement is similar at all orders. Indeed, in all
cases MSHT20 and NNPDF4.0 results agree within uncertainties at
aN$^3$LO, while they do not at NLO for VBF, nor at NLO and NNLO
for associated production.
The impact of using aN$^3$LO PDFs instead of NNLO PDFs at N$^3$LO for NNPDF4.0 is
very moderate for gluon fusion, somewhat more significant for associated
production, and more significant for VBF, in which it is comparable to the 
PDF uncertainty. For MSHT20 instead a significant change from using aN$^3$LO 
instead of NNLO PDFs is also observed for gluon fusion, where
suppression of the cross-sections is seen when replacing NNLO with
aN$^3$LO PDFs. This follows from the behavior of the gluon luminosity
seen in \cref{fig:lumis_NLO_vs_N3LO}.
The impact of MHOUs on the PDFs is generally quite small on the scale
of the PDF uncertainty at all perturbative orders, and essentially 
absent for gluon fusion. For associated production it marginally improves 
perturbative convergence. Interestingly, for NNPDF4.0, for all
Higgs production processes considered, and especially for gluon fusion, 
the estimate of \cref{eq:PDFimpact_xsec_approx} is a substantial underestimate 
of the actual error which is made using NNLO PDFs at N$^3$LO. 
This follows from the fact that for $m_X\sim 100$~GeV the NNLO gluon-gluon 
luminosity is actually closer to the NLO than to the aN$^3$LO 
(see \cref{fig:lumis}), which in turn appears to be an accidental consequence
of the behavior of the gluon PDF for $x\sim 10^{-2}$. 

\section{Summary}
\label{sec:an3lo_summary}
In this chapter we have presented the first aN$^3$LO PDF sets within 
the NNPDF framework, by constructing a full set of approximate N$^3$LO 
splitting functions based on available partial results and known limits, 
approximate massive DIS coefficient functions, and extending to this order 
the FONLL general-mass scheme for DIS coefficient functions. 

The NNPDF4.0 aN$^3$LO PDF sets are available via the \lhapdf{}
interface, 
\begin{center}
    {\bf \url{http://lhapdf.hepforge.org/}~}
\end{center}
and on the NNPDF Collaboration website,
\begin{center}
    {\bf \url{https://nnpdf.mi.infn.it/nnpdf4-0-n3lo/}~}.
\end{center}

In addition to the \lhapdf{} grids themselves, all the results obtained 
in this chapter are reproducible by means of the open-source NNPDF 
code~\cite{NNPDF:2021uiq} and the related suite of theory tools.

We have provided a first assessment of these PDF sets by comparing them to their 
NLO and NNLO counterparts with and without MHOUs. Our main conclusions are the following
\begin{itemize}
    \item For all PDFs good perturbative convergence is observed, with 
    differences decreasing as the perturbative order increases, and the
    aN$^3$LO result always compatible with the NNLO within uncertainties.

    \item For quark PDFs the difference between NNLO and aN$^3$LO 
    results is tiny, suggesting that with current data and 
    methodology the effect of yet higher orders is negligible.
    
    \item For the gluon PDF a more significant shift is observed between
    NNLO and N$^3$LO, thus making the inclusion of N$^3$LO important for 
    precision phenomenology.
    
    \item The inclusion of MHOUs improves perturbative convergence, mostly 
    by shifting central values at each order towards the higher-order result, 
    by an amount that decreases with increasing perturbative order.

    \item Upon inclusion of MHOUs the fit quality becomes all but independent 
    of perturbative order, and PDF uncertainties generally decrease (or remain 
    unchanged) due to the improved data compatibility.

    \item The effect of MHOUs at N$^3$LO is very small for quarks but not negligible 
    for the gluon PDF.
    
    \item Evidence for intrinsic charm is somewhat increased already at NNLO by 
    the inclusion of MHOUs, and somewhat increased again when going from NNLO to N$^3$LO.
    
    \item The impact of N$^3$LO corrections on the total cross-section for Higgs 
    in gluon fusion is very small on the scale of the PDF uncertainty.
\end{itemize}

All in all, these results underline the importance of the inclusion of N$^3$LO corrections 
and MHOUs for precision phenomenology at sub-percent accuracy.

Future NNPDF releases will include by default MHOUs, will be at all orders up to 
aN$^3$LO, and will include a photon PDF. Specifically, we aim to extend to 
aN$^3$LO with MHOUs our recent construction of NNPDF4.0QED PDFs~\cite{NNPDF:2024djq,Barontini:2024dyb}.
Indeed, aN$^3$LO PDFs including a photon PDF (such as those recently released by
MSHT20~\cite{Cridge:2023ryv}) will be a necessary ingredient for theory predictions 
based on state-of-the art QCD and electroweak (EW) corrections. 

Another important line of future development involves the all-order resummation of 
potentially large perturbative contributions in the large $x$ and small $x$ regions
~\cite{Bonvini:2015ira,Ball:2017otu}. This will involve matching
resummed and fixed-order cross-sections and (at small $x$) perturbative evolution 
in the new streamlined NNPDF theory pipeline. Such resummed PDFs will be instrumental 
for precision phenomenology: specifically at small $x$, forward neutrino production 
at the LHC and scattering processes for high-energy astroparticle physics, and at
large $x$, searches for new physics in high-mass final states at the LHC and future 
hadron colliders.


  \chapter{NNLO polarized PDFs with MHOU}
\label{chap:pol}
\begin{center}
\begin{minipage}{1.\textwidth}
    \begin{center}
        \textit{
          This chapter is based my result presented in Refs.~\cite{Cruz-Martinez:2025ahf}.
          In this study my contribution has focused on the development of the 
          theoretical framework needed to extract polarized PDFs and to the data 
          implementation.  
        } 
    \end{center}
\end{minipage}
\end{center}

The interest in helicity dependent (polarized henceforth) PDFs is mainly related 
to the fact that their lowest moments are proportional to the proton axial currents,
which express the fraction of proton spin carried by quarks and
gluons~\cite{Anselmino:1994gn}. In spite of tremendous experimental and
theoretical investigations over the past thirty years, knowledge of
polarized PDFs remains limited in comparison to their unpolarized counterparts,
in particular concerning the distributions of sea quarks and gluons. This fact
hinders the fundamental understanding of proton spin decomposition in the
framework of QCD~\cite{Ji:2020ena}.

The Electron-Ion Collider (EIC)~\cite{AbdulKhalek:2021gbh,AbdulKhalek:2022hcn},
expected to start its operations in the 2030s, is designed to revolutionize
this state of affairs. The EIC will have the possibility to collide polarized
proton and lepton beams, so to measure the polarized inclusive and
semi-inclusive DIS structure functions, to which
polarized PDFs are related through factorization theorem~\cite{Collins:1989gx}.
These measurements are forecast to cover an unprecedented range of proton
momentum fraction $x$ and virtuality $Q^2$, and to attain percent-level precision. 
These facts call for a matching accuracy of the corresponding theoretical predictions, 
which require in turn an improvement in the accuracy of perturbative computations and
of polarized PDF determinations.

As a further step forward into this context, in this paper we present
{\sc NNPDFpol2.0}, a new determination of the proton polarized PDFs based on
the NNPDF methodology~(cf. \cref{chap:methodology}). This determination improves the
previous one, {\sc NNPDFpol1.1}~\cite{Nocera:2014gqa}, in three respects,
which set it apart from~\cite{Bertone:2024taw,Borsa:2024mss}.

\begin{enumerate}

\item We extend the range of fitted datasets. We specifically consider
  measurements of polarized inclusive lepton-nucleon DIS, including legacy
  measurements from HERMES and COMPASS, and measurements of $W$-boson,
  single-inclusive jet, and dijet production from STAR. 

\item We incorporate higher-order corrections in PDF evolution and in the
  hard cross-sections, whenever available, up to NNLO in the strong coupling. 
  We likewise incorporate heavy quark mass corrections in the analysis of
  polarized inclusive DIS structure functions, according to the FONLL
  scheme implemented in~\cite{Hekhorn:2024tqm,Barontini:2024xgu}. We include
  uncertainties due to QCD missing higher-order corrections (MHOUs) by means of
  the methodology developed in~\cite{NNPDF:2019vjt,NNPDF:2019ubu,NNPDF:2024dpb},
  whereby MHOUs are treated through a theory covariance matrix determined by
  scale variations. 

\item We deploy the machine-learning methodology developed
  in~\cite{NNPDF:2021njg}. 
  All aspects of the parametrization and optimization (such as the neural network architecture
  or the choice of minimization algorithm) are now selected through a
  hyperparameter optimization procedure~\cite{Carrazza:2019mzf}, which
  consists in an automated scan of the space of models. 
  
\end{enumerate}

\paragraph{Outline.}

The structure of the chapter is as follows. In \cref{sec:pol_data-theory} 
we review the type of experimental measurements adopted in our fit of polarized
PDFs together with the setting used to compute the corresponding theoretical predictions.
We continue in \cref{sec:pol_methodology} discussing the fitting methodology, 
along with the newer hyperoptimization algorithm adopted to select the optimal 
fit setup. In \cref{sec:pol_results} we conclude highlighting the most relevant 
feature of our new PDFs sets.


\section{Experimental and theoretical input}
\label{sec:pol_data-theory}
In this section, we present the experimental and theoretical input entering
the {\sc NNPDFpol2.0} determination. We first introduce the dataset,
describing the details of each measurement and the computational tools used to
obtain the corresponding theoretical predictions. We then discuss their
perturbative accuracy and specifically the way in which we account for MHOUs.

\subsection{The {\sc NNPDFpol2.0} dataset}
\label{subsec:pol_data}

The {\sc NNPDFpol2.0} parton set is based on measurements of three different
polarized observables: the structure function $g_1$ in polarized inclusive 
lepton-nucleon DIS; the longitudinal single-spin asymmetry $A_L^{W^\pm}$ for
$W^\pm$-boson production in polarized proton-proton collisions; and the
longitudinal double-spin asymmetry $A_{LL}^{1-;2-{\rm jet}}$ for single-inclusive
jet and dijet production in polarized proton-proton collisions. The definition
of these observables can be found, {\it e.g.}, in \cite[Sec.~3]{Ball:2013lla}, 
and in \cite[Sec.~6.2.2]{Nocera:2014vla}.
We review the measurements that we include for each of these observables in turn.

\begin{description}

\item[polarized inclusive DIS structure function.] We include measurements
  performed by the EMC~\cite{EuropeanMuon:1989yki},
  SMC~\cite{SpinMuon:1998eqa,SpinMuon:1999udj},
  and COMPASS~\cite{COMPASS:2015mhb,COMPASS:2016jwv}
  experiments at CERN, by the E142~\cite{E142:1996thl},
  E143~\cite{E143:1998hbs}, E154~\cite{E154:1997xfa},
  and E155~\cite{E155:2000qdr} experiments at SLAC, by the HERMES experiment
  at DESY~\cite{HERMES:1997hjr,HERMES:2006jyl}, and by the
  Hall~A~\cite{JeffersonLabHallA:2016neg,Kramer:2002tt,
    JeffersonLabHallA:2004tea}
  and CLAS~\cite{CLAS:2014qtg,CLAS:2006ozz} experiments at JLab. All of these
  experiments provide data for the polarized inclusive DIS structure function
  $g_1$, reconstructed from the longitudinal double-spin asymmetry (see,
  {\it e.g.} \cite[Sec.~2.1]{Ball:2013lla} for details), except SMC low-$x$,
  E155, Hall A, and CLAS, which instead provide data for $g_1$ normalized to the
  unpolarized inclusive structure function $F_1$. The details of the
  measurements, including their kinematic coverage in the proton momentum
  fraction $x$ and virtuality $Q^2$, are summarized in \cref{tab:DIS_data}.

\begin{table}[!t]
    \scriptsize
    \centering
    \renewcommand{\arraystretch}{1.4}
    \begin{tabularx}{\textwidth}{Xccccc}
  \toprule
  Dataset
  & Ref.
  & $N_{\rm {dat}}$
  & $x$
  & $Q^2$~[GeV$^2$]
  & Theory \\
  \midrule
  EMC $g_1^p$
  & \cite{EuropeanMuon:1989yki}
  & $10 \ (10)$
  & [0.015, 0.466]
  & [3.5, 29.5]
  & {\sc Yadism} \\
  SMC $g_1^p$
  & \cite{SpinMuon:1998eqa}
  & $13 \ (12)$
  & [0.002, 0.48]
  & [0.50, 54.8]
  & {\sc Yadism} \\
  SMC $g_1^d$
  & \cite{SpinMuon:1998eqa}
  & $13 \ (12)$
  & [0.002, 0.48]
  & [0.50, 54.80]
  & {\sc Yadism} \\
  SMC low-$x$ $g_1^p/F_1^p$
  & \cite{SpinMuon:1999udj}
  & $15 \ (8)$
  & [0.00011, 0.121]
  & [0.03, 23.1]
  & {\sc Yadism} \\
  SMC low-$x$ $g_1^d/F_1^d$
  & \cite{SpinMuon:1999udj}
  & $15 \ (8)$
  & [0.00011, 0.121]
  & [0.03, 22.9]
  & {\sc Yadism} \\
  COMPASS $g_1^p$
  & \cite{COMPASS:2015mhb}
  & $17 \ (17)$
  & [0.0036, 0.57]
  & [1.1, 67.4]
  & {\sc Yadism} \\
  COMPASS $g_1^d$
  & \cite{COMPASS:2016jwv}
  & $15 \ (15)$
  & [0.0046, 0.567]
  & [1.1, 60.8]
  & {\sc Yadism}\\
  \midrule
  E142 $g_1^n$
  & \cite{E142:1996thl}
  & $8 \ (8)$
  & [0.035, 0.466]
  & [1.1, 5.5]
  & {\sc Yadism} \\
  E143 $g_1^p$
  & \cite{E143:1998hbs}
  & $28 \ (27)$
  & [0.035, 0.466]
  & [1.27, 9.52]
  & {\sc Yadism} \\
  E143 $g_1^d$
  & \cite{E143:1998hbs}
  & $28 \ (27)$
  & [0.031, 0.749]
  & [1.27, 9.52]
  & {\sc Yadism} \\       
  E154 $g_1^n$
  & \cite{E154:1997xfa}
  & $11 \ (11)$
  & [0.017, 0.024]
  & [1.2, 15.0]
  & {\sc Yadism} \\
  E155 $g_1^p/F_1^p$
  & \cite{E155:2000qdr}
  & $24 \ (24)$
  & [0.015, 0.750]
  & [1.22, 34.72]
  & {\sc Yadism} \\
  E155 $g_1^n/F_1^n$
  & \cite{E155:2000qdr}
  & $24 \ (24)$
  & [0.015, 0.750]
  & [1.22, 34.72]
  & {\sc Yadism} \\
  \midrule 
  HERMES $g_1^n$
  & \cite{HERMES:1997hjr}
  & $9 \ (9)$
  & [0.033, 0.464]
  & [1.22, 5.25]
  & {\sc Yadism} \\
  HERMES $g_1^p$
  & \cite{HERMES:2006jyl}
  & $15 \ (15)$
  & [0.0264, 0.7248]
  & [1.12, 12.21]
  & {\sc Yadism} \\
  HERMES $g_1^d$
  & \cite{HERMES:2006jyl}
  & $15 \ (15)$
  & [0.0264, 0.7248]
  & [1.12, 12.21]
  & {\sc Yadism} \\
  \midrule
  JLAB E06 014 $g_1^n/F_1^n$
  & \cite{JeffersonLabHallA:2016neg}
  & $6 \ (4)$
  & [0.277, 0.548]
  & [3.078, 3.078]
  & {\sc Yadism} \\
  JLAB E97 103 $g_1^n$
  & \cite{Kramer:2002tt}
  & $5 \ (2)$
  & [0.160, 0.200]
  & [0.57, 1.34]
  & {\sc Yadism}\\
  JLAB E99 117 $g_1^n/F_1^n$
  & \cite{JeffersonLabHallA:2004tea}
  & $3 \ (1)$
  & [0.33, 0.60]
  & [2.71, 4.83]
  & {\sc Yadism}\\
  JLAB EG1 DVCS $g_1^p/F_1^p$
  & \cite{CLAS:2014qtg}
  & $47 \ (21)$
  & [0.154, 0.578]
  & [1.064, 4.115]
  & {\sc Yadism} \\
  JLAB EG1 DVCS $g_1^d/F_1^d$
  & \cite{CLAS:2014qtg}
  & $44 \ (19)$
  & [0.158, 0.574]
  & [1.078, 4.666]
  & {\sc Yadism}\\      
  JLAB EG1B $g_1^p/F_1^p$
  & \cite{CLAS:2006ozz}
  & $787 \ (114)$
  & [0.0262, 0.9155]
  & [0.0496, 4.96]
  & {\sc Yadism} \\
  JLAB EG1B $g_1^d/F_1^d$
  & \cite{CLAS:2006ozz}
  & $2465 \ (301)$
  & [0.0295, 0.9337]
  & [0.0496, 4.16]
  & {\sc Yadism} \\
  \bottomrule
\end{tabularx}

    \vspace{0.3cm}
    \caption{The polarized inclusive DIS measurements included in
      {\sc NNPDFpol2.0}. We denote each dataset with a name used throughout
      this paper, and we indicate its reference, number of data points before
      (after) applying kinematic cuts, kinematic coverage, and tool used to
      compute the corresponding theoretical predictions.}
    \label{tab:DIS_data}
\end{table}

We compute the corresponding theoretical predictions with
\yadism, which we have developed to handle the
computation of the polarized structure function $g_1$ and interfaced to
\pineappl. Predictions are accurate up to NNLO and include charm-quark
mass corrections through the FONLL general-mass variable-flavor-number
scheme~\cite{Forte:2010ta}, recently extended to the case of polarized 
structure functions~\cite{Hekhorn:2024tqm}. Target mass corrections are also
included as explained in Appendix~B of~\cite{Hekhorn:2024tqm}.
We nevertheless apply kinematic cuts on the virtuality $Q^2$ and on the
invariant mass of the final state $W^2$, by requiring
$Q^2\geq Q^2_{\rm min}=1.0~{\rm GeV}^2$ and $W^2\geq W^2_{\rm min}=4.0~{\rm GeV}^2$.
These cuts remove, respectively, the region where perturbative QCD becomes
unreliable due to the growth of the strong coupling, and the region where
higher-twist corrections in the factorization of $g_1$ (which we do not
include) may become sizeable. Nuclear corrections affecting experiments that
utilize a deuterium target are neglected. Whereas, in principle, they could be
accounted for as described in~\cite{Ball:2020xqw}, we consider them to be
negligible in comparison to the precision of the experimental measurements.
We therefore model the deuteron as the average of a proton and a neutron,
and relate the PDFs of the latter to the PDFs of the former assuming
isospin symmetry.

\item[$W$-boson longitudinal single-spin asymmetry.] We include the
  measurement of the longitudinal single-spin asymmetry for $W^\pm$-boson
  production in polarized proton-proton collisions, $A_L^{W^\pm}$, performed by
  STAR at a center-of-mass-energy $\sqrt{s}=510$~GeV~\cite{STAR:2018fty}.
  The measurement combines events recorded during the 2011-2012 and 2013 runs,
  and it supersedes the previous one~\cite{STAR:2014afm}. 
  It is given as a differential distribution in the lepton pseudorapidity $\eta_{\ell^\pm}$, which is
  proportional to the $W^\pm$-boson rapidity, and it covers the interval
  $-1.25\leq\eta_{\ell^\pm}\leq +1.25$. The details of the measurement are
  summarized in \cref{tab:DY_data}. 

\begin{table}[!t]
    \scriptsize
    \centering
    \renewcommand{\arraystretch}{1.4}
    \begin{tabularx}{\textwidth}{Xccccc}
  \toprule
  Dataset
  & Ref.
  & $N_{\rm dat}$
  & $\eta_{\ell}$
  & $\sqrt{s}$~[GeV]
  & Theory \\
  \midrule
  STAR $A_L^{W^+}$
  & \cite{STAR:2018fty}
  & 6
  & [$-$1.25, +1.25]
  & $510$
  & {\sc MCFM$^\star$} \\
  STAR $A_L^{W^-}$
  & \cite{STAR:2018fty}
  & 6
  & [$-$1.25, +1.25]
  & $510$
  & {\sc MCFM$^\star$} \\
  \bottomrule
\end{tabularx}

    \vspace{0.3cm}
    \caption{Same as \cref{tab:DIS_data} for $W^\pm$-boson production data. The
    numerical codes used for the computations is an unofficial release of {\tt MCFM}$^\star$ that 
    was presented in~\cite{Boughezal:2021wjw} and modified to produce \pineappl{}
    grids.}
    \label{tab:DY_data}
\end{table}
  
  We compute the corresponding theoretical predictions with a modified version
  of {\tt MCFM}~\cite{Boughezal:2021wjw}, which we interfaced to \pineappl{}
  up to NLO. Given the complexity of the computation, NNLO corrections are
  included, for both the unpolarized and polarized cross-sections entering
  the asymmetry, by means of a $K$-factor, which we determine with the same
  version of {\tt MCFM}. We observe that NNLO corrections are generally small
  (at most of $\mathcal{O}(3\%)$) and that they are relatively
  independent of the lepton rapidity, consistently with \cite[Fig.~2]{Boughezal:2021wjw}. 
  This follows from cancellations occurring between the polarized numerator 
  and the unpolarized denominator in the asymmetry.

\item[Single-inclusive jet and dijet longitudinal double-spin asymmetry.] We
  include measurements of the longitudinal double-spin asymmetry for
  single-inclusive jet and dijet production in polarized proton-proton
  collisions, $A_{\rm LL}^{\text{1-;2-jet}}$, performed by PHENIX~\cite{PHENIX:2010aru}
  at a center-of-mass energy $\sqrt{s}=200$~GeV, and by STAR at
  center-of-mass energies $\sqrt{s}=200$~GeV~\cite{
    STAR:2012hth,STAR:2014wox,STAR:2021mfd} and
  $\sqrt{s}=510$~GeV~\cite{STAR:2019yqm,STAR:2021mqa}. 
  The measurements are given as distributions
  differential in the transverse momentum of the leading jet, $p_T$, in the
  case of single-inclusive jet production, and in the invariant mass of the
  dijet system, $m_{jj}$, in the case of dijet production. For dijet
  production, we consider all the topologies provided. The details of these
  measurements, which include their kinematic coverage, are summarized in
  \cref{tab:JET_data}. 

  We compute the corresponding theoretical predictions with the code
  presented in~\cite{deFlorian:1998qp,Jager:2004jh}, which we interfaced to
  \pineappl, and modified to handle the necessary cuts that define different
  dijet topologies. Predictions are accurate up to NLO, given that NNLO
  corrections are not known yet. We therefore supplement them with a theory
  uncertainty, accounting for missing higher orders, estimated by varying the
  renormalization scale, as we will explain in \cref{subsec:pol_th_covmat}.

\begin{table}[!t]
    \scriptsize
    \centering
    \renewcommand{\arraystretch}{1.4}
    \begin{tabularx}{\textwidth}{Xcccccc}
   \toprule
   Dataset
   & Ref.
   & $N_{\rm {dat}}$
   & $p_{T}$ or $m_{jj}$~[GeV]
   & $\sqrt{s}$~[GeV]
   & Theory \\
   \midrule
   PHENIX $A_{LL}^{\text{1-jet}}$
   & \cite{PHENIX:2010aru}
   & $6$
   & [2.4,10.]
   & 200
   & \cite{deFlorian:1998qp,Jager:2004jh} \\
   \midrule
   STAR $A_{LL}^{\text{1-jet}}$ (2005)
   & \cite{STAR:2012hth}
   & $10$
   & [2.4, 11.]
   & 200
   & \cite{deFlorian:1998qp,Jager:2004jh}\\
   STAR $A_{LL}^{\text{1-jet}}$ (2006)
   & \cite{STAR:2012hth}
   & $9$
   & [8.5, 35.]
   & 200
   & \cite{deFlorian:1998qp,Jager:2004jh}\\
   STAR $A_{LL}^{\text{1-jet}}$ (2009)
   & \cite{STAR:2014wox} 
   & $22$
   & [5.5, 32.]
   & 200
   & \cite{deFlorian:1998qp,Jager:2004jh} \\
   STAR $A_{LL}^{\text{2-jet}}$ (2009)
   & \cite{STAR:2016kpm}
   & $33$
   & [17., 68.]
   & 200
   & \cite{deFlorian:1998qp,Jager:2004jh} \\
   STAR $A_{LL}^{\text{1-jet}}$ (2012)
   & \cite{STAR:2019yqm}
   & $14$
   & [6.8, 55.]
   & 510
   & \cite{deFlorian:1998qp,Jager:2004jh} \\
   STAR $A_{LL}^{\text{2-jet}}$ (2012)
   & \cite{STAR:2019yqm}
   & $42$
   & [20., 110.]
   & 510
   & \cite{deFlorian:1998qp,Jager:2004jh} \\
   STAR $A_{LL}^{\text{1-jet}}$ (2013)
   & \cite{STAR:2021mqa}
   & $14$
   & [8.7, 63.]
   & 510
   & \cite{deFlorian:1998qp,Jager:2004jh} \\
   STAR $A_{LL}^{\text{2-jet}}$ (2013)
   & \cite{STAR:2021mqa}
   & $49$
   & [14., 133.]
   & 510
   & \cite{deFlorian:1998qp,Jager:2004jh} \\
   STAR $A_{LL}^{\text{1-jet}}$ (2015)
   & \cite{STAR:2021mfd} 
   & $22$
   & [5.8, 34.]
   & 200
   & \cite{deFlorian:1998qp,Jager:2004jh} \\
   STAR $A_{LL}^{\text{2-jet}}$ (2015)
   & \cite{STAR:2021mfd}
   & $14$
   & [20., 71.]
   & 200
   & \cite{deFlorian:1998qp,Jager:2004jh} \\
  \bottomrule
\end{tabularx}

    \vspace{0.3cm}
    \caption{Same as \cref{tab:DIS_data} for single-inclusive jet and
      dijet production data. The original numerical codes used for the predictions
      have been modified to allow for the generation of \pineappl{} grids.}
    \label{tab:JET_data}
\end{table}

\end{description}
 
\begin{figure}[!t]
  \centering
  \includegraphics[width=0.8\textwidth]{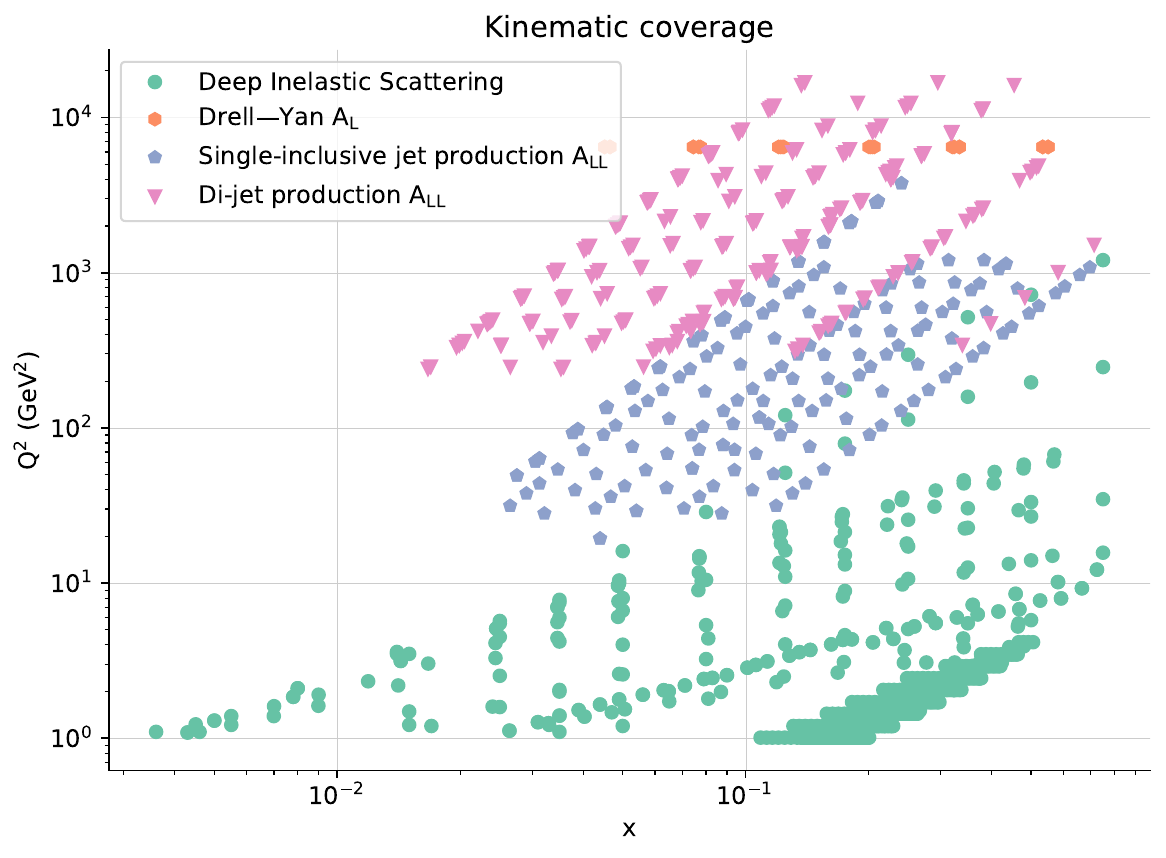}
  \caption{
    The kinematic coverage of the {\sc NNPDFpol2.0} dataset in the $(x, Q^2)$
    plane after applying kinematic cuts.}
  \label{fig:pol_kinplot}
\end{figure}

The total number of data points included in our study, after applying
the aforementioned kinematic cuts, is $N_{\rm dat}=951$, irrespective of the
perturbative accuracy of the determination. The corresponding kinematic
coverage in the $(x,Q^2)$ plane is displayed in \cref{fig:pol_kinplot}. For
$W^\pm$-boson, single-inclusive jet, and dijet production in polarized
proton-proton collisions, LO kinematic relations have been used to determine
$x$ and $Q^2$ from the relevant hadronic variables.

As can be seen from \cref{fig:pol_kinplot}, the largest number of data points
correspond to polarized inclusive DIS measurements. Because of the very moderate
values of the virtuality $Q^2$, the scattering involves the exchange of a
virtual photon. At LO, these measurements are therefore sensitive only to the
singlet PDF flavor combination, whereas the sensitivity to the gluon PDF,
which enters only at higher orders, is suppressed by powers of the strong
coupling. Sensitivity to valence-like PDF flavor combinations is achieved
thanks to $W^\pm$-boson production measurements in polarized proton-proton
collisions, which is a parity-violating process. Complementary to this are
measurements of single-hadron production in DIS, that however we do not
consider because they require the simultaneous knowledge of FFs.
Sensitivity to the gluon PDF is achieved thanks to single-inclusive jet and
dijet production measurements in polarized proton-proton collisions, which 
account for almost all the rest of our dataset. Additional constraints on
the gluon PDF may come from measurements of two other processes:
single-hadron production in polarized proton-proton collisions, which we do not
consider because of the need for the simultaneous knowledge of FFs;
and open-charm production in DIS, which we do not consider because the
available datasets were demonstrated to bring in a negligible amount of
information~\cite{Nocera:2014gqa}.

The complete information on experimental uncertainties, including on their
correlations, is taken into account whenever available from the
{\sc HEPdata} repository~\cite{Maguire:2017ypu} or from the corresponding
publications. Specifically, full covariance matrices are
provided only for the HERMES measurement of~\cite{HERMES:2006jyl} and for
the STAR measurements of~\cite{STAR:2014wox,STAR:2016kpm,STAR:2019yqm,
  STAR:2021mqa,STAR:2021mfd}. Most notably, the latter include correlations
between single-inclusive jet and dijet bins, a fact that allows us to include
all the measurements at the same time in the fit. Information on correlations
is generally not provided by other experiments, except for the highlight of
a multiplicative, fully correlated, uncertainty due to the beam polarization.

\subsection{Perturbative accuracy}
\label{subsec:pol_th_covmat}

The perturbative accuracy of the theoretical predictions corresponding to the
measurements described in \cref{subsec:pol_data} relies on the perturbative
accuracy of matrix elements and of DGLAP splitting functions, which are both
expanded as a series in the strong coupling $a_s$. In this context, this
work pursues two goals: first, to include corrections up to NNLO in both; and,
second, to include MHOUs arising from the truncation of the expansion series
to a finite accuracy.

The first goal is achieved by deploying a set of computational tools, available
as open-source software, that we specifically designed for PDF fitting.
As already mentioned, these include: \yadism~(\cref{sec:yadism})
for the computation of the polarized inclusive structure function $g_1$;
\pineappl~\cite{Carrazza:2020gss}
(interfaced with the private pieces of software
in~\cite{Boughezal:2021wjw,deFlorian:1998qp,Jager:2004jh}
used to compute the polarized proton-proton collision spin asymmetries)
for the construction of PDF-independent interpolation grids;
\eko~(\cref{sec:eko}) for PDF evolution. Each of these
pieces of software has been extended to handle the computation of the polarized
observables at the desired accuracy, including
in the case in which unpolarized and polarized PDFs ought to be used
simultaneously. This amounted to the following.
We implemented in \yadism{} the FONLL general-mass variable-flavor-number
scheme up to NNLO~\cite{Hekhorn:2024tqm,Barontini:2024xgu}, which combines
the massless computation~\cite{Zijlstra:1993sh} with the recent massive
one~\cite{Hekhorn:2018ywm} and its asymptotic
limit~\cite{Bierenbaum:2022biv}. We implemented in \eko{} the
polarized splitting functions, including the known corrections up to
NNLO~\cite{Moch:2014sna,Moch:2015usa,Blumlein:2021enk,Blumlein:2021ryt} and their
matching conditions~\cite{Bierenbaum:2022biv}.
We have extended {\sc PineAPPL} and {\sc pineko} to deal with polarized 
observables, including in the case in which unpolarized and polarized PDFs ought 
to be used simultaneously, such as in the computation of spin asymmetries.

The second goal is achieved following the methodology developed by the NNPDF
collaboration in~\cite{NNPDF:2019vjt,NNPDF:2019ubu,NNPDF:2024dpb}. Specifically,
we supplement the experimental covariance matrix, reconstructed from knowledge
of experimental uncertainties, with a MHOU covariance matrix, constructed from
renormalization and factorization scale variations
\begin{equation}
  \text{cov}^{(\text{tot})}_{ij}
  = \text{cov}^{(\text{exp})}_{ij}
  + \text{cov}^{(\text{MHOU})}_{ij} \, \quad i,j = 1 \dots N_{\text{dat}}.
  \label{eq:tot_cov}
\end{equation}
%
To properly correlate renormalization scale variations, we
therefore define four process categories: neutral current DIS (NC DIS),
corresponding to measurements of $g_1$; charged current Drell-Yan (CC DY),
corresponding to measurements of $A_L^{W^\pm}$; single-inclusive jet production
(JETS), corresponding to measurements of $A_{LL}^{\text{1-jet}}$; and dijet production
(DIJET), corresponding to measurements of $A_{LL}^{\text{2-jet}}$. We thus assume four 
independent renormalization scales $\mu_{r,i}$ and one common factorization 
scale $\mu_{f}$. For each scale we define the ratio
$\rho_k=\mu_k/Q$, where $Q$ denotes the typical scale of the process.
The computation of ${\rm cov}^{(\text{MHOU})}_{ij}$ then follows the scheme B
prescription detailed in~\cite{NNPDF:2019ubu}. Using an approach similar to
that developed in~\cref{sec:an3lo_hadronic_coeff}, we can distinguish two different scale
variation procedures, depending on which MHOU component we want to estimate.

\begin{figure}[!t]
  \centering
  \includegraphics[width=0.48\textwidth]{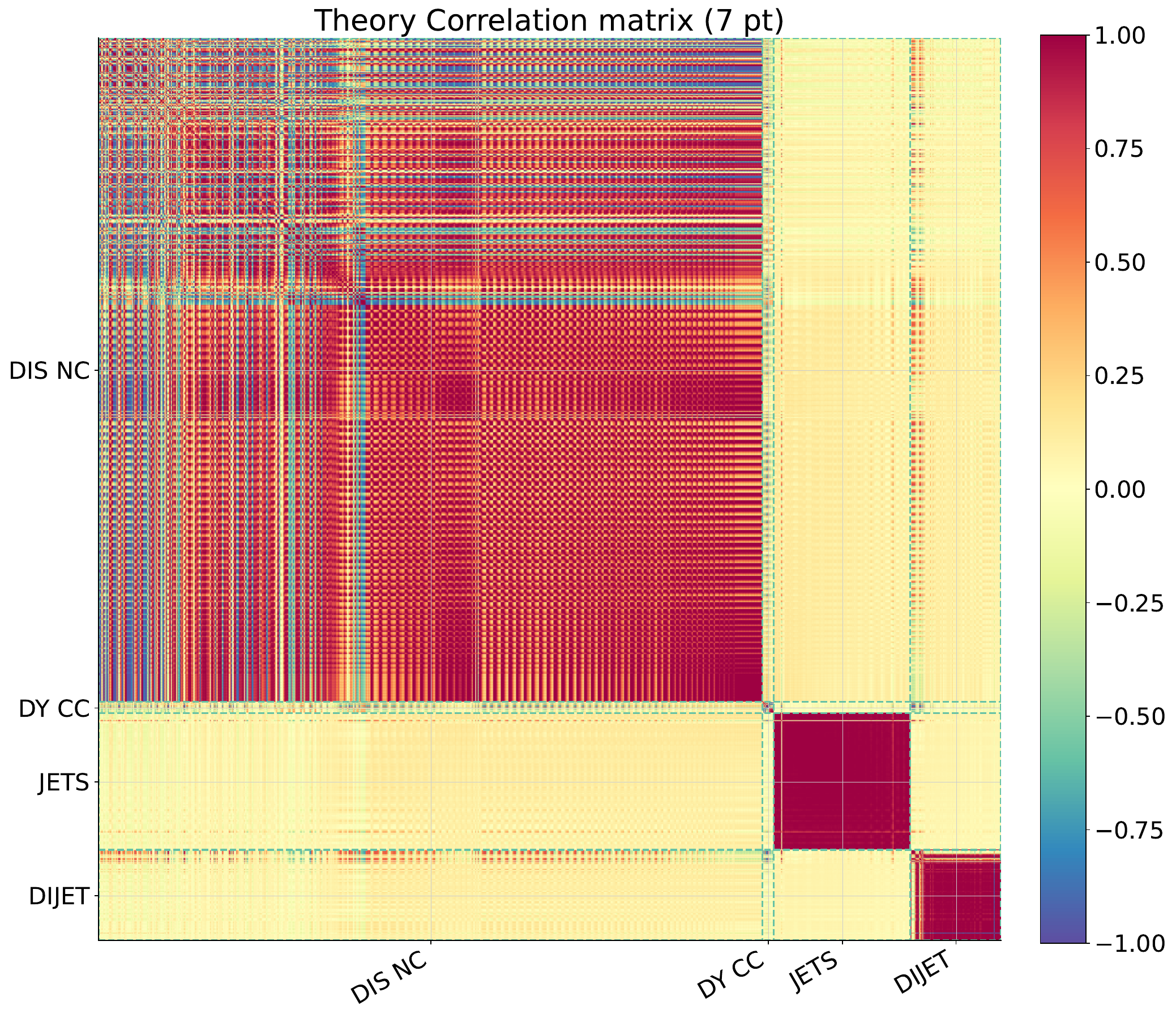}
  \includegraphics[width=0.48\textwidth]{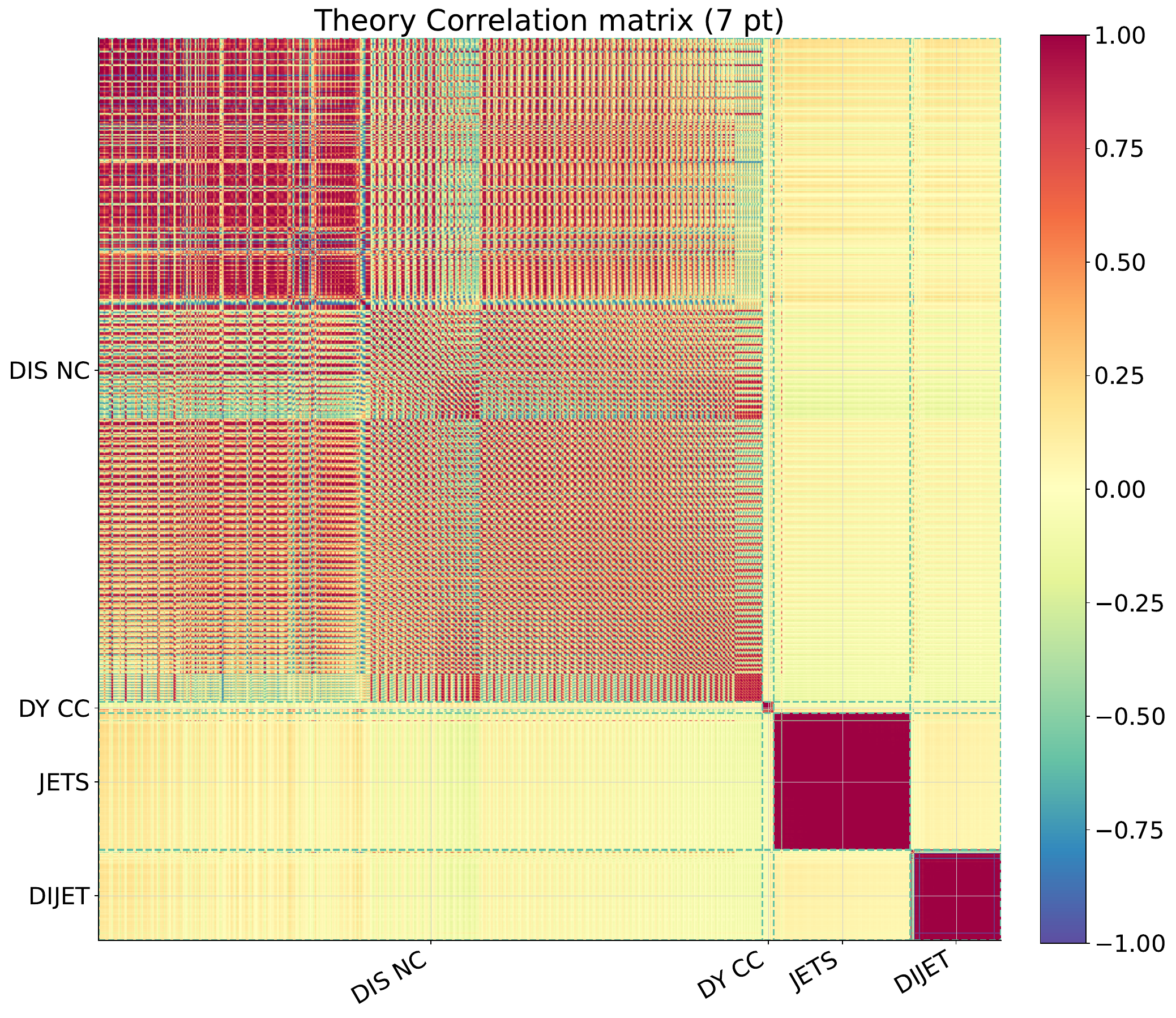}
  \caption{
    The MHOU correlation matrix computed with the 7-point prescription 
    (b) at NLO (left) and NNLO (right).
  }
  \label{fig:pol_7pt_correlation}
\end{figure}

\begin{enumerate}[(a)]

\item We adopt a 3-point renormalization scale variation prescription to
  estimate missing NNLO corrections in the matrix elements of those processes
  for which they are unknown, that is, single-inclusive jet and dijet
  production in polarized proton-proton collisions. We therefore vary the
  ratios $\rho_{r,i}$ of the JETS and DIJET processes in the range
  $\rho_{r,i} \in \{0.5, 1.0, 2.0\}$. 
  
\item We adopt a 7-point renormalization and factorization scale variation
  prescription to estimate the MHOU for the complete dataset. This
  prescription can be applied both at NLO and at NNLO. We therefore consider
  simultaneous variations of the factorization and renormalization scales in
  the range $\rho_k=\{0.5, 1.0, 2.0\}$, and discard the two outermost
  combinations $(\rho_{r,i} = 0.5, \rho_f = 2.0)$, $(\rho_{r,i} = 2.0, \rho_f = 0.5)$.
  The associated correlation matrix is displayed in
  \cref{fig:pol_7pt_correlation}, both at NLO (left) and at NNLO (right).
\end{enumerate}


The two prescriptions are exclusive. Prescription (a) will be adopted only in
the NNLO fits that we will call {\it without MHOUs} in \cref{sec:pol_results}.
This nomenclature puts the emphasis on the fact that MHOUs are
included only partially, and specifically only to account for unknown NNLO
corrections. All the other fits called {\it without MHOUs}, either LO or NLO,
will not use either prescription. Prescription (b) will be adopted
in all the fits, whether LO, NLO or NNLO, that we will call {\it with MHOUs}
in \cref{sec:pol_results}. This nomenclature puts the emphasis on the fact
that MHOUs, beyond the nominal accuracy of the fit, are included on all data
points. Be that as it may, we have checked that MHOUs are generally smaller
than experimental uncertainties, and that they are significantly more
correlated. From \cref{fig:pol_7pt_correlation} we can also appreciate how
the NNLO MHOU correlations are generally smaller that the NLO ones,
indicating a consistent perturbative convergence of theoretical predictions.




\section{Methodology}
\label{sec:pol_methodology}
In this section, we discuss the methodology deployed to determine the
{\sc NNPDFpol2.0} parton set. The methodology, based on parametric regression,
closely follows the one laid out in~\cite{NNPDF:2021njg,NNPDF:2021uiq} for
the determination of the {\sc NNPDF4.0} set of unpolarized PDFs. We review how
aspects of PDF parametrization, optimization, and hyperoptimization are
upgraded and adapted to the polarized case. 

\subsection{Parametrization}
\label{subsec:pol_parametrization}

Parton distribution parametrization entails two choices: first, a choice of
parametrization basis, that is, the set of linearly independent distributions
that are parametrized; second, a choice of parametrization form, that is, the
function that maps the parameters into the elements of the basis.
Concerning the parametrization basis, we choose the set of functions
\begin{equation}
  \Delta f(x,Q_0^2)
  =
  \{\Delta g, \Delta\Sigma, \Delta T_3, \Delta T_8, \Delta V, \Delta V_3\}(x,Q_0^2)\,,
  \label{eq:ev_basis}
\end{equation}
made of the gluon PDF $\Delta g$ and of five independent quark flavor PDF
combinations: the singlet $\Delta\Sigma$, the non-singlet sea $\Delta T_i$,
and the non-singlet valence $\Delta V,\Delta V_3$. 
These PDF combinations are defined as in \cref{eq:evol_basis}.
The parametrization scale is $Q_0^2=1.0$~GeV$^2$; 
PDFs are then evolved to the scale of the physical processes by means of DGLAP 
equations, see \cref{sec:pol_data-theory}. 
Because the available piece of experimental information is sensitive to an asymmetry 
between $\Delta s$ and $\Delta\bar s$ only very mildly, we assume $\Delta s(x,Q_0^2)=\Delta\bar s(x,Q_0^2)$.
%
Differences between $\Delta s$ and $\Delta \bar s$ may occur for $Q^2>Q_0^2$ at NNLO 
and beyond, because higher-order QCD corrections make the two distributions evolve
differently. This effect is however very small. Finally, we assume that charm
is completely generated from gluon splitting through parton evolution,
therefore we set to zero, and do not parametrize, a possible intrinsic charm
component at the parametrization scale $Q_0^2$.

Concerning the parametrization form, we choose a feed-forward neural network
with six output nodes, each of which corresponds to an element of the basis
defined in \cref{eq:ev_basis}.
The neural network architecture and activation function are determined according 
to the hyperparameter optimization procedure delineated in \cref{subsec:pol_hyperoptimization}.
The output of the neural network is then related to the polarized PDFs as
\begin{equation}
  x\Delta f(x,Q_0^2,\boldsymbol\theta)
  =
  A_{\Delta f}
  x^{1-\alpha_{\Delta f}}
  (1-x)^{\beta_{\Delta f}}
  {\rm NN}_{\Delta f}(x,\boldsymbol\theta)\,,
  \label{eq:parametrization}
\end{equation}
where $\Delta f$ denotes each element of the chosen basis, $A_{\Delta f}$ is a
normalization factor, $\alpha_{\Delta f}$ and $\beta_{\Delta f}$ are preprocessing
exponents, and ${\rm NN}_{\Delta f}(x,\boldsymbol\theta)$ is the output of the
neural network, which depends on weights and biases, collectively denoted as
$\boldsymbol\theta$.

The normalization factor $A_{\Delta f}$ is equal to one for all PDFs but the
non-singlet triplet and octet PDF combinations, $\Delta T_3$ and $\Delta T_8$,
for which we define
\begin{equation}
  A_{\Delta T_3}
  =
  a_3 \left[ \int_{x_{\rm min}}^{1} dx \, \Delta T_3 (x, Q_0^2) \right]^{-1}
  \qquad{\rm and}\qquad
  A_{\Delta T_8}
  =
  a_8 \left[ \int_{x_{\rm min}}^{1} dx \, \Delta T_8 (x, Q_0^2) \right]^{-1}\,;
  \label{eq:normalisation}
\end{equation}
here $a_3$ and $a_8$ are the baryon octet decay (scale-independent) constants
whose experimental values are~\cite{ParticleDataGroup:2024cfk}
\begin{equation}
  a_3 = 1.2756\pm 0.0013
  \qquad{\rm and}\qquad
  a_8 = 0.585\pm 0.025\,.
  \label{eq:decay_constants}
\end{equation}
The integrals in \cref{eq:normalisation} are computed each time the
parameters $\boldsymbol\theta$ change, assuming $x_{\rm min}=10^{-4}$. For each
replica, the values of $a_3$ and $a_8$ are random numbers sampled from a
Gaussian distribution with mean value and standard deviation equal to the
corresponding experimental central value and uncertainty,
\cref{eq:decay_constants}. Enforcing \cref{eq:normalisation}
therefore corresponds to requiring that SU(2) and SU(3) flavor symmetries are
exact up to the experimental uncertainties quoted in
\cref{eq:decay_constants}.  

Finally, the preprocessing exponents $\alpha_{\Delta f}$ and $\beta_{\Delta f}$,
which are needed to speed up optimization, are determined by means of an
iterative procedure, firstly introduced in~\cite{Ball:2013lla}. Specifically,
their values are initially random sampled from a flat distribution
which limits are iteratively determined. 


\subsection{Optimization}
\label{subsec:pol_optimization}

Optimization of the neural network parameters $\boldsymbol\theta$ requires
a choice of cost function and of optimization algorithm, including a stopping
criterion. We discuss each of these two choices in turn.

Concerning the cost function, we make considerations that are peculiar 
to the determination of polarized PDFs. For each replica $k$, we define it
as the sum of three terms
\begin{equation}
  \left(\chi^{(k)}(\boldsymbol\theta)\right)^2
  + \Lambda_{\rm int} R_{\rm int}^{(k)}(\boldsymbol\theta)
  + \Lambda_{\rm pos} R_{\rm pos}^{(k)}(\boldsymbol\theta)\,.
  \label{eq:cost_function}
\end{equation}
The first term, $\left(\chi^{(k)}(\boldsymbol\theta)\right)^2$, 
is the usual quadratic loss
\begin{equation}
  \left(\chi^{(k)}(\boldsymbol\theta)\right)^2
  =
  \frac{1}{N_{\rm dat}}\sum_{i,j}^{N_{\rm dat}}
  \left[
  T_i\left(\Delta f^{(k)}(\boldsymbol\theta)\right) - D_i^{(k)}
  \right]
  {\rm Cov}_{ij}^{-1}
  \left[
  T_j\left(\Delta f^{(k)}(\boldsymbol\theta)\right) - D_j^{(k)}
  \right]\,,
  \label{eq:chi2}
\end{equation}
where $i,j$ are indexes that run on the number of data points $N_{\rm dat}$,
${\rm Cov}_{ij}$ is the covariance matrix, $D_{i,j}^{(k)}$ are the $k$-th
experimental pseudodata replicas,
and $T_{i,j}\left(\Delta f^{(k)}(\boldsymbol\theta)\right)$ are the
corresponding theoretical predictions. The covariance matrix is computed as
explained in \cref{sec:pol_data-theory}. Specifically, the $t_0$
prescription~\cite{Ball:2009qv} is used to determine the contribution due to
experimental uncertainties, whereas point prescriptions are used to determine
the MHOU contribution when these are taken into account.
Theoretical predictions are computed as a convolution of the parametrized PDFs 
(specifically of their luminosities $\mathcal{L}$) with fast-kernel interpolating 
tables in \pineappl{} format. These are in turn a convolution of partonic matrix 
elements and evolution kernel operators (EKOs), that evolve PDFs from the 
parametrization scale $Q_0^2$ to the scale $Q^2$ of the physical process 
(see for details \cref{sec:theory_methodology}). 
Some experimental data consist of asymmetries, for which the theoretical 
predictions depend both on polarized PDFs and the unpolarized ones:
\begin{equation}
  T\left(\Delta f(x,Q_0^2,\boldsymbol\theta)\right)
  =
  \frac{T^{\rm pol}\left(\Delta f(x,Q_0^2,\boldsymbol\theta)\right)}{T^{\rm unp}\left(f(x,Q_0^2)\right)}
  =
  \frac{ \Delta FK(x, Q^2 \leftarrow Q_0^2) \otimes \Delta \mathcal{L} \left( \Delta f (x,Q_0^2,\boldsymbol\theta) \right)  }{ FK(x, Q^2 \leftarrow Q_0^2) \otimes \mathcal{L} \left( f(x,Q_0^2) \right) }
  \,.
  \label{eq:asy_theo}
\end{equation}
Both numerator and denominator are computed via FK-tables, with the difference 
that, in the latter, the PDF sets are kept fixed during the optimization and 
parameter optimization enters \cref{eq:asy_theo} only through 
the numerator.
Specifically, we use unpolarized PDFs of the {\sc NNPDF40\_pch} family which assume 
perturbative charm, and have consistent perturbative order with that of the 
polarized PDF we aim to determine.
This way, unpolarized and polarized PDFs are determined with the same values 
of the physical parameters and with the same methodology, ensuring 
perfect consistency between the two.
\\ The second term in \cref{eq:cost_function}, $\Lambda_{\rm int}R_{\rm int}(\boldsymbol\theta)$, 
is a regularization term that enforces the lowest moments of polarized PDFs 
to be finite. This requirement follows from the assumption that the nucleon 
matrix element of the axial current is finite for each parton. Therefore, the
small-$x$ behavior of polarized PDFs must obey
\begin{align}
  \lim _{x \rightarrow 0} x \Delta f(x, Q^2) = 0
  & \qquad\mbox{for $f=g,\Sigma,T_3,T_8$}\,,\\
  \lim _{x \rightarrow 0} x^2 \Delta f(x, Q^2) = 0
  & \qquad\mbox{for $f=V_3,V_8$}\,.
  \label{eq:integrablity}
\end{align}
The first of these two conditions is fulfilled by construction for the polarized
quark triplet and octet PDF combinations, given the choice of normalization
made in their parametrization, see \cref{eq:normalisation}.
The regularization term is therefore
\begin{align}
  \Lambda_{\rm int}R_{\rm int}(\boldsymbol\theta)
  =
  \Lambda_{\rm int} \sum_{f} 
  \left[x \Delta f\left(x_{\rm{int}}, Q_{\rm int}^2,\boldsymbol\theta\right)\right]^2
  & \qquad\mbox{for $f=g,\Sigma$}\,,
  \label{eq:int_reg_1}
  \\
  \Lambda_{\rm int}R_{\rm int}(\boldsymbol\theta)
  =
  \Lambda_{\rm int} \sum_{f} 
  \left[x^2 \Delta f\left(x_{\rm{int}}, Q_{\rm int}^2,\boldsymbol\theta\right)\right]^2
  & \qquad\mbox{for $f=V_3,V_8$}\,,
  \label{eq:int_reg_2}  
\end{align}
where $Q_{\rm int}^2=1$~GeV$^2$ and $x_{\rm{int}}=10^{-5}$. The Lagrange multiplier
$\Lambda_{\rm int}$ grows exponentially during the fit and reaches the maximum
value $\Lambda_{\rm int}=100$ at maximum training length.
\\ The third term in \cref{eq:cost_function}, $\Lambda_{\rm pos} R_{\rm pos}^{(k)}(\boldsymbol\theta)$, 
is enforces PDFs to lead to positive cross-sections. This implies
that polarized PDFs are bound by their unpolarized counterparts for each
parton $f$, for each $x$, and for each $Q^2$~\cite{Altarelli:1998gn}
\begin{equation}
  \vert \Delta f (x, Q^2) \vert \leq f(x, Q^2)\,.
  \label{eq:positivity_pdfs}
\end{equation}
Whereas \cref{eq:positivity_pdfs} is formally valid only at LO, it can be suitably used 
to enforce positivity bounds on polarized PDFs at all orders. 
This is justified by the following observation: NLO corrections to the positivity bounds 
only differ from its lowest order by an amount less than $10\%$ in the small-$x$ regions 
($x \sim 10^{-2}$) while the positivity bounds in \cref{eq:positivity_pdfs} is only significant 
at large-$x$, as for e.g. $g_1/F_1 \sim x$ as $x \to 0$. Higher-order corrections 
to the positivity bounds are negligible in comparison to the size of the PDF uncertainties 
in regions where they are found to provide no constraints.
%
%
The positivity regularization term is
\begin{equation}
  \Lambda_{\rm{pos}} R_{\rm pos}^{(k)}(\boldsymbol\theta)
  =
  \Lambda_{\rm{pos}} \sum_f \sum_{i=1}^{n} 
  \operatorname{ReLU}\left(-\mathcal{C}_f\left(x^i_{\rm pos}, Q^2_{\rm pos},\boldsymbol\theta\right)\right), \qquad \operatorname{ReLU}(t)=\left\{\begin{array}{ll}
  t & \text { if } t>0 \\
  0 & \text { if } t \leq 0
  \end{array}\right.\,,
  \label{eq:positivity-constraints}
\end{equation}
where the function
\begin{align}
  \mathcal{C}_f \left(x^i_{\rm pos}, Q^2_{\rm pos},\boldsymbol\theta\right)
  = f\left(x^i_{\rm pos},Q^2_{\rm pos}\right)
  -\left|\Delta f\left(x^i_{\rm pos},Q^2_{\rm pos},\boldsymbol\theta\right)\right|
  +\sigma_f\left(x^i_{\rm pos},Q^2_{\rm pos}\right),
  \label{eq:c-function}
\end{align}
encodes the positivity condition of \cref{eq:positivity_pdfs}.
In \cref{eq:positivity-constraints,eq:c-function},
$f$ denotes the parton, and $i$ denotes the point at which the function
$\mathcal{C}_f$ is evaluated. In particular, $n=20$ points are sampled
in the range $\left[5\cdot 10^{-7}, 9\cdot 10^{-1}\right]$, half of which are
logarithmically spaced below $10^{-1}$ and half of which are linearly spaced
above. The unpolarized PDF $f$ and its one-$\sigma$ uncertainty $\sigma_f$
are taken from the same PDF set that enters the
computation of theoretical predictions. Finally,
$Q^2_{\rm pos}=5$~GeV$^2$ and the Lagrange multiplier
$\Lambda_{\rm pos}$ grows exponentially during the fit and reaches the maximum
value $\Lambda_{\rm int}=10^{10}$ at maximum training length.

Optimization of the parameters $\boldsymbol\theta$ is achieved through stochastic
gradient descent, as in NNPDF4.0~\cite[Sect.~3.2]{NNPDF:2021njg}. 
The specific optimization algorithm is selected from those that are 
readily available in the {\sc TensorFlow} library~\cite{tensorflow2015:whitepaper} 
through hyperparameter optimization, as discussed in the next section. 
Cross-validation is used to prevent overfitting and design a stopping criterion.
To this purpose, for each pseudodata replica, the data points are split into a
training and a validation set, in a proportion of 60\% and 40\%.
Post-fit checks are finally enforced to exclude parameter configurations that violate 
the positivity constraint, or that have values of $\chi^2$ outside the 4$\sigma$ 
interval of their distribution.

\subsection{Hyperoptimization}
\label{subsec:pol_hyperoptimization}
We now discuss the hyperparameter optimization procedure adopted to determine the 
baseline methodology. The underlying approach follows Ref~\cite{Cruz-Martinez:2024wiu}, 
where the hyperoptimization is performed at the level of the PDF distributions resulting 
from a fit of multiple replicas. This was not accessible in the previous studies of 
\cref{chap:ic,chap:an3lo} due to several limitations, the main one being the inability 
to perform simultaneous fit of multiple replicas at once, which are now evaluated using 
graphics processing units (GPUs). 
Such improvements allow us to distribute the hyperoptimization scans across multiple GPUs, 
enabling an asynchronous search of the parameter space. In principle, the greater the number 
of GPUs utilized, the faster the scan of the hyperparameter space proceeds.

\begin{figure}[!t]
	\centering
	\includegraphics[width=0.80\textwidth]{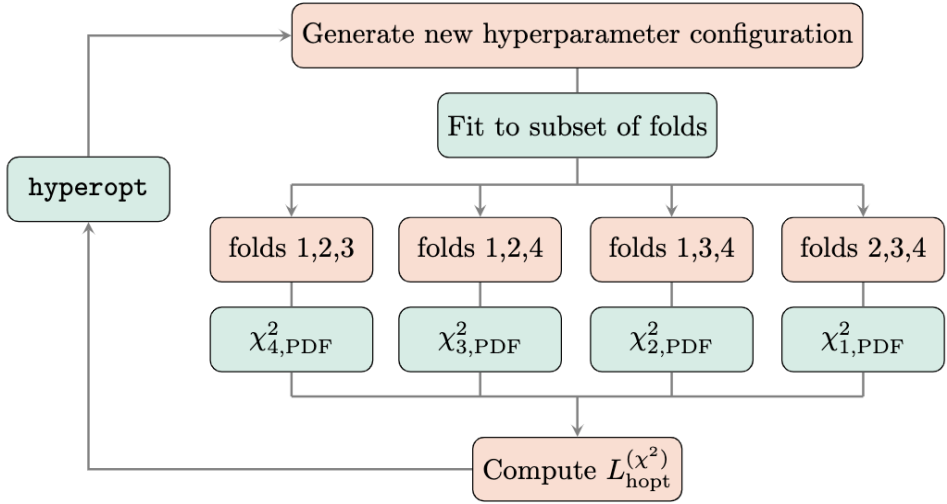}
	\caption{
    Diagrammatic representation of the $K$-fold algorithm used for the hyperparameter 
    optimization. Similar to Ref.~\cite[Fig.~3.5]{NNPDF:2021njg} but now accounting 
    for the PDF replicas distribution, when computing the loss $L_{\rm hopt}^{(\chi^2)}$ 
    of \cref{eq:hyperopt-loss}. See also Ref.~\cite{Cruz-Martinez:2024wiu} for 
    further details.
	}
	\label{fig:kfolds}
\end{figure}

The improved method extends the $K$-fold procedure used in the NNPDF4.0 methodology, and
its diagrammatic representation is shown in \cref{fig:kfolds}. The algorithm starts each 
trial with a selected set of hyperparameters from which $n_{\rm folds}$
folds are constructed. For each subset of folds, the $p$-th fold is left-out and the 
remaining ones are combined into a dataset from which the neural network is optimized 
according to the procedure described above in \cref{subsec:pol_optimization}. 
Each of these fits is performed simultaneously drawing $N_{\rm rep}$ replicas. 
%
%
The hyperoptimization loss function is then defined as
\begin{equation}
\label{eq:hyperopt-loss}
  L_{\rm hopt}^{(\chi^2)} \lp \boldsymbol{\hat{\theta}}\rp =  
  \frac{1}{n_\text{folds}}\sum_{p=1}^{n_{\rm folds}}
  \underset{\boldsymbol{\theta} \in {\boldsymbol{\Theta}}}{\text{ min}^*} \lp \ \la \chi^2_{{\rm PDF},p}\lp \boldsymbol{\theta},\boldsymbol{\hat{\theta}} \rp \ra_{\rm rep} \rp \, ,
\end{equation}
where we distinguish between the model parameters $\boldsymbol{\theta}$ (e.g. network 
weights and biases) and hyperparameters of interest $\boldsymbol{\hat{\theta}}$.
The $^*$ sign indicates that the minimization is regularized with training and validation 
split to avoid overfitting, and the figure of merit $\chi^2_{{\rm PDF},p}$ is evaluated 
on the $p$-fold datasets for all the replicas $N_{\rm rep}$ and averaged.
$\chi^2_{{\rm PDF},p}$ includes contributions from the PDF uncertainties, added in 
quadrature to the experimental covariance matrix. For each replica $k$, it is defined via
\begin{equation}
	\chi^{2(k)}_{{\rm PDF},p}(\boldsymbol{\theta}) = \frac{1}{n_p} \displaystyle\sum_{i,j \in p} \lp D^{(0)}_{i}-T^{(k)}_{i}
	(\boldsymbol{\theta}) \rp\lp {\rm cov}_{\rm (exp)} + {\rm cov}_{\rm (PDF)}  \rp^{-1}_{ij} \lp D^{(0)}_{j}-T^{(k)}_{j}
	(\boldsymbol{\theta})\rp 
	\,,
	\label{eq:chi2definition}
\end{equation}
where we have left the dependence on $\boldsymbol{\hat{\theta}}$ implicit.
\\ The algorithm proceeds iterating over $n_{\rm trials}$ hyperparameter configurations 
ending up with an array of losses computed according to \cref{eq:hyperopt-loss}. In 
principle, one would like to select the optimal hyperparameter set 
$\hat{\boldsymbol{\theta}}^\star$ such that 
\begin{equation}
  \qquad \hat{\boldsymbol{\theta}}^\star = \underset{\hat{\boldsymbol{\theta}} \in \hat{\boldsymbol{\Theta}}}{\arg \min }
	\left( L_{\rm hopt}^{(\chi^2_{\rm pdf})} \lp \boldsymbol{\hat{\theta}}\rp \right) \,,
\end{equation}
however, due to the flexibility of the NN the set of parameters might not be 
unique as there exists different models leading to an equal description of the 
unseen folds. 
Thus, to further discriminate our hyperparameter space we can introduce and additional
loss function. For example, we can evaluate the standard deviation of $\chi^2_{{\rm PDF},p}$ 
over the replica sample in units of the data uncertainty on the left-out folds. 
We then define a second hyperoptimization loss
\begin{equation}
	L_{\rm hopt}^{(\varphi^2)} \lp \boldsymbol{\hat{\theta}}\rp  \equiv \lp  \frac{1}{n_\text{folds}}
	\displaystyle\sum^{n_\text{folds}}_{p=1} \varphi_{\chi^2_p}^2 \lp \boldsymbol{\hat{\theta}}\rp\rp^{-1} \, ,
	\label{eq:hyperoptloss_phi2}
\end{equation}
where the metric that probes the second moment of the PDF distribution is given by
\begin{equation}
	\varphi^2_{\chi^2_p} = \langle \chi^2_p \left[ T \left(\Delta f_{\rm fit}\right), D \right] \rangle_{\rm rep} 
	- \chi^2_p \left[  \langle T \left( \Delta f_{\rm fit} \right) \rangle_{\rm rep},D \right].
	\label{eq:phi2_metric}
\end{equation} 
\cref{eq:phi2_metric} measures the PDF uncertainties on the scale of the data uncertainties,
the preferred extrapolation to the non-fitted $p$-th fold is the one with the 
largest uncertainties, ie. with small values of \cref{eq:hyperoptloss_phi2}.
\\ In summary, given a set of successful models and their corresponding 
$L_{\rm hopt}^{(\chi^2_{\rm pdf})}$, we then select the best one as follows. 
We evaluate the standard deviation, $\Sigma_{\chi^2}$, of $\chi_{{\rm PDF},k}^{2(k)}$
over the replicas corresponding to the fit minimizing \cref{eq:hyperopt-loss}. 
We use this value to define a selection range:
\begin{equation}
	\mathcal{R}: \left[ \hat{\boldsymbol{\theta}}^\star, \hat{\boldsymbol{\theta}}^\star + \Sigma_{\chi^2} \right], \qquad
	\text{with} \qquad \hat{\boldsymbol{\theta}}^\star = \underset{\hat{\boldsymbol{\theta}} \in \hat{\boldsymbol{\Theta}}}{\arg \min }
	\left( L_{\rm hopt}^{(\chi^2_{\rm pdf})} \lp \boldsymbol{\hat{\theta}}\rp \right).
\end{equation}
We select the sought-for optimal set of hyperparameters as the one yielding to the 
lowest value of $L_{\rm hopt}^{(\varphi^2)}$ within the range $\mathcal{R}$.
\footnote{
Let us note that, other model selection criteria are possible. For example, for 
each final PDF replica one can choose a different model among the ones in $\mathcal{R}$, 
or weight their probability to according to $L_{\rm hopt}^{(\varphi^2)}$. We plan to 
investigate the effect of these choices in future studies.
}
The specific values and the found optimal hyperparameter for our polarized fits are listed
in the following paragraph.

\begin{figure}[!t]
	\centering
    \includegraphics[width=0.75\textwidth]{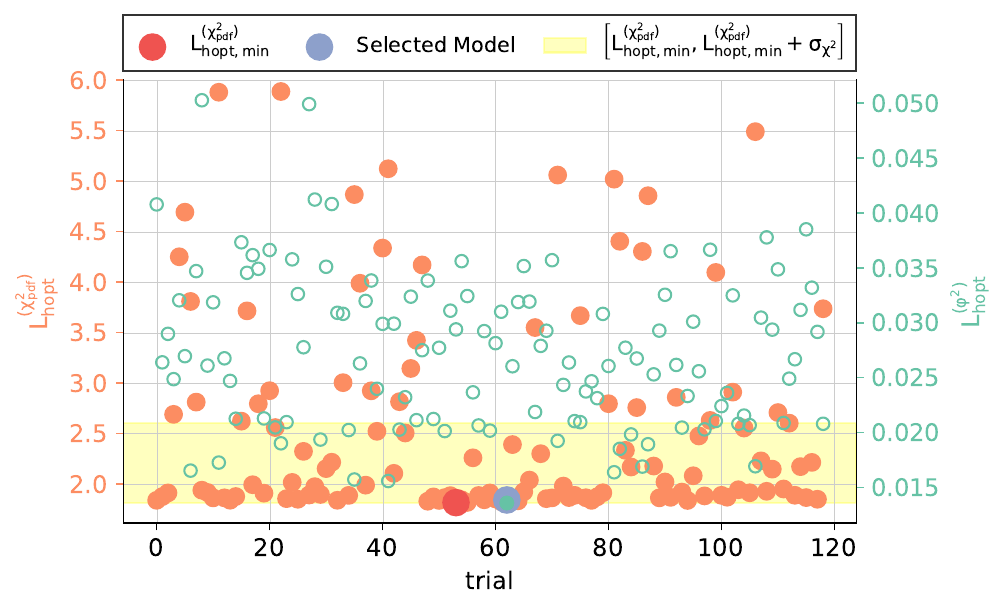}
	\caption{
		The distribution of the hyperoptimization losses as a function of the trials. For each 
    model we show the value of $L_{\rm hopt}^{(\chi^2)}$ (orange dot, left $y$-axis) and 
    of $L_{\rm hopt}^{(\varphi^2)}$ (green maker, right $y$-axis) computed
		on the left-out folds. The yellow shaded band indicate the range within which we select 
    the model with the lowest $L_{\rm hopt}^{(\varphi^2)}$.
	}
	\label{fig:hyperopt-dist}
\end{figure}

\paragraph{Hyperparameters for {\sc NNPDFpol2.0}.}
We perform a scan of $n_{\rm trials}=200$ possible configurations, distributed across four 
A100 Nvidia GPUs. We opt for $N_{\rm rep} = 60$ replicas and $n_{\rm folds} = 4$. 
The dataset partitions are chosen such that each fold is representative of the global 
dataset in terms of both kinematic coverage and process types.
We consider hyperoptimization of different parameters: the NN architecture, the type of 
optimizer, the clipnorm and the value of the learning rate, which are varied as reported 
in \cref{tab:hyper_param}.
The distribution of the loss estimators $L_{\rm hopt}^{(\chi^2)}$ for each of these 
trial models is displayed in \cref{fig:hyperopt-dist}. Many model exhibit a similar 
value of the loss function, closer to the minimum, indicating that there exists many 
different methodology configuration leading to equally performing fits. If this proves 
the flexibility of our fitting methodology, it also implies that an accurate 
hyperoptimization is essential to determine methodological PDF uncertainties in our 
framework.
On contrary, the spread of the second momentum estimator $L_{\rm hopt}^{(\varphi^2)}$ 
is more pronounced suggesting that some of these models, despite achieving good 
description of the data, do not generalize in a conservative way, leading to a poor 
description of the unseen fold. Our selection criteria, based on both estimators aim 
to balance the presence of equally performing hyperparameter configuration selecting 
the optimal model leading to the largest PDF uncertainties given the available data.

\begin{table}[!t]
    \scriptsize
    \centering
    \renewcommand{\arraystretch}{1.4}
    \begin{tabularx}{\textwidth}{Xlll}
\toprule
Parameter                            & \multicolumn{2}{c}{Sampled range}                 & \multicolumn{1}{c}{Optimal model} \\
\cmidrule{2-3}
                                     &  min.                  & max.                     &               \\
\midrule
NN architecture                      & $n_1,n_2,n_3=10$       & $n_1,n_2,n_3=40$         & $n_1=29$,$n_2=12$,$n_3=6$ \\
Number of layers                     & $2$                    & $3$                      & $3$ \\ 
NN initializer                       & \textsc{Glorot\_normal}& \textsc{Glorot\_uniform} & \textsc{Glorot\_uniform} \\
Activation functions                 & \textsc{Tanh}          & \textsc{Sigmoid}         & \textsc{Tanh}  \\
Optimizer                            & \textsc{Nadam}         & \textsc{Adam}            & \textsc{Nadam} \\
Clipnorm                             & $10^{-7}$              & $10^{-4}$                & $2.95 \times 10^{-5}$ \\
Learning rate                        & $10^{-4}$              & $10^{-2}$                & $1.40 \times 10^{-3}$ \\
\midrule
Maximum \# training epochs           & \multicolumn{2}{c}{$17000$}                       & $17000$ \\
Stopping patience                    & \multicolumn{2}{c}{$0.1$}                         & $0.1$   \\
Initial positivity  multiplier       & \multicolumn{2}{c}{$185$}                         & $185$   \\   
Initial integrability multiplier     & \multicolumn{2}{c}{$10$}                          & $10$    \\
\bottomrule
\end{tabularx}

    \vspace{0.3cm}
    \caption{
        The hyperparameter space considered in this study. 
        We scan the internal neural network architecture (number of layers, nodes and activation functions), 
        the $\chi^2$ optimizer, the value of the clipnorm parameter, and the learning rate.
        In the lower part we list also other relevant hyperparameters which are kept fixed during the hyperopt
        and the PDF fit.
    }
    \label{tab:hyper_param}
\end{table}

\section{Results}
\label{sec:pol_results}
\begin{figure}[!t]
  \centering
  \includegraphics[width=0.45\textwidth]{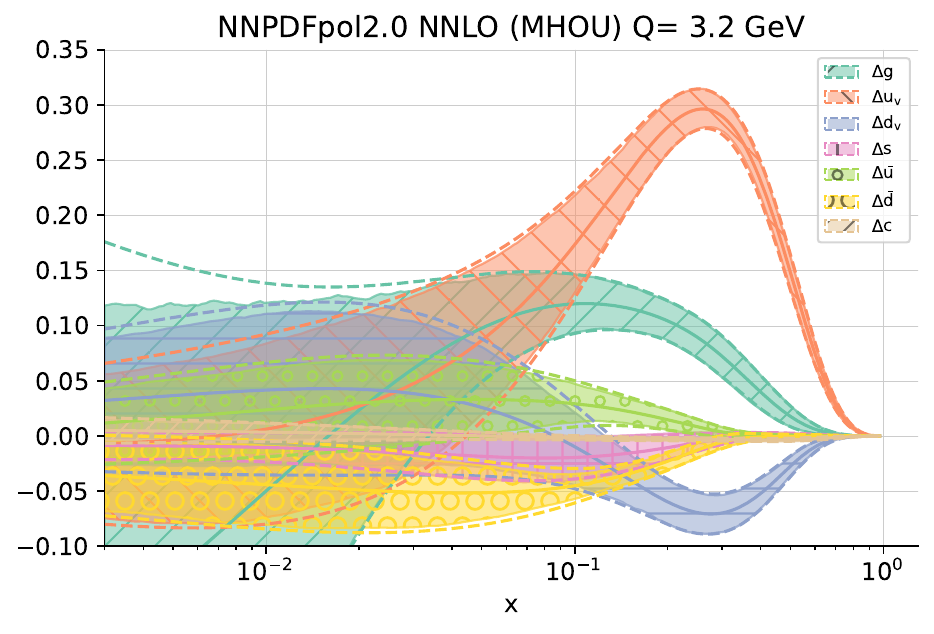}
  \includegraphics[width=0.45\textwidth]{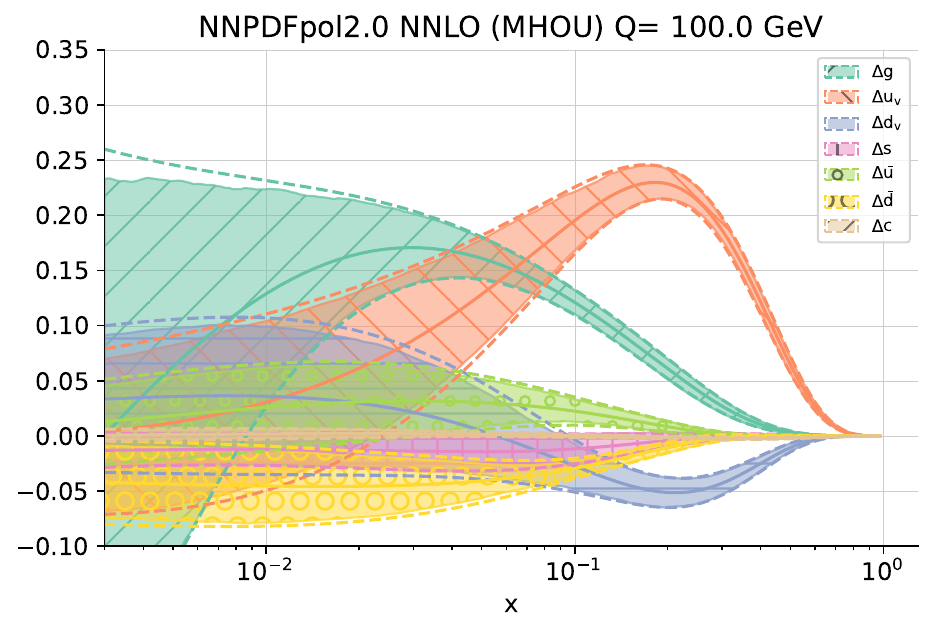}  
  \caption{
        The {\sc NNPDFpol2.0} NNLO MHOUs PDFs at $Q=3.2$~GeV (left) and $Q=10^2$~GeV (right).
  }
  \label{fig:pol_pdg}
\end{figure}

In this section, we present the {\sc NNPDFpol2.0} parton set. The results of the 
NNLO fit are shown in \cref{fig:pol_pdg} at low ($Q=3.2~\text{GeV}$) and a high 
($Q=100~\text{GeV}$) scale.
The quarks polarization is mainly dominated by the valence quarks, which display
a valence-like structure with $\Delta u^{-}$ being positive and $\Delta d^{-}$ 
negative. Overall $\Delta u^{-}$ and $\Delta d^{-}$ are quite well determined 
by the available data, with small uncertainties in the peak region. The contribution 
from $\Delta \bar{u}$ and $\Delta \bar{d}$ are almost identical and opposite at all scales. 
The polarization of the other flavors is suppressed, both at low- and high-$Q$ scales
and affected by larger uncertainties.
The gluon contribution instead, is positive in the large-$x$ region and 
compatible with 0 in the small-$x$ limit. Its magnitude has a stronger dependency
on the scale, with the DGLAP mixing pulling $\Delta g$ in the positive direction
at smaller $x$'s in favor of a depletion of the valence quark polarization for higher-$Q$. 

We now turn to scrutinize the perturbative stability of our results, in terms of fit 
quality and of parton distribution functions, and analyze the differences of our
new determination with respect to the previous NNPDF analysis. We then conclude
examining the implications of our fits on the total proton spin decomposition. 

\subsection{Dependence on theory and dataset variations}
\label{subsec:pol_pert_stability}

We begin the discussion looking at the fit quality. In \cref{tab:chi2_TOTAL_nnlo_mhou} 
we report the number of data points and the $\chi^2$ per data point in the LO, NLO and 
NNLO {\sc NNPDFpol2.0} PDF determinations before and after inclusion of MHOUs. 
Datasets are grouped according to the classification of \cref{sec:pol_data-theory} and 
MHOUs are computed with the 7-point prescription described above. 
All the fits display an overall good quality with the $\chi^2$ being closer to the unity.
As observed in the unpolarized case (cf. \cref{sec:an3lo_fit_settings}), fits which include 
MHOUs are more stable at different perturbative orders. Here, we can notice that the largest 
difference in $\chi^2$ between fits with and without theory uncertainties is indeed visible 
for the LO fits.
While inclusive jet and dijet data are equally described at all orders, the effect of QCD corrections 
is mainly noticeable in the DIS data, especially when going from LO to NLO.

\begin{table}[t!]
    \renewcommand{\arraystretch}{1.4}
    \scriptsize
    \centering
    \begin{tabularx}{\textwidth}{Xrcccccc}
  \toprule
  & 
  & \multicolumn{2}{c}{LO}
  & \multicolumn{2}{c}{NLO}
  & \multicolumn{2}{c}{NNLO} \\
  Dataset
  & $N_{\rm dat}$
  & no MHOU
  & MHOU
  & no MHOU
  & MHOU 
  & no MHOU
  & MHOU \\
  \midrule
  DIS NC
  &  704 & 1.15 & 1.03
         & 0.96 & 0.86
         & 0.95 & 0.92 \\
  DY CC
  &  12 & 1.28 & 0.79
        & 0.88 & 0.73
        & 0.72 & 0.64 \\
  Single-inclusive jets
  &  97 & 1.07 &  1.06
        & 1.07 &  1.07
        & 1.08 &  1.06 \\
  Dijet
  &  138 & 1.01 & 1.01
         & 1.04 & 1.03
         & 1.03 & 1.02 \\
  \midrule
  Total
  & 951 & 1.12 & 1.02
        & 0.97 & 0.90
        & 0.96 & 0.93 \\
\bottomrule
\end{tabularx}

    \vspace{0.3cm}
    \caption{
        The $\chi^2$ per datapoint for the 4 groups of datasets included in the fits, namely
        DIS Neutral current $g_1$ and $g_1/F_1$ (DIS NC), Drell-Yan asymmetries (DY CC), 
        single-inclusive jet and dijet asymmetries.
        We display result at different perturbative orders with and without MHOU computed
        with the methodology discussed in \cref{subsec:pol_th_covmat}. 
    }
    \label{tab:chi2tot}
    \end{table}

\begin{figure}[!t]
    \centering
    \includegraphics[width=0.45\textwidth]{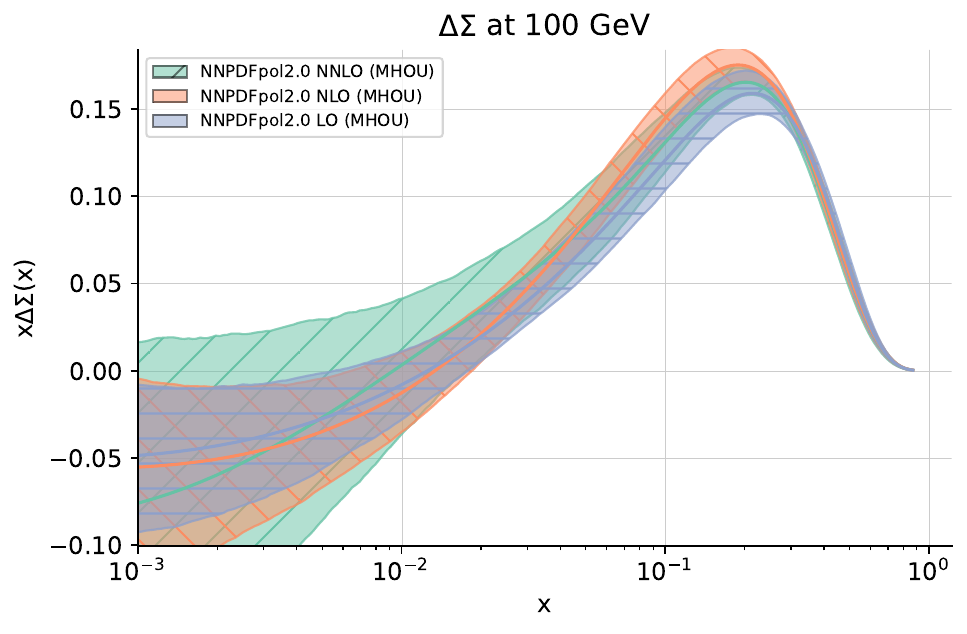}
    \includegraphics[width=0.45\textwidth]{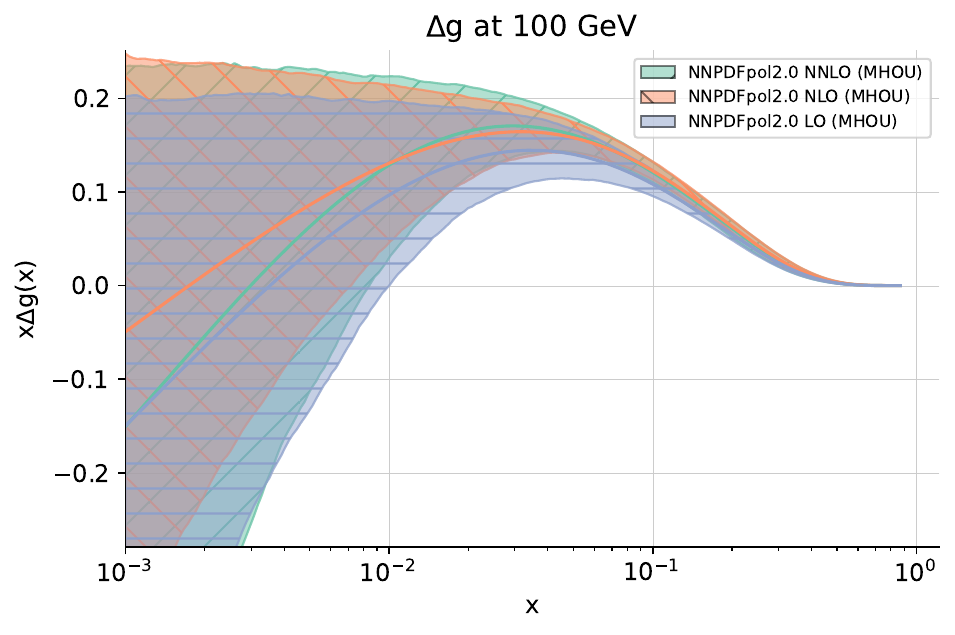}
    \includegraphics[width=0.45\textwidth]{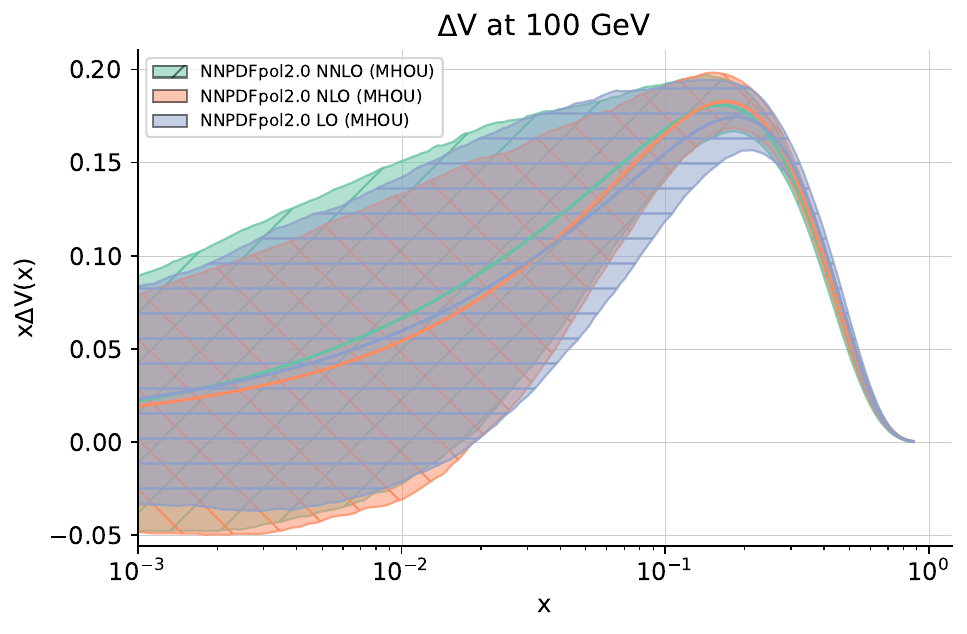}
    \caption{
        The LO, NLO, NNLO {\sc NNPDFpol2.0} PDFs at $Q=100$~GeV. 
        We display the gluon $\Delta g$, the total singlet $\Delta \Sigma$ and total valence $\Delta V$
        PDFs. Error bands correspond to one sigma PDF uncertainties including MHOUs.
    }
    \label{fig:pol_pdf_pert}
\end{figure}

The remarkable stability of the polarized PDF fits at different QCD orders is also visible 
directly on the PDFs, as shown in \cref{fig:pol_pdf_pert}. There we display the gluon $\Delta g$, 
the total singlet $\Delta \Sigma$ and total valence $\Delta V$ at $Q=100$~GeV 
for each perturbative order with the set including MHOUs. All the flavor combinations are compatible 
at the one $\sigma$ level, with the major differences visible only in the size of the uncertainties
bands, for specific kinematic regions. 
In particular, LO fits have broader uncertainties in both in the gluon and the valence-like PDFs 
and have lower central value both for the singlet and gluon PDFs. The size of the uncertainties at 
NNLO and NLO is comparable in most of the cases, except for the small-$x$ singlet, where we see a
quite broad enlargement of the PDF error in the NNLO fits (independently of the presence of MHOUs).
This behavior has been also observed in a similar study~\cite{Borsa:2024mss} and seems to be 
originated by somehow poorer control of the down-like quarks small-$x$ polarization. 
Although the origin of this behavior is not yet clear, it might be interesting to see if this 
effects is an artifact related to the inclusion of NNLO corrections in the $W$-asymmetries via 
$K$-factors.

\begin{figure}[!t]
    \centering
    \includegraphics[width=0.45\textwidth]{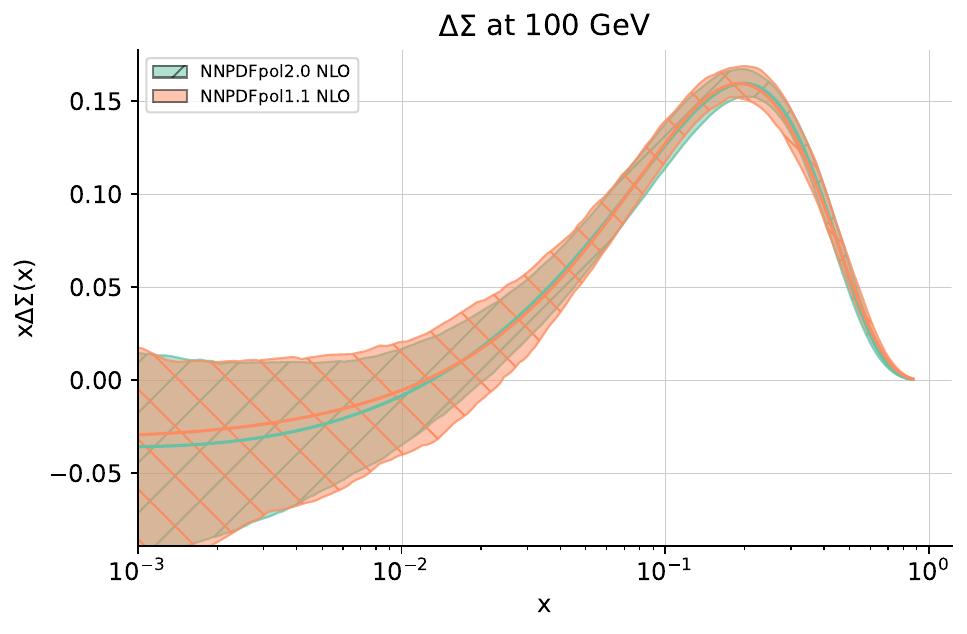}
    \includegraphics[width=0.45\textwidth]{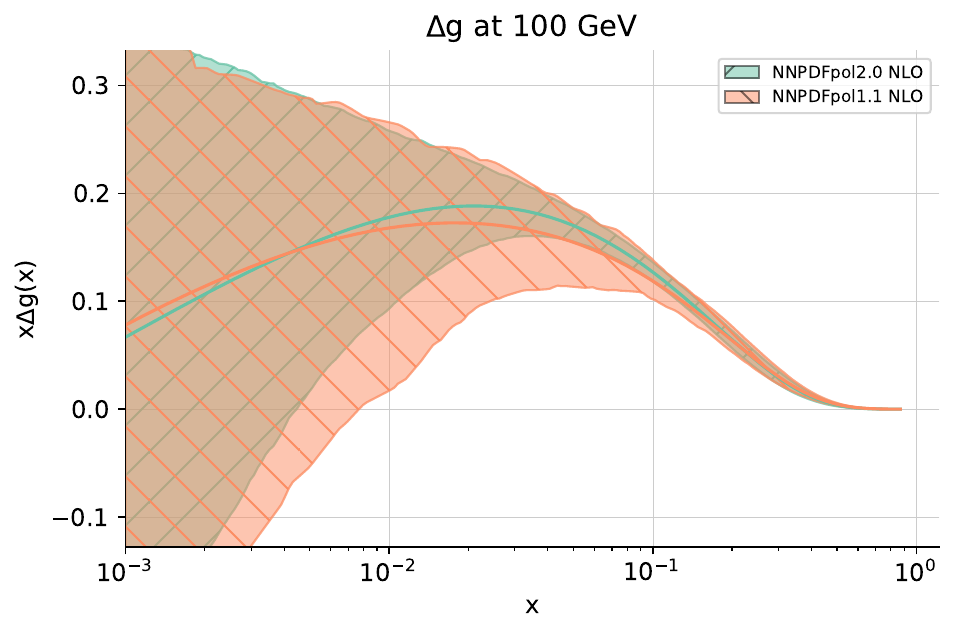}
    \includegraphics[width=0.45\textwidth]{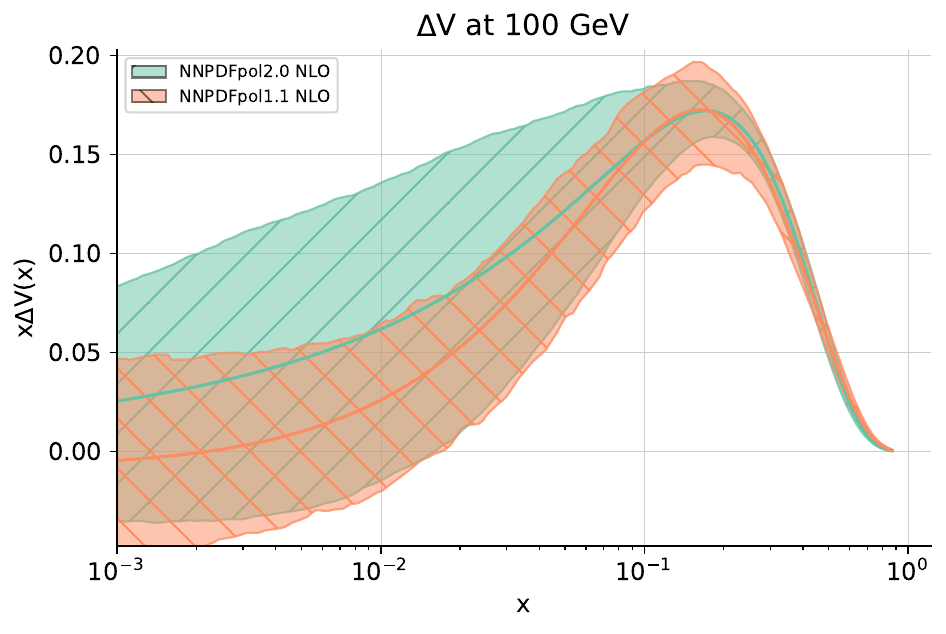}
    \caption{
    }
    \label{fig:pol_11_vs_20}
\end{figure}

By comparing our NLO determination with the previous release of the NNPDF 
collaboration~\cite{Nocera:2014gqa}, we can infer what are the impact of the new methodology 
and of the extended dataset coverage. \cref{fig:pol_11_vs_20} reports the same comparison of 
\cref{fig:pol_pdf_pert}, but now for the NLO set without MHOU and {\sc NNPDFpol1.1}. Overall 
we observe good consistency of the two results.
The singlet PDF is almost identical to the one of the previous study. This is a non-trivial 
finding that can be back traced to having included in the fits a very similar information on 
the DIS structure functions.
Moreover, this suggests that, in the case where large number of datapoints are available, the 
impact of the new hyperopt scan is also limited as verified also in Ref~\cite{Cruz-Martinez:2024wiu}.
The situation is different for the gluon PDF, which is mainly constrained by the jet data.
Indeed, $\Delta g$ displays smaller uncertainties for {\sc NNPDFpol2.0}, in the region $10^-2 \leq x \leq 0.2$,
due to the inclusion of the larger and more accurate STAR measurements of the jets and dijet
asymmetries.
In the case of non-singlet distributions, for e.g. $\Delta V$, we can distinguish a different 
behavior of the new fit in the small- and large-$x$ regions. This is originated from the proper
inclusion of the $W$-asymmetries into the fitting framework and to the addition of the 
valence-like PDFs $\Delta V, \Delta V_3$ as true degree of freedom of our fits.
In {\sc NNPDFpol1.1} the DY asymmetries where included only by means of Bayesian reweighting 
with the starting valence-like PDFs assumed to be equal of a specific prior functional form. 
Thus, we can appreciate that in our new $\Delta V$ determination, uncertainties are larger
in the extrapolation region (small-$x$), while are slightly improved in the peak region
which is the kinematic interval more correlated to the $W$-asymmetries.

In Ref~\cite[Sec.~4]{Cruz-Martinez:2025ahf} we have checked the impact on the fit of dataset variations 
and/or removal of theoretical constraints.
Specifically, we have verified that jets and dijet asymmetry measurements are equally 
constraining the gluon PDF, with the former measurements being slightly more effective.
The removal of small-$Q$ data, as for e.g. the large number of JLAB measurements, has little 
or no impact, mostly due to the magnitude of the experimental uncertainties of these data. 
This enforces confidence on having properly selected the validity of PDF determination. 
Finally, we have studied the impact of the sum rules from the baryon octet decays, 
\cref{eq:normalisation} and, the cross-section positivity constraint of \cref{eq:positivity_pdfs}.
Regarding the former item, a sizeable breaking of SU(3) is advocated in the 
literature~\cite{Flores-Mendieta:1998tfv}, which results in an inflation of the uncertainty 
on $a_8$ up to $30\%$. In order to account for this, and more generally to test the 
sensitivity of the data to SU(3) symmetry breaking, we perform a PDF fit 
that differ for the uncertainty associated to $a_8$, which is enlarged up to 
50\% of the nominal value. This results in PDFs which are fully compatible with the standard 
settings, showing that after training our default $\Delta T_3$ and $\Delta T_8$ are fully 
constrained by the data rather than by these constraints.
Regarding the cross-section positivity, we have performed fits removing completely the 
condition of \cref{eq:positivity_pdfs}. Here we observe that the larger size and more 
constrained PDFs $\Delta \Sigma$ and $\Delta g$ are only mildly affected, with the singlet 
being slightly enhanced in the large-$x$ region. On the other hand, the positivity bound 
has a relevant impact on the suppressed flavor, as $\Delta \bar{u},\Delta \bar{d}$ and $\Delta s$. 
These, once \cref{eq:positivity_pdfs} is not imposed, tend to be unnaturally wiggly in the 
large-$x$ region or slowly converging to $0$ in the $x \to 1$ limit.

\subsection{Implications for the proton spin}
\label{subsec:pol_spin}

By definition, the first moment of the polarized PDF is related to the spin fraction carried 
by the parton inside the original nucleon. In fact, starting from the definition of 
\cref{eq:pdf_def}, one can obtain the polarized PDF by inserting a spin projector.
Given a parton $q$ with spin $S_q$ we define the net-spin fraction $\eta_q$ as
\begin{equation}
    \eta_q = S_q \int_0^1 dz \Delta q(z) \,.
    \label{eq:spin_def}
\end{equation}
The naive parton model, where the proton is described by the quasi-free valence quarks, suggests 
then to decompose the total proton angular momentum $J$ in terms of the quark spins and 
the orbital angular momentum $L_z$.
However, as we have shown in the previous section, also the gluon polarization must be taken
into account, and thus we decompose $J$ as
\begin{equation}
    J = \sum_q \eta_q + \eta_g + L_z =  \frac{1}{2} \,.
    \label{eq:j_decomposition}
\end{equation}
Let us mention that this decomposition is not unique \cite{Deur:2018roz}, but further decomposition
of $L_z$ in terms of gluon and quarks are would not be fully gauge invariant.
The experimental measurements show that parton net-spins contribute only to a fraction, approximately 
30\%, of the total angular momentum, leaving significant room for a non-vanishing $L_z$, 
and a possible explanation of its origin.
Specifically, also the presence of proton's anomalous magnetic moment requires a nonzero 
orbital angular momentum.
Different non-perturbative models, as well as lattice QCD calculations (see \cite{,Deur:2018roz} 
for an extensive review), allow us to predict values of the components of \cref{eq:j_decomposition}, 
and in some cases are even able to relate them to a particular flavor asymmetry.

\begin{figure}[!t]
    \centering
    \includegraphics[width=0.7\textwidth]{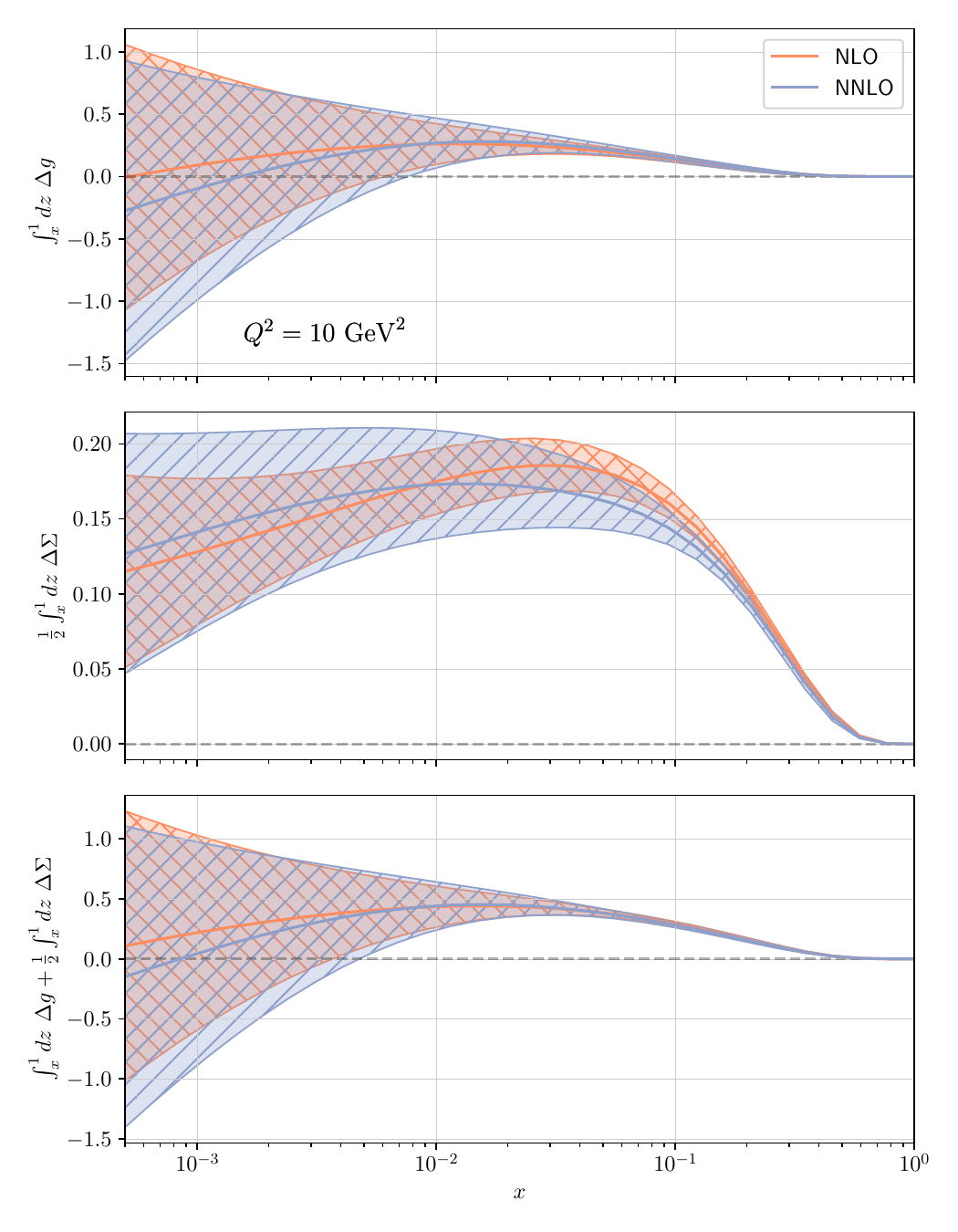}
    \caption{
        The truncated net-spin fraction as a function of $x$ and $Q^2=10~\text{GeV}^2$.
        We display the gluon (top), the total quark singlet (middle) and the 
        combined contributions (bottom). The uncertainty band includes PDF and MHOU
        contributions. 
    }
    \label{fig:spin_fraction}
\end{figure}

Currently, the nucleon spin origin is still an open problem. Our updated determination of polarized
PDFs can be used to infer what is the most likely scenario suggested by the high-energy scattering data.
In \cref{fig:spin_fraction} we display the truncated net-spin fraction
\begin{equation}
    \eta_q^{[x,1]} = S_q \int_x^1 dx \Delta q(x) \,,
    \label{eq:spin_def_tru}
\end{equation}
for the gluon (top), the quark singlet (middle) and the sum of the two contributions (bottom plot) 
at the scale $Q^2=10~\text{GeV}^2$. The values are computed with the NLO and NNLO sets and include MHOUs.
As we probe smaller $x$ values, the quark contribution appears to be convergent and stable.
At $x = 10^{-3}$ the partial net-spin is equal to $0.14 \pm 0.05$. 
Unfortunately, the situation is rather different for the gluon, where the poor accuracy of the PDFs
in small-$x$ region, prevents us to predict a stable estimate of $\eta_g$, which results to be 
compatible with 0 and affected by a large uncertainty.
The poor knowledge of $\Delta g$ affects also the combined value of singlet and gluon.
In this respect our analysis shows that the EIC measurements are still needed to fully resolve the
proton content and its polarization at small-$x$.

\begin{figure}[!t]
    \centering
    \includegraphics[width=0.49\textwidth]{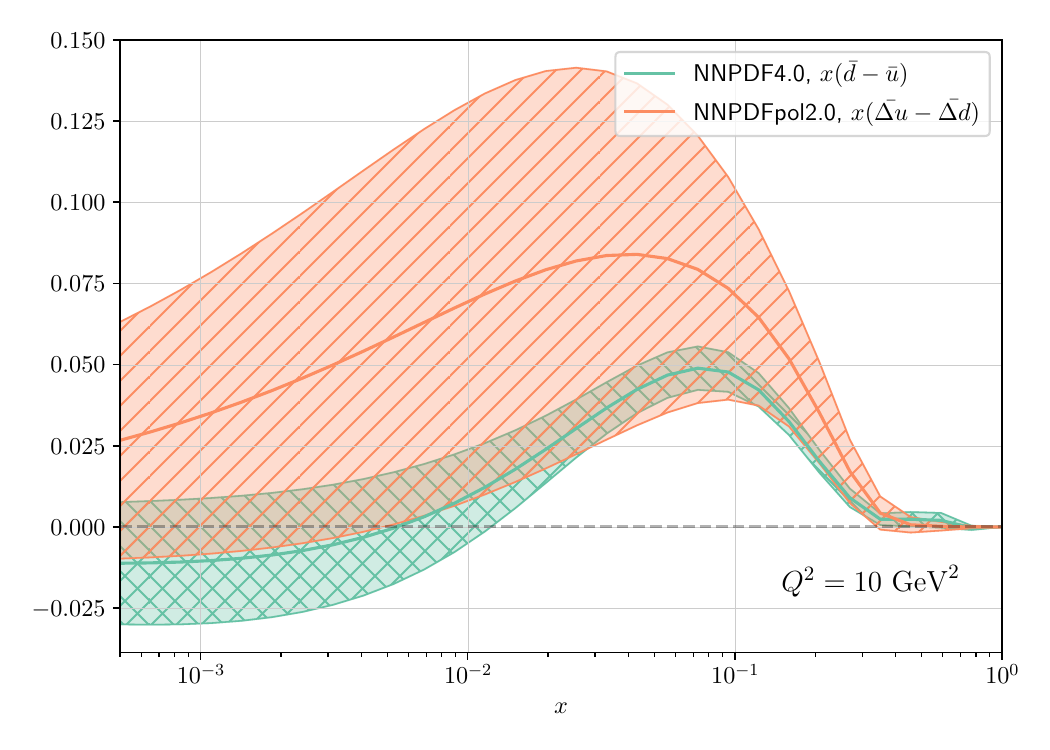}
    \caption{
        Comparison of the polarized ($x\Delta \bar{u} - x\Delta \bar{d}$) and unpolarized 
        ($x\bar{d} - x\bar{d}$) sea quark asymmetry. 
        We display the result for NNLO PDFs with MHOUs at $Q^2=10~\text{GeV}^2$.
    }
    \label{fig:unpo_pol_asy}
\end{figure}

Regarding a possible light flavor polarized sea asymmetry, our determination confirms the finding of
previous studies \cite{Nocera:2014oia} with a $\Delta \bar{u} > 0 > \Delta \bar{d}$, with the 
magnitude of the difference equal or greater than the unpolarized one and 
similar in shape as visible in \cref{fig:unpo_pol_asy}. 

  \chapter{Conclusion and future prospects}
\label{chap:Conclusion}

This thesis collects and summarizes different works regarding 
the topic of Parton Distribution Functions (PDFs). 
PDF uncertainties are currently among the dominant contributions 
to the theoretical error of many LHC observables, such as Higgs boson 
production, Drell-Yan or strong coupling constant measurements. 
Results of the same analysis, carried out with different input PDFs, are 
often not fully consistent, and their combination is therefore non-trivial. 
Thus, to achieve a better accuracy of theoretical predictions we need a 
deeper understanding of both theoretical and methodological
uncertainties arising during the determination of PDFs. 
Moreover, the use of perturbative methods to compute Standard Model
observables entails that to reach higher precision we must include 
higher order corrections, among which QCD are often the largest contribution.

The studies presented here aim to tackle both problems and include, as 
deliverables, different new unpolarized and polarized PDF sets which 
will serve as a useful input for upcoming research in the high energy physics 
phenomenology community.


\section*{Summary}

To begin, in \cref{chap:introduction,chap:methodology} we introduced,
the theoretical and methodological framework of NNPDF4.0, adopted for the following 
sections.

In \cref{chap:ic} we have provided a first evidence of the intrinsic 
charm presence inside the proton, carrying a total momentum fraction less than 
$1~\%$.
The resulting fitted charm PDF displays a characteristic valence-like shape, 
which is difficult to reconcile with a perturbatively generated charm 
from quark and gluon splitting. By inverting PDF matching conditions we have isolated
the intrinsic charm component and analyzed its stability.  
Moreover, we have discussed the possibility of finding a non-vanishing intrinsic 
charm asymmetry, although this is now beyond the level of accuracy 
of the present PDF sets.
Both studies take into account QCD higher order corrections and are 
supplemented by a comparison with present and future LHC and EIC observables, which 
can eventually discriminate this tiny effect.

In \cref{chap:an3lo}, we have shown how approximate N$^3$LO corrections 
and their theoretical uncertainties can be systematically included in a PDF fit, 
obtaining one of the most precise unpolarized PDF determination currently available. 
In particular, we have considered QCD corrections to DGLAP evolution and DIS coefficient 
functions which grasp the largest component of the full N$^3$LO corrections.
The resulting PDF set can be used for a twofold purpose of complementing N$^3$LO
matrix element calculations o compute consistent cross-sections and improve the 
estimate of previous order theory uncertainties. 

Finally, \cref{chap:pol} describes the most recent extraction of helicity dependent PDFs 
at NNLO accuracy. The study accounts for missing higher order uncertainties,
and it shows an excellent perturbative convergence of polarized PDFs. 
The considered experimental data embrace Deep Inelastic measurements 
as well as hadronic data of jet, dijet and Drell-Yan production, which extend the 
kinematic coverage and are coherently fitted for the first time.
We see that NNLO corrections to polarized PDFs are small in almost all kinematic regions,
with impact of MHOUs also being limited. The more consistent treatment of the hadronic 
asymmetries, however, provides better sensitivity in specific kinematic regions with respect 
to previous determinations. This is visible principally for mid-$x$ ranges in the gluon 
and valence-like PDFs. 
On the other hand the enhanced fitting methodology suggests more conservative bounds 
in the extrapolation regions, especially for flavor suppressed components.
Our analysis further suggests that the quarks carry only a fraction of the total proton 
spin (around 30\%), consistent with previous studies. Current sensitivity from the RHIC data 
however is not sufficient to determine accurately the small-$x$ gluon, leading to integrated 
gluon polarization compatible with 0 and large uncertainties.


\section*{Future prospects}

Scientific studies rely on the ability to compare models with data and describe new 
phenomena in a quantitative way both at the level of predictions and uncertainties.
While the former are continuously improved by new ideas and methodological
updates, the latter are more difficult to validate and, the presence of different studies 
based on the same underlying rules, is essential to confirm a scientific discovery 
and rule out possible incidental biases.

Throughout this thesis, to improve the determination of collinear PDF, we have 
applied a consistent inclusion of higher order QCD corrections together with 
theoretical uncertainties in different PDF fits and, studied their impact 
assuming a fixed fitting methodology. 
The ensuing uncertainties are determined within the covariance matrix formalism 
combined with scale or parameter varied predictions. The approach has the advantage 
to be valid for all the scattering processes considered in a PDF fit and to decouple 
the estimate of theoretical errors from the methodological and experimental ones.
However, this procedure contains some arbitrariness, which calls for benchmark studies
against different approaches~\cite{Kassabov:2022orn,Bonvini:2020xeo,McGowan:2022nag}.
Although there is not yet a consensus about the use of (PDF) theory errors,
reaching an agreement is now becoming a more stringent task for the high-energy 
physics community, especially in view of new LHC analysis.
In this respect, our new approximate N$^3$LO PDFs could be helpful to validate 
the goodness of different theoretical uncertainty methodologies, for e.g. 
by comparing N$^3$LO-NNLO cross-section shifts with lower order results 
supplemented with theory uncertainties~\cite{MSHT:2024tdn}.
Future updates on theoretical computations for PDF fits can include the complete 
removal of the NNLO $K$-factors for hadronic observables~\cite{Sharma:2024cnc} which 
will open the possibility for a larger inclusion of exact N$^3$LO partonic matrix elements 
and reduce or correct the current estimate of the incomplete higher order uncertainties.

On the other hand, in this thesis, we have not discussed the effect of fitting methodology 
on PDF uncertainties. Here, a more extensive use of common closure or future tests 
\cite{Harland-Lang:2024kvt} would be beneficial to trim the number of available PDF 
sets to the ones offering more reliable predictions for future observables.
A consistent validation of the current PDF methodologies based on an orthogonal technique, 
such as Gaussian Kernels, would be very interesting \cite{Candido:2024hjt}. This study 
could indeed confirm or reject the current size of the PDF uncertainties.

Regarding the topic of intrinsic charm, a competitive analysis from a different PDF 
fitting group would be highly beneficial to corroborate our finding.
From our side, to further elucidate the origin of the proton intrinsic charm, a more 
comprehensive study of the relationship between non-perturbative models and the PDF extracted 
from high energy data might be needed in the future.
For example, the investigation of higher twist effects neglected in the current 
factorization approach could be a first step in this direction. 
Eventually, the detection of non-vanishing intrinsic charm asymmetry effect at the 
HL-LHC or EIC could be a cornerstone measurement on this long-standing topic.

Turning to the analysis of helicity dependent PDFs, the EIC, expected to start its 
operations in the 2030s, is designed to revolutionize this state of affairs. The 
advanced detector design will allow us to resolve the small-$x$ region down to 
$x \approx 10^{-5}$ reducing the uncertainties on the $g_1$ measurements and 
possibly also on the gluon polarization. 
This should solve the long-standing proton spin puzzle allowing us to determine 
quantitatively the proton orbital angular momentum.
These forthcoming measurements will involve inclusive DIS on nuclear targets but also 
semi-inclusive structure functions, which are then sensitive to light hadron fragmentation 
functions (FF).
With these prospects in mind, our study is a first step towards a simultaneous global QCD 
fit of polarized and unpolarized, proton and nuclear PDFs together with hadronic FFs which 
will allow us to fully take into account all the theoretical and methodological correlations 
originating from the same underlying law.
In spite of the fact that the goal seems quite challenging, we believe the tools
used and developed in this thesis to be flexible and user-friendly enough to be 
extended for such scope.

\end{mainmatter}

\renewcommand{\thechapter}{\Alph{chapter}}
\renewcommand{\thesection}{\thechapter.\arabic{section}}
\renewcommand{\theequation}{\thesection.\arabic{equation}}
\begin{appendices}
  \chapter{An improved aN$^3$LO $P_{gq}^{(3)}$ parametrization}
\label{app:new_an3lo_pgq}

Ref.~\cite{Falcioni:2024xyt} provides an additional number of moments of the 
N$^3$LO splitting function $P_{gq}^{(3)}$ with respect to the one available
when the study of Ref.~\cite{NNPDF:2024nan} was performed. 
This allows us to assess the accuracy of our splitting function approximation 
by comparing results obtained by including increasingly more information.

\begin{figure}[!ht]
    \centering
    \includegraphics[width=0.49\linewidth]{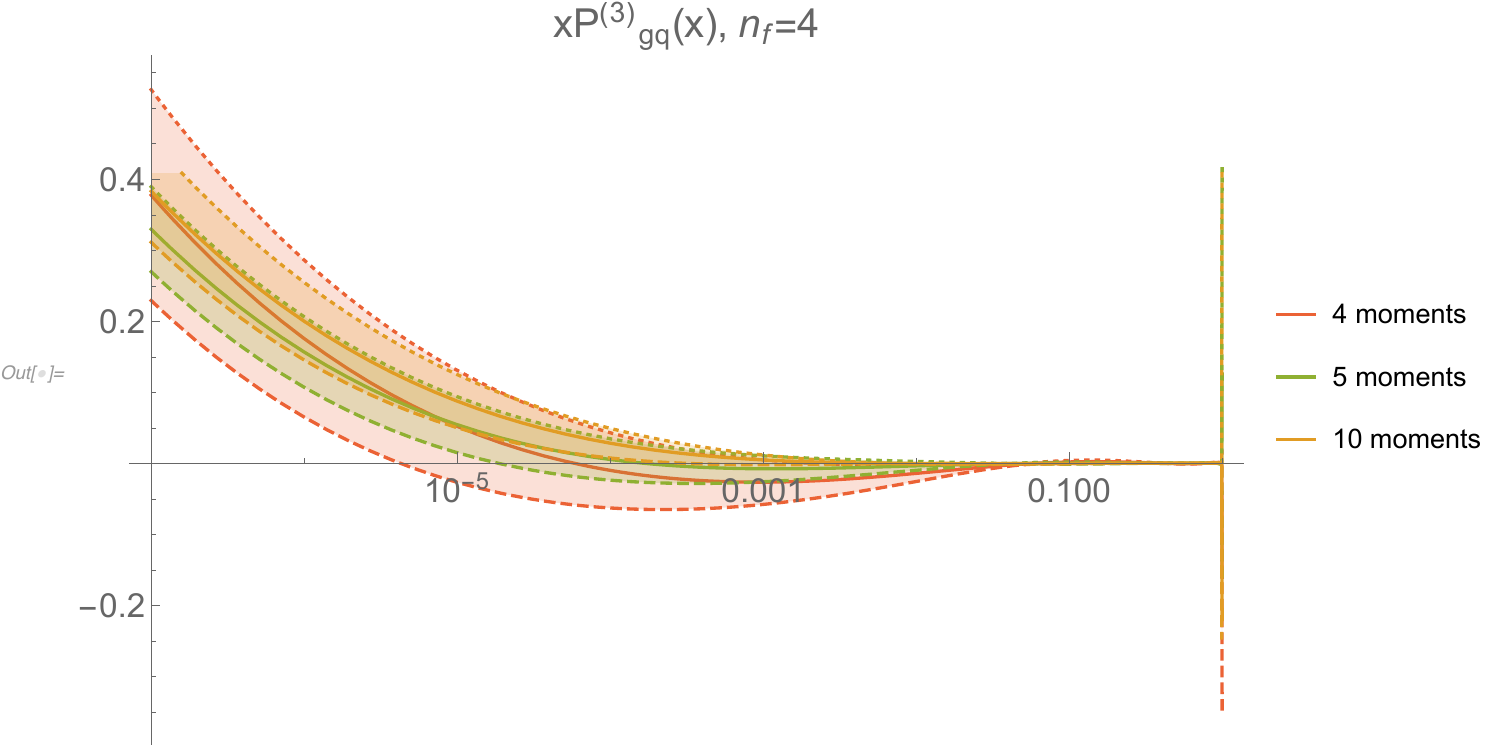}
    \includegraphics[width=0.49\linewidth]{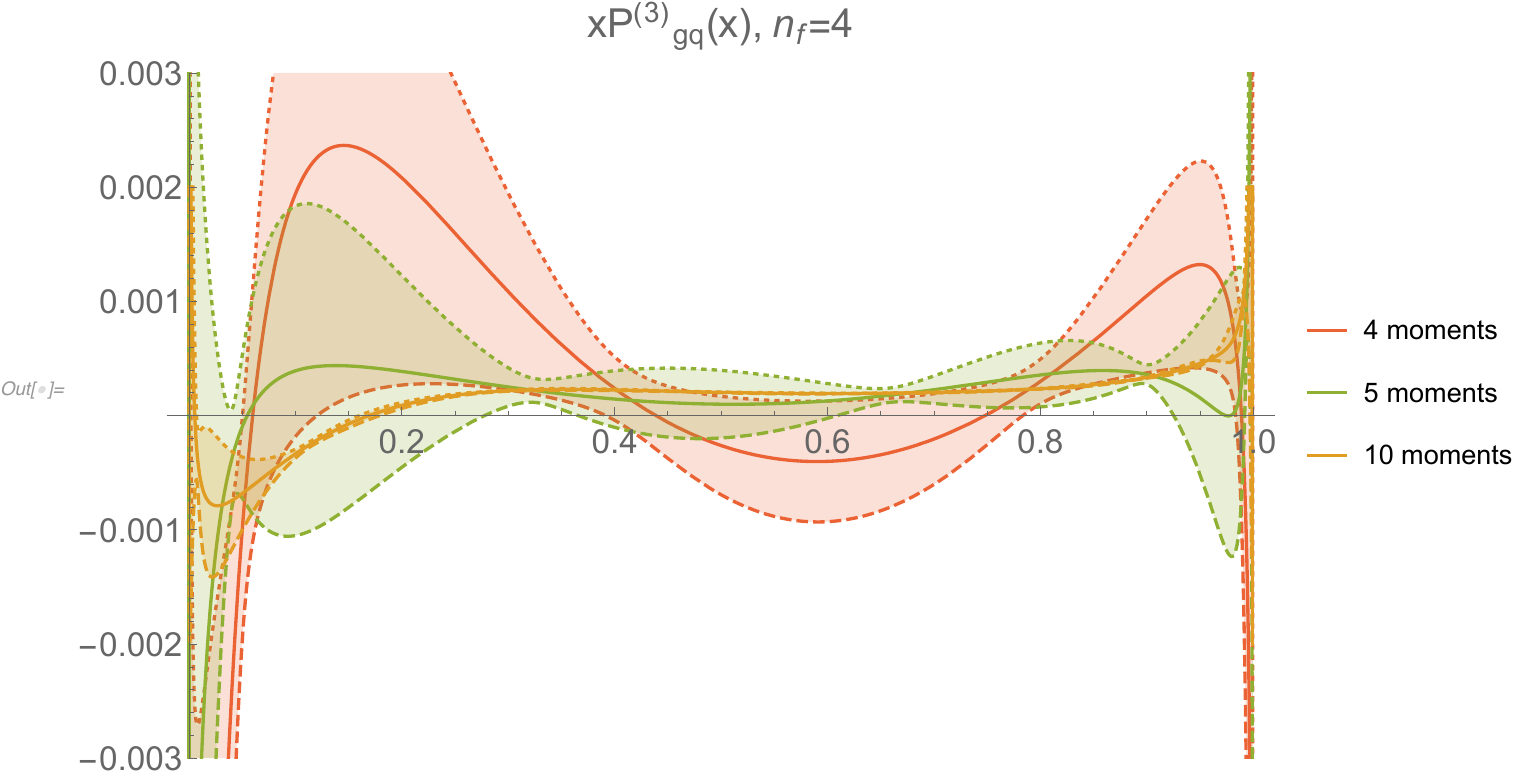}
    \caption{
      The  aN$^3$LO contribution to the gluon-quark splitting function, 
      $xP_{gq}^{(3)}(x)$, together with the corresponding IHOU estimate,
      in logarithmic (left) and linear (right) scale.
      We compare results obtained with the full set of 10 available Mellin moments
      with those where the $xP_{gq}^{(3)}(x)$ parametrization is constrained by a subset
      composed of only 4 or 5 moments.
    }    
    \label{fig:xpg3_moments}
\end{figure}

In order to illustrate this point, we show in \cref{fig:xpg3_moments}
how the $P_{gq}^{(3)}$ approximation obtained with the method of 
\cref{sec:an3lo_general_strategy} with $4$, $5$ or $10$ Mellin moments
are considered. 
As we include further constraints, the uncertainty on the approximation 
becomes smaller and the final parametrization contains fewer oscillations
especially in the large-$x$ region (right plot).
The methodology seems to estimate IHOU correctly with the more
precise result within the error band of the previous in most of the 
$x$-range.
The reduction of IHOU is more prominent for values $x \geq 10^{-4}$, 
with the small-$x$ behavior quite unaffected by the presence of more moments.
This further suggest that our approximate N$^3$LO splitting functions 
are only reliable in a finite small-$x$ region, while a precise 
determination of small-$x$ splitting (and PDFs) requires resummation techniques.

  \chapter{Comparison with MSHT20aN3LO}
\label{app:an3lo_msht20}

In this appendix, we compare the NNPDF4.0 aN$^3$LO PDF set to the only other existing aN$^3$LO
PDF set, MSHT20 aN$^3$LO~\cite{McGowan:2022nag}. As already discussed
in \cref{sec:an3lo_comp}, MSHT20 aN$^3$LO PDFs are determined by fitting
to the data the nuisance parameters that parametrize the IHOU
uncertainty on a prior approximation to splitting functions.
It follows that the ensuing central
value is partly determined by the data, and the IHOU is entirely
data-driven. When comparing NNPDF4.0 and MSHT20 aN$^3$LO PDF sets it should of
course be borne in mind that the sets already differ at NNLO due to
differences in dataset and methodology. The NNLO MSHT20 and NNPDF4.0
PDF sets were compared in Fig.~21 and the corresponding parton
luminosity in Fig.~60 of Ref.~\cite{NNPDF:2021njg}, while a detailed
benchmarking was presented in Ref.~\cite{PDF4LHCWorkingGroup:2022cjn}.

\begin{figure}[!p]
  \centering
  \includegraphics[width=0.45\textwidth]{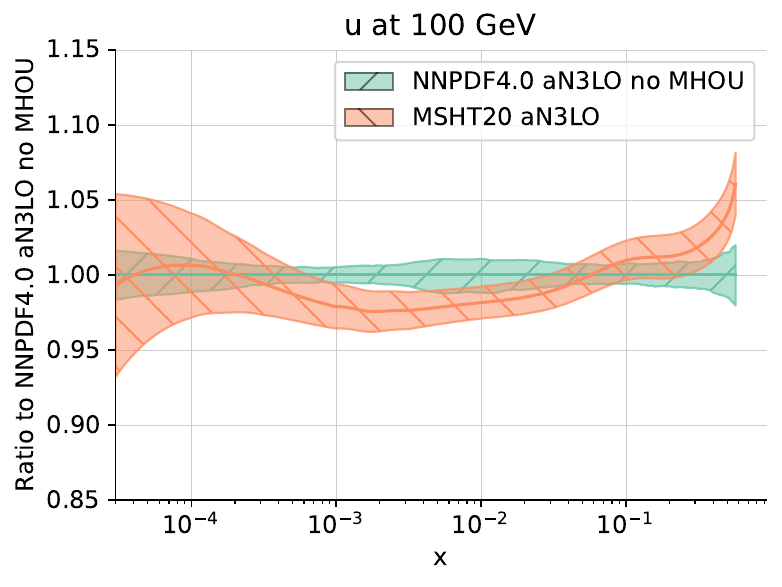}
  \includegraphics[width=0.45\textwidth]{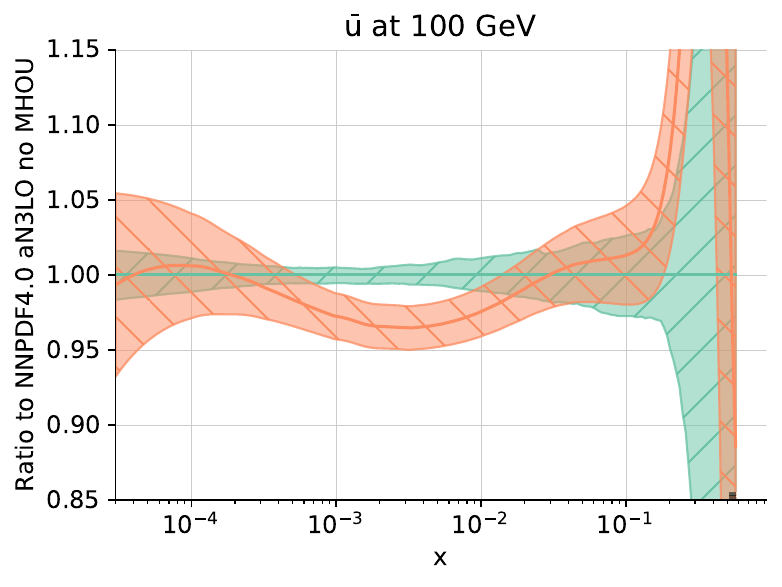}\\
  \includegraphics[width=0.45\textwidth]{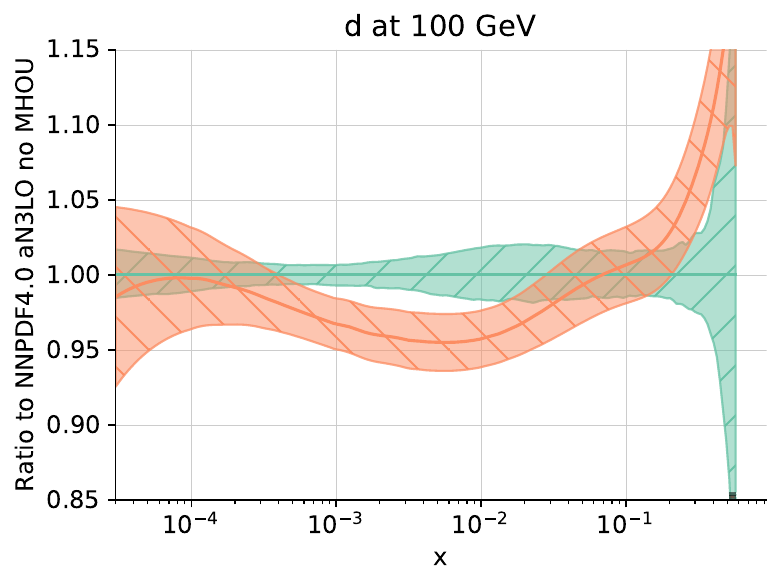}
  \includegraphics[width=0.45\textwidth]{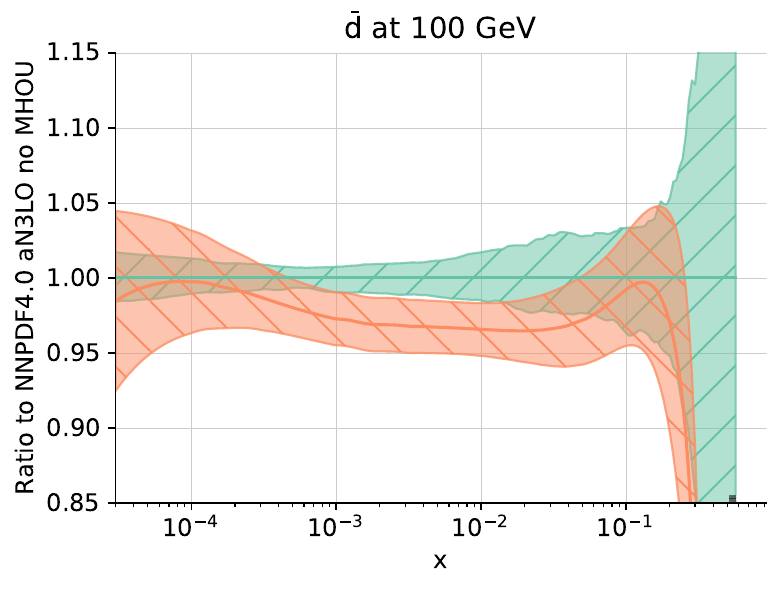}\\
  \includegraphics[width=0.45\textwidth]{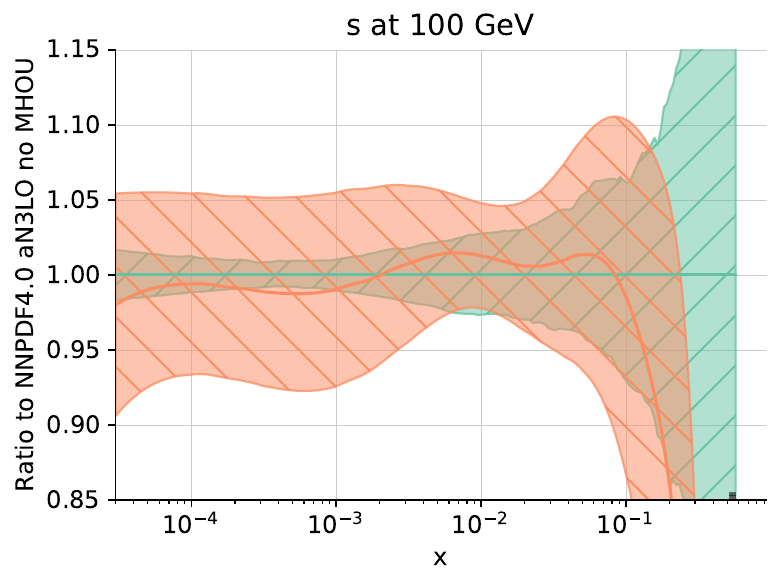}
  \includegraphics[width=0.45\textwidth]{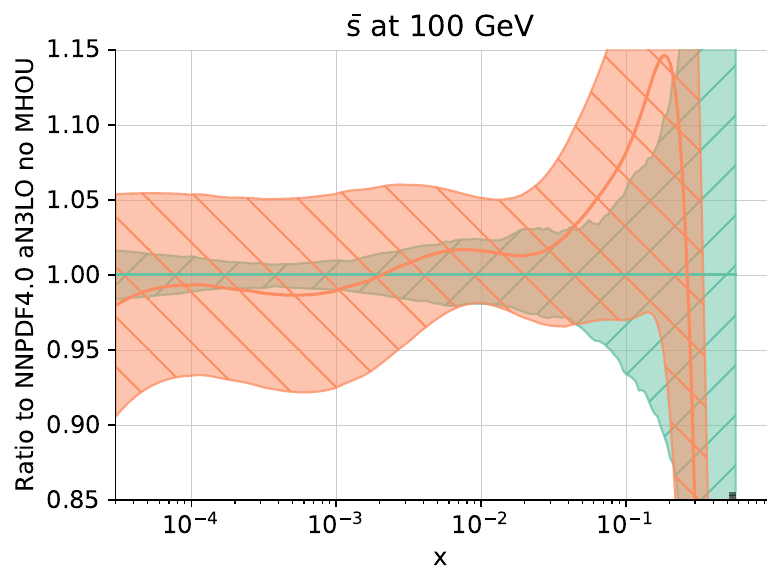}\\
  \includegraphics[width=0.45\textwidth]{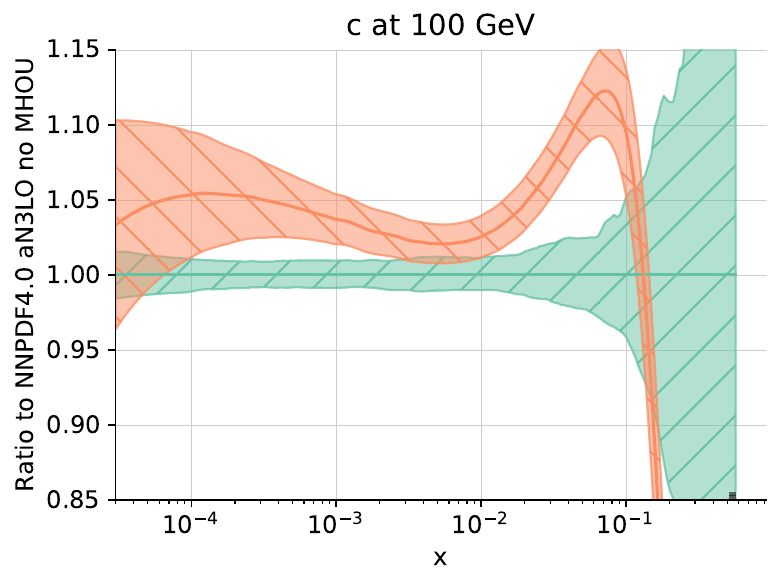}
  \includegraphics[width=0.45\textwidth]{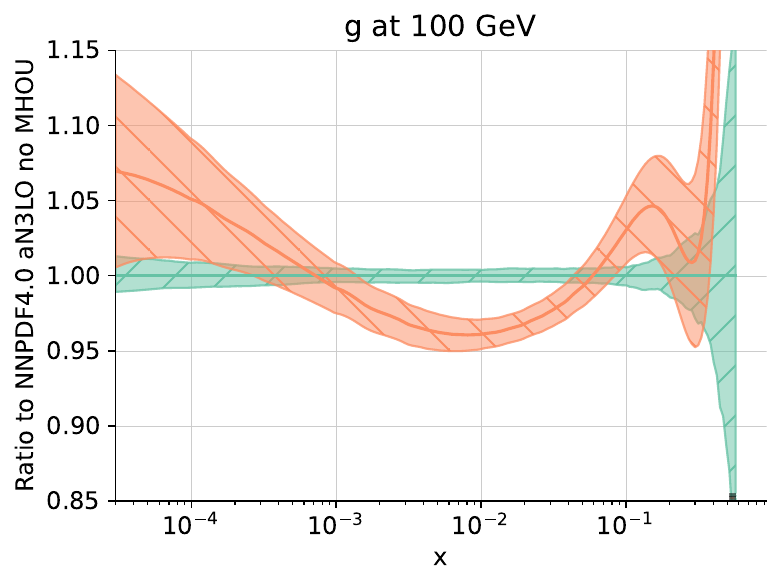}\\
  \caption{Same as
    \cref{fig:pdfs_noMHOU_log}, now
    comparing the NNPDF4.0 aN$^3$LO baseline PDF set without
    MHOUs to  the MSHT20 set recommended as baseline in
    Ref.~\cite{McGowan:2022nag}.}
  \label{fig:PDFs_MSHT20}
\end{figure}

\begin{figure}[!t]
  \centering
  \includegraphics[width=0.45\textwidth]{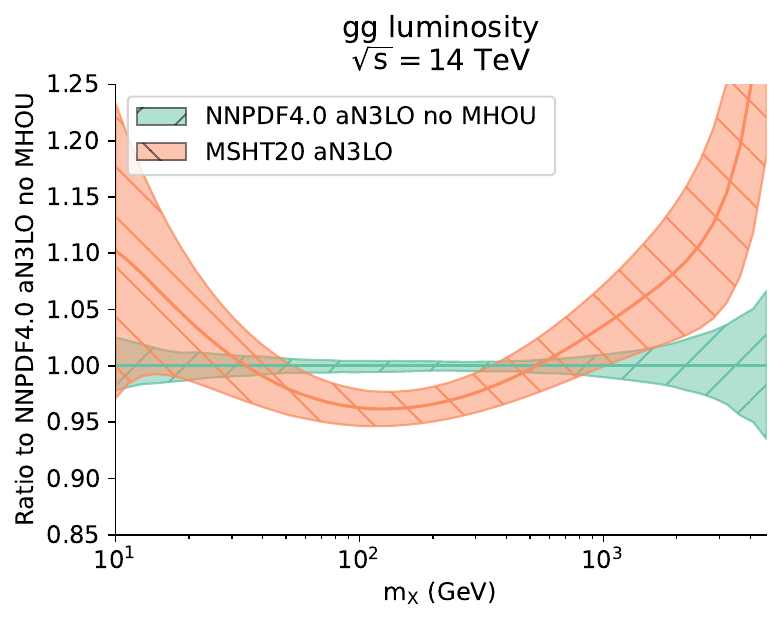}
  \includegraphics[width=0.45\textwidth]{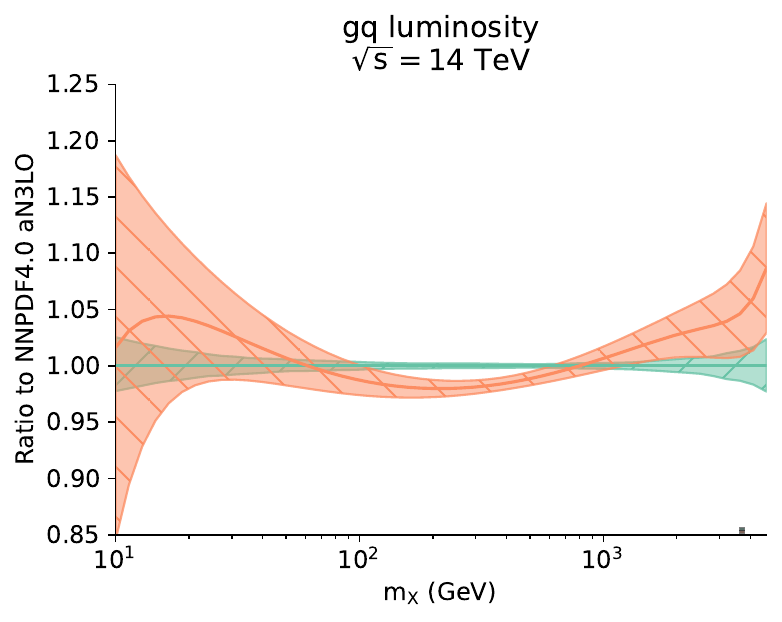}\\
  \includegraphics[width=0.45\textwidth]{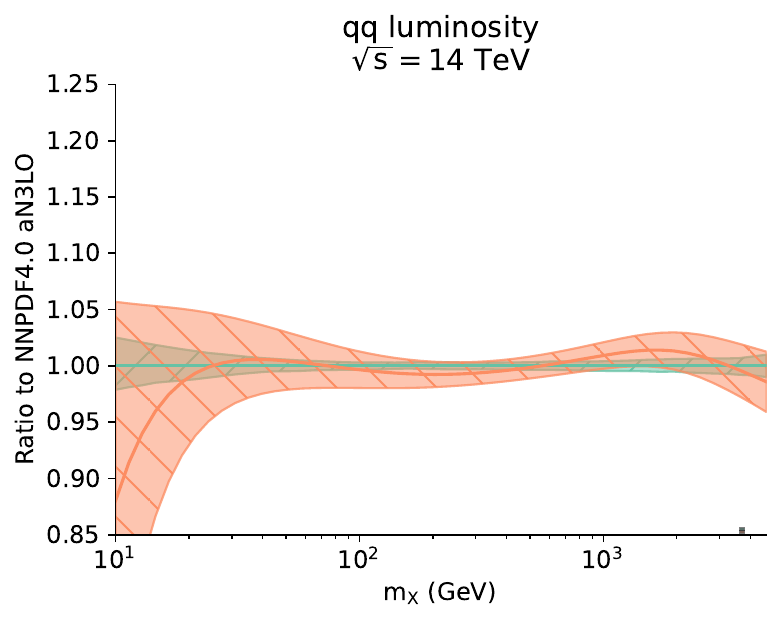}
  \includegraphics[width=0.45\textwidth]{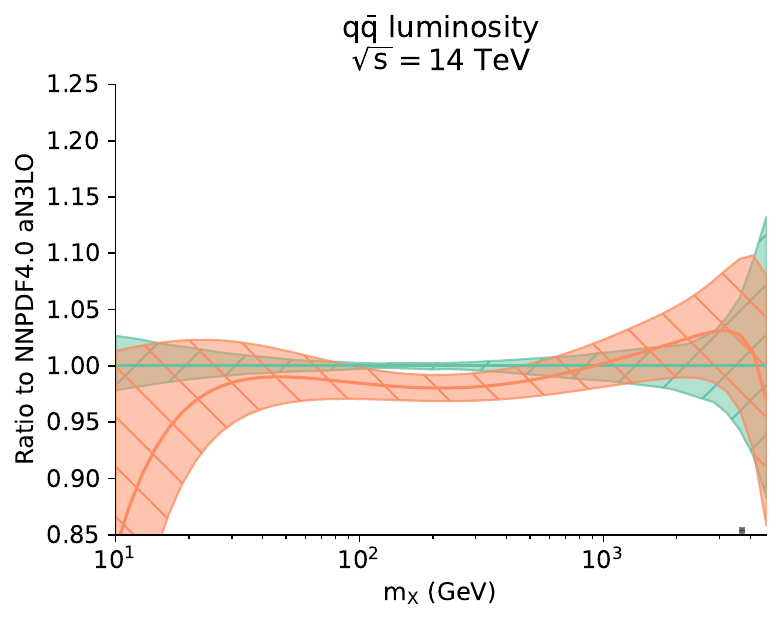}\\
  \caption{Same as \cref{fig:PDFs_MSHT20} for parton luminosities as in
    \cref{fig:lumis}.}
  \label{fig:lumis_MSHT20}
\end{figure}

The comparison of the aN$^3$LO sets is presented in
\cref{fig:PDFs_MSHT20}, where we show the NNPDF4.0
no MHOU set and the  MSHT20 set recommended as baseline in
Ref.~\cite{McGowan:2022nag} at $Q=100$~GeV, normalized to the NNPDF4.0 central
value. All error bands are one sigma uncertainties. 
The dominant differences between the PDF sets are the
same as already observed at NNLO, with  the largest difference
observed for the charm PDF, which is independently parametrized in
NNPDF4.0, but not in MSHT20, where it is determined by perturbative
matching conditions. However, the differences, while remaining qualitatively
similar, are slightly reduced (by $1-2~\%$) when moving from NNLO to aN$^3$LO.
Exceptions are the charm and especially the gluon PDF, which differ more at aN$^3$LO. 
Specifically, the gluon PDF, while reasonably compatible for $x\lesssim 0.07$ at NNLO, 
disagrees at aN$^3$LO, with the MSHT20 result suppressed by $3-4~\%$ in the region
$10^{-3}\lesssim x \lesssim 10^{-1}$, with a PDF uncertainty of $1-2~\%$. 
This suppression of the MSHT20 gluon can likely be traced to the 
behavior of the $P_{gq}$ splitting function seen in \cref{fig:splitting-functions-mhst}.

\begin{figure}[!p]
  \centering
  \includegraphics[width=0.42\textwidth]{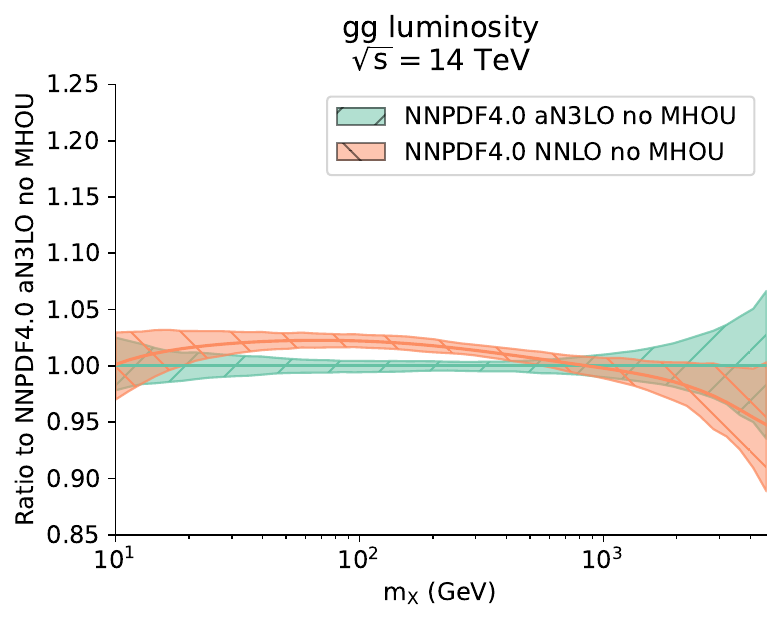}
  \includegraphics[width=0.42\textwidth]{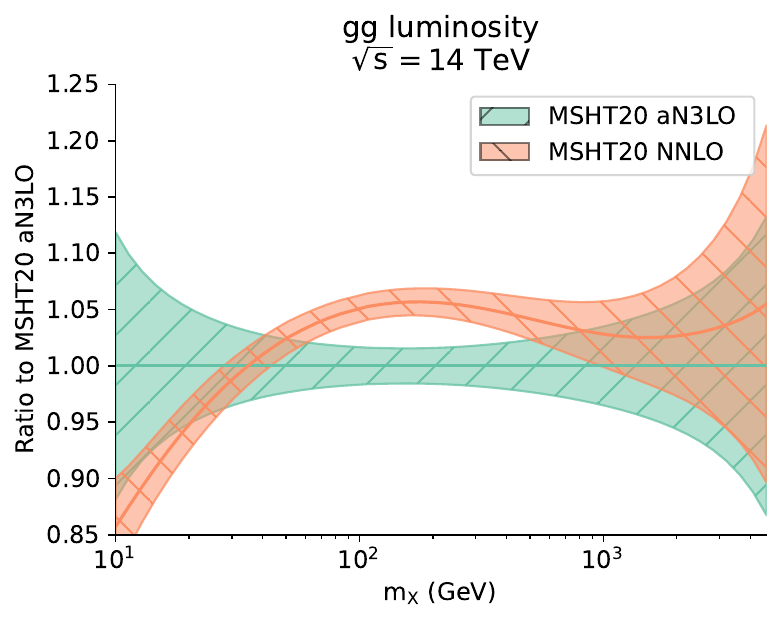}
  \includegraphics[width=0.42\textwidth]{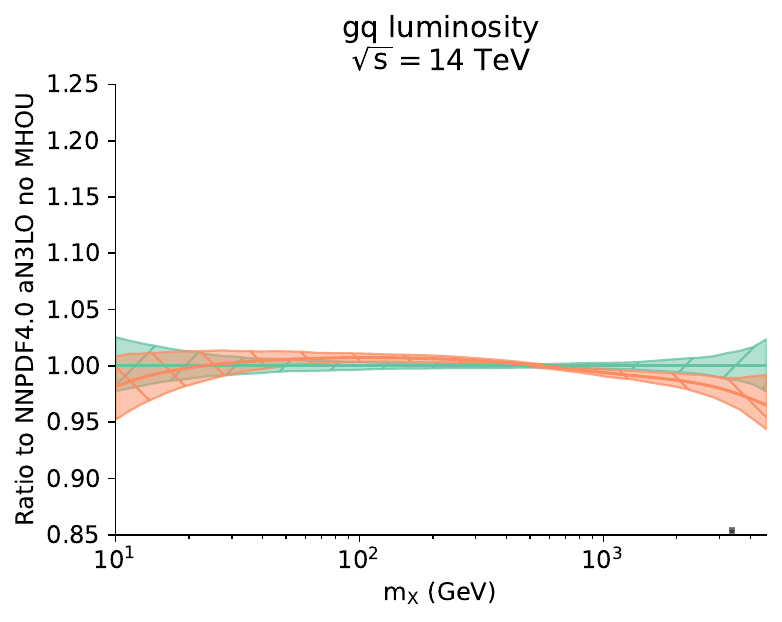}
  \includegraphics[width=0.42\textwidth]{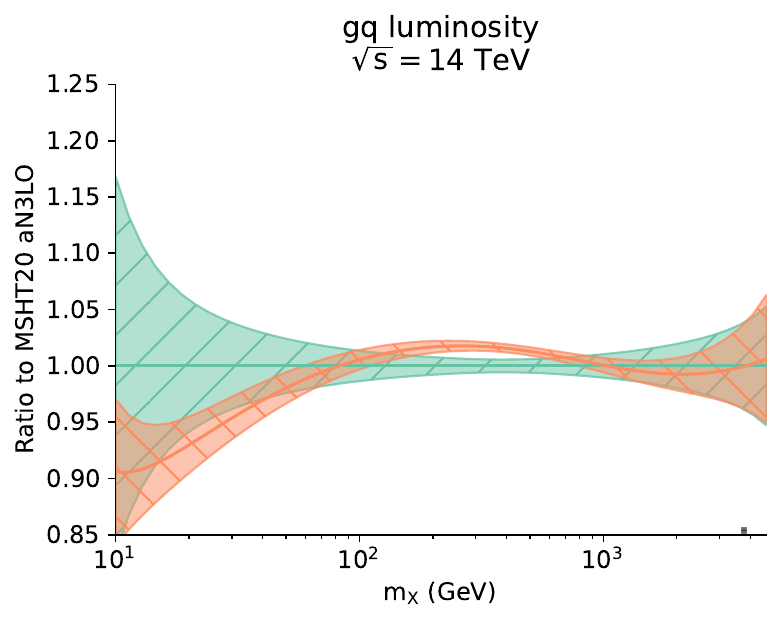}
  \includegraphics[width=0.42\textwidth]{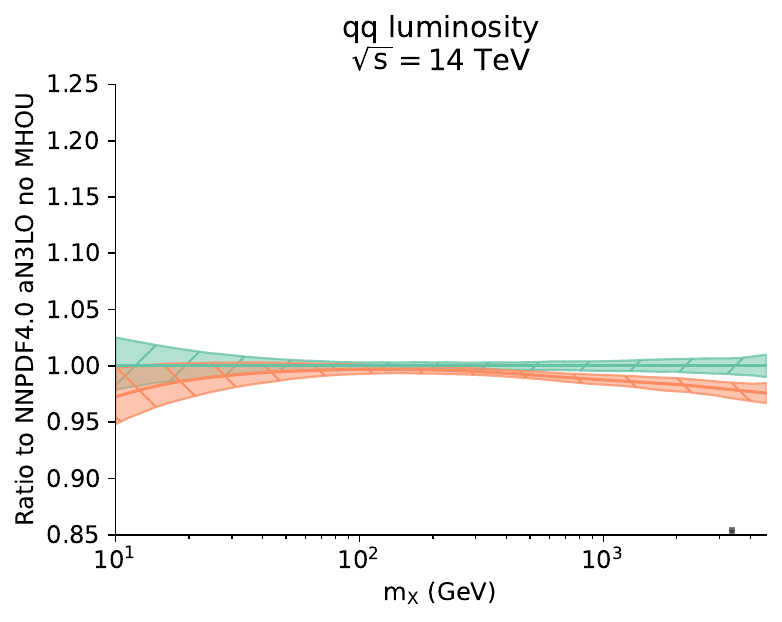}
  \includegraphics[width=0.42\textwidth]{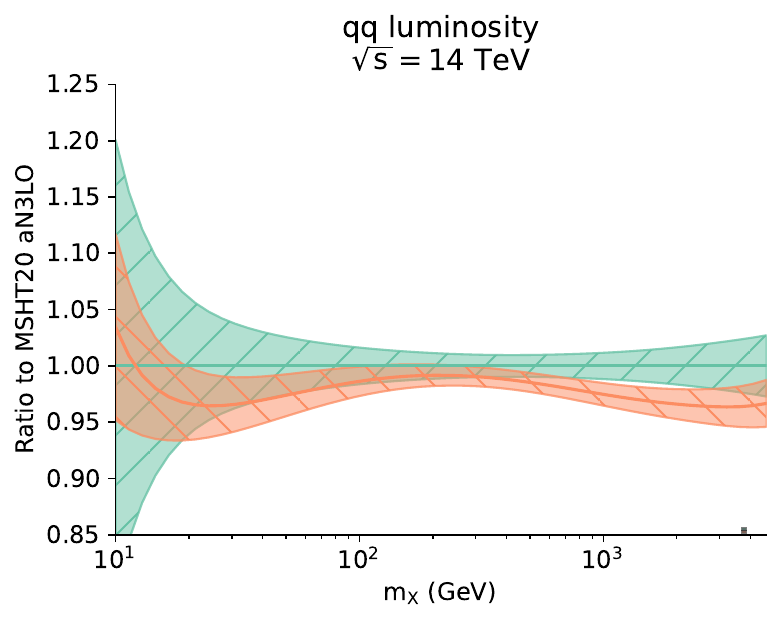}
  \includegraphics[width=0.42\textwidth]{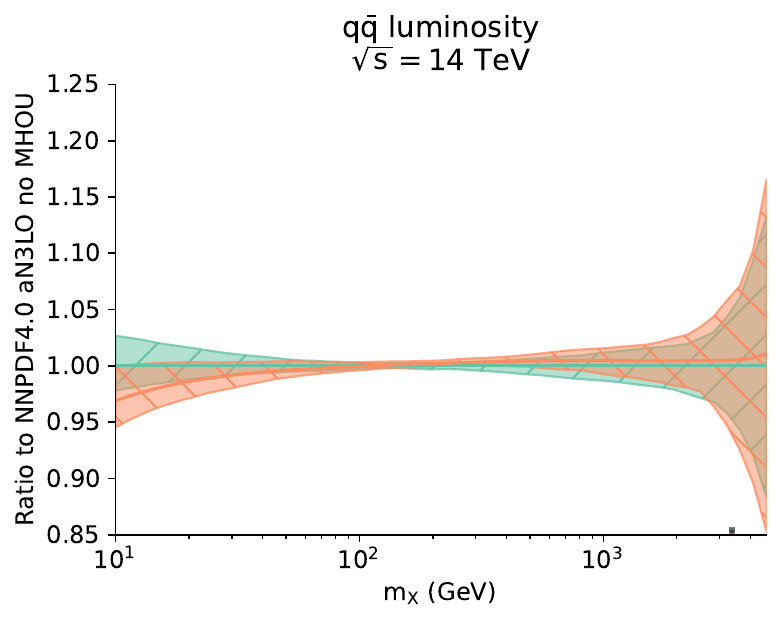}
  \includegraphics[width=0.42\textwidth]{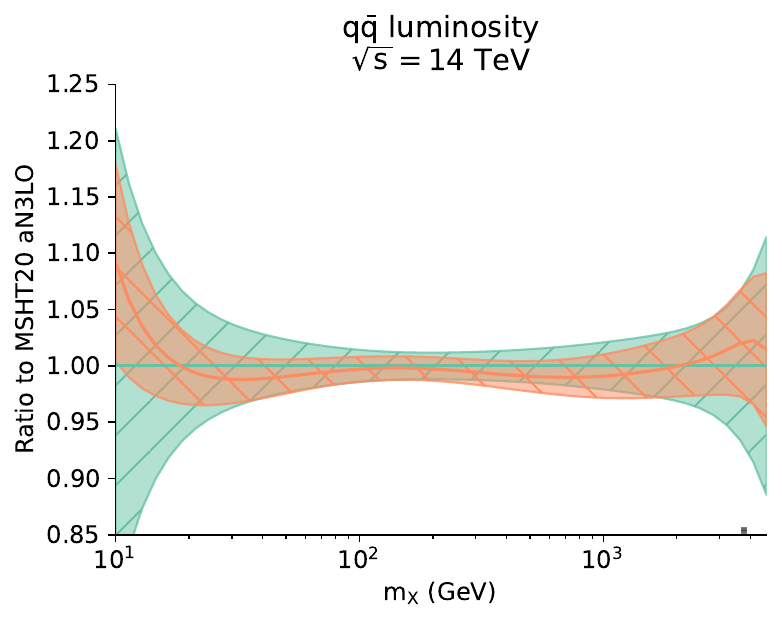}
  \caption{Same as \cref{fig:lumis_MSHT20}, now comparing aN$^3$LO and NNLO
    parton luminosities, separately for the NNPDF4.0 (left) and MSHT20
    (right) PDF sets, normalized to the aN$^3$LO result.}
  \label{fig:lumis_NLO_vs_N3LO}
\end{figure}

Parton luminosities are compared in \cref{fig:lumis_MSHT20}.
Again the pattern is similar to that seen
at NNLO, but now with a considerable suppression of the gluon-gluon and
gluon-quark luminosities in the $M_X\sim100$~GeV region that can be
traced to the behavior of the gluon PDF seen in \cref{fig:PDFs_MSHT20}.
The quark-quark luminosity remains similar in MSHT20 and NNPDF4.0 both at NNLO
and aN$^3$LO. The impact of these effects on the computation of
precision LHC cross-sections is addressed in \cref{sec:an3lo_pheno}.

In order to understand better the comparative impact of aN$^3$LO corrections,
we compare for each set the NNLO and aN$^3$LO luminosities. Results are shown
in \cref{fig:lumis_NLO_vs_N3LO}, normalized to the aN$^3$LO result. 
The qualitative impact of the aN$^3$LO corrections on
either set is similar, but with a stronger aN$^3$LO suppression of gluon
luminosities for MSHT20. In particular the gluon-gluon luminosity is
suppressed for $10^2\lesssim m_X\lesssim 10^3$~GeV  by about $3~\%$ in NNPDF4.0
and $6~\%$ in MSHT20 and the gluon-quark luminosity is suppressed in the same
region by about $1~\%$ in NNPDF4.0 and $3~\%$ in MSHT20. 
In the case of the gluon-gluon luminosity the differences between
NNLO and aN$^3$LO  are larger than the respective PDF uncertainties
(that do not include MHOUs in either case). As already mentioned in
\cref{sec:an3lo_comp}, a dedicated benchmark of aN$^3$LO results
is presented in Ref.~\cite{Cooper-Sarkar:2024crx}.

\end{appendices}

\begin{backmatter}

  \renewcommand{\bibfont}{\scriptsize}
  \printbibliography[heading=bibintoc]

  \begin{summary}
\phantomsection
\addcontentsline{toc}{chapter}{Summary}
\vspace{-0.25cm}

This thesis investigates some fundamental questions about the proton, 
the most abundant object we experience in our world.
Protons are indeed complex systems made of tightly bound elementary particles:
quarks and gluons. Their interaction is among the strongest forces in nature,
making the proton a stable object of which we are all composed.
Understanding how these smaller components interact and distribute 
their energy within the proton is crucial for advancing 
our knowledge of particle physics.

In particular, the proton's substructure becomes relevant during 
high-energy collisions, such as those at the Large Hadron Collider (LHC). 
This inner structure is described by Parton Distribution Functions (PDFs),
which are the main tool adopted in our research. 
PDFs describe how the proton's energy is probabilistically distributed among 
quarks and gluons.

PDFs are thus essential for interpreting data from LHC experiments 
and making predictions about the behavior of particles under extreme conditions.
Like any physical quantity, PDFs are subject to uncertainties, which can 
affect our ability to identify new particles or verify existing theories, such 
as the Standard Model of particle physics.

The main goal of this thesis is thus to improve the current estimate of PDFs 
by adopting state-of-the-art of theoretical calculations combined with
machine learning techniques and a vast amount of high energy data collected 
from various experiments.

Our key findings are presented in \cref{chap:ic,chap:an3lo,chap:pol} and include:

\begin{description}
    \item[Approximate N$^3$LO QCD PDFs:]
    we incorporate the current known next-to-next-to-next-to-leading-order (N$^3$LO) 
    QCD corrections relevant for PDF evolution and Deep Inelastic Scattering,
    within the NNPDF framework. This allows us to refine our knowledge of PDFs 
    accounting for suppressed effects and theoretical errors, which are usually
    neglected but can be relevant in current and forthcoming LHC analyses. 
    \item[Intrinsic Charm:]
    we provide a first evidence of non-vanishing intrinsic charm contribution
    to the proton structure. This is a phenomenon where virtual charm quarks are 
    present also in low energy protons and are not only generated by high energy 
    splitting. This evidence is supported by a comparison with the most up-to-date 
    LHCb data and could be further probed in the upcoming Electron-Ion Collider.
    \item[Polarized PDFs and proton spin:]
    we revise the determination of spin-dependent PDFs by including next-to-next-to-leading-order 
    (NNLO) QCD corrections together with a large number of data from proton-proton 
    scattering not included in previous analyses. 
    Our study further suggests that quarks carry only a fraction of the total proton spin 
    but remains uncertain regarding the gluon contribution.
    In this respect, our study will be beneficial for upcoming experiments from the EIC 
    which are predicted to definitely shed light on the proton spin puzzle. 
\end{description}

By addressing both the theoretical challenges and practical applications of PDFs, 
this research strengthens our knowledge of the proton behavior during high energy
collisions. 
Hopefully, our contribution could pave the way for a more precise interpretation of 
complex experimental measurements, ultimately giving us a better understanding 
of the interaction between elementary particles.

\end{summary}
  \begin{acknowledgements}
\phantomsection
\addcontentsline{toc}{chapter}{Acknowledgments}
\vspace{-0.25cm}

Paris, February 2025.

Now that two months have passed since I left Nikhef and Amsterdam, 
I am starting to better realize how lucky I was to carry out my PhD there.
For sure the place matters, and indeed living in Amsterdam is great, 
but most important are the people you meet. Hoping not to forget 
anyone ...

First, I would like to thank Juan, my supervisor, for having accompanied 
me during the PhD. From the very beginning, you have always been positive 
and encouraging, keeping our motivation high for the work.
Thanks to your help and connections, I was able to meet and collaborate with 
many physicists from all over the world.
Thank you for your trust and sponsorship. 
I am grateful to Mara and Wouter for having supported and monitored my
trajectory at Nikhef. I always appreciate your enthusiasm and passion in 
keeping track of my progress. 
My gratitude also goes to the reading committee—Kjeld, Melissa, Daniel, 
and Tristan—for providing me feedback on this thesis.

Most of the projects I worked on during the PhD were carried out in 
collaboration, which made the work much easier and more enjoyable.
In particular, I want to thank all the members of the NNPDF collaboration: 
Richard, Alessandro, Stefano, Amedeo, Mark, Juan, Luigi, Stefano, Tommaso, 
Felix, Peter, Emanuele, Tanjona, Juan, Tanishq, Roy, Maria, Christopher, Niccolò, 
Andrea, Rabah and Zahari.
Being part of your team has been a pleasure and an honor.
Stefano, Richard, and Maria, you also have been my mentors. I am grateful 
for your patience in correcting my mistakes and reviewing all the details 
of our work. Thank you also for your guidance and lessons in the PDF world. 
As special thanks go to Emanuele (the first person I met at Nikhef) for your 
kindness and precision in guiding me through all the steps and moments where 
something was not clear. Chatting with you is always a pleasure. 
I am immensely grateful to Felix, Tanjona and Tommaso for assisting me in my 
day-to-day work and being good friends. Without your knowledge and help all 
the work would have been impossible. 
In this respect, I would like to thank also Alessandro, Roy and Juan for your 
help and teaching me all the insights of best coding and working practices. 
Andrea and Niccolò, you have been my "PhD mates" and I will always cherish 
our time spent together struggling to understand what we were doing 
debugging code and sharing countless laughs.
Thanks to all NNPDF, juniors and seniors, for the nice time we spent 
during many conferences and workshops.

I am happy to thank also the SMEFiT collaboration: Eleni, Fabio, Juan, 
Alejo, Eugenia, Jaco, Marion and Tommaso. Working with you during the early 
years of my PhD provided me with a broader perspective on the HEP field, 
which was truly valuable.

Next, I would like to thank all the members of the Nikhef theory group: Robin, 
Pieter, Jelle, Max, Jaco, Coenraad, Heleen, Peter, Vaisakh, Andres, 
Tommaso, Tanishq, Maximilian, Sachin, Ankita, Tommaso, Juraj, Johannes, 
Robert, Lucaš, Melissa, Marieke, Eric, Jordy, and Wouter.
Thanks for being always welcoming with me and create an enjoyable and friendly 
environment every day. 
To all the wonderful people I met at Nikhef, with whom I shared evenings and fun moments
—Andrea P., Avanish, Alice, Andrea A., Andrea V., Alex, Antonella, Carlo, 
Enrico, Evridiki, Luca, Lydia, Masha, Roberto, and Vlad—thank you for making 
my time there so enjoyable.

A special thanks go to my flatmates, Tommaso and Masha. Living with you at 
Zeeburgerkade 702 has been amazing. I will always remember all the 
dinners together-the simple ones and the fancy ones-or enjoying the astonishing 
sunset view from our balcony. Thanks for being flexible, supporting and positive. 
From you, I have learned a lot. If coming back home has always been fun, it
is because of you and I am grateful to have shared these years together.
From Amsterdam, I will eagerly remember the time spent playing 
tennis at the ATVD club in Diemen. In particular, I am grateful to Tito, Kenneth, 
Etienne, Jesper and the tireless squad of Federico and Andrea, you are the best!

Living in the Netherlands for the past four years has been an eye-opening experience,
and despite the challenging beginning, each year became better and more rewarding.

Regardless of spending my time in Amsterdam, I cannot forget to thank all the 
friends from Milan. 
Among them, my lifelong group of friends from "Precotto and surroundings" 
(sorry for not listing your names here). I met some of you when we were kids, 
and the fact that we still see each other and share experiences always impresses me.
You always make me feel home and younger when back to Milan, thanks. 
Next, are my physicist friends from university: Alessandro, Andrea, Federica, 
Laura and Tiziano. Sharing the studies and the experience of our PhD around 
Europe has been really helpful and fruitful.
To Laura, thanks for all the nice bike rides and trips around the Netherlands.

Finally, I am grateful to my paranymphs Jaco and Tiziano. You have been loyal
companions during the PhD years, thank you for your support, 
organization and help. 

Last but not least, to my family—from my parents to my cousins—I feel incredibly 
fortunate to have grown up with you. You will always be an example and an inspiration 
in my life. Mom Paola, dad Riccardo and Filippo thanks for all the trust 
and the patience you have from the very beginning.

And to Anna—thank you for walking with me side by side.

\bigskip

\begin{flushright}
With gratitude, \\
Giacomo
\end{flushright}

\end{acknowledgements}



\end{backmatter}

\end{document}